\documentclass[oneside,english]{book}
\usepackage[T1]{fontenc}
\usepackage[latin9]{inputenc}
\usepackage[active]{srcltx}
\usepackage{longtable}
\usepackage{mathrsfs}
\usepackage{bm}
\usepackage{amsmath}
\usepackage{amssymb}
\usepackage{undertilde}
\usepackage{graphicx}
\usepackage{esint}

\usepackage{multind}
\makeindex{SI}

\makeatletter

\newcommand{\noun}[1]{\textsc{#1}}
\providecommand{\tabularnewline}{\\}

\usepackage{amsmath}
\global\newcommand{\dbtilde}[1]{\tilde{\raisebox{-0.1pt}[0.85\height]{$\tilde{#1}$}}}
\newcommand{\dbar}{\mathrm{d}\hspace*{-0.16em}\bar{}\hspace*{0.1em}}

\usepackage{babel}

\makeatother

\usepackage{babel}
\begin{document}

\title{\noun{\huge{}Foundations of}\vspace{2mm}\\
\noun{\huge{} }\textbf{\noun{\huge{}\Huge{Convection}}}{\huge{}}\vspace{2mm}\\
\noun{\huge{} with density stratification}\\
{\huge{} }\textbf{\huge{}~}{\huge{}}\\
{\huge{} }\textbf{\huge{}~}{\huge{}}\\
{\huge{} }}

\author{{\LARGE{}Krzysztof Mizerski}}

\date{}

\maketitle
\newpage{}

~

~

~

~

~

~

~

~

~

~

~

~

~

~

~

~

~

~

~

~

~

~

~

~

~

~

~

~

~

~

~

~

~

~

~

~

~
\noindent \begin{flushright}
\emph{\Large{}To my wife, }\\
\emph{\Large{}my daughter,}\\
\emph{\Large{}and my son}
\par\end{flushright}{\Large \par}

\newpage{}

\tableofcontents{}

\chapter*{Preface}

The phenomenon of thermal and compositional (chemical) convection
is very common in nature and therefore of great importance from the
point of view of understanding of many fundamental aspects of the
environment and universe. It has attracted a great deal of attention
over the last half a century and a significant progress in a detailed
quantitative description of this phenomenon has been made, in various
physical contexts. A number of books have been written on the topic,
such as e.g. the seminal work of Chandrasekhar (1961) on \emph{Hydrodynamic
and Hydromagnetic Stability}, a large portion of which is devoted
to the convective instability near its onset or the outstanding book
of Getling (1998) where systematization of the knowledge on convection
has been continued with a thorough description of the weakly nonlinear
stages. Most of the works, however, considered weakly stratified,
that is the so-called Boussinesq convection.

It is the aim of this book to continue the process of systematization.
It seems important to put the current knowledge on weakly and strongly
stratified convection in order and provide a comprehensive description
of the marginal, weakly nonlinear and fully developed stages of convective
flow in both cases. To that end the book provides a short compendium
of knowledge on the linear and weakly nonlinear limits of the Boussinesq
convection, as a useful reference for a reader and than proceeds with
a review of the theory on fully developed, weakly stratified convection.
The entire third chapter is devoted to a detailed derivation and a
study of the three aforementioned stages of stratified (\emph{anelastic})
convection. The description of stratified convection requires extreme
care, since many aspects have to be considered simultaneously for
full consistency. Detailed and systematic explanations are therefore
provided. It is not the aim to deliver a comprehensive review of findings
on convection, but rather to pinpoint precisely to the relevant works
(even pages and equations) on particular aspects of the dynamics of
convective flows. This book is meant as a textbook for courses on
hydrodynamics and convective flows, for the use of lecturers and students,
however, it may also be of use for the entire scientific community
as a practical reference.

\chapter{The equations of hydrodynamics\label{chap:1}}

The derivation of the fundamental equations for Newtonian fluids has
been provided in many books, e.g. in Chandrasekhar (1961) and many
others. Therefore here we only briefly recall the main points of the
derivation, with the aim to keep the book self-consistent and set
grounds for later chapters.
A reader interested in the historical origins of hydrodynamics and the long process of gradual increase in the rigorousness of the description of dynamical flows since Daniel Bernoulli's \emph{Hydrodynamica} published in 1738 is directed to Darrigol (2005).

\section{General conservation law in a continuous medium \label{sec:General-conservation-law}}

Let us take an extensive quantity $A(t)$ and introduce its density
per unit volume, $a(\mathbf{x},t)$, so that
\begin{equation}
A\left(t\right)=\int_{V}a\left(\mathbf{x},t\right)\mathrm{d}^{3}x.\label{eq:A}
\end{equation}
The variation of $A(t)$ in a volume $V$ can be attributed only to
two phenomena, that is either to sources (or sinks, which will be
thought of as negative sources) of this quantity within the volume,
denoted by $\sigma_{A}(\mathbf{x},t)$ or the flux $\mathbf{j}_{A}(\mathbf{x},t)$
of the quantity $A(t)$, which at least partly, is due to the flow
of the medium. Therefore the total variation of $A(t)$ in a fixed
volume $V$ can be expressed as follows
\begin{equation}
\frac{\mathrm{d}A}{\mathrm{d}t}=\int_{V}\frac{\partial a\left(\mathbf{x},t\right)}{\partial t}\mathrm{d}^{3}x=\int_{V}\sigma_{A}\left(\mathbf{x},t\right)\mathrm{d}^{3}x-\oint_{\partial V}\mathbf{j}_{A}(\mathbf{x},t)\cdot\hat{\mathbf{n}}\mathrm{d}\Sigma,\label{eq:dAbydt}
\end{equation}
where $\hat{\mathbf{n}}$ is the unit normal directed outside of the
surface $\partial V$ enclosing the volume $V$. Therefore a quantity
is conserved in a volume $V$ if there are no sources of this quantity
within the volume and the same amount of flux enters and leaves the
volume. Since $V$ is arbitrary, by the use of the Gauss-Ostrogradsky
divergence theorem,
\begin{equation}
\oint_{\partial V}\mathbf{j}_{A}(\mathbf{x},t)\cdot\hat{\mathbf{n}}\mathrm{d}\Sigma=\int_{V}\nabla\cdot\mathbf{j}_{A}(\mathbf{x},t)\mathrm{d}^{3}\mathbf{x},\label{eq:e1}
\end{equation}
we obtain locally
\begin{equation}
\frac{\partial a\left(\mathbf{x},t\right)}{\partial t}+\nabla\cdot\mathbf{j}_{A}(\mathbf{x},t)=\sigma_{A}\left(\mathbf{x},t\right),\label{eq:gen_evolution_law_local}
\end{equation}
which is a general local evolution law\index{SI}{evolution law} of a certain extensive quantity
$A(t)$ in a fluid.

\section{The continuity equation - mass conservation law\label{sec:The-continuity-equation}}\index{SI}{continuity equation}\index{SI}{mass conservation}

At this point, to derive the law of mass conservation it is enough
to substitute the mass $m(t)$ for $A(t)$ and thus the mass density
$\rho(\mathbf{x},t)$ for $a\left(\mathbf{x},t\right)$ from the previous
section, which yields
\begin{equation}
\frac{\partial\rho\left(\mathbf{x},t\right)}{\partial t}+\nabla\cdot\mathbf{j}_{m}(\mathbf{x},t)=\sigma_{m}\left(\mathbf{x},t\right),\label{eq:e2}
\end{equation}
and since the flux of mass\index{SI}{flux!of mass} in a fluid results solely from the flow
$\mathbf{u}(\mathbf{x},t)$, it takes the form $\mathbf{j}_{m}(\mathbf{x},t)=\rho\left(\mathbf{x},t\right)\mathbf{u}\left(\mathbf{x},t\right)$.
Hence finally the continuity equation reads
\begin{equation}
\frac{\partial\rho\left(\mathbf{x},t\right)}{\partial t}+\nabla\cdot\left[\rho\left(\mathbf{x},t\right)\mathbf{u}\left(\mathbf{x},t\right)\right]=\sigma_{m}\left(\mathbf{x},t\right),\label{eq:mass_evolution_gen}
\end{equation}
which expresses local mass conservation if the sources of matter $\sigma_{m}\left(\mathbf{x},t\right)$
vanish at least locally.

\section{The Navier-Stokes equation - momentum balance\label{sec:The-Navier-Stokes-equation}}\index{SI}{momentum balance}

The general evolution law (\ref{eq:gen_evolution_law_local}) applied
to momentum per unit volume, $\rho\mathbf{u}$, which simply expresses
the Newton's second law of dynamics, takes the form
\begin{equation}
\frac{\partial\rho\mathbf{u}}{\partial t}+\nabla\cdot\left[\rho\mathbf{u}\mathbf{u}+\boldsymbol{\Pi}\right]=\rho\mathbf{F},\label{eq:fundamental_momentum_balance}
\end{equation}
where , $\rho\mathbf{u}\mathbf{u}+\boldsymbol{\Pi}$ is the flux of
momentum\index{SI}{flux!of momentum} (tensorial, since the momentum is vectorial) with $\rho\mathbf{u}\mathbf{u}$
being the advective flux and $\boldsymbol{\Pi}$ the pressure tensor
(pressure and frictional flux); $\mathbf{F}$ is the body force density
per unit mass and the notation with explicit dependence on $\left(\mathbf{x},t\right)$
is dropped form now on for clarity. The pressure tensor $\boldsymbol{\Pi}$
is equal to the negative stress tensor $\boldsymbol{\tau}$, whose
components describe forces per unit surface that form between fluid
elements in a flow. A force per unit surface exerted on a
given fluid element, say element $(1)$, by its neighbour, say element
$(2)$, is given by $\boldsymbol{\tau}\cdot\hat{\mathbf{n}}$, where
$\hat{\mathbf{n}}$ is the unit normal to the surface separating the
two fluid elements, directed from $(1)$ to $(2)$. For standard,
isotropic Newtonian fluids the stress tensor is expressed in terms
of fluid pressure $p$ and velocity gradients in the following way
(cf. e.g. Batchelor  1967, Chandrasekhar 1961)\footnote{The fundamental assumption for Newtonian fluids is that the dissipative
part of the stress tensor, say $\boldsymbol{\tau}_{\mu}$, associated
with frictional effects due to the flow, is linearly related to the
tensor of deformation rate, that is the flow velocity gradient tensor,
since it is the presence of velocity gradients which is necessary
and sufficient for frictional forces to appear. Once this assumption
is made, the final form of the constitutive relation (\ref{eq:Newtonian_stress_tensor})
for an isotropic fluid is simply an outcome of symmetry of the stress
tensor, which follows from the angular momentum balance (cf. the next
section \ref{subsec:The-angular-momentum}) and very basic properties
of tensorial objects known from the group theory. The latter is simply
the Curie's principle 1894 (cf. Chalmers 1970, de Groot and Mazur
1984), namely that in an isotropic system only those tensorial objects,
which at rotations of a system of reference transform according to
the same irreducible (therefore distinct) representations of the rotation
groups can be linearly related. Since the trace and the symmetric
traceless part of a tensor transform differently, one arrives at the
constitutive relations $\mathrm{Tr}\boldsymbol{\tau}_{\mu}=3\mu_{b}\mathrm{Tr}\mathbf{G}=3\mu_{b}\nabla\cdot\mathbf{u}$
and $\boldsymbol{\tau}_{\mu}^{s}-(1/3)(\mathrm{Tr}\boldsymbol{\tau}_{\mu})\mathbf{I}=2\mu(\mathbf{G}^{s}-(1/3)(\nabla\cdot\mathbf{u})\mathbf{I})$,
which are equivalent to (\ref{eq:Newtonian_stress_tensor}). In the
given relations the superfix $s$ denotes a symmetric part of a tensor,
$\mathbf{G}$ is the velocity gradient, as in (\ref{eq:velocity_gradient_def}),
$\mathbf{I}$ is the unitary matrix and the coefficients $\mu>0$
and $\mu_{b}>0$ since they describe frictional, therefore dissipative effects, see section 1.4.1 on the entropy production. },
\begin{equation}
\tau_{ij}=-\Pi_{ij}=-p\delta_{ij}+2\mu\left(G_{ij}^{s}-\frac{1}{3}\nabla\cdot\mathbf{u}\delta_{ij}\right)+\mu_{b}\nabla\cdot\mathbf{u}\delta_{ij},\label{eq:Newtonian_stress_tensor}
\end{equation}
\index{SI}{stress tensor}where $\mu$ is the coefficient of dynamic shear viscosity, $\mu_{b}$
is the bulk viscosity associated with expansion (compression) processes,
the subscripts $ij$ denote the cartesian components of tensors, $\delta_{ij}$
denotes a unitary tensor. Moreover,
\begin{equation}
G_{ij}=\frac{\partial u_{i}}{\partial x_{j}},\qquad G_{ij}^{s}=\frac{1}{2}\left(\frac{\partial u_{i}}{\partial x_{j}}+\frac{\partial u_{j}}{\partial x_{i}}\right),\label{eq:velocity_gradient_def}
\end{equation}
is the velocity gradient tensor\index{SI}{velocity gradient tensor} and its symmetric part; note, that
often kinematic coefficients of shear and bulk viscosities are used,
defined as $\nu=\mu/\rho$ and $\nu_{b}=\mu_{b}/\rho$. The dissipative
part of the stress tensor (\ref{eq:Newtonian_stress_tensor}), that
is $2\mu\left(G_{ij}^{s}-\frac{1}{3}\nabla\cdot\mathbf{u}\delta_{ij}\right)+\mu_{b}\nabla\cdot\mathbf{u}\delta_{ij}$,
describes the viscous stresses in the fluid. Under the assumption,
that there are no mass sources in the entire fluid volume, $\sigma_{m}\left(\mathbf{x},t\right)=0$,
the fundamental momentum balance (\ref{eq:fundamental_momentum_balance})
takes the form
\begin{equation}
\rho\left[\frac{\partial\mathbf{u}}{\partial t}+\left(\mathbf{u}\cdot\nabla\right)\mathbf{u}\right]=\nabla\cdot\boldsymbol{\tau}+\rho\mathbf{F}.\label{eq:gen_momentum_balance}
\end{equation}
After introduction of the stress tensor form for a Newtonian fluid
(\ref{eq:Newtonian_stress_tensor}), we arrive at the well-kown Navier-Stokes
equation\index{SI}{Navier-Stokes equation} with non-uniform viscosity 
\begin{align}
\rho\left[\frac{\partial\mathbf{u}}{\partial t}+\left(\mathbf{u}\cdot\nabla\right)\mathbf{u}\right]= & -\nabla p+\rho\mathbf{F}+\mu\nabla^{2}\mathbf{u}+\left(\frac{\mu}{3}+\mu_{b}\right)\nabla\left(\nabla\cdot\mathbf{u}\right)\nonumber \\
 & \qquad\qquad+2\nabla\mu\cdot\mathbf{G}^{s}+\nabla\left(\mu_{b}-\frac{2}{3}\mu\right)\nabla\cdot\mathbf{u}.\label{NS-Bderiv-2}
\end{align}

\subsection{The angular momentum balance\label{subsec:The-angular-momentum}}\index{SI}{angular momentum balance}

It is straightforward to verify, that the stress tensor $\boldsymbol{\tau}$
for Newtonian fluids given in (\ref{eq:Newtonian_stress_tensor})
is symmetric. As mentioned earlier, its symmetry results from the
balance of angular momentum in a fluid. From the general evolution
law for a continuous medium (\ref{eq:gen_evolution_law_local}) we
obtain for the angular momentum
\begin{equation}
\frac{\partial}{\partial t}\left(\mathbf{x}\times\rho\mathbf{u}\right)+\nabla\cdot\mathbf{j}_{L}=\rho\mathbf{x}\times\mathbf{F},\label{eq:fundamental_angular_momentum_balance}
\end{equation}
where the flux of angular momentum $\mathbf{j}_{L}$ satisfies for
any volume $V$ within the fluid
\begin{equation}
\int_{V}\mathbf{j}_{L}\mathrm{d^{3}\mathbf{x}}=\int_{V}\nabla\cdot\left[\rho\left(\mathbf{x}\times\mathbf{u}\right)\mathbf{u}\right]\mathrm{d^{3}\mathbf{x}}+\oint_{\partial V}\mathbf{x}\times\left(\boldsymbol{\tau}\cdot\hat{\mathbf{n}}\right)\mathrm{d}\Sigma.\label{eq:e3}
\end{equation}
In the above the first term on the right hand side is the advective flux
and the second is the angular momentum generated by pressure and friction.
We can apply the divergence theorem to the latter, which yields
\begin{equation}
\oint_{\partial V}\mathbf{x}\times\left(\boldsymbol{\tau}\cdot\hat{\mathbf{n}}\right)\mathrm{d}\Sigma=\int_{V}\mathbf{x}\times\nabla\cdot\boldsymbol{\tau}\mathrm{d^{3}\mathbf{x}}+\int_{V}\mathbf{a}_{\tau}\mathrm{d^{3}\mathbf{x},}\label{eq:e4}
\end{equation}
where $(\mathrm{a}_{\tau})_{i}=\epsilon_{ijk}\tau_{jk}^{a}$, that
is $\mathbf{a}_{\tau}=2[\tau_{23}^{a},\,-\tau_{13}^{a},\,\tau_{12}^{a}]$
and the superscript $a$ denotes the antisymmetric part of a tensor.
This means that the angular momentum evolution law (\ref{eq:fundamental_angular_momentum_balance})
can be written as follows
\begin{equation}
\frac{\partial}{\partial t}\left(\mathbf{x}\times\rho\mathbf{u}\right)+\nabla\cdot\left[\rho\left(\mathbf{x}\times\mathbf{u}\right)\mathbf{u}\right]=\mathbf{x}\times\nabla\cdot\boldsymbol{\tau}+\rho\mathbf{x}\times\mathbf{F}+\mathbf{a}_{\tau}.\label{eq:fundamental_angular_momentum_balance-1}
\end{equation}
The left hand side of the above equation (\ref{eq:fundamental_angular_momentum_balance-1})
can be manipulated to give
\begin{equation}
\frac{\partial}{\partial t}\left(\mathbf{x}\times\rho\mathbf{u}\right)+\nabla\cdot\left[\rho\left(\mathbf{x}\times\mathbf{u}\right)\mathbf{u}\right]=\mathbf{x}\times\left[\frac{\partial\rho\mathbf{u}}{\partial t}+\nabla\cdot\left(\rho\mathbf{u}\mathbf{u}\right)\right],\label{eq:e5}
\end{equation}
so that the angular momentum balance can be expressed in the following
way
\begin{equation}
\mathbf{x}\times\left[\frac{\partial\rho\mathbf{u}}{\partial t}+\nabla\cdot\left(\rho\mathbf{u}\mathbf{u}-\boldsymbol{\tau}\right)-\rho\mathbf{F}\right]=\mathbf{a}_{\tau}.\label{eq:e6}
\end{equation}
On the other hand the general momentum balance (\ref{eq:fundamental_momentum_balance})
must be satisfied, which yields
\begin{equation}
\mathbf{a}_{\tau}=0\;\Rightarrow\;\boldsymbol{\tau}^{a}=0,\label{eq:e7}
\end{equation}
and hence the antisymmetric part of the stress tensor must vanish or, in
other words, the stress tensor is necessarily symmetric. Simply for the sake of
completeness we may provide a final, general expression for the angular
momentum flux\index{SI}{flux!of angular momentum} $(j_{L})_{im}=\epsilon_{ijk}x_{j}(\rho u_{k}u_{m}+\tau_{km})$.

\section{The energy equation\label{sec:The-energy-equation}}

The total energy density per unit mass\index{SI}{energy per unit mass} in a fluid volume $V$ is
\begin{equation}
e=\frac{1}{2}\mathbf{u}^{2}+\psi+\varepsilon,\label{eq:e8}
\end{equation}
where $\mathbf{u}^{2}/2$ is the kinetic energy density, $\psi$ the
potential energy resulting from presence of conservative body forces,
$\mathbf{F}=-\nabla\psi$, which we will assume stationary, and $\varepsilon$
denotes the internal energy of the fluid. The general local evolution
law (\ref{eq:gen_evolution_law_local}) for the total energy reads
\begin{equation}
\frac{\partial\rho e}{\partial t}+\nabla\cdot\left(\rho e\mathbf{u}+\mathbf{j}_{\mathrm{mol}}\right)=Q,\label{eq:energy_evolution_gen}
\end{equation}
where $\rho e\mathbf{u}$ is the flux due to energy advection by the
flow, $\mathbf{j}_{\mathrm{mol}}$ is the flux of energy from molecular
mechanical and thermal effects and the energy sources, here denoted
by $Q\,(=\sigma_{E})$, describe heating processes (absorbed heat
per unit volume per unit time) such as e.g. the radioactive heating,
thermal radiation, etc. To establish the formula for the molecular
energy flux $\mathbf{j}_{\mathrm{mol}}$ we must realize the effects
that lead to variation of the total energy. The total change of the
energy $\rho e$ in the volume $V$, in the absence of energy sources
$Q$, results solely from two factors, that is the heat transfer between
the volume and the rest of the fluid and the total work done on the
volume by the stresses described by the stress tensor $\boldsymbol{\tau}$,
\begin{equation}
\oint_{\partial V}\mathbf{u}\cdot\boldsymbol{\tau}\cdot\hat{\mathbf{n}}\mathrm{d}\Sigma=\int\nabla\cdot\left(\boldsymbol{\tau}\cdot\mathbf{u}\right)\mathrm{d}V.\label{eq:e9}
\end{equation}
The body forces are assumed conservative hence their work
\begin{equation}
\int\rho\mathbf{u}\cdot\mathbf{F}\mathrm{d}V=-\int\rho\mathbf{u}\cdot\nabla\psi,\label{eq:e10}
\end{equation}
simply expresses the total change of the potential energy, which is
due to advection only, since the body forces are also assumed stationary.
Therefore the molecular flux $\mathbf{j}_{\mathrm{mol}}$ can be decomposed
into two contributions. The first one comes from thermal effects and
is described by the Fourier's law,\index{SI}{Fourier's law}
\begin{equation}
\mathbf{j}_{T}=-k\nabla T,\label{eq:e11}
\end{equation}
stating, that the thermal heat flux is proportional to its cause,
that is the temperature gradient\footnote{which is yet another manifestation of the Curie's principle 1894 (cf.
de Groot and Mazur 1984), stating that fluxes are linear in ``thermodynamic
forces'', which is a term used to describe the physical causative
factors of the fluxes.} and the coefficient of proportionality $k$ is called the thermal
conduction coefficient. The second contribution to the molecular energy
flux is of mechanical nature and for now will be denoted by $\mathbf{j}_{\mathrm{mech}}$.
This implies the following form of the total energy flux
\begin{equation}
\mathbf{j}_{\rho e}=\rho e\mathbf{u}+\mathbf{j}_{\mathrm{mech}}-k\nabla T,\label{eq:e12}
\end{equation}
which allows to rewrite the energy evolution equation (\ref{eq:energy_evolution_gen})
in the form
\begin{equation}
e\left[\frac{\partial\rho}{\partial t}+\nabla\cdot\left(\rho\mathbf{u}\right)\right]+\rho\left(\frac{\partial e}{\partial t}+\mathbf{u}\cdot\nabla e\right)=-\nabla\cdot\mathbf{j}_{\mathrm{mech}}+\nabla\cdot\left(k\nabla T\right)=Q,\label{eq:energy_evolution_gen-1}
\end{equation}
and in the absence of mass sources within the volume the term in the
square brackets on the left hand side of (\ref{eq:fundamental_angular_momentum_balance-1})
vanishes. However, as said, in the absence of energy sources $Q$,
the energy change in a volume $V$ can result only from the heat exchanged
with the surroundings and the total work done by the stresses on the
fluid volume, therefore must be equal to 
\begin{equation}
\nabla\cdot\left(k\nabla T\right)+\nabla\cdot\left(\boldsymbol{\tau}\cdot\mathbf{u}\right),\label{eq:e13}
\end{equation}
which implies
\begin{equation}
\mathbf{j}_{\mathrm{mech}}=-\boldsymbol{\tau}\cdot\mathbf{u}.\label{eq:e14}
\end{equation}
The total, advective time derivative\index{SI}{advective time derivative} of the kinetic energy 
\begin{equation}
\rho\left(\frac{\partial}{\partial t}+\mathbf{u}\cdot\nabla\right)\frac{1}{2}\mathbf{u}^{2}=\rho\mathbf{u}\cdot\left(\frac{\partial}{\partial t}+\mathbf{u}\cdot\nabla\right)\mathbf{u},\label{eq:e15}
\end{equation}
with the use of the momentum balance (\ref{eq:gen_momentum_balance})
can be rearranged into
\begin{equation}
\rho\mathbf{u}\cdot\left(\frac{\partial}{\partial t}+\mathbf{u}\cdot\nabla\right)\mathbf{u}=\mathbf{u}\cdot\nabla\cdot\boldsymbol{\tau}+\rho\mathbf{u}\cdot\mathbf{F}=\nabla\cdot\left(\boldsymbol{\tau}\cdot\mathbf{u}\right)-\boldsymbol{\tau}:\mathbf{G}+\rho\mathbf{u}\cdot\mathbf{F},\label{eq:e16}
\end{equation}
where $\mathbf{G}$ is the velocity gradient tensor defined in (\ref{eq:velocity_gradient_def})
and double dot denotes contraction over both indices, $\boldsymbol{\tau}:\mathbf{G}=\tau_{ij}G_{ij}$.
Note, that because the stress tensor is symmetric, we can also substitute
only the symmetric part of the velocity gradient tensor into $\boldsymbol{\tau}:\mathbf{G}=\boldsymbol{\tau}:\mathbf{G}^{s}$.
On the other hand, as already remarked, the advection of the potential
energy per unit mass, stationary by assumption, is easily expressed
by
\begin{equation}
\rho\left(\frac{\partial}{\partial t}+\mathbf{u}\cdot\nabla\right)\psi=-\rho\mathbf{u}\cdot\mathbf{F},\label{eq:e17}
\end{equation}
therefore the total energy balance yields
\begin{equation}
\rho\left(\frac{\partial\varepsilon}{\partial t}+\mathbf{u}\cdot\nabla\varepsilon\right)=\nabla\cdot\left(k\nabla T\right)+\boldsymbol{\tau}:\mathbf{G}^{s}+Q.\label{eq:energy_evolution_gen-1-1}
\end{equation}
The above equation (\ref{eq:energy_evolution_gen-1-1}) expresses
the first law of thermodynamics\index{SI}{first law of thermodynamics}, that is the change in the internal
energy of a fluid in a volume $V$ is due to the heat exchanged between
the volume and its surroundings, $\nabla\cdot\left(k\nabla T\right)$,
and the work done by the fluid flow on the volume unbalanced by the
kinetic energy change, $\boldsymbol{\tau}:\mathbf{G}^{s}$. We can
now write down the formula for the total energy flux\index{SI}{flux!of total energy}
\begin{equation}
\mathbf{j}_{\rho e}=\rho e\mathbf{u}-\boldsymbol{\tau}\cdot\mathbf{u}-k\nabla T,\label{eq:e18}
\end{equation}
with $Q$ being the volume sources of the total energy, whereas the
flux of the internal energy\index{SI}{flux!of internal energy} and its volume sources can be defined
as follows
\begin{equation}
\mathbf{j}_{\rho\varepsilon}=\rho\varepsilon\mathbf{u}-k\nabla T,\qquad\sigma_{\rho\varepsilon}=\boldsymbol{\tau}:\mathbf{G}^{s}+Q.\label{eq:e19}
\end{equation}
By the use of the formula for the stress tensor in Newtonian
fluids (\ref{eq:Newtonian_stress_tensor})
\begin{equation}
\boldsymbol{\tau}:\mathbf{G}^{s}=-p\nabla\cdot\mathbf{u}+2\mu\mathbf{G}^{s}:\mathbf{G}^{s}+\left(\mu_{b}-\frac{2}{3}\mu\right)\left(\nabla\cdot\mathbf{u}\right)^{2},\label{eq:e20}
\end{equation}
which includes the viscous heating\index{SI}{viscous heating}, that is $2\mu\mathbf{G}^{s}:\mathbf{G}^{s}+(\mu_{b}-2\mu/3)(\nabla\cdot\mathbf{u})^{2}$.
The equation for internal energy can be transformed either into an
equation for the fluid's entropy $s$ or temperature $T$. The first
and second laws of thermodynamics yield\footnote{It is well known from the second law of thermodynamics, that the differential of the
internal energy for a single component system takes the form\index{SI}{second law of thermodynamics}
\[
\mathrm{d}\mathcal{E}=T\mathrm{d}S-p\mathrm{d}V,
\]
where $\mathcal{E}$, $S$ and $V$ are the ``canonical'' thermodynamic
variables, that is the actual internal energy, the entropy and the
volume of the system (as opposed to the mass densities $\varepsilon$,
$s$ and $\rho$); division by the total mass $M=m_{m}N$, where $m_{m}$
denotes the molecular mass of the fluid particles and $N$ the number
of particles, allows to transform the above into the differential
for the mass density of the internal energy, which takes the form (\ref{eq:int_en_differential}).}
\begin{equation}
\mathrm{d}\varepsilon=T\mathrm{d}s-p\mathrm{d}\left(\frac{1}{\rho}\right),\label{eq:int_en_differential}
\end{equation}
which implies for the total, advective derivatives, that
\begin{equation}
\frac{\partial\varepsilon}{\partial t}+\mathbf{u}\cdot\nabla\varepsilon=T\left(\frac{\partial s}{\partial t}+\mathbf{u}\cdot\nabla s\right)+\frac{p}{\rho^{2}}\left(\frac{\partial\rho}{\partial t}+\mathbf{u}\cdot\nabla\rho\right)=T\left(\frac{\partial s}{\partial t}+\mathbf{u}\cdot\nabla s\right)-\frac{p}{\rho}\nabla\cdot\mathbf{u},\label{eq:e22}
\end{equation}
where the continuity equation (\ref{eq:mass_evolution_gen}) without
mass sources was used. Therefore the entropy equation reads\index{SI}{entropy!equation}
\begin{equation}
\rho T\left(\frac{\partial s}{\partial t}+\mathbf{u}\cdot\nabla s\right)=\nabla\cdot\left(k\nabla T\right)+2\mu\mathbf{G}^{s}:\mathbf{G}^{s}+\left(\mu_{b}-\frac{2}{3}\mu\right)\left(\nabla\cdot\mathbf{u}\right)^{2}+Q.\label{Energy_eq_s}
\end{equation}
Next, we parametrize the fluid's entropy with temperature $T$ and
pressure $p$, that is take $s=s(T,p)$ and write
\begin{equation}
\mathrm{d}s=\left(\frac{\partial s}{\partial T}\right)_{p}\mathrm{d}T+\left(\frac{\partial s}{\partial p}\right)_{T}\mathrm{d}p=\frac{c_{p}}{T}\mathrm{d}T-\frac{\alpha}{\rho}\mathrm{d}p.\label{eq:ds}
\end{equation}
We adopt the standard convention of denoting by $(\partial A/\partial B)_{C}$
a derivative of quantity $A$ with respect to quantity $B$ at constant
quantity $C$. In the above equation (\ref{eq:ds}) $c_{p}=T(\partial s/\partial T)_{p}$
is the heat capacity at constant pressure and 
\begin{equation}
\alpha=-\frac{1}{\rho}\left(\frac{\partial\rho}{\partial T}\right)_{p}\label{eq:alpha}
\end{equation}
 is the thermal expansion coefficient\index{SI}{thermal expansion coefficient}, which by the Maxwell identity
$\alpha=-\rho(\partial s/\partial p)_{T}$. Standard fluids expand
with increasing temperature and therefore we will be considering only
fluids with $\alpha>0$. With the use of (\ref{eq:ds}) applied to
the advective derivatives in (\ref{Energy_eq_s}), we can obtain the
temperature equation in the form\index{SI}{temperature equation}
\begin{align}
\rho c_{p}\left(\frac{\partial T}{\partial t}+\mathbf{u}\cdot\nabla T\right)-\alpha T\left(\frac{\partial p}{\partial t}+\mathbf{u}\cdot\nabla p\right)= & \nabla\cdot\left(k\nabla T\right)+2\mu\mathbf{G}^{s}:\mathbf{G}^{s}\nonumber \\
 & +\left(\mu_{b}-\frac{2}{3}\mu\right)\left(\nabla\cdot\mathbf{u}\right)^{2}+Q.\label{Energy_eq1}
\end{align}
On the other hand one can also parametrize the entropy with temperature
$T$ and density $\rho$, which yields
\begin{equation}
\mathrm{d}s=\left(\frac{\partial s}{\partial T}\right)_{\rho}\mathrm{d}T+\left(\frac{\partial s}{\partial\rho}\right)_{T}\mathrm{d}\rho.\label{eq:ds-1}
\end{equation}
The first term on right hand side of the latter equation can be easily
expressed by the heat capacity at constant volume, which is defined
by $c_{v}=T(\partial s/\partial T)_{\rho}$. To find an expression for the entropy
derivative with respect to density at constant temperature we can
first use the Maxwell identity $\rho^{2}(\partial s/\partial\rho)_{T}=-(\partial p/\partial T)_{\rho}$
and then the implicit function theorem to get $(\partial p/\partial T)_{\rho}=-(\partial\rho/\partial T)_{p}/(\partial\rho/\partial p)_{T}$,
so that
\begin{equation}
\mathrm{d}s=\frac{c_{v}}{T}\mathrm{d}T-\frac{\alpha}{\rho^{2}\beta}\mathrm{d}\rho,\label{eq:ds-1-1}
\end{equation}
and we have defined the isothermal compressibility coefficient\index{SI}{isothermal compressibility coefficient}
\begin{equation}
\beta=\frac{1}{\rho}\left(\frac{\partial\rho}{\partial p}\right)_{T}.\label{eq:beta}
\end{equation}
Finally, by the use of (\ref{eq:ds-1-1}) and the mass conservation
law (\ref{eq:mass_evolution_gen}) without mass sources, $\sigma_{m}=0$,
we obtain yet another form of the temperature equation
\begin{equation}
\rho c_{v}\left(\frac{\partial T}{\partial t}+\mathbf{u}\cdot\nabla T\right)+\frac{\alpha T}{\beta}\nabla\cdot\mathbf{u}=\nabla\cdot\left(k\nabla T\right)+2\mu\mathbf{G}^{s}:\mathbf{G}^{s}+\left(\mu_{b}-\frac{2}{3}\mu\right)\left(\nabla\cdot\mathbf{u}\right)^{2}+Q,\label{Energy_eq2}
\end{equation}
which does not involve any other time derivatives except for time
derivative of the temperature. 

\subsection{Production of the total entropy\label{subsec:Total-entropy-production}}\index{SI}{entropy!production}

According to the second law of thermodynamics, in the absence of the
volume heat sources/sinks, $Q=0$, the total entropy production in
the system must be greater than or equal to zero, when the adiabatic
insulation
\begin{equation}
\left.-k\nabla T\cdot\hat{\mathbf{n}}\right|_{\partial V}=0,\label{eq:adiabatic_insulation-1}
\end{equation}
is assumed. In the above $\hat{\mathbf{n}}$ is the unit normal to
the boundary of the entire fluid region $\partial V$. Introducing
the following notation for the traceless and symmetric part of the
velocity gradient tensor
\begin{equation}
\utilde{\mathbf{G}}^{s}=\mathbf{G}^{s}-\frac{1}{3}\left(\nabla\cdot\mathbf{u}\right)\mathbf{I},\label{eq:e23}
\end{equation}
the term describing the viscous heating in the energy balance (\ref{Energy_eq_s})
can be easily rearranged into the following form
\begin{equation}
2\mu\mathbf{G}^{s}:\mathbf{G}^{s}+\left(\mu_{b}-\frac{2}{3}\mu\right)\left(\nabla\cdot\mathbf{u}\right)^{2}=2\mu\utilde{\mathbf{G}}^{s}:\utilde{\mathbf{G}}^{s}+\mu_{b}\left(\nabla\cdot\mathbf{u}\right)^{2},\label{eq:e24}
\end{equation}
which is clearly positive definite, since a straight forward calculation
yields
\begin{equation}
\utilde{\mathbf{G}}^{s}:\utilde{\mathbf{G}}^{s}=\left(\utilde{G}_{11}^{s}\right)^{2}+\left(\utilde{G}_{22}^{s}\right)^{2}+\left(\utilde{G}_{33}^{s}\right)^{2}+2\left[\left(\utilde{G}_{12}^{s}\right)^{2}+\left(\utilde{G}_{13}^{s}\right)^{2}+\left(\utilde{G}_{23}^{s}\right)^{2}\right].\label{eq:e25}
\end{equation}
On dividing the energy equation (\ref{Energy_eq_s}) by $T$ with
a little bit of algebra one obtains the entropy equation\index{SI}{entropy!equation}
\begin{equation}
\frac{\partial}{\partial t}\left(\rho s\right)+\nabla\cdot\left(\rho\mathbf{u}s\right)=\nabla\cdot\left(\frac{k}{T}\nabla T\right)+\frac{k}{T^{2}}\left(\nabla T\right)^{2}+2\frac{\mu}{T}\utilde{\mathbf{G}}^{s}:\utilde{\mathbf{G}}^{s}+\frac{\mu_{b}}{T}\left(\nabla\cdot\mathbf{u}\right)^{2},\label{Energy_Aderiv-1-1-3-1}
\end{equation}
where the continuity equation (\ref{eq:mass_evolution_gen}) in the
absence of the sources of matter, $\sigma_{m}=0$\footnote{The assumption of vanishing sources of matter is necessary here, since
we assume adiabatic insulation and it is thermodynamically impossible
to introduce a non-zero flux of matter without simultaneous heat
transfer (cf. Guminski 1974 and section \ref{sec:Compositional-and-heat-fluxes} of this book).}, was used. Assuming periodic boundary conditions in the horizontal
directions with some periods $L_{x}$, $L_{y}$, and by the use of
the adiabatic (\ref{eq:adiabatic_insulation-1}) and impermeability
$u_{z}(z=0,\,L)=0$ conditions on the top and bottom boundaries, integration
of the two ``divergence'' terms in the entropy equation (\ref{Energy_Aderiv-1-1-3-1})
over the entire horizontally periodic volume $V=(0,\,L_{x})\times(0,\,L_{y})\times(0,\,L)$
leads to
\begin{equation}
\int_{V}\nabla\cdot\left(\rho\mathbf{u}s\right)\mathrm{d}^{3}x=\int_{\partial V}\rho s\mathbf{u}\cdot\hat{\mathbf{n}}\mathrm{d}\Sigma=0,\label{eq:e26}
\end{equation}
\begin{equation}
\int_{V}\nabla\cdot\left[\frac{k}{T}\nabla T\right]\mathrm{d}^{3}x=\int_{\partial V}\frac{k}{T}\nabla T\cdot\hat{\mathbf{n}}\mathrm{d}\Sigma=0.\label{eq:e27}
\end{equation}
These terms are therefore, easily eliminated from the global entropy
balance, which takes the form
\begin{equation}
\frac{\partial}{\partial t}\int_{V}\rho s\mathrm{d}^{3}x=\int_{V}\frac{k}{T^{2}}\left(\nabla T\right)^{2}\mathrm{d}^{3}x+2\int_{V}\frac{\mu}{\tilde{T}}\mathbf{\utilde{G}}^{s}:\utilde{\mathbf{G}}^{s}\mathrm{d}^{3}x+\int_{V}\frac{\mu_{b}}{\tilde{T}}\left(\nabla\cdot\mathbf{u}\right)^{2}\mathrm{d}^{3}x\geq0.\label{eq:e28}
\end{equation}
The latter inequality is, of course, true only when the coefficients
$\mu$, $\mu_{b}$ and $k$ associated with irreversible processes
are positive. The presented calculation demonstrates in fact, that
the second law of thermodynamics demands positivity of those physical
parameters of the fluid.

This way we have demonstrated, that when adiabatic insulation of the
fluid is assumed and $\mu>0$, $\mu_{b}>0$ and $k>0$, the total entropy production in the fluid is positive
or null. This verifies the agreement of the derived energy balance
for fluids with the second law of thermodynamics.

\section{Fundamental ideas in theoretical description of the phenomenon of
convection\label{sec:Fundamental-ideas-in-conv}}

\subsection{Adiabatic gradient\label{subsec:Adiabatic-gradient}}\index{SI}{adiabatic gradient}

\begin{figure}
\begin{centering}
\includegraphics[scale=0.2]{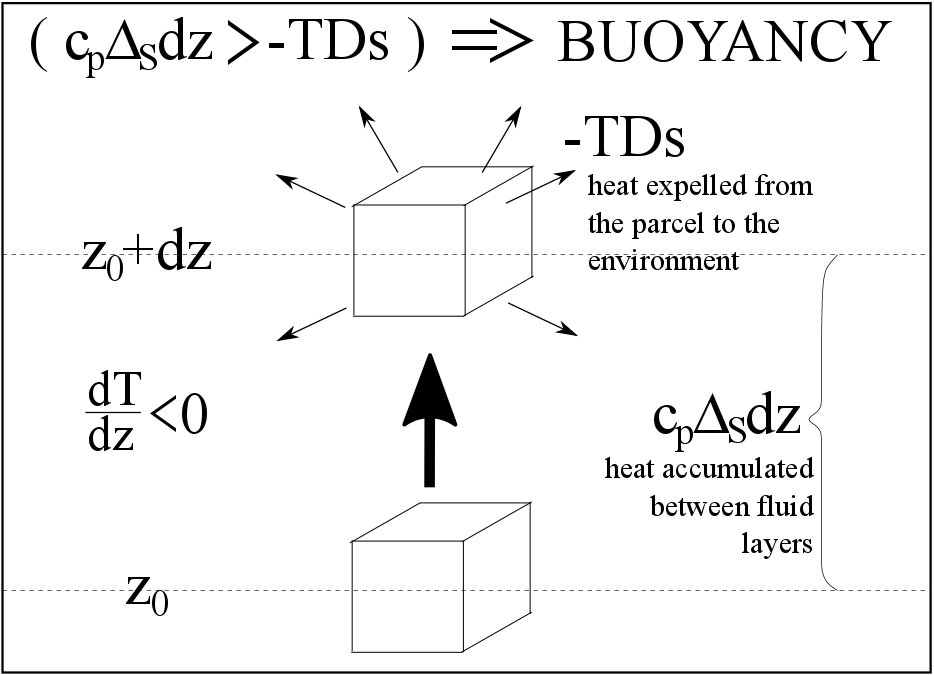}\caption{\label{fig:rising_bubble}{\footnotesize{}A schematic representation
of physical conditions required for appearance of buoyancy in a fluid
with negative vertical gradient of temperature. When a fluid parcel
is shifted upwards between two infinitesimally distant horizontal layers,
it releases heat to the environment and thus its temperature decreases,
causing the density to increase. When the total heat accumulated between
the layers $z_{0}$ and $z_{0}+\mathrm{d}z$ starts to exceed the
total heat released from the fluid element to its surroundings, this
lost heat can not account for an increase in density of the parcel
significant enough for stability, thus the parcel's density remains
smaller then that of the surroundings and buoyancy is created.}}
\par\end{centering}
\end{figure}
To illustrate the importance of the adiabatic state in convective
flows we will introduce now a simple and basic picture of a blob rise,
which we will later come back to, in order to explain fundamental
aspects of convection. In a static state, i.e. when no motions are
present and the thermodynamics fields $\rho$, $T$, $p$ and $s$
are stationary but height-dependent one can study the stability of
a fluid layer, heated from below, by considering two infinitesimally
spaced horizontal (perpendicular to gravity) fluid layers situated
at heights $z_{0}$ and $z_{0}+\mathrm{d}z$. If a fluid volume of a
unitary mass $V=1/\rho$ is taken from the lower level $z_{0}$ and
placed slightly higher at $z_{0}+\mathrm{d}z$ it will experience
a thermodynamic transformation, since its temperature and pressure
will start adjusting to the environment at the higher level (cf. figure
\ref{fig:rising_bubble}). If the fluid volume after the transformation
becomes denser than the surroundings, the gravity acts to put it
back at the original level $z_{0}$ and then the situation is stable;
in the opposite case the buoyancy is non-zero and the system looses
stability. Let us take the pressure $p(z)$ and the entropy $s(z)$
as the system parameters, then the fluid volume, initially 
\[
V\left(p(z_{0}),s(z_{0})\right),
\]
after the shift and the thermodynamic transformation which adjusts
the pressure to the value $p(z_{0}+\mathrm{d}z)$ of the surrounding fluid at
the higher level changes to
\[
V\left(p(z_{0}+\mathrm{d}z),s(z_{0})+\mathrm{D}s\right),
\]
where $s(z_{0})+\mathrm{D}s$ denotes the entropy after the thermodynamic
process thus $\mathrm{D}s$ is the entropy change in the process.
In order for the static state to be stable we must require
\begin{equation}
V\left(p(z_{0}+\mathrm{d}z),s(z_{0})+\mathrm{D}s\right)<V\left(p(z_{0}+\mathrm{d}z),s(z_{0}+\mathrm{d}z_{0})\right),\label{eq:e29}
\end{equation}
that the new volume of a mass unit is smaller than that of the surroundings
at the new level $z_{0}+\mathrm{d}z$. Since the pressure is the same
on both sides of the above inequality, we may expand both sides in
the entropy alone about the values at the level $z_{0}$ to get
\begin{equation}
V\left(p(z_{0}+\mathrm{d}z),s(z_{0})\right)+\left(\frac{\partial V}{\partial s}\right)_{p}\mathrm{D}s<V\left(p(z_{0}+\mathrm{d}z),s(z_{0})\right)+\left(\frac{\partial V}{\partial s}\right)_{p}\mathrm{d}s,\label{eq:e30}
\end{equation}
where $\mathrm{d}s=s(z_{0}+\mathrm{d}z_{0})-s(z_{0})$. Since
\begin{equation}
\left(\frac{\partial V}{\partial s}\right)_{p}=\frac{\left(\frac{\partial V}{\partial T}\right)_{p}}{\left(\frac{\partial s}{\partial T}\right)_{p}}=\frac{T}{c_{p}}\left(\frac{\partial V}{\partial T}\right)_{p}>0\label{eq:e31}
\end{equation}
for all standard fluids,where $(\partial V/\partial T)_{p}$ is the
coefficient of heat expansion, it is obvious now, that the stability
condition demands\index{SI}{stability criterion}
\begin{equation}
\mathrm{D}s<\mathrm{d}s,\label{eq:e32}
\end{equation}
and hence in the case of decreasing temperature with height, when
the fluid volume moved upwards looses heat so that $\mathrm{D}s\leq0$,
the strongest stability restriction is imposed by an adiabatic transformation
requiring
\begin{equation}
\mathrm{D}s=0\;\Rightarrow\;\mathrm{d}s>0,\label{eq:e33}
\end{equation}
for stability. This means that
\begin{equation}
0<\frac{\mathrm{d}s}{\mathrm{d}z}=\left(\frac{\partial s}{\partial T}\right)_{p}\frac{\mathrm{d}T}{\mathrm{d}z}+\left(\frac{\partial s}{\partial p}\right)_{T}\frac{\mathrm{d}p}{\mathrm{d}z}=\frac{c_{p}}{T}\frac{\mathrm{d}T}{\mathrm{d}z}+\alpha g,\label{eq:e34}
\end{equation}
where the Maxwell identity $(\partial s/\partial p)_{T}=-(\partial V/\partial T)_{p}$
and the hydrostatic balance $\mathrm{d}p/\mathrm{d}z=-g/V$ were used.
As a result we obtain the following sufficient (but \emph{not necessary})
condition for stability\index{SI}{adiabatic gradient}
\begin{equation}
-\frac{\mathrm{d}T}{\mathrm{d}z}<\frac{g\alpha T}{c_{p}},\label{eq:e35}
\end{equation}
that is convection does not develop when the temperature gradient
is below the adiabatic one and buoyancy forces may only start to appear
when the temperature gradient exceeds that of the adiabatic profile. 

In real fluids the dissipative effects are present, such as viscous
friction and molecular heat conduction and therefore the thermodynamic
transformation of a raised fluid element is not adiabatic and the
loss of heat imposes $\mathrm{D}s<0$. This relaxes the stability
condition and the convection threshold is shifted to higher temperature
gradients. Therefore a precise condition that must be satisfied for
the system to be convectively stable is expressed as follows\index{SI}{stability criterion}
\begin{equation}
-T\mathrm{D}s>-c_{p}\left(\frac{\mathrm{d}T}{\mathrm{d}z}+\frac{\alpha Tg}{c_{p}}\right)\mathrm{d}z,\label{eq:GENERAL_STABILITY_CONDITION}
\end{equation}
where $\mathrm{D}s<0$ and $-\mathrm{d}T/\mathrm{d}z-\alpha Tg/c_{p}$
becomes positive once the temperature gradient exceeds the adiabatic
one and can then be called the superadiabatic temperature gradient. 

~

\emph{The physical meaning of the latter inequality in terms of the
convective instability threshold is that the system becomes unstable
as soon as the total heat per unit mass accumulated between two infinitesimally
distant fluid layers starts to exceed the total heat per unit mass
released to the environment by a rising fluid element on the infinitesimal
distance between the layers; the fluid element, therefore,
does not loose enough heat (hence is hotter than its surroundings)
and the resulting energy excess is transformed into the work of the
buoyancy. }

~

In other words immediately above the convection threshold the heat
flux through the system becomes large enough, that the molecular mechanisms
are no longer capable of sustaining it and the system looses stability,
since perturbed fluid elements do not loose enough heat as they rise,
hence are hotter than surroundings and therefore buoyant. Note, however,
that this is not the only possibility for appearance of thermally
induced motion, since a time oscillatory flow may appear under some
circumstances even when the condition (\ref{eq:GENERAL_STABILITY_CONDITION})
is still satisfied; we postpone the discussion of such cases until
the end of this section.

The general stability condition (\ref{eq:GENERAL_STABILITY_CONDITION})
can be investigated further to yield a general expression for the
threshold value of the superadiabatic temperature gradient below which
the system is convectively stable and above which the instability
develops. Denoting
\begin{equation}
\Delta_{S}=-\frac{\mathrm{d}T}{\mathrm{\mathrm{d}z}}-\frac{g\alpha T}{c_{p}},\label{eq:e36}
\end{equation}
the marginal (critical) state at convection threshold, by the use
of (\ref{eq:GENERAL_STABILITY_CONDITION}) is described by\index{SI}{stability criterion}
\begin{equation}
\frac{\mathrm{min}\left(-T\frac{\mathrm{D}s}{\mathrm{d}t}\right)}{u_{z}}=c_{p}\Delta_{S\,crit},\label{eq:general_marginal_condition}
\end{equation}
where the minimum of $-T\mathrm{D}s/\mathrm{d}t$ on the left hand
side is taken over all possible convective states in the vicinity
of convection threshold at any point in space and $\Delta_{S\,crit}$
denotes the critical value of the superadiabatic temperature gradient
$\Delta_{S}$ exactly at threshold.\footnote{In cases, when the hydrostatic temperature gradient at threshold is
height-dependent this determination of critical temperature jump
across the fluid layer involves taking a vertical average of (\ref{eq:general_marginal_condition}).} In the latter equation $\mathrm{D}/\mathrm{d}t$ denotes the total
advective derivative with respect to time\index{SI}{advective time derivative}
\begin{equation}
\frac{\mathrm{D}}{\mathrm{d}t}=\frac{\partial}{\partial t}+\mathbf{u}\cdot\nabla\,,\label{eq:convective_derivative}
\end{equation}
which describes the total change of a quantity (such as the entropy,
temperature etc.) in a fluid element during its evolution in the flow
and $\Delta_{S\,crit}$ denotes the critical value of the superadiabatic
gradient at the threshold. We therefore make use of the energy equation
expressed in terms of the entropy (\ref{Energy_eq_s}) to rewrite
the marginal relation (\ref{eq:general_marginal_condition}) in the
form
\begin{equation}
\Delta_{S\,crit}=\frac{\mathrm{min}\left[-\frac{1}{\rho}\left(\nabla\cdot\left(k\nabla T\right)+Q_{visc}+Q\right)\right]}{c_{p}u_{z}}.\label{eq:e37}
\end{equation}
The entire viscous heating term
\begin{equation}
Q_{visc}=2\mu\mathbf{G}^{s}:\mathbf{G}^{s}+\left(\mu_{b}-\frac{2}{3}\mu\right)\left(\nabla\cdot\mathbf{u}\right)^{2},\label{eq:e38}
\end{equation}
is quadratically nonlinear in the velocity field $\mathbf{u}$, thus at the threshold
is very weak; consequently all the quadratic terms in the velocity can be
neglected in the marginal state at leading order. This yields the following expression
for the critical value of the superadiabatic temperature gradient
at the convection threshold
\begin{equation}
\Delta_{S\,crit}=\frac{\mathrm{min}\left[-\frac{1}{\rho}\left(\nabla\cdot\left(k\nabla T\right)+Q\right)\right]}{c_{p}u_{z}}.\label{eq:general_condition_on_Delta_c}
\end{equation}
It must be emphasized at this stage, that the above general criterion
for convective stability (\ref{eq:GENERAL_STABILITY_CONDITION}) and
the resulting relation (\ref{eq:general_condition_on_Delta_c}) are
relevant only to the cases, when instability sets in without time
oscillations, that is when the marginal state at convection threshold
is stationary. In other words, the above reasoning and results can
be directly applied to the cases, when the so-called \emph{principle
of the exchange of stabilities} formulated in Chandrasekhar (1961)
pp. 24-26 is valid, that is when the growth rate of small perturbations
in the vicinity of convection threshold is purely real and passes
through zero from negative to positive values exactly at the threshold
as the superadiabatic temperature gradient increases. But a situation,
when oscillations appear before the superadiabatic gradient (\ref{eq:general_condition_on_Delta_c})
is reached is, in general, not excluded. In such a case the flow takes form of
a travelling wave, which may exist even for superadiabatic gradient values
significantly lower than that defined by the heat balance (\ref{eq:general_marginal_condition}).
It is enough then if only the advected heat per unit time and per
unit mass $T\mathbf{u}\cdot\nabla s$ released by a rising fluid parcel
becomes equal to $c_{p}\Delta_{S\,crit}$ at threshold, since the flow
is then generated at the instances when the oscillatory component
of the released heat $T\partial_{t}s$ vanishes (at other times the
heat released by the parcel exceeds the one accumulated between fluid
layers, thus the parcel is not buoyant). According to (\ref{eq:convective_derivative})
we have
\begin{equation}
\mathbf{u}\cdot\nabla s=\frac{\mathrm{D}s}{\mathrm{d}t}-\frac{\partial s}{\partial t}\,,\label{eq:e39}
\end{equation}
therefore the final, most general expression for the critical superadiabatic
gradient at convection threshold takes the form\index{SI}{stability criterion!general}
\begin{equation}
\Delta_{S\,crit}=\frac{\mathrm{min}\left[-T\left(\frac{\mathrm{D}s}{\mathrm{d}t}-\frac{\partial s}{\partial t}\right)\right]}{u_{z}}\approx\frac{\mathrm{min}\left[T\frac{\partial s}{\partial t}-\frac{1}{\rho}\left(\nabla\cdot\left(k\nabla T\right)+Q\right)\right]}{c_{p}u_{z}}.\label{eq:general_condition_on_Delta_c-1}
\end{equation}
Note also, that near the marginal state, when the velocity field and
the temperature perturbation are weak $T\mathbf{u}\cdot\nabla s\approx-c_{p}u_{z}\Delta_{S}$,
since the term quadratic in the perturbations to the hydrostatic state
is negligible. The issue of convection threshold will be further enlightened
in later chapters, namely chapters \ref{sec:Linear-stability-analysis}
and \ref{sec:Linear-stability-A} concerned with linear analysis of
the convective instability.

\subsection{Filtering sound waves\label{subsec:Filtering-sound-waves}}\index{SI}{filtration of sound waves}

It is useful and often possible to filter sound waves from the wave
spectrum of the system of dynamical equations, since these are very
fast waves and the typical time scales of convective phenomena in
many stellar and planetary interiors, including the Earth's core,
likewise in many laboratory experiments are much longer than those associated
with sound propagation. From the technical point of view it is very
useful, because filtering sound waves is equivalent to an assumption,
that the sound velocity is infinite and hence pressure spreads infinitely
fast. This allows to obtain a stationary Poisson-like problem for
the pressure. In numerical simulations the sound waves are difficult
to resolve because of their high frequencies, therefore an approximated
set of equations with the sound waves eliminated from the wave spectrum
allows for much more efficient simulations.

It will be demonstrated now, how from the technical point of view
the sound waves can be removed from the spectrum of the dynamical
equations, i.e. precisely which term is responsible for their presence
or absence. Let us assume for simplicity, that there is no radiative
heating, $Q=0$, and the heat conduction coefficient likewise the
dynamical viscosity are uniform, $k=\mathrm{const}$, $\mu=\mathrm{const}$.
In such a case the full system of equations describing the dynamics
of a fluid consists of the momentum balance (the Navier-Stokes equation),
the continuity and energy equations and the equation of state, which
can be cast in the following form \begin{subequations}
\begin{equation}
\rho\left[\frac{\partial\mathbf{u}}{\partial t}+\left(\mathbf{u}\cdot\nabla\right)\mathbf{u}\right]=-\nabla p+\rho\mathbf{g}+\mu\nabla^{2}\mathbf{u}+\left(\frac{\mu}{3}+\mu_{b}\right)\nabla\left(\nabla\cdot\mathbf{u}\right),\label{eq:e40}
\end{equation}
\begin{equation}
\frac{\partial\rho}{\partial t}+\mathbf{u}\cdot\nabla\rho+\rho\nabla\cdot\mathbf{u}=0,\label{eq:e41}
\end{equation}
\begin{align}
\rho c_{v}\left(\frac{\partial T}{\partial t}+\mathbf{u}\cdot\nabla T\right)-\frac{T}{\rho}\left(\frac{\partial p}{\partial T}\right)_{\rho}\left(\frac{\partial\rho}{\partial t}+\mathbf{u}\cdot\nabla\rho\right)= & k\nabla^{2}T+2\mu\mathbf{G}^{s}:\mathbf{G}^{s}\nonumber \\
 & +\left(\mu_{b}-\frac{2}{3}\mu\right)\left(\nabla\cdot\mathbf{u}\right)^{2},\label{eq:Energy_SW}
\end{align}
\begin{equation}
\rho=\rho(p,T),\label{eq:e42}
\end{equation}
\end{subequations} 
Let us assume a static, spatially uniform equilibrium
\begin{equation}
\rho_{0}=\mathrm{const},\quad T_{0}=\mathrm{const},\quad p_{0}=\mathrm{const},\quad\mathbf{u}_{0}=0,\label{eq:e43}
\end{equation}
and introduce small perturbations upon the equilibrium \begin{subequations}
\begin{equation}
\mathbf{u}\left(\mathbf{x},t\right),\label{eq:e44}
\end{equation}
\begin{equation}
\rho\left(\mathbf{x},t\right)=\rho_{0}+\rho'\left(\mathbf{x},t\right),\;T\left(\mathbf{x},t\right)=T_{0}+T'\left(\mathbf{x},t\right),\;p\left(\mathbf{x},t\right)=p_{0}+p'\left(\mathbf{x},t\right).\label{eq:e45}
\end{equation}
 \end{subequations} 
Linearisation of the equations with respect to the small perturbations
leads to 
\begin{subequations}
\begin{equation}
\rho_{0}\frac{\partial\mathbf{u}}{\partial t}+\left(\frac{\partial p}{\partial\rho}\right)_{T}\nabla\rho'+\left(\frac{\partial p}{\partial T}\right)_{\rho}\nabla T'-\mu\nabla^{2}\mathbf{u}-\left(\frac{\mu}{3}+\mu_{b}\right)\nabla\left(\nabla\cdot\mathbf{u}\right)=0,\label{eq:NS_soundwaves_pert}
\end{equation}
\begin{equation}
\frac{\partial\rho'}{\partial t}+\rho_{0}\nabla\cdot\mathbf{u}=0,\label{eq:cont_soundwaves_pert}
\end{equation}
\begin{equation}
\rho_{0}c_{v}\frac{\partial T'}{\partial t}-\frac{T}{\rho}\left(\frac{\partial p}{\partial T}\right)_{\rho}\frac{\partial\rho'}{\partial t}-k\nabla^{2}T'=0,\label{eq:Energy_soundwaves_pert}
\end{equation}
\end{subequations} 
It is now possible to decompose the perturbations
into Fourier modes and consider single wave vector $\boldsymbol{\mathcal{K}}$,
\begin{equation}
\rho'\left(\mathbf{x},t\right)=\hat{\rho}\mathrm{e}^{\mathrm{i}\left(\boldsymbol{\mathcal{K}}\cdot\mathbf{x}-\omega t\right)},\quad T'\left(\mathbf{x},t\right)=\hat{T}\mathrm{e}^{\mathrm{i}\left(\boldsymbol{\mathcal{K}}\cdot\mathbf{x}-\omega t\right)},\quad\mathbf{u}\left(\mathbf{x},t\right)=\hat{\mathbf{u}}\mathrm{e}^{\mathrm{i}\left(\boldsymbol{\mathcal{K}}\cdot\mathbf{x}-\omega t\right)}.\label{eq:e46}
\end{equation}
Introduction of the Fourier forms of the perturbations into the linearised
equations ($\ref{eq:NS_soundwaves_pert}$-c) allows to obtain the
characteristic equation, since non-zero solutions can only exist if
the determinant
\begin{equation}
\left|\begin{array}{ccccc}
-\mathcal{W}_{1} & 0 & 0 & \mathcal{K}_{x}\mathcal{W}_{2} & \left(\frac{\partial p}{\partial T}\right)_{\rho}\mathcal{K}_{x}\\
0 & -\mathcal{W}_{1} & 0 & \mathcal{K}_{y}\mathcal{W}_{2} & \left(\frac{\partial p}{\partial T}\right)_{\rho}\mathcal{K}_{y}\\
0 & 0 & -\mathcal{W}_{1} & \mathcal{K}_{z}\mathcal{W}_{2} & \left(\frac{\partial p}{\partial T}\right)_{\rho}\mathcal{K}_{z}\\
\rho_{0}\mathcal{K}_{x} & \rho_{0}\mathcal{K}_{y} & \rho_{0}\mathcal{K}_{z} & -\omega & 0\\
0 & 0 & 0 & \frac{\omega T_{0}}{\rho_{0}}\left(\frac{\partial p}{\partial T}\right)_{\rho} & -\left(\rho_{0}c_{v}\omega+\mathrm{i}k\mathcal{K}^{2}\right)
\end{array}\right|=0\label{eq:e46andhalf}
\end{equation}
vanishes, where 
\begin{subequations}
\begin{equation}
\mathcal{W}_{1}=\rho_{0}\omega+\mathrm{i}\mu\mathcal{K}^{2},\label{eq:e47}
\end{equation}
\begin{equation}
\mathcal{W}_{2}=\left(\frac{\partial p}{\partial\rho}\right)_{T}-\mathrm{i}\left(\frac{\mu}{3}+\mu_{b}\right)\frac{\omega}{\rho_{0}}.\label{eq:e48}
\end{equation}
\end{subequations} 
This yields
\begin{align}
\left(\rho_{0}\omega+\mathrm{i}\mu\mathcal{K}^{2}\right)^{2}\left\{ \omega T_{0}\mathcal{K}^{2}\left(\frac{\partial p}{\partial T}\right)_{\rho}^{2}+\left(\rho_{0}c_{v}\omega+\mathrm{i}k\mathcal{K}^{2}\right)\left[\rho_{0}\mathcal{K}^{2}\left(\left(\frac{\partial p}{\partial\rho}\right)_{T}\right.\right.\right.\nonumber \\
\left.\left.\left.-\mathrm{i}\frac{\omega}{\rho_{0}}\left(\frac{\mu}{3}+\mu_{b}\right)\right)-\omega\left(\rho_{0}\omega+\mathrm{i}\mu\mathcal{K}^{2}\right)\right]\right\}  & =0\label{eq:dispersion_SW_gen}
\end{align}
The modes $\omega=-\mathrm{i}\mu\mathcal{K}^{2}/\rho_{0}$ are purely
diffusive and decaying and do not involve sound waves, therefore we
will concentrate only on the part of the above dispersion relation
given in the braces. The presence of sound waves in the spectrum can
be most easily shown in the following way. Since the speed of sound\index{SI}{speed of sound}
is
\begin{eqnarray}
C^{2}=\left(\frac{\partial p}{\partial\rho}\right)_{s} & = & \left(\frac{\partial p}{\partial\rho}\right)_{T}+\left(\frac{\partial p}{\partial T}\right)_{\rho}\left(\frac{\partial T}{\partial\rho}\right)_{s}\nonumber \\
 & = & \left(\frac{\partial p}{\partial\rho}\right)_{T}+\left(\frac{\partial p}{\partial T}\right)_{\rho}^{2}\frac{T}{\rho^{2}c_{v}},\label{eq:e49}
\end{eqnarray}
which is satisfied by virtue of the implicit function theorem which
implies $(\partial T/\partial\rho)_{s}=-T(\partial s/\partial\rho)_{T}/c_{v}$
and the Maxwell relation $\rho^{2}\left(\partial s/\partial\rho\right)_{T}=-(\partial p/\partial T)_{\rho}$,
the dispersion relation in the limit of vanishing diffusion, $\mu\rightarrow0$,
$\mu_{b}\rightarrow0$ and $k\rightarrow0$, can be simplified to
\begin{equation}
\rho_{0}^{2}c_{v}\omega^{3}-\rho_{0}^{2}c_{v}C^{2}\mathcal{K}^{2}\omega=0.\label{eq:e50}
\end{equation}
Thus
\begin{equation}
\omega=\pm C\mathcal{K},\label{eq:e60}
\end{equation}
so that the phase velocity of waves is equal to the speed of sound\footnote{\label{fn:damping_of_sound_waves}Note, that in the general dissipative
case, $\mu\neq0$, $\mu_{b}\neq0$ and $k\neq0$ the sound waves are
damped by the all three dissipative processes. In particular in the
long-wavelength limit $\mathcal{K}\ll\rho_{0}C/\mu$ the dispersion
relation (\ref{eq:dispersion_SW_gen}) can be solved by means of asymptotic
expansions in the wave number and the sound modes are characterized
by 
\[
\omega=\pm C\mathcal{K}-\mathrm{i}\frac{1}{2\rho_{0}}\left[\frac{4}{3}\mu+\mu_{b}+k\left(\frac{1}{c_{v}}-\frac{1}{c_{p}}\right)\right]\mathcal{K}^{2}.
\]
The dissipative damping of sound waves is also often formulated in
terms of spatial absorption, that is diminishing of the waves intensity
as it travels a certain distance in the fluid; in other words such a
formulation involves complex wave vector and real frequency, but the
absorption coefficient is essentially the same and involves all three
dissipation coefficients (cf. Landau and Lifschitz 1987, ch. 79 on
``\emph{Absorption of sound}'', eq. (79.6)).}. The sound waves can be eliminated from the spectrum through elimination
of the term $\partial\rho'/\partial t$ from the continuity equation\index{SI}{continuity equation}
($\ref{eq:cont_soundwaves_pert}$), which could be achieved e.g. by
assuming\index{SI}{filtration of sound waves}
\begin{equation}
\frac{\rho'}{\rho_{0}}\sim\frac{T'}{T_{0}}\ll1,\quad\omega L\sim\left\Vert \mathbf{u}\right\Vert \sim\left(\frac{T'}{T_{0}}\right)^{1/2}.\label{eq:scalings_SWfiltration}
\end{equation}
Such an assumption allows to neglect the term $\partial\rho'/\partial t$
with respect to $\rho_{0}\nabla\cdot\mathbf{u}$ with all the other
terms in the set of perturbation equations ($\ref{eq:NS_soundwaves_pert}$-c)
retained. The dispersion relation ($\ref{eq:dispersion_SW_gen}$)
is then modified to
\begin{equation}
\frac{\omega T_{0}}{\rho_{0}}\left(\frac{\partial p}{\partial T}\right)_{\rho}^{2}+\left(\rho_{0}c_{v}\omega+\mathrm{i}k\mathcal{K}^{2}\right)\left[\left(\frac{\partial p}{\partial\rho}\right)_{T}-\mathrm{i}\frac{\omega}{\rho_{0}}\left(\frac{\mu}{3}+\mu_{b}\right)\right]=0,\label{eq:e61}
\end{equation}
and it can be easily seen, that in the non-diffusive limit $\mu\rightarrow0$,
$\mu_{b}\rightarrow0$ and $k\rightarrow0$ we get
\begin{equation}
\rho_{0}c_{v}C^{2}\omega=0,\label{eq:e62}
\end{equation}
and hence the sound waves are eliminated from the spectrum. Finally,
it is of interest to note that neglection of the density time derivative in the energy
equation ($\ref{eq:Energy_SW}$) only, through assumption of $\left(\partial p/\partial T\right)_{\rho}=0$
does not lead to elimination of sound waves from the spectrum, since
then $\left(\partial p/\partial\rho\right)_{T}=C^{2}$ and sound wave
dispersion relation $\omega=\pm C\mathcal{K}$ is the only non-zero
and non-diffusive in nature root of the dispersion relation ($\ref{eq:dispersion_SW_gen}$).

The two most frequently used approximations for description of convection
dynamics utilize the idea presented in this section, in particular
the scalings ($\ref{eq:scalings_SWfiltration}$) to filter the sound
waves. These are the well-known Oberbeck-Boussinesq and anelastic
approximations, which will be discussed in detail in the following
chapters.

\section*{Review exercises}

{\textbf{Exercise 1.}} \\
Which fundamental physical laws are responsible for symmetry of the stress tensor, $\tau_{ij}=\tau_{ji}$?
\\

\noindent{\textbf{Exercise 2.}} \\
In the Cattaneo (1948) model the heat flux possesses a correction with respect to the Fourier's law (\ref{eq:e11}) and takes the form $\mathbf{j}_T=-k\nabla T+\sigma\nabla\partial_t T$. This leads to the following hyperbolic temperature equation (cf. Straughan 2011)
\[
\frac{\partial^2 T}{\partial t^2}+\frac{1}{\tau}\frac{\partial T}{\partial t}-\frac{k}{\rho c_p\tau}\nabla^2 T=0,
\]
with $\tau=\sigma/k>0$. Demonstrate that under periodic boundary conditions in all three spatial directions (periodic box) the solutions can take form of damped waves, which are termed \emph{damped heat waves}.
\\

\noindent{\textbf{Exercise 3.}} \\
For the problem of Ex. 2 calculate the group velocity of the damped heat waves $\mathbf{v}_g$. Then calculate the time average over the time $\tau$ of the heat flux $\mathbf{j}_T$ associated with a single standing heat wave, assuming $T(t=0)=T_0\cos(\boldsymbol{\mathcal{K}}\cdot\mathbf{x})$ and $\partial_t T(t=0)=-T(t=0)/2\tau$.

\noindent \emph{Hint}: Assuming the phasor of the temperature oscillations in the form $\exp[\mathrm{i}(\boldsymbol{\mathcal{K}}\cdot\mathbf{x}-\omega t)]$ the group velocity is defined as $\mathbf{v}_g=\nabla_{\boldsymbol{\mathcal{K}}}(\Re\mathfrak{e}\omega)$, i.e. the gradient of the real part of the complex frequency $\Re\mathfrak{e}\omega$ with respect to the wave vector $\boldsymbol{\mathcal{K}}$.
\\

\noindent{\textbf{Exercise 4.}} \\
Derive an expression for the vertical temperature gradient when the entropy density per unit mass $s$ remains uniform.

\noindent\emph{Hint}: cf. equation (\ref{eq:e34}).
\\

\noindent{\textbf{Exercise 5.}} \\
Consider a long wavelength limit $\mathcal{K}\ll\rho_0 C/\mu$ and demonstrate, that sound waves are damped by viscous effects.

\noindent\emph{Hint}: cf. equation (\ref{eq:dispersion_SW_gen}) and footnote \ref{fn:damping_of_sound_waves}.

\chapter{The Oberbeck-Boussinesq convection\label{chap:The-Boussinesq-convection}}

The simplest and at the same time the most classic approach to mathematical
description of convection is the Oberbeck-Boussinesq approximation,
whose applicability covers systems with low density variations that
result only from variations of temperature\footnote{This will be made precise in the following section.}.
At the turn of the nineteenth and twentieth centuries a German physicist
Anton Oberbeck and a French mathematician Joseph Valentin Boussinesq
worked independently on a rigorous mathematical description of buoyancy
driven flows. The equations which are now known as the Oberbeck-Boussinesq
approximation were first obtained by Oberbeck (1879), who derived
them through formal expansion in power series in the thermal expansion
coefficient $\alpha$ (the equations appear as the leading order approximation)
and utilized them to describe convection in spherical geometry. Two
years later Lorenz (1881) published a study of heat transfer by a
free convection in a cartesian geometry using the Oberbeck equations.
Quite independently Boussinesq (1903, p. 174 of that book) obtained
essentially the same equations by making a series of assumptions,
which led to buoyancy force expressed solely in terms of temperature
and otherwise constant density in all the equations, in particular
the solenoidal constraint for the velocity field. As commonly done,
we will often refer in short to the flow of fluid described by the
Oberbeck-Boussinesq equations simply as Boussinesq convection.

It is of interest to give a glimpse on some of the most important
literature concerning significant milestones in understanding of Boussinesq
convection and systematization of knowledge, so that interested readers
can broaden their horizons. The earliest experiments involving thermally
driven flow in a horizontal layer of fluid heated from below date
back to Thomson (1882). A more detailed and comprehensive experimental
study was done later by B$\acute{\textrm{e}}$nard (1900)\footnote{however, B$\acute{\textrm{e}}$nard studied in fact a problem significantly
influenced by a temperature-dependent surface tension, which is now
known as the B$\acute{\textrm{e}}$nard-Maragoni convection.}. In sixteen years time from that seminal experimental work Lord Rayleigh
(1916) was the first one to provide analytic results concerning convective
instability threshold; Pellew and Sothwell (1940) developed his theory
to study the influence of boundary conditions on the conditions of
stability breakdown. After the two main pioneers the problem of thermal
convection in a horizontal layer heated from below is now commonly
referred to as the Rayleigh-B$\acute{\textrm{e}}$nard problem. A
survey of findings from those
early stages concerning the phenomenon of convection can be found in Ostrach (1957).

A noteworthy derivation of the Oberbeck-Boussinesq equations came
from Spiegel and Veronis (1960), with some later corrections regarding
the magnitude of viscous heating in Veronis (1962). This derivation
was based on the assumption of large scale heights with respect to
the layer's depth, i.e. $L\ll\left|\rho/\mathrm{d}_{z}\rho\right|$,
$L\ll\left|T/\mathrm{d}_{z}T\right|$, $L\ll\left|p/\mathrm{d}_{z}p\right|$,
and formal expansions in the small magnitude of density stratification.
Next step in mathematical formalization and obtaining full mathematical
rigour in the derivation of the Boussinesq system of equations was
due to Mihaljan (1962), who explicitly included the assumption of
smallness of diffusivities and derived the Boussinesq equations through
two-parameter expansions in $\alpha\Delta T$ and $\kappa^{2}/c_{v}L^{2}\Delta T$;
he also included an analysis of the energetics within the Oberbeck-Boussinesq
approximation. Later Cordon and Velarde (1975) and Velarde and Cordon
(1976) utilized similar expansion parameters, with the particular
emphasis on the viscous dissipation and large-gap effects.

A systematization of knowledge about Boussinesq convection has continued
throughout a substantial collection of books on the topic. A quick
review of some of the most important contributions includes:

(1) Chandrasekhar (1961) - a comprehensive analysis of the linearised
Boussinesq equations in the very weak amplitude regime close to convection
onset; the effect of background rotation and magnetic field have been
thoroughly considered and the case of convection onset in spheres
and spherical shells has been analysed.

(2) Gershuni and Zhukhovitskii (1972, eng. trans. 1976) - a general
book on convective instability, with a strong focus on the influence
of various physical effects on the conditions and flow structure at
the threshold of convection. This includes e.g. the effects of geometry,
consideration of binary mixtures, internal heat sources, effects of
longitudinal temperature gradient, the study of convection in porous
medium saturated with fluid (convective filtration) or the thermocapillary
effect.

Here, for the sake of a quick reference of readers and self-consistency
and completeness of this book, in section \ref{sec:Linear-stability-analysis}
we provide a compendium of some of the most important, known results
for the linear regime at convection onset in different physical configurations.

(3) Joseph (1976) - a fully rigorous mathematical description of convection
under the Oberbeck-Boussinesq approximation including the bifurcation
theory and the global stability analysis through a general energy
theory of stability; the effects of geometry, the magnetic field and
a study of turbulent convection in porous materials are included.

(4) Straughan (2004) - mathematically fully rigorous application of
the energy method to Boussinesq convection. Various physical circumstances
are considered, such as e.g. convection in porous media, internal
heating, surface tension, the micropolar model of suspensions, electric
and magnetic fields and temperature dependent viscosity. The pattern
selection problem for Boussinesq convection is also considered.

(5) Getling (1998) - a seminal work containing a very comprehensive
study and review of the weakly nonlinear convection and pattern selection
near the onset

The primary goal of this chapter is to provide a consistent framework
and reference for later chapters and gather knowledge on the development
of Boussinesq convection from linear stages, through the weakly nonlinear
ones up to a fully nonlinear regime. Let us recall, that the early
Malkus' (1954) experiments, confirmed by many later studies clearly
show that development of convection with increasing driving occurs
through a sequence of consecutive instabilities, which build upon
successive, gradually more complicated flows until a fully turbulent
state is reached. The aforementioned Getling's (1998) book is a wonderful
reference for description of the problem of pattern selection by the
convective flow near onset, i.e. the changes in stable flow structure
when the driving is continuously enhanced, but weak. A number of significant
contributions in this field come from a German scientist Friedrich
Busse. We only mention a few out of a large collection of his works,
that is the influential studies Busse (1970) and Busse and Cuong (1977)
on the onset of rotating convection in spherical geometry and Busse
(1969, 1978), Busse and Riahi (1980) and Clever and Bussse (1994)
on nonlinear effects and heat transfer estimates. We conclude the
short review by recalling the distinguished work of Grossmann and
Lohse (2000), later updated in Stevens\emph{ et al}. (2013), who gathered
the experimental, numerical and theoretical results concerning fully
developed, nonlinear Boussinesq convection, systematized them and
constructed a consistent dynamical picture together with a theory
of heat transfer (including estimates of the magnitude of the flow)
for Boussinesq convection with strong driving.

In this chapter we attempt to present a general picture of Boussinesq
convection and explain the approaches undertaken in the description
at different stages of convection development. We start from the rigorous
derivation of the Boussinesq system of equations and a study of general
energetic properties of Boussinesq convection, with general definitions
of the Rayleigh and Nusselt numbers. Next the linear regime near the
onset is described under various physical conditions and we proceed
to explain the weakly nonlinear approach and major approaches to the
pattern selection problem near onset. Finally we review the Grossmann
and Lohse (2000) theory of fully developed convection. A comprehensive
but short summary is offered in the last section of this chapter.

\section{Derivation of the Oberbeck-Boussinesq equations\label{sec:Derivation-of-the_B-eqs}}

With the aim of derivation of one of the most standard approximations
of the hydrodynamic equations constructed to describe convective flows,
that is the so-called Boussinesq approximation for convection, we first restate the
system of Navier-Stokes, mass continuity and energy equations, supplied
by the equation of state\index{SI}{hydrodynamic equations (general)}
\begin{subequations}
\begin{eqnarray}
\rho\left[\frac{\partial\mathbf{u}}{\partial t}+\left(\mathbf{u}\cdot\nabla\right)\mathbf{u}\right] & = & -\nabla p+\rho\mathbf{g}+\mu\nabla^{2}\mathbf{u}+\left(\frac{\mu}{3}+\mu_{b}\right)\nabla\left(\nabla\cdot\mathbf{u}\right)\nonumber \\
 &  & \qquad\qquad+2\nabla\mu\cdot\mathbf{G}^{s}+\nabla\left(\mu_{b}-\frac{2}{3}\mu\right)\nabla\cdot\mathbf{u},\qquad\label{NS-Bderiv}
\end{eqnarray}
\begin{equation}
\frac{\partial\rho}{\partial t}+\mathbf{u}\cdot\nabla\rho+\rho\nabla\cdot\mathbf{u}=0,\label{Cont_Bderiv}
\end{equation}
\begin{equation}
\rho c_{v}\left(\frac{\partial T}{\partial t}+\mathbf{u}\cdot\nabla T\right)+\frac{\alpha T}{\beta}\nabla\cdot\mathbf{u}=\nabla\cdot\left(k\nabla T\right)+2\mu\mathbf{G}^{s}:\mathbf{G}^{s}+\left(\mu_{b}-\frac{2}{3}\mu\right)\left(\nabla\cdot\mathbf{u}\right)^{2}+Q,\label{Energy_Bderiv}
\end{equation}
\begin{equation}
\rho=\rho(p,T),\label{State_eq_Bderiv}
\end{equation}
\end{subequations} with $G_{ij}^{s}=(\partial_{j}u_{i}+\partial_{i}u_{j})/2$
being the symmetric part of the velocity field gradient tensor, $\mathbf{G}^{s}:\mathbf{G}^{s}=G_{ij}^{s}G_{ij}^{s}$
and $Q$ denoting heat sources, e.g. due to radiation. All the coefficients
such as $\mu$, $\mu_{b}$, $c_{v}$ and $k$ have been assumed nonuniform
for generality, with the exception that the horizontal variations
of the thermal conductivity coefficient $k$ (i.e. the variations
in the plane perpendicular to gravity acceleration) are neglected and
thus $k$ is assumed to be a function of height only,
\begin{equation}
k=k(z).\label{eq:e63}
\end{equation}
The approach of Spiegel and Veronis (1960) will be generally followed
with some changes, in particular a general equation of state will
be considered and a more detailed explanation of all the assumptions
made will be provided. Let us assume the $z$-axis of the coordinate
system in the vertical direction, so that the gravitational acceleration points in the negative
$z$-direction, i.e. $\mathbf{g}=-g\hat{\mathbf{e}}_{z}$. The thermodynamic
variables, such as density, temperature, pressure and entropy in the
considered case of time-independent boundary conditions can be split
into a hydrostatic part, which is $z$-dependent only and a small but
fully spatially and time-dependent fluctuation. The former will be
denoted by a single upper tilde, i.e.
\[
\tilde{\rho},\;\tilde{T},\;\tilde{p},\;\tilde{s}\quad\textrm{denote hydrostatic parts},
\]
and the fluctuations will be denoted by primes, 
\[
\rho',\;T',\;p',\;s',\quad\textrm{denote fluctuations}.
\]
The hydrostatic part can be further split into a mean, $\bar{T}$,
and a correction, which involves the variations in the hydrostatic
state; the latter one will be denoted by an upper double tilde in the following, $\dbtilde{T}$,
and in the case of temperature it satisfies the imposed
inhomogeneous boundary conditions, which drive the convective flow.
Consequently homogeneous boundary conditions are imposed on the temperature
fluctuation $T'$. Hence the thermodynamic variables are represented
as follows 
\begin{subequations}
\begin{equation}
\rho(\mathbf{x},t)=\bar{\rho}+\dbtilde{\rho}(z)+\rho'(\mathbf{x},t),\label{eq:rho_gen_B_deriv}
\end{equation}
\begin{equation}
T(\mathbf{x},t)=\bar{T}+\dbtilde{T}(z)+T'(\mathbf{x},t),\label{eq:T_gen_B_deriv}
\end{equation}
\begin{equation}
p(\mathbf{x},t)=\bar{p}+\dbtilde{p}(z)+p'(\mathbf{x},t),\label{eq:p_gen_B_deriv}
\end{equation}
\end{subequations} 
where quantities marked by the upper bar, that
is $\bar{\rho}$, $\bar{T}$ and $\bar{p}$ are space averages (upper
bar will denote the full spatial average in the following) of $\rho(\mathbf{x},t)$,
$T(\mathbf{x},t)$ and $p(\mathbf{x},t)$ respectively; the quantities
with an upper double tilde correspond to hydrostatic state variation
thus in the absence of motion ($z$-dependent only, since $k=k(z)$
is a sole function of height by assumption) and primed variables denote
the fluctuations resulting from motion. The central point of the Boussinesq
approximation is the \emph{first simplifying assumption} that the
scale heights\index{SI}{scale heights}
\begin{equation}
D_{\rho}=\left|\frac{1}{\bar{\rho}}\frac{\mathrm{d}\dbtilde{\rho}}{\mathrm{d}z}\right|^{-1},\quad D_{T}=\left|\frac{1}{\bar{T}}\frac{\mathrm{d}\dbtilde{T}}{\mathrm{d}z}\right|^{-1},\quad D_{p}=\left|\frac{1}{\bar{p}}\frac{\mathrm{d}\dbtilde{p}}{\mathrm{d}z}\right|^{-1},\label{eq:scale_heights_def}
\end{equation}
are all much larger than the thickness $L$ of the layer of fluid,
which undergoes convection\index{SI}{Boussinesq!assumption (1)}
\begin{equation}
L\ll\min\left\{ D_{\rho},\,D_{T},\,D_{p}\right\} ,\label{eq:LvsScaleHeights_B}
\end{equation}
everywhere in the fluid domain. Using this assumption we define small
parameter\index{SI}{Boussinesq!small parameter}
\begin{equation}
\epsilon=\frac{\Delta\dbtilde{\rho}}{\bar{\rho}}\ll1,\label{eq:epsilon_def_B}
\end{equation}
which is small by virtue of integration of relation $L/D_{\rho}\ll1$
from the level of minimal to the level of maximal density within the
fluid layer and the density jump between these layers is denoted by
$\Delta\dbtilde{\rho}$. 

A \emph{second crucial restriction} that
we have to impose by assumption is that the fluctuations of thermodynamic
variables induced by the convective motions are of the same order
of magnitude or smaller than the static variation, i.e.
\begin{equation}
\left|\frac{\rho'}{\Delta\dbtilde{\rho}}\right|\lesssim\mathcal{O}(1),\quad\left|\frac{T'}{\Delta\dbtilde{T}}\right|\lesssim\mathcal{O}(1),\quad\left|\frac{p'}{\Delta\dbtilde{p}}\right|\lesssim\mathcal{O}(1),\label{eq:e64}
\end{equation}
which, in turn leads to\index{SI}{Boussinesq!assumption (2)}
\begin{equation}
\left|\frac{\rho'}{\bar{\rho}}\right|\lesssim\mathcal{O}(\epsilon),\quad\left|\frac{T'}{\bar{T}}\right|\lesssim\mathcal{O}(\epsilon),\quad\left|\frac{p'}{\bar{p}}\right|\lesssim\mathcal{O}(\epsilon).\label{fluct_magnitude_B}
\end{equation}
This assumption means that (\ref{eq:rho_gen_B_deriv})-(\ref{eq:p_gen_B_deriv})
could be understood as formal power series in $\epsilon$, most possibly
asymptotic in nature. Therefore mathematically the first few terms of the
expansion, hence the leading order of the fluctuating parts as well,
can be expected to provide a satisfactory approximation of the full
solution, since $\epsilon$ can always be chosen small enough. However,
in modeling real systems, in which the parameter $\epsilon=\Delta\dbtilde{\rho}/\bar{\rho}$
is set, care must be taken to verify \emph{a posteriori} consistency
with assumption (\ref{fluct_magnitude_B}), although there does not
seem to be any experimental nor numerical evidence for fluctuations
ever to significantly exceed the static variation.

The static, motionless state is described with the use of the dynamical
equations (\ref{NS-Bderiv}) and (\ref{Energy_Bderiv}) by \index{SI}{reference (basic) state}
\begin{subequations}
\begin{equation}
\frac{\partial\dbtilde{p}}{\partial x}=\frac{\partial\dbtilde{p}}{\partial y}=0,\quad\frac{\partial\dbtilde{p}}{\partial z}=-\bar{\rho}g-\dbtilde{\rho}g,\label{eq:Static_NS}
\end{equation}
\begin{equation}
\nabla\cdot\left(k\nabla\dbtilde{T}\right)=-\dbtilde{Q}.\label{eq:Static_energy}
\end{equation}
\end{subequations} 
Furthermore, we expand the equation of state (\ref{State_eq_Bderiv})
in Taylor series to obtain
\begin{equation}
\rho=\bar{\rho}\left[1-\bar{\alpha}\left(T-\bar{T}\right)+\bar{\beta}\left(p-\bar{p}\right)+\mathcal{O}\left(\epsilon^{2}\right)\right],\label{eq:Taylor_exp_rho_B}
\end{equation}
where
\begin{equation}
\bar{\alpha}=-\frac{1}{\bar{\rho}}\left.\left(\frac{\partial\rho}{\partial T}\right)\right|_{\left(T,\,p\right)=\left(\bar{T},\,\bar{p}\right)},\quad\bar{\beta}=\frac{1}{\bar{\rho}}\left.\left(\frac{\partial\rho}{\partial p}\right)\right|_{\left(T,\,p\right)=\left(\bar{T},\,\bar{p}\right)},\label{eq:e65}
\end{equation}
which by the use of (\ref{eq:rho_gen_B_deriv}-c) implies for both
the static density and the density fluctuation 
\begin{subequations}
\begin{equation}
\frac{\dbtilde{\rho}}{\bar{\rho}}=\bar{\beta}\dbtilde{p}-\bar{\alpha}\dbtilde{T}+\mathcal{O}\left(\epsilon^{2}\right),\label{eq:rho_0_B}
\end{equation}
\begin{equation}
\frac{\rho'}{\bar{\rho}}=\bar{\beta}p'-\bar{\alpha}T'+\mathcal{O}\left(\epsilon^{2}\right),\label{eq:rho_prime_B}
\end{equation}
\end{subequations}
since the static variation and fluctuations due
to the flow have to be balanced separately. In the above $\bar{\alpha}$
is the thermal expansion coefficient and $\bar{\beta}$ the isothermal
compressibility coefficient related to the isothermal speed of sound
$C_{T}$\index{SI}{speed of sound!isothermal}
\begin{equation}
\bar{\beta}=\frac{1}{\bar{\rho}C_{T}^{2}}.\label{eq:e66}
\end{equation}
Furthermore, making use of (\ref{eq:rho_gen_B_deriv}) and (\ref{eq:epsilon_def_B})
in the mass conservation equation (\ref{Cont_Bderiv}) we obtain the
first significant result of the Boussinesq approximation for convective
flows, namely that the flow divergence is negligibly small in terms
of any natural frequency in the system, such as e.g. $k/\bar{\rho}c_{p}L^{2}$,
$\mu/\bar{\rho}L^{2}$, $\mu_{b}/\bar{\rho}L^{2}$ or $\sqrt{g/L}$,\index{SI}{continuity equation}
\begin{eqnarray}
\nabla\cdot\mathbf{u} & = & -\left(\frac{\partial}{\partial t}+\mathbf{u}\cdot\nabla\right)\frac{\dbtilde{\rho}+\rho'}{\bar{\rho}}+\mathcal{O}\left(\epsilon^{2}\frac{\mathscr{U}}{L}\right)\nonumber \\
 & = & -\epsilon\left(\frac{\partial}{\partial t}+\mathbf{u}\cdot\nabla\right)\frac{\dbtilde{\rho}+\rho'}{\Delta\dbtilde{\rho}}+\mathcal{O}\left(\epsilon^{2}\frac{\mathscr{U}}{L}\right)\nonumber \\
 & = & 0+\mathcal{O}\left(\epsilon\frac{\mathscr{U}}{L}\right),\label{eq:divu_B}
\end{eqnarray}
where $\mathscr{U}$ is the velocity scale defined later in (\ref{eq:vel_and_time_scales}).

\subsection{Momentum balance\label{subsec:Momentum-balance_B}}

On inserting the static state equation (\ref{eq:Static_NS}) into
the Navier-Stokes equation (\ref{NS-Bderiv}) and utilizing the negligibility
of the flow divergence (\ref{eq:divu_B}) one obtains
\begin{equation}
\rho\left[\frac{\partial\mathbf{u}}{\partial t}+\left(\mathbf{u}\cdot\nabla\right)\mathbf{u}\right]=-\nabla p'-\rho'g\hat{\mathbf{e}}_{z}+\mu\nabla^{2}\mathbf{u}+2\nabla\mu\cdot\mathbf{G}^{s}+\mathcal{O}\left(\mu\epsilon\frac{\mathscr{U}}{L^{2}}\right).\label{NS-Bderiv-1}
\end{equation}
Next, dividing the latter equation by $\bar{\rho}$ and using the
definition of the small parameter $\epsilon=\Delta\dbtilde{\rho}/\bar{\rho}$,
as in (\ref{eq:epsilon_def_B}), and keeping only the leading order
terms the momentum balance simplifies to
\begin{equation}
\frac{\partial\mathbf{u}}{\partial t}+\left(\mathbf{u}\cdot\nabla\right)\mathbf{u}=-\frac{1}{\bar{\rho}}\nabla p'-\epsilon\frac{\rho'}{\Delta\dbtilde{\rho}}g\hat{\mathbf{e}}_{z}+\nu\nabla^{2}\mathbf{u}+2\nabla\nu\cdot\mathbf{G}^{s},\label{NS-Bderiv-1-1}
\end{equation}
where $\nu=\mu/\bar{\rho}$ is the kinematic viscosity. The buoyancy
force $\epsilon g\rho'/\Delta\dbtilde{\rho}\hat{\mathbf{e}}_{z}$
drives the motions and thus the flow acceleration must be of the order
$\epsilon g$, i.e. necessarily much smaller than the acceleration
of gravity. This implies the following for the convective velocity
and time scales\index{SI}{Boussinesq!velocity scale}\index{SI}{Boussinesq!time scale}
\begin{equation}
\mathscr{U}\sim\epsilon^{1/2}\sqrt{gL},\qquad\mathscr{T}\sim\epsilon^{-1/2}\sqrt{\frac{L}{g}},\label{eq:vel_and_time_scales}
\end{equation}
and immediately for the viscosity\index{SI}{Boussinesq!viscosity scales}
\begin{equation}
\mu_{b}/\bar{\rho}\lesssim\nu\sim\epsilon^{1/2}\sqrt{gL}L\label{eq:visc_scale}
\end{equation}
(the bulk viscosity does not exceed shear viscosity in order of magnitude).
Furthermore, the buoyancy term together with vertical pressure gradient
in (\ref{NS-Bderiv-1-1}) can be rearranged with the aid of the relation
between density, temperature and pressure fluctuations (\ref{eq:rho_prime_B})
and the definition of $\epsilon$ in (\ref{eq:epsilon_def_B}) to give
\begin{eqnarray}
-\frac{1}{\bar{\rho}}\frac{\partial p'}{\partial z}-\frac{\rho'}{\bar{\rho}}g & = & -\frac{1}{\bar{\rho}}\frac{\partial p'}{\partial z}-\bar{\beta}p'g+\bar{\alpha}T'g\nonumber \\
 & = & -\frac{1}{\bar{\rho}}\left(\frac{\partial p'}{\partial z}+\frac{p'}{H}\right)+\bar{\alpha}T'g,\label{eq:Buoyancy_simpl_B}
\end{eqnarray}
where
\begin{equation}
H=\frac{1}{g\bar{\rho}\bar{\beta}}=\frac{C_{T}^{2}}{g}\label{eq:e67}
\end{equation}
has a dimension of length. In the case of a perfect gas equation of
state, i.e. $p=\rho RT$\footnote{$R=k_{B}/m_{m}$ is the specific gas constant; $m_{m}$ is the molecular
mass, $k_{B}$ the Boltzmann constant.} the compressibility coefficient $\bar{\beta}=1/\bar{p}$ and the
quantity $H$ is a thickness defined by a hydrostatic pressure balance
in a system with uniform mean density $\bar{\rho}$. On the other
hand the definition of the pressure scale height supplied by the hydrostatic
balance (\ref{eq:Static_NS}) implies
\begin{equation}
D_{p}=\left[\frac{g\bar{\rho}}{\bar{p}}\left(1+\frac{\dbtilde{\rho}}{\bar{\rho}}\right)\right]^{-1}=\frac{\bar{p}}{g\bar{\rho}}\frac{1}{1+\epsilon\frac{\dbtilde{\rho}}{\Delta\dbtilde{\rho}}},\label{eq:e68}
\end{equation}
therefore
\begin{equation}
H=D_{p}+\mathcal{O}\left(\epsilon\right)\gg L.\label{eq:HvsDp_B}
\end{equation}
The above relation (\ref{eq:HvsDp_B}) is, in fact, more general and
does not apply solely to fluids whose thermodynamic properties are
described by the perfect gas equation of state. All that is necessary
is that in terms of the small parameter $\epsilon$ being a measure
of weak density stratification, $\partial\rho/\partial p\sim\rho/p$
or equivalently $C_{T}^{2}\sim p/\rho$; this then implies that $H$
and $D_{p}$ are of the same order of magnitude. It is therefore necessary
from the point of view of derivation of the Boussinesq equations to
make a \emph{third assumption} concerning the fluid thermodynamic
properties, namely that\index{SI}{Boussinesq!assumption (3)}
\begin{equation}
\frac{\sqrt{gL}}{C_{T}}=\mathcal{O}\left(\epsilon^{1/2}\right),\quad\textrm{or equivalently}\quad\frac{1}{g\bar{\rho}\bar{\beta L}}=\mathcal{O}\left(\epsilon^{-1}\right),\label{eq:third_assumption}
\end{equation}
obviously satisfied by a perfect gas by virtue of the assumption (\ref{eq:epsilon_def_B}); of course this means that the Mach number, 
\begin{equation}
Ma=\frac{\mathcal{U}}{C}=\mathcal{O}(\epsilon)\label{Mach_B}
\end{equation} 
is small and scales linearly with $\epsilon$. It has to be emphasized, however,
that the third assumption is by no means as fundamental as the previous
two expressed in (\ref{eq:epsilon_def_B}) and (\ref{fluct_magnitude_B}),
but in fact reflects a rather typical experimental situation. As a consequence
the term $p'/H$ in equation (\ref{eq:Buoyancy_simpl_B}) is necessarily
much smaller than the vertical variation rate of the pressure fluctuation
\begin{equation}
\frac{p'}{H}\ll\frac{\partial p'}{\partial z},\label{eq:e69}
\end{equation}
and thus finally the leading order momentum balance (\ref{NS-Bderiv-1-1})
takes the form
\begin{equation}
\frac{\partial\mathbf{u}}{\partial t}+\left(\mathbf{u}\cdot\nabla\right)\mathbf{u}=-\frac{1}{\bar{\rho}}\nabla p'+g\bar{\alpha}T'\hat{\mathbf{e}}_{z}+\nu\nabla^{2}\mathbf{u}+2\nabla\nu\cdot\mathbf{G}^{s}.\label{NS-Bderiv-1-1-1}
\end{equation}
Relation (\ref{eq:HvsDp_B}) allowed to simplify the buoyancy force
through neglection of the small pressure fluctuation and retaining
only the temperature fluctuation. This in turn allows to simplify
also the relation (\ref{eq:rho_prime_B}) between the fluctuations
of thermodynamic variables to
\begin{equation}
\frac{\rho'}{\bar{\rho}}=-\bar{\alpha}T'+\mathcal{O}\left(\epsilon^{2}\right).\label{eq:rho_prime_B-1}
\end{equation}
A straightforward consequence of the considerations leading to derivation
of the final form of the Navier-Stokes equation under the Boussinesq
approximation is\index{SI}{Boussinesq!scalings}
\begin{subequations}
\begin{equation}
\frac{\dbtilde{\rho}}{\bar{\rho}}=\mathcal{O}\left(\epsilon\right),\qquad\frac{\rho'}{\bar{\rho}}=\mathcal{O}\left(\epsilon\right),\label{eq:rho_scaling_B}
\end{equation}
\begin{equation}
\frac{\dbtilde{T}}{\bar{T}}=\mathcal{O}\left(\epsilon\right),\qquad\frac{T'}{\bar{T}}=\mathcal{O}\left(\epsilon\right),\label{eq:T_scaling_B}
\end{equation}
\begin{equation}
\frac{\dbtilde{p}}{\bar{p}}=\mathcal{O}\left(\epsilon\right),\qquad\frac{p'}{\bar{p}}=\mathcal{O}\left(\epsilon^{2}\right),\label{eq:p_scaling_B}
\end{equation}
\begin{equation}
\frac{\bar{\rho}gL}{\bar{p}}=\mathcal{O}\left(\epsilon\right).\label{eq:p_vs_rho_B}
\end{equation}
\end{subequations} 
In particular the last relations (\ref{eq:p_vs_rho_B})
and (\ref{eq:p_scaling_B}) imply that the pressure unperturbed by
fluid motion in Boussinesq systems is very strong. 

Finally, we provide
an example of a specific solution of (\ref{eq:Static_NS}) and (\ref{eq:Static_energy})
for the hydrostatic reference state\index{SI}{reference (basic) state} in Boussinesq convection for the
case of an ideal gas, $p=\rho RT$, with $Q=0$, $\kappa=\mathrm{const}$, which can now be
fully understood from the point of view of orderings of terms,
\begin{subequations}
\begin{equation}
\frac{\dbtilde{T}}{\bar{T}}=\frac{\Delta T}{\bar{T}}\left(\frac{1}{2}-\frac{z}{L}\right),\label{eq:T0_example}
\end{equation}
\begin{equation}
\frac{\dbtilde{\rho}}{\bar{\rho}}=\left(\frac{\bar{\rho}gL}{\bar{p}}-\frac{\Delta T}{\bar{T}}\right)\left[\left(\frac{1}{2}-\frac{z}{L}\right)-\frac{\bar{\rho}gL}{2\bar{p}}\frac{z}{L}\left(1-\frac{z}{L}\right)\right]+\mathcal{O}\left(\epsilon^{3}\right),\label{eq:rho0_example}
\end{equation}
\begin{equation}
\frac{\dbtilde{p}}{\bar{p}}=\frac{\bar{\rho}gL}{\bar{p}}\left(\frac{1}{2}-\frac{z}{L}\right)-\frac{\bar{\rho}gL}{2\bar{p}}\left(\frac{\bar{\rho}gL}{\bar{p}}-\frac{\Delta T}{\bar{T}}\right)\frac{z}{L}\left(1-\frac{z}{L}\right)+\mathcal{O}\left(\epsilon^{3}\right),\label{eq:p0_example}
\end{equation}
\end{subequations}
where $\Delta T>0$ is the total temperature jump
across the fluid layer, positive by definition ($\Delta T=T_{B}-T_{T}$).
All the three variables $\dbtilde{T}/\bar{T}$, $\dbtilde{\rho}/\bar{\rho}$
and $\dbtilde{p}/\bar{p}$ are of course of the order $\mathcal{O}(\epsilon)$
and are all linear functions of $z$ at leading order. The density
and pressure in the hydrostatic state possess higher order corrections
and in particular the quadratic corrections
\begin{equation}
-\frac{\bar{\rho}gL}{2\bar{p}}\left(\frac{\bar{\rho}gL}{\bar{p}}-\frac{\Delta T}{\bar{T}}\right)\frac{z}{L}\left(1-\frac{z}{L}\right)=\mathcal{O}\left(\epsilon^{2}\right),\label{eq:correction_example}
\end{equation}
are still important as the pressure quadratic correction contributes
to the hydrostatic balance (\ref{eq:Static_NS}) at the order $\mathcal{O}(\epsilon)$.
We observe, that in many experimental situations $gL/R\Delta T$, which in terms of the small parameter is of the order $\epsilon^0$, is in fact
likely to be significantly less than unity, since the specific gas constant
$R$ for laboratory liquids is of the order $10^{2}\,J/kgK$ and the
typical temperature gradients in laboratory are at the order $10\,K/m$.
This implies, that as a result of the fluid's thermal expansion the
density gradient in the hydrostatic state,
\begin{equation}
L\frac{\mathrm{d}}{\mathrm{d}z}\frac{\dbtilde{\rho}}{\bar{\rho}}=\frac{\Delta T}{\bar{T}}-\frac{\bar{\rho}gL}{\bar{p}}+\mathcal{O}\left(\epsilon^{2}\right)=\mathcal{O}(\epsilon),\label{eq:density_grad_example}
\end{equation}
is likely to be positive in vicinity of convection threshold, although small. However,
if the Boussinesq system remains close to adiabatic, that is $\Delta T/L-g/c_{p}\ll\Delta T/L$,
then it is a simple matter to demonstrate, that $gL/R\Delta T\approx c_{p}/R>1$
and thus the density in the hydrostatic reference state decreases slightly 
with height.

\subsection{Energy balance\label{subsec:Energy-balance_B}}

Utilizing the static state equation (\ref{eq:Static_energy}) and
the general energy balance (\ref{Energy_Bderiv}) the equation for
the temperature fluctuation can be written in the form
\begin{equation}
\rho c_{v}\left(\frac{\partial T'}{\partial t}+\mathbf{u}\cdot\nabla T\right)+\frac{\alpha T}{\beta}\nabla\cdot\mathbf{u}=\nabla\cdot\left(k\nabla T'\right)+2\mu\mathbf{G}^{s}:\mathbf{G}^{s}+\left(\mu_{b}-\frac{2}{3}\mu\right)\left(\nabla\cdot\mathbf{u}\right)^{2}+Q',\label{Energy_Bderiv-1}
\end{equation}
where
\begin{equation}
Q'=Q-\dbtilde{Q}.\label{eq:e70}
\end{equation}
Next we introduce the state equations (\ref{eq:rho_0_B}) and (\ref{eq:rho_prime_B-1})
into the flow divergence estimate (\ref{eq:divu_B}) to obtain
\begin{equation}
\frac{\alpha T}{\beta}\nabla\cdot\mathbf{u}=\frac{\bar{\alpha}\bar{T}}{\bar{\beta}}\left(\frac{\partial}{\partial t}+\mathbf{u}\cdot\nabla\right)\left[\bar{\alpha}\left(\dbtilde{T}+T'\right)-\bar{\beta}\dbtilde{p}\right]+\mathcal{O}\left(\bar{p}\epsilon^{2}\frac{\mathscr{U}}{L}\right).\label{eq:e71}
\end{equation}
This can be further simplified by making use of the hydrostatic balance
in (\ref{eq:Static_NS}), which allows to write
\begin{equation}
-\bar{\alpha}\bar{T}\left(\frac{\partial}{\partial t}+\mathbf{u}\cdot\nabla\right)\dbtilde{p}=u_{z}g\bar{\rho}\bar{\alpha}\bar{T}+\mathcal{O}\left(\bar{p}\epsilon^{2}\frac{\mathscr{U}}{L}\right),\label{eq:e72}
\end{equation}
and therefore
\begin{equation}
\frac{\alpha T}{\beta}\nabla\cdot\mathbf{u}=\frac{\bar{\alpha}^{2}\bar{T}}{\bar{\beta}}\left(\frac{\partial}{\partial t}+\mathbf{u}\cdot\nabla\right)\left(\dbtilde{T}+T'\right)+u_{z}g\bar{\rho}\bar{\alpha}\bar{T}+\mathcal{O}\left(\bar{p}\epsilon^{2}\frac{\mathscr{U}}{L}\right).\label{eq:div_u_energy_B}
\end{equation}
Inserting the latter expression (\ref{eq:div_u_energy_B}) into the
energy balance (\ref{Energy_Bderiv-1}), at the leading order yields
\begin{align}
\left(\bar{\rho}\bar{c}_{v}+\frac{\bar{\alpha}^{2}\bar{T}}{\bar{\beta}}\right)\left(\frac{\partial T'}{\partial t}+\mathbf{u}\cdot\nabla T\right)+u_{z}g\bar{\rho}\bar{\alpha}\bar{T}= & \nabla\cdot\left(k\nabla T'\right)+2\mu\mathbf{G}^{s}:\mathbf{G}^{s}\nonumber \\
 & +\left(\mu_{b}-\frac{2}{3}\mu\right)\left(\nabla\cdot\mathbf{u}\right)^{2}+Q'.\label{Energy_Bderiv-1-1}
\end{align}
Now we recall a thermodynamic identity\index{SI}{specific heats difference}
\begin{eqnarray}
c_{p}-c_{v} & = & T\left[\left(\frac{\partial s}{\partial T}\right)_{p}-\left(\frac{\partial s}{\partial T}\right)_{\rho}\right]=T\left[\left(\frac{\partial s}{\partial\rho}\right)_{T}\left(\frac{\partial\rho}{\partial T}\right)_{p}\right]\nonumber \\
 & = & -\frac{T}{\rho^{2}}\left(\frac{\partial p}{\partial T}\right)_{\rho}\left(\frac{\partial\rho}{\partial T}\right)_{p}=\frac{\alpha^{2}T}{\rho\beta},\label{eq:cp-cv_B}
\end{eqnarray}
where we have used the Maxwell's identity $\rho^{2}\left(\partial s/\partial\rho\right)_{T}=-(\partial p/\partial T)_{\rho}$
and the implicit function theorem $(\partial p/\partial T)_{\rho}=\alpha/\beta$. First of all this allows to simplify the factor in front of the
temperature time derivative,
\begin{equation}
\bar{\rho}\bar{c}_{v}+\frac{\bar{\alpha}^{2}\bar{T}}{\bar{\beta}}=\bar{\rho}\bar{c}_{p}.\label{eq:e73}
\end{equation}
Secondly, since according to (\ref{eq:p_vs_rho_B}) we can estimate
the order of magnitude of the right hand side in (\ref{eq:cp-cv_B})
as $\bar{\alpha}^{2}\bar{T}/\bar{\beta}\bar{\rho}\sim\bar{\alpha}^{2}\bar{T}\bar{p}/\bar{\rho}\sim\epsilon^{-1}gL\bar{\alpha}^{2}\bar{T}\sim\epsilon^{-1}gL/\bar{T}$,
it follows that
\begin{equation}
\bar{c}_{p}\sim\bar{c}_{v}\sim\epsilon^{-1}\frac{gL}{\bar{T}}.\label{eq:e74}
\end{equation}
Hence comparison of the orders of magnitude of the advective term
with viscous heating\index{SI}{viscous heating} in equation (\ref{Energy_Bderiv-1-1}) 
\begin{subequations}
\begin{equation}
\left|2\mu\mathbf{G}^{s}:\mathbf{G}^{s}+\left(\mu_{b}-\frac{2}{3}\mu\right)\left(\nabla\cdot\mathbf{u}\right)^{2}\right|\sim\epsilon^{3/2}\bar{\rho}\frac{\left(gL\right)^{3/2}}{L},\label{eq:e75}
\end{equation}
\begin{equation}
\left|\left(\bar{\rho}\bar{c}_{v}+\frac{\bar{\alpha}^{2}\bar{T}}{\bar{\beta}}\right)\left(\frac{\partial T'}{\partial t}+\mathbf{u}\cdot\nabla T\right)\right|\sim\left|u_{z}g\bar{\rho}\bar{\alpha}\bar{T}\right|\sim\epsilon^{1/2}\bar{\rho}\frac{\left(gL\right)^{3/2}}{L},\label{eq:e76}
\end{equation}
\end{subequations} 
leading to
\begin{equation}
\frac{\left|2\mu\mathbf{G}^{s}:\mathbf{G}^{s}+\left(\mu_{b}-\frac{2}{3}\mu\right)\left(\nabla\cdot\mathbf{u}\right)^{2}\right|}{\left|\left(\bar{\rho}\bar{c}_{v}+\frac{\bar{\alpha}^{2}\bar{T}}{\bar{\beta}}\right)\left(\frac{\partial T'}{\partial t}+\mathbf{u}\cdot\nabla T\right)\right|}\sim\epsilon,\label{eq:visc_heat_negligible_B}
\end{equation}
allows to conclude that the viscous heating provides a negligible
contribution to the energy balance of Boussinesq systems. On the other hand, the
process of molecular heat transfer has to be included in the energy
balance, since it is a crucial process of temperature relaxation, which
requires that
\begin{equation}
k\sim\epsilon^{-1/2}\frac{\bar{\rho}}{\bar{T}}\left(gL\right)^{3/2}L,\label{eq:e77}
\end{equation}
and the thermal diffusivity\index{SI}{Boussinesq!thermal diffusivity scale}
\begin{equation}
\kappa=\frac{k}{\bar{\rho}\bar{c}_{p}}\sim\epsilon^{1/2}\sqrt{gL}L,\label{eq:e78}
\end{equation}
in consistency with the estimate of viscous diffusivity in (\ref{eq:visc_scale}).
Therefore finally the leading order form of the temperature equation
reads\index{SI}{temperature equation!Boussinesq}
\begin{equation}
\frac{\partial T'}{\partial t}+\mathbf{u}\cdot\nabla T'+u_{z}\left(\frac{\mathrm{d}\dbtilde{T}}{\mathrm{d}z}+\frac{g\bar{\alpha}\bar{T}}{\bar{c}_{p}}\right)=\nabla\cdot\left(\kappa\nabla T'\right)+\frac{Q'}{\bar{\rho}\bar{c}_{p}},\label{Energy_Bderiv-1-1-1}
\end{equation}
where $-g\bar{\alpha}\bar{T}/\bar{c}_{p}$ is the adiabatic gradient,
i.e. temperature gradient corresponding to constant entropy per unit
mass.\footnote{We note, that the final form of the temperature equation (\ref{Energy_Bderiv-1-1-1})
could also be derived directly from the second form of the energy
equation given in (\ref{Energy_eq1}) and the following estimate 
$-\bar{\alpha}\bar{T}\left(\partial_{t}p'+\mathbf{u}\cdot\nabla\dbtilde{p}+\mathbf{u}\cdot\nabla p'\right)=u_{z}g\bar{\rho}\bar{\alpha}\bar{T}+\mathcal{O}\left(\bar{p}\epsilon^{2}\mathscr{U}/L\right)$, obtained by
virtue of the imposed scalings likewise the hydrostatic balance (\ref{eq:Static_NS}).} This completes the derivation of the Boussinesq system of equations
given in (\ref{NS-Bderiv-1-1-1}), (\ref{eq:divu_B}) and (\ref{Energy_Bderiv-1-1-1})
and supplied by the static state equations (\ref{eq:Static_NS}-b)
and (\ref{eq:rho_0_B}). We note, that the isobaric heat capacity
$c_{p}$ for standard laboratory liquids is of the order of $10^{3}\,J/kgK$,
thus for $L\sim1\,m$ and $\left|\mathrm{d}\dbtilde{T}/\mathrm{d}z\right|=\left|\Delta T/L\right|\sim10\,K/m$
one obtains that $gL/\Delta Tc_{p}\sim10^{-3}$ is very small. Taking
into account that typically $\bar{\alpha}\bar{T}\lesssim1$, the term
involving the adiabatic gradient in equation (\ref{Energy_Bderiv-1-1-1})
is typically negligible. However, it is not necessarily the case in
natural convective systems, since e.g. in the Earth's core the adiabatic
gradient is huge, of the order $10^{-4}\,K/m$ and very close to
the static state gradient. This results from the fact, that vigorous
convection is so efficient in transporting heat, that after billions of years the established
temperature gradient which drives the convection
in natural large-scale systems such as planetary and stellar interiors can not
exceed too much the adiabatic gradient.

Finally, it has to be stressed, that the two assumptions stated in
(\ref{eq:epsilon_def_B}) and (\ref{fluct_magnitude_B}) lead to much
smaller kinetic energy $\sim\epsilon\bar{\rho}gL$ than the internal
(or thermal) energy in the system $\sim\epsilon^{-1}\bar{\rho}gL$
with the difference of two orders of magnitude in $\epsilon$. Even
when considering only the varying part of the thermal energy, it is
still of the order of $c_{v}\dbtilde{T}\sim\bar{\rho}gL$, which $\epsilon^{-1}$
times greater than the kinetic energy. An already stated consequence
of this is the negligibility of viscous heating in the energy balance.

For the sake of completeness the entropy equation can now be easily
written down
\begin{equation}
\bar{\rho}\bar{T}\left(\frac{\partial s}{\partial t}+\mathbf{u}\cdot\nabla s\right)=\nabla\cdot\left(k\nabla T'\right)+Q'.\label{Entropy_Bderiv}
\end{equation}
Expanding about the mean state, as in (\ref{eq:Taylor_exp_rho_B})
\begin{equation}
s=\bar{s}-\bar{\alpha}\frac{p-\bar{p}}{\bar{\rho}}+\bar{c}_{p}\frac{T-\bar{T}}{\bar{T}}+\mathcal{O}\left(\epsilon\frac{gL}{\bar{T}}\right),\label{eq:e79}
\end{equation}
where the Maxwell's identity $\rho^{2}(\partial s/\partial p)_{T}=(\partial\rho/\partial T)_{p}=-\rho\alpha$
was used, one obtains by virtue of (\ref{eq:p_scaling_B})
\begin{equation}
\dbtilde{s}=-\frac{\bar{\alpha}}{\bar{\rho}}\dbtilde{p}+\bar{c}_{p}\frac{\dbtilde{T}}{\bar{T}}+\mathcal{O}\left(\epsilon\frac{gL}{\bar{T}}\right),\qquad s'=\bar{c}_{p}\frac{T'}{\bar{T}}+\mathcal{O}\left(\epsilon\frac{gL}{\bar{T}}\right),\label{eq:entropy_temperature_in_Boussinesq}
\end{equation}
so that under the Boussinesq approximation the entropy fluctuation
is the same as the fluctuation of temperature up to a constant factor.
This allows to simplify the entropy equation, at leading order, to
\begin{equation}
\bar{\rho}\bar{T}\left(\frac{\partial s'}{\partial t}+\mathbf{u}\cdot\nabla s'\right)+\bar{\rho}\bar{c}_{p}u_{z}\left(\frac{\mathrm{d}\dbtilde{T}}{\mathrm{d}z}+\frac{g\bar{\alpha}\bar{T}}{\bar{c}_{p}}\right)=\nabla\cdot\left(k\nabla T'\right)+Q'.\label{Entropy_Bderiv-1}
\end{equation}
or
\begin{equation}
\frac{\partial s'}{\partial t}+\mathbf{u}\cdot\nabla s'+\frac{\bar{c}_{p}}{\bar{T}}u_{z}\left(\frac{\mathrm{d}\dbtilde{T}}{\mathrm{d}z}+\frac{g\bar{\alpha}\bar{T}}{\bar{c}_{p}}\right)=\nabla\cdot\left(\kappa\nabla s'\right)+\frac{Q'}{\bar{\rho}\bar{T}}.\label{Entropy_Bderiv-1-1}
\end{equation}

\subsection{Energetic properties of Boussinesq systems\label{subsec:Energetic-properties-of_B}}

Gathering all the necessary equations the complete and closed system
of approximate Boussinesq equations at leading order reads\index{SI}{Boussinesq!equations, general} \begin{subequations}
\begin{equation}
\frac{\partial\mathbf{u}}{\partial t}+\left(\mathbf{u}\cdot\nabla\right)\mathbf{u}=-\frac{1}{\bar{\rho}}\nabla p'+g\bar{\alpha}T'\hat{\mathbf{e}}_{z}+\nu\nabla^{2}\mathbf{u}+2\nabla\nu\cdot\mathbf{G}^{s},\label{eq:NS_B_final}
\end{equation}
\begin{equation}
\nabla\cdot\mathbf{u}=0,\label{eq:Mass_cons_B_final}
\end{equation}
\begin{equation}
\frac{\partial T'}{\partial t}+\mathbf{u}\cdot\nabla T'+u_{z}\left(\frac{\mathrm{d}\dbtilde{T}}{\mathrm{d}z}+\frac{g\bar{\alpha}\bar{T}}{\bar{c}_{p}}\right)=\nabla\cdot\left(\kappa\nabla T\right)+\frac{Q}{\bar{\rho}\bar{c}_{p}},\label{eq:Energy_B_final}
\end{equation}
\end{subequations} 
where the static temperature equation $\nabla\cdot(k\nabla\dbtilde{T})=-\dbtilde{Q}$
has been incorporated back into the energy balance. Let us consider
a plane layer of fluid of thickness $L$, say periodic in the horizontal
directions with periods $L_{x}$, $L_{y}$ and take the $z$-axis
vertical aligned with the gravity $\mathbf{g}=-g\hat{\mathbf{e}}_{z}$.
Applying a horizontal average
\begin{equation}
\left\langle \cdot\right\rangle _{h}=\frac{1}{L_{x}L_{y}}\int_{-L_{x}/2}^{L_{x}/2}\int_{-L_{y}/2}^{L_{y}/2}\left(\cdot\right)\mathrm{d}x\mathrm{d}y,\label{eq:e80}
\end{equation}
to the energy equation one obtains
\begin{equation}
\frac{\partial}{\partial t}\left\langle \bar{\rho}\bar{c}_{p}T'\right\rangle _{h}=-\frac{\partial}{\partial z}\left[\left\langle \bar{\rho}\bar{c}_{p}u_{z}T'\right\rangle _{h}-k\frac{\partial\left\langle T\right\rangle _{h}}{\partial z}\right]+\left\langle Q\right\rangle _{h},\label{eq:e81}
\end{equation}
since horizontal average of the continuity equation (\ref{eq:Mass_cons_B_final})
results in 
\begin{equation}\left\langle u_{z}\right\rangle _{h}=0.\label{mean_uz_null}\end{equation} 
Therefore the
rate of change of the thermal energy is governed by vertical variation
of the total heat flux (and radiative heat sources). A stationary
state requires that
\begin{equation}
\left\langle \bar{\rho}\bar{c}_{p}u_{z}T'\right\rangle _{h}-k\frac{\partial\left\langle T\right\rangle _{h}}{\partial z}-\left\langle Q\right\rangle L=-\left.k\frac{\partial\left\langle T\right\rangle _{h}}{\partial z}\right|_{z=0},\label{eq:stationary_flux_B}
\end{equation}
which was obtained by integration of the stationary equation (\ref{eq:e81})
along $'z'$ from $0$ to $L$, where 
\begin{equation}
\left\langle Q\right\rangle =\frac{1}{L_{x}L_{y}L}\int_{-L_{x}/2}^{L_{x}/2}\int_{-L_{y}/2}^{L_{y}/2}\int_{0}^{L}Q\mathrm{d}z\mathrm{d}y\mathrm{d}x\label{eq:e81andhalf}
\end{equation}
is the radiative heating averaged over the entire periodic fluid domain.
In the absence of heat sources $Q=0$, equation (\ref{eq:stationary_flux_B})
states that the total, horizontally averaged heat flux\index{SI}{heat flux!total} entering the
system at the bottom is the same at every horizontal plane (independent
of height), which is a crucial feature of Boussinesq systems; in particular
the total heat flux entering at the bottom, in a stationary state
is equal to the total flux released through the top boundary,\index{SI}{heat flux balance}
\begin{equation}
-\left.k\frac{\partial\left\langle T\right\rangle _{h}}{\partial z}\right|_{z=0}=-\left.k\frac{\partial\left\langle T\right\rangle _{h}}{\partial z}\right|_{z=L}.\label{eq:e82}
\end{equation}
When the radiative heating is negligible with respect to advection of heat and conduction, i.e. 
\begin{equation}
\left\langle Q\right\rangle L \ll\left\langle \bar{\rho}\bar{c}_{p}u_{z}T'-k\partial_{z}T\right\rangle _{h},\label{eq:e82andhalf}
\end{equation}
the following definition of the Nusselt number, being a measure of
effectiveness of heat transfer by convection seems most practically
useful\index{SI}{Nusselt number!Boussinesq}
\begin{equation}
Nu=\frac{\bar{\rho}\bar{c}_{p}\left\langle u_{z}T\right\rangle _{h}-k\partial_{z}\left\langle T\right\rangle _{h}-\bar{k}g\bar{\alpha}\bar{T}/\bar{c}_{p}}{\left\langle k\Delta_{S}\right\rangle },\label{eq:Nu_def_B}
\end{equation}
where
\begin{equation}
\Delta_{S}=-\frac{\mathrm{d}\dbtilde{T}}{\mathrm{\mathrm{d}z}}-\frac{g\bar{\alpha}\bar{T}}{\bar{c}_{p}}>0\label{eq:e83}
\end{equation}
is the reference temperature gradient excess with respect to the adiabatic
gradient (superadiabatic gradient excess of the static state) and
\begin{equation}
\left\langle \cdot\right\rangle =\frac{1}{L_{x}L_{y}L}\int_{-L_{x}/2}^{L_{x}/2}\int_{-L_{y}/2}^{L_{y}/2}\int_{0}^{L}\left(\cdot\right)\mathrm{d}z\mathrm{d}y\mathrm{d}x,\label{eq:e84}
\end{equation}
is the full spatial average (the same as the upper bar). In such a way the Nusselt number is a ratio of the total superadiabatic heat flux in the convective state to the total superadiabatic heat flux in the hydrostatic reference state. The term $\bar{\rho}\bar{c}_{p}\left\langle u_{z}T\right\rangle _{h}$ represents the flux contribution from advection of temperature whereas $-k\partial_{z}\left\langle T\right\rangle _{h}$ from molecular conduction.  Note, that
the Nusselt number $Nu$ in (\ref{eq:Nu_def_B}) is in principle a
function of $z$ and $t$ and becomes a constant in a stationary state,
due to (\ref{eq:stationary_flux_B}).

\subsubsection{Convection driven by a fixed temperature difference between top and
bottom boundaries, in the absence of radiative heat sources, $Q=0$\label{subsec:fixed_DeltaT}}

Accompanied by the following definition of the Rayleigh number\footnote{When $Q=0$ and $\kappa=\mathrm{const}$ the static temperature distribution
is linear in the vertical coordinate $z$ and the definition of the
Rayleigh number corresponds to the standard one $Ra=g\bar{\alpha}\Delta T_{S}L^{3}/\kappa\nu$,
where $\Delta T_{S}=\Delta T-g\bar{\alpha}\bar{T}L/c_{p}$, $\Delta T=T_{bottom}-T_{top}$
is the temperature difference between bottom and top boundaries, $g\bar{\alpha}\bar{T}L/c_{p}$
expresses such temperature difference in the adiabatic state and in
most experimental situations $\Delta T_{S}\approx\Delta T$, since
the adiabatic gradient is negligible compared to the static one (as
argued below equation (\ref{Energy_Bderiv-1-1-1})).}\index{SI}{Rayleigh number!Boussinesq}
\begin{equation}
Ra=\frac{g\bar{\alpha}\left\langle \kappa\Delta_{S}\right\rangle L^{4}}{\bar{\kappa}^{2}\bar{\nu}},\label{eq:Ra_def_B}
\end{equation}
measuring the relative strength of the buoyancy forces (controlled
by the static state temperature difference between top and bottom
plates) with respect to diffusive effects, the above considerations allow to express the time variation
of the mean kinetic energy (thorough taking the dot-product of the
Navier-Stokes equation and $\mathbf{u}$) in a simple form
\begin{equation}
\frac{\partial}{\partial t}\left(\frac{1}{2}\mathbf{u}^{2}\right)=-\nabla\cdot\left[\left(\frac{1}{2}\mathbf{u}^{2}+\frac{p'}{\bar{\rho}}\right)\mathbf{u}\right]+g\bar{\alpha}T'u_{z}+\mathbf{u}\cdot\left[\nabla\cdot\left(2\nu\mathbf{G}^{s}\right)\right],\label{eq:e85}
\end{equation}
and so\footnote{Note, that the expression for the mean viscous energy dissipation is
the same as that for the mean viscous heating in the energy equation.}
\begin{equation}
\frac{\partial}{\partial t}\left\langle \frac{1}{2}\mathbf{u}^{2}\right\rangle =\frac{\bar{\kappa}^{2}\bar{\nu}}{L^{4}}Ra\left(\frac{1}{L}\int_{0}^{L}Nu(z,t)\mathrm{d}z-1\right)+g\bar{\alpha}\left\langle \kappa\frac{\partial T'}{\partial z}\right\rangle -2\left\langle \nu\mathbf{G}^{s}:\mathbf{G}^{s}\right\rangle .\label{eq:kin_E_variation_B}
\end{equation}
Of course in a stationary state $Nu=\mathrm{const}$ and the first term on the right hand side becomes $\bar{\kappa}^{2}\bar{\nu}Ra(Nu-1)/L^4$. In obtaining (\ref{eq:kin_E_variation_B}) we have used the boundary conditions that the component
of the velocity field normal to the boundaries, likewise either the
tangent viscous stresses or the tangent velocity components at the
boundaries are either null or periodic, i.e. either
\begin{subequations}
\begin{equation}
\left.\mathbf{u}\cdot\mathbf{n}\right|_{\partial V}=0,\label{eq:impermeable}
\end{equation}
\begin{equation}
\left.\mathbf{n}\times\left(2\nu\mathbf{G}^{s}\cdot\mathbf{n}\right)\right|_{\partial V}=0\quad\textrm{or}\quad\left.\mathbf{n}\times\mathbf{u}\right|_{\partial V}=0,\label{eq:stress-free_or_no-slip}
\end{equation}
\end{subequations} 
or the values at the boundaries are periodic (not
necessarily zero), where $\mathbf{n}$ is the normal unit vector at
the boundaries. This implies
\begin{equation}
\left\langle \nabla\cdot\left[\left(\frac{1}{2}\mathbf{u}^{2}+\frac{p'}{\bar{\rho}}\right)\mathbf{u}\right]\right\rangle =0,\quad\left\langle \nabla\cdot\left(2\nu\mathbf{u}\cdot\mathbf{G}^{s}\right)\right\rangle =0,\label{eq:e86}
\end{equation}
and
\begin{equation}
\left\langle \mathbf{u}\cdot\left[\nabla\cdot\left(2\nu\mathbf{G}^{s}\right)\right]\right\rangle =\left\langle \nabla\cdot\left(2\nu\mathbf{u}\cdot\mathbf{G}^{s}\right)\right\rangle -2\left\langle \nu\mathbf{G}:\mathbf{G}^{s}\right\rangle =-2\left\langle \nu\mathbf{G}^{s}:\mathbf{G}^{s}\right\rangle ,\label{eq:e87}
\end{equation}
since the double contraction of the antisymmetric part of the velocity
gradient tensor $G_{ij}=\partial u_{i}/\partial x_{j}$ with the symmetric
part is necessarily zero therefore $\mathbf{G}:\mathbf{G}^{s}=\mathbf{G}^{s}:\mathbf{G}^{s}$.
It follows, that either the tangent stress-free or no-slip boundary
conditions work for derivation of the kinetic energy evolution
equation (\ref{eq:kin_E_variation_B}). Moreover, the mean work of
the buoyancy force per unit mass has been expressed with the aid of
the definitions of the Nusselt (\ref{eq:Nu_def_B}) and Rayleigh numbers
(\ref{eq:Ra_def_B}) in the following way
\begin{align}
\frac{g\bar{\alpha}}{L}\int_{0}^{L}\mathrm{d}z\left\langle T'u_{z}\right\rangle _{h}= & g\bar{\alpha}\left\langle \kappa\Delta_{S}\right\rangle Nu+\frac{g^{2}\bar{\alpha}^{2}\bar{T}\bar{\kappa}}{\bar{c}_{p}}+\frac{g\bar{\alpha}}{L}\int_{0}^{L}\mathrm{d}z\left\langle \kappa\frac{\partial}{\partial z}\left(\dbtilde{T}+T'\right)\right\rangle _{h}\nonumber \\
= & \frac{\bar{\nu}\bar{\kappa}^{2}}{L^{4}}Ra\left(Nu-1\right)+g\bar{\alpha}\left\langle \kappa\frac{\partial T'}{\partial z}\right\rangle ,\label{eq:e88}
\end{align}
since $\kappa\mathrm{d}_{z}\dbtilde{T}$ is uniform (independent of $\mathbf{x}$)
by virtue of the static state balance and we have assumed stationarity, i.e. $Nu=\mathrm{const}$. The last term in the latter
equation is necessarily zero in the case when $\kappa$ is spatially
uniform, since $\left.\left\langle T'\right\rangle \right|_{z=0,L}=0$,
and then equation (\ref{eq:kin_E_variation_B}), in a stationary state
reads
\begin{equation}
0=\frac{\kappa^{2}\bar{\nu}}{L^{4}}Ra\left(Nu-1\right)-2\left\langle \nu\mathbf{G}^{s}:\mathbf{G}^{s}\right\rangle .\label{eq:kin_E_variation_B-1}
\end{equation}
However, if the thermal diffusivity $\kappa$ is depth-dependent this
term is, in general non-zero and in particular in the fully nonlinear,
turbulent regime analysed in section \ref{sec:Fully-developed-convection}
it can be estimated as 
\begin{equation}
\frac{g\bar{\alpha}}{L}\int_{0}^{L}\kappa\partial_{z}\left\langle T'\right\rangle _{h}\mathrm{d}z\sim-\frac{1}{2}g\bar{\alpha}\Delta_{S}(\kappa_{B}+\kappa_{T})=\frac{\bar{\kappa}^{2}\bar{\nu}Ra(\kappa_{B}+\kappa_{T})\Delta_{S}}{2\left\langle \kappa\Delta_{S}\right\rangle L^{4}},\label{eq:e89}
\end{equation}
where the subscripts $B$ and $T$ denote values of $\kappa$ taken
at the bottom and top boundaries respectively.

Taking now the full spatial average of the temperature equation\index{SI}{temperature equation!Boussinesq} (\ref{eq:Energy_B_final})
with excluded static state contribution and multiplied by $T'$
\begin{equation}
\frac{\partial}{\partial t}\left\langle \frac{1}{2}T'^{2}\right\rangle +\left\langle \nabla\cdot\left(\frac{1}{2}\mathbf{u}T'^{2}\right)\right\rangle -\left\langle u_{z}T'\Delta_{S}\right\rangle =\left\langle T'\nabla\cdot\left(\kappa\nabla T'\right)\right\rangle.\label{eq:Energy_B_final-2}
\end{equation}
For a stationary state one obtains in a straight forward manner
\begin{equation}
\frac{\partial}{\partial t}\left\langle \frac{1}{2}T'^{2}\right\rangle =0,\quad\left\langle \nabla\cdot\left(\frac{1}{2}\mathbf{u}T'^{2}\right)\right\rangle =0,\label{eq:e90}
\end{equation}
where the latter comes form the assumed periodicity in horizontal
directions and impermeable top and bottom boundaries; also by virtue
of integration by parts and application of fixed temperature boundary
conditions $T'(z=0,\,L)=0$,
\begin{equation}
\left\langle T'\nabla\cdot\left(\kappa\nabla T'\right)\right\rangle =-\left\langle \kappa\left(\nabla T'\right)^{2}\right\rangle .\label{eq:e91}
\end{equation}
Moreover, using the Nusselt number definition (\ref{eq:Nu_def_B})
the last term on the left hand side of (\ref{eq:Energy_B_final-2}),
in a stationary state, can be cast as follows
\begin{eqnarray}
\left\langle u_{z}T'\Delta_{S}\right\rangle  & = & \frac{1}{L}\int_{0}^{L}\mathrm{d}z\Delta_{S}\left[\left\langle \kappa\Delta_{S}\right\rangle Nu+\left\langle \kappa\frac{\partial}{\partial z}\left(\dbtilde{T}+T'\right)\right\rangle _{h}+\frac{g\bar{\alpha}\bar{T}\bar{\kappa}}{\bar{c}_{p}}\right]\nonumber \\
 & = & \left\langle \kappa\Delta_{S}\right\rangle \left\langle \Delta_{S}\right\rangle Nu+\left\langle \Delta_{S}\right\rangle \frac{g\bar{\alpha}\bar{T}\bar{\kappa}}{\bar{c}_{p}}+\left\langle \Delta_{S}\kappa\frac{\mathrm{d}\dbtilde{T}}{\mathrm{d}z}\right\rangle +\left\langle \Delta_{S}\kappa\frac{\partial T'}{\partial z}\right\rangle \nonumber \\
 &  & \left\langle \kappa\Delta_{S}\right\rangle \left\langle \Delta_{S}\right\rangle \left(Nu-1\right)+\left\langle \Delta_{S}\kappa\frac{\partial T'}{\partial z}\right\rangle \label{eq:e92}
\end{eqnarray}
where, again, we have used $\kappa\mathrm{d}_{z}\dbtilde{T}=\mathrm{const}$,
a consequence of the static state balance. Therefore a stationary
convective state implies
\begin{equation}
\left\langle \kappa\left(\nabla T'\right)^{2}\right\rangle =\left\langle \kappa\Delta_{S}\right\rangle \left\langle \Delta_{S}\right\rangle \left(Nu-1\right)+\left\langle \Delta_{S}\kappa\frac{\partial T'}{\partial z}\right\rangle ,\label{eq:mean_gradT_squared_B}
\end{equation}
and hence\index{SI}{Nusselt number!Boussinesq}
\begin{equation}
Nu=\frac{\left\langle \kappa\left(\nabla T'\right)^{2}\right\rangle -\left\langle \Delta_{S}\kappa\frac{\partial T'}{\partial z}\right\rangle }{\left\langle \kappa\Delta_{S}\right\rangle \left\langle \Delta_{S}\right\rangle }+1,\label{eq:Nu_by_grad_squared}
\end{equation}
In cases, when the thermal diffusivity $\kappa$ can be considered
uniform we get $\left\langle \Delta_{S}\kappa\frac{\partial T'}{\partial z}\right\rangle =\Delta_{S}\kappa\left\langle \frac{\partial T'}{\partial z}\right\rangle =0$
and the above relation simplifies to
\begin{equation}
Nu=\frac{\kappa\left\langle \left(\nabla T'\right)^{2}\right\rangle }{\kappa\Delta_{S}^{2}}+1,\label{eq:thermal_diss_vs_Nusselt}
\end{equation}
an expression for the Nusselt number in terms of a quadratic form
in $\nabla T'$ - the heat flow stimulus. 

Finally, by virtue of the
Nusselt number estimate $Nu\leq\sqrt{Ra}/4-1$ valid for $Ra\geq64$,
derived in Doering and Constantin (1996) for Boussinesq equations
with $Q=0$, $\mu=\mathrm{const},$ $k=\mathrm{const}$ and neglection
of the adiabatic gradient with respect to $\Delta T/L$ one obtains
\begin{equation}
\kappa\left\langle \left(\nabla T'\right)^{2}\right\rangle \leq\left(\frac{\sqrt{Ra}}{4}-2\right)\kappa\Delta_{S}^{2}.\label{eq:Nu_est_1}
\end{equation}
An alternative upper bound on the Nusselt number under the same assumptions
$Q=0$, $\mu=\mathrm{const},$ $k=\mathrm{const}$ and neglection
of the adiabatic gradient, valid for $Ra\gg1$ was obtained by Nobili
(2015) and Choffrut \emph{et al}. (2016), which we recall here
\begin{equation}
Nu=\frac{\kappa\left\langle \left(\nabla T'\right)^{2}\right\rangle }{\kappa\Delta_{S}^{2}}+1\lesssim\begin{cases}
Ra^{1/3}\left(\ln Ra\right)^{1/3} & \textrm{ for }Pr\geq Ra^{1/3}\left(\ln Ra\right)^{1/3}\\
Ra^{1/2}\left(\frac{\ln Ra}{Pr}\right)^{1/2} & \textrm{ for }Pr\leq Ra^{1/3}\left(\ln Ra\right)^{1/3}
\end{cases}.\label{eq:Nu_est_2}
\end{equation}
The estimates (\ref{eq:Nu_est_1}) and (\ref{eq:Nu_est_2}), however,
are also valid in the same form when the adiabatic gradient is included
in the equations with definitions of the Nusselt and Rayleigh numbers
as in (\ref{eq:Nu_def_B}) and (\ref{eq:Ra_def_B}). This is easily seen, since inclusion of the adiabatic gradient in the driving parameter $\Delta_S$
in the case of uniform fluid properties and absence of the radiative
sources can be effectively interpreted as a decrease of the temperature
difference $\Delta T\rightarrow\Delta T-g\bar{\alpha}\bar{T}L/\bar{c}_{p}$.

\subsubsection{Convection driven by a fixed heat flux at the boundaries, in the
absence of radiative heat sources, $Q=0$\label{subsec:fixed_flux}}

The case when the convection is driven by a fixed heat flux at the
boundaries requires $\left.\partial_{z}T'\right|_{z=0,L}=0$, therefore
in the absence of radiative heat sources the equation (\ref{eq:stationary_flux_B})
corresponding to a stationary state, takes the form
\begin{equation}
\left\langle \bar{\rho}\bar{c}_{p}u_{z}T'\right\rangle _{h}-k\frac{\partial\left\langle T\right\rangle _{h}}{\partial z}=-\left.k\frac{\partial\dbtilde{T}}{\partial z}\right|_{z=0}=-k\frac{\partial\dbtilde{T}}{\partial z}=\mathrm{const},\label{eq:stationary_flux_B-1}
\end{equation}
where the static state equation $\partial_{z}(k\partial_{z}\dbtilde{T})=0$
was used, and then the Nusselt number (\ref{eq:Nu_def_B}), in
a stationary state, is simply unity. Furthermore, as a consequence
\begin{equation}
\left\langle \bar{\rho}\bar{c}_{p}u_{z}T'\right\rangle _{h}-k\frac{\partial\left\langle T'\right\rangle _{h}}{\partial z}=0,\label{eq:stationary_flux_B-1-1}
\end{equation}
which implies that in a stationary state the mean advective heat flux
is balanced by the mean fluctuation molecular flux. The definition
of the Rayleigh number (\ref{eq:Ra_def_B}) is sustained\index{SI}{Rayleigh number!Boussinesq}
\begin{equation}
Ra=\frac{g\bar{\alpha}\left\langle \kappa\Delta_{S}\right\rangle L^{4}}{\bar{\kappa}^{2}\bar{\nu}},\label{eq:e93}
\end{equation}
since $\left\langle \kappa\Delta_{S}\right\rangle $ is the heat flux
excess with respect to the average adiabatic flux, which drives the
flow. The stationary mean kinetic energy equation, by the use of (\ref{eq:stationary_flux_B-1-1}),
is now
\begin{equation}
0=g\bar{\alpha}\left\langle \kappa\frac{\partial T'}{\partial z}\right\rangle -2\left\langle \nu\mathbf{G}^{s}:\mathbf{G}^{s}\right\rangle .\label{eq:e94}
\end{equation}
The following new Nusselt number definition, as a measure of advective heat flux
over the superadiabatic static flux turns out useful in description
of convection driven by fixed heat flux on boundaries\index{SI}{Nusselt number!Boussinesq}
\begin{equation}
Nu_{Q}=\frac{\left\langle \bar{\rho}\bar{c}_{p}u_{z}T'\right\rangle }{\left\langle k\Delta_{S}\right\rangle },\label{eq:Nu_Q}
\end{equation}
which allows to express the stationary balance between the work of
the buoyancy and viscous heating in a simple way
\begin{equation}
0=\frac{\bar{\kappa}^{2}\bar{\nu}}{L^{4}}RaNu_{Q}-2\left\langle \nu\mathbf{G}^{s}:\mathbf{G}^{s}\right\rangle .\label{eq:e95}
\end{equation}
The work of the buoyancy force in the case of uniform $\kappa$ is
$g\bar{\alpha}\kappa\Delta\left\langle T'\right\rangle _{h}/L$, with
$\Delta\left\langle T'\right\rangle _{h}$ denoting the difference
of the mean temperature fluctuation between top and bottom boundaries
and then the Nusselt number is simply 
\begin{equation}
Nu_{Q}=\frac{\Delta\left\langle T'\right\rangle _{h}}{\Delta_{S}L}.\label{eq:e95andhalf}
\end{equation}
Furthermore, the equation (\ref{eq:mean_gradT_squared_B}) in the
case of a fixed heat flux at the boundaries takes the form
\begin{equation}
\left\langle \kappa\left(\nabla T'\right)^{2}\right\rangle =\left\langle \Delta_{S}\kappa\frac{\partial T'}{\partial z}\right\rangle ,\label{eq:e96}
\end{equation}
and at uniform $\kappa$ one obtains 
\begin{equation}
\left\langle \Delta_{S}\kappa\frac{\partial T'}{\partial z}\right\rangle =\kappa\Delta_{S}\left\langle \frac{\partial T'}{\partial z}\right\rangle =\kappa\Delta_{S}\frac{\Delta\left\langle T'\right\rangle _{h}}{L}=\kappa\Delta_{S}^{2}Nu_{Q}\label{eq:e96andhalf}
\end{equation}
and therefore
\begin{equation}
Nu_{Q}=\frac{\left\langle \kappa\left(\nabla T'\right)^{2}\right\rangle }{\kappa\Delta_{S}^{2}}.\label{eq:e97}
\end{equation}
Summarizing, the Boussinesq convection, either driven by a fixed temperature
difference or a fixed heat flux at boundaries, is characterized by the thermal energy being
much stronger than the kinetic one with the latter governed by a balance
between the work of the buoyancy force and the viscous dissipation.
The thermal energy remains uninfluenced by those effects at leading order.
Moreover, the total, horizontally averaged heat flux flowing through
the system is constant, i.e. independent of height.

\subsection{Conservation of mass and values of the mean pressure at boundaries\label{subsec:Conservarion-of-mass}}

At the heart of the Boussinesq approximation lies the solenoidal constraint
$\nabla\cdot\mathbf{u}=0$, which implies sound-proof dynamics and
therefore the pressure spreads with infinite velocity. Consequently,
the pressure fluctuation is determined by a Poisson-type, elliptic
equation
\begin{equation}
\nabla^{2}p'=g\bar{\rho}\bar{\alpha}\frac{\partial T'}{\partial z}+\bar{\rho}\nabla\nu\cdot\nabla^{2}\mathbf{u}-\bar{\rho}\nabla\cdot\left[\nabla\cdot\left(\mathbf{u}\mathbf{u}\right)-2\nabla\nu\cdot\mathbf{G}^{s}\right],\label{eq:Mass_cons_pressure_B_1}
\end{equation}
obtained by taking a divergence of the Navier-Stokes equation (\ref{eq:NS_B_final}).
However, in order to fully resolve the dynamics of the equations of
convection (\ref{eq:NS_B_final}-c) one must keep the total mass of
the fluid conserved. This means, that if we assume, that the total
mass is contained in the reference state $\bar{\rho}+\dbtilde{\rho}$,
we must impose\index{SI}{mass conservation}
\begin{equation}
\left\langle \rho'\right\rangle =0\quad\textrm{at all times}.\label{eq:Mass_cons_pressure_B_3}
\end{equation}
The latter is \emph{not} guaranteed by the system of equations (\ref{eq:NS_B_final}-c)
throughout the entire evolution, if only the initial condition is
chosen to satisfy (\ref{eq:Mass_cons_pressure_B_3}), and therefore
constitutes and additional constraint, which must be imposed at every
instant. 

If we further make an additional assumption that the viscosity is
allowed to be a function of $'z'$ only, i.e. $\nu=\nu(z)$, and average
the $z$-component of the Navier-Stokes equation (\ref{eq:NS_B_final})
over the entire periodic domain, substituting $T'=-\rho'/\bar{\rho}\bar{\alpha},$
we get
\begin{equation}
\left\langle p'\right\rangle _{h}(z=L)-\left\langle p'\right\rangle _{h}(z=0)=-gL\left\langle \rho'\right\rangle \label{eq:Mass_cons_pressure_B_2}
\end{equation}
which by (\ref{eq:Mass_cons_pressure_B_3}) implies\index{SI}{pressure boundary conditions}
\begin{equation}
\left\langle p'\right\rangle _{h}(z=L)=\left\langle p'\right\rangle _{h}(z=0)\quad\textrm{at all times}.\label{eq:Mass_cons_pressure_B_4}
\end{equation}
This constitutes a boundary condition, which must be imposed on the
pressure field at every moment in time, used in tandem with the elliptic
equation (\ref{eq:Mass_cons_pressure_B_1}). Let us note, that whether
or not the condition of null pressure fluctuation jump across the
layer is imposed, the convective velocity field remains uninfluenced,
since a shift in pressure fluctuation which is time-dependent only
corresponds to a simple gauge transformation. Nevertheless, the condition
(\ref{eq:Mass_cons_pressure_B_4}) is important in order to fully
resolve the dynamics of the temperature field.

An important consequence of the mass conservation constraint (\ref{eq:Mass_cons_pressure_B_3})
and the Boussinesq relation between density and temperature fluctuations
$\rho'=-\bar{\rho}\bar{\alpha}T'$, is that 
\begin{equation}
\left\langle T'\right\rangle =0\quad\textrm{at all times}\label{eq:Mass_cons_pressure_B_5}
\end{equation}
must also be satisfied throughout the evolution of the system. Therefore
averaging of the temperature equation\index{SI}{temperature equation!Boussinesq} (\ref{eq:Energy_B_final}) over
the entire fluid domain leads to
\begin{equation}
\left.-k\frac{\partial\left\langle T\right\rangle _{h}}{\partial z}\right|_{z=0}+\left.k\frac{\partial\left\langle T\right\rangle _{h}}{\partial z}\right|_{z=L}+\left\langle Q\right\rangle L=0\quad\textrm{at all times},\label{eq:Mass_cons_pressure_B_6}
\end{equation}
so that in the absence of volume heating sources, $Q=0$, the total
heat flux entering the system at the bottom must equal the total heat
flux leaving the system at the top at every instant in time (consistently with (\ref{eq:e82}) obtained for a stationary state).

\subsection{Boussinesq up-down symmetries\label{subsec:Boussinesq-up-down-symmetries}}\index{SI}{Boussinesq!up-down symmetries}

For the purpose of clarity we restate here the Boussinesq approximated equations with uniform thermal and viscous
diffusivity coefficients, $\nu=\mathrm{const}$, $\kappa=\mathrm{const}$,
and no radiative heat sources $Q=0$, in a more explicit
form  
\begin{subequations}
\begin{equation}
\frac{\partial\mathbf{u}}{\partial t}+\left(\mathbf{u}\cdot\nabla_{h}\right)\mathbf{u}+u_{z}\frac{\partial\mathbf{u}}{\partial z}=-\frac{1}{\bar{\rho}}\nabla p'+g\bar{\alpha}T'\hat{\mathbf{e}}_{z}+\nu\left(\nabla_{h}^{2}\mathbf{u}+\frac{\partial^{2}\mathbf{u}}{\partial z^{2}}\right),\label{eq:NS_B_final-1}
\end{equation}
\begin{equation}
\nabla_{h}\cdot\mathbf{u}+\frac{\partial u_{z}}{\partial z}=0,\label{eq:Mass_cons_B_final-1}
\end{equation}
\begin{equation}
\frac{\partial T'}{\partial t}+\mathbf{u}\cdot\nabla_{h}T'+u_{z}\left(\frac{\partial T'}{\partial z}-\Delta_{S}\right)=\kappa\left(\nabla_{h}^{2}T'+\frac{\partial^{2}T'}{\partial z^{2}}\right),\label{eq:Energy_B_final-1}
\end{equation}
\end{subequations} 
where $\nabla_{h}$ denotes the horizontal component
of the $\nabla$, operator\footnote{in Cartesian geometry $\nabla_{h}=(\partial_{x},\,\partial_{y})$}.
This system of equations possesses an up-down symmetry with respect to the mid plane, i.e. under
the transformation
\begin{equation}
z\rightarrow L-z,\label{eq:e98}
\end{equation}
the velocity field components and the temperature and pressure fields
transform in the following way 
\begin{subequations}
\begin{equation}
u_{z}\left(L-z\right)=-u_{z}\left(z\right),\quad\mathbf{u}_{h}\left(L-z\right)=\mathbf{u}_{h}\left(z\right),\label{eq:symmetry_nonlin_1}
\end{equation}
\begin{equation}
T'\left(L-z\right)=-T'\left(z\right),\quad p'\left(L-z\right)=p'\left(z\right).\label{eq:symmetry_nonlin_2}
\end{equation}
\end{subequations} 
In other words, when the top and bottom boundary
conditions are of the same type (e.g. impermeable and either no-slip or stress-free and fixed temperature or fixed heat flux) and the initial conditions satisfy the mid-plane symmetries (\ref{eq:symmetry_nonlin_1},b), the full nonlinear Boussinesq equations
imply, that the vertical velocity and temperature perturbation are
antisymmetric whereas the horizontal velocity and pressure perturbation
are symmetric with respect to the mid-plane\footnote{Physically it is likely, that the initial conditions do not satisfy the mid-plane symmetries (\ref{eq:symmetry_nonlin_1},b), or perturbations are introduced which do not satisfy them. However, the total thermal flux at the top and bottom of the domain needs to be symmetric with respect to mid-plane (cf. (\ref{eq:Mass_cons_pressure_B_6})), thus only the antisymmetric mode of the mean temperature can create/experience heat flux on the boundaries. Nevertheless, the nonlinear interactions between modes can still seed the modes with opposite symmetry at least for some time. Summarizing, the modes satisfying the up-down symmetries (\ref{eq:symmetry_nonlin_1},b) can last infinitely long in the absence of non-symmetric perturbations, whereas the modes with opposite symmetries can not survive alone.}. This will be called the
\emph{nonlinear up-down symmetry} or the \emph{up-down symmetry of
developed convection}. Interestingly, the linearised set of Boussinesq
equations, under the assumption that the convective flow and pressure
and temperature perturbations are weak, i.e. 
\begin{subequations}
\begin{equation}
\frac{\partial\mathbf{u}}{\partial t}=-\frac{1}{\bar{\rho}}\nabla p'+g\bar{\alpha}T'\hat{\mathbf{e}}_{z}+\nu\left(\nabla_{h}^{2}\mathbf{u}+\frac{\partial^{2}\mathbf{u}}{\partial z^{2}}\right),\label{eq:NS_B_final-1-1}
\end{equation}
\begin{equation}
\nabla_{h}\cdot\mathbf{u}+\frac{\partial u_{z}}{\partial z}=0,\label{eq:Mass_cons_B_final-1-1}
\end{equation}
\begin{equation}
\frac{\partial T'}{\partial t}-u_{z}\Delta_{S}=\kappa\left(\nabla_{h}^{2}T'+\frac{\partial^{2}T'}{\partial z^{2}}\right),\label{eq:Energy_B_final-1-1}
\end{equation}
\end{subequations} 
possesses an additional opposite up-down symmetry,
that is 
\begin{subequations}
\begin{equation}
u_{z}\left(L-z\right)=u_{z}\left(z\right),\quad\mathbf{u}_{h}\left(L-z\right)=-\mathbf{u}_{h}\left(z\right),\label{eq:symmetry_lin_1}
\end{equation}
\begin{equation}
T'\left(L-z\right)=T'\left(z\right),\quad p'\left(L-z\right)=-p'\left(z\right),\label{eq:symmetry_lin_2}
\end{equation}
\end{subequations} 
so that the vertical velocity and temperature
perturbation are symmetric whereas the horizontal velocity and pressure
perturbation are antisymmetric with respect to the mid-plane. This
symmetry corresponds to large-scale convective rolls, that is rolls
on the entire scale $L$ of the fluid layer, which set in first in
the weakly overcritical regime, i.e. slightly above convection threshold
(c.f. next section on linear regime). This will be called the \emph{linear
up-down symmetry} or the \emph{up-down symmetry near convection
threshold}.

It can be seen easily, that in the case when both the top and bottom
boundary conditions are the same, as the Rayleigh number keeps increasing
hence the driving force for convection is magnified, the system passes
from a linear regime to a nonlinear one, which can be associated with
a change of the solution symmetry and therefore a radical change in
the form of the flow and temperature distribution. When the boundary
conditions break the up-down symmetry, that is the top and bottom
boundaries are physically different, the solutions in general do not
possess any of the above symmetries.

\section{Linear stability analysis at convection threshold - compendium of
results for different physical conditions\label{sec:Linear-stability-analysis}}\index{SI}{threshold of convection}

The aim of this and two following sections is to describe and explain
the development of the convective instability from its onset, through
the weakly nonlinear stage with various types of convective patterns
to the fully developed turbulent convection. We start with the linear
analysis at convection threshold and derivation of the critical Rayleigh
number for convection onset. Such a linear regime is known as the Rayleigh-B$\acute{\textrm{e}}$nard
problem, which has been thoroughly explained in the books of Chandrasekhar
(1961), Gershuni and Zhukhovitskii (1976), Getling (1998) and it is now considered a part of a very standard knowledge on thermal convection.
Therefore the derivations in the linear regime will be only briefly
recalled here and the results for different types of boundary conditions
with inclusion of the effects of background rotation and radiative
heating will be summarized in a compact way. Attention will be focused
on the discussion of the flow properties at threshold and explanation
of the convective instability on physical grounds. It will be assumed,
that the transport coefficients such as the viscosity $\nu$ and thermal
conductivity $\kappa$, likewise the heat capacity $c_{p}$ are constant.

At the initial stage of convection development, that is close to convection
threshold, the perturbations to the hydrostatic basic state are weak,
\begin{equation} 
\frac{\mathbf{u}}{\epsilon^{1/2}\sqrt{gL}}\ll1,\quad\textrm{and}\quad \frac{T'}{\dbtilde{T}}\ll1,\label{weakness_linear}
\end{equation}
and thus the dynamical equations can be linearised to yield the set
of equations (\ref{eq:NS_B_final-1-1}-c). The linear problem can
be solved in terms of decomposition of the small perturbation fields
into normal, Fourier-type modes
\begin{equation}
\mathbf{u}\left(x,y,z,t\right)=\Re\mathfrak{e}\;\,\hat{\mathbf{u}}\left(z\right)\mathrm{e}^{\sigma t}\mathrm{e}^{\mathrm{i}\left(\mathcal{K}_{x}x+\mathcal{K}_{y}y\right)},\quad T'\left(x,y,z,t\right)=\Re\mathfrak{e}\;\,\hat{T}\left(z\right)\mathrm{e}^{\sigma t}\mathrm{e}^{\mathrm{i}\left(\mathcal{K}_{x}x+\mathcal{K}_{y}y\right)},\label{eq:linear_perturbations_forms}
\end{equation}
where $\sigma$ is the growth rate, thus the system becomes convectively
unstable as soon as there appears at least one mode with $\Re\mathfrak{e}\sigma>0$.
The most general form of the solution at threshold
consists of a superposition of Fourier modes of the type
\begin{equation}
\mathbf{u}\left(x,y,z,t\right)=\Re\mathfrak{e}\underset{\mathcal{K}=\mathcal{K}_{crit}}{\sum_{\boldsymbol{\mathcal{K}}}}\hat{\mathbf{u}}_{\boldsymbol{\mathcal{K}}}\left(z\right)\mathrm{e}^{\sigma_{\boldsymbol{\mathcal{K}}}t}\mathrm{e}^{\mathrm{i}\left(\mathcal{K}_{x}x+\mathcal{K}_{y}y\right)},\label{eq:Fourier_decomposition_linear_B}
\end{equation}
where $\mathcal{K}_{crit}$ is the value of the marginal wave number,
however, the planform of the solution remains unknown within the scope
of linear theory (note, that when the system is homogeneous and isotropic
in the horizontal directions, there is no preferation for any horizontal
direction and the amplitudes $\hat{\mathbf{u}}_{\boldsymbol{\mathcal{K}}}\left(z\right)$
and the growth rates $\sigma_{\boldsymbol{\mathcal{K}}}$ are the
same for each Fourier mode). Taking a double curl of the Navier-Stokes
equation (\ref{eq:NS_B_final-1-1}) allows to separate the problem
for the vertical velocity component and temperature fluctuation from
the equations for horizontal velocity components. The latter can be
simply calculated \emph{a posteriori} once $\hat{u}_{z}(z)$ and $\hat{T}(z)$
are known. We will first consider the case without the radiative heating.
On introducing the forms of the perturbations (\ref{eq:linear_perturbations_forms})
into the equations (\ref{eq:Energy_B_final-1-1}) with $Q=0$ and
the equation for $u_{z}$, obtained from double curl of (\ref{eq:NS_B_final-1-1})
we get 
\begin{subequations}
\begin{equation}
\left[\sigma\left(\frac{\mathrm{d}^{2}}{\mathrm{d}z^{2}}-\mathcal{K}^{2}\right)-\nu\left(\frac{\mathrm{d}^{2}}{\mathrm{d}z^{2}}-\mathcal{K}^{2}\right)^{2}\right]\hat{u}_{z}=-g\bar{\alpha}\mathcal{K}^{2}\hat{T},\label{eq:linear_eqs_Boussinesq_NS}
\end{equation}
\begin{equation}
\left[\sigma-\kappa\left(\frac{\mathrm{d}^{2}}{\mathrm{d}z^{2}}-\mathcal{K}^{2}\right)\right]\hat{T}=\Delta_{S}\hat{u}_{z}.\label{eq:linear_eqs_Boussinesq_Th}
\end{equation}
\end{subequations} 
It can be rigorously shown (cf. Chandrasekhar
1961, \S II.11, pp. 24--26), that the instability in this case occurs as a direct mode with
$\sigma\in\mathbb{R}$ passing through zero. This is often called
the \emph{Principle of the exchange of stabilities}, which is satisfied
in this case. Therefore the marginal state, exactly at threshold,
is characterized by $\sigma=0$, which by the use of (\ref{eq:linear_eqs_Boussinesq_NS},b)
allows to write down the vertical velocity amplitude equation in a
compact form
\begin{equation}
\left(\frac{\mathrm{d}^{2}}{\mathrm{d}z^{2}}-\mathcal{K}^{2}\right)^{3}\hat{u}_{z}=-\frac{Ra}{L^{4}}\mathcal{K}^{2}\hat{u}_{z}.\label{eq:uz_equation_lin_B}
\end{equation}
This equation is subject to boundary conditions and the natural choice
is that the boundaries are isothermal, impermeable and either rigid
or stress-free. Below, we will quickly summarize the results for each
case separately, however, before this, let us first examine the general
expression for the critical superadiabatic gradient at threshold (\ref{eq:general_condition_on_Delta_c});
in the Boussinesq case, cf. equations (\ref{Entropy_Bderiv})-(\ref{Entropy_Bderiv-1-1}),
with constant viscosity, thermal conductivity and heat capacity and
without radiation it is easily reduced to\footnote{The condition (\ref{eq:Boussinesq_Delta_crit}) directly corresponds
to the linearised energy equation (\ref{eq:linear_eqs_Boussinesq_Th})
with $\sigma=0$, where the right hand side corresponds to the full
advective derivative of the entropy per unit mass, since in the Boussinesq
case $s'\approx\bar{c}_{p}T'/\bar{T}$ is equivalent to the temperature
fluctuation up to the constant factor $\bar{c}_{p}/\bar{T}$, and
$\mathrm{d}s_{0}/\mathrm{d}z=-c_{p}\Delta_{S}/\bar{T}.$ }
\begin{equation}
\Delta_{S\,crit}=\kappa\frac{\mathrm{min}\left[-\nabla^{2}T'\right]}{u_{z}},\quad\textrm{at any }z.\label{eq:Boussinesq_Delta_crit}
\end{equation}
By the use of (\ref{eq:linear_eqs_Boussinesq_NS}) we can further
simplify this expression to
\begin{equation}
Ra_{crit}=\frac{L^{4}\mathrm{min}\left[-\frac{1}{\mathcal{K}^{2}}\left(\frac{\mathrm{d}^{2}}{\mathrm{d}z^{2}}-\mathcal{K}^{2}\right)^{3}u_{z}\right]}{u_{z}},\quad\textrm{at any }z\label{eq:Boussinesq_general_condition_for_Rac}
\end{equation}
where in the current case\index{SI}{Rayleigh number!critical}
\begin{equation}
Ra_{crit}=\frac{g\bar{\alpha}\Delta_{S\,crit}L^{4}}{\kappa\nu},\label{eq:e99}
\end{equation}
is the critical Rayleigh number for convection threshold. Thus the
heat per unit mass released by a fluid parcel rising on an infinitesimal
distance $\mathrm{d}z$ in a time unit in the marginal state is
\begin{equation}
-T\frac{\mathrm{D}s}{\mathrm{d}t}=-\bar{c}_{p}\kappa\nabla^{2}T'=\bar{c}_{p}\frac{\kappa\nu}{g\bar{\alpha}L^{4}}Ra_{crit}u_{z},\label{eq:e100}
\end{equation}
which is equal to the total heat per unit mass accumulated between
the infinitesimally distant (by $u_{z}\mathrm{d}t$) horizontal fluid
layers, between which the perturbed parcel travels, per unit time of the rise in the marginal state, $\bar{c}_{p}\Delta_{S\,crit}u_{z}$; when $-T\mathrm{D}s/\mathrm{d}t$
falls below $\bar{c}_{p}\Delta_{S}u_{z}$, the system becomes unstable.
This result will be used to compare the physical nature of the convective
instability trigger between various specific cases, to which we turn
now.

\subsection{Two isothermal, stress-free boundaries, $Q=0$.\label{subsec:Two-isothermal-stress-free}}

In the case of two stress-free boundaries at $z=0,\,L$ the boundary
conditions for the vertical velocity amplitude yield
\begin{equation}
\hat{u}_{z}=0,\quad\frac{\mathrm{d}^{(2m)}\hat{u}_{z}}{\mathrm{d}z^{(2m)}}=0,\label{eq:BCs_Boussinesq_case1}
\end{equation}
for all natural numbers $m\in\mathbb{N}$, where the equations (\ref{eq:linear_eqs_Boussinesq_NS},b)
with $\sigma=0$ (at threshold) and $\hat{T}(z=0,\,L)=0$ were used.
The solution of the problem (\ref{eq:uz_equation_lin_B}) can be sought
in the form of a superposition of modes of the type $\mathrm{Const}_{1}\sin qz+\mathrm{Const}_{2}\cos qz$,
where $q$ corresponds to a set of constant coefficients determined
by the boundary conditions. This allows to conclude, that the only
possible solution in this case takes the form
\begin{equation}
\hat{u}_{z}(z)=A\sin\left(\frac{n\pi z}{L}\right),\label{eq:uz_for_2SF_isoth_Q0}
\end{equation}
where $n\in\mathbb{N}$ and $A$ is a constant amplitude, undetermined
within the scope of linear theory. Substitution of the above form
of solution into equation (\ref{eq:uz_equation_lin_B}) gives $Ra=(n^{2}\pi^{2}+\mathcal{K}^{2}L^{2})^{3}/\mathcal{K}^{2}L^{2}$,
which for a given $\mathcal{K}$ takes the minimal value at $n=1$,
thus
\begin{equation}
Ra=\frac{\left(\pi^{2}+\mathcal{K}^{2}L^{2}\right)^{3}}{\mathcal{K}^{2}L^{2}}.\label{eq:Ra_crit_of_k}
\end{equation}
This already allows to draw a conclusion about the symmetry of the
flow at convection threshold; since $n=1$ at threshold, in the case
of two stress-free boundaries the marginal solutions possesses the \emph{linear
up-down symmetry} defined in (\ref{eq:symmetry_lin_1},b), which corresponds
to large-scale rolls. Minimization of the expression (\ref{eq:Ra_crit_of_k})
over all possible values of $\mathcal{K}$ leads to the critical Rayleigh
number for development of convective instability in the current case,
\begin{equation}
Ra_{crit}=\frac{27}{4}\pi^{4}\approx657.5,\quad\textrm{achieved at}\quad\mathcal{K}_{crit}=\frac{\pi}{\sqrt{2}L}.\label{eq:e101}
\end{equation}
The ratio of the horizontal to vertical thickness of a single convective
roll, as depicted on figure \ref{fig:compendium_of_LinBsq}a, is $\pi/\mathcal{K}_{crit}L=\sqrt{2},$
thus the rolls are slightly flattened. The result for $Ra_{crit}$
can be, of course, obtained from (\ref{eq:Boussinesq_general_condition_for_Rac})
and (\ref{eq:uz_for_2SF_isoth_Q0}) in a straightforward way.
\begin{figure}
\begin{centering}
a)\includegraphics[scale=0.17]{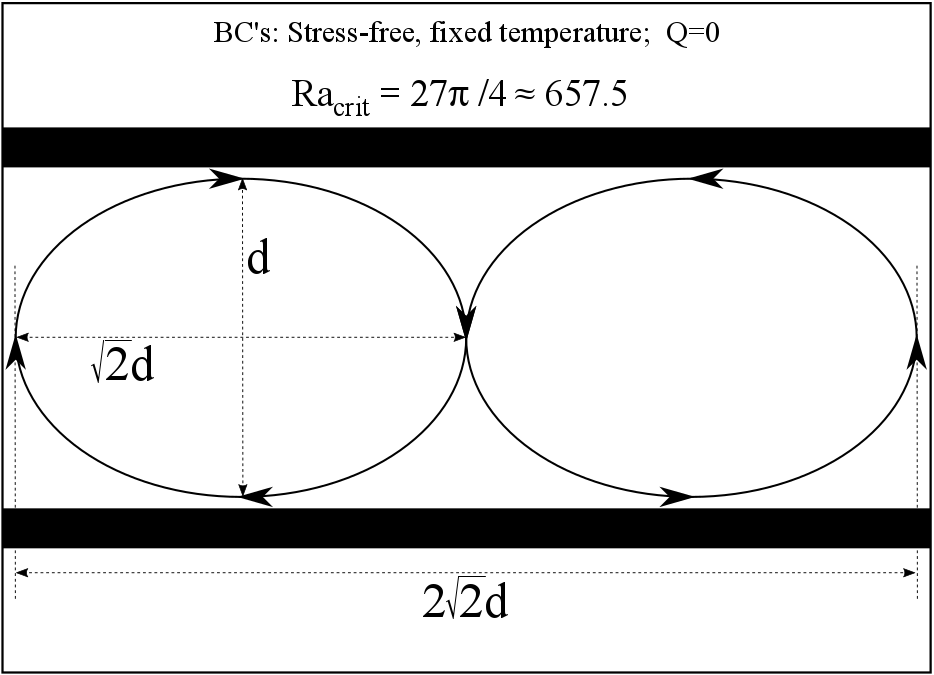}~~b)\includegraphics[scale=0.17]{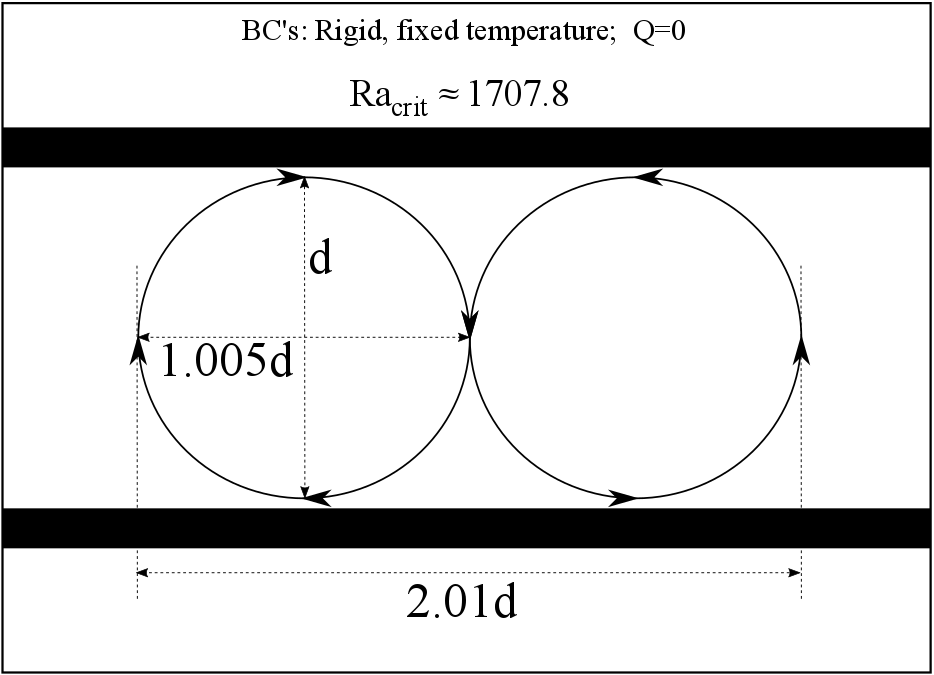}
\par\end{centering}
\begin{centering}
~
\par\end{centering}
\begin{centering}
~
\par\end{centering}
\begin{centering}
c)\includegraphics[scale=0.17]{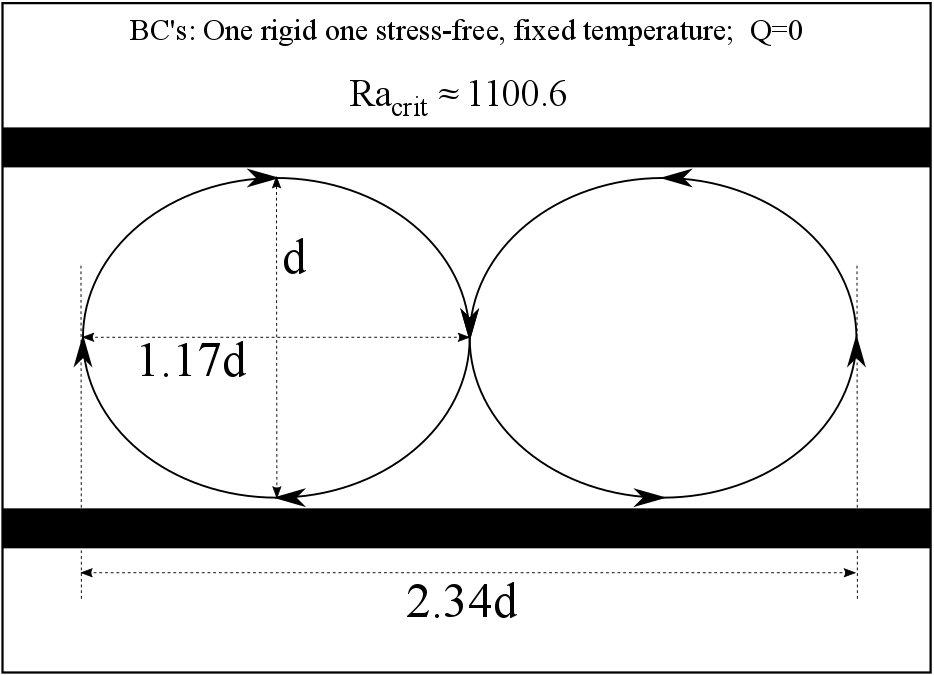}~~d)\includegraphics[scale=0.17]{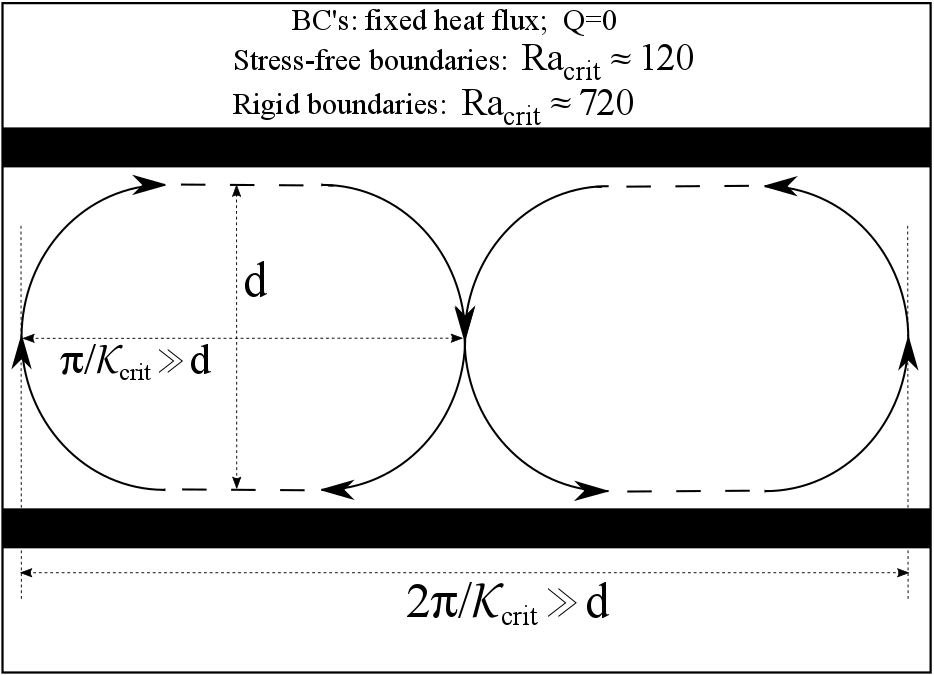}
\par\end{centering}
\begin{centering}
~
\par\end{centering}
\begin{centering}
~
\par\end{centering}
\begin{centering}
e)\includegraphics[scale=0.17]{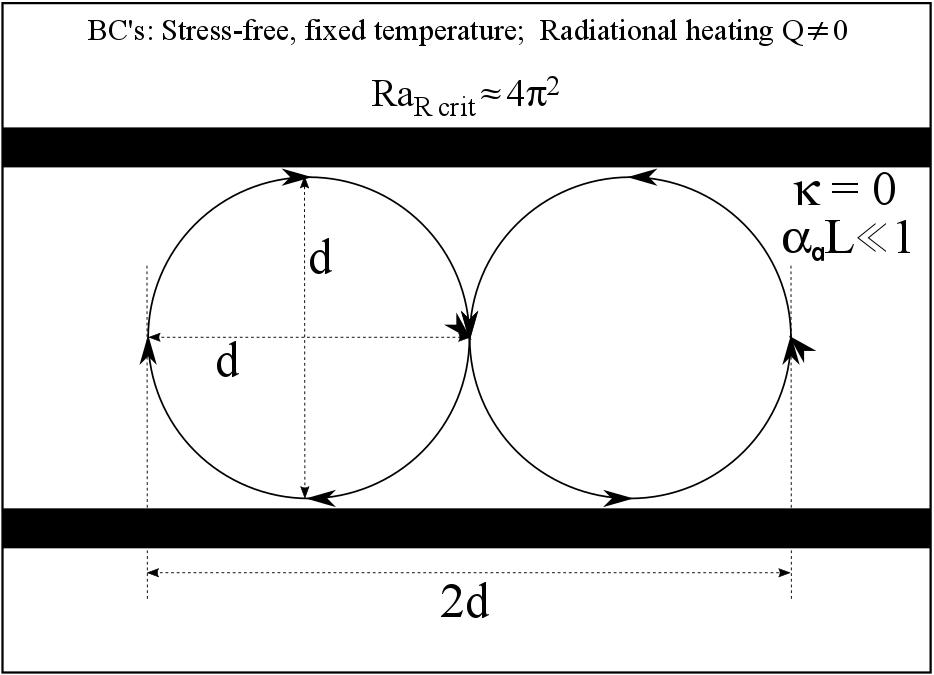}~~f)\includegraphics[scale=0.17]{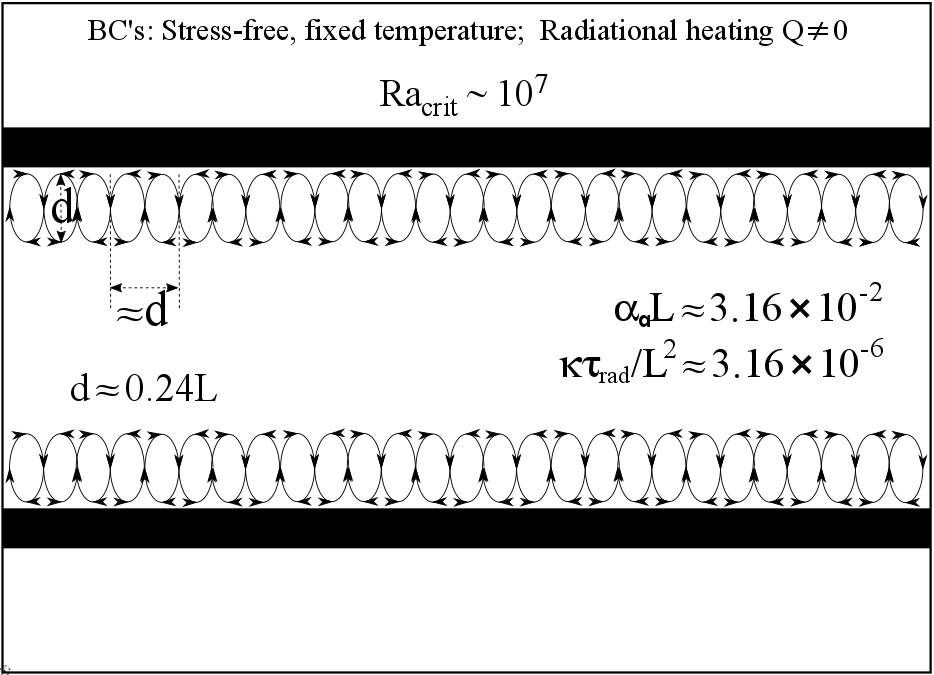}
\par\end{centering}
\begin{centering}
~
\par\end{centering}
\begin{centering}
~
\par\end{centering}
\begin{centering}
g)\includegraphics[scale=0.17]{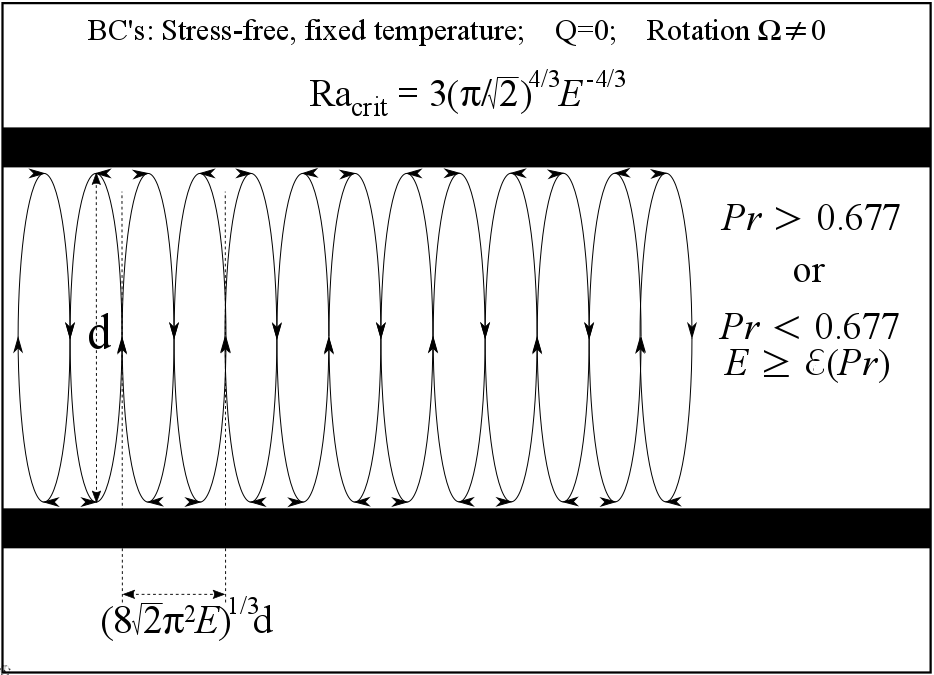}~~h)\includegraphics[scale=0.17]{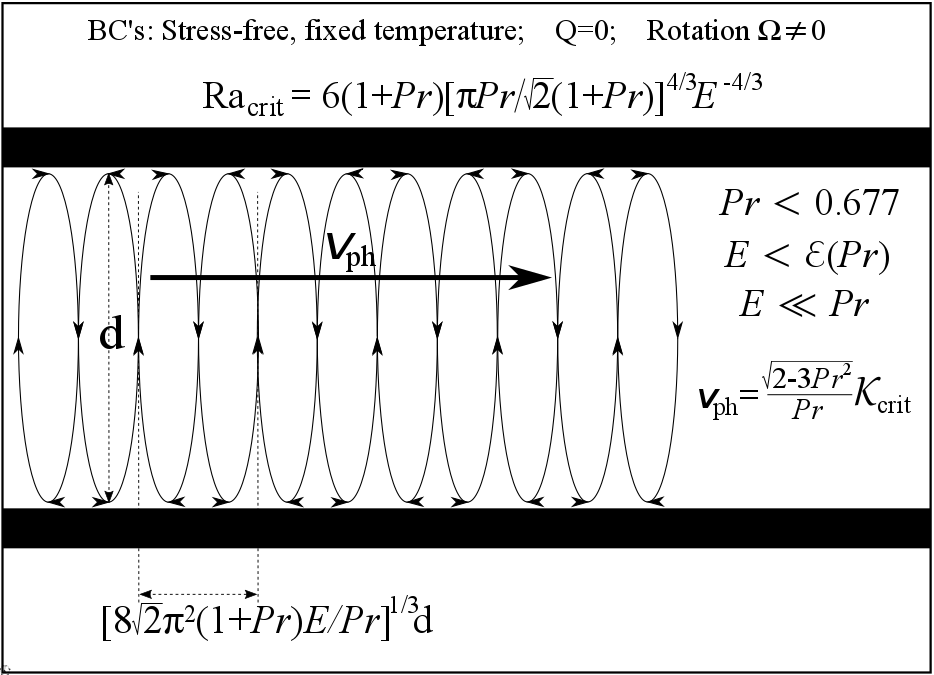}
\par\end{centering}
~

~

~
\centering{}\emph{\footnotesize{}figure caption on next page $\Longrightarrow$}{\footnotesize \par}
\end{figure}
 
\begin{figure}
\centering{}\caption{{\footnotesize{}\label{fig:compendium_of_LinBsq}Schematic picture
of convection rolls in the marginal state for Boussinesq convection
in 8 different physical situations: a) stress-free, isothermal boundaries,
$Q=0$; b) rigid, isothermal boundaries, $Q=0$; c) one rigid one
stress-free boundary, both isothermal, $Q=0$; d) heat flux held fixed
at the boundaries (either rigid or stress-free), $Q=0$ - the rolls
are strongly elongated in this case; e) and f) demonstrate the possible
effects of radiational heating $Q\protect\neq0$ for $\kappa=0$ and
$\kappa\protect\neq0$ when the boundaries are isothermal and stress-free;
g) and h) demonstrate the effect of background rotation for isothermal
and stress-free boundaries and $Q=0$ for large and small Prandtl
numbers, $Pr=\nu/\kappa$. The vertical cross-sections along the plane determined by
the $z$-axis and the horizontal wave vector $\boldsymbol{\mathcal{K}}$
are shown.}}
\end{figure}

\subsection{Two isothermal rigid boundaries, $Q=0$.\label{subsec:Two-isothermal-rigid}}

Next we consider the case of two rigid boundaries with no-slip conditions
at $z=0,\,L$
\begin{equation}
\hat{u}_{z}=0,\quad\frac{\mathrm{d}\hat{u}_{z}}{\mathrm{d}z}=0.\label{eq:BCs_rigrig}
\end{equation}
Because the boundaries are assumed isothermal the equation (\ref{eq:linear_eqs_Boussinesq_NS})
implies additionally
\begin{equation}
\left(\frac{\mathrm{d}^{2}}{\mathrm{d}z^{2}}-\mathcal{K}^{2}\right)^{2}\hat{u}_{z}=0,\label{eq:BCs_rigrig_1}
\end{equation}
at $z=0,\,L$. On introducing the general form of the solution $\mathrm{Const}_{1}\sin qz+\mathrm{Const}_{2}\cos qz$
into the equation (\ref{eq:uz_equation_lin_B}) it can be easily seen,
that the boundary conditions (\ref{eq:BCs_rigrig}) and (\ref{eq:BCs_rigrig_1})
imply that there are two distinct classes of solutions, one satisfying
the ``linear'' up-down symmetry (\ref{eq:symmetry_lin_1},b) and
second one obeying the ``developed'' symmetry (\ref{eq:symmetry_nonlin_1},b).
As demonstrated by Chandrasekhar (1961) the marginal state corresponds
to the large-scale flow, with $\hat{u}_{z}(z)$ and $\hat{T}(z)$
even with respect to the mid-plane\footnote{This is clear from a general approach based on expansion of $\hat{u}_{z}(z)$
and $\hat{T}(z)$ in eigenfunctions of the problem (\ref{eq:linear_eqs_Boussinesq_NS},b)
with boundary conditions either of Dirichlet or Neumann type, which
can be easily shown to form a complete set. It was demonstrated in
Chandrasekhar (1961), chapter 2, section 13(a), pp. 27-31, that the
eigen mode of the lowest order, that is with the largest possible
wavelength in the vertical direction corresponds to the lowest $Ra$-the
eigenvalue of the problem. When the top and bottom boundaries are
symmetric, this mode must be symmetric with respect to the mid-plane.}; the $z$-dependent amplitude of the vertical velocity $\hat{u}_{z}$
at threshold takes the form
\begin{align}
\hat{u}_{z}\left(z\right)= & A\cos\left(q_{0}\frac{2z-L}{2L}\right)-A\,0.0615\cosh\left(q_{1}\frac{2z-L}{2L}\right)\cos\left(q_{2}\frac{2z-L}{2L}\right)\nonumber \\
 & +A\,0.1039\sinh\left(q_{1}\frac{2z-L}{2L}\right)\sin\left(q_{2}\frac{2z-L}{2L}\right),\label{eq:uz_rigid_rigid_B}
\end{align}
where
\begin{equation}
q_{0}\approx3.9736,\quad q_{1}\approx5.1952,\quad q_{2}\approx2.1261,\label{eq:e102}
\end{equation}
and $A$ is an undetermined constant. The critical Rayleigh number
and the wave number of the marginal mode in the case of two rigid
and isothermal boundaries are
\begin{equation}
Ra_{crit}\approx1707.76,\qquad\mathcal{K}_{crit}=\frac{3.12}{L},\label{eq:Ra_crit_2rigid_2isoth}
\end{equation}
and hence the horizontal to vertical single roll thicknesses ratio is $\pi/\mathcal{K}_{crit}L\approx1.01$,
i.e. the rolls are very close to circular (cf. figure \ref{fig:compendium_of_LinBsq}b).
In this case the heat per unit mass released by a fluid parcel rising
on an infinitesimal distance $\mathrm{d}z$ in a time unit in the
marginal state (which is equal to the maximal heat per unit mass,
per unit time that can be accumulated between fluid layers before
convection starts), that is $\bar{c}_{p}\kappa\nu Ra_{crit}u_{z}/g\bar{\alpha}L^{4}$,
is higher than in the previous case. This is because the viscous stresses
on the rigid walls do not allow the simple-type solutions $\hat{u}_{z}\sim\sin(n\pi z)$
to develop and the allowed solutions have a more complex structure
(\ref{eq:uz_rigid_rigid_B}), which turns out to require higher superadiabatic
temperature gradients.

\subsection{One rigid and one stress-free boundary, both isothermal, $Q=0$.\label{subsec:One-rigid-one_SF_isothermal}}

As remarked above, in the case of two rigid and isothermal boundaries
there is a class of solutions with odd symmetry of $\hat{u}_{z}(z)$
and $\hat{T}(z)$ with respect to the mid-plane. These solutions necessarily
satisfy the stress-free conditions $\hat{u}_{z}=0$, $\mathrm{d}_{z}^{2}\hat{u}_{z}=0$
and $(\mathrm{d}_{z}^{2}-\mathcal{K}^{2})^{2}\hat{u}_{z}=0$ at $z=L/2$,
thus the solution for the current case of one rigid and one stress-free
boundary, both isothermal can be retrieved from the odd solution for
the previous case, provided by Chandrasekhar (1961). The critical
parameter values for convection threshold do not depend on whether
the top boundary is assumed stress free and the bottom one rigid or
the other way around. In the former case the solution takes the form
\begin{align}
\hat{u}_{z}\left(z\right)= & A\sin\left(q_{0}\frac{z-L}{2L}\right)-A\,0.0171\sinh\left(q_{1}\frac{z-L}{2L}\right)\cos\left(q_{2}\frac{z-L}{2L}\right)\nonumber \\
 & +A\,0.00346\cosh\left(q_{1}\frac{z-L}{2L}\right)\sin\left(q_{2}\frac{z-L}{2L}\right),\label{eq:e103}
\end{align}
where
\begin{equation}
q_{0}\approx7.1379,\quad q_{1}\approx9.1108,\quad q_{2}\approx3.7893.\label{eq:e104}
\end{equation}
and $A$ is an undetermined constant. The critical Rayleigh number
and the wave number of the marginal mode are in this case
\begin{equation}
Ra_{crit}\approx1100.65,\qquad\mathcal{K}_{crit}=\frac{2.68}{L},\label{eq:e105}
\end{equation}
and hence the horizontal to vertical single roll thicknesses ratio is $\pi/\mathcal{K}_{crit}L\approx1.17$,
i.e. the rolls are only slightly flattened (cf. figure \ref{fig:compendium_of_LinBsq}c).
In terms of the value of the critical Rayleigh number, it is perhaps
no surprise, that this case lies in between the two previous cases.

There are, however, other types of boundary conditions and important
physical effects, that have been widely considered due to relevance
to particular physical settings, which have a profound effect on the
stability characteristics and structure of the flow at convection
threshold. E.g. instead of fixing the temperature at the top and bottom
boundary the heat flux can be fixed, which is a condition well corresponding
to situations found in astrophysics, where the heat flux at interfaces
between layers in planetary and stellar interiors can be reasonably
considered fixed for long periods of time. Another important effect
in planetary mantles and stellar interiors is the radiative heating.
We will first show the influence of the boundaries with fixed heat
flux on the onset of convection. 

\subsection{Two stress-free, fixed-thermal-flux boundaries, $Q=0$.\label{subsec:Two-stress-free-fixed-flux}}

The case when top and bottom boundaries are stress-free and the heat
flux on both of them is fixed, that is
\begin{equation}
\left.\frac{\partial T'}{\partial z}\right|_{z=0,L}=0,\label{eq:thermally_insulating_BCs}
\end{equation}
has been thoroughly investigated and the linear analysis can be found
e.g. in Jakeman (1968) and Park and Sirovich (1991). The convection
threshold in this case is achieved at 
\begin{equation}
Ra_{crit}=120,\qquad\mathcal{K}_{crit}=0,\label{eq:e106}
\end{equation}
with the trivial solution 
\begin{equation}
\hat{u}_{z}=0,\quad\textrm{and}\quad\hat{T}=\mathrm{const}.\label{eq:e107}
\end{equation}
This means, that when the Rayleigh number only slightly exceeds the
critical value, convection develops in the form of very wide (strongly
flattened) large-scale roll; since the imposed boundary conditions
are symmetric with respect to $z=L/2$, $\hat{u}_{z}(z)$ and $\hat{T}(z)$
just above the threshold possess the even symmetry with respect to
the mid-plane (linear up-down symmetry), as depicted on figure \ref{fig:compendium_of_LinBsq}d.
The condition of fixed thermal flux physically corresponds to a situation,
when the thermal conductivity of the bounding solid walls at $z=0,\,L$
is much smaller than that of the fluid (as opposed to the isothermal
walls, which can be physically realized when their thermal conductivity
is much larger than the fluid's, to allow for quick temperature relaxation).
The pathological structure of the critical solution can be interpreted
in the context of physically realizable situations, when the thermal
conductivity of the solid boundaries is always finite, which again,
leads to large-scale rolls of very long wavelength at the threshold and/or in its vicinity.

Despite the triviality of the marginal solution (\ref{e107}), the situation in this case does not differ very significantly
from the previous cases in terms of general qualitative description of fluid's behaviour,
since the amplitude of convection is known to scale with a positive
power of departure from convection threshold, $Ra-Ra_{crit}$ (see
next section \ref{sec:WNT} for weakly nonlinear estimates), thus
exactly at threshold there are no motions in all the considered cases.

Furthermore, when the superadiabatic gradient is slightly above the
threshold value, which in the current case implies $\mathcal{K}\ll1$,
the heat per unit mass, per unit time released by a rising fluid parcel
on an infinitesimal vertical distance, which is equal to $\bar{c}_{p}\kappa\nu Ra_{crit}u_{z}/g\bar{\alpha}L^{4}$,
is significantly smaller than in the cases with isothermal boundaries,
(and at the same time also smaller than the heat per unit mass
per unit time accumulated between two horizontal planes at distance
$u_{z}\mathrm{d}t$, therefore convective flow is triggered). This
means that when the boundaries are held at fixed thermal flux, to
drive a convection it is necessary to accumulate much less heat between
horizontal fluid layers by rising the superadiabatic gradient, than
in the cases when the boundaries are held at fixed temperatures. The
reason for this, is that in the case at hand the heat excess coming
from perturbations cannot leave the system through boundaries thus
it is easier to achieve the critical heat balance (\ref{eq:general_marginal_condition})
at convection threshold as opposed to the previous cases when the
heat could freely leave through isothermal boundaries. 

\subsection{Two rigid, fixed-thermal-flux boundaries, $Q=0$.\label{subsec:Two-rigid-fixed-flux}}

The next case is again symmetric about the mid-plane with the heat
flux fixed at both top and bottom boundaries, but this time both these
boundaries are rigid (boundary conditions (\ref{eq:thermally_insulating_BCs})
and (\ref{eq:BCs_rigrig})). This problem has been addressed e.g. in
Sparrow \emph{et al}. (1964), Jakeman (1968), Gershuni and Zhukhovitskii
(1976) and Cerisier \emph{et al}. (1998), who showed, that for such a
choice of boundary conditions the critical Rayleigh number and the
associated wave number for convection are
\begin{equation}
Ra_{crit}\approx720,\qquad\mathcal{K}_{crit}=0,\label{eq:Ra_crit_2rigid_2insul}
\end{equation}
with the trivial solution
\begin{equation}
\hat{u}_{z}=0,\quad\textrm{and}\quad\hat{T}=\mathrm{const},\label{eq:e108}
\end{equation}
thus, again, slightly above the threshold the temperature gradients
are very small, $\hat{u}_{z}(z)$ and $\hat{T}(z)$ are symmetric
with respect to the mid-plane (linear up-down symmetry) and the flow
is organized in a pattern with large horizontal wavelength (cf. figure
\ref{fig:compendium_of_LinBsq}d). As expected, the rigid boundaries
make the critical balance (\ref{eq:general_marginal_condition}) for
convection threshold harder to achieve. Still, at the same time $Ra_{crit}$
in the current case (\ref{eq:Ra_crit_2rigid_2insul}) is much smaller
than in the case when both boundaries are isothermal and rigid (\ref{eq:Ra_crit_2rigid_2isoth}),
indicating the influence of fixed-thermal-flux boundaries, which keep
the heat excess resulting from fluctuations in the system.

We now proceed to study the influence of two different physical
effects, very common in nature, such as radiative heating and background
rotation on the threshold of convection. We start
with the effect of radiation.

\subsection{The effect of radiative heating, $Q\protect\neq0$, at two stress-free,
isothermal boundaries.\label{subsec:The-effect-of-RDH}}\index{SI}{radiative!heating}

We start by introducing the model of thermal radiation following Goody
(1956), Goody and Yung (1989) and Goody (1995)\footnote{The details of the linear stability analysis for convective systems
with radiative heating can be found in Goody (1956), Getling (1998),
Goody and Yung (1989) and Goody (1995), with perhaps most comprehensive
treatment by Getling (1980) and Larson (2001).}. Denoting the radiative energy flux by $\mathbf{j}_{rad}$ we can
express the heat per unit volume delivered to the system in a time
unit by thermal radiation as
\begin{equation}
Q=-\nabla\cdot\mathbf{j}_{rad},\label{eq:e109}
\end{equation}
with the following equation governing $\mathbf{j}_{rad}$
\begin{equation}
\nabla\frac{1}{\alpha_{a}}\nabla\cdot\mathbf{j}_{rad}-3\alpha_{a}\mathbf{j}_{rad}=4\sigma_{rad}\nabla\left(T^{4}\right),\label{eq:radiative_flux_governing_eq}
\end{equation}
where $\alpha_{a}$ is the coefficient of absorption of radiation
per unit volume and $\sigma_{rad}$ is the Stefan-Boltzmann constant.
The absorption coefficient, just as $\mu$, $\kappa$ and $\bar{c}_{p}$,
will be assumed uniform in what follows. Let us first linearise the
thermal source term on the right hand side of (\ref{eq:radiative_flux_governing_eq}),
taking into account, that under the Boussinesq approximation $T=\bar{T}+\dbtilde{T}+T'$
and $(\dbtilde{T}+T')/\bar{T}=\mathcal{O}\left(\epsilon\right)\ll1$
according to (\ref{eq:T_scaling_B}), which yields
\begin{equation}
\nabla T^{4}=4\bar{T}^{3}\nabla\left(\dbtilde{T}+T'\right)+\mathcal{O}\left(\epsilon\frac{\bar{T}^{4}}{L}\right),\label{eq:e110}
\end{equation}
and hence the equation (\ref{eq:radiative_flux_governing_eq}) simplifies
to
\begin{equation}
\nabla\left(\nabla\cdot\mathbf{j}_{rad}\right)-3\alpha_{a}^{2}\mathbf{j}_{rad}=16\alpha_{a}\sigma_{rad}\bar{T}^{3}\nabla\left(\dbtilde{T}+T'\right).\label{eq:radiative_flux_governing_eq-1}
\end{equation}
For the basic, hydrostatic state dependent on $z$ alone, $\tilde T(z) = \bar T + \dbtilde{T}(z)$ and $\dbtilde{\mathbf{j}}_{rad}(z)$ (note that $\bar{\mathbf{j}}_{rad}=0$), the latter equation and the energy equation (\ref{eq:Energy_B_final})
take the form 
\begin{subequations}
\begin{equation}
\kappa\frac{\mathrm{d}^{2}\dbtilde{T}}{\mathrm{d}z^{2}}-\frac{1}{\bar{\rho}\bar{c}_{p}}\frac{\mathrm{d}\dbtilde{j}_{rad\,z}}{\mathrm{d}z}=0,\label{eq:T_basic_RadB}
\end{equation}
\begin{equation}
\frac{\mathrm{d}^{2}\dbtilde{j}_{rad\,z}}{\mathrm{d}z^{2}}-3\alpha_{a}^{2}\dbtilde{j}_{rad\,z}=16\alpha_{a}\sigma_{rad}\bar{T}^{3}\frac{\mathrm{d}\dbtilde{T}}{\mathrm{d}z},\label{eq:j_basic_radB}
\end{equation}
\begin{equation}
\dbtilde{j}_{rad\,x}=\dbtilde{j}_{rad\,y}=0.\label{eq:jxjy_basic_rad_B}
\end{equation}
\end{subequations} 
Furthermore, let us consider the limit of a transparent
fluid, when the absorption coefficient is very small, $\alpha_{a}\ll L^{-1}$.
In such a case the term quadratic in $\alpha_{a}^{2}$ can be neglected
in the equation (\ref{eq:radiative_flux_governing_eq-1}). This allows
for a simplification of the equation for the fluctuation of the radiative
flux $\mathbf{j}_{rad}^{\prime}$, by integration of the equation
once with respect to $\mathbf{x}$,
\begin{equation}
\nabla\cdot\mathbf{j}_{rad}^{\prime}\approx16\alpha_{a}\sigma_{rad}\bar{T}^{3}T',\label{eq:divj_vs_T_linear_B_rad}
\end{equation}
where the constant of integration is zero, since when the temperature
perturbation vanishes $T'=0$ the perturbation to the radiative heat
transfer must vanish as well. Hence the linearised thermal energy
equation can now be written solely in terms of the temperature fluctuation
\begin{equation}
\frac{\partial T'}{\partial t}-u_{z}\Delta_{S}=\kappa\nabla^{2}T'-16\alpha_{a}\sigma_{rad}\bar{T}^{3}T'.\label{eq:e111}
\end{equation}
This equation together with (\ref{eq:NS_B_final-1-1},b) are subject
to boundary conditions, which will now be specified. We apply the
Fourier decomposition as in (\ref{eq:Fourier_decomposition_linear_B}),
thus the rigid, isothermal boundaries imply for the fluctuations
\begin{equation}
\hat{u}_{z}(z=0,\,L)=0,\quad\left.\frac{\partial^{2}\hat{u}_{z}}{\partial z^{2}}\right|_{z=0,\,L}=0,\quad\hat{T}(z=0,\,L)=0.\label{eq:e112}
\end{equation}
To specify the superadiabatic gradient $\Delta_{S}$ the basic state
solution of (\ref{eq:T_basic_RadB}-c) is necessary. The boundary
conditions on radiative flux have been derived by Goody (1956) and
thus in the case when both the top and bottom boundaries are black
bodies the hydrostatic state must satisfy 
\begin{subequations}
\begin{equation}
\left.\left(\frac{\mathrm{d}\dbtilde{j}_{rad\,z}}{\mathrm{d}z}-2\alpha_{a}\dbtilde{j}_{rad\,z}\right)\right|_{z=0}=0,\quad\left.\left(\frac{\mathrm{d}\dbtilde{j}_{rad\,z}}{\mathrm{d}z}+2\alpha_{a}\dbtilde{j}_{rad\,z}\right)\right|_{z=L}=0,\label{eq:BC_hydrostatic_radB_1}
\end{equation}
\begin{equation}
\dbtilde{T}(z=0)=\dbtilde{T}_{B},\quad\dbtilde{T}(z=L)=\dbtilde{T}_{T},\label{eq:BC_hydrostatic_radB_2}
\end{equation}
\end{subequations} 
where according to our definition $\tilde T(z) = \bar T + \dbtilde{T}(z)$ we have $\dbtilde{T}_B=\Delta T/2$ and $\dbtilde{T}_T=-\Delta T/2$. The solution for the hydrostatic balance (\ref{eq:T_basic_RadB}-c)
with the boundary conditions (\ref{eq:BC_hydrostatic_radB_1},b) is
provided e.g. in Larson (2001) in his equation (31), but note the
shift of the reference system origin to the mid-plane; since the general
formulae are quite cumbersome, they will not be repeated here. Particularly
interesting is the limit of radiatively transparent fluid (i.e. $L^{-1}\gg\alpha_{a}$
) and small thermal diffusivity, defined by 
\begin{equation}
L^{-1}\gg\alpha_{a}\gg(2\pi)^{2}\frac{\kappa\tau_{rad}}{3L^{3}},\label{eq:radiative_regime_def_B}
\end{equation}
where 
\begin{equation}
\tau_{rad}=3\bar{\rho}\bar{c}_{p}L/16\sigma_{rad}\bar{T}^{3}\label{eq:tau_rad}
\end{equation}
is the radiative cooling time-scale\index{SI}{radiative!cooling time-scale}. In this limit the basic temperature
profile adopts a form with strong gradients near the boundaries, indicating
possible formation of boundary layers in the convective flow. The
system then exhibits two distinct regimes for convective threshold:
regime (I) with typical large-scale convective rolls, and regime (II)
with small scale convective rolls located near boundaries. The former
is achieved at small and extremely small values of $\kappa\tau_{rad}/L^{2}$
(at a given $\tau_{rad}$ and $\alpha_{a}L\ll1$), say for $\kappa_{2}<\kappa\ll L^{2}/\tau_{rad}$ or $\kappa<\kappa_{1}\ll L^{2}/\tau_{rad}$,
where $\kappa_{1}$ and
$\kappa_{2}$ are dependent on system parameters. The regime (II)
is achieved for intermediate values of $\kappa\tau_{rad}/L^{2}$ (at
a given $\tau_{rad}$ and $\alpha_{a}L\ll1$), in-between the limiting
values for the regime (I), i.e. for $\kappa_{1}<\kappa<\kappa_{2}$.
Both regimes satisfy the linear up-down symmetry (\ref{eq:symmetry_lin_1},b).
The dependence of the radiative Rayleigh number at threshold on the
wavenumber of perturbations, $Ra_{R}(\mathcal{K})$, can be obtained
from a numerical solution of the linear eigenvalue problem at threshold,
cf. Getling (1998, chapter 7.3.2) and Larson (2001). Under the conditions
corresponding to regime (II) the curve $Ra_{R}(\mathcal{K})$ possesses
two minima. The first minimum, around $\mathcal{K}L=\pi$, corresponds
to the large-scale convective rolls extending over the entire depth
of the fluid layer, whereas the second one, with larger $\mathcal{K}L$,
to the small-scale rolls near the boundaries, as depicted on figures
\ref{fig:compendium_of_LinBsq}e and f (in fact the figure e) depicts
the situation for $\kappa=0$, which qualitatively resembles that
of regime (I); figure f) is based on parameter values calculated in
chapter 7.3.2 of Getling (1998)). The position of the second minimum,
that is the critical wave number for the regime (II) increases with
the value of the parameter $\alpha_{a}L^{3}/\kappa\tau_{rad}$, so
that when the latter parameter increases, the convection rolls in
the regime (II) are horizontally thinning. Depending on the system
parameter values either one of the minima can become a global minimum
corresponding to the critical Rayleigh number for convection.

When $\kappa$ is finite the $Ra_{R}(\mathcal{K})$ dependence, where
$Ra_{R}$ denotes the Rayleigh number in the presence of radiative
heating, possesses only one minimum around $\mathcal{K}L=\pi$ and
system also forms large-scale rolls of similar type as in the regime
(I). Furthermore, when the thermal diffusivity approaches exactly
zero, $\kappa=0$, in the limit of a transparent fluid $\alpha_{a}L\ll1$
the linear problem becomes fully analytically tractable with the results
at convection threshold summarized as follows
\begin{equation}
\hat{T}(z)\sim\hat{u}_{z}(z)\sim\sin\left(\pi\frac{z}{L}\right),\label{eq:e113}
\end{equation}
and
\begin{equation}
Ra_{R\,crit}=4\pi^{2},\qquad\mathcal{K}_{crit}L=\pi,\label{eq:e114}
\end{equation}
where the radiative Rayleigh number is defined in the following way
\begin{equation}
Ra_{R}=\frac{g\bar{\alpha}\Delta_{S}L^{4}}{3(\alpha_{a}L)(L^{2}/\tau_{rad})\nu},\label{eq:radiative_Ra}
\end{equation}
\begin{equation}
\Delta_{S}=\frac{3\alpha_{a}L}{4+3\alpha_{a}L}\frac{\Delta T}{L}-\frac{g}{\bar{c}_{p}},\qquad\tau_{rad}=\frac{3\bar{\rho}\bar{c}_{p}L}{16\sigma_{rad}\bar{T}^{3}}.\label{eq:Delta_S_rad_tau_rad}
\end{equation}
The parameter $\Delta_{S}$ denotes the interior superadiabatic temperature
gradient, since in the limit $\kappa=0$, when no thermal conduction
is possible the temperature of the fluid adjacent to the boundaries,
say $T_{b}$ (bottom) and $T_{t}$ (top), differs from the temperatures
of the boundaries denoted as usually by $T_{B}$ and $T_{T}$, so
that there is a discontinuity of temperature at the boundaries $T_{B}>T_{b}$
and $T_{T}<T_{t}$. This is simply a manifestation of a radiative
thermal boundary layer, which forms in a realistic situation of small,
but finite diffusion, $\kappa\ll L^{2}/\tau_{rad}$, due to joint
action of radiation and thermal diffusivity, where the temperature
decays exponentially to the value at a boundary; such a boundary layer
is shrank to a discontinuity at a boundary in the limit $\kappa=0$.

Furthermore, the $\tau_{rad}$ depends explicitly on the mean temperature
$\bar{T}$, which in turn implies $Ra_{R}\sim\bar{T}^{-3}$, thus
a radiative medium can be thermally stabilized by an increase of the
mean temperature of the system. This is in contrast to all the previous, non-radiative
cases considered, when only the basic temperature gradient played
a role in triggering the convective instability and the value of the
mean temperature had no effect on stability.

Also in this case we can estimate the heat per unit mass released
by a rising fluid parcel on an infinitesimal vertical distance in
a time unit in the marginal state, which for $\kappa=0$ is equal
to $\bar{c}_{p}3(\alpha_{a}L)(L^{2}/\tau_{rad})\nu Ra_{R\,crit}u_{z}/g\bar{\alpha}L^{4}$.
It includes the effects of thermal radiation and absorption, but note
that absorption by assumption is weak, $\alpha_{a}\ll L^{-1}$. Comparison
with previous non-radiative cases can be made when the thermal diffusivity
of a non-radiative system $\kappa$ is assumed comparable with the
``radiative diffusivity'' parameter $3(\alpha_{a}L)(L^{2}/\tau_{rad})$
of a radiative, but thermally insulating ($\kappa=0$) system. Then
the maximal heat per unit mass which can accumulate between infinitesimally
distant fluid layers before convection starts, $\bar{c}_{p}3(\alpha_{a}L)(L^{2}/\tau_{rad})\nu Ra_{R\,crit}\mathrm{d}z/g\bar{\alpha}L^{4}$,
is significantly smaller than in all the previous, non-radiative cases
($Ra_{R\,crit}$ is only about $40$), which indicates that thermal
radiation has a destabilizing effect in this sense (in terms of the
interior temperature gradient, since in fact the temperature difference
between the top and bottom plates, i.e. $\Delta T=T_{B}-T_{T}$ may
be larger at threshold in the radiative case, since $\alpha_{a}L\ll1$).

When both, the thermal diffusivity $\kappa\neq0$ and the radiative
effects are included the heat per unit mass released by a rising fluid
parcel on an infinitesimal vertical distance in a time unit in the
marginal state can be most conveniently expressed by $\bar{c}_{p}u_{z}\Delta_{S\,crit}(z)$,
since in such a case the basic state temperature gradient depends
on depth.

\subsection{Constraints from rapid rotation at two stress-free, isothermal boundaries, $Q=0$.\label{subsec:Constraints-from-rapid_rotation}}\index{SI}{rapid rotation}

We will now briefly comment on the effect of the Coriolis force, which
is a common body force in many natural convective systems such as
the planetary and stellar atmospheres and cores. When the system rotates
at a uniform rate about vertical axis $\bm{\Omega}=\Omega\hat{\mathbf{e}}_{z}$
(perpendicular to the planes of the parallel boundaries) the linearised
Navier-Stokes equation under the Boussinesq approximation has to include
the ``non-inertiality'' of the system of reference, namely the Coriolis
force
\begin{equation}
\frac{\partial\mathbf{u}}{\partial t}=-\nabla\frac{p'}{\bar{\rho}}+g\bar{\alpha}T'\hat{\mathbf{e}}_{z}-2\Omega\hat{\mathbf{e}}_{z}\times\mathbf{u}+\nu\nabla^{2}\mathbf{u};\label{eq:Coriolis_NS_linearised}
\end{equation}
the centrifugal force $-\bar{\rho}\bm{\Omega}\times(\bm{\Omega}\times\mathbf{x})$,
which is potential, only modifies the hydrostatic pressure distribution
$\mathrm{d}_{z}\dbtilde{p}=-(\bar{\rho}+\dbtilde{\rho})g+\bar{\rho}(\bm{\Omega}\times\mathbf{x})^{2}/2$,
where (\ref{eq:Static_energy}) and (\ref{eq:rho_0_B}) still hold.
Thorough analysis of the linear regime at convection threshold for
rotating systems together with the derivation of the form of the dynamical
equations in rotating systems can be found in $\S$III of Chandrasekhar
(1961). One can easily see, that by taking a curl of the equation
(\ref{eq:Coriolis_NS_linearised}) and letting $\nu=0$ one obtains
$\partial_{z}u_{z}=0$, thus for impermeable boundaries, $u_{z}(z=0,\,L)=0$,
rapid rotation leading to domination of the Coriolis force over the
viscous friction inhibits convection. More precisely, when the so-called
Ekman number
\begin{equation}
E=\nu/2\Omega L^{2}\label{eq:e115}
\end{equation}
is small $E\ll1$, the rotational effects tend to suppress the vertical
convective motions.\footnote{This is a manifestation of the well-known Taylor-Proundman theorem.}
When the background rotation is present the system can no longer be
sufficiently described by the vertical velocity component and the
temperature only, since the horizontal components of velocity can
not be separated anymore. Introducing the vorticity $\bm{\zeta}=\nabla\times\mathbf{u}$
the linearised equations for the $z$-dependent amplitudes defined as in (\ref{eq:linear_perturbations_forms})
take the form 
\begin{subequations}
\begin{equation}
\left[\sigma\left(\frac{\mathrm{d}^{2}}{\mathrm{d}z^{2}}-\mathcal{K}^{2}\right)-\nu\left(\frac{\mathrm{d}^{2}}{\mathrm{d}z^{2}}-\mathcal{K}^{2}\right)^{2}\right]\hat{u}_{z}+2\Omega\frac{\mathrm{d}\hat{\zeta}_{z}}{\mathrm{d}z}=-g\bar{\alpha}\mathcal{K}^{2}\hat{T},\label{eq:linear_eqs_Boussinesq_NS-1}
\end{equation}
\begin{equation}
\left[\sigma-\nu\left(\frac{\mathrm{d}^{2}}{\mathrm{d}z^{2}}-\mathcal{K}^{2}\right)\right]\hat{\zeta}_{z}=2\Omega\frac{\mathrm{d}\hat{u}_{z}}{\mathrm{d}z},\label{eq:linear_eqs_Boussinesq_NS-vort}
\end{equation}
\begin{equation}
\left[\sigma-\kappa\left(\frac{\mathrm{d}^{2}}{\mathrm{d}z^{2}}-\mathcal{K}^{2}\right)\right]\hat{T}=\Delta_{S}\hat{u}_{z}.\label{eq:linear_eqs_Boussinesq_Th-1}
\end{equation}
\end{subequations} 
Two cases can be distinguished. In the first one
the instability sets in as stationary convection with $\Im\mathfrak{m}\sigma=0$
and $\sigma$ passing through zero (thus the principle of exchange
of stabilities holds). This is always the case, when the Prandtl number $Pr=\nu/\kappa$\index{SI}{Prandtl number} satisfies $Pr>\mathscr{P}r\approx0.677$;
for $Pr<\mathscr{P}r$ there always
exists a finite value of the Ekman number $E=\mathscr{E}(Pr)$, such
that for $E\geq\mathcal{E}(Pr)$ convection still sets in as a stationary
flow. The second case is obtained for $E<\mathcal{E}(Pr)$, when the flow is oscillatory at the onset
with $\Im\mathfrak{m}\sigma\neq0$ and $\Re\mathfrak{e}\sigma$ passing
through zero. In the first case ($Pr>\mathscr{P}r$) or ($Pr<\mathscr{P}r$
and $E\geq\mathcal{E}(Pr)$), when the marginal flow is stationary
\begin{equation}
u_{z}=\Re\mathfrak{e}\;\,A\sin\left(\pi\frac{z}{L}\right)\mathrm{e}^{\mathrm{i}\left(\mathcal{K}_{x}x+\mathcal{K}_{y}y\right)},\label{eq:e116}
\end{equation}
where $A$ is an undetermined constant and in the limit of rapid rotation
$E\ll1$ one obtains
\begin{equation}
Ra_{crit}\approx3\left(\frac{\pi}{\sqrt{2}E}\right)^{4/3},\qquad\mathcal{K}_{crit}L\approx\left(\frac{\pi}{\sqrt{2}E}\right)^{1/3}.\label{eq:Ra_crit_rotation_1}
\end{equation}
In the second case ($Pr<\mathscr{P}r$ and $E<\mathcal{E}(Pr)$) the
purely oscillatory marginal state\footnote{it should be noted, however, that in all the cases studied above exactly
at threshold the amplitude of the marginal states vanishes, and the
flow in the form of linear solutions is observed only slightly above
the threshold value of the temperature gradient, cf. the next section
\ref{sec:WNT} on weakly nonlinear analysis.} in the rapidly rotating limit at finite $Pr$, that is when $EPr^{-1}\ll1$,
is described by
\begin{equation}
u_{z}=\Re\mathfrak{e}\;\,A\sin\left(\pi\frac{z}{L}\right)\mathrm{e}^{\mathrm{i}\omega t}\mathrm{e}^{\mathrm{i}\left(\mathcal{K}_{x}x+\mathcal{K}_{y}y\right)},\label{eq:e117}
\end{equation}
where $A$ is an undetermined constant, and
\begin{equation}
\omega=\frac{\sqrt{2-3Pr^{2}}}{Pr}\left(\frac{Pr}{1+Pr}\frac{\pi}{\sqrt{2}E}\right)^{2/3},\label{eq:e118}
\end{equation}
\begin{equation}
Ra_{crit}=6\left(1+Pr\right)\left(\frac{Pr}{1+Pr}\frac{\pi}{\sqrt{2}E}\right)^{4/3},\qquad\mathcal{K}_{crit}L=\left(\frac{Pr}{1+Pr}\frac{\pi}{\sqrt{2}E}\right)^{1/3}.\label{eq:Ra_crit_rotation_2}
\end{equation}
In both cases the state satisfies the linear up-down symmetry (\ref{eq:symmetry_lin_1},b).
Moreover, the results for both regimes (\ref{eq:Ra_crit_rotation_1})
and (\ref{eq:Ra_crit_rotation_2}) indicate the stabilizing role of
rotation, since $Ra_{crit}\rightarrow\infty$ as $E\rightarrow0$,
thus strong rotation inhibits vertical convective heat transfer. Because
in the limit $E\ll1$ the threshold value of the temperature gradient
is very high, as indicated by $Ra_{crit}\gg1$ the heat per unit mass
released by a rising fluid parcel on an infinitesimal vertical distance
in a time unit in the marginal state, $\bar{c}_{p}\kappa\nu Ra_{R\,crit}u_{z}/g\bar{\alpha}L^{4}$,
is extremely high. In other words rapid background rotation allows
for very high values of molecular heat flux before the vertically
perturbed fluid parcels become buoyant. The form of the convection
rolls in marginal rapidly rotating convection is schematically depicted
on figures \ref{fig:compendium_of_LinBsq}g and h.

\subsection{Summary\label{subsec:Summary_Linear_B}}

We can quickly summarize the effects of different types of boundaries,
radiation and background rotation as follows. The rigid walls stabilize
the system by reducing its ability to develop flow near the boundaries.
Thermally insulating walls exert a destabilizing effect by keeping
the perturbation heat flux in the system, thus making it easier for
a vertically perturbed fluid parcel to become buoyant. Furthermore,
the effect of thermal radiation is, in general, complex, but for a
radiatively transparent fluid, $\alpha_{a}\ll L^{-1}$, in the limit
of weak thermal diffusion, $\kappa\ll L^{2}/\tau_{rad}$, it can be
concluded, that radiation convectively destabilizes the system in
the sense, that smaller interior superadiabatic temperature gradients
are sufficient for convection threshold; this however, may correspond
to larger temperature difference $\Delta T=T_{B}-T_{T}$ between the
top and bottom boundaries, since the temperature jumps in thin radiative
boundary layers in the hydrostatic state are finite, but the interior
temperature gradient is proportional to $\alpha_{a}L\ll1$. Finally
the effect of rapid rotation is to inhibit the vertical flow, thus
is strongly stabilizing.

A fairly general relation, that may come useful for straightforward
calculation of the critical Rayleigh number for convection once the
vertical structure of the flow close to threshold is known, valid
for $\kappa=\mathrm{const}$ and isothermal, impermeable and either
stress-free or no-slip boundaries can be obtained from equations (\ref{eq:linear_eqs_Boussinesq_NS-1}-b)
and the temperature equation with heat sources
\begin{equation}
\left[\sigma-\kappa\left(\frac{\mathrm{d}^{2}}{\mathrm{d}z^{2}}-\mathcal{K}^{2}\right)\right]\hat{T}=\Delta_{S}\hat{u}_{z}+\frac{\hat{Q}}{\bar{\rho}\bar{c}_{p}},\label{eq:linear_thermal_eq_Boussinesq_gen}
\end{equation}
in a similar manner to Chandrasekhar's (1961) eq. (253), p. 125 in
that book. In the above it has been assumed, that just as for all
the perturbations the heat source perturbation satisfies $Q'=\hat{Q}(z)\exp\left[\mathrm{i}\left(\bm{\mathcal{K}}\cdot\mathbf{x}_{h}+\omega t\right)\right]$.
Introducing $\sigma=\mathrm{i}\omega$, $\omega\in\mathbb{R}$ at
threshold and
\begin{equation}
F(z)=\left(\frac{\mathrm{d}^{2}}{\mathrm{d}z^{2}}-\mathcal{K}^{2}\right)\left(\frac{\mathrm{d}^{2}}{\mathrm{d}z^{2}}-\mathcal{K}^{2}-\mathrm{i}\frac{\omega}{\nu}\right)\hat{u}_{z}-\frac{2\Omega}{\nu}\frac{\mathrm{d}\hat{\zeta}}{\mathrm{d}z}=\frac{g\bar{\alpha}\mathcal{K}^{2}}{\nu}\hat{T},\label{eq:e119}
\end{equation}
the two equations (\ref{eq:linear_eqs_Boussinesq_NS-1}) and (\ref{eq:linear_thermal_eq_Boussinesq_gen})
can be reduced to
\begin{equation}
\left(\frac{\mathrm{d}^{2}}{\mathrm{d}z^{2}}-\mathcal{K}^{2}-\frac{\mathrm{i}\omega}{\kappa}\right)F=-\frac{g\bar{\alpha}\Delta_{S}}{\nu\kappa}\mathcal{K}^{2}\hat{u}_{z}-\frac{g\bar{\alpha}}{\nu\kappa}\mathcal{K}^{2}\frac{\hat{Q}}{\bar{\rho}\bar{c}_{p}}.\label{eq:F_function}
\end{equation}
Multiplying the latter equation by $F$ and integrating with respect
to the vertical variable $z$ from $0$ to $L$, after integration
by parts of the left hand side one obtains the Chandrasekhar's expression
for the critical Rayleigh number modified here as to include the effect
of heat sources (e.g. thermal radiation, radioactivity etc.)\footnote{Note, that alternatively equation (\ref{eq:F_function}) could be
multiplied simply by $\hat{u}_{z}$ and integrated over the range of
$z$, which gives 
\[
Ra=\int_{0}^{L}\left[\hat{u}_{z}\left(\frac{\mathrm{i}\omega}{\kappa}+\mathcal{K}^{2}-\frac{\mathrm{d}^{2}}{\mathrm{d}z^{2}}\right)F\right]\mathrm{d}z\left/\mathcal{K}^{2}\int_{0}^{L}\left(\frac{\Delta_{S}}{\left\langle \Delta_{S}\right\rangle }\frac{\hat{u}_{z}^{2}}{L^{4}}+\frac{\hat{Q}\hat{u}_{z}}{\bar{\rho}\bar{c}_{p}\left\langle \Delta_{S}\right\rangle L^{4}}\right)\mathrm{d}z\right.
\]
.},
\begin{equation}
Ra=\frac{g\bar{\alpha}\left\langle \Delta_{S}(z)\right\rangle L^{4}}{\kappa\nu}=\frac{\int_{0}^{L}\left[\left(\frac{\mathrm{d}F}{\mathrm{d}z}\right)^{2}+\left(\mathcal{K}^{2}+\mathrm{i}\frac{\omega}{\kappa}\right)F^{2}\right]\mathrm{d}z}{\mathcal{K}^{2}\int_{0}^{L}\left(\frac{\Delta_{S}}{\left\langle \Delta_{S}\right\rangle }\frac{\hat{u}_{z}}{L^{4}}+\frac{\hat{Q}}{\bar{\rho}\bar{c}_{p}\left\langle \Delta_{S}\right\rangle L^{4}}\right)F\mathrm{d}z}.\label{eq:Rayleigh_from_variational_principle-1}
\end{equation}
With the use of the general expression on $\Delta_{S\,crit}$ (\ref{eq:general_condition_on_Delta_c-1})
it is clear, that to obtain the actual value of the critical Rayleigh
number at the convection threshold $Ra_{crit}$, the above result
(\ref{eq:Rayleigh_from_variational_principle-1}) has to be minimized
over all possible values of the horizontal wave number $\mathcal{K}$,
the inverse vertical variation scale of perturbations, say $q$, and
the real frequency of oscillations $\omega(\mathcal{K},q)$. The latter
expression for the Rayleigh number corresponds exactly to the general
definition (\ref{eq:Ra_def_B}) for $\kappa=\mathrm{const}$. On the
other hand, when additionally $Q=0$ and $\bm{\Omega}=0$ the expression
for the Rayleigh number simplifies to
\begin{equation}
Ra=\frac{g\bar{\alpha}\Delta_{S}L^{4}}{\kappa\bar{\nu}}=\frac{L^{4}\int_{0}^{L}\left[\left(\frac{\mathrm{d}F}{\mathrm{d}z}\right)^{2}+\mathcal{K}^{2}F^{2}\right]\mathrm{d}z}{\mathcal{K}^{2}\int_{0}^{L}\hat{u}_{z}F\mathrm{d}z}=\frac{L^{4}\int_{0}^{L}\left[-\hat{u}_{z}\left(\frac{\mathrm{d}^{2}}{\mathrm{d}z^{2}}-\mathcal{K}^{2}\right)^{3}\hat{u}_{z}\right]\mathrm{d}z}{\mathcal{K}^{2}\int_{0}^{L}\hat{u}_{z}^{2}\mathrm{d}z}.\label{eq:Rayleigh_from_variational_principle}
\end{equation}
At this stage, we recall here the Chandrasekhar's (1961) statement
from $\S$II, p. 34, below equation (185) about the physical conditions
for convective instability trigger in non-rotating, Boussinesq systems
(for which by construction the thermal energy greatly exceeds the kinetic one): 

~

\emph{Instability occurs at the minimum temperature gradient at which
a balance can be steadily maintained between the kinetic energy dissipated
by viscosity and the work done by the buoyancy force, which
in general both include the effect of heating sources }$Q$.

~

Finally, it is of interest to provide a general relation, which can
be used to calculate the growth rate\index{SI}{growth rate} of perturbations for the case
when heat sources $Q$ and background rotation $\bm{\Omega}$ are
present. By the use of equations (\ref{eq:linear_eqs_Boussinesq_NS-1},b)
and (\ref{eq:linear_thermal_eq_Boussinesq_gen}) one obtains
\begin{eqnarray}
\left(\frac{\mathrm{d}^{2}}{\mathrm{d}z^{2}}-\mathcal{K}^{2}\right)\left[\sigma-\kappa\left(\frac{\mathrm{d}^{2}}{\mathrm{d}z^{2}}-\mathcal{K}^{2}\right)\right]\left[\sigma-\nu\left(\frac{\mathrm{d}^{2}}{\mathrm{d}z^{2}}-\mathcal{K}^{2}\right)\right]^{2}\hat{u}_{z}\qquad\qquad\nonumber \\
+4\Omega^{2}\frac{\mathrm{d}^{2}}{\mathrm{d}z^{2}}\left[\sigma-\kappa\left(\frac{\mathrm{d}^{2}}{\mathrm{d}z^{2}}-\mathcal{K}^{2}\right)\right]\hat{u}_{z}\qquad\qquad\nonumber \\
=-g\bar{\alpha}\mathcal{K}^{2}\left[\sigma-\nu\left(\frac{\mathrm{d}^{2}}{\mathrm{d}z^{2}}-\mathcal{K}^{2}\right)\right]\left(\Delta_{S}\hat{u}_{z}+\frac{\hat{Q}}{\bar{\rho}\bar{c}_{p}}\right). &  & \qquad\label{eq:grate_gen_B}
\end{eqnarray}
Of course a relation between $\hat{Q}$ and $\hat{T}$ and thus effectively
between $\hat{Q}$ and $\hat{u}_{z}$ must be specified by the physical
properties of the system, such as e.g. in the case of thermal radiation
by (\ref{eq:divj_vs_T_linear_B_rad}). However, if there are no volume
heat sources, $Q=0$, and the boundaries are assumed stress-free and
isothermal, the solution for the class of most unstable modes near
threshold takes the form $\hat{u}_{z}\sim\sin\left(m\pi\frac{z}{L}\right)\mathrm{e}^{\sigma t}$
with $m=1$ and the relation for the growth rate greatly simplifies,
\begin{eqnarray}
\left(\pi^{2}+\mathcal{K}^{2}L^{2}\right)\left[\frac{\sigma L^{2}}{\kappa}+\left(\pi^{2}+\mathcal{K}^{2}L^{2}\right)\right]\left[\frac{\sigma L^{2}}{\nu}+\left(\pi^{2}+\mathcal{K}^{2}L^{2}\right)\right]^{2}\qquad\qquad\nonumber \\
+\left(\frac{\pi}{E}\right)^{2}\left[\frac{\sigma L^{2}}{\kappa}+\left(\pi^{2}+\mathcal{K}^{2}L^{2}\right)\right]\qquad\qquad\nonumber \\
=Ra\mathcal{K}^{2}L^{2}\left[\frac{\sigma L^{2}}{\nu}+\left(\pi^{2}+\mathcal{K}^{2}L^{2}\right)\right]. &  & \qquad\label{eq:grate_no_Q_B}
\end{eqnarray}
Furthermore, if the rotation is neglected, $\bm{\Omega}=\mathbf{0}$,
the dispersion relation takes the form
\begin{equation}
\left(\pi^{2}+\mathcal{K}^{2}L^{2}\right)\left[\frac{\sigma L^{2}}{\kappa}+\left(\pi^{2}+\mathcal{K}^{2}L^{2}\right)\right]\left[\frac{\sigma L^{2}}{\nu}+\left(\pi^{2}+\mathcal{K}^{2}L^{2}\right)\right]-Ra\mathcal{K}^{2}L^{2}=0,\label{eq:disp_rel_BCisothSF_B}
\end{equation}
which leads to\index{SI}{growth rate}
\begin{equation}
\frac{\sigma L^{2}}{\kappa}=-\frac{1+Pr}{2}\left(\pi^{2}+\mathcal{K}^{2}L^{2}\right)+\sqrt{\left(\frac{1-Pr}{2}\right)^{2}\left(\pi^{2}+\mathcal{K}^{2}L^{2}\right)^{2}+RaPr\frac{\mathcal{K}^{2}L^{2}}{\pi^{2}+\mathcal{K}^{2}L^{2}}},\label{eq:grate_BCisothSF_B}
\end{equation}
where only the positive root was provided. The latter expression allows
to calculate the growth rate of unstable modes in the vicinity of
threshold, $Ra-Ra_{crit}\ll Ra_{crit}$ and $\mathcal{K}L-\mathcal{K}_{crit}L\ll1$
for systems with stress-free and isothermal boundaries.

As remarked at the beginning of this section the horizontal planform
of the solutions remains undetermined in the linear regime. However,
in the case of a horizontally infinite layer for which neither direction
nor point in a horizontal plane is preferred, one can say, based purely on this symmetry
and geometrical considerations, that the only possible cell patterns
consist of contiguous parallel stripes (rolls), triangles, squares
or hexagons filling and fitting into the entire $xy$ plane. Indeed,
because the convection cells are all contiguous and fill the entire
horizontal plane, the cells are either rolls (stripes) or regular
polygons and in the latter case the angle at an apex of an $n$-sided
regular polygon, $\pi-2\pi/n$ must be equal to $2\pi/m$, where $m$
is an integer, since in such a configuration the apex is shared by
$m$ exactly the same regular polygons; the relation $1-2/n=2/m$,
where $m$ and $n$ are integers can only be satisfied for pairs $(n=3,\,m=6)$
or $(n=4,\,m=4)$or $(n=6,\,m=3)$. The linear analysis does not allow
to determine, which one of those cell patterns is stable. We will
address the issue of horizontal pattern selection\index{SI}{pattern selection} by convective flow
near threshold in the following section.

\section{A word on weakly nonlinear estimates\label{sec:WNT}}\index{SI}{weakly nonlinear theory}

There have been many works concerned with the weakly nonlinear analysis of
convection near threshold involving determination of the amplitude
of convection rolls and pattern selection problems. Most of them were based
on similar approaches as the one developed by Schl$\ddot{\textrm{u}}$ter
\emph{et al}. (1965), that is perturbative expansions in the amplitude
of the convective flow, assumed weak. A most comprehensive review
of the literature and results on the topic for Rayleigh-B$\acute{\textrm{e}}$nard problem
with isothermal and either stress-free or rigid boundaries is provided
in the excellent book of Getling (1998). A thorough analysis of the
pattern selection problem for the case of fixed heat flux at boundaries
can be found in the seminal paper of Knobloch (1990). A brief compendium
of the results on nearly marginal convection and a comprehensive picture
of patterns developed by convection as the Rayleigh number departs
from threshold is provided in this section. We start by considering
the classical model of Rayleigh-B$\acute{\textrm{e}}$nard convection without heat sources,
$Q=0$, with uniform diffusivities $\mu=\textrm{const}.$, $\kappa=\textrm{const}.$
and stress-free, isothermal boundaries to demonstrate how the amplitude
of two-dimensional rolls near the threshold can be established.

It is useful to introduce the dynamical equations in non-dimensional
form. With the choice of $L$, $\kappa/L$, $L^{2}/\kappa$, $\bar{\rho}\kappa^{2}/L^{2}$
and $L\Delta_{S}$ as units of length, velocity, time, pressure fluctuation
and temperature fluctuation respectively
\begin{equation}
\mathbf{x}=L\mathbf{x}^{\sharp},\quad\mathbf{u}=\frac{\kappa}{L}\mathbf{u}^{\sharp},\quad t=\frac{L^{2}}{\kappa}t^{\sharp},\quad p'=\bar{\rho}\left(\frac{\kappa}{L}\right)^{2}p^{\sharp},\quad T'=L\Delta_{S}T^{\sharp},\label{eq:e120}
\end{equation}
the set of dynamical of equations takes the following form 
\begin{subequations}
\begin{equation}
\frac{\partial\mathbf{u}^{\sharp}}{\partial t^{\sharp}}+\left(\mathbf{u}^{\sharp}\cdot\nabla^{\sharp}\right)\mathbf{u}^{\sharp}=-\nabla^{\sharp}p^{\sharp}+RaPrT^{\sharp}\hat{\mathbf{e}}_{z}+Pr\nabla^{\sharp2}\mathbf{u}^{\sharp},\label{eq:NS_eq_WNT}
\end{equation}
\begin{equation}
\nabla^{\sharp}\cdot\mathbf{u}^{\sharp}=0,\label{eq:div_eq_WNT}
\end{equation}
\begin{equation}
\frac{\partial T^{\sharp}}{\partial t^{\sharp}}+\mathbf{u}^{\sharp}\cdot\nabla^{\sharp}T^{\sharp}-u_{z}^{\sharp}=\nabla^{\sharp2}T^{\sharp}.\label{eq:T_eq_WNT}
\end{equation}
\end{subequations}
We start by considering the simplest case of solutions
in the form of two-dimensional rolls, and since $\nabla^{\sharp}\cdot\mathbf{u}^{\sharp}=0$
we introduce the stream function $\psi=\psi(x^{\sharp},z^{\sharp})$,
$\mathbf{u}^{\sharp}=\nabla\times\left(\psi\hat{e}_{y}\right)$, so
that
\begin{equation}
u_{x}^{\sharp}=-\frac{\partial\psi}{\partial z^{\sharp}},\quad u_{y}^{\sharp}=0,\quad u_{z}^{\sharp}=\frac{\partial\psi}{\partial x^{\sharp}}.\label{eq:e121}
\end{equation}
By taking a curl of the momentum equation (\ref{eq:NS_eq_WNT}) and
taking its $y$-component one obtains 
\begin{subequations}
\begin{equation}
\frac{\partial}{\partial t^{\sharp}}\nabla^{\sharp2}\psi+\mathcal{J}\left(\zeta^{\sharp},\,\psi\right)=RaPr\frac{\partial T^{\sharp}}{\partial x^{\sharp}}+Pr\nabla^{\sharp4}\psi,\label{eq:psi_eq_general_WNT}
\end{equation}
\begin{equation}
-\frac{\partial T^{\sharp}}{\partial t^{\sharp}}+\mathcal{J}\left(T^{\sharp},\,\psi\right)=-\frac{\partial\psi}{\partial x^{\sharp}}-\nabla^{\sharp2}T^{\sharp},\label{eq:T_eq_general_WNT}
\end{equation}
\end{subequations} 
where 
\begin{equation}
\zeta^{\sharp}=-\nabla^{\sharp2}\psi\label{eq:zeta_eq_gen_WNT}
\end{equation}
is the $y$-component of the flow vorticity and 
\begin{equation}
\mathcal{J}\left(f,\,g\right)=\frac{\partial f}{\partial x^{\sharp}}\frac{\partial g}{\partial z^{\sharp}}-\frac{\partial f}{\partial z^{\sharp}}\frac{\partial g}{\partial x^{\sharp}}\label{eq:e122}
\end{equation}
denotes the Jacobian. As we know from the linear considerations, in
the problem at hand of classical Rayleigh-B$\acute{\textrm{e}}$nard convection with stress-free,
isothermal boundaries, the instability sets in through a stationary mode
and thus near convection threshold the time scale of flow evolution
is slow. To establish the time scale of evolution of the amplitude
of convection we expand the growth rate $\sigma^{\sharp}=\sigma L^{2}/\kappa$
given in (\ref{eq:grate_BCisothSF_B}) in powers of the departure
from threshold defined in the following way 
\begin{equation}
\eta=\frac{Ra-Ra_{crit}}{Ra_{crit}}\ll1,\label{eq:eta_def_WNT}
\end{equation}
and $\delta\mathcal{K}^{\sharp}=\mathcal{K}L-\mathcal{K}_{crit}L$ to obtain
(cf. also Getling 1998 or Fauve 2017)
\begin{equation}
\sigma^{\sharp}=\frac{3\pi^{2}Pr}{2\left(1+Pr\right)}\left(\eta-\frac{8}{3\pi^{2}}\delta\mathcal{K}^{\sharp 2}\right).\label{eq:sigma_approximate_WNT}
\end{equation}
It is evident from (\ref{eq:sigma_approximate_WNT}), that the time
scale of evolution of the most unstable mode with $\mathcal{K}=\mathcal{K}_{crit}$,
thus $\delta\mathcal{K}^{\sharp}=0$ is slow, of the order $\eta^{-1}L^{2}/\kappa$.
Hence we introduce $\tau=\eta\,t^{\sharp}$ for the slow time-scale
corresponding to the dynamics of the amplitude of convection. The
growth rate of the most unstable mode, at the initial stage of evolution,
when the dynamics is governed by the linear equations is
\begin{equation}
\sigma_{0}^{\sharp}=\frac{\eta}{\tau_{0}},\quad\textrm{where}\quad\frac{1}{\tau_{0}}=\frac{3\pi^{2}Pr}{2\left(1+Pr\right)}.\label{eq:MUM_grate_B}
\end{equation}
The main task of the weakly nonlinear analysis is to establish the
equation for time evolution of the amplitude of fluctuations, which
includes the effect of nonlinearities and thus allows for saturation
after the initial stage of exponential amplification. 

The next step is the expansion of all the dependent variables in powers
of the departure from threshold, which we assume in the following
form 
\begin{subequations}
\begin{equation}
\psi=\eta^{r}\psi_{1}+\eta^{2r}\psi_{2}+\eta^{3r}\psi_{3}+\dots,\quad\zeta^{\sharp}=\eta^{r}\zeta_{1}+\eta^{2r}\zeta_{2}+\eta^{3r}\zeta_{3}+\dots,\label{eq:e123}
\end{equation}
\begin{equation}
T^{\sharp}=\eta^{r}T_{1}+\eta^{2r}T_{2}+\eta^{3r}T_{3}+\dots,\label{eq:e124}
\end{equation}
\end{subequations} 
where $r$ is a positive rational number. Introducing
the above expansions, $\tau=\eta\,t^{\sharp}$ and (\ref{eq:eta_def_WNT})
into the equations (\ref{eq:psi_eq_general_WNT},b) and (\ref{eq:zeta_eq_gen_WNT})
and dividing by $\eta^{r}$ leads to 
\begin{subequations}
\begin{align}
 & \eta\frac{\partial}{\partial\tau}\left(\nabla^{\sharp2}\psi_{1}+\eta^{r}\nabla^{\sharp2}\psi_{2}\right)+\eta^{r}\mathcal{J}\left(\zeta_{1},\,\psi_{1}\right)+\eta^{2r}\left[\mathcal{J}\left(\zeta_{1},\,\psi_{2}\right)+\mathcal{J}\left(\zeta_{2},\,\psi_{1}\right)\right]\qquad\qquad\nonumber \\
 & \qquad\qquad\qquad\qquad=Ra_{crit}\left(1+\eta\right)Pr\frac{\partial}{\partial x^{\sharp}}\left(T_{1}+\eta^{r}T_{2}+\eta^{2r}T_{3}\right)\nonumber \\
 & \qquad\qquad\qquad\qquad\quad+Pr\nabla^{\sharp4}\left(\psi_{1}+\eta^{r}\psi_{2}+\eta^{2r}\psi_{3}\right)+\mathcal{O}\left(\eta^{3r}\right),\label{eq:psi_eq_general_WNT-1}
\end{align}
\begin{align}
 & -\eta\frac{\partial}{\partial\tau}\left(T_{1}+\eta^{r}T_{2}\right)+\eta^{r}\mathcal{J}\left(T_{1},\,\psi_{1}\right)+\eta^{2r}\left[\mathcal{J}\left(T_{1},\,\psi_{2}\right)+\mathcal{J}\left(T_{2},\,\psi_{1}\right)\right]\qquad\qquad\qquad\nonumber \\
 & \qquad\qquad\qquad\qquad=-\frac{\partial}{\partial x^{\sharp}}\left(\psi_{1}+\eta^{r}\psi_{2}+\eta^{2r}\psi_{3}\right)-\nabla^{\sharp2}\left(T_{1}+\eta^{r}T_{2}+\eta^{2r}T_{3}\right)\nonumber \\
 & \qquad\qquad\qquad\qquad\quad+\mathcal{O}\left(\eta^{3r}\right),\label{eq:T_eq_general_WNT-1}
\end{align}
\begin{equation}
\zeta_{n}=-\nabla^{\sharp2}\psi_{n}\quad\textrm{for all }n=1,\,2,\,3,\,\dots\label{eq:zeta_eq_gen_WNT-1}
\end{equation}
\end{subequations} 
Balancing the leading order terms (order unity
terms) gives the linear balance which allows to obtain the critical
Rayleigh number and the spatial structure of $\psi_{1}$ and $T_{1}$,
but not the amplitude 
\begin{subequations}
\begin{equation}
Ra_{crit}Pr\frac{\partial T_{1}}{\partial x^{\sharp}}+Pr\nabla^{\sharp4}\psi_{1}=0,\label{eq:psi_1_WNT}
\end{equation}
\begin{equation}
-\frac{\partial\psi_{1}}{\partial x^{\sharp}}-\nabla^{\sharp2}T_{1}=0,\label{eq:T_1_WNT}
\end{equation}
\begin{equation}
\zeta_{1}=-\nabla^{\sharp2}\psi_{1},\label{eq:zeta_eq_gen_WNT-1-1}
\end{equation}
\end{subequations} 
The solution for the most unstable mode takes
the form (cf. section \ref{subsec:Two-isothermal-stress-free} on the relevant linear problem) 
\begin{subequations}
\begin{eqnarray}
\psi_{1} & = & A\left(\tau\right)\sin\left(\pi z^{\sharp}\right)\cos\left(\mathcal{K}_{crit}^{\sharp}x^{\sharp}\right),\label{eq:psi_1_sol}\\
T_{1} & = & -\frac{\mathcal{K}_{crit}^{\sharp}}{\pi^{2}+\mathcal{K}_{crit}^{\sharp2}}A\left(\tau\right)\sin\left(\pi z^{\sharp}\right)\sin\left(\mathcal{K}_{crit}^{\sharp}x^{\sharp}\right),\label{eq:T_1_sol}\\
\zeta_{1} & = & \left(\pi^{2}+\mathcal{K}_{crit}^{\sharp2}\right)\psi_{1}.\label{eq:zeta_1_sol}
\end{eqnarray}
\end{subequations} 
with
\begin{equation}
Ra_{crit}=\frac{\left(\pi^{2}+\mathcal{K}_{crit}^{\sharp2}\right)^{3}}{\mathcal{K}_{crit}^{\sharp2}}=\frac{27}{4}\pi^{4},\quad\mathcal{K}_{crit}^{\sharp}=\frac{\pi}{\sqrt{2}},\label{eq:Ra_c_and_K_c}
\end{equation}
and for the sake of clarity only the $\cos\left(\mathcal{K}_{crit}^{\sharp}x^{\sharp}\right)$
mode was chosen. The next order balance involves terms of the order
$\eta^{r}$ and since at this stage we do not know yet the value of
$r$, we must also include in the balance the terms of the order $\eta$;
supplied by $\mathcal{J}(\zeta_{1},\,\psi_{1})=(\pi^{2}+\mathcal{K}_{crit}^{\sharp2})\mathcal{J}(\psi_{1},\,\psi_{1})=0$
and $\mathcal{J}(T_{1},\,\psi_{1})=-\mathcal{K}_{crit}^{\sharp2}\pi A^{2}\sin(2\pi z^{\sharp})/2(\pi^{2}+\mathcal{K}_{crit}^{\sharp2})$
this yields 
\begin{subequations}
\begin{equation}
Ra_{crit}Pr\frac{\partial T_{2}}{\partial x^{\sharp}}+Pr\nabla^{\sharp4}\psi_{2}=\eta^{1-r}\frac{\partial}{\partial\tau}\left(\nabla^{\sharp2}\psi_{1}\right)-\eta^{1-r}Ra_{crit}Pr\frac{\partial T_{1}}{\partial x^{\sharp}},\label{eq:psi_2_WNT}
\end{equation}
\begin{equation}
-\frac{\partial\psi_{2}}{\partial x^{\sharp}}-\nabla^{\sharp2}T_{2}=-\eta^{1-r}\frac{\partial T_{1}}{\partial\tau}-\frac{\pi\mathcal{K}_{crit}^{\sharp2}}{2\left(\pi^{2}+\mathcal{K}_{crit}^{\sharp2}\right)}A^{2}\sin\left(2\pi z^{\sharp}\right),\label{eq:T_2_WNT}
\end{equation}
\begin{equation}
\zeta_{2}=-\nabla^{\sharp2}\psi_{2},\label{eq:zeta_eq_gen_WNT-1-2}
\end{equation}
\end{subequations} 
At this stage the standard procedure in the approach
based on perturbative expansions involves application of the Fredholm
Alternative Theorem, which in the case at hand means, that non-trivial
solutions of the latter set of equations can only exist if the solution
to the linear problem on the left hand side is orthogonal to the non-homogenity
on the right hand side (cf. Korn \& Korn 1961). First, however, we
must introduce a certain inner product on the space of solutions and
show, that the linear operator on the left hand side of (\ref{eq:psi_2_WNT},b)
is self-adjoint. We write down the equations (\ref{eq:psi_2_WNT},b)
in the form
\begin{equation}
\bm{\mathfrak{L}}\mathbf{v}=\bm{\mathfrak{N}},\label{eq:e125}
\end{equation}
where
\begin{equation}
\mathbf{v}=\left[\begin{array}{c}
\psi_{2}\\
T_{2}
\end{array}\right],\quad\bm{\mathfrak{L}}=\left[\begin{array}{cc}
Pr\nabla^{\sharp4} & Ra_{crit}Pr\frac{\partial}{\partial x^{\sharp}}\\
-\frac{\partial}{\partial x^{\sharp}} & -\nabla^{\sharp2}
\end{array}\right],\label{eq:e126}
\end{equation}
\begin{equation}
\bm{\mathfrak{N}}=\left[\begin{array}{c}
\eta^{1-r}\frac{\partial}{\partial\tau}\left(\nabla^{\sharp2}\psi_{1}\right)-\eta^{1-r}Ra_{crit}Pr\frac{\partial T_{1}}{\partial x^{\sharp}}\\
-\eta^{1-r}\frac{\partial T_{1}}{\partial\tau}-\frac{\pi\mathcal{K}_{crit}^{\sharp2}}{2\left(\pi^{2}+\mathcal{K}_{crit}^{\sharp2}\right)}A^{2}\sin\left(2\pi z^{\sharp}\right)
\end{array}\right],\label{eq:e127}
\end{equation}
and define the inner product of $\mathbf{v}_{\alpha}=[\psi_{\alpha},\,T_{\alpha}]$
and $\mathbf{v}_{\beta}=[\psi_{\beta},\,T_{\beta}]$ in the following
way\footnote{Note, that the linear operator $\bm{\mathfrak{L}}$ is the same as
in the leading order problem (\ref{eq:psi_1_WNT},b). }
\begin{equation}
\left\langle \mathbf{v}_{\alpha},\,\mathbf{v}_{\beta}\right\rangle =\int_{0}^{2\pi/\mathcal{K}_{crit}^{\sharp}}\mathrm{d}x^{\sharp}\int_{0}^{1}\mathrm{d}z^{\sharp}\left(\psi_{\alpha}\psi_{\beta}+Ra_{crit}PrT_{\alpha}T_{\beta}\right).\label{eq:e128}
\end{equation}
Indeed, it is now easy to verify, that with such a definition the
linear operator $\bm{\mathfrak{L}}$ is self-adjoint, that is $\left\langle \bm{\mathfrak{L}}\mathbf{v}_{\alpha},\,\mathbf{v}_{\beta}\right\rangle =\left\langle \mathbf{v}_{\alpha},\,\bm{\mathfrak{L}}\mathbf{v}_{\beta}\right\rangle $
is satisfied. We can, therefore utilize the Fredholm Alternative Theorem
and write
\begin{align}
\eta^{1-r}\int_{0}^{2\pi/\mathcal{K}_{crit}^{\sharp}}\mathrm{d}x^{\sharp}\int_{0}^{1}\mathrm{d}z^{\sharp}\left[\frac{\partial}{\partial\tau}\left(\nabla^{\sharp2}\psi_{1}\right)-Ra_{crit}Pr\frac{\partial T_{1}}{\partial x^{\sharp}}\right]\sin\left(\pi z^{\sharp}\right)\cos\left(\mathcal{K}_{crit}^{\sharp}x^{\sharp}\right)\nonumber \\
+Ra_{crit}Pr\frac{\mathcal{K}_{crit}^{\sharp}}{\pi^{2}+\mathcal{K}_{crit}^{\sharp2}}\int_{0}^{2\pi/\mathcal{K}_{crit}^{\sharp}}\mathrm{d}x^{\sharp}\int_{0}^{1}\mathrm{d}z^{\sharp}\Bigg[\eta^{1-r}\frac{\partial T_{1}}{\partial\tau}\quad\qquad\qquad\qquad\qquad\nonumber \\
+\frac{\pi\mathcal{K}_{crit}^{\sharp2}}{2\left(\pi^{2}+\mathcal{K}_{crit}^{\sharp2}\right)}A^{2}\sin\left(2\pi z^{\sharp}\right)\Bigg]\sin\left(\pi z^{\sharp}\right)\sin\left(\mathcal{K}_{crit}^{\sharp}x^{\sharp}\right)=0.\label{eq:solvability_order_2}
\end{align}
Since
\begin{equation}
\int_{0}^{2\pi/\mathcal{K}_{crit}^{\sharp}}\mathrm{d}x^{\sharp}\sin\left(\mathcal{K}_{crit}^{\sharp}x^{\sharp}\right)\int_{0}^{1}\mathrm{d}z^{\sharp}\sin\left(2\pi z^{\sharp}\right)\sin\left(\pi z^{\sharp}\right)=0,\label{eq:e129}
\end{equation}
we are left only with terms of the order $\eta^{1-r}$. The value of $r$ can be established as follows. If we first assume tentatively,
that $r=1$, so that $\eta^{1-r}=1$, then the solvability
condition involves terms from the right hand side $\bm{\mathfrak{N}}$ which are all of
the same order as the left hand side (originally $\eta^{2r}$) and thus from
(\ref{eq:solvability_order_2}) one obtains
\begin{equation}
\frac{\mathrm{d}A}{\mathrm{d}\tau}=\frac{1}{\tau_{0}}A.\label{eq:e130}
\end{equation}
The latter amplitude equation does not involve any contribution from
nonlinear terms in the dynamical equations and is fully linear, leading
to exponential amplification of the amplitude, without the possibility
for saturation. This means that the choice $r=1$ was wrong. Therefore
the contributions from the buoyancy force and time-derivatives in
the dynamical equations, both involving terms proportional to $\eta$,
do not enter the second order balance (\ref{eq:psi_2_WNT},b), but are required to enter the balance at the next and higher orders so that saturation can be achieved. Consequently the only reasonable choice is $r=1/2$. This establishes
the well known feature of nearly marginal convection, that the amplitude
of the flow\index{SI}{amplitude of convection} scales like square root of the departure from threshold,
$\sqrt{(Ra-Ra_{crit})/Ra_{crit}}$.

Now, with $r=1/2$ we solve the second-order equations (\ref{eq:psi_2_WNT},b),
which take the form 
\begin{subequations}
\begin{equation}
Ra_{crit}Pr\frac{\partial T_{2}}{\partial x^{\sharp}}+Pr\nabla^{\sharp4}\psi_{2}=0\label{eq:psi_2_WNT-1}
\end{equation}
\begin{equation}
-\frac{\partial\psi_{2}}{\partial x^{\sharp}}-\nabla^{\sharp2}T_{2}=-\frac{\pi\mathcal{K}_{crit}^{\sharp2}}{2\left(\pi^{2}+\mathcal{K}_{crit}^{\sharp2}\right)}A^{2}\sin\left(2\pi z^{\sharp}\right),\label{eq:T_2_WNT-1}
\end{equation}
\end{subequations} 
and since the right hand side is independent of
$x^{\sharp}$ the solution takes the form\footnote{It is enough to take only the particular solution of the inhomogeneous
problem since the solution of the homogeneous problem provides only
an order $\eta$ correction to the amplitude.}
\begin{equation}
\psi_{2}=0,\quad T_{2}=-\frac{\mathcal{K}_{crit}^{\sharp2}}{8\pi\left(\pi^{2}+\mathcal{K}_{crit}^{\sharp2}\right)}A^{2}\sin\left(2\pi z^{\sharp}\right).\label{eq:e131}
\end{equation}
Next we gather the terms of the order $\eta$ in the equations (\ref{eq:psi_eq_general_WNT-1}-c)
to obtain 
\begin{subequations}
\begin{equation}
Ra_{crit}Pr\frac{\partial T_{3}}{\partial x^{\sharp}}+Pr\nabla^{\sharp4}\psi_{3}=\frac{\partial}{\partial\tau}\nabla^{\sharp2}\psi_{1}-Ra_{crit}Pr\frac{\partial T_{1}}{\partial x^{\sharp}},\label{eq:psi_3_WNT}
\end{equation}
\begin{equation}
-\frac{\partial\psi_{3}}{\partial x^{\sharp}}-\nabla^{\sharp2}T_{3}=-\frac{\partial T_{1}}{\partial\tau}+\mathcal{J}\left(T_{2},\,\psi_{1}\right)\label{eq:T_3_WNT}
\end{equation}
\begin{equation}
\zeta_{3}=-\nabla^{\sharp2}\psi_{3},\label{eq:zeta_eq_gen_WNT-1-3}
\end{equation}
\end{subequations} 
where $\psi_{2}=0$ has been substituted. The
solvability condition for the system of equations (\ref{eq:psi_3_WNT},b)
yields
\begin{align}
\int_{0}^{2\pi/\mathcal{K}_{crit}^{\sharp}}\mathrm{d}x^{\sharp}\int_{0}^{1}\mathrm{d}z^{\sharp}\left[\frac{\partial}{\partial\tau}\left(\nabla^{\sharp2}\psi_{1}\right)-Ra_{crit}Pr\frac{\partial T_{1}}{\partial x^{\sharp}}\right]\sin\left(\pi z^{\sharp}\right)\cos\left(\mathcal{K}_{crit}^{\sharp}x^{\sharp}\right)\quad\nonumber \\
+Ra_{crit}Pr\frac{\mathcal{K}_{crit}^{\sharp}}{\pi^{2}+\mathcal{K}_{crit}^{\sharp2}}\int_{0}^{2\pi/\mathcal{K}_{c}^{\sharp}}\mathrm{d}x^{\sharp}\int_{0}^{1}\mathrm{d}z^{\sharp}\bigg[\frac{\partial T_{1}}{\partial\tau}\qquad\qquad\qquad\qquad\qquad\quad\nonumber \\
-\mathcal{J}\left(T_{2},\,\psi_{1}\right)\bigg]\sin\left(\pi z^{\sharp}\right)\sin\left(\mathcal{K}_{crit}^{\sharp}x^{\sharp}\right)=0.\label{eq:solvability_order_3}
\end{align}
which on introduction of the first-order solutions (\ref{eq:psi_1_sol},b)
and (\ref{eq:Ra_c_and_K_c}) and 
\begin{equation}
\mathcal{J}\left(T_{2},\,\psi_{1}\right)=-\frac{\mathcal{K}_{crit}^{\sharp3}}{4\left(\pi^{2}+\mathcal{K}_{crit}^{\sharp2}\right)}A^{3}\cos\left(2\pi z^{\sharp}\right)\sin\left(\pi z^{\sharp}\right)\sin\left(\mathcal{K}_{crit}^{\sharp}x^{\sharp}\right)\label{eq:e132}
\end{equation}
simplifies to\index{SI}{amplitude (Landau) equation}
\begin{equation}
\frac{\mathrm{d}A}{\mathrm{d}\tau}=\frac{1}{\tau_{0}}A-\frac{1}{24\tau_{0}}A^{3}\label{eq:Landau_eq_WNT_B}
\end{equation}
where
\begin{equation}
\int_{0}^{2\pi/\mathcal{K}_{crit}^{\sharp}}\mathrm{d}x^{\sharp}\sin^{2}\left(\mathcal{K}_{crit}^{\sharp}x^{\sharp}\right)\int_{0}^{1}\mathrm{d}z^{\sharp}\cos\left(2\pi z^{\sharp}\right)\sin^{2}\left(\pi z^{\sharp}\right)=-\frac{1}{8}\label{eq:e133}
\end{equation}
was used. Equation (\ref{eq:Landau_eq_WNT_B}) is called the Landau
equation and it governs the time evolution of the amplitude of nearly
marginal solutions in the form of two-dimensional rolls. A stationary
solution of this type can be easily obtained and takes the form
\begin{equation}
A=\pm\sqrt{24},\label{eq:e134}
\end{equation}
and hence with the accuracy up to order $\mathcal{O}(\eta^{3/2})$
\begin{subequations}
\begin{align}
\psi= & \pm\eta^{1/2}\sqrt{24}\sin\left(\pi z^{\sharp}\right)\cos\left(\frac{\pi}{\sqrt{2}}x^{\sharp}\right)+\mathcal{O}\left(\eta^{3/2}\right),\label{eq:psi_sol_WNT_B}\\
T^{\sharp}= & \mp\eta^{1/2}\frac{4}{\sqrt{3}\pi}\sin\left(\pi z^{\sharp}\right)\sin\left(\frac{\pi}{\sqrt{2}}x^{\sharp}\right)-\eta\frac{1}{\pi}\sin\left(2\pi z^{\sharp}\right)+\mathcal{O}\left(\eta^{3/2}\right).\label{eq:T_sol_WNT_B}
\end{align}
\end{subequations} 
This allows to calculate the Nusselt number in
such a stationary state (cf. equation (\ref{eq:Nu_def_B})),
\begin{eqnarray}
Nu & = & \frac{\bar{\rho}\bar{c}_{p}\left\langle u_{z}T\right\rangle _{h}-k\partial_{z}\left\langle T\right\rangle _{h}-\bar{k}g\bar{\alpha}\bar{T}/\bar{c}_{p}}{k\Delta_{S}}=1+\left\langle \frac{\partial\psi}{\partial x^{\sharp}}T^{\sharp}\right\rangle _{h}-\partial_{z^{\sharp}}\left\langle T^{\sharp}\right\rangle _{h}\nonumber \\
 & = & 1+2\eta+\mathcal{O}\left(\eta^{3/2}\right)\approx2\frac{Ra}{Ra_{crit}}-1.\label{eq:e135}
\end{eqnarray}
Note, that the associated horizontally averaged superadiabatic temperature
gradient 
\begin{figure}
\centering{}\includegraphics[scale=0.3]{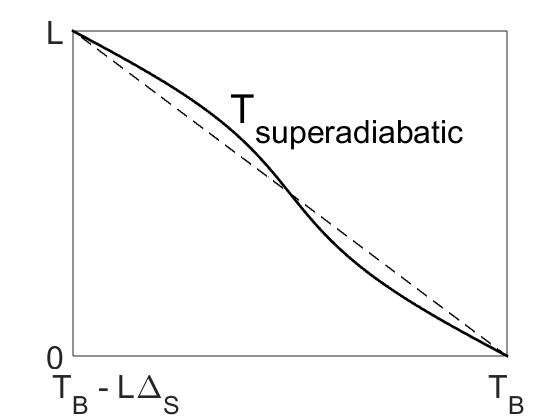}\caption{\label{fig:WNT_profile_Boussinesq}{\footnotesize{}Vertical profile
of the mean superadiabatic temperature $T_{\mathrm{superadiabatic}}=\tilde{T}+\left\langle T'\right\rangle _{h}+g\bar{\alpha}\bar{T}z/\bar{c}_{p}$,
in the weakly nonlinear regime close to convection threshold in the
absence of heat sources, $Q=0$, obtained from equation (\ref{eq:e136})
for $\eta=0.2$; the superadiabatic temperature profile in the hydrostatic
reference state $\tilde{T}+g\bar{\alpha}\bar{T}z/\bar{c}_{p}$ is
plotted with a dashed line for reference. The total mean temperature
profile becomes closer to adiabatic in the middle of the fluid domain
and its gradient sharpens near the boundaries.}}
\end{figure}
\begin{equation}
\left\langle \frac{\mathrm{d}T}{\mathrm{d}z}\right\rangle _{h}+\frac{g\bar{\alpha}\bar{T}}{\bar{c}_{p}}=-\Delta_{S}\left[1+2\eta\cos\left(\frac{2\pi z}{L}\right)\right]\label{eq:e136}
\end{equation}
is weakest near the mid point and strongest near the boundaries, thus
the temperature profile becomes closer to adiabatic in the middle
and the gradients become sharper as the boundaries are approached
(see figure \ref{fig:WNT_profile_Boussinesq}).

\subsection{Short introduction into the pattern selection problem near threshold
\label{subsec:Pattern_selection}}\index{SI}{pattern selection}

The selection of the horizontal planform by the convective system
is typically described by stability analysis of a stationary (or more generally oscillatory) solution with a given horizontal pattern, such
as e.g. the stationary two-dimensional roll solution (\ref{eq:psi_sol_WNT_B},b)
with the amplitude determined by the equation (\ref{eq:Landau_eq_WNT_B})
obtained via the weakly nonlinear analysis. In practice the stationary
solution is perturbed $\mathbf{u}_{st}^{\sharp}+\delta\mathbf{u}^{\sharp}$,
$T_{st}^{\sharp}+\delta T^{\sharp}$ and its stability with respect
to infinitesimal perturbations is studied by consideration of an eigenvalue
problem for the growth rate $s$ of the perturbations. By the use
of the following decomposition of the velocity perturbation
involving vertically irrotational and rotational parts of the horizontal component\footnote{This can be viewed as the standard poloidal-toroidal decomposition of solenoidal fields, $\delta\mathbf{u}^{\sharp}=\nabla^{\sharp}\times\left(\varPsi\hat{\mathbf{e}}_{z}\right)+\nabla^{\sharp}\times\nabla^{\sharp}\times\left(\varPsi_{pol}\hat{\mathbf{e}}_{z}\right)$, where the poloidal component is separated into a horizontal part $\nabla_{h}^{\sharp}\varPhi$ and a vertical part $\delta u_{z}^{\sharp}\hat{\mathbf{e}}_{z}$, by introduction of new variables $\varPhi=\partial_z^{\sharp}\varPsi_{pol}$ and $\delta u_{z}^{\sharp}=-\nabla_h^{\sharp 2}\varPsi_{pol}$. The decomposition (\ref{eq:decomposition_of_uh_WNT_B}) is allowed (and mathematically equivalent to the poloidal-toroidal decomposition) once the divergence-free condition on the original velocity field $\nabla^{\sharp}\cdot\delta\mathbf{u}^{\sharp}=0$ expressed in (\ref{eq:div_eq_WNT-1}) is imposed.}
\begin{equation}
\delta\mathbf{u}^{\sharp}=\nabla_{h}^{\sharp}\varPhi+\nabla^{\sharp}\times\left(\varPsi\hat{\mathbf{e}}_{z}\right)+\delta u_{z}^{\sharp}\hat{\mathbf{e}}_{z},\label{eq:decomposition_of_uh_WNT_B}
\end{equation}
the eigenvalue problem can be expressed in the following way 
\begin{subequations}
\begin{align}
s\nabla^{\sharp2}\delta u_{z}^{\sharp}-\hat{\mathbf{e}}_{z}\cdot\nabla^{\sharp}\times\nabla^{\sharp}\times\left[\left(\delta\mathbf{u}^{\sharp}\cdot\nabla^{\sharp}\right)\mathbf{u}_{st}^{\sharp}+\left(\mathbf{u}_{st}^{\sharp}\cdot\nabla^{\sharp}\right)\delta\mathbf{u}^{\sharp}\right]\qquad\qquad\nonumber \\
=Ra_{crit}\left(1+\eta\right)Pr\nabla_{h}^{\sharp2}\delta T^{\sharp}+Pr\nabla^{\sharp4}\delta u_{z}^{\sharp},\quad\label{eq:NS_eq_WNT-1}
\end{align}
\begin{equation}
s\nabla_{h}^{\sharp2}\varPsi-\hat{\mathbf{e}}_{z}\cdot\nabla^{\sharp}\times\left[\left(\delta\mathbf{u}^{\sharp}\cdot\nabla^{\sharp}\right)\mathbf{u}_{st}^{\sharp}+\left(\mathbf{u}_{st}^{\sharp}\cdot\nabla^{\sharp}\right)\delta\mathbf{u}^{\sharp}\right]=Pr\nabla^{\sharp2}\nabla_{h}^{\sharp2}\varPsi,\label{eq:NS_eq_WNT-1-1}
\end{equation}
\begin{equation}
\nabla_{h}^{\sharp2}\varPhi=-\frac{\partial\delta u_z^{\sharp}}{\partial z},\label{eq:div_eq_WNT-1}
\end{equation}
\begin{equation}
s\delta T^{\sharp}+\delta\mathbf{u}^{\sharp}\cdot\nabla^{\sharp}T_{st}^{\sharp}+\mathbf{u}_{st}^{\sharp}\cdot\nabla^{\sharp}\delta T^{\sharp}-\delta u_{z}^{\sharp}=\nabla^{\sharp2}\delta T^{\sharp}.\label{eq:T_eq_WNT-1}
\end{equation}
\end{subequations} 
An approach of this type has been undertaken e.g. by Schl$\ddot{\textrm{u}}$tter
\emph{et al}. (1965), Busse (1967)\footnote{Busse (1967) utilizes a simplifying assumption of infinite $Pr$.}
with the results summarized in Busse (1978) and Getling (1998) for
three types of planform: two-dimensional rolls, squares and hexagons,
were the boundaries were assumed isothermal, impermeable and either rigid or stress-free. It has
been shown, that for purely symmetric systems with respect to the
mid-plane the first stable solution takes the form of two-dimensional
rolls when the Rayleigh number slightly exceeds the critical value for convection threshold. These rolls become unstable when the Rayleigh number is increased
even further via a variety of instability types, such as e.g. the
\emph{Zigzag instability}, which leads to sinusoidal curving of the
rolls, the \emph{Cross-roll instability} which forms a new system
of rolls perpendicular to the initial ones which eventually take over
or the \emph{Eckhaus instability}. The latter instability type is
the only one that does not break the two-dimensionality of the flow
and leads to phase alteration, that is compression and expansion of
groups of rolls along their wave vector. Which instability sets in
depends on the value of the Rayleigh number and the set of wave numbers
in a wave packet of the perturbation. For a comprehensive list of the
types of instabilities of 2D roll structures see $\S$\textbf{6} of
Getling (1998). However, it was also reported, that when the up-down
symmetry is broken by either (i) allowing for a weak temperature variation
of the expansion coefficient $\alpha(T)$, of the viscosity $\nu(T)$
or of the thermal diffusivity $\kappa(T)$, (ii) imposing different boundary
conditions at the top and bottom boundaries or (iii) including the effect
of free surface curvature, the hexagonal pattern is preferred.

Utilizing the same decomposition as in (\ref{eq:decomposition_of_uh_WNT_B}),
Manneville (1983) has derived an equation for the pattern structure
function $\check{u}_{1z}$ (convection amplitude) near convection
onset for the case of stress-free, isothermal, impermeable boundaries
by assuming the following expansions of variables 
\begin{subequations}
\begin{equation}
\delta u_{z}^{\sharp}=\sum_{n}\check{u}_{nz}\left(x^{\sharp},y^{\sharp},t^{\sharp}\right)\sin\left(n\pi z^{\sharp}\right),\label{eq:e137}
\end{equation}
\begin{equation}
\delta T^{\sharp}=\sum_{n}\check{T}_{nz}\left(x^{\sharp},y^{\sharp},t^{\sharp}\right)\sin\left(n\pi z^{\sharp}\right),\label{eq:e138}
\end{equation}
\begin{equation}
\varPsi=\sum_{n}\check{\varPsi}_{nz}\left(x^{\sharp},y^{\sharp},t^{\sharp}\right)\cos\left(n\pi z^{\sharp}\right),\label{eq:e139}
\end{equation}
\end{subequations} 
and applying the Galerkin method\footnote{The Galerkin method is essentially based on expansion of variables
in a complete set of basis functions, introduction of the expansions
into the equations and equating their inner product with each of the
basis functions to zero.} within the scope of weakly nonlinear theory. This study was concerned
with the primary instabilities, thus $\mathbf{u}_{st}^{\sharp}=0$
and the nonlinear terms $\left(\delta\mathbf{u}^{\sharp}\cdot\nabla^{\sharp}\right)\delta\mathbf{u}^{\sharp}$
and $\delta\mathbf{u}^{\sharp}\cdot\nabla^{\sharp}\delta T^{\sharp}$
likewise explicit time derivatives were included in the equations
(\ref{eq:NS_eq_WNT-1}-d) to yield \footnote{Note the different temperature scale assumed by Manneville (1983),
which corresponds to $L\Delta_{S}/Ra$.}
\begin{align}
\tau_{0}\frac{\partial\check{u}_{1z}}{\partial t^{\sharp}}= & \left[\eta-\frac{4}{3\pi^{4}}\left(\nabla_{h}^{\sharp2}+\mathcal{K}_{crit}^{\sharp2}\right)^{2}\right]\check{u}_{1z}-\frac{1}{6\pi^{4}}\check{u}_{1z}\left[\left(\nabla_{h}^{\sharp}\check{u}_{1z}\right)^{2}+\mathcal{K}_{crit}^{\sharp2}\check{u}_{1z}^{2}\right]\nonumber \\
 & -\tau_{0}\left(\check{u}_{0x}\frac{\partial}{\partial x^{\sharp}}+\check{u}_{0y}\frac{\partial}{\partial y^{\sharp}}\right)\check{u}_{1z},\label{eq:e140}
\end{align}
\begin{equation}
\left(\frac{\partial}{\partial t^{\sharp}}-Pr\nabla_{h}^{\sharp2}\right)\nabla_{h}^{\sharp2}\check{\varPsi}_{0}=\frac{1}{\mathcal{K}_{crit}^{\sharp2}}\left[\frac{\partial\check{u}_{1z}}{\partial y^{\sharp}}\frac{\partial\left(\nabla_{h}^{\sharp2}\check{u}_{1z}\right)}{\partial x^{\sharp}}-\frac{\partial\check{u}_{1z}}{\partial x^{\sharp}}\frac{\partial\left(\nabla_{h}^{\sharp2}\check{u}_{1z}\right)}{\partial y^{\sharp}}\right],\label{eq:e141}
\end{equation}
where $\mathcal{K}_{crit}^{\sharp}=2/\sqrt{\pi}$ and
\begin{equation}
\check{u}_{0x}=\frac{\partial\check{\varPsi}_{0}}{\partial y^{\sharp}},\qquad\check{u}_{0y}=-\frac{\partial\check{\varPsi}_{0}}{\partial x^{\sharp}}\label{eq:e143}
\end{equation}
so that $\check{\varPsi}_{0}$ is the stream function of the horizontal flow $(\check{u}_{0x},\,\check{u}_{0y})$, z-independent by definition (\ref{eq:e137}-c). After solving
the amplitude equations the flow and temperature are given by 
\begin{subequations}
\begin{equation}
u_{x}^{\sharp}=\check{u}_{0x}+\frac{2}{\pi}\frac{\partial\check{u}_{1z}}{\partial x}\cos\left(\pi z\right)-\frac{3\left(8+3Pr\right)}{64Pr\pi^{4}}\frac{\partial}{\partial x}\left[\left(\nabla_{h}\check{u}_{1z}\right)^{2}+\mathcal{K}_{crit}^{2}\check{u}_{1z}^{2}\right]\cos\left(2\pi z\right),\qquad\label{eq:e144}
\end{equation}
\begin{equation}
u_{y}^{\sharp}=\check{u}_{0y}+\frac{2}{\pi}\frac{\partial\check{u}_{1z}}{\partial y}\cos\left(\pi z\right)-\frac{3\left(8+3Pr\right)}{64Pr\pi^{4}}\frac{\partial}{\partial y}\left[\left(\nabla_{h}\check{u}_{1z}\right)^{2}+\mathcal{K}_{crit}^{2}\check{u}_{1z}^{2}\right]\cos\left(2\pi z\right),\qquad\label{eq:e145}
\end{equation}
\begin{equation}
u_{z}^{\sharp}\approx\check{u}_{1z}\sin\left(\pi z\right)+\frac{3\left(8+3Pr\right)}{128Pr\pi^{5}}\nabla_{h}^{2}\left[\left(\nabla_{h}\check{u}_{1z}\right)^{2}+\mathcal{K}_{crit}^{2}\check{u}_{1z}^{2}\right]\sin\left(2\pi z\right),\qquad\label{eq:e146}
\end{equation}
\begin{align}
T^{\sharp}= & \frac{2}{3\pi^{2}}\check{u}_{1z}\left(x,y\right)\sin\left(\pi z\right)\nonumber \\
 & +\frac{1}{6\pi^{5}}\left[\frac{9\left(8+3Pr\right)}{256Pr\pi^{2}}\nabla_{h}^{2}-1\right]\left[\left(\nabla_{h}\check{u}_{1z}\right)^{2}+\mathcal{K}_{crit}^{2}\check{u}_{1z}^{2}\right]\sin\left(2\pi z\right),\label{eq:e147}
\end{align}
\end{subequations} 
with the accuracy
up to terms of the order $\mathcal{O}(\eta^{3/2})$ (see (\ref{eq:eta_def_WNT}) for the definition of $\eta$). A similar equation for the pattern structure function,
but for the case of both rigid boundaries held at a fixed heat flux,
when the nearly marginal flow varies on very large horizontal length
scales (as known from the linear theory, cf. section \ref{subsec:Two-rigid-fixed-flux}) was
derived by Proctor (1981). It was later generalized by Knobloch (1990) to include
asymmetric top-bottom boundary conditions and temperature variation
of fluid properties (such as $\nu$, $\alpha$ and $\kappa$). They
assumed doubly periodic lattice and utilized perturbative expansions
in square of the inverse length scale of horizontal variation, which
is equivalent to expansions in square of the wave number of the planform
$\mathcal{K}^{\sharp2}=\mathcal{K}_{x}^{\sharp2}+\mathcal{K}_{y}^{\sharp2}\ll1$
\begin{equation}
T^{\sharp}\left(x,y,z,t\right)=T_{0}\left(x,y,t\right)+\mathcal{K}^{2}T_{2}\left(x,y,z,t\right)+\dots,\label{eq:e148}
\end{equation}
where the leading order term for temperature is independent of $z$
by virtue of the temperature equation. For symmetric, fixed-heat flux
boundary conditions and uniform fluid properties it has been shown,
that $\partial_{t}\sim\mathcal{K}^{\sharp4}$ and $Ra-Ra_{crit}\sim\mathcal{\mathcal{K}}^{\sharp2}$
and the pattern structure function obeys the following equation
\begin{equation}
\frac{\partial T_{0}}{\partial t}=-\eta\nabla_{h}^{2}T_{0}-\frac{34}{231}\nabla_{h}^{4}T_{0}+\frac{10}{7}\nabla_{h}\cdot\left(\left|\nabla_{h}T_{0}\right|^{2}\nabla_{h}T_{0}\right),\label{eq:e149}
\end{equation}
where $\eta=(Ra-Ra_{crit})/Ra_{crit}$ denotes the departure from
threshold, as before. Proctor (1981) showed, that small departures
from constant heat flux at the boundaries lead to stability of square-cell
solutions over the entire range of validity of the approximation. Knobloch
(1990), however, demonstrated that temperature variation of fluid
properties and asymmetric boundary conditions can result in stable
patterns in the forms of rolls, squares and hexagons, depending on
values of system parameters.

Finally we note an interesting result concerning upper bounds on heat
transport by nearly marginal convection which extends to turbulent
regime, obtained for rigid and isothermal boundaries, which was reported
by Busse (1969). It was conjectured, that the dynamics of convection
is controlled by stability of the boundary layers and importance of
the horizontal scale of convection has been emphasized. Busse (1969)
generalized the results of Howard (1963) for a variational problem
of finding a maximum of the convective heat transport $\left\langle u_{z}T'\right\rangle$ at
a given Rayleigh number, by introducing a structure of successive
boundary layers to adjust the horizontal length scale of variation
of the convective velocities and temperature from a boundary to the
interior value. The latter was assumed comparable with the interior
vertical variation length scale which allowed to minimize dissipation
and maximize the heat transport. This resulted in a sequence of upper
bounds for the Nusselt number realized for different ranges of the
Rayleigh number, well approximated by the following three formulae
\begin{eqnarray}
Nu\leq1+0.1252\,Ra^{3/8} & \;\textrm{for}\; & Ra<2.642\times10^{4},\nonumber \\
Nu\leq1+0.0482\,Ra^{15/32} & \;\textrm{for}\; & 2.642\times10^{4}\leq Ra<4.644\times10^{5},\qquad\quad\label{eq:Nusselt_estimates_Busse}\\
Nu\leq1+0.0311\,Ra^{1/2} & \;\textrm{for}\; & Ra\geq4.644\times10^{5}.\nonumber 
\end{eqnarray}
The latter upper bound on the Nusselt number obtained by Busse (1969),
namely $Nu\leq1+0.0311\,Ra^{1/2}$ is indeed satisfied for all $Ra\geq4.644\times10^{5}$,
even in a fully turbulent regime for extremely high values of the Rayleigh
number. This can be verified through comparison of this estimate with
heat flux estimates obtained from the theory of Grossman and Lohse
(2000) and Stevens $\textit{et al.}$ (2013) for fully developed turbulent
convection. However, the theory of Grossman and Lohse, based on data
from a large number of laboratory and numerical experiments, provides
far more precise estimates than (\ref{eq:Nusselt_estimates_Busse}),
and therefore more useful within the ranges of their validity, which
depend on both, the Rayleigh number and the Prandtl number.

\section{Fully developed convection\label{sec:Fully-developed-convection}}

The fully nonlinear, developed convection at very high Rayleigh numbers
is a very common in nature, but at the same time extremely complex
phenomenon, for which a complete mathematical description is not achievable.
The same is of course true for many turbulent flows of a different
physical origin than convection, which results from the fact, that
a mathematically rigorous and complete description of a fully developed
turbulence, despite intensive attempts from over a century still remains
beyond the grasp of the present-day fluid mechanics. Nevertheless, the
physical picture of developed convection, obtained through a mixture
of application of fully rigorously derived relations, some scaling
arguments and experimental/numerical data was obtained in the comprehensive
study of Grossman and Lohse (2000), later updated in Stevens $\textit{et al.}$
(2013). It is the intention of this section to utilize their theory
to explain the physics and major dynamical aspects of developed convection;
an interested reader is referred to the seminal work of Grossman and
Lohse (2000) for all the detailed information about scaling laws for
the convective heat flux (measured by the Nusselt number) and the
magnitude of the convective velocities (measured by the Reynolds number)
with the Rayleigh number, and their thorough derivation.

After Grossamn and Lohse (2000) let us introduce the following simplified
physical setting. The system we consider is a layer of fluid described
by the perfect gas equation, confined between two flat parallel
boundaries at distance $L$, with spatially uniform vertical gravity
$\mathbf{g}=-g\hat{\mathbf{e}}_{z}$ pointing downwards and no radiative
heat sources, $Q=0$. This is the configuration often used in numerical
experiments. All the physical properties of the fluid, such as $k$
(heat conductivity), $\mu$ (viscosity), $\alpha$ (thermal expansion)
and the heat capacities $c_{p}$, $c_{v}$ are assumed uniform. Standard
no-slip conditions are imposed at the boundaries, which are impermeable.
The fluid is heated from below and the boundaries are held at constant
temperatures. The equations governing the evolution of such system,
under the Boussinesq approximation are given in (\ref{eq:NS_B_final}-c)
with $Q=0$; note, that the adiabatic gradient in the energy equation
is typically negligible in comparison with the temperature gradient
of the static state in experimental situations but very large in the
case of many natural systems such as the Earth's and planetary cores,
stellar interiors, etc. The Nusselt and Rayleigh numbers are therefore
as defined in (\ref{eq:Nu_def_B}) and (\ref{eq:Ra_def_B}), but since
in the absence of heat sources the static temperature profile is linear
in $z$, the expressions simplify to\index{SI}{Nusselt number!Boussinesq}\index{SI}{Rayleigh number!Boussinesq}
\begin{equation}
Nu=\frac{\bar{\rho}c_{p}\left\langle u_{z}T\right\rangle _{h}-k\partial_{z}\left\langle T\right\rangle _{h}-kg/c_{p}}{k\Delta_{S}},\label{eq:Nu_def_B-1}
\end{equation}
\begin{equation}
Ra=\frac{g\alpha\Delta_{S}L^{4}}{\kappa\nu},\label{eq:Ra_def_B-1}
\end{equation}
where $\kappa=k/\bar{\rho}c_{p}=\mathrm{const}$ and the superadiabatic
gradient is now uniform,
\begin{equation}
\Delta_{S}=\frac{\Delta T}{L}-\frac{g}{\bar{c}_{p}}>0,\label{eq:e150}
\end{equation}
and $\Delta T=T_{B}-T_{T}>0.$ Although Grossman and Lohse (2000)
do not include the adiabatic gradient in their considerations, their
entire analysis and results can also be applied to the more general
situation considered here, but with the static state temperature profile
replaced by the superadiabatic profile $\bar T+\dbtilde{T}(z)+gz/c_{p}$, thus effectively
$\Delta T$ replaced by $L\Delta_{S}$. 
\begin{figure}
\centering{}a)\includegraphics[scale=0.17]{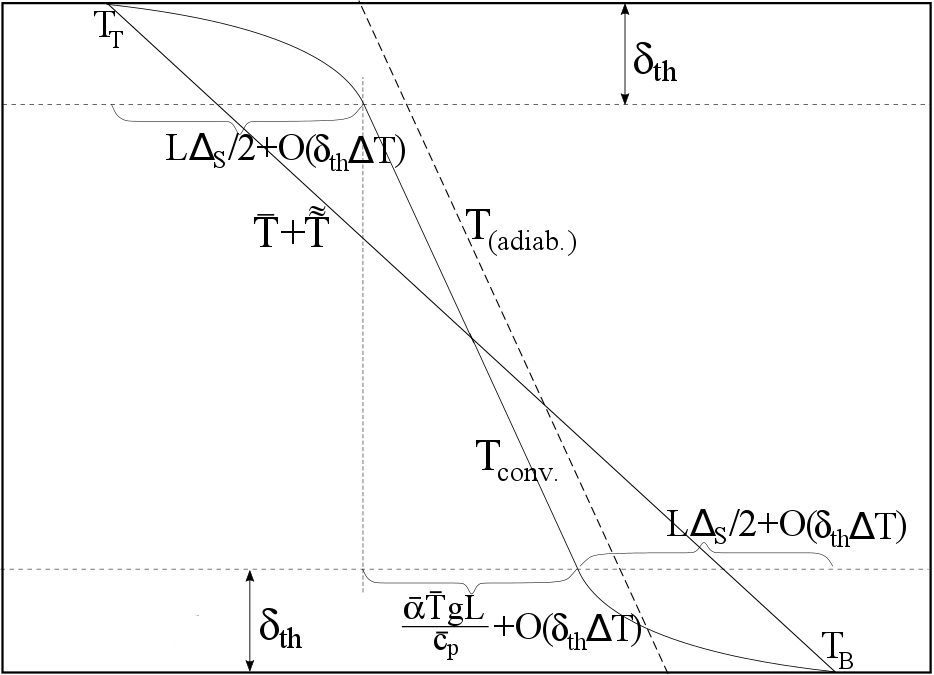}~~~b)\includegraphics[scale=0.17]{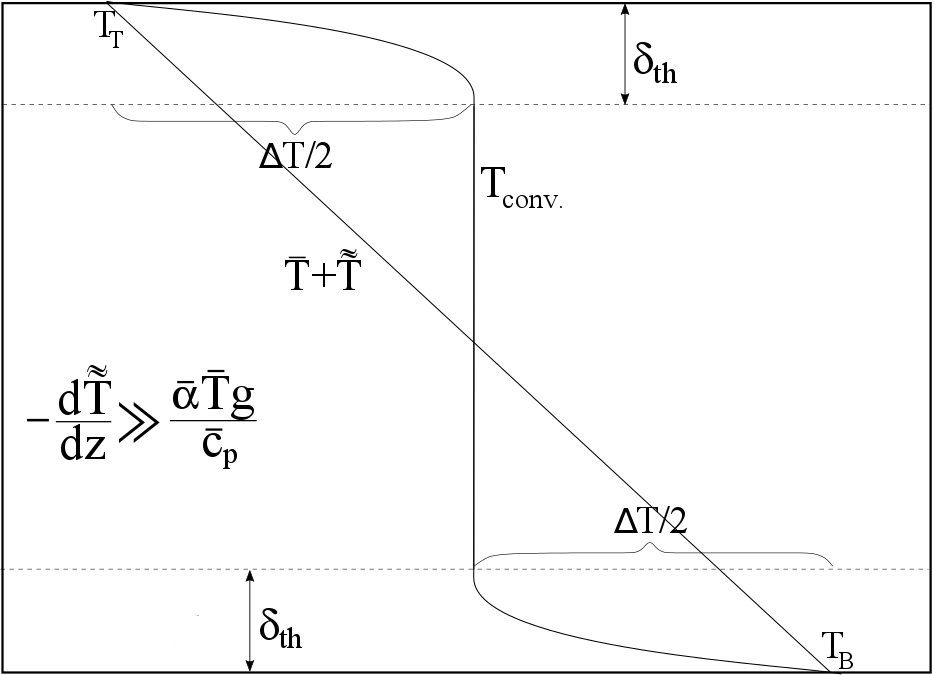}\caption{\label{fig:Boussinesq_profiles_GL2000}{\footnotesize{}A schematic
picture of vertical profiles of the mean temperature $T_{\mathrm{conv.}}=\tilde{T}+\left\langle T'\right\rangle _{h}$,
the hydrostatic reference state temperature $\tilde{T}$ and the adiabatic
profile $T_{(\mathrm{adiab.})}=-gz/c_{p}+\mathrm{const}$ in fully
developed convection with fixed temperature at boundaries and $Q=0$.
Figure a) depicts the general case, whereas figure b) was constructed
to depict the experimentally most common situation, when the adiabatic
gradient $g/c_{p}$ is negligibly small compared to the basic state
temperature gradient $\Delta T/L$. The boundary layers are marked,
which have the same thicknesses at top and bottom. The temperature
jumps across the boundary layers are also the same at top and bottom
and are approximately equal to half of the superadiabatic temperature
jump across the entire fluid layer (up to corrections of order $\mathcal{O}(\delta_{th}\Delta T)$).
Note, that figure b) could also be interpreted as a profile of superadiabatic
temperature $\tilde{T}+\left\langle T'\right\rangle _{h}+gz/c_{p}=T_{B}-\Delta_{S}z+\left\langle T'\right\rangle _{h}$,
with the temperature jumps across the boundary layers $\Delta T/2$
replaced by $L\Delta_{S}/2+\mathcal{O}(\delta_{th}\Delta T)$ and
the top temperature $T_{T}$ replaced by $T_{B}-L\Delta_{S}$.}}
\end{figure}

The physical picture of fully developed convection which emerges from
a large number of theoretical, numerical and experimental studies
can be summarised as follows. The average temperature profile in developed
convection for the considered physical setting is depicted on figure
\ref{fig:Boussinesq_profiles_GL2000} along with the adiabatic temperature
profile and that of the hydrostatic, diffusive state. Both latter
profiles are linear whereas in developed convection the system separates
into three different regions - two thermal boundary layers and the
bulk. The convective temperature profile in the bulk is bound to be
very close to the adiabatic one because of the efficient mixing by
the vigorous convective flow, with a gradient only slightly above
the adiabatic one, which nevertheless is enough to drive vigorous
convection. In other words conduction, $-k\partial_{z}\left\langle T'\right\rangle _{h}$,
is negligible with respect to advection, $\bar{\rho}c_{p}\left\langle u_{z}T\right\rangle _{h}$,
in the bulk\footnote{\label{fn:B_cond_may_balance_adv_in_bulk}Nevertheless, the conductive
and advective fluxes associated with fluctuations about the horizontal
means may be in balance, as happens in the case when the total thermal
dissipation is dominated by its bulk contribution considered below
among other cases; the fluctuations about the horizontal means are
small compared to the means in a well-mixed, turbulent bulk.}. Hence the temperature is advected by the vigorous convective flow in the bulk
with very little heat losses, i.e. like an almost conserved quantity
following the motion of fluid parcels. This is why the bulk has been
called a thermal shortcut in the literature (cf. Grossman and Lohse
2000). The top and bottom boundary layers adjust the almost adiabatic
bulk profile to the boundary conditions, that is $T_{T}$ and $T_{B}$
respectively, which results in formation of large temperature gradients
(see figure \ref{fig:Boussinesq_profiles_GL2000}). In the boundary
layers the flow is not as efficient as in the bulk and heat advection
is balanced by the enhanced diffusion, since the horizontally averaged
vertical temperature gradient can significantly exceed the adiabatic
one. The temperature gradient must decrease with the distance from
the boundary, therefore the thickness of the boundary layer is established
by the distance from the boundary on which conduction, $-k\partial_{z}\left\langle T\right\rangle _{h}$,
is strong enough to balance (or overcome) advection, $\bar{\rho}c_{p}\left\langle u_{z}T\right\rangle _{h}$.
Bearing in mind the Boussinesq up-down symmetry this allows to estimate
the total, horizontally averaged superadiabatic convective heat flux as $-kL\Delta_{S}/2\delta_{th}$
and thus the thickness of thermal boundary layers as\index{SI}{boundary layer thickness!thermal boundary layer}
\begin{equation}
\frac{\delta_{th}}{L}=\frac{1}{2Nu},\label{eq:thermal_thickness}
\end{equation}
where the factor of a half results from the fact, that the temperature
jumps across the two top and bottom boundary layers must be the same,
thus equal to $L\Delta_{S}/2$ and both the boundary layers must have
the same thicknesses. 

\begin{figure}
\centering{}\includegraphics[scale=0.2]{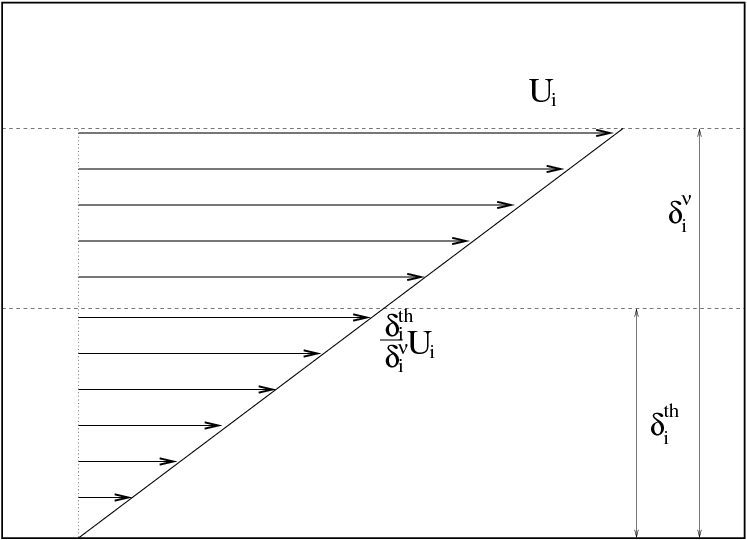}\caption{\label{fig:nesting_of_BLs_B_GL2000}{\footnotesize{}A schematic picture
of nested boundary layers. When the thermal layer is nested in the
viscous one, which occurs for $Pr>1$, the velocity magnitude for
the thermal layer needs to be rescaled with a ratio of the boundary
layers thicknesses, $\delta_{th}/\delta_{\nu}<1$, due to the approximately
linear velocity profile in the boundary layers which adjusts the wind
of turbulence $\mathscr{U}$ to the zero velocity at boundary.}}
\end{figure}

The central idea of the Grossmann and Lohse (2000) theory for convection between rigid, isothermal boundaries introduced
on the basis of a vast experimental and numerical evidence involves
the existence of a mean convective flow termed the ``wind of turbulence'',
that is a large-scale convection roll, for which the velocity scale
will be denoted by $\mathscr{U}$. This scale is used to estimate
the advective terms in the viscous boundary layers of the Blasius
type, which in turn allows to use the standard estimate for the viscous,
laminar boundary layer thickness, \index{SI}{boundary layer thickness!Blasius boundary layer}
\begin{equation}
\frac{\delta_{\nu}}{L}=Re^{-1/2},\qquad Re=\frac{\mathscr{U}L}{\nu},\label{eq:Reynolds_number}
\end{equation}
characterized
by the Reynolds number $Re$ which measures the ratio of inertia to viscous diffusion. Similar
type estimates of temperature advection and thermal diffusion can
be made for the thermal boundary layers, which means that the thickness
of the thermal boundary layer can be also characterized by the P$\acute{\textrm{e}}$clet
number, $Pe=\mathscr{U}_{th}L/\kappa$, measuring the ratio of heat
advection to heat conduction, thus implying a relation between $Pe$
and $Nu$. There is a subtle difference, however, since the velocity
scale $\mathscr{U}_{th}$ used to estimate heat advection is not necessarily
the same as the scale$\mathscr{U}$ in the estimate of the advection
of momentum, the latter being the mean large-scale convective flow (the wind
of turbulence) which stirs the bulk and is responsible for creation of viscous
boundary layers. The scale $\mathscr{U}_{th}$ is related to $\mathscr{U}$
and it depends on whether the viscous layer is nested inside the thermal
one, $\delta_{\nu}<\delta_{th}$, or the thermal layer is nested inside
the viscous one, $\delta_{th}<\delta_{\nu}$ as shown on figure \ref{fig:nesting_of_BLs_B_GL2000}.
In the latter case, which corresponds to large Prandtl numbers $Pr=\nu/\kappa$,
the wind of turbulence has to be scaled with the boundary layer thicknesses
ratio $\delta_{th}/\delta_{\nu}<1$, to obtain the thermal layer velocity
scale in the form $\mathscr{U}_{th}=\mathscr{U}\delta_{th}/\delta_{\nu}$.
On the other hand, when $\delta_{\nu}<\delta_{th}$ there is no need
for rescaling the velocity in the thermal layer, as the layer is directly
influenced by the wind, so that $\mathscr{U}_{th}=\mathscr{U}$ in
this case. Moreover, as it will become evident from the following
analysis, a simple estimate of the thermal layer thickness as $Pe^{-1/2}$
leading to a similar advection-diffusion balance as in the viscous
layer is not always valid, which suggests a significant role of large
horizontal gradients of temperature in the dynamics of thermal layers
in some cases.

The main objective now is the derivation of the scaling laws for the
Nusselt (\ref{eq:Nu_def_B-1}) and Reynolds, $Re=\mathscr{U}L/\nu$
numbers with $Ra$ (\ref{eq:Ra_def_B-1}) in developed convection.
The form of the scaling laws crucially depends on whether the thermal
and viscous dissipation take place predominantly in the boundary layers
or in the bulk of the convection. At a very high Rayleigh number the strongly
turbulent convection becomes dominated by a vast number of small-scale
structures with strong gradients spread across the entire bulk, in
which case the dissipation takes place in the entire fluid domain.
Grossmann and Lohse (2000) estimate it to be the case at $Ra\gtrsim10^{14}$.
However, before such a regime is reached, that is at $10^{7}\lesssim Ra\lesssim10^{14}$,
the turbulence is not as strong yet and the dissipation may be
dominated by the large gradients in the top and bottom boundary layers,
despite the fact, that they occupy only a small fraction of the entire
fluid volume. First, we provide estimates of the bulk contributions
to the viscous and thermal dissipation rates. When strong dissipation
takes place in the entire fluid domain, with the large-scale wind
of turbulence stirring the fluid, the Kolmogorov picture of turbulent
energy cascade can be applied to estimate the viscous and thermal
dissipation with the magnitude of the relevant nonlinear term in the
Navier-Stokes and energy equations respectively 
\begin{subequations}
\begin{equation}
2\nu\left\langle \mathbf{G}^{s}:\mathbf{G}^{s}\right\rangle \approx\frac{\mathscr{U}^{3}}{L}=\frac{\nu^{3}}{L^{4}}Re^{3},\label{eq:visc_diss_estimate}
\end{equation}
\begin{align}
\kappa\left\langle \left(\nabla T'\right)^{2}\right\rangle  & \approx\mathscr{U}\frac{\left(L\Delta_{S}\right)^{2}}{L}=\kappa\Delta_{S}^{2}Pe=\kappa\Delta_{S}^{2}RePr,\nonumber \\
 & \qquad\qquad\qquad\qquad\qquad\qquad\textrm{when}\quad\delta_{\nu}<\delta_{th},\;Pr\textrm{-small},\label{eq:therm_diss_estimate_1}
\end{align}
\begin{align}
\kappa\left\langle \left(\nabla T'\right)^{2}\right\rangle  & \approx\frac{\delta_{th}}{\delta_{\nu}}\mathscr{U}\frac{\left(L\Delta_{S}\right)^{2}}{L}=\kappa\Delta_{S}^{2}Pe=\frac{1}{2}\kappa\Delta_{S}^{2}Nu^{-1}Re^{3/2}Pr,\nonumber \\
 & \qquad\qquad\qquad\qquad\qquad\qquad\textrm{when}\quad\delta_{\nu}>\delta_{th},\;Pr\textrm{-large},\label{eq:therm_diss_estimate_2}
\end{align}
\end{subequations} 
where
\begin{equation}
Pr=\frac{\nu}{\kappa},\quad Pe=\frac{\mathscr{U}_{th}L}{\kappa},\quad\delta_{\nu}=LRe^{-1/2},\label{eq:parameters_Pe_Pr_deltanu}
\end{equation}
and the thermal boundary layer thickness is given in (\ref{eq:thermal_thickness}).
By the use of (\ref{eq:thermal_diss_vs_Nusselt}) we can already see,
that when the total thermal dissipation can be estimated by the bulk contributions
we have $Nu-1\approx Pe$; in turbulent convection the Nusselt number is typically much greater than unity, therefore $Nu-1\approx Nu$. Consequently the thickness of the thermal boundary layers can be estimated with the use of the P$\acute{\textrm{e}}$clet number as $\delta_{th}\approx1/2Pe$ for all
values of the Prandtl number (that is for both cases of $\delta_{\nu}<\delta_{th}$
and $\delta_{\nu}>\delta_{th}$)\footnote{\label{ft:TS}Note, that the temperature equation\index{SI}{temperature equation!Boussinesq} (\ref{eq:Energy_B_final}) can in fact be rewritten in terms of the superadiabatic temperature $T_S=\tilde T + T' +gz/c_p= T_B - \Delta_S z+T'$, to yield
\[
\frac{\partial T_S}{\partial t}+\mathbf{u}\cdot\nabla T_S=\kappa\nabla^2 T_S,
\]
where we have used the current assumptions $Q=0$, $\kappa=\mathrm{const}$ and $\alpha T=1$, the latter being a general property of a perfect gas. It follows, that the estimate $\left\langle \left(\nabla T_S\right)^{2}\right\rangle\approx \Delta_S^2 Pe$, analogous to (\ref{eq:therm_diss_estimate_1},c) still holds for the case of bulk-dominated thermal dissipation and since for isothermal boundaries we have $\left\langle \left(\nabla T'\right)^{2}\right\rangle =\left\langle \left(\nabla T_S\right)^{2}\right\rangle - \Delta_S^2$, inspection of (\ref{eq:thermal_diss_vs_Nusselt}) allows to obtain the estimate $Nu\approx Pe$ in an even more straightforward way. We stress again, that it is valid only for the case when thermal dissipation takes place predominantly in the bulk.}. This suggests that in the case of bulk-dominated thermal dissipation the horizontal length scales of temperature variation in the thermal boundary layers are
similar to the vertical scale of temperature variation (thus very small, of the order $\delta_{th}$), since advection has to balance diffusion
in the boundary layers.

Next we turn to the case, when the total viscous and thermal dissipation
are dominated by the contributions from boundary layers. Estimating
the magnitude of velocity gradients in the boundary layers by $\mathscr{U}/\delta_{\nu}$
a straightforward integration of the expression for the total viscous
dissipation gives
\begin{equation}
2\nu\left\langle \mathbf{G}^{s}:\mathbf{G}^{s}\right\rangle \approx\nu\frac{\mathscr{U}^{2}}{\delta_{\nu}^{2}}\frac{\delta_{\nu}}{L}\approx\frac{\nu^{3}}{L^{4}}Re^{5/2},\label{eq:visc_diss_BL}
\end{equation}
where the factor of $\delta_{\nu}/L$ results from vertical integration
and the fraction of the total volume occupied by the boundary layers
gives the dominant contribution to dissipation. A similar estimate
for the thermal dissipation leads to a tautology, that is
\begin{equation}
\kappa\left\langle \left(\nabla T'\right)^{2}\right\rangle \approx\kappa\frac{\left(L\Delta_{S}\right)^{2}}{\delta_{th}^{2}}\frac{\delta_{th}}{L}\approx\kappa\Delta_{S}^{2}Nu,\label{eq:Tdiss_estimate_BL}
\end{equation}
which is equivalent to the rigorous relation (\ref{eq:thermal_diss_vs_Nusselt}) (up to neglection of unity with respect to $Nu\gg1$)\footnote{See also the footnote (\ref{ft:TS}) for $\left\langle \left(\nabla T_S\right)^{2}\right\rangle \approx\Delta_{S}^{2}Nu$, which is exactly equivalent to (\ref{eq:thermal_diss_vs_Nusselt}).},
thus no new information is provided. However, it is worth noting,
that the above estimate in (\ref{eq:Tdiss_estimate_BL}) can always
be applied solely to the boundary layers, that is with the total average
replaced by the horizontal one and vertical integration over the thermal
boundary layers only, $0<z<\delta_{th}$ and $L-\delta_{th}<z<L$.
This means, that in fact the thermal dissipation in the thermal boundary
layers is always of the same order of magnitude in terms of the non-dimensional
numbers $Nu$ or $Ra$ as the total thermal dissipation. Hence the formerly considered
regime of strong bulk dissipation in (\ref{eq:visc_diss_estimate}-c)
corresponds to at most a moderate domination of the bulk contribution
to the total dissipation over contributions from boundary layers;
the equations (\ref{eq:visc_diss_estimate}-c) are nevertheless valid,
even if the boundary layer dissipation is comparable to the one in
the bulk. 

It is however possible for the boundary layer contribution
to the total thermal dissipation to dominate over the bulk one. In
such a case, to obtain the necessary relations between the numbers $Nu$,
$Re$ and $Ra$, $Pr$ one needs to consider the advection-diffusion
balance within the thermal boundary layers. Estimating the horizontal
temperature gradients by $\Delta_{S}$ and the diffusive term by $\kappa L\Delta_{S}/\delta_{th}^{2}$
one obtains 
\begin{subequations}
\begin{equation}
\mathscr{U}\Delta_{S}=\frac{\kappa L\Delta_{S}}{\delta_{th}^{2}}\;\Rightarrow\;Nu\approx\frac{1}{2}Re^{1/2}Pr^{1/2},\qquad\textrm{when}\quad\delta_{\nu}<\delta_{th},\;Pr\textrm{-small},\label{eq:therm_diss_estimate_1_BL}
\end{equation}
\begin{equation}
\frac{\delta_{th}}{\delta_{\nu}}\mathscr{U}\Delta_{S}=\frac{\kappa L\Delta_{S}}{\delta_{th}^{2}}\;\Rightarrow\;Nu\approx\frac{1}{2}Re^{1/2}Pr^{1/3},\qquad\textrm{when}\quad\delta_{\nu}>\delta_{th},\;Pr\textrm{-large}.\label{eq:therm_diss_estimate_2_BL}
\end{equation}
\end{subequations}
It is noteworthy, that the above estimates (\ref{eq:therm_diss_estimate_1_BL},b)
in conjunction with the exact equation (\ref{eq:thermal_diss_vs_Nusselt})
allow to estimate the thickness of the thermal boundary layers with
the P$\acute{\textrm{e}}$clet number as $\delta_{th}\approx Pe^{-1/2}$
for all values of the Prandtl number. This resembles the Blasius layer
relation $\delta_{\nu}=Re^{-1/2}$, the reason for that being that
the estimates of horizontal and vertical gradients ($\nabla_h\sim1/L$, $\partial_z\sim1/\delta_{th}$) utilized here are of same type as in the standard Blasius boundary layer theory, hence lead
to the same type of final distinguished balance. We recall, that
in the formerly considered regime of extremely large Rayleigh numbers
we obtained $\delta_{th}\approx(2Pe)^{-1}$ (see discussion below
(\ref{eq:parameters_Pe_Pr_deltanu}) and the footnote (\ref{ft:TS})), which suggests, that in that
regime the boundary layer balance on the far left of relations (\ref{eq:therm_diss_estimate_1_BL},b)
has to be modified by a new estimate of the horizontal gradients,
as large as $L\Delta_{S}/\delta_{th}$.

\begin{figure}
\begin{centering}
\includegraphics[scale=0.35]{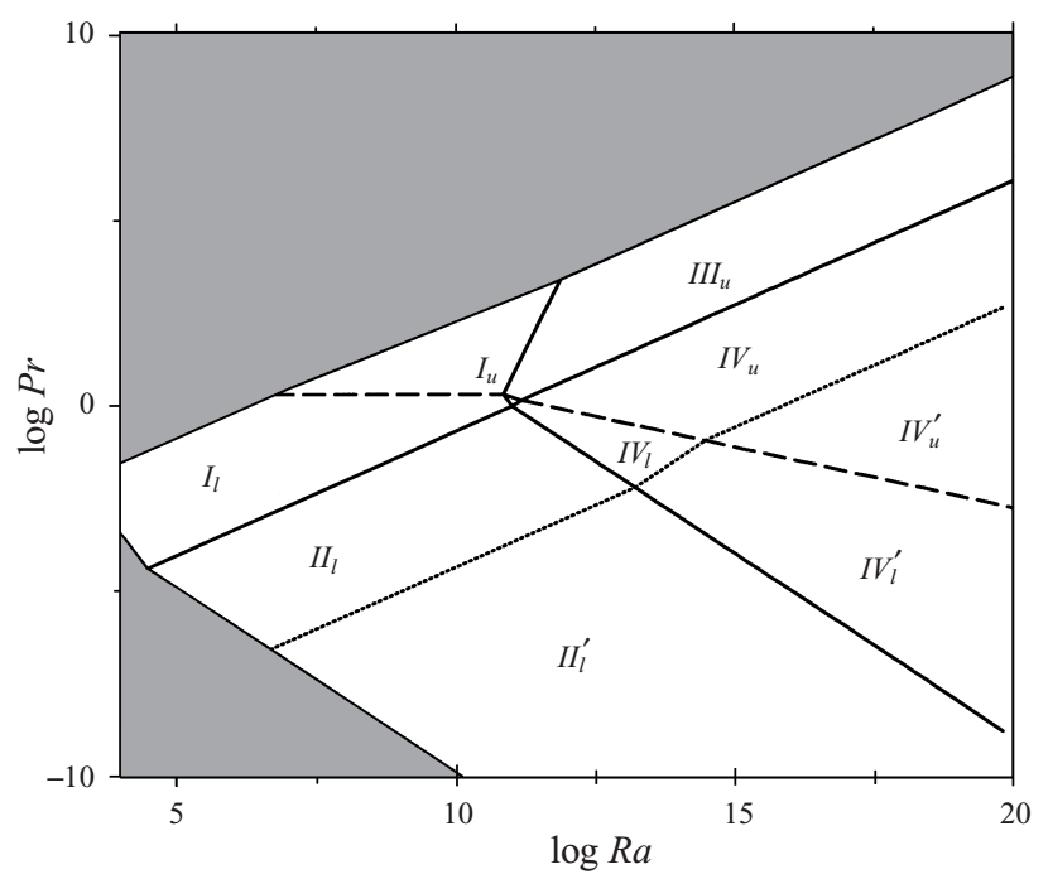}\caption{\label{fig:PrRa_plane_diagram_GL2000}{\footnotesize{}A diagram of
distinct dynamical regimes for developed Boussinesq convection with
fixed temperature on boundaries and no heat sources $Q=0.$ The different
regimes are determined by different combinations of dominant contributions
to viscous and thermal dissipation (cf. (\ref{eq:visc_diss_estimate}-c)
for bulk estimates and (\ref{eq:visc_diss_BL}) and (\ref{eq:therm_diss_estimate_1_BL}-b)
for boundary layer estimates). The dashed line, obtained by equating
$\delta_{th}=LNu^{-1}$ and $\delta_{\nu}=LRe^{-1/2}$ divides the
parameter space into regions with low and high $Pr$ denoted by subscripts
$l$ and $u$ respectively. The gray regions correspond to regimes
of not fully developed turbulence (after Grossmann and Lohse 2000).}}
\par\end{centering}
\end{figure}

Having now the complete set of relations between the parameters $Nu$
and $Re$, which describe the response of the system to driving described
by $Ra$ with system properties included in $Pr$, we can equate the
exact expressions for viscous and thermal dissipation in (\ref{eq:thermal_diss_vs_Nusselt})
and (\ref{eq:kin_E_variation_B-1}) with their estimates for different
dynamical regimes in (\ref{eq:visc_diss_estimate}-c), (\ref{eq:visc_diss_BL})
and (\ref{eq:therm_diss_estimate_1_BL}-b). This results in four different
basic regimes of turbulent convection, in which the dominant contributions
to total viscous or thermal dissipation come either from the bulk
or the boundary layers, which sums up to four cases symbolically denoted
as $\nu_{BL}\kappa_{BL}$ (regime I), $\nu_{bulk}\kappa_{BL}$ (regime
II), $\nu_{BL}\kappa_{bulk}$ (regime III) and \emph{$\nu_{bulk}\kappa_{bulk}$,}
(regime IV), where \emph{BL} stands for boundary layer. Figure \ref{fig:PrRa_plane_diagram_GL2000}
taken from Grossmann and Lohse (2000) depicts the regimes on the $Pr$-$Ra$
plane. Each of the regimes is further divided into a ``lower'' regime
achieved at small values of the Prandtl number when $\delta_{\nu}<\delta_{th}$
and an ``upper'' regime obtained at large $Pr$ when $\delta_{\nu}>\delta_{th}$.
The scaling relations\index{SI}{scaling laws} $Nu(Ra,\,Pr)$ and $Re(Ra,\,Pr)$ obtained from
equating the exact expressions and estimates of dissipation are therefore
distinct in all the regimes $I_{l}$, $I_{u}$, $II_{l}$,... $IV_{u}$.
E.g. in the bulk-dominated regime at small Prandtl number one obtains
$Nu\sim Ra^{1/2}Pr^{1/2}$ and $Re\sim Ra^{1/2}Pr^{-1/2}$ whereas
in the regime dominated by dissipation in boundary layers at small
Prandtl number $Nu\sim Ra^{1/4}Pr^{1/8}$ and $Re\sim Ra^{1/2}Pr^{-3/4}$.
Furthermore, the boundaries between the basic regimes $I$, $II$,
$III$ and $IV$ are calculated by equating the relevant dissipation
estimates, e.g. the boundaries between $I$ and $II$ and between
$III$ and $IV$ are obtained by equating the viscous dissipation
estimates (\ref{eq:visc_diss_BL}) and (\ref{eq:visc_diss_estimate})
but with the Reynolds numbers expressed in terms of the $Ra$ and
$Pr$ numbers through the corresponding scalings; the boundary between $I_{u}$
and $III_{u}$ is calculated by equating (\ref{eq:therm_diss_estimate_2_BL}) and
(\ref{eq:therm_diss_estimate_2}), again, expressed by $Ra$ and $Pr$
with the use of the corresponding scalings, etc. The dashed line on
figure \ref{fig:PrRa_plane_diagram_GL2000}, obtained by equating
$\delta_{th}=LNu^{-1}$ and $\delta_{\nu}=LRe^{-1/2}$ divides the
parameter space into regions with low and high $Pr$ denoted by subscripts
$l$ and $u$ respectively. The primed regimes marked below the dotted
line, reached at high enough values of $Ra$ are characterized by
already turbulent boundary layers and therefore are all bulk-dominated.
In other words the scalings in regime $II_{l}^{\prime}$ are the same
as in $IV_{l}$. A complete derivation and discussion of the scaling
laws for all different possible dynamical regimes is provided in Grossmann
and Lohse (2000), where also the prefactors for scaling relations
are obtained based on comparison with laboratory and numerical experiments,
later updated in Stevens \emph{et al}. (2013).

\section{Validity of the approximation and summary\label{sec:Validity-B}}

Let us briefly recall all the necessary assumptions for validity of
the Boussinesq approximation. \emph{Firstly} it must be required,
that the density, temperature and pressure scale heights satisfy
\begin{equation}
L\ll\min\left\{ D_{\rho},\,D_{T},\,D_{p}\right\} ,\label{eq:LvsScaleHeights_B-1}
\end{equation}
everywhere in the fluid domain. Using this assumption we define the small
parameter
\begin{equation}
\epsilon=\frac{\Delta\dbtilde{\rho}}{\bar{\rho}}\ll1,\label{eq:epsilon_def_B-1}
\end{equation}
which is small by virtue of integration of relation $L/D_{\rho}\ll1$
from the level of minimal to the level of maximal density within the
fluid layer and the density jump between these layers is denoted by
$\Delta\dbtilde{\rho}$. A \emph{second restriction} concerns the
magnitude of the fluctuations of thermodynamic variables induced by
the convective motions, which effectively implies
\begin{equation}
\frac{\rho'}{\bar{\rho}}=\mathcal{O}\left(\epsilon\right),\quad\frac{T'}{\bar{T}}=\mathcal{O}\left(\epsilon\right),\quad\frac{p'}{\bar{p}}=\mathcal{O}\left(\epsilon^{2}\right),\label{eq:e151}
\end{equation}
\begin{equation}
\frac{\dbtilde{\rho}}{\bar{\rho}}=\mathcal{O}\left(\epsilon\right),\quad\frac{\dbtilde{T}}{\bar{T}}=\mathcal{O}\left(\epsilon\right),\quad\frac{\dbtilde{p}}{\bar{p}}=\mathcal{O}\left(\epsilon\right),\quad\frac{\bar{\rho}gL}{\bar{p}}=\mathcal{O}\left(\epsilon\right),\label{fluct_magnitude_B-1}
\end{equation}
(an assumption, which in principle needs to be checked for consistency
\emph{a posteriori}) and for the magnitude of convective velocity $\mathscr{U}$, the
time scales $\mathscr{T}$, the diffusivities and the heat capacity
\begin{equation}
\mathscr{U}\sim\epsilon^{1/2}\sqrt{gL},\qquad\mathscr{T}\sim\epsilon^{-1/2}\sqrt{\frac{L}{g}},\label{eq:vel_and_time_scales-2}
\end{equation}
\begin{equation}
\mu_{b}/\bar{\rho}\lesssim\nu\sim\epsilon^{1/2}\sqrt{gL}L,\quad\kappa=\frac{k}{\bar{\rho}\bar{c}_{p}}\sim\epsilon^{1/2}\sqrt{gL}L,\quad\bar{c}_{p}\sim\bar{c}_{v}\sim\epsilon^{-1}\frac{gL}{\bar{T}}.\label{eq:visc_scale-2}
\end{equation}
A \emph{third weakest restriction} concerns the fluid thermodynamic
properties,
\begin{equation}
C_{T}\gg\sqrt{gL},\quad\textrm{or equivalently}\quad\frac{1}{g\bar{\rho}\bar{\beta}}\gg L.\label{eq:third_assumption-1}
\end{equation}
The latter assumption is easily satisfied for most of fluids by virtue of the first assumption (\ref{eq:epsilon_def_B-1}), in particular obviously satisfied for a weakly stratified perfect gas. The set of above assumptions
leads to the following set of momentum, energy and mass balance supplied
by the equation of state, under the Boussinesq approximation at leading
order in $\epsilon$ 
\begin{subequations}
\begin{equation}
\frac{\partial\mathbf{u}}{\partial t}+\left(\mathbf{u}\cdot\nabla\right)\mathbf{u}=-\frac{1}{\bar{\rho}}\nabla p'+g\bar{\alpha}T'\hat{\mathbf{e}}_{z}+\nu\nabla^{2}\mathbf{u}+2\nabla\nu\cdot\mathbf{G}^{s},\label{eq:NS_B_final-2}
\end{equation}
\begin{equation}
\nabla\cdot\mathbf{u}=0,\label{eq:Mass_cons_B_final-2}
\end{equation}
\begin{equation}
\frac{\partial T'}{\partial t}+\mathbf{u}\cdot\nabla T'+u_{z}\left(\frac{\mathrm{d}\dbtilde{T}}{\mathrm{d}z}+\frac{g\bar{\alpha}\bar{T}}{\bar{c}_{p}}\right)=\nabla\cdot\left(\kappa\nabla T'\right)+\frac{Q'}{\bar{\rho}\bar{c}_{p}},\label{eq:Energy_B_final-3}
\end{equation}
\[
\rho=\bar{\rho}+\dbtilde{\rho}+\rho'=\bar{\rho}\left[1-\bar{\alpha}\dbtilde{T}+\bar{\beta}\dbtilde{p}-\bar{\alpha}T'\right].
\]
\end{subequations} 
The system (\ref{eq:NS_B_final-2}-c) is a closed
set of equations for the velocity field $\mathbf{u}$, temperature
fluctuation $T'$ and the pressure fluctuation $p'$, whereas the state
equation can be used for calculation of the density fluctuation $\rho'=-\bar{\alpha}T'$.
The equations describing the hydrostatic state (denoted by the subscript
$0$) are given in (\ref{eq:Static_NS},b) and (\ref{eq:rho_0_B}). 

Finally, in what follows a short summary of main aspects of Boussinesq
convection is provided. The general, useful definition of the Rayleigh
number is given in the form
\begin{equation}
Ra=\frac{g\bar{\alpha}\left\langle \kappa\Delta_{S}\right\rangle L^{4}}{\bar{\kappa}^{2}\bar{\nu}},\label{eq:Ra_def_B-2}
\end{equation}
which is a measure of the magnitude of the superadiabatic gradient,
thus of the driving force for convective flow. Two Boussinesq systems
are dynamically equivalent when the Rayleigh numbers and the Prandtl
numbers $Pr=\nu/\kappa$ are the
same for both systems (likewise in the case when some additional
effects are present any other nondimensional number describing the
relative strength of those effects, such as e.g. the Ekman number
$E=\nu/2\Omega L^{2}$ describing the effect of rotation); in such a case the response of the systems
to driving must be the same. One possible measure of the dynamical
response of a convective system to driving, in the absence of heat
sources $Q=0$ is provided by the Nusselt number, which is defined
as a ratio of the total superadiabatic convective heat flux over the
superadiabatic molecular heat flux in a corresponding hydrostatic
state\footnote{Such a hydrostatic state is, of course, unstable, since it corresponds
to the same driving $\Delta_{S}$ as that in the convective
state.} in the following way
\begin{equation}
Nu=\frac{\bar{\rho}\bar{c}_{p}\left\langle u_{z}T\right\rangle _{h}-k\partial_{z}\left\langle T\right\rangle _{h}-\bar{k}g\bar{\alpha}\bar{T}/\bar{c}_{p}}{\left\langle k\Delta_{S}\right\rangle }=\frac{L}{2\delta_{th}},\label{eq:Nu_def_B-2}
\end{equation}
for isothermal boundaries and
\begin{equation}
Nu_{Q}=\frac{\left\langle \bar{\rho}\bar{c}_{p}u_{z}T'\right\rangle }{\left\langle k\Delta_{S}\right\rangle }=\frac{1}{2}\left(\frac{L}{\delta_{th}}-1\right)\approx\frac{L}{2\delta_{th}},\label{eq:e152}
\end{equation}
when the heat flux at boundaries is held fixed. A characteristic feature
of Boussinesq systems is that when $Q=0$ the total, horizontally
averaged heat flux through every horizontal plane is the same, in
other words the total heat flux is height independent. $\delta_{th}$
denotes the thickness of the thermal boundary layers which form at
the top and bottom in fully developed convection with large temperature
gradients to account for the large convective heat flux in the bulk
of the flow. 

We also recall here the results of section \ref{subsec:Conservarion-of-mass}.
The conservation of mass implies, that in order for the density and
thus temperature fluctuations to be correctly resolved, the jump of
the mean pressure fluctuation across the depth of the fluid layer
must vanish at all times, i.e. $\left\langle p'\right\rangle _{h}(z=L)-\left\langle p'\right\rangle _{h}(z=0)=0$.
This constitutes a boundary condition, which must be imposed on the
pressure field.

Furthermore, there are two possible types of up-down symmetry in Boussinesq
convection with symmetric top-bottom boundary conditions: the \emph{nonlinear
up-down symmetry}, the only one allowed by the nonlinear equations,
defined by
\begin{align}
u_{z}\left(L-z\right) & =-u_{z}\left(z\right),\quad\mathbf{u}_{h}\left(L-z\right)=\mathbf{u}_{h}\left(z\right),\label{eq:symmetry_nonlin_1-1}\\
T'\left(L-z\right) & =-T'\left(z\right),\quad p'\left(L-z\right)=p'\left(z\right),\label{eq:symmetry_nonlin_1-1b}
\end{align}
and corresponding to developed convection, and the \emph{linear up-down
symmetry}, opposite to the previous one
\begin{align}
u_{z}\left(L-z\right) & =u_{z}\left(z\right),\quad\mathbf{u}_{h}\left(L-z\right)=-\mathbf{u}_{h}\left(z\right),\label{eq:symmetry_lin_1-1}\\
T'\left(L-z\right) & =T'\left(z\right),\quad p'\left(L-z\right)=-p'\left(z\right),\label{eq:symmetry_lin_1-1b}
\end{align}
allowed only by the linearised dynamical equations and corresponding
to large-scale convective rolls at convection threshold. The critical
Rayleigh number for convection, in the absence of heat sources $Q=0$,
for $\nu=\mathrm{const.}$, $\kappa=\mathrm{const.}$ and for the
case when the instability sets in as stationary flow can be calculated
from the following formula
\begin{equation}
Ra_{crit}=\frac{L^{4}\mathrm{min}\left[-\frac{1}{\mathcal{K}^{2}}\left(\frac{\mathrm{d}^{2}}{\mathrm{d}z^{2}}-\mathcal{K}^{2}\right)^{3}\hat{u}_{z}\right]}{\hat{u}_{z}},\quad\textrm{at any }z\label{eq:Boussinesq_general_condition_for_Rac-1}
\end{equation}
once the flow at threshold is obtained as a solution of the dynamical
equations with the growth rate $\sigma=0$. A more general formula
for $Q\neq0$, for $\nu=\nu(z)$, $\kappa=\kappa(z)$ and $\bm{\Omega}\neq0$,
thus including the possibility of oscillatory flow at convection threshold
can be formulated to yield
\begin{align}
Ra_{crit}= & \frac{\bar{\alpha}g\langle\kappa\Delta_{S}\rangle L^{4}}{\bar{\kappa}^{2}\bar{\nu}}\nonumber \\
= & \mathrm{min}\,\Re\mathfrak{e}\,\left\langle \frac{\kappa(z)}{\bar{\kappa}}\frac{\left[\mathrm{i}\omega-\kappa(z)\left(\frac{\mathrm{d}^{2}}{\mathrm{d}z^{2}}-\mathcal{K}^{2}\right)\right]\left(\frac{\bar{\alpha}gL^{4}\hat{T}}{\bar{\kappa}\bar{\nu}}\right)-\frac{\bar{\alpha}gL^{4}\hat{Q}}{\bar{\rho}\bar{c}_{p}\bar{\kappa}\bar{\nu}}}{\hat{u}_{z}}\right\rangle ,\label{eq:Boussinesq_Delta_crit-1}
\end{align}
where the frequency of oscillations at threshold $\omega$ as a function
of the horizontal wave number $\mathcal{K}$ and the inverse vertical
variation scale (vertical ``wave number'') of perturbations $q$
(say), $\omega=\omega(\mathcal{K},q)$ can be determined by equating
the imaginary part of the expression in the angular brackets on the
right hand side of the latter equation to zero,
\begin{equation}
\Im\mathfrak{m}\,\left\{ \frac{\kappa(z)}{\bar{\kappa}}\frac{\left[\mathrm{i}\omega-\kappa(z)\left(\frac{\mathrm{d}^{2}}{\mathrm{d}z^{2}}-\mathcal{K}^{2}\right)\right]\left(\frac{\bar{\alpha}gL^{4}\hat{T}}{\bar{\kappa}\bar{\nu}}\right)-\frac{\bar{\alpha}gL^{4}\hat{Q}}{\bar{\rho}\bar{c}_{p}\bar{\kappa}\bar{\nu}}}{\hat{u}_{z}}\right\} =0.\label{eq:e153}
\end{equation}
In the marginal state the heat per unit mass released by a rising
fluid parcel on an infinitesimal vertical distance in a time unit
can be expressed by $\bar{c}_{p}\kappa\nu Ra_{crit}u_{z}/g\bar{\alpha}L^{4}$.
For growth rates of perturbations to the hydrostatic state in the
vicinity of convection threshold, i.e for $Ra-Ra_{crit}\ll Ra_{crit}$, $\mathcal{K}L-\mathcal{K}_{crit}L\ll1$,
see the general relation (\ref{eq:grate_gen_B}) and equations (\ref{eq:grate_BCisothSF_B})
and (\ref{eq:sigma_approximate_WNT}) for the particular case of stress-free
and isothermal boundaries.

\section*{Review exercises}

{\textbf{Exercise 1.}} \\
Formulate the fundamental assumptions which lead to the Boussinesq system of equations and derive the elliptic Poisson-type equation for pressure.

\noindent\emph{Hint}: The assumptions are formulated in (\ref{eq:LvsScaleHeights_B}) and (\ref{fluct_magnitude_B}) (or \ref{eq:rho_scaling_B}-d), supplied by (\ref{eq:visc_scale}), (\ref{eq:e74}) and (\ref{eq:e77}).
\\

\noindent{\textbf{Exercise 2.}} \\
Calculate the mean value of the entropy fluctuation in the Boussinesq convection, $\langle s'\rangle$.

\noindent\emph{Hint}: cf. the equation (\ref{eq:entropy_temperature_in_Boussinesq}) and section \ref{subsec:Conservarion-of-mass}.
\\

\noindent{\textbf{Exercise 3.}} \\
Consider Boussinesq fluid of viscosity 
\[
\nu=\nu_0\left(1-\frac{z^2}{2L^2}\right),
\] convectively driven by a fixed temperature difference between bottom and top plates $\Delta T$, under uniform gravity $g$. The mean temperature $\bar{T}$, mean density $\bar{\rho}$ and the mean specific heat $\bar{c}_p$ are given; the fluid is described by the equation of state of the perfect gas. The heat flux of a hydrostatic state is also given and is denoted by $F_0$. The thermal diffusivity is uniform. Given the total viscous dissipation rate $D_{\nu}$ calculate the total convective heat flux in the system $F_{conv}$. Then calculate the Rayleigh number. 

\noindent \emph{Hint}: utilize the results of section \ref{subsec:fixed_DeltaT} to show that $F_{conv}=\langle k\Delta_S\rangle (Nu-1)= \bar{\rho}\bar{c}_p\bar{T}D_{\nu}/g$ and 
\[
Ra=\frac{6g\Delta T^2L^2}{5\bar{T}F_0\nu_0}\left(1-\frac{gL}{\bar{c}_p\Delta T}\right).
\]
\\

\noindent{\textbf{Exercise 4.}} \\
Calculate the growth rate of convection near threshold in a system rotating at the angular velocity $\boldsymbol{\Omega}=\Omega \hat{\mathbf{e}}_z$.

\noindent\emph{Hint}: calculate roots of the cubic equation (\ref{eq:grate_no_Q_B}).
\\

\noindent{\textbf{Exercise 5.}} \\
By the use of the weakly nonlinear theory derive the relation between the Nusselt and Rayleigh numbers near convection threshold.

\noindent\emph{Hint}: cf. section \ref{sec:WNT}.

\chapter{Anelastic convection\label{chap:Anelastic-convection}}

The Boussinesq approximation is applicable to thin layers of fluid,
where the density variation is weak. However, in most astrophysical
applications such a thin layer approximation is not satisfactory,
because the typical scale heights in the system, $D_{p}$, $D_{\rho}$
and $D_{T}$ (cf. equation (\ref{eq:scale_heights_def})) are comparable
or even significantly smaller than the depth of the convective layer.
Still, the phenomenon of convection is very common in natural systems
and it is in fact a crucial factor in the dynamics of stellar and
planetary interiors and atmospheres, since the convectively driven
flow transports energy and angular momentum. Moreover, convection
in electrically conducting domains within the stellar and planetary
interiors (cores) is responsible for the hydromagnetic dynamo effect,
in other words generation of magnetic fields of those astrophysical
bodies (see, e.g. Soward 1991, Tobias and Weiss 2007a,b). This results
in a strong need for an accurate description of convection in systems
with strong density stratification. The fully compressible models
are very cumbersome due to inclusion of the dynamics of fast sound
waves, which needs to be thoroughly resolved, hence it is desirable,
that the mathematical description of planetary and stellar convection
allows to filter out the sound waves, similarly as in the case of
the Boussinesq approximation. In order to satisfy the needs for a
sound-proof description of strongly stratified convection the anelastic
approximation was formulated by Ogura and Phillips (1962) and Gough
(1969). Ogura and Phillips (1962) also invented the name \emph{anelastic}
approximation, based on the fact, that what they called an 'elastic'
part of the internal energy of the fluid can be neglected. From a
more general point of view the distinction of the 'elastic energy'
is not necessarily strictly definite, nevertheless the name anelastic
approximation has been established through a wide use over many decades. 

The anelastic approximation was later even more thoroughly explained
and generalized to the magnetohydrodynamic case by Lantz and Fan (1999).
The main idea underlying this approach is that guided by the observation
that natural large-scale convective systems in their long-time evolution
develop states that are nearly adiabatic, and only a slight excess
above the adiabatic gradient drives a very vigorous convective flow,
a fundamental assumption of weak superadiabaticity of the dynamical
system is put forward. This implies, that the Mach number, i.e. the
ratio of convective velocity to the speed of sound is small, but the
temperature and density stratification can be arbitrary. There is
a large amount of scientific literature on the properties of anelastic
convection. Although we do not intend here to provide a complete review
of the developments on various dynamical aspects of convection under
the anelastic approximation, it is without a doubt of interest to
direct an interested reader to some of the most important findings.
In a series of papers Gilman and Glatzmaier (1981) and Glatzmaier
and Gilman (1981a, b) have studied the influence of various physical
effects and conditions, such as the dissipative effects, the boundary
conditions and zone depth on the dynamics of anelastic convection.
They have also studied the linear convection onset in spherical shells,
which was further investigated by Drew \emph{et al}. (1995). It is
important to point out at this stage, that recently Calkins \emph{et
al}. (2015) and Verhoeven and Glatzmaier (2018) have demonstrated,
that in rapidly rotating low-Prandtl number systems the anelastic
approximation breaks down. More precisely when the rate of background
rotation is fast enough, that the value of $\Omega L$ starts to exceed the
mean speed of sound $\bar{C}$ and for $\nu/\kappa\ll1$ the growth
rate of the convective instability at threshold is greatly enhanced
and the time derivative of the density fluctuation in the continuity
equation ceases to be negligible. In turn the sound waves start to
play a dynamical role (cf. (\ref{eq:scalings_SWfiltration}) and the
discussion below) and thus the evolution obtained from the anelastic
system of equations does not capture a crucial factor of the dynamics;
therefore in this limit the anelastic approximation is clearly not
applicable. Nevertheless, for non-rotating systems or systems for
which the background rotation is not as rapid, i.e. satisfies $\Omega L\ll\bar{C}$
the anelastic approximation was never shown to break down and the restriction
$\Omega L\ll\bar{C}$ for its validity is by no means strong.

Furthermore, a notable development concerning adaptation of the anelastic
formulation to the dynamics of the Earth's core, together with a comprehensive
discussion of the dynamics of geophysical convection was done in a
seminal paper by Braginsky and Roberts (1995). Even further developments
on the topic, including comparison of the anelastic and Boussinesq
approaches in the Earth's core context have been performed by Anufriev
\emph{et al.} (2005); a more general comparison of the two sound-proof
approaches can be found in Lilly (1996).

From the point of view of dynamics of compressible atmospheres the
issue of stratification in the hydrostatic reference state is an important
one. Noteworthy models developed for description of dynamics of convection
in compressible atmospheres include e.g. Wilhelmson and Ogura (1972),
Lipps and Henler (1982) and Durran (1989). The latter was further
analysed and explained by Durran (2008), Klein (2009), Klein \emph{et
al}. (2010) and Klein and Puluis (2012). A survey of different approaches
to stratified convection in compressible atmospheres was provided
by Bannon (1996). 

In the following sections of this chapter we consider a general case
of an arbitrary equation of state and provide a thorough discussion
of all undertaken steps necessary to derive the anelastic equations;
possible further simplifications in special cases are proposed. The
linear convection close to convection threshold likewise the general
energetic characteristics of anelastic convection and nonlinear heat
transfer at fully developed state are described. Finally a comparison
of different approaches with an adiabatic and non-adiabatic reference
temperature profiles is provided and a relation between the anelastic
and Boussinesq approximations is explained.

\section{Derivation of the anelastic equations\label{sec:Derivation-A}}

The anelastic liquid approximation is commonly applied to the study
of planetary and stellar interiors where convection takes place in
large and heavy fluid regions. Those regions typically have comparable
mass with the entire celestial body that they are a part of. Therefore
typically gravity also significantly varies with depth. Moreover,
convection disturbs the mass distribution in the fluid layers providing
further corrections to the acceleration of gravity. Consequently  variation of gravity should, in general, be included (allowed) in the mathematical description of anelastic convection. The full set of dynamical equations expressing
the physical laws of conservation of momentum, mass and energy, supplied
by the equations of state and the gravitational potential equation takes
the form 
\begin{subequations}
\begin{align}
\rho\left[\frac{\partial\mathbf{u}}{\partial t}+\left(\mathbf{u}\cdot\nabla\right)\mathbf{u}\right]= & -\nabla p+\rho\mathbf{g}+\mu\nabla^{2}\mathbf{u}+\left(\frac{\mu}{3}+\mu_{b}\right)\nabla\left(\nabla\cdot\mathbf{u}\right)\nonumber \\
 & \qquad\qquad+2\nabla\mu\cdot\mathbf{G}^{s}+\nabla\left(\mu_{b}-\frac{2}{3}\mu\right)\nabla\cdot\mathbf{u},\label{NS-Aderiv}
\end{align}
\begin{equation}
\frac{\partial\rho}{\partial t}+\nabla\cdot\left(\rho\mathbf{u}\right)=0,\label{Cont_Aderiv}
\end{equation}
\begin{equation}
\nabla^{2}\psi=4\pi G\rho,\qquad\mathbf{g}=-\nabla\psi,\label{eq:grav_pot_aderiv}
\end{equation}
\begin{equation}
\rho T\left(\frac{\partial s}{\partial t}+\mathbf{u}\cdot\nabla s\right)=\nabla\cdot\left(k\nabla T\right)+2\mu\mathbf{G}^{s}:\mathbf{G}^{s}+\left(\mu_{b}-\frac{2}{3}\mu\right)\left(\nabla\cdot\mathbf{u}\right)^{2}+Q,\label{Energy_Aderiv}
\end{equation}
\begin{equation}
\rho=\rho(p,T),\quad s=s(p,T),\label{State_eq_Aderiv}
\end{equation}
\end{subequations} 
where $G$ is the gravitational constant and $\psi$
is the gravitational potential. Once again let us assume that the
shear dynamical viscosity $\mu$, the bulk viscosity $\mu_{b}$, the
specific heat $c_{v}$ and thermal conduction $k$ are nonuniform
for generality, however, the latter is a function of depth only. Furthermore,
we take the $z$-axis of the coordinate system so that it is perpendicular
to the plates at $z=0,\,L$. For time-independent boundary conditions
we decompose the thermodynamic variables into the static (reference)\index{SI}{reference (basic) state}
state contributions, denoted by an upper tilde (now incorporating both,
the mean and the vertically varying static departure from mean, previously
denoted by double tilde) and fluctuations induced by convective
flow denoted by prime, 
\begin{subequations}
\begin{equation}
\rho(\mathbf{x},t)=\tilde{\rho}(z)+\rho'(\mathbf{x},t),\label{eq:rho_gen_A_deriv}
\end{equation}
\begin{equation}
T(\mathbf{x},t)=\tilde{T}(z)+T'(\mathbf{x},t),\label{eq:T_gen_A_deriv}
\end{equation}
\begin{equation}
p(\mathbf{x},t)=\tilde{p}(z)+p'(\mathbf{x},t),\label{eq:p_gen_A_deriv}
\end{equation}
\begin{equation}
s(\mathbf{x},t)=\tilde{s}(z)+s'(\mathbf{x},t),\label{eq:p_gen_A_deriv-1}
\end{equation}
\begin{equation}
\psi(\mathbf{x},t)=\tilde{\psi}\left(z\right)+\psi'(\mathbf{x},t).\label{eq:psi_gen_A_deriv}
\end{equation}
\end{subequations} 
The gravitational acceleration, therefore, is
also decomposed in a similar way
\begin{equation}
\mathbf{g}(\mathbf{x},t)=\tilde{\mathbf{g}}(z)+\mathbf{g}'(\mathbf{x},t)=-\tilde{g}(z)\hat{\mathbf{e}}_{z}+\mathbf{g}'(\mathbf{x},t),\label{eq:e154}
\end{equation}
and since the density distribution in the hydrostatic state is depth-dependent
only, so is the gravitational acceleration, and it is along the $z$-direction;
we have chosen the direction of the $z$-axis, so that it points vertically
upwards, thus $\tilde{\mathbf{g}}(z)=-\tilde{g}(z)\hat{\mathbf{e}}_{z}$.
In consequence, from this point onwards we will call the $z$-direction
vertical. Note, that this implies, that the density of the bottom
and top bodies occupying volumes at $z<0$ and at $z>L$ is by assumption
homogeneous in the horizontal directions, so that they only generate
vertical gravity. 

For the sake of simplicity, let us now make an assumption, that the
top body at $z>L$ is light enough not to influence the total gravity.
Such an assumption corresponds to the situation in the planetary and
stellar interiors, where because of the roughly spherical geometry
the exterior regions of the body do not influence gravitationally
the inner ones in a significant way. This allows to say, that $\tilde{g}(z)>0$
(if the fluid layer is not heavier than the bottom body) and therefore
for the convection to take place the bottom plate must be hotter than
the upper one.

Next, the equation (\ref{eq:grav_pot_aderiv}) written separately
for $\tilde{\mathbf{g}}(z)$ and $\mathbf{g}'$ in the fluid region
$0<z<L$ takes the form
\begin{equation}
-\nabla\cdot\tilde{\mathbf{g}}=4\pi G\tilde{\rho},\qquad-\nabla\cdot\mathbf{g}'=4\pi G\rho'.\label{eq:e155}
\end{equation}
Since $\tilde{\mathbf{g}}$ must be continuous across the top and
bottom interfaces at $z=0,\,L$, the gravitational fluctuation $\mathbf{g}'$
must vanish at the top and bottom
\begin{equation}
\left.g'\right|_{z=0,\,L}=0,\label{eq:e156}
\end{equation}
which is consistent with the equation $-\nabla\cdot\mathbf{g}'=4\pi G\rho'$
and the fact, that the mass conservation equation $\partial_{t}(\tilde{\rho}+\rho')+\nabla\cdot[(\tilde{\rho}+\rho')\mathbf{u}]=0$
implies $\left\langle \rho'\right\rangle =0$. Consequently, the gravity
fluctuation $\mathbf{g}'$ is of the same order of magnitude as $G\rho'L$.
Integration of equation $-\nabla\cdot\tilde{\mathbf{g}}=4\pi G\tilde{\rho}$
over a cuboid $V_{f}(z_{cm};z)=\left\{ (x,y,z'):\,|x|\leq L_{x},\,|y|\leq L_{y},\,z_{cm}<z'\leq z\right\} $
bounded by horizontal planes at $z=z_{cm}$, where $z_{cm}$ denotes
the position of the centre of mass for the fluid layer and the bottom
body, and at the height $z$, yields 
\begin{equation}
\tilde{g}(z)L_{x}L_{y}=4\pi GL_{x}L_{y}\left[\int_{-z_{cm}}^{0}\rho_{b}(z')\mathrm{d}z'+\int_{0}^{z}\tilde{\rho}(z')\mathrm{d}z'\right]=4\pi G\left[M_{bc}+M_{f}(z)\right].\label{eq:gravity_acceleration_A}
\end{equation}
In the above we have denoted by $M_{f}(z)$ the mass of the fluid
occupying the volume $V_{f}(0;z)$ and by $M_{bc}$ the mass of the part of the bottom
body contained within the region $z_{cm}<z<0$ between the level of the centre
of mass and the bottom of the fluid layer ($z_{cm}$ is expected to
lie within the bottom body, so that $\tilde{\mathbf{g}}$ is directed
downwards in the fluid region); $\rho_{b}$ denotes the density of
the bottom body. It is of interest to comment on the limit, when the
bottom body is much heavier than the fluid layer. When the total mass
of the fluid $M_{f}=M_{f}(L)$ is negligible with respect to the mass
of the bottom body $M_{b}$, i.e. $M_{f}/M_{b}\ll1$, then $M_{bc}\approx M_{b}/2$\footnote{Note, that $z_{cm}=-M_{b}L_{b}/2M_{tot}+L_{x}L_{y}\int_{0}^{L}z\tilde{\rho}(z)\mathrm{d}z/M_{tot}$,
where $M_{tot}=M_{b}+M_{f}$ is the total mass of the system fluid-bottom
body and $L_{b}$ is the vertical span of the bottom body. Integration
by parts leads to $\int_{0}^{L}z\tilde{\rho}(z)\mathrm{d}z=LM_{f}/L_{x}L_{y}-\int_{0}^{L}M_{f}(z)\mathrm{d}z/L_{x}L_{y}>0$,
hence in the limit $M_{b}\gg M_{f}$ the term $L_{x}L_{y}\int_{0}^{L}z\tilde{\rho}(z)\mathrm{d}z/M_{tot}$
becomes negligibly small with respect to $M_{b}L_{b}/2M_{tot}$ and
we obtain $z_{cm}\approx-L_{b}/2$; this implies $M_{bc}\approx M_{b}/2$.} 
and at any $0<z\leq L$ we get $M_{f}(z)/M_{bc}\ll1$. It is then
predominantly the contribution $4\pi GM_{bc}/L_{x}L_{y}$ to the gravity
acceleration in (\ref{eq:gravity_acceleration_A}) which is responsible
for creation of buoyancy in the fluid and which drives the fluid motions
when the bottom boundary is heated. Therefore in such a case it is
legitimate to neglect the effect of the fluid on the gravity and assume,
that the gravitational force, which acts on the fluid is generated
solely by the presence of the bottom body, $\mathbf{g}\approx\tilde{\mathbf{g}}\approx2\pi GM_{b}/L_{x}L_{y}\hat{\mathbf{e}}_{z}$ (the gravitational fluctuation is then of the order $g'\sim G M_f \rho'/L_x L_y\tilde{\rho}$, thus its influence is also negligible, $\tilde{\rho}g'/\rho'\tilde{g}\sim M_f/M_b$). In the flat geometry this implies, that $\mathbf{g}$ is approximately constant,
but in the spherical case, corresponding to the convective zones
in celestial bodies, $\mathbf{g}\sim1/r^{2}$, where $r$ measures
the radial distance from the centre of the body.

The general equations of the hydrostatic equilibrium take the following
form \index{SI}{reference (basic) state}
\begin{subequations}
\begin{equation}
\frac{\mathrm{d}\tilde{p}}{\mathrm{d}z}=-\tilde{\rho}\tilde{g},\label{eq:hydrostatic_eq_A_1}
\end{equation}
\begin{equation}
\frac{\mathrm{d}^{2}\tilde{\psi}}{\mathrm{d}z^{2}}=4\pi G\tilde{\rho},\qquad\tilde{\mathbf{g}}=-\frac{\mathrm{d}\tilde{\psi}}{\mathrm{d}z}\hat{\mathbf{e}}_{z},\label{eq:grav_pot_aderiv-1}
\end{equation}
\begin{equation}
\frac{\mathrm{d}}{\mathrm{d}z}\left(k\frac{\mathrm{d}\tilde{T}}{\mathrm{d}z}\right)=-\tilde{Q},\label{eq:hydrostatic_eq_A_2}
\end{equation}
\begin{equation}
\tilde{\rho}=\rho(\tilde{p},\tilde{T}),\quad\tilde{s}=s(\tilde{p},\tilde{T}).\label{eq:hydrostatic_eq_A_3}
\end{equation}
\end{subequations} 
We elaborate on the possible forms of the static
reference state later. We now introduce the two fundamental assumptions
of the anelastic approximation. \emph{Firstly}\index{SI}{anelastic!assumption (1)}, guided by the observations
that the natural large-scale convective systems such as e.g. planetary
and stellar interiors or atmospheres in their long-time evolution
develop states that are nearly adiabatic and only a slight excess
above the adiabatic gradient drives a very vigorous convective flow
we assume that
\begin{equation}
0<\delta\equiv\left\langle \frac{L}{\tilde{T}}\Delta_{S}\right\rangle =-\left\langle \frac{L}{\tilde{T}}\left(\frac{\mathrm{d}\tilde{T}}{\mathrm{\mathrm{d}z}}+\frac{\tilde{g}\tilde{\alpha}\tilde{T}}{\tilde{c}_{p}}\right)\right\rangle \ll1,\label{eq:e157}
\end{equation}
i.e. thermal gradient in the fluid, which undergoes convection is
only weakly superadiabatic. Note, that the general relation
\begin{equation}
\frac{\partial s}{\partial z}=\frac{c_{p}}{T}\frac{\partial T}{\partial z}-\frac{\alpha}{\rho}\frac{\partial p}{\partial z},\label{eq:dsbydz_A_general}
\end{equation}
obtained with the aid of heat capacity definition $c_{p}=T(\partial s/\partial T)_{p}$
and the Maxwell relation $\rho^{2}(\partial s/\partial p)_{T}=(\partial\rho/\partial T)_{p}=-\rho\alpha$
allows to write
\begin{equation}
\frac{\mathrm{d}\tilde{s}}{\mathrm{\mathrm{d}z}}=\frac{\tilde{c}_{p}}{\tilde{T}}\left(\frac{\mathrm{d}\tilde{T}}{\mathrm{\mathrm{d}z}}+\frac{\tilde{g}\tilde{\alpha}\tilde{T}}{\tilde{c}_{p}}\right),\label{eq:dsbydz}
\end{equation}
thus
\begin{equation}
\delta=-\left\langle \frac{L}{\tilde{T}}\left(\frac{\mathrm{d}\tilde{T}}{\mathrm{\mathrm{d}z}}+\frac{\tilde{g}\tilde{\alpha}\tilde{T}}{\tilde{c}_{p}}\right)\right\rangle =-\left\langle \frac{L}{\tilde{c}_{p}}\frac{\mathrm{d}\tilde{s}}{\mathrm{\mathrm{d}z}}\right\rangle \ll1.\label{eq:delta_definition}
\end{equation}
\emph{Secondly}\index{SI}{anelastic!assumption (2)}, similarly as in the case of Boussinesq approximation
we assume that the fluctuations of thermodynamic variables are much
smaller than their static profiles. Since it is the departure from
adiabatic state that drives the convection it is natural to assume
that it is also a measure of the relative magnitude of fluctuations
\begin{equation}
\left|\frac{\rho'}{\tilde{\rho}}\right|\sim\left|\frac{T'}{\tilde{T}}\right|\sim\left|\frac{p'}{\tilde{p}}\right|\sim\left|\frac{s'}{\tilde{s}}\right|\sim\mathcal{O}(\delta)\ll1,\label{fluct_magnitude_A}
\end{equation}
an assumption justified to a large extent by experimental and numerical evidence
which, however, needs to be verified for consistency \emph{a posteriori} in each particular case\footnote{\label{fn:fluct_magnitude_note}Some developments concerning the magnitude
of fluctuations for sound-proof equations turned out possible for
the case of weak solutions. It has been shown, that either in infinite
space or for boundaries absorbing the energy of acoustic waves the
time-dependent solutions remain in proximity to the initial conditions in a sense
of certain integral bounds; in other words the solutions do not departure
too far from the initial state - cf. e.g. the book on this topic by
Feireisl and Novotn$\acute{\textrm{y}}$ (2017). These estimates,
however, have not yet benefited from taking into account the damping
of acoustic waves by dissipative processes, such as viscosities and
thermal conduction (cf. footnote \ref{fn:damping_of_sound_waves}).}. Note, that the equilibrium entropy $\tilde{s}$ consists of two
contributions, $\tilde{s}=\tilde{s}_{0}+\dbtilde{s}(z)$, where the constant entropy $\tilde{s}_{0}=\mathrm{const}$
corresponds to an adiabatic state of uniform entropy. The superadiabatic vertical variation of the entropy in the
hydrostatic equilibrium, denoted here by
$\dbtilde{s}(z)=\mathcal{O}(\delta\tilde{s}_0)$, constitutes only a weak correction to the
mean.

As a consequence the equations of state (\ref{State_eq_Aderiv}) are
now expanded about the hydrostatic equilibrium at every height 
\begin{subequations}
\begin{equation}
\rho=\tilde{\rho}\left[1-\tilde{\alpha}\left(T-\tilde{T}\right)+\tilde{\beta}\left(p-\tilde{p}\right)+\mathcal{O}\left(\delta^{2}\right)\right],\label{eq:Taylor_exp_rho_A}
\end{equation}
\begin{equation}
s=\tilde{s}-\tilde{\alpha}\frac{p-\tilde{p}}{\tilde{\rho}}+\tilde{c}_{p}\frac{T-\tilde{T}}{\tilde{T}}+\mathcal{O}\left(\tilde{c}_{p}\delta^{2}\right),\label{eq:Taylor_exp_entropy_A}
\end{equation}
\end{subequations} 
(where, again the Maxwell relation $\rho^{2}(\partial s/\partial p)_{T}=(\partial\rho/\partial T)_{p}=-\rho\alpha$
was used) which results in 
\begin{subequations}
\begin{equation}
\frac{\rho'}{\tilde{\rho}}=-\tilde{\alpha}T'+\tilde{\beta}p'+\mathcal{O}\left(\delta^{2}\right),\label{eq:Taylor_exp_rho_A-1}
\end{equation}
\begin{equation}
s'=-\tilde{\alpha}\frac{p'}{\tilde{\rho}}+\tilde{c}_{p}\frac{T'}{\tilde{T}}+\mathcal{O}\left(\tilde{c}_{p}\delta^{2}\right).\label{eq:Taylor_exp_entropy_A-1}
\end{equation}
\end{subequations} 
Furthermore, a similar type of argument as the one used to
establish the convective flow magnitude and time scales in the Boussinesq fluid
can also be applied here. The buoyancy force in the momentum balance
(\ref{NS-Aderiv}) i.e. $-\tilde{g}\rho'\hat{\mathbf{e}}_{z}-\tilde{\rho}\nabla\psi'$
drives the fluid motion which implies the flow acceleration of the
order $g\rho'/\tilde{\rho}\sim\delta g$, thus much smaller than the
acceleration of gravity. This, in turn, results in the following convective
velocity\index{SI}{anelastic!velocity scale} and time scales\index{SI}{anelastic!time scale}
\begin{equation}
\mathscr{U}\sim\delta^{1/2}\sqrt{\bar{g}L},\qquad\mathscr{T}\sim\delta^{-1/2}\sqrt{\frac{L}{\bar{g}}},\label{eq:vel_and_time_scales-1}
\end{equation}
and hence the viscosity scales\index{SI}{anelastic!viscosity scales} also have to be small for consistency,
\begin{equation}
\mu_{b}\lesssim\mu\sim\delta^{1/2}\bar{\rho}\sqrt{\bar{g}L}L.\label{eq:visc_scale-1}
\end{equation}
Similarly, the thermal conductivity\index{SI}{anelastic!thermal conductivity scale} coefficient must satisfy
\begin{equation}
k\sim\delta^{1/2}\bar{\rho}\bar{c}_{p}\sqrt{\bar{g}L}L,\label{eq:kappa_scale}
\end{equation}
(note, that contrary to the Boussinesq case, in the anelastic approximation
the thermal conductivity $k$ and the thermal diffusion $\kappa$
are both of the same order of magnitude in terms of the small parameter,
$\delta$). It follows, that by assumption\index{SI}{anelastic!heat source scales}
\begin{equation}
\tilde{Q}\sim k\frac{\bar{T}}{L^{2}}\sim\delta^{1/2}\bar{c}_{p}\bar{\rho}\bar{T}\sqrt{\frac{\bar{g}}{L}},\qquad\frac{Q'}{\tilde{Q}}\sim\delta,\label{eq:radiogenic_A}
\end{equation}
has to be satisfied in order for the anelastic approximation to be
valid. In other words the anelastic approximation can be applied only,
when the radiogenic heating is weak enough not drive the system too
far away from the adiabatic state; i.e. $\tilde{Q}/\bar{c}_{p}\bar{\rho}\bar{T}\sqrt{\bar{g}/L}\sim\delta^{1/2}$
allows the fundamental assumption (\ref{eq:delta_definition}) to
be satisfied.

An interesting consequence is that the Mach number squared, which
is the ratio of the convective velocity scale to the mean speed of
sound $\bar{C}$
\begin{equation}
Ma^{2}=\frac{\mathscr{U}^{2}}{\bar{C}^{2}}=\frac{\mathscr{U}^{2}}{\left\langle \left(\frac{\partial p}{\partial\rho}\right)_{s}\right\rangle }=\mathcal{O}\left(\delta\bar{g}L\left\langle \tilde{\rho}\tilde{\beta}\right\rangle \right)=\mathcal{O}\left(\delta\right)\ll1\label{eq:e158}
\end{equation}
must be of the order $\delta$, hence small. In the above we have
used the estimate $\Delta\tilde{p}\sim\tilde{\rho}gL$ of the hydrostatic
pressure jump across the fluid layer and the thermodynamic identity
\begin{eqnarray}
C^{2}=\left(\frac{\partial p}{\partial\rho}\right)_{s} & = & \left(\frac{\partial p}{\partial\rho}\right)_{T}+\left(\frac{\partial p}{\partial T}\right)_{\rho}\left(\frac{\partial T}{\partial\rho}\right)_{s}\nonumber \\
 & = & \frac{1}{\rho\beta}+\left(\frac{\partial p}{\partial T}\right)_{\rho}^{2}\frac{T}{\rho^{2}c_{v}}\nonumber \\
 & = & \frac{1}{\rho\beta}\left(1+\frac{\alpha^{2}T}{c_{v}\beta\rho}\right)=\frac{\gamma}{\rho\beta},\label{eq:speed_of_sound}
\end{eqnarray}
satisfied by virtue of the implicit function theorem which implies
$(\partial p/\partial T)_{\rho}=\alpha/\beta$ and $(\partial T/\partial\rho)_{s}=-T(\partial s/\partial\rho)_{T}/c_{v}$
and the Maxwell relation $\rho^{2}\left(\partial s/\partial\rho\right)_{T}=-(\partial p/\partial T)_{\rho}$;
the last equality in (\ref{eq:speed_of_sound}) is obtained on the
basis of (\ref{eq:cp-cv_B}) and $\gamma=c_{p}/c_{v}$. Of course
the above estimates correspond, in fact, to the \emph{third assumption}\index{SI}{anelastic!assumption (3)},
which was made in the process of derivation of the Boussinesq equations (\ref{eq:third_assumption}),
which is typically satisfied by fluids and therefore is a weak assumption,
not introducing strong restrictions. From the point of view of the
state equations it means we assume, that in terms of the small parameter
$\delta$ one can estimate $\partial\rho/\partial p\sim\rho/p$ etc. Note, however, that in the Boussinesq case the Mach number scales linearly with the small parameter of perturbative expansions $\epsilon$, whereas in the anelastic case the Mach number scales like square root of the small parameter $\delta$.

It is also important to note the physical consequence of the assumption
(\ref{eq:vel_and_time_scales-1}) concerning the velocity and time
scales of evolution. In the anelastic case the scale heights in the system are comparable
with the system's vertical span $L$, i.e.
\begin{equation}
\frac{D_{p}}{L}=\left(-\frac{L}{\tilde{p}}\frac{\mathrm{d}\tilde{p}}{\mathrm{d}z}\right)^{-1}=\frac{\tilde{g}L}{R\tilde{T}}=\gamma\frac{\tilde{g}L}{\tilde{C}^{2}}=\mathcal{O}\left(1\right),\label{eq:e159}
\end{equation}
(similarly $-L\mathrm{d}_{z}\tilde{\rho}/\tilde{\rho}=\mathcal{O}(1)$
and $-L\mathrm{d}_{z}\tilde{T}/\tilde{T}=\mathcal{O}(1)$), where
we have used the hydrostatic pressure balance for the reference state
$\mathrm{d}_{z}\tilde{p}=-\tilde{\rho}g$. Therefore the value of $\sqrt{\bar{g}L}$ which has the dimension of velocity is of comparable magnitude as the mean speed of sound $\bar{C}$.
It follows, that the assumed time scale $\mathscr{T}\sim\delta^{-1/2}\sqrt{L/\bar{g}}$
is of the same order of magnitude as the inertial time scale\index{SI}{inertial time scale}
\begin{equation}
\mathscr{T}_{\mathrm{inertial}}=\frac{L}{\mathscr{U}}=Ma^{-1}\frac{L}{\bar{C}}\sim\delta^{-1/2}\frac{L}{\bar{C}}\sim\mathscr{T}.\label{eq:e160}
\end{equation}
Moreover, time scale associated with internal gravity waves\index{SI}{internal-gravity-waves time scale}
\begin{equation}
\mathscr{T}_{\mathrm{grav.}}=\left(\frac{g}{\tilde{T}}\Delta_{S}\right)^{-1/2}\sim\delta^{-1/2}\sqrt{\frac{L}{g}}=\mathscr{T}\label{eq:e161}
\end{equation}
is also comparable with the chosen dynamical time scale, but all the
three time scales $\mathscr{T}_{\mathrm{inertial}}$, $\mathscr{T}_{\mathrm{grav.}}$
and $\mathscr{T}$ are all much longer than the fast acoustic time
scale $L/\bar{C}$. Therefore the following relation is satisfied
\begin{equation}
\mathscr{T}\sim\mathscr{T}_{\mathrm{inertial}}\sim\mathscr{T}_{\mathrm{grav.}}\gg\frac{L}{\bar{C}},\label{eq:e162}
\end{equation}
so that both the inertial and buoyancy effects are included in the
dynamics, but the acoustic effects are filtered out.

We are now ready to write down the dynamic equations for fluctuations,
i.e. with subtracted hydrostatic balance (\ref{eq:hydrostatic_eq_A_1},b)
\begin{subequations}
\begin{align}
\left(\tilde{\rho}+\rho'\right)\left[\frac{\partial\mathbf{u}}{\partial t}+\left(\mathbf{u}\cdot\nabla\right)\mathbf{u}\right]= & -\nabla p'+\rho'\tilde{\mathbf{g}}-\tilde{\rho}\nabla\psi'+\mu\nabla^{2}\mathbf{u}+\left(\frac{\mu}{3}+\mu_{b}\right)\nabla\left(\nabla\cdot\mathbf{u}\right)\nonumber \\
 & +2\nabla\mu\cdot\mathbf{G}^{s}+\nabla\left(\mu_{b}-\frac{2}{3}\mu\right)\nabla\cdot\mathbf{u},\label{NS-Aderiv-1}
\end{align}
\begin{equation}
\frac{\partial\rho'}{\partial t}+\nabla\cdot\left[\left(\tilde{\rho}+\rho'\right)\mathbf{u}\right]=0,\label{Cont_Aderiv-1}
\end{equation}
\begin{equation}
\nabla^{2}\psi'=4\pi G\rho',\label{eq:grav_pot_aderiv-2}
\end{equation}
\begin{align}
\left(\tilde{\rho}+\rho'\right)\left(\tilde{T}+T'\right)\left[\frac{\partial s'}{\partial t}+\mathbf{u}\cdot\nabla\left(\tilde{s}+s'\right)\right]= & \nabla\cdot\left(k\nabla T'\right)+2\mu\mathbf{G}^{s}:\mathbf{G}^{s}\nonumber \\
 & +\left(\mu_{b}-\frac{2}{3}\mu\right)\left(\nabla\cdot\mathbf{u}\right)^{2}+Q',\label{Energy_Aderiv-1}
\end{align}
\begin{equation}
\frac{\rho'}{\tilde{\rho}}=-\tilde{\alpha}T'+\tilde{\beta}p'+\mathcal{O}\left(\delta^{2}\right),\qquad s'=-\tilde{\alpha}\frac{p'}{\tilde{\rho}}+\tilde{c}_{p}\frac{T'}{\tilde{T}}+\mathcal{O}\left(\tilde{c}_{p}\delta^{2}\right),\label{State_eq_Aderiv-1}
\end{equation}
\end{subequations} 
and with the aid of (\ref{fluct_magnitude_A}),
(\ref{eq:vel_and_time_scales-1}), (\ref{eq:visc_scale-1}) and (\ref{eq:radiogenic_A})
take their leading order form\index{SI}{anelastic!equations, general}
\begin{subequations}
\begin{eqnarray}
\tilde{\rho}\left[\frac{\partial\mathbf{u}}{\partial t}+\left(\mathbf{u}\cdot\nabla\right)\mathbf{u}\right] & = & -\nabla p'+\rho'\tilde{\mathbf{g}}-\tilde{\rho}\nabla\psi'+\mu\nabla^{2}\mathbf{u}+\left(\frac{\mu}{3}+\mu_{b}\right)\nabla\left(\nabla\cdot\mathbf{u}\right)\nonumber \\
 &  & +2\nabla\mu\cdot\mathbf{G}^{s}+\nabla\left(\mu_{b}-\frac{2}{3}\mu\right)\nabla\cdot\mathbf{u},\label{NS-Aderiv-1-1}
\end{eqnarray}
\begin{equation}
\nabla\cdot\left(\tilde{\rho}\mathbf{u}\right)=0,\label{Cont_Aderiv-1-1}
\end{equation}
\begin{equation}
\nabla^{2}\psi'=4\pi G\rho',\label{eq:grav_pot_aderiv-2-1}
\end{equation}
\begin{equation}
\tilde{\rho}\tilde{T}\left[\frac{\partial s'}{\partial t}+\mathbf{u}\cdot\nabla\left(\tilde{s}+s'\right)\right]=\nabla\cdot\left(k\nabla T'\right)+2\mu\mathbf{G}^{s}:\mathbf{G}^{s}+\left(\mu_{b}-\frac{2}{3}\mu\right)\left(\nabla\cdot\mathbf{u}\right)^{2}+Q',\label{Energy_Aderiv-1-1}
\end{equation}
\begin{equation}
\frac{\rho'}{\tilde{\rho}}=-\tilde{\alpha}T'+\tilde{\beta}p',\qquad s'=-\tilde{\alpha}\frac{p'}{\tilde{\rho}}+\tilde{c}_{p}\frac{T'}{\tilde{T}}.\label{State_eq_Aderiv-1-1}
\end{equation}
\end{subequations} 
Note, that according to \ref{subsec:Filtering-sound-waves} (cf. eq. (\ref{eq:scalings_SWfiltration})) the anelastic continuity equation\index{SI}{continuity equation} filters out sound waves. Making use of (\ref{eq:dsbydz}) the energy equation
can be also written in the form
\begin{align}
\tilde{\rho}\tilde{T}\left(\frac{\partial s'}{\partial t}+\mathbf{u}\cdot\nabla s'\right)-\tilde{\rho}\tilde{c}_{p}u_{z}\Delta_{S}= & \nabla\cdot\left(k\nabla T'\right)+2\mu\mathbf{G}^{s}:\mathbf{G}^{s}\nonumber \\
 & +\left(\mu_{b}-\frac{2}{3}\mu\right)\left(\nabla\cdot\mathbf{u}\right)^{2}+Q'.\label{eq:Energy_eq_A_entropy_A}
\end{align}
For the sake of completeness we also provide the temperature equation\index{SI}{temperature equation}
within the anelastic approximation (cf. (\ref{Energy_eq2}))
\begin{align}
\left(\tilde{\rho}+\rho'\right)c_{v}\left[\frac{\partial T'}{\partial t}+\mathbf{u}\cdot\nabla\left(\tilde{T}+T'\right)\right]+\frac{\alpha}{\beta}\left(\tilde{T}+T'\right)\nabla\cdot\mathbf{u}\qquad\qquad\nonumber \\
=\nabla\cdot\left(k\nabla T'\right)+2\mu\mathbf{G}^{s}:\mathbf{G}^{s}+\left(\mu_{b}-\frac{2}{3}\mu\right)\left(\nabla\cdot\mathbf{u}\right)^{2} & +Q'\quad\label{Energy_Aderiv-1-2}
\end{align}
which at the leading order can be written in the following form
\begin{align}
\tilde{\rho}\tilde{c}_{v}\left(\frac{\partial T'}{\partial t}+\mathbf{u}\cdot\nabla T'\right)+\left(\tilde{\rho}+\rho'\right)c_{v}u_{z}\frac{\mathrm{d}\tilde{T}}{\mathrm{d}z}+\frac{\alpha}{\beta}\left(\tilde{T}+T'\right)\nabla\cdot\mathbf{u}\nonumber \\
=\nabla\cdot\left(k\nabla T'\right)+2\mu\mathbf{G}^{s}:\mathbf{G}^{s}+\left(\mu_{b}-\frac{2}{3}\mu\right)\left(\nabla\cdot\mathbf{u}\right)^{2} & +Q'.\label{Energy_Aderiv-1-2-1}
\end{align}
Note, that $c_{v}$ in the term proportional to $\mathrm{d}\tilde{T}/\mathrm{d}z$
and $\alpha/\beta$ in the term proportional to the flow divergence
are not taken at the hydrostatic equilibrium (not marked with an
upper tilde), but $\alpha$ and $\beta$ and in particular also their
fluctuations $\alpha'$, $\beta'$ can be obtained once the equations
of state are known; the specific heat $c_{v}$
is an input parameter in the theory of fluid mechanics resulting from
kinetic or phenomenological models, but it is known, that the specific
heat often significantly depends on the temperature; once the relation $c_v(T)$ is known $\tilde{c}_v$ and $c_v^{\prime}$ can be established, if necessary. We stress, that
it is important to keep the terms which involve the fluctuations of
the fluid's thermodynamic properties, i.e. $\tilde{\rho}c_{v}^{\prime}u_{z}\mathrm{d}_{z}\tilde{T}$
and $\alpha'\tilde{T}\nabla\cdot\mathbf{u}/\tilde{\beta}$ and $-\tilde{\alpha}\beta'\tilde{T}\nabla\cdot\mathbf{u}/\tilde{\beta}^{2}$
in the energy equation, since they contribute to the leading order
balance obtained at the order $\mathcal{O}\left(\delta\mathcal{U}\bar{\rho}\bar{g}\right)=\mathcal{O}\left(\delta^{3/2}\bar{\rho}\bar{g}\sqrt{\bar{g}L}\right)$
in the energy equation. Furthermore, although it is obvious that at the
held degree of accuracy the terms $\frac{\alpha}{\beta}T'\nabla\cdot\mathbf{u}$,
$\alpha'\tilde{T}\nabla\cdot\mathbf{u}/\tilde{\beta}$ and $-\tilde{\alpha}\beta'\tilde{T}\nabla\cdot\mathbf{u}/\tilde{\beta}^{2}$
can only involve the leading order expression for the flow divergence
$\nabla\cdot\mathbf{u}=-\mathrm{d}_{z}\tilde{\rho}u_{z}/\tilde{\rho}+\mathcal{O}(\delta\mathcal{U}/L)$,
because $\tilde{\alpha}T'\mathcal{U}/\tilde{\beta}L$, $\alpha'\tilde{T}\mathcal{U}/\tilde{\beta}L$
and $\tilde{\alpha}\beta'\tilde{T}\mathcal{U}/L\tilde{\beta}^{2}$
are already of the required order of magnitude $\mathcal{O}\left(\delta\mathcal{U}\bar{\rho}\bar{g}\right)$,
on the contrary the term $\frac{\alpha}{\beta}\tilde{T}\nabla\cdot\mathbf{u}$
must include the entire expression $\nabla\cdot\mathbf{u}=-[\nabla(\tilde{\rho}+\rho')\cdot\mathbf{u}+\partial_{t}\rho']/(\tilde{\rho}+\rho')$,
resulting from the law of mass conservation; the corrections $\mathcal{O}(\delta\mathcal{U}/L)$
in the expression for the flow divergence are vital for this term, since $(\tilde{\alpha}\tilde{T}/\tilde{\beta})\mathcal{O}(\delta\mathcal{U}/L)$
is of the required order of magnitude. We note, however, that there
are also terms such as $\tilde{\rho}\tilde{c}_{v}u_{z}\mathrm{d}_{z}\tilde{T}$
and 
\begin{equation}
\frac{\tilde{\alpha}\tilde{T}}{\tilde{\beta}}\nabla\cdot\mathbf{u}=-\frac{\tilde{\alpha}\tilde{T}}{\tilde{\rho}\tilde{\beta}}u_{z}\frac{\mathrm{d}\tilde{\rho}}{\mathrm{d}z}+\mathcal{O}\left(\delta\mathcal{U}\bar{\rho}\bar{g}\right),\label{eq:e163}
\end{equation}
which alone are $\delta^{-1}$ times stronger than the rest of the
terms in the energy equation, all of the order $\mathcal{O}\left(\delta\mathcal{U}\bar{\rho}\bar{g}\right)$.
We will demonstrate now, that the sum of the two unfitting terms appearing
in the equation is in fact of the required order of magnitude $\mathcal{O}\left(\delta\mathcal{U}\bar{\rho}\bar{g}\right)$.
Namely with the use of the continuity equation $\partial_{t}\rho'+\nabla\cdot((\tilde{\rho}+\rho')\mathbf{u})=0$
the sum of the second and third terms on the left hand side of (\ref{Energy_Aderiv-1-2-1})
can be rearranged to give
\begin{align}
\left(\tilde{\rho}+\rho'\right)c_{v}u_{z}\frac{\mathrm{d}\tilde{T}}{\mathrm{d}z}+\frac{\alpha}{\beta}\left(\tilde{T}+T'\right)\nabla\cdot\mathbf{u}= & u_{z}\left(\tilde{\rho}c_{v}\frac{\mathrm{d}\tilde{T}}{\mathrm{d}z}-\frac{\alpha}{\beta}\frac{\tilde{T}}{\tilde{\rho}}\frac{\mathrm{d}\tilde{\rho}}{\mathrm{d}z}\right)+\mathcal{O}\left(\delta\mathcal{U}\bar{\rho}\bar{g}\right)\nonumber \\
= & \tilde{\rho}\tilde{c}_{v}u_{z}\left(\frac{\mathrm{d}\tilde{T}}{\mathrm{d}z}-\frac{\tilde{\alpha}}{\tilde{c}_{v}\tilde{\beta}}\frac{\tilde{T}}{\tilde{\rho}}\frac{\mathrm{d}\tilde{\rho}}{\mathrm{d}z}\right)+\mathcal{O}\left(\delta\mathcal{U}\bar{\rho}\bar{g}\right).\label{eq:term_in_general_T_eq_A1}
\end{align}
Furthermore, the $z$-derivative of the basic density $\tilde{\rho}(\tilde{p},\,\tilde{T})$
reads
\begin{equation}
\frac{\mathrm{d}\tilde{\rho}}{\mathrm{d}z}=\tilde{\rho}\tilde{\beta}\frac{\mathrm{d}\tilde{p}}{\mathrm{d}z}-\tilde{\rho}\tilde{\alpha}\frac{\mathrm{d}\tilde{T}}{\mathrm{d}z},\label{eq:rho_derivative_in_z_A}
\end{equation}
and from the definition of $\delta$ in (\ref{eq:delta_definition})
and the hydrostatic balance $\mathrm{d}\tilde{p}/\mathrm{d}z=-\tilde{\rho}\tilde{g}$
we obtain an estimate of the vertical gradient of density in the hydrostatic state
\begin{align}
\frac{\mathrm{d}\tilde{\rho}}{\mathrm{d}z}= & -\tilde{\rho}^{2}\tilde{\beta}\tilde{g}-\tilde{\rho}\tilde{\alpha}\frac{\mathrm{d}\tilde{T}}{\mathrm{d}z}=-\left(\frac{\tilde{\rho}^{2}\tilde{\beta}}{\tilde{\alpha}\tilde{T}}\right)\left[\frac{\tilde{c}_{v}\tilde{\alpha}\tilde{T}\tilde{g}}{\tilde{c}_{p}}-\left(\tilde{c}_{p}-\tilde{c}_{v}\right)\frac{\tilde{T}}{L}\frac{L\Delta_{S}}{\tilde{T}}\right]\nonumber \\
= & -\frac{\tilde{c}_{v}\tilde{\beta}\tilde{\rho}^{2}\tilde{g}}{\tilde{c}_{p}}+\mathcal{O}\left(\tilde{\rho}\delta/L\right)=\frac{\tilde{c}_{v}\tilde{\beta}\tilde{\rho}^{2}}{\tilde{\alpha}\tilde{T}}\frac{\mathrm{d}\tilde{T}}{\mathrm{d}z}+\mathcal{O}\left(\tilde{\rho}\delta/L\right).\label{eq:drhobydz_A}
\end{align}
The latter can be utilized to get 
\begin{align}
\left(\tilde{\rho}+\rho'\right)c_{v}u_{z}\frac{\mathrm{d}\tilde{T}}{\mathrm{d}z}+\frac{\alpha}{\beta}\left(\tilde{T}+T'\right)\nabla\cdot\mathbf{u}= & \,\tilde{\rho}\tilde{c}_{v}u_{z}\left(\frac{\mathrm{d}\tilde{T}}{\mathrm{d}z}+\frac{\tilde{\alpha}\tilde{T}\tilde{g}}{\tilde{c}_{p}}\right)+\mathcal{O}\left(\delta\mathcal{U}\bar{\rho}\bar{g}\right)\nonumber \\
= & -\tilde{\rho}c_{v}u_{z}\Delta_{S}+\mathcal{O}\left(\delta\mathcal{U}\bar{\rho}\bar{g}\right)=\mathcal{O}\left(\delta\mathcal{U}\bar{\rho}\bar{g}\right).\label{eq:term_in_general_T_eq_A}
\end{align}
This proves consistency for the temperature equation (\ref{Energy_Aderiv-1-2-1})
which constitutes a balance between terms at the order $\mathcal{O}\left(\delta\mathcal{U}\bar{\rho}\bar{g}\right)$.
However, we recall that when deriving (\ref{eq:term_in_general_T_eq_A1})
we have in fact used the full (not only leading order) continuity
equation, 
\begin{equation}
\left(\tilde{\rho}+\rho'\right)\nabla\cdot\mathbf{u}=-\mathbf{u}\frac{\mathrm{d}\tilde{\rho}}{\mathrm{d}z}-\frac{\partial\rho'}{\partial t}-\mathbf{u}\cdot\nabla\rho',\label{eq:e164}
\end{equation}
since the terms 
\begin{equation}
-\frac{\alpha\tilde{T}}{\beta}\left(\frac{\partial\rho'}{\partial t}+\mathbf{u}\cdot\nabla\rho'\right)=\mathcal{O}\left(\delta\mathcal{U}\bar{\rho}\bar{g}\right)\label{eq:e165}
\end{equation}
also contribute to the leading order temperature balance. This means
that application of the temperature balance expressed as in (\ref{Energy_Aderiv-1-2-1})
along with the leading order mass conservation law $\nabla\cdot(\tilde{\rho}\mathbf{u})=0$
is \emph{not} possible, since it leads to neglection of vital terms
in the temperature balance and hence the full continuity equation
would have to be used. This is an undesirable situation, since in
such a case the quick sound waves are not filtered out and the entire
point of anelastic approximation is lost. The situation can be rectified
by rearranging the temperature equation (\ref{Energy_Aderiv-1-2-1})
with the use of the full mass conservation law (cf. (\ref{Energy_eq1})),
which yields\index{SI}{temperature equation!anelastic, general}
\begin{align}
\tilde{\rho}\tilde{c}_{p}\left(\frac{\partial T'}{\partial t}+\mathbf{u}\cdot\nabla T'\right)-\tilde{\alpha}\tilde{T}\left(\frac{\partial p'}{\partial t}+\mathbf{u}\cdot\nabla p'\right)+\rho c_{p}u_{z}\frac{\mathrm{d}\tilde{T}}{\mathrm{d}z}-\alpha Tu_{z}\frac{\mathrm{d}\tilde{p}}{\mathrm{d}z}\qquad\nonumber \\
=\nabla\cdot\left(k\nabla T'\right)+2\mu\mathbf{G}^{s}:\mathbf{G}^{s}+\left(\mu_{b}-\frac{2}{3}\mu\right)\left(\nabla\cdot\mathbf{u}\right)^{2} & +Q'.\quad\label{Energy_eq1-1}
\end{align}
This equation no longer suffers from issues associated with higher
order corrections to the flow divergence and can be used along with
the simplified law of mass conservation $\nabla\cdot(\tilde{\rho}\mathbf{u})=0$.
The sum of the terms $\rho c_{p}u_{z}\mathrm{d}\tilde{T}/\mathrm{d}z-\alpha Tu_{z}\mathrm{d}\tilde{p}/\mathrm{d}z$ in equation (\ref{Energy_eq1-1})
is also of the order $\mathcal{O}\left(\delta\mathcal{U}\bar{\rho}\bar{g}\right)$,
however, a simpler form is not easily obtained in the general case
of an unspecified equation of state. Some simplification is possible
under the assumption that the fluid satisfies the perfect gas equation
of state, $p=\rho RT$, where $R=k_{B}/m_{m}$ is the specific gas
constant, which depends on the type of gas/fluid; $k_{B}$ is the
Boltzmann constant and $m_{m}$ is the molecular mass of the fluid
particles. Then $\alpha=1/T$ and $\beta=1/p$ and since in such a
case $c_{p}-c_{v}=R=\mathrm{const}$ and $c_{p}/c_{v}=\gamma=\mathrm{const}$\footnote{Of course in a real fluid/gas the specific heat ratio depends on temperature
since the amount of degrees of freedom of fluid/gas particles increases
with temperature. However, since the case of a general equation of
state is presented in detail therefore whenever the perfect gas equation
of state will be considered here, it will also be assumed for simplicity,
that the specific heats at constant volume and pressure are uniform
(and then a generalisation to nonuniform $c_{v}$ and $c_{p}$ in
the case of a perfect gas state equation is fairly simple).} we can assume $\tilde{c}_{v}=c_{v}=\mathrm{const.}$ and $\tilde{c}_{p}=c_{p}=\mathrm{const.}$
Therefore by virtue of $\mathrm{d}\tilde{p}/\mathrm{d}z=-\tilde{\rho}\tilde{g}$
and (\ref{State_eq_Aderiv-1-1}) the energy equation at leading order
takes the form\index{SI}{temperature equation!anelastic, perfect gas}
\begin{align}
\tilde{\rho}c_{p}\left(\frac{\partial T'}{\partial t}+\mathbf{u}\cdot\nabla T'\right)-\left(\frac{\partial p'}{\partial t}+\mathbf{u}\cdot\nabla p'\right)+\tilde{\rho}c_{p}\frac{\mathrm{d}\tilde{T}}{\mathrm{d}z}u_{z}\left(\frac{p'}{\tilde{p}}-\frac{T'}{\tilde{T}}\right)-\tilde{\rho}c_{p}u_{z}\Delta_{S}\quad\nonumber \\
=\nabla\cdot\left(k\nabla T'\right)+2\mu\mathbf{G}^{s}:\mathbf{G}^{s}+\left(\mu_{b}-\frac{2}{3}\mu\right)\left(\nabla\cdot\mathbf{u}\right)^{2}+Q'.\label{Energy_eq1-1-1}
\end{align}
Such an equation does not involve the flow divergence and can be used
along with $\nabla\cdot(\tilde{\rho}\mathbf{u})=0$, however, the
lagrangian derivative of pressure needs to be kept, since it provides
a contribution of the same order of magnitude as all the other terms
in the above equation (which is, of course, also true for the general
temperature equation (\ref{Energy_eq1-1})). The temperature equation\index{SI}{temperature equation} (\ref{Energy_eq1-1}) or in the case of a perfect gas (\ref{Energy_eq1-1-1}) can replace the entropy equation\footnote{To be precise the correct terminology for these equations is the energy equation expressed in terms of either the entropy or the temperature.} in the closed system of equations (\ref{NS-Aderiv-1-1}-e) describing the anelastic convection.

However, in the case of a perfect gas under uniform gravity, $\mathbf{g}=-g\hat{\mathbf{e}}_z$, $g=\mathrm{const}$, the temperature equation (\ref{Energy_eq1-1-1}) can be combined
with a rearranged Navier-Stokes equation
\begin{align}
\frac{1}{\tilde{T}}\left[\frac{\partial\mathbf{u}}{\partial t}+\left(\mathbf{u}\cdot\nabla\right)\mathbf{u}\right]= & -\nabla\left(\frac{p'}{\tilde{\rho}\tilde{T}}\right)-\frac{T'}{\tilde{T}^{2}}\mathbf{g}+\frac{\mu}{\tilde{\rho}\tilde{T}}\nabla^{2}\mathbf{u}+\left(\frac{\mu}{3\tilde{\rho}\tilde{T}}+\frac{\mu_{b}}{\tilde{\rho}\tilde{T}}\right)\nabla\left(\nabla\cdot\mathbf{u}\right)\nonumber \\
 & +\frac{2}{\tilde{\rho}\tilde{T}}\nabla\mu\cdot\mathbf{G}^{s}+\frac{1}{\tilde{\rho}\tilde{T}}\nabla\left(\mu_{b}-\frac{2}{3}\mu\right)\nabla\cdot\mathbf{u},\label{NS-Aderiv-1-1-2-1-1}
\end{align}
so that the buoyancy is expressed solely by the temperature fluctuation.
This way both, the velocity field equation and the Poisson-type problem
for pressure obtained from the Navier-Stokes equation supplied by
$\nabla\cdot(\tilde{\rho}\mathbf{u})=0$ (by multiplying (\ref{NS-Aderiv-1-1-2-1-1}) by $\tilde{\rho}\tilde{T}$ and taking its divergence) depend on the temperature
fluctuation $T'$, which depends on the lagrangian pressure derivative.
However, as shown in section \ref{subsec:Filtering-sound-waves} the
sound waves are indeed filtered out within such an approach, despite
the presence of the lagrangian pressure derivative in the energy equation. Therefore the temperature formulation\index{SI}{temperature formulation} of the anelastic equations for a perfect gas under uniform gravity consists
of equations (\ref{Energy_eq1-1-1}), (\ref{NS-Aderiv-1-1-2-1-1})
and $\nabla\cdot(\tilde{\rho}\mathbf{u})=0$.

Finally, the anelastic mass conservation equation,
\begin{equation}
\nabla\cdot(\tilde{\rho}\mathbf{u})=0,\label{eq:mass_conserv_AN}
\end{equation}
which filtrates the fast sound waves, can be cast in a somewhat different
form by the use of (\ref{eq:rho_derivative_in_z_A}), the hydrostatic
force balance $\mathrm{d}_{z}\tilde{p}=-\tilde{\rho}\tilde{g}$ and
the definition of $\Delta_{S}=-\mathrm{d}_{z}\tilde{T}-\tilde{\alpha}\tilde{T}\tilde{g}/\tilde{c}_{p}$.
Namely the latter two imply
\begin{equation}
\frac{\mathrm{d}\tilde{T}}{\mathrm{d}z}=-\Delta_{S}-\frac{\tilde{\alpha}\tilde{T}\tilde{g}}{\tilde{c}_{p}}=-\Delta_{S}+\frac{\tilde{\alpha}\tilde{T}}{\tilde{c}_{p}\tilde{\rho}}\frac{\mathrm{d}\tilde{p}}{\mathrm{d}z},\label{eq:e166}
\end{equation}
which can be introduced into (\ref{eq:rho_derivative_in_z_A}) to
obtain
\begin{align}
\frac{1}{\tilde{\rho}}\frac{\mathrm{d}\tilde{\rho}}{\mathrm{d}z}= & \frac{\tilde{\beta}}{\tilde{c}_{p}}\left(\tilde{c}_{p}-\frac{\tilde{\alpha}^{2}\tilde{T}}{\tilde{\rho}\tilde{\beta}}\right)\frac{\mathrm{d}\tilde{p}}{\mathrm{d}z}+\tilde{\alpha}\Delta_{S}\nonumber \\
= & \frac{\tilde{\beta}}{\tilde{\gamma}}\frac{\mathrm{d}\tilde{p}}{\mathrm{d}z}+\tilde{\alpha}\Delta_{S},\label{eq:rho_derivative_in_z_A-1}
\end{align}
and we have used the thermodynamic identity (\ref{eq:cp-cv_B}) and
$\gamma=c_{p}/c_{v}$. This allows to transform the mass conservation
equation (\ref{eq:mass_conserv_AN}) into
\begin{equation}
\nabla\cdot\mathbf{u}+\frac{\tilde{\beta}}{\tilde{\gamma}}\frac{\mathrm{d}\tilde{p}}{\mathrm{d}z}u_{z}=-\tilde{\alpha}\Delta_{S}u_{z}=\mathcal{O}\left(\delta\sqrt{\frac{\bar{g}}{L}}\right).\label{eq:e167}
\end{equation}
For a perfect gas $\tilde{p}=\tilde{\rho}R\tilde{T}$, thus the mass
conservation equation simplifies to
\begin{equation}
\nabla\cdot\mathbf{u}+\frac{1}{\gamma\tilde{p}}\frac{\mathrm{d}\tilde{p}}{\mathrm{d}z}u_{z}=0,\label{eq:div_PI_1}
\end{equation}
which is often expressed in one the following equivalent forms
\begin{equation}
\nabla\cdot\left(\tilde{p}^{1/\gamma}\mathbf{u}\right)=0,\label{eq:div_PI_2}
\end{equation}
\begin{equation}
\nabla\cdot\mathbf{u}-\frac{\tilde{g}}{\tilde{C}^{2}}u_{z}=0,\label{eq:div_PI_3}
\end{equation}
where $\tilde{C}=\sqrt{\gamma R\tilde{T}}$ is the speed of sound
in the static reference state. In the literature the set of equations
(\ref{NS-Aderiv-1-1},c,d,e) for a perfect gas with either (\ref{eq:div_PI_1})
or (\ref{eq:div_PI_2}) or (\ref{eq:div_PI_3}) substituted for the
mass conservation equation (\ref{Cont_Aderiv-1-1}), i.e. 
\begin{subequations}
\begin{align}
\tilde{\rho}\left[\frac{\partial\mathbf{u}}{\partial t}+\left(\mathbf{u}\cdot\nabla\right)\mathbf{u}\right]= & -\nabla p'+\rho'\tilde{\mathbf{g}}-\tilde{\rho}\nabla\psi'+\mu\nabla^{2}\mathbf{u}+\left(\frac{\mu}{3}+\mu_{b}\right)\nabla\left(\nabla\cdot\mathbf{u}\right)\nonumber \\
 & +2\nabla\mu\cdot\mathbf{G}^{s}+\nabla\left(\mu_{b}-\frac{2}{3}\mu\right)\nabla\cdot\mathbf{u},\label{NS-PI}
\end{align}
\begin{align}
\tilde{\rho}\tilde{T}\left(\frac{\partial s'}{\partial t}+\mathbf{u}\cdot\nabla s'\right)-\tilde{\rho}\tilde{c}_{p}u_{z}\Delta_{S}= & \nabla\cdot\left(k\nabla T'\right)+2\mu\mathbf{G}^{s}:\mathbf{G}^{s}\nonumber \\
 & +\left(\mu_{b}-\frac{2}{3}\mu\right)\left(\nabla\cdot\mathbf{u}\right)^{2}+Q',\label{Energy_PI}
\end{align}
\begin{equation}
\nabla^2\psi'=4\pi G\rho',\qquad\nabla\cdot\left(\tilde{p}^{1/\gamma}\mathbf{u}\right)=0,\label{Div-PI}
\end{equation}
\begin{equation}
\frac{\rho'}{\tilde{\rho}}=-\frac{T'}{\tilde{T}}+\frac{p'}{\tilde{p}},\qquad s'=-R\frac{p'}{\tilde{p}}+c_{p}\frac{T'}{\tilde{T}},\label{State_eq_PI}
\end{equation}
\end{subequations} 
is called the \emph{pseudo-incompressible approximation}\index{SI}{pseudo-incompressible approximation}
(cf. Durran 1989, Durran 2008, Klein 2009, Klein \emph{et al}. 2010,
Klein and Puluis 2012 and Vasil \emph{et al}. 2013\footnote{The pseudo-incompressible approximation is particularly useful in the absence of thermal effects, as it allows for a sound-proof treatment of gravity waves without the constraint of nearly adiabatic dynamics present in the anelastic approximation. The low-Mach number theory for
convective flows is currently being further developed based on a variational approach by Toby Wood and his group at the University of Newcastle, UK. The aim of this approach is to allow for
relaxation of the assumption of small superadiabaticity of convection required by the
anelastic approximation and demand only the smallness of the pressure fluctuation, 
$p'/p\ll1$, with the density and temperature fluctuations comparable in magnitude to their reference state values}). The entropy equation (\ref{Energy_PI}) can of course be replaced
by the temperature equation (\ref{Energy_eq1-1-1}).

\subsection{Production of the total entropy within the anelastic approximation\label{subsec:Entropy-production}}\index{SI}{entropy!production}

It is of fundamental importance, that the derived anelastic energy
equation (\ref{Energy_Aderiv-1-1}) is consistent with the second
law of thermodynamics, i.e. in the case when the system is adiabatically
insulated the production of the total entropy must be positive or
null. This is not immediately obvious, since we have shown in section
\ref{subsec:Total-entropy-production}, that the entropy production
consists of the viscous heating (positive definite) and the volume
integral of the square of the temperature gradient. However, in the
process of derivation of the anelastic energy balance expressed in
terms of the entropy variations (\ref{Energy_Aderiv-1-1}) we have
neglected on the left hand side of this equation the fluctuation $T'$
with respect to the reference temperature $\tilde{T}$, leaving only
$\tilde{\rho}\tilde{T}\mathrm{D}_{t}s$. This implies appearance of
terms of the type $\nabla\tilde{T}\cdot\nabla T$ in the equation for the entropy per unit volume $\tilde{\rho}s$.
To demonstrate, that the anelastic approximation does not violate
the second law of thermodynamics we must show, that in the absence of any heat sources and adiabatic insulation on boundaries the system of anelastic equation does not allow the total entropy to decrease. Let us first assume adiabatic insulation
\begin{equation}
\left.-k\frac{\partial}{\partial z}\left(\tilde{T}+T'\right)\right|_{z=0,\,L}=0,\label{eq:adiabatic_insulation}
\end{equation}
and no radiogenic heat sources
\begin{equation}
Q=0.\label{eq:e168}
\end{equation}
However, we must realize, that the assumption of adiabatic insulation implies, that there is no energy input neither on the boundaries nor within the fluid volume, since also $Q=0$.  It follows from (\ref{eq:adiabatic_insulation}), that the reference state gradient $\mathrm{d}_z\tilde{T}$ is either asymptotically small of the order of $\partial_z T'=\mathcal{O}(\delta T_B/L)$ or null. This puts the system in the Boussinesq limit (cf. section \ref{subsec:Boussinesq-limit-A}). Since the assumption of small departure from adiabaticity, $\delta\ll1$, must always hold within the anelastic approximation, we conclude, that adiabatic insulation within the anelastic approximation can only be assumed for systems characterized by very small adiabatic gradient, $\tilde{\alpha}\tilde{T}g/\tilde{c}_p=\mathcal{O}(\delta T_B/L)$, i.e. within the Boussinesq limit. In other words, the anelastic approximation, useful and distinct from the Boussinesq approximation only at strong stratification, is constructed for systems which are not adiabatically insulated. This is, of course, a desired situation, since convective flow has to be driven by a thermal energy flux at the bottom (and/or non-zero $Q$), and naturally convective systems, such as stellar and planetary interiors or atmospheres are obviously not insulated. Nevertheless, we still must demonstrate, that the derived system of anelastic equations does not violate the second law of thermodynamics and is consistent with it, even if this effectively implies consideration of the Boussinesq limit.

The viscous heating\index{SI}{viscous heating} term in (\ref{Energy_Aderiv-1-1}) has a standard
form and has been shown to be positive definite in section \ref{subsec:Total-entropy-production}.
On dividing the energy equation (\ref{Energy_Aderiv-1-1}) by $\tilde{T}$
one obtains the equation for the entropy per unit volume $\tilde{\rho}s$, which can be cast in the form
\begin{align}
\frac{\partial}{\partial t}\left[\tilde{\rho}\left(\tilde{s}+s'\right)\right]+\nabla\cdot\left[\tilde{\rho}\mathbf{u}\left(\tilde{s}+s'\right)\right]= & \nabla\cdot\left[\frac{k}{\tilde{T}}\nabla\left(\tilde{T}+T'\right)\right]+\frac{k}{\tilde{T}^{2}}\frac{\mathrm{d}\tilde{T}}{\mathrm{d}z}\frac{\partial}{\partial z}\left(\tilde{T}+T'\right)\nonumber \\
 & +2\frac{\mu}{\tilde{T}}\utilde{\mathbf{G}}^{s}:\utilde{\mathbf{G}}^{s}+\frac{\mu_{b}}{\tilde{T}}\left(\nabla\cdot\mathbf{u}\right)^{2},\label{Energy_Aderiv-1-1-3}
\end{align}
where the anelastic continuity equation was used $\nabla\cdot(\tilde{\rho}\mathbf{u})=0$
(recall, that $\utilde{\mathbf{G}}^{s}$ denotes the traceless part
of the symmetric velocity gradient tensor). Next, if periodic boundary
conditions in the horizontal directions with some periods $L_x$ and $L_y$
are assumed, the adiabatic (\ref{eq:adiabatic_insulation}) and impermeability
$u_{z}(z=0,\,L)=0$ conditions on the top and bottom boundaries allow
to easily demonstrate, that the two ``divergence'' terms, $\nabla\cdot\left[\tilde{\rho}\mathbf{u}\left(\tilde{s}+s'\right)\right]$
and $\nabla\cdot\left[k\nabla\left(\tilde{T}+T'\right)/\tilde{T}\right]$,
do not contribute to the global entropy balance 
\begin{equation}
\int_{V}\nabla\cdot\left[\tilde{\rho}\mathbf{u}\left(\tilde{s}+s'\right)\right]\mathrm{d}^{3}x=\int_{\partial V}\tilde{\rho}\left(\tilde{s}+s'\right)\mathbf{u}\cdot\hat{\mathbf{n}}\mathrm{d}\Sigma=0,\label{eq:nonlin_entropy_null}
\end{equation}
\begin{equation}
\int_{V}\nabla\cdot\left[\frac{k}{\tilde{T}}\nabla\left(\tilde{T}+T'\right)\right]\mathrm{d}^{3}x=\int_{\partial V}\frac{k}{\tilde{T}}\nabla\left(\tilde{T}+T'\right)\cdot\hat{\mathbf{n}}\mathrm{d}\Sigma=0,\label{eq:diff_entropy_null}
\end{equation}
where $V=(0,\,L_x)\times(0,\,L_y)\times(0,\,L)$ is
the total periodic fluid volume. Furthermore, the adiabatic boundary
condition (\ref{eq:adiabatic_insulation}) can always be imposed in such
a way, that
\begin{equation}
\left.-k\frac{\mathrm{d}\tilde{T}}{\mathrm{d}z}\right|_{z=0,\,L}=0,\qquad\left.-k\frac{\partial T'}{\partial z}\right|_{z=0,\,L}=0,\label{eq:e169}
\end{equation}
which simply corresponds to a rather natural choice of the reference
state in this case. This, however, in conjunction with the reference
state equation
\begin{equation}
\frac{\mathrm{d}}{\mathrm{d}z}\left(k\frac{\mathrm{d}\tilde{T}}{\mathrm{d}z}\right)=0\;\Rightarrow\;k\frac{\mathrm{d}\tilde{T}}{\mathrm{d}z}=\mathrm{const},\label{eq:e170}
\end{equation}
implies, that
\begin{equation}
k\frac{\mathrm{d}\tilde{T}}{\mathrm{d}z}=0,\label{eq:ref_state_ad}
\end{equation}
and hence the entire term 
\begin{equation}
\frac{k}{\tilde{T}^{2}}\frac{\mathrm{d}\tilde{T}}{\mathrm{d}z}\frac{\partial}{\partial z}\left(\tilde{T}+T'\right)=0,\label{eq:e171}
\end{equation}
from the entropy equation vanishes. Therefore integration of the entropy
balance (\ref{Energy_Aderiv-1-1-3}) over the entire periodic fluid
volume $V$ leads to
\begin{equation}
\frac{\partial}{\partial t}\int_{V}\tilde{\rho}s\mathrm{d}^{3}x=2\int_{V}\frac{\mu}{\tilde{T}}\mathbf{\utilde{\mathbf{G}}}^{s}:\utilde{\mathbf{G}}^{s}\mathrm{d}^{3}x+\int_{V}\frac{\mu_{b}}{\tilde{T}}\left(\nabla\cdot\mathbf{u}\right)^{2}\mathrm{d}^{3}x\geq0.\label{eq:e172}
\end{equation}
This way we have demonstrated, that the anelastic approximation is
consistent with the second law of thermodynamics. In other words,
we have shown that the production of the total entropy in anelastic,
adiabatically insulated systems, calculated from the entropy equation
(\ref{Energy_Aderiv-1-1}) is positive definite.

\subsection{Conservation of mass and values of the mean pressure at boundaries\label{subsec:Conservarion-of-mass-A}}\index{SI}{pressure boundary conditions}

The solenoidal constraint $\nabla\cdot(\tilde{\rho}\mathbf{u})=0$,
obtained by neglection of the density time variation implies sound-proof
dynamics and therefore the pressure spreads infinitely fast. Consequently,
the pressure fluctuation is determined by a Poisson-type, elliptic
equation
\begin{align}
\nabla^{2}p'= & \frac{\partial}{\partial z}\left(\rho'\tilde{g}\right)-\nabla\cdot\left[\nabla\cdot\left(\tilde{\rho}\mathbf{u}\mathbf{u}\right)+\tilde{\rho}\nabla\psi'\right]-\left(\frac{4}{3}\mu+\mu_{b}\right)\nabla^{2}\left(\frac{1}{\tilde{\rho}}\frac{\mathrm{d}\tilde{\rho}}{\mathrm{d}z}u_{z}\right)\nonumber \\
 & +2\frac{\mathrm{d}\mu}{\mathrm{d}z}\nabla^{2}u_{z}-\frac{\mathrm{d}}{\mathrm{d}z}\left(\frac{2}{3}\mu+2\mu_{b}\right)\frac{\partial}{\partial z}\left(\frac{1}{\tilde{\rho}}\frac{\mathrm{d}\tilde{\rho}}{\mathrm{d}z}u_{z}\right)\nonumber \\
 & +2\frac{\mathrm{d}^{2}\mu}{\mathrm{d}z^{2}}\frac{\partial u_{z}}{\partial z}-2\frac{\mathrm{d}^{2}}{\mathrm{d}z^{2}}\left(\mu_{b}-\frac{2}{3}\mu\right)\frac{\partial}{\partial z}\left(\frac{1}{\tilde{\rho}}\frac{\mathrm{d}\tilde{\rho}}{\mathrm{d}z}u_{z}\right),\label{eq:Mass_cons_pressure_A_1}
\end{align}
obtained by taking a divergence of the Navier-Stokes equation (\ref{NS-Aderiv-1-1}),
utilizing $\nabla\cdot(\tilde{\rho}\mathbf{u})=0$ and a simplifying
assumption that $\mu=\mu(z)$ and $\mu_{b}=\mu_{b}(z)$ are functions
of height only. However, neglection of the time variation of density
implies also, that the total mass is \emph{not} conserved in the dynamics
described by the anelastic system of equations (\ref{NS-Aderiv-1-1}-e),
and thus conservation of the total mass must be imposed additionally.
This means, that if we assume, that the total mass is contained in
the reference state $\tilde{\rho}$, we must impose\index{SI}{mass conservation}
\begin{equation}
\left\langle \rho'\right\rangle =0\quad\textrm{at all times}.\label{eq:Mass_cons_pressure_A_2}
\end{equation}
The latter constraint must be imposed
at every instant. This is achieved in a very simple way in the general
case of non-uniform gravity, when because of
\begin{equation}
\frac{\mathrm{d}^{2}\left\langle \psi'\right\rangle _{h}}{\mathrm{d}z^{2}}=4\pi G\left\langle \rho'\right\rangle _{h}\label{eq:MCP_A}
\end{equation}
the mass conservation is simply ensured by continuity of the gravitational
field at top and bottom (cf. (\ref{eq:e156})),
\begin{equation}
\left.\frac{\mathrm{d}\left\langle \psi'\right\rangle _{h}}{\mathrm{d}z}\right|_{z=L}=\left.\frac{\mathrm{d}\left\langle \psi'\right\rangle _{h}}{\mathrm{d}z}\right|_{z=0}=0.\label{eq:MCP_A_1}
\end{equation}
It follows, that if we average the $z$-component of the Navier-Stokes
equation (\ref{NS-Aderiv-1-1}) (with the viscosities $z$-dependent
only) over the entire periodic domain, we get
\begin{equation}
\left\langle p'\right\rangle _{h}(z=L)-\left\langle p'\right\rangle _{h}(z=0)=-L\left\langle \rho'\tilde{g}\right\rangle -L\left\langle \tilde{\rho}\frac{\partial\psi'}{\partial z}\right\rangle =-2L\left\langle \rho'\tilde{g}\right\rangle ,\label{eq:Mass_cons_pressure_A_3-1}
\end{equation}
where we have used (\ref{eq:grav_pot_aderiv-1}) and (\ref{eq:MCP_A_1})
to get the final equality. 

When the gravity is uniform the right hand side of the latter equation
needs to vanish, since then
$\tilde{g}=\mathrm{const}$ and $\left\langle \rho'\right\rangle$ needs to vanish
due to the mass conservation constraint. In such a case the gravitational
potential $\psi'$ drops out of the dynamical equations (cf. the discussion
below (\ref{eq:gravity_acceleration_A})) thus there is no need to
solve for $\psi'$ and as a result the condition (\ref{eq:MCP_A_1})
is never applied. Therefore when gravity is uniform the simplest way
to achieve the total mass conservation, is to impose a null mean pressure
fluctuation jump across the layer,\index{SI}{pressure boundary conditions}
\begin{equation}
\left\langle p'\right\rangle _{h}(z=L)=\left\langle p'\right\rangle _{h}(z=0)\quad\textrm{at all times},\label{eq:Mass_cons_pressure_A_4}
\end{equation}
similarly as in the case of Boussinesq convection (cf. section \ref{subsec:Conservarion-of-mass}).
This constitutes a boundary condition, which must be imposed on the
pressure field at every moment in time, used in tandem with the elliptic
equation (\ref{eq:Mass_cons_pressure_A_1}) at $\tilde{g}=\mathrm{const}$
and $\nabla\psi'=0$. We note, however, that whether or not the condition
(\ref{eq:Mass_cons_pressure_A_4}) is imposed, the convective velocity
field remains uninfluenced, since a shift in pressure fluctuation
which is time-dependent only, corresponds to a simple gauge transformation.
Nevertheless, the condition (\ref{eq:Mass_cons_pressure_A_4}) is
important in order to fully resolve the dynamics of the thermodynamic
fluctuations.

An important consequence of the mass conservation constraint (\ref{eq:Mass_cons_pressure_A_2})
and (\ref{State_eq_Aderiv-1-1}), is that 
\begin{equation}
-\left\langle \tilde{\rho}\tilde{\alpha}T'\right\rangle +\left\langle \tilde{\rho}\tilde{\beta}p'\right\rangle =0\quad\textrm{at all times}\label{eq:Mass_cons_pressure_A_5}
\end{equation}
or equivalently
\begin{equation}
-\left\langle \tilde{\rho}\tilde{\alpha}\tilde{T}\frac{s'}{\tilde{c}_{p}}\right\rangle +\left\langle \frac{\tilde{c}_{v}}{\tilde{c}_{p}}\tilde{\rho}\tilde{\beta}p'\right\rangle =0\quad\textrm{at all times}\label{eq:Mass_cons_pressure_A_6}
\end{equation}
must also be satisfied throughout the evolution of the system. 

\subsection{Boussinesq limit\label{subsec:Boussinesq-limit-A}}\index{SI}{Boussinesq!limit}

Perhaps the easiest link with the Boussinesq approximation is obtained through the energy equation expressed in terms of entropy (\ref{eq:Energy_eq_A_entropy_A}), by assuming that the scale
heights associated with density, temperature and pressure are large
compared to the fluid layer thickness, i.e. $\mathrm{d}\tilde{\rho}/\mathrm{d}z\ll\bar{\rho}/L$,
$\mathrm{d}\tilde{T}/\mathrm{d}z\ll\bar{T}/L$ and $\mathrm{d}\tilde{p}/\mathrm{d}z\ll\bar{p}/L$.
This implies $\delta\lesssim\mathcal{O}(\epsilon)$, $\tilde{\rho}\approx\bar{\rho}$,
$\tilde{T}\approx\bar{T}$, $\tilde{p}\approx\bar{p}$, $\mathrm{d}\tilde{\rho}/\mathrm{d}z\approx\mathrm{d}\dbtilde{\rho}/\mathrm{d}z$,
$\mathrm{d}\tilde{T}/\mathrm{d}z\approx\mathrm{d}\dbtilde{T}/\mathrm{d}z$,
$\mathrm{d}\tilde{p}/\mathrm{d}z\approx\mathrm{d}\dbtilde{p}/\mathrm{d}z$
where $\epsilon=\Delta\dbtilde{\rho}/\bar{\rho}$ and the hydrostatic
contributions to thermodynamic fields, just as in the previous chapter
are decomposed into the mean and a hydrostatic vertically varying correction,
e.g. $\tilde{T}=\bar{T}+\dbtilde{T}$. Note also, that in the Boussinesq
limit $L$ becomes so small (or alternatively the averaged temperature
$\bar{T}$ so large) that by the use of general formula (\ref{eq:dsbydz_A_general}),
resulting directly from the definition of $c_{p}$ and thermodynamic
identities, and with the aid of the hydrostatic balance we get $c_{p}\sim-\bar{\alpha}\bar{T}gL/(L\mathrm{d}_{z}\dbtilde{T})\sim\epsilon^{-1}gL/\bar{T}$.
This transforms the above entropy equation (\ref{eq:Energy_eq_A_entropy_A})
into the Boussinesq one (\ref{Entropy_Bderiv-1}) and the continuity
equation into $\nabla\cdot\mathbf{u}\approx0$ in a straightforward
way (the term proportional to $\Delta_{S}\sim-\mathrm{d}\tilde{T}/\mathrm{d}z\approx-\mathrm{d}\dbtilde{T}/\mathrm{d}z\sim\mathcal{O}(\bar{T}\epsilon/L)$
remains in the entropy equation (\ref{eq:Energy_eq_A_entropy_A})
since it is of the same order $\mathcal{O}(\epsilon^{1/2})$ as the
rest of the terms in that equation, but in the continuity equation
$u_{z}\mathrm{d}\tilde{\rho}/\mathrm{d}z\sim\mathcal{O}(\bar{\rho}\sqrt{g/L}\epsilon^{3/2})$
is negligible compared to $\bar{\rho}\nabla\cdot\mathbf{u}\sim\mathcal{O}(\bar{\rho}\sqrt{g/L}\epsilon^{1/2})$).
Moreover, as demonstrated in the previous chapter (cf. discussion below
(\ref{eq:Buoyancy_simpl_B})) the assumption of large scale heights implies also $\left|\tilde{\beta}p'\right|\ll\left|\tilde{\alpha}T'\right|$, which allows to simplify the buoyancy
force in the Navier-Stokes equation (\ref{NS-Aderiv-1-1}) to $-\tilde{\alpha}T'\mathbf{g}$
thus obtaining the Boussinesq momentum balance (\ref{NS-Bderiv-1-1-1}).
A consequence of $\delta\lesssim\mathcal{O}(\epsilon)$ is that the
departure from adiabatic state in the Boussinesq limit is always small
but in the sense, that the total temperature (and density, pressure)
variation is very weak in comparison with its mean value; the actual
difference $(L/\bar{T})(\mathrm{d}\dbtilde{T}/\mathrm{d}z+g\bar{\alpha}\bar{T}/\bar{c}_{p})$
can be even smaller, but it is not a necessary requirement; in other
words under the Boussinesq approximation both contributions from $(L/\bar{T})(\mathrm{d}\dbtilde{T}/\mathrm{d}z)$
and $(L/\bar{T})(g\bar{\alpha}\bar{T}/\bar{c}_{p})$ are small, of
the order $\mathcal{O}(\epsilon)$, but it is allowed that $g\bar{\alpha}\bar{T}/\bar{c}_{p}\ll\mathrm{d}\dbtilde{T}/\mathrm{d}z$ (cf. section \ref{subsec:Anelastic-vs-Boussinesq}).

Taking the Boussinesq limit of the temperature equation\index{SI}{temperature equation} (\ref{Energy_eq1-1})
is achieved by neglection of the entire lagrangian pressure derivative
on the grounds of $\left|\tilde{\beta}p'\right|\ll\left|\tilde{\alpha}T'\right|$
and
\begin{align}
\rho c_{p}u_{z}\frac{\mathrm{d}\tilde{T}}{\mathrm{d}z}-\alpha Tu_{z}\frac{\mathrm{d}\tilde{p}}{\mathrm{d}z}= & \,\bar{\rho}\bar{c}_{p}u_{z}\frac{\mathrm{d}\dbtilde{T}}{\mathrm{d}z}-\bar{\alpha}\bar{T}u_{z}\frac{\mathrm{d}\dbtilde{p}}{\mathrm{d}z}+\mathcal{O}\left(\epsilon^{2}\bar{c}_{p}\mathcal{U}\bar{\rho}\bar{T}/L\right)\nonumber \\
\approx & -\bar{\rho}c_{p}u_{z}\Delta_{S}+\mathcal{O}\left(\epsilon^{2}\bar{c}_{p}\mathcal{U}\bar{\rho}\bar{T}/L\right).\label{eq:e173}
\end{align}

\section{The reference state\label{subsec:The-reference-state}}\index{SI}{reference (basic) state}

We will focus for a moment on the static reference state, which by
assumption is close to an adiabatic state, to give an idea of the
possible formulations within the anelastic approximation. There are
basically two options - either, as in all the previous sections, the
reference state is not adiabatic but its departure from adiabaticity
is small by assumption and given by (\ref{eq:delta_definition}) or
the reference state itself can be chosen to be adiabatic and then
the flow is driven by the boundary conditions which only slightly
departure from those corresponding to the adiabatic state. The latter
case requires an alternative definition of the parameter $\delta$
involving then the boundary conditions. Both formulations are equivalent
and can be transformed into each other. The case of the adiabatic
reference state is postponed until the end of the current section
and we will start with the case, when the hydrostatic reference state
slightly departures from adiabatic and satisfies the boundary conditions,
which drive convection. From the mathematical point of view this corresponds
simply to a well-established trick in the theory of differential equations
of subtracting a stationary state which satisfies the non-homogeneous
boundary conditions, to allow for homogeneous boundary conditions for
the remainder (fluctuation). The hydrostatic state equations (\ref{eq:hydrostatic_eq_A_1})-(\ref{eq:hydrostatic_eq_A_3})
for a perfect gas in the absence of the radiative heat sources and
at uniform gravity take the form\footnote{Note, that the expression for the entropy of an ideal gas results
directly from the definition of the specific heat at constant volume
$c_{v}=T(\partial_{T}s)_{\rho}$, the Maxwell relation $\rho^{2}(\partial_{\rho}s)_{T}=(\partial_{T}p)_{\rho}$
and the equation of state $p=\rho RT$.}
\begin{equation}
\frac{\mathrm{d}\tilde{p}}{\mathrm{d}z}=-\tilde{\rho}g,\quad\frac{\mathrm{d}}{\mathrm{d}z}\left(k\frac{\mathrm{d}\tilde{T}}{\mathrm{d}z}\right)=0,\quad\tilde{p}=\tilde{\rho}R\tilde{T},\quad\tilde{s}=c_{v}\ln\frac{\tilde{p}}{\tilde{\rho}^{\gamma}}+\mathrm{const}.\label{eq:hydrostatic_eq_A_pg}
\end{equation}
Let us consider an illustrative example when the thermal conductivity
$k$ is assumed uniform, in which case the reference state is polytropic
with the reference temperature $\tilde{T}$ being a simple linear
function of $z$ and\index{SI}{reference (basic) state}
\begin{subequations}
\begin{equation}
\tilde{T}=T_{B}\left(1-\theta\frac{z}{L}\right),\qquad\tilde{\rho}=\rho_{B}\left(1-\theta\frac{z}{L}\right)^{m},\qquad\tilde{p}=\frac{gL\rho_{B}}{\theta\left(m+1\right)}\left(1-\theta\frac{z}{L}\right)^{m+1},\label{eq:BS1}
\end{equation}
\begin{equation}
\tilde{s}=c_{p}\frac{m+1-\gamma m}{\gamma}\ln\left(1-\theta\frac{z}{L}\right)+\mathrm{const},\label{eq:BS2}
\end{equation}
\begin{equation}
m=\frac{gL}{R\Delta T}-1.\label{eq:m_exp1}
\end{equation}
\end{subequations} 
where $\theta=\Delta T/T_{B}$ is the
magnitude of the basic temperature gradient and a measure of compressibility
of the fluid; $T_{B}$ and $\rho_{B}$ denote here the values of temperature and
density at the bottom of the layer in the reference state respectively, $\Delta T=T_{B}-T_{T}>0$
is the temperature jump across the layer in the reference state and $0\leq\theta<1$. We
note, that for the case when the temperature is held constant at boundaries
the values $T_{B}$ and $T_{T}$ simply constitute the thermodynamic
boundary conditions, however, in the cases of specified heat flux
at boundaries or isentropic boundaries, the actual top and bottom
values of the temperature in the convective state are generally shifted with respect to $T_T$ and $T_B$
by corrections of the order $\delta$, namely $T'(x,y,z=0,t)$ and
$T'(x,y,z=L,t)$. On the other hand the bottom value of density in the reference state $\rho_B$
is established by the total mass of the fluid in the considered domain
and thus in general is also altered by order $\delta$ corrections
in a convective state. 

Since in the adiabatic state $\tilde{\rho}\sim\tilde{T}^{1/(\gamma-1)}$,
here the departure from adiabaticity is manifested by the small difference
between the adiabatic exponent $1/(\gamma-1)$ and the polytropic
index $m$, i.e. $\delta\sim1/(\gamma-1)-m>0$. More precisely, in
this case 
\begin{equation}
\delta=\left(\frac{1}{\gamma-1}-m\right)\frac{\gamma-1}{\gamma}\ln\frac{1}{1-\theta}=\frac{L}{T_{B}}\left(\frac{\Delta T}{L}-\frac{g}{c_{p}}\right)\frac{1}{\theta}\ln\frac{1}{1-\theta}>0;\label{eq:delta_intermsof_epsan}
\end{equation}
note, that the non-dimensional expression
\begin{equation}
\frac{L}{T_{B}}\left(\frac{\Delta T}{L}-\frac{g}{c_{p}}\right)\label{eq:alternative_delta}
\end{equation}
is often utilized as an alternative definition of the small anelastic
parameter $\delta$.

Let us expand the reference state about the adiabatic neglecting terms
of the order $\mathcal{O}(c_{p}^{2}\Delta_{S}^{2}/g^{2})$ and higher;
this yields 
\begin{subequations}
\begin{equation}
\tilde{T}=T_{B}\left(1-\frac{gz}{c_{p}T_{B}}\right)-\Delta_{S}z,\label{eq:Ttilde_expanded_RS}
\end{equation}
\begin{align}
\tilde{\rho}\approx & \rho_{B}\left(1-\frac{gz}{c_{p}T_{B}}-\frac{\Delta_{S}}{T_{B}}z\right)^{\frac{1}{\gamma-1}-\frac{\Delta_{S}c_{p}}{g(\gamma-1)}}\approx\rho_{B}\left(1-\frac{gz}{c_{p}T_{B}}\right)^{\frac{1}{\gamma-1}}\nonumber \\
 & -\frac{\rho_{B}c_{p}\Delta_{S}}{g\left(\gamma-1\right)}\left[\frac{gz}{c_{p}T_{B}}\left(1-\frac{gz}{c_{p}T_{B}}\right)^{\frac{2-\gamma}{\gamma-1}}+\left(1-\frac{gz}{c_{p}T_{B}}\right)^{\frac{1}{\gamma-1}}\ln\left(1-\frac{gz}{c_{p}T_{B}}\right)\right],\label{eq:rhotilde_expanded_RS}
\end{align}
\begin{align}
\tilde{p}\approx & \rho_{B}RT_{B}\left(1-\frac{gz}{c_{p}T_{B}}-\frac{\Delta_{S}}{T_{B}}z\right)^{\frac{\gamma}{\gamma-1}-\frac{\Delta_{S}c_{p}}{g(\gamma-1)}}\approx\rho_{B}RT_{B}\left(1-\frac{gz}{c_{p}T_{B}}\right)^{\frac{\gamma}{\gamma-1}}\nonumber \\
 & -\frac{\rho_{B}T_{B}c_{v}c_{p}\Delta_{S}}{g}\left[\frac{\gamma gz}{c_{p}T_{B}}\left(1-\frac{gz}{c_{p}T_{B}}\right)^{\frac{1}{\gamma-1}}+\left(1-\frac{gz}{c_{p}T_{B}}\right)^{\frac{\gamma}{\gamma-1}}\ln\left(1-\frac{gz}{c_{p}T_{B}}\right)\right],\label{eq:ptilde_expanded_RS}
\end{align}
\begin{equation}
\tilde{s}\approx\frac{c_{p}^{2}\Delta_{S}}{g}\ln\left(1-\frac{gz}{c_{p}T_{B}}\right)+\mathrm{const}.\label{eq:BS2-1}
\end{equation}
\end{subequations} 
The order $\mathcal{O}(c_{p}\Delta_{S}/g)$ corrections
are vital, since they allow to satisfy the boundary conditions, which
drive the flow. It is this order correction to the constant adiabatic
entropy profile in (\ref{eq:BS2-1}) which is responsible for non-zero
entropy gradient and driving in the energy equation (\ref{Energy_Aderiv-1-1}),
i.e. $\tilde{\rho}\tilde{T}u_{z}\mathrm{d}\tilde{s}/\mathrm{d}z\neq0$.
It is, therefore, important to realize, that the $\mathcal{O}(c_{p}\Delta_{S}/g)$
corrections to the adiabatic profile in the reference state need to
be established precisely, therefore the reference state must be obtained
from the full equations, not those approximated, with some $\mathcal{O}(c_{p}\Delta_{S}/g)$
order terms already neglected.

As mentioned, an often used alternative formulation is based on taking
the reference state hydrostatic and adiabatic and application of non-homogeneous
boundary conditions to the fluctuations. In such a case the Navier-Stokes and continuity
equations (\ref{NS-Aderiv-1-1})-(\ref{Cont_Aderiv-1-1}) remain unchanged
but in the energy equation (\ref{Energy_Aderiv-1-1}) the basic entropy
gradient vanishes (there is no term proportional to $\Delta_{S}$
in (\ref{eq:Energy_eq_A_entropy_A})), so that it takes on a simpler
form
\begin{equation}
\rho_{ad}T_{ad}\left(\frac{\partial s'}{\partial t}+\mathbf{u}\cdot\nabla s'\right)=\nabla\cdot\left(k\nabla T'\right)+2\mu\mathbf{G}^{s}:\mathbf{G}^{s}+\left(\mu_{b}-\frac{2}{3}\mu\right)\left(\nabla\cdot\mathbf{u}\right)^{2}+Q',\label{eq:Energy_eq_A_entropy_A-2}
\end{equation}
where the subscript $ad$ denotes the static, adiabatic reference state and the boundary conditions on $T'$ are non-zero in this case. In the above example
of a perfect gas with uniform thermal conductivity and gravity and
no radiative heat sources the adiabatic static state can take the
form \index{SI}{reference (basic) state}
\begin{subequations}
\begin{align}
T_{ad} & =T_{ad\,B}\left(1-\frac{gz}{c_{p}T_{ad\,B}}\right),\quad\rho_{ad}=\rho_{ad\,B}\left(1-\frac{gz}{c_{p}T_{ad\,B}}\right)^{\frac{1}{\gamma-1}},\label{eq:adiabatic_profile_RS}\\
p_{ad} & =\rho_{ad\,B}RT_{ad\,B}\left(1-\frac{gz}{c_{p}T_{B}}\right)^{\frac{\gamma}{\gamma-1}},\quad s_{ad}=\mathrm{const}.,\label{eq:adiabatic_profile_RS_1}
\end{align}
\end{subequations} 
which is of the same form as (\ref{eq:Ttilde_expanded_RS}-d)
but without the $\mathcal{O}(c_{p}\Delta_{S}/g)$ corrections, and
satisfies the hydrostatic dynamical equations (the quantities $\rho_{ad\,B}$
and $T_{ad\,B}$ are simply the bottom values of density and temperature
in the adiabatic state). The small parameter measuring the departure
from adiabaticity introduced by the boundary conditions, either with fixed temperature or fixed heat flux, which imply $\Delta T=T_{B}-T_{T}>gL/c_{p}$
is conveniently defined as
\begin{equation}
\delta_{ad}=\frac{1}{T_{B}}\left(\Delta T-\frac{gL}{c_{p}}\right),\label{eq:e174}
\end{equation}
and in the current case of constant thermal conductivity when the
temperature profile is linear it is simply equivalent to $\delta_{ad}=\delta=L\Delta_{S}/T_{B}$.
Note, that the total mass (per horizontal unit surface) contained
in the fluid domain can be expressed in two ways
\begin{subequations}
\begin{equation}
\int_{0}^{L}\tilde{\rho}\mathrm{d}z=\frac{R\rho_{B}T_{B}}{g}\left[1-\left(1-\frac{\Delta T}{T_{B}}\right)^{\frac{gL}{R\Delta T}}\right],\label{eq:M_cond}
\end{equation}
\begin{equation}
\int_{0}^{L}\rho_{ad}\mathrm{d}z=\frac{R\rho_{ad\,B}T_{B}}{g}\left[1-\left(1-\frac{gL}{c_{p}T_{B}}\right)^{\frac{\gamma}{\gamma-1}}\right],\label{eq:M_ad}
\end{equation}
\end{subequations}
where $T_{ad\,B}=T_{B}$ has been assumed and thus
a full correspondence between the two formulations with the hydrostatic conduction
and adiabatic reference states is obtained by assuming
\begin{equation}
\rho_{ad\,B}=\rho_{B}\frac{1-\left(1-\frac{\Delta T}{T_{B}}\right)^{\frac{gL}{R\Delta T}}}{1-\left(1-\frac{gL}{c_{p}T_{B}}\right)^{\frac{\gamma}{\gamma-1}}}=\rho_{B}+\mathcal{O}\left(\delta\rho_{B}\right),\label{eq:rho_rel}
\end{equation}
which means that the bottom values of the density in the hydrostatic
conduction reference state and in the hydrostatic adiabatic state
are slightly shifted by an order $\mathcal{O}(\delta\rho_{B})$ value.
This way the total mass of the fluid is contained in the reference
states, $\tilde{\rho}$ and $\rho_{ad}$, and therefore the mass conservation
implies that the spatial average over the entire fluid domain of the
density fluctuation is always zero, both in the case of fluctuation
about the hydrostatic conduction reference state, $\left\langle \rho-\tilde{\rho}\right\rangle =0$
and in the case of fluctuation about the hydrostatic adiabatic reference
state, $\left\langle \rho-\rho_{ad}\right\rangle =0$. Recall, that in practice this is obtained by imposing condition (\ref{eq:Mass_cons_pressure_A_4}) on the mean pressure fluctuation.

To illustrate full correspondence between the conduction reference state formulation and the formulation with the adiabatic reference state let us consider a perfect gas, characterized by constant dynamic viscosities $\mu=\mathrm{const}$, $\mu_b=\mathrm{const}$, thermal conductivity $k=\mathrm{const}$, specific heats $c_v=\mathrm{const}$, $c_p=\mathrm{const}$ and gravity $g=\mathrm{const}$ and the total mass $M_f$, which is driven by keeping the temperature fixed at the boundaries and $\Delta T=T_{B}-T_{T}>gL/c_{p}$. Of course the system physically responds to the driving regardless of our choice of mathematical description, i.e. no matter which formulation we pick, the final results for the velocity field and the total temperature $T(\mathbf{x},t)$, total pressure $p(\mathbf{x},t)$, total density $\rho(\mathbf{x},t)$, and the total entropy $s(\mathbf{x},t)$ must be the same. The total thermodynamic variables are expressed in the following way
\begin{subequations}\label{full_therm_vars}
\begin{equation}
T(\mathbf{x},t)=\tilde{T}(z)+T'(\mathbf{x},t)=T_{ad}(z)+T_S(\mathbf{x},t),
\end{equation}
\begin{equation}
\rho(\mathbf{x},t)=\tilde{\rho}(z)+\rho'(\mathbf{x},t)=\rho_{ad}(z)+\rho_S(\mathbf{x},t),
\end{equation}
\begin{equation}
p(\mathbf{x},t)=\tilde{p}(z)+p'(\mathbf{x},t)=p_{ad}(z)+p_S(\mathbf{x},t),
\end{equation}
\begin{equation}
s(\mathbf{x},t)=\tilde{s}(z)+s'(\mathbf{x},t)=\mathrm{const}+s_S(\mathbf{x},t),
\end{equation}
\end{subequations}
where the subscript $S$ denotes the superadiabatic fluctuation about the adiabatic state. The latter expressions define transformations between the two formulations of the type
\begin{equation}
T_S(\mathbf{x},t)=T'(\mathbf{x},t)+\tilde{T}(z)-T_{ad}(z)=T'(\mathbf{x},t)-\Delta_S z,\label{TS_Tprime_transform}
\end{equation}
where (\ref{eq:Ttilde_expanded_RS}) and (\ref{eq:adiabatic_profile_RS}) have been used and $T_{ad\,B}=T_{B}$. The transformation between the entropy fluctuations $s_S(\mathbf{x},t)=s'(\mathbf{x},t)+\tilde{s}(z)+\mathrm{const}$ is also rather simple, since the inhomogeneous correction involves only the vertical variation of $\tilde{s}(z)$, cf. (\ref{eq:BS2-1}). On the other hand the transformations for the density and pressure fluctuations, which result from (\ref{full_therm_vars}b,c), are slightly more complicated, because they must involve the bottom values of the density in both formulations, i.e. $\rho_B$ and $\rho_{ad\,B}$. Therefore if we assume, that the total mass of the fluid is contained in the references states (which is a natural assumption, see section \ref{subsec:Conservarion-of-mass-A}), and utilize the relation (\ref{eq:rho_rel}), the transformations for the density and pressure fluctuations between the two anelastic formulations about the conduction and adiabatic reference states become well defined. In particular for the density fluctuations we provide the explicit transformation formula (cf. (\ref{eq:rhotilde_expanded_RS}) and (\ref{eq:adiabatic_profile_RS}))
\begin{align}
\rho_S(\mathbf{x},t)= & \,\,\rho'(\mathbf{x},t)+\tilde{\rho}(z)-\rho_{ad}(z)\nonumber\\
= & \,\,\rho'(\mathbf{x},t)+\rho_{ad\,B}\left(1-\frac{gz}{c_{p}T_{B}}\right)^{\frac{1}{\gamma-1}}\Bigg\{\frac{\rho_{B}}{\rho_{ad\,B}}-1\Bigg .\nonumber \\
& \qquad\qquad\left .-\frac{c_{p}\Delta_{S}}{g\left(\gamma-1\right)}\left[\frac{\frac{gz}{c_{p}T_{B}}}{1-\frac{gz}{c_{p}T_{B}}}+\ln\left(1-\frac{gz}{c_{p}T_{B}}\right)\right]\right\}+\mathcal{O}\left(\rho_{ad\,B}\delta^2\right), \label{rho_transform}
\end{align}
where by the use of $\delta=L\Delta_S /T_B=\theta-gL/c_p T_B$ and (\ref{eq:rho_rel})  the ratio $\rho_{B}/\rho_{ad\,B}$ is given by
\begin{equation}
\frac{\rho_B}{\rho_{ad\,B}}=1-\frac{c_{p}\Delta_{S}}{g\left(\gamma-1\right)}\frac{\gamma\left(1-\frac{gL}{c_{p}T_{B}}\right)^{\frac{\gamma}{\gamma-1}}}{1-\left(1-\frac{gL}{c_{p}T_{B}}\right)^{\frac{\gamma}{\gamma-1}}}\left[\frac{\frac{gL}{c_{p}T_{B}}}{1-\frac{gL}{c_{p}T_{B}}}+\ln\left(1-\frac{gL}{c_{p}T_{B}}\right)\right]+\mathcal{O}\left(\delta^2\right).\label{rho_ratio_formulations}
\end{equation}
Of course the correction to $\rho'(\mathbf{x},t)$ in equation (\ref{rho_transform}) is of the same order of magnitude as the density fluctuation, i.e. $\mathcal{O}(\rho_{ad\,B}\delta)$. Similar transformation formula can be obtained for the pressure fluctuation. We conclude, that the results obtained with one formulation, say for fluctuations about a conduction reference state, can be easily transformed into fluctuations about the adiabatic state in the same physical setting using (\ref{full_therm_vars}a-d). In practice, it is often necessary to compare results obtained from numerical simulations utilizing the two different formulations. The above recipe allows to do this, but of course the physical situation modelled with the two approaches {\it{must}} be the same in order for the results to correspond directly to each other. Therefore such comparisons must be done with great care. In the considered example the two formulations necessarily produce the same results, as long as the aforementioned set of physical parameters, i.e. $\mu$, $\mu_b$, $k$, $c_p$, $c_v$, $g$, the total mass of the fluid $M_f$ and the driving bottom-top temperature difference $\Delta T$ is the same in both approaches. We elaborate on the issue of how to compare the results of the conduction reference state formulation with the results of the formulation with the adiabatic reference state in section \ref{subsec:Comparison-of-anelastic}.

Finally we note, that in numerical modelling of anelastic convection
an often undertaken approach is to utilize time-dependent basic states.
All the variables such as $\rho$, $T$, $p$ and $\mathbf{u}$ are
horizontally averaged at each time step, so that the basic state $\left\langle \rho\right\rangle _{h}(z,t)$,
$\left\langle T\right\rangle _{h}(z,t)$, $\left\langle p\right\rangle _{h}(z,t)$
and $\left\langle \mathbf{u}\right\rangle _{h}(z,t)$ depends only
on height and time. Often a time average is applied as well,
\begin{equation}
\left\langle \rho\right\rangle _{h,t}=\frac{1}{t}\int_{0}^{t}\left\langle \rho\right\rangle _{h}(z,s)\mathrm{d}s,\label{eq:e175}
\end{equation}
which implies that the time dependence of the basic state is slow.
Such a state naturally satisfies all the boundary conditions responsible
for driving the flow; if the fluid flow is thermally driven this implies
that the temperature fluctuation satisfies homogeneous boundary conditions.
If, however, convection is driven by fixed entropy at boundaries,
the entropy fluctuation has to vanish there. The small parameter $\delta$ is then naturally defined by the boundary conditions, as in (\ref{eq:e174}). 

\section{Simplifications through entropy formulations\label{subsec:Simplifications-through-entropy}}\index{SI}{entropy!formulation}

The aim of this section is to further simplify the full system of
dynamical anelastic equations and express them solely in terms of
two thermodynamic variables, the entropy $s'$ and the pressure $p'$
with the latter appearing only in the Navier-Stokes equation under
the $\nabla$ operator, thus being easily removable by taking its
curl. By making use of equations (\ref{State_eq_Aderiv-1-1}) which
result from the equation of state, one can eliminate the density and
temperature fluctuations and express them by the entropy and pressure
fluctuations 
\begin{subequations}
\begin{equation}
\rho'=-\frac{\tilde{\alpha}\tilde{T}\tilde{\rho}}{\tilde{c}_{p}}s'+\tilde{\rho}\tilde{\beta}\frac{\tilde{c}_{v}}{\tilde{c}_{p}}p',\label{eq:rho_prime_A}
\end{equation}
\begin{equation}
T'=\frac{\tilde{T}}{\tilde{c}_{p}}s'+\frac{\tilde{\alpha}\tilde{T}}{\tilde{c}_{p}\tilde{\rho}}p',\label{eq:T_prime_A}
\end{equation}
\end{subequations} 
where the thermodynamic identity
\begin{equation}
c_{p}-c_{v}=\frac{\alpha^{2}T}{\beta\rho},\label{eq:cp-cv_A}
\end{equation}
justified in (\ref{eq:cp-cv_B}) was used in obtaining the above expression
for $\rho'$. Introduction of (\ref{eq:T_prime_A}) into the energy
balance (\ref{Energy_Aderiv-1-1}) leads to
\begin{align}
\tilde{\rho}\tilde{T}\left[\frac{\partial s'}{\partial t}+\mathbf{u}\cdot\nabla\left(s'+\tilde{s}\right)\right]= & \nabla\cdot\left[k\nabla\left(\frac{\tilde{T}}{\tilde{c}_{p}}s'\right)\right]+\nabla\cdot\left[k\nabla\left(\frac{\tilde{\alpha}\tilde{T}}{\tilde{c}_{p}}\frac{p'}{\tilde{\rho}}\right)\right]\nonumber \\
 & +2\mu\mathbf{G}^{s}:\mathbf{G}^{s}+\left(\mu_{b}-\frac{2}{3}\mu\right)\left(\nabla\cdot\mathbf{u}\right)^{2}+Q'.\label{eq:Energy_eq_A_entropy_A-1}
\end{align}
On the other hand the pressure term together with the buoyancy force
in the Navier-Stokes equation (\ref{NS-Aderiv-1-1}), divided by $\tilde{\rho}$,
with the aid of (\ref{eq:rho_prime_A}) can be expressed in the following
way
\begin{align}
-\frac{1}{\tilde{\rho}}\nabla p'+\frac{\rho'}{\tilde{\rho}}\tilde{\mathbf{g}}-\nabla\psi'= & -\frac{1}{\tilde{\rho}}\nabla p'+\left(\frac{\tilde{\alpha}\tilde{T}}{\tilde{c}_{p}}s'-\tilde{\beta}\frac{\tilde{c}_{v}}{\tilde{c}_{p}}p'\right)\tilde{g}\hat{\mathbf{e}}_{z}-\nabla\psi'\nonumber \\
= & -\nabla\left(\frac{p'}{\tilde{\rho}}+\psi'\right)+\frac{\tilde{\alpha}\tilde{T}}{\tilde{c}_{p}}s'\tilde{g}\hat{\mathbf{e}}_{z}-\left(\tilde{g}\tilde{\beta}\frac{\tilde{c}_{v}}{\tilde{c}_{p}}+\frac{1}{\tilde{\rho}^{2}}\frac{\mathrm{d}\tilde{\rho}}{\mathrm{d}z}\right)p'\hat{\mathbf{e}}_{z}.\label{eq:Buoyancy_entropy_A_1}
\end{align}
Next, since the $z$-derivative of the static density distribution
is given by (\ref{eq:rho_derivative_in_z_A}) and, again, making use
of $c_{p}-c_{v}=\alpha^{2}T/\beta\rho$ (cf. (\ref{eq:cp-cv_A})),
one obtains
\begin{eqnarray}
-\frac{1}{\tilde{\rho}}\nabla p'+\frac{\rho'}{\tilde{\rho}}\tilde{\mathbf{g}}-\nabla\psi' & = & -\nabla\left(\frac{p'}{\tilde{\rho}}+\psi'\right)+\frac{\tilde{\alpha}\tilde{T}}{\tilde{c}_{p}}s'\tilde{g}\hat{\mathbf{e}}_{z}-\tilde{\alpha}\frac{p'}{\tilde{\rho}}\Delta_{S}\hat{\mathbf{e}}_{z}\nonumber \\
 & = & -\nabla\left(\frac{p'}{\tilde{\rho}}+\psi'\right)+\frac{\tilde{\alpha}\tilde{T}}{\tilde{c}_{p}}s'\tilde{g}\hat{\mathbf{e}}_{z}+\mathcal{O}\left(\bar{g}\delta^{2}\right),\label{eq:Buoyancy_entropy_A_2}
\end{eqnarray}
since $\Delta_{S}=\mathcal{O}(\bar{T}\delta/L)$ and $p'/\tilde{p}=\mathcal{O}(\delta)$. 

This allows to rewrite the system of dynamic equations under the anelastic
approximation in the form 
\begin{subequations}
\begin{align}
\frac{\partial\mathbf{u}}{\partial t}+\left(\mathbf{u}\cdot\nabla\right)\mathbf{u}= & -\nabla\left(\frac{p'}{\tilde{\rho}}+\psi'\right)+\frac{\tilde{\alpha}\tilde{T}}{\tilde{c}_{p}}s'\tilde{g}\hat{\mathbf{e}}_{z}+\frac{\mu}{\tilde{\rho}}\nabla^{2}\mathbf{u}+\left(\frac{\mu}{3\tilde{\rho}}+\frac{\mu_{b}}{\tilde{\rho}}\right)\nabla\left(\nabla\cdot\mathbf{u}\right)\nonumber \\
 & +\frac{2}{\tilde{\rho}}\nabla\mu\cdot\mathbf{G}^{s}+\frac{1}{\tilde{\rho}}\nabla\left(\mu_{b}-\frac{2}{3}\mu\right)\nabla\cdot\mathbf{u},\label{NS-Aderiv-1-1-1}
\end{align}
\begin{equation}
\nabla^{2}\psi'=4\pi G\frac{\tilde{\rho}}{\tilde{c}_{p}}\left(\tilde{\beta}\tilde{c}_{v}p'-\tilde{\alpha}\tilde{T}s'\right)\label{eq:psi_prime_A_ent}
\end{equation}
\begin{equation}
\nabla\cdot\left(\tilde{\rho}\mathbf{u}\right)=0,\label{Cont_Aderiv-1-1-1}
\end{equation}
\begin{align}
\tilde{\rho}\tilde{T}\left(\frac{\partial s'}{\partial t}+\mathbf{u}\cdot\nabla s'\right)-\tilde{\rho}\tilde{c}_{p}u_{z}\Delta_{S}= & \nabla\cdot\left[k\nabla\left(\frac{\tilde{T}}{\tilde{c}_{p}}s'\right)\right]+\nabla\cdot\left[k\nabla\left(\frac{\tilde{\alpha}\tilde{T}}{\tilde{c}_{p}}\frac{p'}{\tilde{\rho}}\right)\right]\nonumber \\
 & +2\mu\mathbf{G}^{s}:\mathbf{G}^{s}+\left(\mu_{b}-\frac{2}{3}\mu\right)\left(\nabla\cdot\mathbf{u}\right)^{2}+Q',\label{eq:Energy_eq_A_entropy_A-1-1}
\end{align}
\end{subequations}
which constitute a closed problem for two thermodynamic
variables $s'$ and $p'$ and the velocity field $\mathbf{u}$. However,
up to now we have merely substituted for the density and temperature
fluctuations from (\ref{eq:rho_prime_A},b), but the pressure fluctuation
still appears in the energy equation and in the equation for the gravitational
potential. Therefore we will now assume, that the gravitational acceleration
is constant (cf. the discussion below (\ref{eq:gravity_acceleration_A}))
\begin{equation}
\textrm{assume:}\quad g\approx\tilde{g}\approx\mathrm{const},\quad\left|\nabla\psi'\right|\ll\frac{\rho'}{\tilde{\rho}}g,\label{eq:e176}
\end{equation}
which allows to eliminate the fluctuation $\mathbf{g}'=-\nabla\psi'$
from the momentum balance, as it is negligibly small in comparison
with the term $\tilde{\alpha}\tilde{T}s'g\hat{\mathbf{e}}_{z}/\tilde{c}_{p}$,
and hence the equation (\ref{eq:psi_prime_A_ent}) can also be removed. 

Next, multiplying the equation (\ref{NS-Aderiv-1-1-1}) by $\tilde{\rho}$
and taking its divergence, by the use of the continuity equation (\ref{Cont_Aderiv-1-1-1})
we obtain a stationary Poisson-type problem for $p'/\tilde{\rho}$,
at the leading order in $\delta$,
\begin{equation}
\nabla\cdot\left(\tilde{\rho}\nabla\frac{p'}{\tilde{\rho}}\right)=\nabla\cdot\left[\nabla\cdot\left(2\mu\mathbf{G}^{s}-\tilde{\rho}\mathbf{u}\mathbf{u}\right)\right]+\nabla^{2}\left[\left(\mu_{b}-\frac{2}{3}\mu\right)\nabla\cdot\mathbf{u}\right]+\frac{\partial}{\partial z}\left(\frac{\tilde{\alpha}\tilde{T}\tilde{\rho}g}{\tilde{c}_{p}}s'\right),\label{eq:p'_eq_2}
\end{equation}
where 
\begin{equation}
\nabla\cdot\left[\nabla\cdot\left(2\mu\mathbf{G}^{s}-\tilde{\rho}\mathbf{u}\mathbf{u}\right)\right]=\partial_{i}\partial_{j}\left(2\mu G_{ij}^{s}-\tilde{\rho}u_{i}u_{j}\right),\label{eq:e177}
\end{equation}
and with the aid of the estimate of the static density vertical variation
in (\ref{eq:drhobydz_A}) we get
\begin{equation}
\nabla\cdot\left(\tilde{\rho}\nabla\frac{p'}{\tilde{\rho}}\right)=\tilde{\rho}\nabla^{2}\frac{p'}{\tilde{\rho}}-\frac{\tilde{c}_{v}\tilde{\beta}\tilde{\rho}^{2}g}{\tilde{c}_{p}}\frac{\partial}{\partial z}\frac{p'}{\tilde{\rho}}.\label{eq:e178}
\end{equation}
The Poisson-type, stationary problem for the pressure perturbation is a
manifestation of the fact, that under the anelastic approximation
all the terms of the order $\mathcal{O}(Ma^{2})$ are neglected and
thus pressure spreads infinitely fast. Note, that up to now no additional
assumptions have been made except for $g=\mathrm{const}$
and the standard anelastic assumption of small system departure from
the adiabatic state. If we assume, that the fluid satisfies the equation
of state of a perfect gas (with constant specific heats) the above
system of dynamical equations reduces to 
\begin{subequations}
\begin{align}
\frac{\partial\mathbf{u}}{\partial t}+\left(\mathbf{u}\cdot\nabla\right)\mathbf{u}= & -\nabla\frac{p'}{\tilde{\rho}}+\frac{gs'}{c_{p}}\hat{\mathbf{e}}_{z}+\frac{\mu}{\tilde{\rho}}\nabla^{2}\mathbf{u}+\left(\frac{\mu}{3\tilde{\rho}}+\frac{\mu_{b}}{\tilde{\rho}}\right)\nabla\left(\nabla\cdot\mathbf{u}\right)\nonumber \\
 & +\frac{2}{\tilde{\rho}}\nabla\mu\cdot\mathbf{G}^{s}+\frac{1}{\tilde{\rho}}\nabla\left(\mu_{b}-\frac{2}{3}\mu\right)\nabla\cdot\mathbf{u},\label{NS-Aderiv-1-1-1-1}
\end{align}
\begin{equation}
\nabla\cdot\left(\tilde{\rho}\mathbf{u}\right)=0,\label{Cont_Aderiv-1-1-1-1}
\end{equation}
\begin{align}
\tilde{\rho}\tilde{T}\left(\frac{\partial s'}{\partial t}+\mathbf{u}\cdot\nabla s'\right)-\tilde{\rho}c_{p}u_{z}\Delta_{S}= & \nabla\cdot\left[\kappa\tilde{\rho}\nabla\left(\tilde{T}s'\right)\right]+\nabla\cdot\left[\kappa\tilde{\rho}\nabla\left(\frac{p'}{\tilde{\rho}}\right)\right]\nonumber \\
 & +2\mu\mathbf{G}^{s}:\mathbf{G}^{s}+\left(\mu_{b}-\frac{2}{3}\mu\right)\left(\nabla\cdot\mathbf{u}\right)^{2}+Q',\label{eq:Energy_eq_A_entropy_A-1-1-1}
\end{align}
\end{subequations} 
where we have introduced the thermal diffusivity
\begin{equation}
\kappa=\frac{k}{\tilde{\rho}c_{p}}.\label{eq:e179}
\end{equation}
The Poisson-type problem for the pressure is given by (\ref{eq:p'_eq_2}),
or more explicitly
\begin{align}
\nabla^{2}\frac{p'}{\tilde{\rho}}-\frac{g}{c_{p}\left(\gamma-1\right)\tilde{T}}\frac{\partial}{\partial z}\frac{p'}{\tilde{\rho}}= & \frac{1}{\tilde{\rho}}\nabla\cdot\left[\nabla\cdot\left(2\mu\mathbf{G}^{s}-\tilde{\rho}\mathbf{u}\mathbf{u}\right)\right]+\frac{1}{\tilde{\rho}}\nabla^{2}\left[\left(\mu_{b}-\frac{2}{3}\mu\right)\nabla\cdot\mathbf{u}\right]\nonumber \\
 & +\frac{1}{c_{p}\tilde{\rho}}\frac{\partial}{\partial z}\left(\tilde{\rho}gs'\right),\label{eq:p'overrho_perfectgas_A}
\end{align}
with $\gamma=c_{p}/c_{v}=\mathrm{const}$ being the specific heat
ratio.

\subsection{Boundary conditions for the entropy\label{subsec:Boundary-conditions-for-s}}

Formulation in terms of the entropy density per unit mass $s'$ requires specification
of boundary conditions for this variable. These naturally depend on
a particular problem and may differ from one application to another.
However, on general grounds we can say, that by the use of (\ref{eq:Taylor_exp_entropy_A-1})
the fixed temperature boundary conditions $T'|_{z=0,L}=0$ correspond
to 
\begin{equation}
s'+\frac{\tilde{\alpha}}{\tilde{\rho}}p'=0\quad\textrm{at}\;z=0,\,L\label{eq:BC_entropy_0L_fT}
\end{equation}
whereas fixed heat flux at boundaries $\partial_{z}T'|_{z=0,L}=0$
is equivalent to
\begin{equation}
\frac{\partial}{\partial z}\left(\frac{\tilde{T}}{\tilde{c}_{p}}s'+\frac{\tilde{\alpha}\tilde{T}}{\tilde{c}_{p}\tilde{\rho}}p'\right)=0\quad\textrm{at}\;z=0,\,L.\label{eq:BC_entropy_0L_fF}
\end{equation}
Either one of the boundary conditions, i.e. fixed temperature or fixed
heat flux is supplied by the general mass conservation law $\partial_{t}(\tilde{\rho}+\rho')+\nabla\cdot[(\tilde{\rho}+\rho')\mathbf{u}]=0$
which implies $\left\langle \tilde{\rho}+\rho'\right\rangle =\mathrm{const}$.
Under standard initial conditions not introducing additional mass
into the system it follows that $\left\langle \rho'\right\rangle =0$, and consequently
\begin{equation}
\left\langle \frac{\tilde{\alpha}\tilde{T}\tilde{\rho}}{\tilde{c}_{p}}s'\right\rangle =\left\langle \frac{\tilde{c}_{v}\tilde{\beta}\tilde{\rho}}{\tilde{c}_{p}}p'\right\rangle .\label{eq:entropy_pressure_mean_rel}
\end{equation}
Since there are two boundary conditions required for the entropy field
$s'$ and one for the pressure $p'$ the three relations: two boundary
conditions at $z=0,\,L$ either (\ref{eq:BC_entropy_0L_fT}) or (\ref{eq:BC_entropy_0L_fF})
and the relation between spatial means (\ref{eq:entropy_pressure_mean_rel})
constitute a necessary and sufficient set of conditions, allowing
to fully determine the fields. However, this set of conditions is
rather cumbersome in terms of applicability in numerical simulations,
because not only that it couples the entropy and pressure but also
involves computation of means in order to fully determine the entropy
and pressure fields.

\subsubsection{The case of isentropic boundary conditions\label{subsec:isentropic_BCs_temp_shift}}

Turbulent convection is often modelled with application of isentropic
boundary conditions. Although keeping the entropy fixed at boundaries
does not seem physical, since physically one can not control the pressure
and hence also the entropy at boundaries, it is a common practice due
to simplicity achieved when the entropy formulations are used in tandem
with fixed entropy boundary conditions. In such a case $s'|_{z=0,L}=0$
corresponds to 
\begin{equation}
-\frac{p'}{c_{p}\tilde{\rho}}+T'=0\quad\textrm{at}\;z=0,\,L\label{eq:fixed_entropy_BCs_forT}
\end{equation}
whereas the conservation of the total mass, $\left\langle \rho'\right\rangle =0$,
implies
\begin{equation}
\left\langle \tilde{\rho}s'\right\rangle =\frac{c_{v}}{R}\left\langle \frac{p'}{\tilde{T}}\right\rangle ,\label{eq:entropy_pressure_mean_rel-1}
\end{equation}
cf. equations (\ref{eq:Taylor_exp_rho_A-1},b). 

The following calculation reveals some interesting features of convective
flows. First we note, that by the mass conservation law the horizontally
averaged vertical velocity must satisfy
\begin{equation}
\frac{\partial\left\langle u_{z}\right\rangle _{h}}{\partial z}=-\frac{\left\langle u_{z}\right\rangle _{h}}{D_{\rho}},\label{eq:e193}
\end{equation}
where $D_{\rho}=-\tilde{\rho}/\mathrm{d}_{z}\tilde{\rho}$ is the
density scale height, therefore impermeability conditions at the boundaries
at $z=0,\,L$ imply
\begin{equation}
\left\langle u_{z}\right\rangle _{h}=0.\label{eq:mean_h_uz_null_A}
\end{equation}
A horizontal average of the $z$-component of the stationary Navier-Stokes
equation (\ref{NS-Aderiv-1-1}) with constant gravity yields
\begin{equation}
\frac{\partial}{\partial z}\left\langle \tilde{\rho}u_{z}^{2}\right\rangle _{h}=-\frac{\partial\left\langle p'\right\rangle _{h}}{\partial z}-g\left\langle \rho'\right\rangle _{h},\label{eq:eq_ad1}
\end{equation}
therefore the mean pressure fluctuation $\left\langle p'\right\rangle _{h}$
in convection satisfies a non-hydrostatic balance, influenced by inertia.
Next, integration of the latter equality over $z$ from $0$ to $L$
shows\index{SI}{pressure boundary conditions}
\begin{equation}
\Delta\left\langle p'\right\rangle _{h}=\left\langle p'\right\rangle _{h,B}-\left\langle p'\right\rangle _{h,T}=0,\label{eq:no_pressure_jump}
\end{equation}
which means that the mean pressure fluctuation is the same at both,
top and bottom boundaries. Consequently the boundary conditions (\ref{eq:fixed_entropy_BCs_forT})
can be expressed in the following way
\begin{subequations}
\begin{equation}
\left\langle T'\right\rangle _{h,B}=\frac{\left\langle p'\right\rangle _{h,B}}{c_{p}\tilde{\rho}_{B}},\label{eq:BC_entropy_0L_fT-1}
\end{equation}
\begin{equation}
\left\langle T'\right\rangle _{h,T}=\frac{\left\langle p'\right\rangle _{h,B}}{c_{p}\tilde{\rho}_{T}}=\frac{\tilde{\rho}_{B}}{\tilde{\rho}_{T}}\left\langle T'\right\rangle _{h,B},\label{eq:BC_entropy_0L_fT-1-1}
\end{equation}
\end{subequations}
and of course at leading order in $\delta$ the
values of the reference state density at top and bottom $\tilde{\rho}_{T}$
and $\tilde{\rho}_{B}$ could be simply replaced by the top and bottom
values of the total density $\rho(z=0)$ and $\rho(z=L)$, which in
general differ by order $\mathcal{O}(\delta)$ corrections. Since
density decreases with height, $\tilde{\rho}_{B}>\tilde{\rho}_{T}$, it is clear from (\ref{eq:BC_entropy_0L_fT-1-1}),
that when the boundaries are isentropic, the mean temperature fluctuation
is of the same sign at the top and bottom boundaries and its magnitude
is significantly greater at the top (note that similar calculation
could be done for the case of isothermal boundaries, with analogous
results for the magnitudes and signs of the mean entropy fluctuation
at boundaries).

\subsection{Constant thermal diffusivity formulation for an ideal gas with non-vanishing
heat sink\label{subsec:Constant-thermal-diffusivity}}

\footnote{This section follows the derivation of Mizerski (2017).}Holding
the assumption, that the fluid satisfies the equation of state of
a perfect gas (with constant specific heats) and $g=\mathrm{const}$
a further simplification can be achieved by elimination of the pressure
term from the energy equation. The pressure term in that equation
can be easily expressed in the form
\begin{equation}
\nabla\cdot\left[\kappa\tilde{\rho}\nabla\left(\frac{p'}{\tilde{\rho}}\right)\right]=\kappa\nabla\cdot\left[\tilde{\rho}\nabla\left(\frac{p'}{\tilde{\rho}}\right)\right]+\tilde{\rho}\frac{\mathrm{d}\kappa}{\mathrm{d}z}\frac{\partial}{\partial z}\left(\frac{p'}{\tilde{\rho}}\right).\label{eq:e180}
\end{equation}
Therefore by virtue of (\ref{eq:p'_eq_2}) the energy equation can
be rewritten in the form
\begin{align}
\tilde{\rho}\tilde{T}\left(\frac{\partial s'}{\partial t}+\mathbf{u}\cdot\nabla s'\right)-\tilde{\rho}c_{p}u_{z}\Delta_{S}= & \nabla\cdot\left[\kappa\tilde{\rho}\nabla\left(\tilde{T}s'\right)\right]+\kappa\frac{g}{c_{p}}\frac{\partial}{\partial z}\left(\tilde{\rho}s'\right)\nonumber \\
 & +\tilde{\rho}\frac{\mathrm{d}\kappa}{\mathrm{d}z}\frac{\partial}{\partial z}\frac{p'}{\tilde{\rho}}+\mathcal{J}+Q',\label{eq:energy_by_entropy_deriv_A}
\end{align}
where the term
\begin{align}
\mathcal{J}= & \kappa\nabla\cdot\left[\nabla\cdot\left(2\mu\mathbf{G}^{s}-\tilde{\rho}\mathbf{u}\mathbf{u}\right)\right]+\kappa\nabla^{2}\left[\left(\mu_{b}-\frac{2}{3}\mu\right)\nabla\cdot\mathbf{u}\right]\nonumber \\
 & +2\mu\mathbf{G}^{s}:\mathbf{G}^{s}+\left(\mu_{b}-\frac{2}{3}\mu\right)\left(\nabla\cdot\mathbf{u}\right)^{2},\label{eq:J_def_A}
\end{align}
gathers all terms nonlinear in the velocity field and all associated
with viscous diffusion. Next, the two terms on the right hand side
of (\ref{eq:energy_by_entropy_deriv_A}) involving the entropy derivatives,
on the basis of the definition of the parameter $\delta$ in (\ref{eq:delta_definition})
yielding
\begin{equation}
\frac{\mathrm{d}\tilde{T}}{\mathrm{d}z}=-\frac{g}{c_{p}}+\mathcal{O}\left(\delta\frac{\bar{T}}{L}\right),\label{eq:e181}
\end{equation}
can be manipulated to give at leading order
\begin{align}
\nabla\cdot\left[\kappa\tilde{\rho}\nabla\left(\tilde{T}s'\right)\right] & +\kappa\frac{1}{c_{p}}\frac{\partial}{\partial z}\left(\tilde{\rho}gs'\right)\nonumber \\
= & \,\frac{\partial}{\partial z}\left(\kappa\tilde{\rho}\frac{\mathrm{d}\tilde{T}}{\mathrm{d}z}s'\right)+\nabla\cdot\left(\kappa\tilde{\rho}\tilde{T}\nabla s'\right)+\kappa\frac{g}{c_{p}}\frac{\partial}{\partial z}\left(\tilde{\rho}s'\right)\nonumber \\
= & -\frac{g}{c_{p}}\frac{\mathrm{d}\kappa}{\mathrm{d}z}\tilde{\rho}s'+\nabla\cdot\left(\kappa\tilde{\rho}\tilde{T}\nabla s'\right)+\mathcal{O}\left(\delta^{5/2}\bar{\rho}g\sqrt{gL}\right).\label{eq:e182}
\end{align}
Finally the energy equation can be cast in the following form
\begin{equation}
\tilde{\rho}\tilde{T}\left(\frac{\partial s'}{\partial t}+\mathbf{u}\cdot\nabla s'\right)-\tilde{\rho}c_{p}u_{z}\Delta_{S}=\nabla\cdot\left(\kappa\tilde{\rho}\tilde{T}\nabla s'\right)+\tilde{\rho}\frac{\mathrm{d}\kappa}{\mathrm{d}z}\left[\frac{\partial}{\partial z}\frac{p'}{\tilde{\rho}}-\frac{g}{c_{p}}s'\right]+\mathcal{J}+Q'.\label{eq:energy_by_entropy_deriv_A-1}
\end{equation}
It is clear now, that the assumption of uniform thermal diffusivity
$\kappa$ allows to simplify the energy equation by removing the entire
term proportional to the $z$-derivative of $\kappa$, which is the
only term involving the pressure fluctuation and then the full system
of anelastic equations reads\index{SI}{entropy!formulation} 
\begin{subequations}
\begin{align}
\frac{\partial\mathbf{u}}{\partial t}+\left(\mathbf{u}\cdot\nabla\right)\mathbf{u}= & -\nabla\frac{p'}{\tilde{\rho}}+\frac{gs'}{c_{p}}\hat{\mathbf{e}}_{z}+\frac{\mu}{\tilde{\rho}}\nabla^{2}\mathbf{u}+\left(\frac{\mu}{3\tilde{\rho}}+\frac{\mu_{b}}{\tilde{\rho}}\right)\nabla\left(\nabla\cdot\mathbf{u}\right)\nonumber \\
 & +\frac{2}{\tilde{\rho}}\nabla\mu\cdot\mathbf{G}^{s}+\frac{1}{\tilde{\rho}}\nabla\left(\mu_{b}-\frac{2}{3}\mu\right)\nabla\cdot\mathbf{u},\label{NS-Aderiv-1-1-1-1-1}
\end{align}
\begin{equation}
\nabla\cdot\left(\tilde{\rho}\mathbf{u}\right)=0,\label{Cont_Aderiv-1-1-1-1-1}
\end{equation}
\begin{equation}
\tilde{\rho}\tilde{T}\left(\frac{\partial s'}{\partial t}+\mathbf{u}\cdot\nabla s'\right)-\tilde{\rho}c_{p}u_{z}\Delta_{S}=\kappa\nabla\cdot\left(\tilde{\rho}\tilde{T}\nabla s'\right)+\mathcal{J}+Q',\label{eq:energy_final_constkappa}
\end{equation}
 \end{subequations} 
where $\mathcal{J}$ is given in (\ref{eq:J_def_A}).

This constitutes a closed system of equations for an ideal gas, obtained
by no additional assumptions other than $g=\mathrm{const}$
and $\kappa=\mathrm{const}$, expressed solely in terms of the velocity
field and the pressure and entropy fluctuations. If the pressure
distribution is required, it can be calculated from the Poisson-type problem for pressure fluctuation provided in (\ref{eq:p'overrho_perfectgas_A}). However, the pressure fluctuation
is entirely eliminated from the energy equation and it appears only
in the momentum balance under the gradient operator. Therefore the
pressure problem, which as mentioned involves some quite significant
complications for computational implementations can be easily avoided
by taking curl of the Navier-Stokes equation and eliminating the pressure
fluctuations from the full set of the dynamical equations. The significant
advantage is that physically one can not control pressure at the boundaries
and therefore boundary conditions on pressure are computationally
cumbersome and often have to involve application of spatial averages.
Hence elimination of pressure is desired. Such an approach typically
involves introduction of some potentials, such as e.g. the toroidal
(say $\mathcal{T}$) and poloidal (say $\mathcal{P}$) potentials
\begin{equation}
\tilde{\rho}\mathbf{u}=\nabla\times\left(\mathcal{T}\hat{\mathbf{e}}_{z}\right)+\nabla\times\nabla\times\left(\mathcal{P}\hat{\mathbf{e}}_{z}\right)\label{eq:e183}
\end{equation}
or the vector potential $\tilde{\rho}\mathbf{u}=\nabla\times\mathbf{A}$
(with some gauge conditions for $\mathcal{T}$ and $\mathcal{P}$
or for $\mathbf{A}$) to satisfy the solenoidal constraint $\nabla\cdot(\tilde{\rho}\mathbf{u})=0$,
thus a reduction of the number of variables. It is often accompanied
by an arbitrary \emph{ad hoc} but greatly simplifying assumption that
the entropy is constant and known on the boundaries, even though physically
one cannot control the entropy nor pressure on boundaries. The boundary
conditions for the potentials ($\mathcal{T}$ and $\mathcal{P}$ or
$\mathbf{A}$) then need to be derived on the basis of the physical
conditions assumed for the velocity field. 

We have started the derivation of the final simplified energy equation
(\ref{eq:energy_final_constkappa}) from the equation (\ref{Energy_Aderiv-1-1})
which is equivalent to (\ref{eq:Energy_eq_A_entropy_A}). Although
the latter two are already simplified with respect to the full energy equation
(\ref{Energy_Aderiv}), we have demonstrated in section \ref{subsec:Entropy-production}
that this simplification does not affect the sign of total entropy
production\index{SI}{entropy!production}, 
\begin{equation}
\frac{\partial}{\partial t}\int\tilde{\rho}s'\mathrm{d}^{3}x,\label{eq:e184}
\end{equation}
which in accordance with the second law of thermodynamics is positive,
if the boundaries are assumed adiabatically insulating. In the process
of derivation of  (\ref{eq:energy_final_constkappa}) we have neglected
some terms of the order $\mathcal{O}(\delta^{5/2})$, however, careful
considerations show, that all the neglected terms in the energy equation
can be gathered to yield a correction of the form
\begin{align}
\kappa\frac{\partial}{\partial z}\left[\frac{\Delta_{S}}{\tilde{T}}\left(p'+\tilde{\rho}\tilde{T}s'\right)\right]= & \,\kappa\frac{\partial}{\partial z}\left[\frac{c_{p}\Delta_{S}}{g}\left(\frac{\gamma-1}{\gamma}\frac{p'}{\tilde{p}}+\frac{s'}{c_{p}}\right)\tilde{\rho}g\right]\nonumber \\
= & \,\mathcal{O}\left(\delta^{5/2}\bar{\rho}g\sqrt{gL}\right).\label{eq:e185}
\end{align}
These terms, divided by $\tilde{T}$ and integrated over the entire
fluid volume provide only a small, $\mathcal{O}(\delta^{5/2})$ order
correction to the total entropy production which is of the order $\mathcal{O}(\delta^{3/2})$.
Therefore the total entropy production in  (\ref{eq:energy_final_constkappa})
is necessarily positive.

\subsubsection{The reference state at constant thermal diffusivity\label{subsec:constant-kappa-reference-state}}

Let us focus now on the possible forms of the hydrostatic reference
state in the case, when the thermal diffusivity $\kappa=k/\tilde{\rho}c_{p}$ is uniform\footnote{We stress, that the above entropy formulation is valid only when the
parameter $\kappa=k/\tilde{\rho}c_{p}$ is uniform, which is the thermal
diffusivity defined with the basic density profile $\tilde{\rho}$,
not the 'full' thermal diffusivity $k/c_{p}(\tilde{\rho}+\rho')$.
This corresponds to an assumption of a particular vertical profile
of the thermal conduction coefficient $k\sim\tilde{\rho}$.}. We consider an ideal gas at uniform gravity $g=\mathrm{const}$.
The reference state equations (\ref{eq:hydrostatic_eq_A_1}-d) in
this case simplify to
\begin{equation}
\frac{\mathrm{d}\tilde{p}}{\mathrm{d}z}=-\tilde{\rho}g,\qquad\frac{\mathrm{d}}{\mathrm{d}z}\left(\tilde{\rho}\frac{\mathrm{d}\tilde{T}}{\mathrm{d}z}\right)=-\frac{\tilde{Q}}{\kappa c_{p}},\qquad\tilde{p}=\tilde{\rho}R\tilde{T}.\label{eq:hydrostatic_eq_A_1-3}
\end{equation}
Of course, once $\tilde{p}$, $\tilde{\rho}$ and $\tilde{T}$ are
known the basic state entropy can be found from
\begin{equation}
\tilde{s}=c_{v}\ln\frac{\tilde{p}}{\tilde{\rho}^{\gamma}}+\mathrm{const}.\label{eq:e186}
\end{equation}
Elimination of the pressure from equations (\ref{eq:hydrostatic_eq_A_1-3})
leaves
\begin{equation}
\frac{\mathrm{d}\tilde{T}}{\mathrm{d}z}+\frac{1}{\tilde{\rho}}\frac{\mathrm{d}\tilde{\rho}}{\mathrm{d}z}\tilde{T}+\frac{g}{R}=0,\qquad\frac{\mathrm{d}}{\mathrm{d}z}\left(\tilde{\rho}\frac{\mathrm{d}\tilde{T}}{\mathrm{d}z}\right)=-\frac{\tilde{Q}}{\kappa c_{p}}.\label{eq:ref_st_kappa}
\end{equation}
The above reference state equations must be satisfied simultaneously
with the fundamental assumption $\delta\ll1$, which implies
\begin{equation}
-\frac{L}{\bar{T}}\left(\frac{\mathrm{d}\tilde{T}}{\mathrm{d}z}+\frac{g}{c_{p}}\right)=\mathcal{O}\left(\delta\right)\ll1.\label{eq:departure_kappa}
\end{equation}
Note, that under the current assumptions the adiabatic gradient $g/c_{p}$
is uniform, therefore the latter equality can only be satisfied when
either both, the constant adiabatic gradient and vertically varying
gradient of the reference temperature profile are small compared to
the mean temperature of the fluid divided by the layer's thickness
$\bar{T}/L$ or the temperature gradient in the reference state is
uniform at leading order. The former case implies only a weak stratification,
thus effectively reduces the model to the Boussinesq approximation
and it is only the latter case which is of interest. When the reference
state gradient is (at least approximately) uniform the energy equation
becomes $\tilde{Q}=-\kappa c_{p}\mathrm{d}_{z}\tilde{T}\mathrm{d}_{z}\tilde{\rho}$
and since both the gradients $\mathrm{d}_{z}\tilde{\rho}$ and $\mathrm{d}_{z}\tilde{T}$
are negative this necessarily implies heat sinks, $\tilde{Q}<0$,
i.e. cooling processes in the fluid volume. Note, that the nearly
adiabatic temperature gradient $\mathrm{d}_{z}\tilde{T}=-g/c_{p}+\mathcal{O}(\delta\Delta T/L)$
together with the hydrostatic force balance $\mathrm{d}_{z}\tilde{p}=-\tilde{\rho}g$
and the state equation $\tilde{p}=\tilde{\rho}R\tilde{T}$ imply the
reference state in the form (\ref{eq:BS1}-c) and consequently the
volumetric heat sink is given by
\begin{align}
\tilde{Q}= & -\kappa c_{p}\frac{\mathrm{d}\tilde{T}}{\mathrm{d}z}\frac{\mathrm{d}\tilde\rho}{\mathrm{d}z}=-\frac{g^{2}\kappa}{\gamma R}\frac{\tilde{\rho}}{\tilde{T}}+\mathcal{O}\left(\delta^{3/2}\rho_{B}\sqrt{g^{3}L}\right)\nonumber \\
= & -\frac{g^{2}\kappa\rho_{B}}{\gamma RT_{B}}\left(1-\theta\frac{z}{L}\right)^{m-1}+\mathcal{O}\left(\delta^{3/2}\rho_{B}\sqrt{g^{3}L}\right).\label{eq:Q_EntrForm}
\end{align}
In other words the mathematical
description of anelastic convection formulated in terms of the entropy fluctuation in (\ref{NS-Aderiv-1-1-1-1-1}-c),
although certainly attractive from the point of view of modelling,
becomes applicable only when the modelled system is being cooled at
every height by some processes, with cooling rate in the form (\ref{eq:Q_EntrForm}).
Such processes can involve natural cooling, e.g. through thermal
radiation or chemical reactions or laboratory induced cooling, e.g.
through the laser cooling process. The latter is a very effective
mechanism of cooling of gases based on interactions of the gas molecules
with a unidirectional ensemble of laser rays and the Doppler effect,
which result in homogenization of the velocity distribution of particles
and thus temperature decrease (cf. Phillips 1998). Let us provide
a simple example of a natural cooling process through thermal radiation.
As in section \ref{subsec:The-effect-of-RDH} we adopt the model of
Goody (1956) and Goody and Yung (1989), but we do not assume that
both the boundaries have radiational properties of a black body, in
particular the intensity of radiation from boundaries can be set to
zero (as in the case of vacuum). 

The heat per unit volume emitted from the system in a time unit by
thermal radiation can be expressed by the radiative energy flux,
\begin{equation}
Q=-\nabla\cdot\mathbf{j}_{rad},\label{eq:Q_def}
\end{equation}
which under the so-called double-stream Milne-Eddington approximation
has to satisfy the following equation
\begin{equation}
\nabla\frac{1}{\alpha_{a}}\nabla\cdot\mathbf{j}_{rad}-3\alpha_{a}\mathbf{j}_{rad}=4\sigma_{rad}\nabla\left(T^{4}\right),\label{eq:radiative_flux_governing_eq-2}
\end{equation}
where $\alpha_{a}$ is the coefficient of absorption of radiation
per unit volume and $\sigma_{rad}$ is the Stefan-Boltzmann constant;
the absorption coefficient is a function of temperature and pressure,
$\alpha_{a}=\alpha_{a}(T,\,p)$. Let us assume, that the boundaries
are kept isothermal. The general radiative boundary conditions derived
from the radiative transfer equation (cf. pages 429\textendash 430
in Goody 1956), supplied by the thermal boundary conditions, yield for the
reference state (which is $z$-dependent only)
\begin{subequations}
\begin{equation}
\left.\left(\frac{\mathrm{d}\tilde{j}_{rad\,z}}{\mathrm{d}z}-2\alpha_{a}\tilde{j}_{rad\,z}\right)\right|_{z=0}=4\alpha_{a}\beta_{B},\quad\left.\left(\frac{\mathrm{d}\tilde{j}_{rad\,z}}{\mathrm{d}z}+2\alpha_{a}\tilde{j}_{rad\,z}\right)\right|_{z=L}=4\alpha_{a}\beta_{T},\label{eq:BC_hydrostatic_radB_1-1}
\end{equation}
\begin{equation}
\tilde{T}(z=0)=T_{B},\quad\tilde{T}(z=L)=T_{T},\label{eq:BC_hydrostatic_radB_2-1}
\end{equation}
\end{subequations} 
where $\beta_{B}=B(T_{B})-I_{B}^{+}$ is the difference
between the Planck intensity of black body radiation at temperature
$T_{B}$ denoted by $B(T_{B})=\sigma_{rad}T_{B}^{4}$ and the upward
radiation intensity from the bottom boundary $I_{B}^{+}$; analogously
$\beta_{T}=B(T_{T})-I_{T}^{-}$ with $I_{T}^{-}$ denoting the downward
radiation intensity from the top boundary and $B(T_{T})=\sigma_{rad}T_{T}^{4}$.
Naturally, in the case when both boundaries are black bodies we get
$\beta_{T}=\beta_{B}=0$. Note, that when the radiative flux is $z$-dependent
only the equation (\ref{eq:radiative_flux_governing_eq-2}) implies
\begin{equation}
\tilde{j}_{rad\,x}=\tilde{j}_{rad\,y}=0.\label{eq:e187}
\end{equation}
To provide a simple example of a possible form of the reference state
let us assume that the absorption coefficient $\alpha_{a}$ is constant
and consider the limit of a transparent fluid, when the mean free
path of photons is much larger than $L$ thus the absorption coefficient
is very small, $\alpha_{a}\ll L^{-1}$ . In such a case the smallest
term $3\alpha_{a}\mathbf{j}_{rad}$ in the radiative flux equation
(\ref{eq:radiative_flux_governing_eq-2}) can be neglected and the
simplified equation reads
\begin{equation}
\frac{\mathrm{d}^{2}\tilde{j}_{rad\,z}}{\mathrm{d}z^{2}}=16\alpha_{a}\sigma_{rad}\tilde{T}^{3}\frac{\mathrm{d}\tilde{T}}{\mathrm{d}z}.\label{eq:radiative_flux_governing_eq-2-1}
\end{equation}
The latter has to be solved together with the energy balance in the
hydrostatic reference state at constant $\kappa=k/c_{p}\tilde{\rho}$,
cf. the second equation in (\ref{eq:ref_st_kappa}) and (\ref{eq:Q_def})
\begin{equation}
\frac{\mathrm{d}}{\mathrm{d}z}\left(\tilde{\rho}\frac{\mathrm{d}\tilde{T}}{\mathrm{d}z}\right)=\frac{1}{\kappa c_{p}}\frac{\mathrm{d}\tilde{j}_{rad\,z}}{\mathrm{d}z},\label{eq:energy_eq_BS_rad}
\end{equation}
subject to the boundary conditions (\ref{eq:BC_hydrostatic_radB_1-1},b).
Since in the case at hand the adiabatic gradient $g/c_{p}$ is constant
the temperature profile must be linear in $z$,
\begin{equation}
\tilde{T}=T_{B}\left(1-\theta\frac{z}{L}\right),\label{eq:e188}
\end{equation}
where $\theta=\Delta T/T_{B}$. Consequently the first equation in
(\ref{eq:ref_st_kappa}) and the perfect gas law $\tilde{p}=\tilde{\rho}R\tilde{T}$
imply the polytropic form of the reference state provided in (\ref{eq:BS1}-c).
Therefore by the use of (\ref{eq:energy_eq_BS_rad}) we get
\begin{equation}
\frac{\mathrm{d}\tilde{j}_{rad\,z}}{\mathrm{d}z}=\frac{m\kappa c_{p}\rho_{B}T_{B}\theta^{2}}{L^{2}}\left(1-\theta\frac{z}{L}\right)^{m-1},\label{eq:jz_derivative}
\end{equation}
and finally (\ref{eq:radiative_flux_governing_eq-2-1}) implies
\begin{equation}
\frac{m\left(m-1\right)\kappa c_{p}\rho_{B}\theta^{2}}{L^{2}T_{B}^{m-2}}\tilde{T}^{m-2}=16\alpha_{a}\sigma_{rad}\tilde{T}^{3}.\label{eq:189}
\end{equation}
The latter can only be satisfied when $m=5$ (and then by $\delta\ll1$
we must have $\gamma=6/5+\mathcal{O}(\delta)$) and when the following
relation between $\rho_{B}$, $T_{B}$ and $\theta$ is satisfied
\begin{equation}
\kappa c_{p}\rho_{B}\theta^{2}=\frac{4}{5}\alpha_{a}\sigma_{rad}T_{B}^{3}L^{2}.\label{eq:condition_rad_1}
\end{equation}
The solution for the radiative flux, obtained from (\ref{eq:jz_derivative})
and (\ref{eq:BC_hydrostatic_radB_1-1}) reads
\begin{align}
\tilde{j}_{rad\,z}= & \frac{\kappa c_{p}\rho_{B}T_{B}\theta}{L}\left[\frac{5\theta}{2\alpha_{a}L}+1-\left(1-\theta\frac{z}{L}\right)^{5}\right]-2\beta_{B}\nonumber \\
= & 4\sigma_{rad}T_{B}^{4}\left\{ \frac{1}{2}+\frac{\alpha_{a}L}{5\theta}\left[1-\left(1-\theta\frac{z}{L}\right)^{5}\right]\right\} -2\beta_{B}\label{eq:e190}
\end{align}
where on top of the relation (\ref{eq:condition_rad_1}) between the
system parameters also
\begin{equation}
\sigma_{rad}T_{B}^{4}\left\{ 1+\left(1-\theta\right)^{4}+\frac{2\alpha_{a}L}{5\theta}\left[1-\left(1-\theta\right)^{5}\right]\right\} =\beta_{T}+\beta_{B},\label{eq:condition_rad_2}
\end{equation}
must be satisfied; the latter, however, does not involve $\rho_{B}$.
We note, that when radiation from the bottom boundary can be neglected,
$I_{B}^{+}\approx0$, and the top boundary can be assumed to have
the radiational properties of a black body, $\beta_{T}=0$, and in
the limit of strong stratification when $T_{B}/T_{T}\gg1$ thus $1-\theta\ll1$
the term proportional to $\alpha_{a}L/\theta$ can be neglected since
$\alpha_{a}L$ was assumed small, and then the relation (\ref{eq:condition_rad_2})
is naturally satisfied.

It should be stressed again, that due to the requirement, that the
temperature gradient in the reference state must be close to adiabatic,
this entropy formulation with constant thermal diffusivity can be
applied only to a class of systems with non-vanishing volume cooling, e.g. in the form (\ref{eq:e190}) with (\ref{eq:condition_rad_1}) and (\ref{eq:condition_rad_2}).

\subsection{Braginsky and Roberts formulation for turbulent convection\label{subsec:Braginsky-and-Roberts}}

The turbulent flow is produced by superadiabatic gradient and hence
in case of a fully developed convectively driven turbulence it is
natural to assume that the heat flux at large scales, produced by
small-scale turbulent fluctuations is proportional to the entropy,
not the temperature gradient. Note, that the turbulent fluctuations in
this case are not the same as the thermodynamic fluctuations denoted
by primes. The turbulent fluctuations are obtained by separating the
thermodynamic fluctuations into large-scale means and small-scale
corrections, the latter being the turbulent fluctuations. Their action
on the means is often modelled by introduction of turbulent transport
coefficients, such as turbulent viscosity and thermal diffusivity.
In the following we will utilize the concept of turbulent transport
coefficients and therefore consider only the mean thermodynamic fluctuations
without making a distinction in notation between the mean and full
thermodynamic fluctuations. In other words the primed variables will
simply denote the mean thermodynamic fluctuations.

Barginsky and Roberts (1995) have proposed a simple ansatz for a fully
developed, well-mixed turbulence in convective flow, that the molecular
heat transport in such a case is much smaller than that generated
by small scale turbulence. Therefore in the evolution of large scale
turbulent components (that is means) the dominant contribution to
diffusive heat flux comes from turbulent diffusivity and can be modeled
by $-\tilde{\rho}\tilde{T}\boldsymbol{\kappa}_{t}\cdot\nabla s'=-\tilde{T}\boldsymbol{k}_{t}\cdot\nabla s'/\tilde{c}_{p}$,
where $\boldsymbol{\kappa}_{t}$ and $\boldsymbol{k}_{t}=\tilde{\rho}\tilde{c}_{p}\boldsymbol{\kappa}_{t}$
are now the turbulent diffusivity and conductivity respectively, which
are typically non-isotropic and hence for generality should be kept
in tensorial form. This allows to neglect the term $\nabla\cdot\left(k\nabla T'\right)$
describing the molecular heat transport in the general energy equation
(\ref{eq:Energy_eq_A_entropy_A}) with respect to the flux associated
with turbulent diffusivity, typically at least few orders of magnitude
larger. Therefore we can write down the set of dynamical equations
in a form much resembling that from the previous section devoted to
constant molecular thermal diffusivity formulation, but with turbulent
transport tensors $\boldsymbol{\kappa}_{t}$, $\boldsymbol{\mu}_{t}$
and $\boldsymbol{\mu_{b}}_{t}$ \index{SI}{entropy!formulation}
\begin{subequations}
\begin{equation}
\frac{\partial\mathbf{u}}{\partial t}+\left(\mathbf{u}\cdot\nabla\right)\mathbf{u}=-\nabla\frac{p'}{\tilde{\rho}}+\frac{gs'}{c_{p}}\hat{\mathbf{e}}_{z}+\nabla\cdot\left[2\boldsymbol{\mu}_{t}\cdot\mathbf{G}^{s}+\left(\boldsymbol{\mu_{b}}_{t}-\frac{2}{3}\boldsymbol{\mu}_{t}\right)\nabla\cdot\mathbf{u}\right]\label{eq:NS_BR1995}
\end{equation}
\begin{equation}
\nabla\cdot\left(\tilde{\rho}\mathbf{u}\right)=0,\label{Cont_entropy_BR_A}
\end{equation}
\begin{align}
\tilde{\rho}\tilde{T}\left(\frac{\partial s'}{\partial t}+\mathbf{u}\cdot\nabla s'\right)-\tilde{\rho}c_{p}u_{z}\Delta_{S}\nonumber \\
=\nabla\cdot\left(\tilde{\rho}\tilde{T}\boldsymbol{\kappa}_{t}\cdot\nabla s'\right)+ & \left[2\boldsymbol{\mu}_{t}\cdot\mathbf{G}^{s}+\left(\boldsymbol{\mu_{b}}_{t}-\frac{2}{3}\boldsymbol{\mu}_{t}\right)\nabla\cdot\mathbf{u}\right]:\mathbf{G}+Q'.\label{eq:Energy_BR1995}
\end{align}
\end{subequations} 
In such a way one obtains a closed system of
equations expressed solely in terms of the velocity field, pressure
and the entropy fluctuations, however, applicable only to a fully developed
turbulence in convective flows. The pressure problem is easily avoidable
by taking curl of the Navier-Stokes equation.

Such a formulation, however, suffers from having the capability to
violate the second law of thermodynamics, i.e. the production of the
total entropy in adiabatically insulated system may turn out negative.
In the case of no-slip boundary conditions the flow near the boundaries
can not be turbulent and therefore the heat flux at boundaries reduces
to the molecular one. Consequently the adiabatic insulation implies
\begin{equation}
\left.-k\frac{\partial}{\partial z}\left(\tilde{T}+T'\right)\right|_{z=0,\,L}=0.\label{eq:adiabatic_insulation-2}
\end{equation}
In the absence of heat sources, $Q=0$, very similar manipulations
to those done in section \ref{subsec:Entropy-production} (cf. equations
(\ref{eq:nonlin_entropy_null}) and (\ref{eq:ref_state_ad})) allow
to derive from (\ref{eq:Energy_BR1995}) the following formula for
the production of the total entropy
\begin{align}
\frac{\partial}{\partial t}\int_{V}\tilde{\rho}s\mathrm{d}^{3}x= & \int_{\partial V}\tilde{\rho}\left(\boldsymbol{\kappa}_{t}\cdot\nabla s'\right)\cdot\hat{\mathbf{n}}\mathrm{d}\Sigma\nonumber \\
 & +\int_{V}\frac{1}{\tilde{T}}\left[2\boldsymbol{\mu}_{t}\cdot\mathbf{G}^{s}+\left(\boldsymbol{\mu_{b}}_{t}-\frac{2}{3}\boldsymbol{\mu}_{t}\right)\nabla\cdot\mathbf{u}\right]:\mathbf{G}\mathrm{d}^{3}x,\label{eq:entropy_production_BR95}
\end{align}
where we have made a natural assumption, that the molecular heat conduction
coefficient $k$ is non-zero everywhere in the fluid volume. Since
the boundary conditions (\ref{eq:adiabatic_insulation-2}) involve
the temperature only, the term
\begin{equation}
\int_{\partial V}\tilde{\rho}\left(\boldsymbol{\kappa}_{t}\cdot\nabla s'\right)\cdot\hat{\mathbf{n}}\mathrm{d}\Sigma\label{eq:e191}
\end{equation}
is in general neither zero nor positive definite and hence the entropy
production can, in principle, be negative. The possible lack of consistency
with the second law of thermodynamics, resulting from application
of the concept of turbulent transport coefficients, and therefore
poor control over the sign of the production of the total entropy
is a significant disadvantage of this formulation.

\subsubsection{The hydrostatic reference state\label{subsec:basic_state_BR1995}}

In this case the hydrostatic reference state can either be adiabatic
or close to adiabatic, but as in the previous cases has to satisfy
the standard static ($\partial_{t}\equiv0$ and $\mathbf{u}\equiv0$)
equations, i.e. with molecular, not turbulent, diffusion coefficients,
and in particular
\begin{equation}
\frac{\mathrm{d}}{\mathrm{d}z}\left(k\frac{\mathrm{d}\tilde{T}}{\mathrm{d}z}\right)=-\tilde{Q}.\label{eq:heat_balance_ref_BR95}
\end{equation}
The turbulent diffusion coefficients, as previously the molecular
ones, now have to satisfy (for all $i=1,\,2,\,3$ and $j=1,\,2,\,3$)
\begin{equation}
\left(\mu_{b_{t}}\right)_{ij}\lesssim\left(\mu_{t}\right)_{ij}\sim\delta^{1/2}\bar{\rho}\sqrt{\bar{g}L}L.\label{eq:visc_scale-1-2}
\end{equation}
\begin{equation}
\left(\kappa_{t}\right)_{ij}\sim\delta^{1/2}\sqrt{\bar{g}L}L,\label{eq:kappa_scale-1}
\end{equation}
for consistency of the anelastic formulation. The molecular transport
coefficients $k$, $\mu$ and $\mu_{b}$ are by assumption negligibly
small compared to the turbulent ones. This, of course, requires the
heating term $\tilde{Q}$ to be small in order to balance the weak
molecular diffusion in the hydrostatic state, as in (\ref{eq:heat_balance_ref_BR95}).

\section{Energetic properties of anelastic systems\label{subsec:Energetic-properties-of-A}}

Just as in the Boussinesq case we now proceed to describe the mean
physical and in particular energetic properties of anelastic systems.
For simplicity and clarity it will be assumed that the system is periodic
in horizontal directions, there are no radiative heat sources, thus
$Q=0$ and the acceleration of gravity is constant $g=\mathrm{const}$.
The evolution of compressible convection under the anelastic approximation
is described by the set of the Navier-Stokes (\ref{NS-Aderiv-1-1-1}),
mass conservation and energy equations (\ref{Cont_Aderiv-1-1})-(\ref{State_eq_Aderiv-1-1}),
i.e. \index{SI}{anelastic!equations, uniform gravity, no Q}
\begin{subequations}
\begin{align}
\frac{\partial\mathbf{u}}{\partial t}+\left(\mathbf{u}\cdot\nabla\right)\mathbf{u}= & -\nabla\frac{p'}{\tilde{\rho}}+\frac{\tilde{\alpha}\tilde{T}}{\tilde{c}_{p}}s'g\hat{\mathbf{e}}_{z}+\frac{\mu}{\tilde{\rho}}\nabla^{2}\mathbf{u}+\left(\frac{\mu}{3\tilde{\rho}}+\frac{\mu_{b}}{\tilde{\rho}}\right)\nabla\left(\nabla\cdot\mathbf{u}\right)\nonumber \\
 & +\frac{2}{\tilde{\rho}}\nabla\mu\cdot\mathbf{G}^{s}+\frac{1}{\tilde{\rho}}\nabla\left(\mu_{b}-\frac{2}{3}\mu\right)\nabla\cdot\mathbf{u},\label{NS-Aderiv-1-1-1-2}
\end{align}
\begin{equation}
\nabla\cdot\left(\tilde{\rho}\mathbf{u}\right)=0,\label{Cont_Aderiv-1-1-2}
\end{equation}
\begin{equation}
\tilde{\rho}\tilde{T}\left[\frac{\partial s'}{\partial t}+\mathbf{u}\cdot\nabla\left(\tilde{s}+s'\right)\right]=\nabla\cdot\left(k\nabla T\right)+2\mu\mathbf{G}^{s}:\mathbf{G}^{s}+\left(\mu_{b}-\frac{2}{3}\mu\right)\left(\nabla\cdot\mathbf{u}\right)^{2},\label{Energy_Aderiv-1-1-1}
\end{equation}
\begin{equation}
\frac{\rho'}{\tilde{\rho}}=-\tilde{\alpha}T'+\tilde{\beta}p',\qquad s'=-\tilde{\alpha}\frac{p'}{\tilde{\rho}}+\tilde{c}_{p}\frac{T'}{\tilde{T}}.\label{State_eq_Aderiv-1-1-1}
\end{equation}
\end{subequations} 
Understanding the energy transfer and production
in convective flow is the key to understanding the physics of compressible
convection. Therefore we derive now a few exact relations which allow
to describe some general aspects of the dynamics of developed compressible
convection. By multiplying the Navier-Stokes equation (\ref{NS-Aderiv-1-1-1-2})
by $\tilde{\rho}\mathbf{u}$ and averaging over the entire periodic volume (periodicity referring
to the '$x$' and '$y$' directions) we obtain the following relation
\begin{equation}
\frac{\partial}{\partial t}\left\langle \frac{1}{2}\tilde{\rho}\mathbf{u}^{2}\right\rangle =\left\langle \frac{g\tilde{\alpha}\tilde{T}\tilde{\rho}}{\tilde{c}_{p}}u_{z}s'\right\rangle -2\left\langle \mu\mathbf{G}^{s}:\mathbf{G}^{s}\right\rangle +\left\langle \left(\frac{2}{3}\mu-\mu_{b}\right)\left(\nabla\cdot\mathbf{u}\right)^{2}\right\rangle ,\;\label{eq:W_BF_and_VD}
\end{equation}
where the impermeable and either no-slip or stress-free
boundary conditions, as in (\ref{eq:impermeable}) and (\ref{eq:stress-free_or_no-slip})
were used. The latter relation states, that changes in the total kinetic
energy are due to the total work per unit volume of the buoyancy force
averaged over the horizontal directions and the total viscous dissipation
in the fluid volume; in a (statistically) stationary state the work
of the buoyancy force and the viscous dissipation are equal
\begin{equation}
\left\langle \frac{g\tilde{\alpha}\tilde{T}\tilde{\rho}}{\tilde{c}_{p}}u_{z}s'\right\rangle =2\left\langle \mu\mathbf{G}^{s}:\mathbf{G}^{s}\right\rangle -\left\langle \left(\frac{2}{3}\mu-\mu_{b}\right)\left(\nabla\cdot\mathbf{u}\right)^{2}\right\rangle .\label{eq:e192}
\end{equation}
Next we derive expressions for the total, superadiabatic
heat flux in the system at every $z$. First we recall, that the impermeability
conditions at boundaries together with the horizontally averaged mass
conservation law imply
\begin{equation}
\left\langle u_{z}\right\rangle _{h}=0,\label{eq:avh_uz}
\end{equation}
cf. equation (\ref{eq:mean_h_uz_null_A}). Now, averaging the heat
equation (\ref{Energy_Aderiv-1-1-1}) over a horizontal plane and
integrating from $0$ to $z$ leads to,
\begin{align}
\frac{\partial}{\partial t}\int_{0}^{z}\left\langle \tilde{\rho}\tilde{T}s'\right\rangle _{h}\mathrm{d}z= & \int_{0}^{z}\tilde{\rho}\frac{\mathrm{d}\tilde{T}}{\mathrm{d}z}\left\langle u_{z}s'\right\rangle _{h}\mathrm{d}z-\tilde{\rho}\tilde{T}\left\langle u_{z}s'\right\rangle _{h}-\left.k\frac{\partial\left\langle T\right\rangle _{h}}{\partial z}\right|_{z=0}+k\frac{\partial\left\langle T\right\rangle _{h}}{\partial z}\nonumber \\
 & +2\int_{0}^{z}\left\langle \mu\mathbf{G}^{s}:\mathbf{G}^{s}\right\rangle _{h}\mathrm{d}z-\int_{0}^{z}\left\langle \left(\frac{2}{3}\mu-\mu_{b}\right)\left(\nabla\cdot\mathbf{u}\right)^{2}\right\rangle _{h}\mathrm{d}z.\label{eq:ThE}
\end{align}
In a stationary state the left hand side of the latter equation vanishes.
Thus setting the upper limit of the vertical integration in (\ref{eq:ThE})
to $z=L$ (i.e. integrating over the entire fluid volume), applying
the boundary conditions of impermeability $u_{z}(z=0,\,L)=0$ and
utilizing (\ref{eq:W_BF_and_VD}) allows to obtain
\begin{equation}
L\left\langle \frac{g\tilde{\alpha}\tilde{T}\tilde{\rho}}{\tilde{c}_{p}}u_{z}s'\right\rangle +L\left\langle \tilde{\rho}\frac{\mathrm{d}\tilde{T}}{\mathrm{d}z}u_{z}s'\right\rangle =-\left.k\frac{\partial\left\langle T\right\rangle _{h}}{\partial z}\right|_{z=L}+\left.k\frac{\partial\left\langle T\right\rangle _{h}}{\partial z}\right|_{z=0},\quad\label{eq:e194}
\end{equation}
and since $\mathrm{d}\tilde{T}/\mathrm{d}z=-g\tilde{\alpha}\tilde{T}/\tilde{c}_{p}+\mathcal{O}(\delta\tilde{T}/L)$
it is clear that the two terms on the left hand side cancel at leading
order. This leads to the expectable conclusion, that in a stationary
state\index{SI}{heat flux balance}
\begin{equation}
-\left.k\frac{\partial\left\langle T\right\rangle _{h}}{\partial z}\right|_{z=0}=-\left.k\frac{\partial\left\langle T\right\rangle _{h}}{\partial z}\right|_{z=L},\label{eq:heat_flux_enter_exit}
\end{equation}
hence the total, horizontally averaged heat flux entering the system
at the bottom equals the flux, which leaves the system at the top.
Furthermore, on denoting the total, horizontally averaged heat flux
by $F_{total}(z)$, the expression (\ref{eq:ThE}) can be used to
derive a formula for the heat flux $F_{total}(z=0)$ which enters
the system at the bottom in a stationary state
\begin{align}
F_{total}\left(z=0\right)= & -k\left.\frac{\mathrm{d}}{\mathrm{d}z}\left(\tilde{T}+\left\langle T'\right\rangle _{h}\right)\right|_{z=0}\nonumber \\
= & -k\frac{\mathrm{d}}{\mathrm{d}z}\left(\tilde{T}+\left\langle T'\right\rangle _{h}\right)\nonumber \\
 & +\tilde{\rho}\tilde{T}\left\langle u_{z}s'\right\rangle _{h}-\int_{0}^{z}\tilde{\rho}\frac{\mathrm{d}\tilde{T}}{\mathrm{d}z}\left\langle u_{z}s'\right\rangle _{h}\mathrm{d}z\nonumber \\
 & -2\int_{0}^{z}\left\langle \mu\mathbf{G}^{s}:\mathbf{G}^{s}\right\rangle _{h}\mathrm{d}z+\int_{0}^{z}\left\langle \left(\frac{2}{3}\mu-\mu_{b}\right)\left(\nabla\cdot\mathbf{u}\right)^{2}\right\rangle _{h}\mathrm{d}z,\label{eq:F_conv_total_1}
\end{align}
which we will refer to as \emph{the first formula for total heat flux}\index{SI}{heat flux!total}.
Clearly, the heat flux 
\begin{equation}
F_{total}\left(z\right)=-k\frac{\mathrm{d}}{\mathrm{d}z}\left(\tilde{T}+\left\langle T'\right\rangle _{h}\right)+\tilde{\rho}\tilde{T}\left\langle u_{z}s'\right\rangle _{h},\label{eq:F_total}
\end{equation}
which consists of the conductive $-k\mathrm{d}_{z}\left(\tilde{T}+\left\langle T\right\rangle _{h}\right)$
and advective $\tilde{\rho}\tilde{T}\left\langle u_{z}s'\right\rangle _{h}$
parts, contrary to the Boussinesq case is not constant at every $z$,
because the work of the buoyancy force, likewise the viscous entropy
production substantially modify the total flux for $0<z<1$, i.e.
\begin{align}
F_{total}\left(z\right)= & F_{total}\left(z=0\right)-\int_{0}^{z}\frac{\tilde{\alpha}\tilde{T}g\tilde{\rho}}{\tilde{c}_{p}}\left\langle u_{z}s'\right\rangle _{h}\mathrm{d}z\nonumber \\
 & +2\int_{0}^{z}\left\langle \mu\mathbf{G}^{s}:\mathbf{G}^{s}\right\rangle _{h}\mathrm{d}z-\int_{0}^{z}\left\langle \left(\frac{2}{3}\mu-\mu_{b}\right)\left(\nabla\cdot\mathbf{u}\right)^{2}\right\rangle _{h}\mathrm{d}z,\;\label{eq:heta_flux_balance_BULK_AN}
\end{align}
where we have used $\mathrm{d}_{z}\tilde{T}=-\tilde{\alpha}\tilde{T}g/\tilde{c}_{p}+\mathcal{O}(\delta g/\bar{c}_{p})$. Note, that $Q=0$ implies $k\mathrm{d}_{z}\tilde{T}=\mathrm{const}$,
hence the terms $-\left.k\mathrm{d}_{z}\tilde{T}\right|_{z=0}$ and
$-k\mathrm{d}_{z}\tilde{T}$ are equal and can be cancelled on the
both sides of (\ref{eq:F_conv_total_1}) leaving an expression for
the convective heat flux only\index{SI}{heat flux!convective}
\begin{subequations}
\begin{equation}
F_{conv.}=F_{total}+k\mathrm{d}_{z}\tilde{T}=-k\left.\frac{\mathrm{d}}{\mathrm{d}z}\left\langle T'\right\rangle _{h}\right|_{z=0}=-k\frac{\mathrm{d}}{\mathrm{d}z}\left\langle T'\right\rangle _{h}+\tilde{\rho}\tilde{T}\left\langle u_{z}s'\right\rangle _{h},\label{eq:F_conv}
\end{equation}
\begin{equation}
F_{conv.}(z=0)=F_{total}(z=0)+k\mathrm{d}_{z}\tilde{T}=-k\left.\frac{\mathrm{d}}{\mathrm{d}z}\left\langle T'\right\rangle _{h}\right|_{z=0}.\label{eq:F_conv_z0}
\end{equation}
\end{subequations} 
Moreover, the formula (\ref{eq:F_conv_total_1})
can be used to derive \emph{the first formula for superadiabatic heat
flux}\index{SI}{heat flux!superadiabatic}
\begin{align}
F_{S}(z=0)= & -k\left.\frac{\mathrm{d}}{\mathrm{d}z}\left(\tilde{T}+\left\langle T'\right\rangle _{h}-T_{ad}\right)\right|_{z=0}\nonumber \\
= & -k\frac{\mathrm{d}}{\mathrm{d}z}\left(\tilde{T}+\left\langle T'\right\rangle _{h}-T_{ad}\right)-\left(k\frac{\mathrm{d}T_{ad}}{\mathrm{d}z}-\left.k\frac{\mathrm{d}T_{ad}}{\mathrm{d}z}\right|_{z=0}\right)\nonumber \\
 & +\tilde{\rho}\tilde{T}\left\langle u_{z}s'\right\rangle _{h}-\int_{0}^{z}\tilde{\rho}\frac{\mathrm{d}\tilde{T}}{\mathrm{d}z}\left\langle u_{z}s'\right\rangle _{h}\mathrm{d}z\nonumber \\
 & -2\int_{0}^{z}\left\langle \mu\mathbf{G}^{s}:\mathbf{G}^{s}\right\rangle _{h}\mathrm{d}z+\int_{0}^{z}\left\langle \left(\frac{2}{3}\mu-\mu_{b}\right)\left(\nabla\cdot\mathbf{u}\right)^{2}\right\rangle _{h}\mathrm{d}z,\label{eq:F_conv_superadiab_1}
\end{align}
Note, that since the basic state profile is assumed close to the adiabatic
the term $-(k\mathrm{d}_{z}T_{ad}-\left.k\mathrm{d}_{z}T_{ad}\right|_{z=0})$
which describes the heat flux jump along the adiabat is of the order
$\mathcal{O}(\delta k\tilde{T}/L)=\mathcal{O}(\delta^{3/2}\tilde{\rho}\tilde{c}_{p}\sqrt{\bar{g}L}\tilde{T})$,
which is consistent with the order of magnitude of the entire formula
(\ref{eq:F_conv_superadiab_1}). The adiabatic temperature profile
is in general curvilinear, however if e.g. one considers the case
of an ideal gas with uniform gravity $g=\mathrm{const}$, then the
adiabatic gradient becomes uniform and the term $-(k\mathrm{d}_{z}T_{ad}-\left.k\mathrm{d}_{z}T_{ad}\right|_{z=0})$
vanishes; in such a case setting $z=L$ in (\ref{eq:F_conv_superadiab_1})
gives $-k\left.\mathrm{d}_{z}(\tilde{T}+\left\langle T'\right\rangle _{h}-T_{ad})\right|_{z=0}=-k\left.\mathrm{d}_{z}(\tilde{T}+\left\langle T'\right\rangle _{h}-T_{ad})\right|_{z=L}$,
so that the superadiabatic heat flux entering the system at $z=0$
is equal to the superadiabatic heat flux leaving the system at $z=L$.
Most generally, however, when the adiabatic gradient cannot be assumed
uniform the vertical variation of the heat flux conducted down the
adiabat, $-(\left.k\mathrm{d}_{z}T_{ad}\right|_{z=L}-\left.k\mathrm{d}_{z}T_{ad}\right|_{z=0})$
contributes to the superadiabatic heat flux balance at the top and
bottom boundaries.

Next we consider the balance of entropy per unit volume, that is the
heat equation (\ref{Energy_Aderiv-1-1-1}) divided by $\tilde{T}$.
If we average the entropy equation over the horizontal planes and
integrate it from bottom to arbitrary height $z$, then for a stationary
state and impermeable and either stress-free or no-slip boundaries
we obtain \emph{the second formula for the total heat flux}\index{SI}{heat flux!total}
\begin{align}
F_{total}\left(z=0\right)= & -k\left.\frac{\mathrm{d}}{\mathrm{d}z}\left(\tilde{T}+\left\langle T'\right\rangle _{h}\right)\right|_{z=0}\nonumber \\
= & -k\frac{\tilde{T}_{B}}{\tilde{T}}\frac{\mathrm{d}}{\mathrm{d}z}\left(\tilde{T}+\left\langle T'\right\rangle _{h}\right)-k\frac{\mathrm{d}\tilde{T}}{\mathrm{d}z}\left[1-\frac{\tilde{T}_{B}}{\tilde{T}}\right]\nonumber \\
 & -\tilde{T}_{B}k\frac{\mathrm{d}\tilde{T}}{\mathrm{d}z}\left(\frac{\left\langle T'\right\rangle _{h}}{\tilde{T}^{2}}-\frac{\left.\left\langle T'\right\rangle _{h}\right|_{z=0}}{\tilde{T}_{B}^{2}}\right)-2\tilde{T}_{B}k\frac{\mathrm{d}\tilde{T}}{\mathrm{d}z}\int_{0}^{z}\frac{\left\langle T'\right\rangle _{h}}{\tilde{T}^{3}}\frac{\mathrm{d}\tilde{T}}{\mathrm{d}z}\mathrm{d}z\nonumber \\
 & +\tilde{\rho}\tilde{T}_{B}\left\langle u_{z}s'\right\rangle _{h}\nonumber \\
 & -2\tilde{T}_{B}\int_{0}^{z}\left\langle \frac{\mu}{\tilde{T}}\mathbf{G}^{s}:\mathbf{G}^{s}\right\rangle _{h}\mathrm{d}z+\tilde{T}_{B}\int_{0}^{z}\left\langle \frac{2\mu-3\mu_{b}}{3\tilde{T}}\left(\nabla\cdot\mathbf{u}\right)^{2}\right\rangle _{h}\mathrm{d}z.\label{eq:F_conv_total_2}
\end{align}
The latter implies\index{SI}{heat flux!superadiabatic}
\begin{align}
F_{S}\left(z=0\right)= & -k\left.\frac{\mathrm{d}}{\mathrm{d}z}\left(\tilde{T}+\left\langle T'\right\rangle _{h}-T_{ad}\right)\right|_{z=0}\nonumber \\
= & -k\frac{\tilde{T}_{B}}{\tilde{T}}\frac{\mathrm{d}}{\mathrm{d}z}\left(\tilde{T}+\left\langle T'\right\rangle _{h}-T_{ad}\right)\nonumber \\
 & -\left[\left.k\frac{\mathrm{d}\left(\tilde{T}-T_{ad}\right)}{\mathrm{d}z}\right|_{z=0}-\frac{\tilde{T}_{B}}{\tilde{T}}k\frac{\mathrm{d}\left(\tilde{T}-T_{ad}\right)}{\mathrm{d}z}\right]\nonumber \\
 & -\tilde{T}_{B}k\frac{\mathrm{d}\tilde{T}}{\mathrm{d}z}\left(\frac{\left\langle T'\right\rangle _{h}}{\tilde{T}^{2}}-\frac{\left.\left\langle T'\right\rangle _{h}\right|_{z=0}}{\tilde{T}_{B}^{2}}\right)-2\tilde{T}_{B}k\frac{\mathrm{d}\tilde{T}}{\mathrm{d}z}\int_{0}^{z}\frac{\left\langle T'\right\rangle _{h}}{\tilde{T}^{3}}\frac{\mathrm{d}\tilde{T}}{\mathrm{d}z}\mathrm{d}z\nonumber \\
 & +\tilde{\rho}\tilde{T}_{B}\left\langle u_{z}s'\right\rangle _{h}\nonumber \\
 & -2\tilde{T}_{B}\int_{0}^{z}\left\langle \frac{\mu}{\tilde{T}}\mathbf{G}^{s}:\mathbf{G}^{s}\right\rangle _{h}\mathrm{d}z+\tilde{T}_{B}\int_{0}^{z}\left\langle \frac{2\mu-3\mu_{b}}{3\tilde{T}}\left(\nabla\cdot\mathbf{u}\right)^{2}\right\rangle _{h}\mathrm{d}z.\label{eq:F_conv_superadiabatic_2}
\end{align}
which will be referred to as \emph{the second formula for superadiabatic
heat flux}\index{SI}{heat flux!superadiabatic}. In the above we have utilized the fact, that for $Q=0$
one obtains $k\mathrm{d}_{z}\tilde{T}=\mathrm{const}$. Note, that
the distinction between the bottom value of the reference temperature
$\tilde{T}_{B}$ and the bottom value of the total temperature $\tilde{T}_B+T'(\mathbf{x},t)|_{z=0}$ is
kept here for clarity, since the boundaries are not necessarily isothermal (recall, that the corrections from the temperature fluctuation $T'(\mathbf{x},t)|_{z=0}$ are of the order $\mathcal{O}(\delta)$).
However, this distinction is not really necessary in the equation
(\ref{eq:F_conv_superadiabatic_2}) for the superadiabatic flux, since
all the terms in this expression are of the order $\mathcal{O}(\delta^{3/2})$,
thus $\mathcal{O}(\delta)$ corrections to $\tilde{T}_{B}$ are irrelevant,
as they produce negligible $\mathcal{O}(\delta^{5/2})$ corrections
to the superadiabatic heat flux.

Equation (\ref{eq:F_conv_total_2}) taken at $z=L$, i.e. the heat
equation divided by $\tilde{T}$ integrated over the entire volume
from $z$ to $L$, with the aid of (\ref{eq:heat_flux_enter_exit})
gives the following relation between the convective heat flux in the
system and the viscous dissipation with corrections proportional to
the basic state heat flux\index{SI}{heat flux!convective}
\begin{align}
F_{conv.}\left(z=0\right)\left(\tilde{\Gamma}-1\right)= & \,2\tilde{T}_{B}L\left\langle \frac{\mu}{\tilde{T}}\mathbf{G}^{s}:\mathbf{G}^{s}\right\rangle -\tilde{T}_{B}L\left\langle \frac{2\mu-3\mu_{b}}{3\tilde{T}}\left(\nabla\cdot\mathbf{u}\right)^{2}\right\rangle \nonumber \\
+k\frac{\mathrm{d}\tilde{T}}{\mathrm{d}z} & \left[\frac{\tilde{\Gamma}^{2}\left.\left\langle T'\right\rangle _{h}\right|_{z=L}-\left.\left\langle T'\right\rangle _{h}\right|_{z=0}}{\tilde{T}_{B}}+2\tilde{T}_{B}\int_{0}^{L}\frac{\left\langle T'\right\rangle _{h}}{\tilde{T}^{3}}\frac{\mathrm{d}\tilde{T}}{\mathrm{d}z}\mathrm{d}z\right],\label{eq:rel_flux_VD}
\end{align}
where we have introduced the bottom to top temperature ratio
\begin{equation}
\tilde{\Gamma}=\frac{\tilde{T}_{B}}{\tilde{T}_{T}}=\frac{1}{1-\tilde{\theta}}>1,\label{eq:e195}
\end{equation}
and $\tilde{\theta}=\Delta\tilde{T}/\tilde{T}_{B}=(\tilde{T}_{B}-\tilde{T}_{T})/\tilde{T}_{B}$.
Of course, in the case, when both boundaries are held at constant
temperature we have $\tilde{T}_{B}=T_{B}$, $\tilde{T}_{T}=T_{T}$,
$\Delta\tilde{T}=\Delta T$ and the term proportional to $\Gamma^{2}\left.\left\langle T'\right\rangle _{h}\right|_{z=L}-\left.\left\langle T'\right\rangle _{h}\right|_{z=0}$
vanishes. 

One of the possible approaches is to define the Nusselt number simply
as the ratio of the convective heat flux to the total flux conducted
by the basic state and the Rayleigh number with the use of the basic
state gradient $-k\mathrm{d}_{z}\tilde{T}/\bar{k}$; this leads to
a simple relation between the Nusselt and Rayleigh numbers via (\ref{eq:rel_flux_VD}).
We will, however, consider now a simplified case of a fluid governed
by the ideal gas equation of state with constant specific heats $c_{p}$
and $c_{v}$, constant gravity $g$, viscosity $\mu$ and thermal
conductivity $k$. This will allow for fairly simple relations between
the Nusselt and Rayleigh numbers defined in a similar manner as in
the Boussinesq case. We separate the two inherently distinct problems
of isothermal (thermally perfectly conducting) and fixed heat flux (thermally insulating)
boundaries.

\subsection{Isothermal boundaries, perfect gas, uniform fluid properties, $g=\mathrm{const}.$\label{subsec:Isothermal-boundaries-IDG}}

Under the current assumptions the equation (\ref{eq:F_conv_superadiabatic_2})
taken at $z=L$ supplied by (\ref{eq:heat_flux_enter_exit}) provide
an expression for the superadiabatic heat flux in the form\index{SI}{heat flux!superadiabatic}
\begin{align}
\left[F_{S}\left(z=0\right)-k\Delta_{S}\right]\left(\Gamma-1\right)= & \,\,2T_{B}L\mu\left\langle \frac{1}{\tilde{T}}\mathbf{G}^{s}:\mathbf{G}^{s}\right\rangle \nonumber \\
 & -T_{B}L\left(\frac{2}{3}\mu-\mu_{b}\right)\left\langle \frac{\left(\nabla\cdot\mathbf{u}\right)^{2}}{\tilde{T}}\right\rangle \nonumber \\
 & +2T_{B}Lk\left(\frac{\Delta T}{L}\right)^{2}\left\langle \frac{T'}{\tilde{T}^{3}}\right\rangle ,\label{eq:rel_flux_VD-1}
\end{align}
On defining the Nusselt number $Nu$\index{SI}{Nusselt number!anelastic} and the Rayleigh number $Ra$\index{SI}{Rayleigh number!anelastic} in
the following way
\begin{equation}
Nu=\frac{F_{S}\left(z=0\right)}{k\Delta_{S}}=\frac{-k\left.\frac{\mathrm{d}}{\mathrm{d}z}\left(\tilde{T}+\left\langle T'\right\rangle _{h}-T_{ad}\right)\right|_{z=0}}{k\Delta_{S}},\label{eq:Nu_def_anapp-1}
\end{equation}
\begin{equation}
Ra=\frac{g\bigtriangleup_{S}L^{4}\rho_{B}^{2}c_{p}}{T_{B}\mu k},\label{eq:Ra_def-1}
\end{equation}
where $k\Delta_{S}=k(\Delta T/L-g/c_{p})$ is the superadiabatic conductive
heat flux in the hydrostatic basic state and $1/T_{B}=\tilde{\alpha}_{B}$,
the relation (\ref{eq:rel_flux_VD-1}) can be rewritten to yield
\begin{eqnarray}
\frac{k^{2}}{\rho_{B}^{2}c_{p}^{2}L^{4}}Ra\left(Nu-1\right)\left(\Gamma-1\right) & = & 2\Delta T\left\langle \frac{1}{\tilde{T}}\mathbf{G}^{s}:\mathbf{G}^{s}\right\rangle \nonumber \\
 &  & -\Delta T\left(\frac{2}{3}-\frac{\mu_{b}}{\mu}\right)\left\langle \frac{\left(\nabla\cdot\mathbf{u}\right)^{2}}{\tilde{T}}\right\rangle \nonumber \\
 &  & +2c_{p}Pr^{-1}L\left(\frac{\Delta T}{L}\right)^{3}\left\langle \frac{T'}{\tilde{T}^{3}}\right\rangle ,\qquad\label{eq:rel_flux_VD-1-1}
\end{eqnarray}
where we have used $\Delta T/L=g/c_{p}+\mathcal{O}(\delta\tilde{T}/L)$.
In the above
\begin{equation}
Pr=\frac{\mu c_{p}}{k},\label{eq:e197}
\end{equation}
is the Prandtl number\index{SI}{Prandtl number}. We note, that since in the case at hand the
basic state is given by the equations (\ref{eq:BS1}a-c) we get
\begin{equation}
\Delta_{S}=\frac{\Delta T}{L}-\frac{g}{c_{p}}=-\frac{T_{B}}{c_{p}}\left.\frac{\mathrm{d}\tilde{s}}{\mathrm{d}z}\right|_{z=0}=\frac{\Delta T\Delta\tilde{s}}{c_{p}L\ln\Gamma},\label{eq:e198}
\end{equation}
where
\begin{equation}
\Delta\tilde{s}=\tilde{s}_{B}-\tilde{s}_{T}=c_{p}\frac{m+1-\gamma m}{\gamma}\ln\Gamma\label{eq:e199}
\end{equation}
is the entropy jump across the fluid layer in the basic hydrostatic
state. Thus we arrive at an alternative expression for the Rayleigh
number\index{SI}{Rayleigh number!anelastic}
\begin{equation}
Ra=\frac{g\Delta T\Delta\tilde{s}L^{3}\rho_{B}^{2}}{\ln\Gamma T_{B}\mu k}.\label{eq:Ra_def-1-1}
\end{equation}
Another useful relation comes from averaging the first formula for
total heat flux (\ref{eq:F_conv_total_1}) over the vertical fluid
gap $0\leq z\leq L$, which under the current assumptions can be cast
in the following form
\begin{align}
F_{conv.}\left(z=0\right)= & 2\left\langle \tilde{\rho}\tilde{T}u_{z}s'\right\rangle -2\mu L\left\langle \left(\frac{\tilde{T}}{\Delta T}-1\right)\mathbf{G}^{s}:\mathbf{G}^{s}\right\rangle \nonumber \\
 & +\left(\frac{2}{3}\mu-\mu_{b}\right)L\left\langle \left(\frac{\tilde{T}}{\Delta T}-1\right)\left(\nabla\cdot\mathbf{u}\right)^{2}\right\rangle ,\label{eq:e200}
\end{align}
where 
\begin{equation}
\frac{\tilde{T}}{\Delta T}-1=\frac{1}{\Gamma-1}-\frac{z}{L}.\label{eq:e201}
\end{equation}
An alternative expression for the convective flux at the bottom is
provided by (\ref{eq:rel_flux_VD}), which in the case at hand implies
\begin{align}
F_{conv.}\left(z=0\right)\left(\Gamma-1\right)= & 2T_{B}L\left\langle \frac{\mu}{\tilde{T}}\mathbf{G}^{s}:\mathbf{G}^{s}\right\rangle -T_{B}L\left\langle \frac{2\mu-3\mu_{b}}{3\tilde{T}}\left(\nabla\cdot\mathbf{u}\right)^{2}\right\rangle \nonumber \\
 & +2T_{B}Lk\left(\frac{\Delta T}{L}\right)^{2}\left\langle \frac{T'}{\tilde{T}^{3}}\right\rangle ,\label{eq:rel_flux_VD-2}
\end{align}

\subsection{Boundaries held at constant heat flux, perfect gas, uniform fluid
properties, $g=\mathrm{const}.$\label{subsec:constant-flux-IDG}}

When the heat flux is held constant at the boundaries we have
\begin{equation}
\left.\frac{\mathrm{d}}{\mathrm{d}z}\left\langle T'\right\rangle _{h}\right|_{z=0}=\left.\frac{\mathrm{d}}{\mathrm{d}z}\left\langle T'\right\rangle _{h}\right|_{z=L}=0,\label{eq:BCs_const_heat_flux}
\end{equation}
and the basic state gradient is uniform, as in the previous case $\tilde{T}=\tilde{T}_{B}-\Delta\tilde{T}z/L$,
and $\Delta\tilde{T}=\tilde{T}_{B}-\tilde{T}_{T}$. This means that
the definition of the Nusselt number (\ref{eq:Nu_def_anapp-1}) utilized
in the case of isothermal boundaries becomes useless in the current
case, since the expression $-k\left.\mathrm{d}_{z}(\tilde{T}+\left\langle T'\right\rangle _{h}-T_{ad})\right|_{z=0}/k\Delta_{S}$
is exactly unity (cf. eq. (\ref{eq:stationary_flux_B-1}) and the
comment below for a similar result in the Boussinesq case). Therefore
an alternative definition of the Nusselt number is required. Contrary
to the Boussinesq case the total convective heat flux (\ref{eq:F_conv})
is non-zero away from the boundaries (cf. (\ref{eq:stationary_flux_B-1-1}) for the Boussinesq case), because of the substantial influence
of viscous heating and the work done by buoyancy effects. Nevertheless,
it seems reasonable to define the Nusselt number only by the mean
advective contribution to the total convective flux (\ref{eq:F_conv}),\index{SI}{Nusselt number!anelastic}
\begin{equation}
Nu_{Q}=\frac{\left\langle \tilde{\rho}\tilde{T}u_{z}s'\right\rangle }{k\Delta_{S}}\label{eq:Nu_Q_anelastic}
\end{equation}
since the mean conductive contribution is expected to be much smaller
in strongly developed convection. Averaging the first formula for
total heat flux (\ref{eq:F_conv_total_1}) over $0\leq z\leq L$ and
taking into account (\ref{eq:BCs_const_heat_flux}) leads to
\begin{align}
0= & -k\frac{\Delta\left\langle T'\right\rangle _{h}}{L}+2\left\langle \tilde{\rho}\tilde{T}u_{z}s'\right\rangle -2\mu L\left\langle \left(\frac{\tilde{T}}{\Delta T}-1\right)\mathbf{G}^{s}:\mathbf{G}^{s}\right\rangle \nonumber \\
 & +\left(\frac{2}{3}\mu-\mu_{b}\right)L\left\langle \left(\frac{\tilde{T}}{\Delta T}-1\right)\left(\nabla\cdot\mathbf{u}\right)^{2}\right\rangle ,\label{eq:e202}
\end{align}
where\footnote{Although it does not play an important role in the following
analysis, we can observe that in the case at hand it is expected,
that convection tends to equalize the total top and bottom temperatures,
which implies $\Delta\left\langle T'\right\rangle _{h}>0$.}
\begin{equation}
\Delta\left\langle T'\right\rangle _{h}=\left.\left\langle T'\right\rangle _{h}\right|_{z=L}-\left.\left\langle T'\right\rangle _{h}\right|_{z=0},\label{eq:e203}
\end{equation}
is the temperature fluctuation jump across the layer. Therefore with
the use of the Nusselt number definition we get the following relation
\begin{align}
\frac{k^{2}}{\tilde{\rho}_{B}^{2}c_{p}^{2}L^{4}}Ra\left(Nu_{Q}-\frac{k\frac{\Delta\left\langle T'\right\rangle _{h}}{2L}}{k\Delta_{S}}\right)\frac{\Gamma}{\Gamma-1}= & \left\langle \left(\frac{\tilde{T}}{\Delta T}-1\right)\mathbf{G}^{s}:\mathbf{G}^{s}\right\rangle \nonumber \\
 & -\left(\frac{1}{3}-\frac{\mu_{b}}{2\mu}\right)\left\langle \left(\frac{\tilde{T}}{\Delta T}-1\right)\left(\nabla\cdot\mathbf{u}\right)^{2}\right\rangle .\label{eq:e204}
\end{align}
As mentioned above, the term $k\Delta\left\langle T'\right\rangle _{h}/2Lk\Delta_{S}$
describing the ratio of the mean conductive heat flux resulting from
convection to the superadiabatic heat flux in the basic state is expected
to be negligible with respect to $Nu_{Q}$ at high $Ra$, when convection
is strongly turbulent and the Nusselt number could be much greater
than unity. If, however this is not the case and the mean conductive
flux is significant it is more convenient to include it in the definition
of the Nusselt number, since the values of the temperature at the
top and bottom boundaries, $\left.\left\langle T'\right\rangle _{h}\right|_{z=L}$
and $\left.\left\langle T'\right\rangle _{h}\right|_{z=0}$ are not
known. Thus in such a case it is natural to define the Nusselt number
in the following way\index{SI}{Nusselt number!anelastic}
\begin{equation}
Nu_{Q}^{\prime}=\frac{\left\langle \tilde{\rho}\tilde{T}u_{z}s'\right\rangle -k\frac{\Delta\left\langle T'\right\rangle _{h}}{2L}}{k\Delta_{S}}.\label{eq:e205}
\end{equation}
Note however, that the constant heat flux boundary conditions (\ref{eq:BCs_const_heat_flux})
imply, that $F_{conv.}\left(z=0\right)=-k\left.\mathrm{d}_{z}\left\langle T'\right\rangle _{h}\right|_{z=0}=0$,
thus (\ref{eq:rel_flux_VD}) allows to express the mean fluctuation
temperature at one boundary (say at the top $\left.\left\langle T'\right\rangle _{h}\right|_{z=L}$)
by the value of the mean fluctuation temperature at the other boundary
(at the bottom $\left.\left\langle T'\right\rangle _{h}\right|_{z=0}$)
and mean quantities.

\section{Linear stability analysis\protect \\
(ideal gas, $Q=0$, isothermal, stress-free and impermeable boundaries,
constant $\nu$, $k$, $\mathbf{g}$ and $c_{p}$)\label{sec:Linear-stability-A}}

It is of interest to study the properties of marginal (linear) anelastic
convection near threshold, to demonstrate the influence of density
stratification on the critical Rayleigh number, convective flow and
the temperature distribution at convection threshold. This will be
achieved via formal expansions in parameter $\theta=\Delta T/T_{B}$,
which is a measure of compressibility/stratification of the fluid. Since
the calculations in the compressible case are cumbersome, we limit
ourselves here only to the simplest case of stress-free, impermeable
and isothermal boundaries, thus we assume
\begin{equation}
u_{z}\left(z=0,\,L\right)=0,\quad\left.\frac{\partial\mathbf{u}_{h}}{\partial z}\right|_{z=0,\,L}=0\label{eq:impermeability_stress_free}
\end{equation}
and
\begin{equation}
T'\left(z=0,\,L\right)=0.\label{eq:isothermal}
\end{equation}
At convection threshold and slightly above it the magnitude of perturbations
to the hydrostatic basic state can be safely assumed small, $\mathbf{u}/\delta^{1/2}\sqrt{gL}\ll1$
and $T'/\delta\tilde{T}\ll1$, etc. which allows to linearize the
full set of dynamical equations (\ref{NS-Aderiv-1-1-1-2}-d) to get\index{SI}{anelastic!equation, perfect gas, linear}
\begin{subequations}
\begin{align}
\frac{\partial\mathbf{u}}{\partial t}= & -\nabla\frac{p'}{\tilde{\rho}}+\frac{g}{c_{p}}s'\hat{\mathbf{e}}_{z}+\nu\nabla^{2}\mathbf{u}+\frac{\nu}{3}\nabla\left(\nabla\cdot\mathbf{u}\right)\nonumber \\
 & +\frac{\nu}{\tilde{\rho}}\frac{\mathrm{d}\tilde{\rho}}{\mathrm{d}z}\left(\nabla u_{z}+\frac{\partial\mathbf{u}}{\partial z}\right)+\frac{2}{3}\nu\left(\frac{1}{\tilde{\rho}}\frac{\mathrm{d}\tilde{\rho}}{\mathrm{d}z}\right)^{2}u_{z}\hat{\mathbf{e}}_{z},\label{NS-Aderiv-1-1-1-2-1}
\end{align}
\begin{equation}
\nabla\cdot\left(\tilde{\rho}\mathbf{u}\right)=0,\label{Cont_Aderiv-1-1-2-1}
\end{equation}
\begin{equation}
\tilde{\rho}\tilde{T}\left(\frac{\partial s'}{\partial t}+\frac{\mathrm{d}\tilde{s}}{\mathrm{d}z}u_{z}\right)=k\nabla^{2}T,\label{Energy_Aderiv-1-1-1-1}
\end{equation}
\begin{equation}
\frac{\rho'}{\tilde{\rho}}=-\frac{T'}{\tilde{T}}+\frac{p'}{\tilde{p}},\qquad\tilde{T}s'=-\frac{p'}{\tilde{\rho}}+c_{p}T',\label{State_eq_Aderiv-1-1-1-1}
\end{equation}
\end{subequations} 
where for simplicity we have assumed that the
fluid satisfies the equation of state of an ideal gas $p=\rho RT$
and the heat capacities $c_{p}$ and $c_{v}$ are constants. Moreover,
it has been assumed, that the kinematic viscosity $\nu=\mu/\tilde{\rho}$
and the thermal conductivity coefficient $k$ are uniform. Under the
current assumptions the hydrostatic reference state is given by equations
(\ref{eq:BS1}-c), hence we can compute the inverse density scale
height\index{SI}{scale heights}, which appears explicitly in the equation (\ref{NS-Aderiv-1-1-1-2-1}),
\begin{equation}
\frac{1}{\tilde{\rho}}\frac{\mathrm{d}\tilde{\rho}}{\mathrm{d}z}=-\frac{1}{L}\frac{m\theta}{1-\theta\frac{z}{L}},\label{eq:inverse_density_scale_height}
\end{equation}
and $\mathrm{d}_{z}\tilde{s}=-c_{p}\Delta_{S}/\tilde{T}$ by the use
of the general formula (\ref{eq:dsbydz}).

To derive one equation for the $z$-dependent amplitude of the vertical
velocity $\hat{u}_{z}$, analogous to (\ref{eq:linear_eqs_Boussinesq_NS}),
we first decompose the perturbation fields into Fourier modes 
\begin{subequations}
\begin{equation}
\mathbf{u}\left(x,y,z,t\right)=\Re\mathfrak{e}\;\,\hat{\mathbf{u}}\left(z\right)\mathrm{e}^{\sigma t}\mathrm{e}^{\mathrm{i}\left(\mathcal{K}_{x}x+\mathcal{K}_{y}y\right)},\label{eq:linear_perturbations_forms-A_u}
\end{equation}
\begin{equation}
T'\left(x,y,z,t\right)=\Re\mathfrak{e}\;\,\hat{T}\left(z\right)\mathrm{e}^{\sigma t}\mathrm{e}^{\mathrm{i}\left(\mathcal{K}_{x}x+\mathcal{K}_{y}y\right)},\label{eq:linear_perturbations_forms-A_T}
\end{equation}
\begin{equation}
s'\left(x,y,z,t\right)=\Re\mathfrak{e}\;\,\hat{s}\left(z\right)\mathrm{e}^{\sigma t}\mathrm{e}^{\mathrm{i}\left(\mathcal{K}_{x}x+\mathcal{K}_{y}y\right)},\label{eq:linear_perturbations_forms-A_s}
\end{equation}
\begin{equation}
p'\left(x,y,z,t\right)=\Re\mathfrak{e}\;\,\hat{p}\left(z\right)\mathrm{e}^{\sigma t}\mathrm{e}^{\mathrm{i}\left(\mathcal{K}_{x}x+\mathcal{K}_{y}y\right)},\label{eq:linear_perturbations_forms-A_p}
\end{equation}
\end{subequations} 
where $\sigma$ is the growth rate, and adopt
the following procedure. We introduce the above form of the perturbation
fields into the equations (\ref{NS-Aderiv-1-1-1-2-1}-d), and take
a $z$-component of the double curl of the equation (\ref{NS-Aderiv-1-1-1-2-1}),
to eliminate the pressure and relate the entropy amplitude $\hat{s}$
to $\hat{u}_{z}$. Next we take a divergence of the equation (\ref{NS-Aderiv-1-1-1-2-1})
to obtain an equation for the pressure amplitude. With the aid of
(\ref{Cont_Aderiv-1-1-2-1}) this yields \begin{subequations}
\begin{equation}
-\sigma\mathfrak{D}^{2}\hat{u}_{z}=\frac{g}{c_{p}}\mathcal{K}^{2}\hat{s}-\nu\mathfrak{D}^{2}\mathfrak{D}^{2}\hat{u}_{z}+\frac{2}{3}\nu\mathcal{K}^{2}\frac{m+3}{m}\left(\frac{1}{\tilde{\rho}}\frac{\mathrm{d}\tilde{\rho}}{\mathrm{d}z}\right)^{2}\hat{u}_{z},\label{eq:NS-A_linear_uz}
\end{equation}
\begin{align}
\nabla^{2}\frac{\hat{p}}{\tilde{\rho}}= & \,\,\sigma\frac{1}{\tilde{\rho}}\frac{\mathrm{d}\tilde{\rho}}{\mathrm{d}z}\hat{u}_{z}+\frac{g}{c_{p}}\frac{\mathrm{d}\hat{s}}{\mathrm{d}z}-\frac{4}{3}\nu\mathfrak{D}^{2}\left(\frac{1}{\tilde{\rho}}\frac{\mathrm{d}\tilde{\rho}}{\mathrm{d}z}\hat{u}_{z}\right)+\nu\frac{1}{\tilde{\rho}}\frac{\mathrm{d}\tilde{\rho}}{\mathrm{d}z}\mathfrak{D}^{2}\hat{u}_{z}\nonumber \\
 & -\frac{2\nu}{m}\left(\frac{1}{\tilde{\rho}}\frac{\mathrm{d}\tilde{\rho}}{\mathrm{d}z}\right)^{2}\frac{\mathrm{d}\hat{u}_{z}}{\mathrm{d}z}-\frac{2\nu}{m}\left(\frac{1}{\tilde{\rho}}\frac{\mathrm{d}\tilde{\rho}}{\mathrm{d}z}\right)^{3}\hat{u}_{z},\label{eq:pressure_A_linear}
\end{align}
\begin{equation}
\sigma\tilde{\rho}\tilde{T}\hat{s}-c_{p}\Delta_{S}\tilde{\rho}\hat{u}_{z}=k\nabla^{2}\hat{T},\label{Energy_A_linear}
\end{equation}
\begin{equation}
\frac{\hat{p}}{\tilde{\rho}}=c_{p}\hat{T}-\tilde{T}\hat{s},\label{State_eq_A_linear}
\end{equation}
\end{subequations} 
where
\begin{equation}
\nabla^{2}=-\mathcal{K}^{2}+\frac{\mathrm{d}^{2}}{\mathrm{d}z^{2}},\qquad\mathfrak{D}^{2}\hat{u}_{z}=\nabla^{2}\hat{u}_{z}+\frac{\mathrm{d}}{\mathrm{d}z}\left(\frac{1}{\tilde{\rho}}\frac{\mathrm{d}\tilde{\rho}}{\mathrm{d}z}\hat{u}_{z}\right),\label{eq:e206}
\end{equation}
and we have used (\ref{eq:inverse_density_scale_height}) to write
$\mathrm{d}_{z}(\mathrm{d}_{z}\tilde{\rho}/\tilde{\rho})=-(1/m)(\mathrm{d}_{z}\tilde{\rho}/\tilde{\rho})^{2}$.

We now introduce $\hat{p}/\tilde{\rho}$ from (\ref{State_eq_A_linear})
into (\ref{eq:pressure_A_linear}), substitute for the temperature
$\hat{T}$ from (\ref{Energy_A_linear}) and for the entropy $\hat{s}$
from (\ref{eq:NS-A_linear_uz}) to obtain the final equation for one
variable $\hat{u}_{z}$, which reads
\begin{align}
-\frac{\sigma^{2}}{\kappa_{B}\nu}\frac{\tilde{\rho}}{\rho_{B}}\tilde{T}\mathfrak{D}^{2}\hat{u}_{z}+\sigma\left[\frac{1}{\kappa_{B}}\frac{\tilde{\rho}}{\rho_{B}}\tilde{T}\mathfrak{D}^{2}\mathfrak{D}^{2}\hat{u}_{z}+\frac{1}{\nu}\nabla^{2}\left(\tilde{T}\mathfrak{D}^{2}\hat{u}_{z}\right)+\frac{g}{c_{p}\nu}\frac{\mathrm{d}}{\mathrm{d}z}\mathfrak{D}^{2}\hat{u}_{z}\right]\qquad\quad\nonumber \\
-\sigma\mathcal{K}^{2}\hat{u}_{z}\left(\frac{1}{\tilde{\rho}}\frac{\mathrm{d}\tilde{\rho}}{\mathrm{d}z}\right)\left[\frac{2}{3}\frac{m+3}{m}\frac{1}{\kappa}\frac{\tilde{\rho}}{\rho_{B}}\tilde{T}\left(\frac{1}{\tilde{\rho}}\frac{\mathrm{d}\tilde{\rho}}{\mathrm{d}z}\right)+\frac{g}{c_{p}\nu}\right]\qquad\quad\qquad\qquad\qquad\qquad\qquad\nonumber \\
=\,\,T_{B}\frac{\mathcal{K}^{2}}{L^{4}}Ra\frac{\tilde{\rho}}{\rho_{B}}\hat{u}_{z}+\nabla^{2}\left(\tilde{T}\mathfrak{D}^{2}\mathfrak{D}^{2}\hat{u}_{z}\right)+\frac{g}{c_{p}}\frac{\mathrm{d}}{\mathrm{d}z}\mathfrak{D}^{2}\mathfrak{D}^{2}\hat{u}_{z}\qquad\quad\qquad\qquad\nonumber \\
-\frac{4}{3}\frac{g}{c_{p}}\mathcal{K}^{2}\mathfrak{D}^{2}\left(\frac{1}{\tilde{\rho}}\frac{\mathrm{d}\tilde{\rho}}{\mathrm{d}z}\hat{u}_{z}\right)+\frac{g}{c_{p}}\mathcal{K}^{2}\frac{1}{\tilde{\rho}}\frac{\mathrm{d}\tilde{\rho}}{\mathrm{d}z}\mathfrak{D}^{2}\hat{u}_{z}\qquad\qquad\qquad\qquad\quad\,\,\qquad\nonumber \\
-\frac{2}{3}\mathcal{K}^{2}\frac{m+3}{m}\nabla^{2}\left[\tilde{T}\left(\frac{1}{\tilde{\rho}}\frac{\mathrm{d}\tilde{\rho}}{\mathrm{d}z}\right)^{2}\hat{u}_{z}\right]-\frac{2}{3}\frac{g}{c_{p}}\mathcal{K}^{2}\frac{m+3}{m}\frac{\mathrm{d}}{\mathrm{d}z}\left[\left(\frac{1}{\tilde{\rho}}\frac{\mathrm{d}\tilde{\rho}}{\mathrm{d}z}\right)^{2}\hat{u}_{z}\right]\nonumber \\
-\frac{2}{m}\frac{g}{c_{p}}\mathcal{K}^{2}\left[\left(\frac{1}{\tilde{\rho}}\frac{\mathrm{d}\tilde{\rho}}{\mathrm{d}z}\right)^{2}\frac{\mathrm{d}\hat{u}_{z}}{\mathrm{d}z}+\left(\frac{1}{\tilde{\rho}}\frac{\mathrm{d}\tilde{\rho}}{\mathrm{d}z}\right)^{3}\hat{u}_{z}\right],\qquad\qquad\qquad\qquad\;\,\,\,\qquad\label{eq:u_z_equation_full_A}
\end{align}
where
\begin{equation}
\kappa_{B}=\frac{k}{\rho_{B}c_{p}}.\label{eq:e207}
\end{equation}
We will now assume, that the \emph{Principle of the exchange of stabilities}
holds in the case at hand, later to verify this hypothesis \emph{a
posteriori}. Therefore the growth rate $\sigma$ at the onset of convection
is now assumed to be purely real and the system becomes convectively
unstable as soon as there appears at least one mode with $\sigma>0$,
whereas exactly at threshold $\sigma=0$. This allows to set the entire
left hand side of (\ref{eq:u_z_equation_full_A}) to zero in the marginal
state and on multiplying this equation by $L^{6}T_{B}^{-1}$ and introducing
non-dimensional coordinates $\mathbf{x}^{\sharp}=\mathbf{x}/L$ the
equation for $\hat{u}_{z}$ reduces to
\begin{align}
0= & \,\,\mathcal{K}^{\sharp2}Ra\left(1-\theta z^{\sharp}\right)^{m}\hat{u}_{z}+\nabla^{\sharp2}\left[\left(1-\theta z^{\sharp}\right)\mathfrak{D}^{\sharp2}\mathfrak{D}^{\sharp2}\hat{u}_{z}\right]+\theta\frac{\mathrm{d}}{\mathrm{d}z^{\sharp}}\mathfrak{D}^{\sharp2}\mathfrak{D}^{\sharp2}\hat{u}_{z}\nonumber \\
 & +\frac{4}{3}m\theta^{2}\mathcal{K}^{\sharp2}\mathfrak{D}^{\sharp2}\left(\frac{\hat{u}_{z}}{1-\theta z^{\sharp}}\right)-m\theta^{2}\mathcal{K}^{\sharp2}\frac{1}{1-\theta z^{\sharp}}\mathfrak{D}^{\sharp2}\hat{u}_{z}\nonumber \\
 & -\frac{2}{3}m\left(m+3\right)\theta^{2}\mathcal{K}^{\sharp2}\nabla^{\sharp2}\left(\frac{\hat{u}_{z}}{1-\theta z^{\sharp}}\right)-\frac{2}{3}m\left(m+3\right)\theta^{3}\mathcal{K}^{\sharp2}\frac{\mathrm{d}}{\mathrm{d}z^{\sharp}}\left[\frac{\hat{u}_{z}}{\left(1-\theta z^{\sharp}\right)^{2}}\right]\nonumber \\
 & -2m\theta^{3}\mathcal{K}^{\sharp2}\frac{1}{\left(1-\theta z^{\sharp}\right)^{2}}\left(\frac{\mathrm{d}\hat{u}_{z}}{\mathrm{d}z^{\sharp}}-m\theta\frac{\hat{u}_{z}}{1-\theta z^{\sharp}}\right),\label{eq:u_z_equation_full_A-1}
\end{align}
where $\mathcal{K}^{\sharp}=\mathcal{K}L$ and
\begin{equation}
Ra=\frac{g\Delta_{S}L^{4}}{T_{B}\kappa_{B}\nu},\label{eq:e208}
\end{equation}
\begin{equation}
\nabla^{\sharp2}=-\mathcal{K}^{\sharp2}+\frac{\mathrm{d}^{2}}{\mathrm{d}z^{\sharp2}},\qquad\mathfrak{D}^{\sharp2}\hat{u}_{z}=\nabla^{\sharp2}\hat{u}_{z}-m\theta\frac{\mathrm{d}}{\mathrm{d}z^{\sharp}}\left(\frac{\hat{u}_{z}}{1-\theta z^{\sharp}}\right).\label{eq:e209}
\end{equation}
This equation is subject to boundary conditions, that is the impermeability
of boundaries
\begin{equation}
\hat{u}_{z}\left(z^{\sharp}=0,\,1\right)=0,\label{eq:impermeability}
\end{equation}
and the stress-free boundary condition (\ref{eq:impermeability_stress_free})
for the horizontal velocity components, which by taking the $z$-derivative
of the mass conservation equation $\nabla_{h}^{\sharp}\cdot\hat{\mathbf{u}}_{h}+\partial_{z^{\sharp}}\hat{u}_{z}=m\theta\hat{u}_{z}/(1-\theta z^{\sharp})$
can be easily transformed to boundary conditions involving only $\hat{u}_{z}$
\begin{subequations}
\begin{equation}
\left.\frac{\partial^{2}\hat{u}_{z}}{\partial z^{\sharp2}}\right|_{z^{\sharp}=0}=m\theta\left.\frac{\partial\hat{u}_{z}}{\partial z^{\sharp}}\right|_{z^{\sharp}=0},\label{eq:stress_free_0}
\end{equation}
\begin{equation}
\left.\frac{\partial^{2}\hat{u}_{z}}{\partial z^{\sharp2}}\right|_{z^{\sharp}=1}=\frac{m\theta}{1-\theta}\left.\frac{\partial\hat{u}_{z}}{\partial z^{\sharp}}\right|_{z^{\sharp}=1}.\label{eq:stress_free_1}
\end{equation}
\end{subequations} 
This problem of the sixth order differential equation (\ref{eq:u_z_equation_full_A-1})
with four boundary conditions (\ref{eq:impermeability}) and (\ref{eq:stress_free_0},b)
must be supplied with the temperature equation (\ref{Energy_A_linear}),
which now reads
\begin{equation}
\nabla^{\sharp2}\frac{\hat{T}}{T_{B}}=-\frac{\nu}{gL^{2}}Ra\left(1-\theta z^{\sharp}\right)^{m}\hat{u}_{z},\label{eq:temperature_eq_A_nondim}
\end{equation}
with two boundary conditions
\begin{equation}
\hat{T}\left(z^{\sharp}=0,\,1\right)=0.\label{eq:BC_temp_A}
\end{equation}
The full problem defined by equations (\ref{eq:u_z_equation_full_A-1})
and (\ref{eq:temperature_eq_A_nondim})\footnote{Still with some aid of (\ref{eq:NS-A_linear_uz}-d) to obtain self-consistency, i.e. eliminate the solution of the homogeneous temperature problem $\nabla^{\sharp 2}\hat T/T_B = 0$.} with the boundary conditions
(\ref{eq:impermeability}), (\ref{eq:stress_free_0},b) and (\ref{eq:BC_temp_A})
is rather cumbersome and perhaps even unsolvable analytically. Therefore
to get some insight into the effect of density stratification on the
convection threshold\index{SI}{threshold of convection} we will assume now, that the stratification is
weak through imposing
\begin{equation}
\theta=1-\frac{1}{\Gamma}\ll1.\label{eq:e210}
\end{equation}
This allows to expand all the dependent variables in perturbation
series, which under additional simplifying assumption, that at threshold
the convective flow takes the form of two-dimensional rolls leads to
the following form of the velocity and temperature fields 
\begin{subequations}
\begin{equation}
u_{z}=\left(\hat{u}_{0z}+\theta\hat{u}_{1z}+\theta^{2}\hat{u}_{2z}+\dots\right)\cos\left(\mathcal{K}x\right),\label{eq:e211}
\end{equation}
\begin{equation}
u_{x}=\left(\hat{u}_{0x}+\theta\hat{u}_{1x}+\theta^{2}\hat{u}_{2x}+\dots\right)\sin\left(\mathcal{K}x\right),\label{eq:e212}
\end{equation}
\begin{equation}
T'=\left(\hat{T}_{0}+\theta\hat{T}_{1}+\theta^{2}\hat{T}_{2}+\dots\right)\cos\left(\mathcal{K}x\right).\label{eq:e213}
\end{equation}
\end{subequations} 
The expansion of dependent variables must be accompanied
by expansion of the critical Rayleigh number and the critical wave number, since these
parameters are expected to be influenced by stratification,
\begin{equation}
Ra_{crit}=Ra_{0}+\theta Ra_{1}+\theta^{2}Ra_{2}+\dots,\quad\mathcal{K}_{crit}^{\sharp}=\mathcal{K}_{0}^{\sharp}+\theta\mathcal{K}_{1}^{\sharp}+\theta^{2}\mathcal{K}_{2}^{\sharp2}+\dots.\label{eq:e214}
\end{equation}
Introducing the above expansions into the equations and gathering
terms at the leading order $\theta^{0}$ one obtains 
\begin{subequations}
\begin{equation}
-\mathcal{K}_{0}^{\sharp2}Ra_{0}\hat{u}_{0z}=\left(-\mathcal{K}_{0}^{\sharp2}+\frac{\mathrm{d}^{2}}{\mathrm{d}z^{\sharp2}}\right)^{3}\hat{u}_{0z},\label{eq:uz0_A}
\end{equation}
\begin{equation}
\left(-\mathcal{K}_{0}^{\sharp2}+\frac{\mathrm{d}^{2}}{\mathrm{d}z^{\sharp2}}\right)\frac{\hat{T}_{0}}{T_{B}}=-\frac{\nu}{gL^{2}}Ra_{0}\hat{u}_{0z},\label{eq:T0_A}
\end{equation}
\end{subequations} 
which is, of course, an equivalent set of equations
to that obtained in the Boussinesq case (\ref{eq:uz_equation_lin_B})
and (\ref{eq:linear_eqs_Boussinesq_Th}). It is, however, not immediately
obvious, that the boundary conditions for the velocity $\hat{u}_{0z}$
can be expressed in the same form as (\ref{eq:BCs_Boussinesq_case1}),
so as to extract only the $\sin\pi z^{\sharp}$ type solutions out
of the class of six independent solutions of the sixth order equation
(\ref{eq:uz0_A}). This is because the anelastic linear problem (\ref{eq:NS-A_linear_uz}-d),
in contrast to the Boussinesq case, involves the entropy and pressure
fields, for which the boundary conditions are not specified. In particular
the entropy appears in the buoyancy force, thus complicating derivation
of a full set of boundary conditions solely in terms of the vertical
velocity component $\hat{u}_{z}$. However, the state equation (\ref{State_eq_A_linear}),
with the aid of the basic state formulae (\ref{eq:BS1}) can be easily
rearranged into the following non-dimensional form
\begin{equation}
\theta\frac{\hat{p}}{gL\rho_{B}\left(1-\theta z^{\sharp}\right)^{m}}=\frac{\hat{T}}{T_{B}}-\left(1-\theta z^{\sharp}\right)\frac{\hat{s}}{c_{p}},\label{eq:e215}
\end{equation}
where the fact, that $m=1/(\gamma-1)+\mathcal{O}(\delta)$ was utilized.
This implies, that at the leading order in $\theta$ the standard Boussinesq consonance
of the temperature and entropy fluctuations occurs, i.e. $\hat{s}_{0}=c_{p}\hat{T}_{0}/T_{B}$
and therefore the leading order problem here corresponds directly
to the threshold of Boussinesq convection. It follows that the leading order solution corresponds
to that from section \ref{subsec:Two-isothermal-stress-free} and
reads
\begin{equation}
\hat{u}_{0z}=A\sin\pi z^{\sharp},\qquad\hat{T}_{0}=T_{B}\frac{\nu}{gL^{2}}A\frac{9}{2}\pi^{2}\sin\left(\pi z^{\sharp}\right),\label{eq:0_order_A_UT}
\end{equation}
\begin{equation}
Ra_{0}=\frac{27}{4}\pi^{4},\qquad\mathcal{K}_{0}^{2}=\frac{\pi^{2}}{2}.\label{eq:0_order_A_RaK}
\end{equation}
Gathering now all the terms next to the first power of $\theta$ in
the equations (\ref{eq:u_z_equation_full_A-1}) and (\ref{eq:temperature_eq_A_nondim})
and in the boundary conditions (\ref{eq:impermeability}), (\ref{eq:stress_free_0},b)
and (\ref{eq:BC_temp_A}) and utilizing (\ref{eq:0_order_A_UT},b),
allows to write down the set of equations determining the first order
corrections to the Boussinesq solutions (\ref{eq:0_order_A_UT},b) resulting from compressibility that is $\hat{u}_{1}$, $\hat{T}_{1}$,
$Ra_{1}$ and $\mathcal{K}_{1}$
in the following simple form 
\begin{subequations}
\begin{align}
\frac{27}{8}\pi^{6}\hat{u}_{1z}+\left(-\frac{\pi^{2}}{2}+\frac{\mathrm{d}^{2}}{\mathrm{d}z^{\sharp2}}\right)^{3}\hat{u}_{1z}= & \,\,\frac{27}{8}\pi^{6}\left(\left(m-1\right)z^{\sharp}-\frac{Ra_{1}}{Ra_{0}}\right)\sin\pi z^{\sharp}\nonumber \\
 & +\frac{9}{4}\pi^{5}\left(2m+1\right)\cos\pi z^{\sharp},\label{eq:uz1_A}
\end{align}
\begin{equation}
\left(-\mathcal{K}_{0}^{\sharp2}+\frac{\mathrm{d}^{2}}{\mathrm{d}z^{\sharp2}}\right)\frac{\hat{T}_{1}}{T_{B}}=\frac{\nu}{gL^{2}}\frac{27}{4}\pi^{4}\left[-\hat{u}_{1z}+\sin\pi z^{\sharp}\left(mz^{\sharp}-\frac{Ra_{1}}{Ra_{0}}+\frac{4}{3\pi^{2}}\mathcal{K}_{0}\mathcal{K}_{1}\right)\right],\label{eq:T1_A}
\end{equation}
\end{subequations} 
with the boundary conditions 
\begin{subequations}
\begin{equation}
\left.\frac{\partial^{2}\hat{u}_{1z}}{\partial z^{\sharp2}}\right|_{z^{\sharp}=0}=m\pi A,\qquad\left.\frac{\partial^{2}\hat{u}_{1z}}{\partial z^{\sharp2}}\right|_{z^{\sharp}=1}=-m\pi A,\label{eq:BCs_order_1_A_1}
\end{equation}
\begin{equation}
\hat{u}_{z}\left(z^{\sharp}=0,\,1\right)=0,\qquad\hat{T}\left(z^{\sharp}=0,\,1\right)=0.\label{eq:BCs_order1_A_2}
\end{equation}
\end{subequations} 
The next step is to consider the solvability condition
for the latter set of equations, i.e. apply the Fredholm Alternative
Theorem (cf. Korn \& Korn 1961). This requires the knowledge of the
solution to the homogeneous problem (\ref{eq:uz1_A}) with adequate
boundary conditions, which must be orthogonal to the non-homogeneity
on the right hand side of (\ref{eq:uz1_A}). The homogeneous problem,
however, corresponds directly to the leading order one, i.e. the Boussinesq
problem, both in terms of the equations and boundary conditions, thus
its solution is simply $\sin\pi z^{\sharp}$ and the solvability condition
yields
\begin{align}
0= & \frac{27}{8}\pi^{6}\int_{0}^{1}\mathrm{d}z\left(\left(m-1\right)z^{\sharp}-\frac{Ra_{1}}{Ra_{0}}\right)\sin^{2}\pi z^{\sharp}\nonumber \\
 & +\frac{9}{4}\pi^{5}\left(2m+1\right)\int_{0}^{1}\mathrm{d}z\cos\pi z^{\sharp}\sin\pi z^{\sharp}.\label{eq:e216}
\end{align}
This implies, that the first order correction to the critical Rayleigh
number at threshold takes the form
\begin{equation}
Ra_{1}=\frac{1}{2}\left(m-1\right)Ra_{0}=\frac{27}{8}\pi^{4}\left(m-1\right).\label{eq:e217}
\end{equation}
The latter result means, that when $m>1$, which roughly corresponds
to fluids characterized by $\gamma<2$ (see discussion below (\ref{eq:BS1}-c)\footnote{The definition of the parameter $\delta$ in (\ref{eq:delta_definition})
and the basic state formulae (\ref{eq:BS1}-c) imply, that $m\approx1/(\gamma-1)$
up to terms as small as $\mathcal{O}(\delta)$.}), the critical Rayleigh number in the case at hand\index{SI}{Rayleigh number!critical}
\begin{equation}
Ra_{crit}\approx\frac{27}{4}\pi^{4}\left[1+\frac{1}{2}\theta\left(m-1\right)\right],\label{eq:Ra_crit_A}
\end{equation}
is increased by the presence of stratification. As a result the total
heat per unit mass accumulated in the entire fluid layer (between top and
bottom boundaries) in the marginal state, which according to (\ref{eq:GENERAL_STABILITY_CONDITION})
equals
\begin{equation}
-\int_{0}^{L}c_{p}\left(\frac{\mathrm{d}\tilde{T}}{\mathrm{d}z}+\frac{g}{c_{p}}\right)\mathrm{d}z=\frac{c_{p}\kappa_{B}\nu T_{B}}{gL^{3}}Ra_{crit},\label{eq:e218}
\end{equation}
(and is equal to the total heat per unit mass released in the marginal
state by a fluid parcel rising from the bottom to the top of the fluid
layer), where $Ra_{crit}$ is given in (\ref{eq:Ra_crit_A}) is also
greater in the stratified case, than in the Boussinesq one.

This, however, is by no means a rule and not every anelastic system
is characterized by a greater critical Rayleigh number than its Boussinesq
analogue. In fact our choice of constant $\nu$ and $k$, thus $z$-dependent
$\mu=\tilde{\rho}\nu$ and $\kappa=k/\tilde{\rho}c_{p}$ allowed to
simplify the algebraic manipulations, but also influenced the stability
characteristics of the system. Mizerski and Tobias (2011) have studied
the marginal anelastic convection in the limit of rapid background
rotation $E\ll1$ (with $Pr=\nu/\kappa_{B}$ significantly exceeding
the Boussinesq transitional value $\mathscr{P}r$ between stationary
and oscillatory marginal convection) and weak stratification\footnote{note the different sign in definition of $\theta$ in Mizerski and
Tobias (2011).} $\theta\ll1$, and they have shown, that when $\nu$ and $\kappa$ are assumed constant
instead of $\nu$ and $k$, the presence of stratification decreases
the critical Rayleigh number to $Ra_{crit}=3(\pi/\sqrt{2}E)^{4/3}\left[1-\theta\gamma/2(\gamma-1)\right]$.
Moreover, the efficiency of heat transport near the marginal state
was also shown to depend strongly on whether the thermal conductivity
coefficient $k$ or the thermal diffusivity $\kappa$ are assumed
constant. In particular they have demonstrated for rapidly rotating
convective systems, that when $k=\textrm{const}$ and $\gamma<2$
the Nusselt number is decreased by the presence of compressibility
and the opposite is true, when $\kappa=\textrm{const}$. However,
the specific heat ratio can achieve values greater than two,
as e.g. in the case of dry air at high pressure ($\sim200\,\mathrm{atm}$)
and low temperature ($200\,\mathrm{K}$), as reported by Perry\emph{
et al.} (1997) and Kamari \emph{et al.} (2014). Therefore it is also possible,
that in the case of constant $\nu$ and $k$ studied above the critical
Rayleigh number (\ref{eq:Ra_crit_A}) may be less than the Boussinesq
value $27\pi^{4}/4$, when the polytropic index $m$ falls below unity. 

Once the correction $Ra_{1}$ to the critical Rayleigh number is known,
the correction to the critical wave number $\mathcal{K}_{1}^{\sharp}$
can be established. Due to the nature of the analysis in the asymptotic
regime $\theta\ll1$ and the perturbation expansions it is, in fact,
rather simple to predict, that the Rayleigh number at convection threshold\index{SI}{threshold of convection}
can only depend on $\mathcal{K}^{\sharp}$ through the $\theta^{2}$
correction of the type $+\mathrm{Const}\theta^{2}\mathcal{K}_{1}^{\sharp2}$,
where $\mathrm{Const}>0$ and thus minimization of $Ra(\mathcal{K}^{\sharp})$
over all possible wave numbers leads to $\mathcal{K}_{1\,crit}^{\sharp}=0$;
no correction proportional to $\mathcal{K}_{1}^{\sharp}$ in the first power is
possible, since the Rayleigh number at threshold can not depend on
the sign of the wave number. This indeed, can be verified by consideration
of the $\theta^{2}$ order terms in the $\hat{u}_{z}$ equation (\ref{eq:u_z_equation_full_A-1}),
even without solving yet for the $\hat{u}_{1z}$ correction to the
vertical velocity \footnote{This still requires the assumption formulated above, that the \emph{Principle
of the exchange of stability} is valid, which will be later verified
for consistency at least up to the order $\theta$, by a straightforward
calculation of the growth rate with its order $\theta$ correction.}. From the solvability condition at the order $\theta^{2}$ all the
terms proportional to first and second powers of $\mathcal{K}_{1}^{\sharp}$
and first power of $\mathcal{K}_{2}^{\sharp}$ can be gathered, which
leads to cancelling of all the terms proportional to $\mathcal{K}_{1}^{\sharp}$
and $\mathcal{K}_{2}^{\sharp}$ in the first powers and the correction
$Ra_{2}$ to the Rayleigh number at threshold takes the form $Ra_{2}=4\mathcal{K}_{1}^{\sharp2}/3+\mathrm{const}$,
where $\mathrm{const}$ is entirely independent of the wave number.
This allows to verify, that the minimal value of the Rayleigh number
at convection threshold, in other words the critical Rayleigh number\index{SI}{Rayleigh number!critical}
for weakly stratified, anelastic convection with constant $\nu$ and
$k$ is achieved by modes with $\mathcal{K}_{1}^{\sharp}=\mathcal{K}_{1\,crit}^{\sharp}=0$.

\begin{figure}
\begin{centering}
a)\includegraphics[scale=0.14]{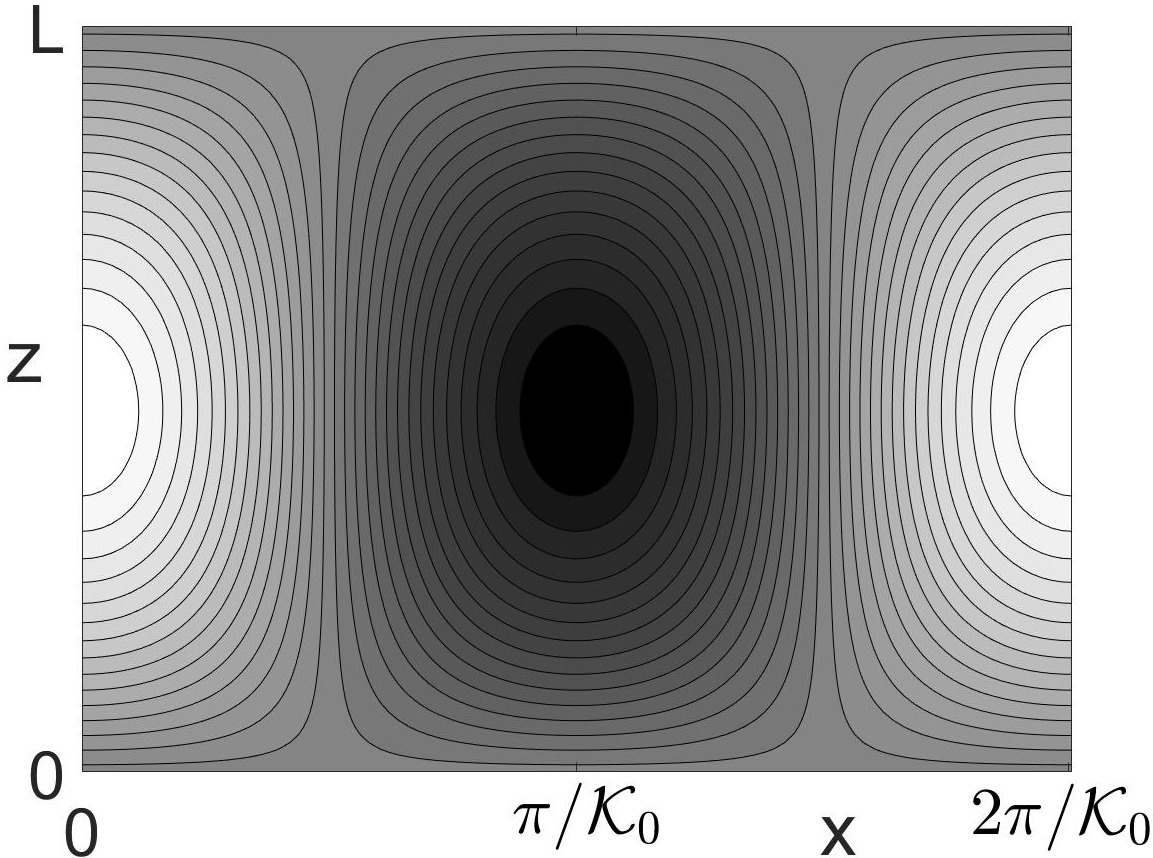}~~~b)\includegraphics[scale=0.14]{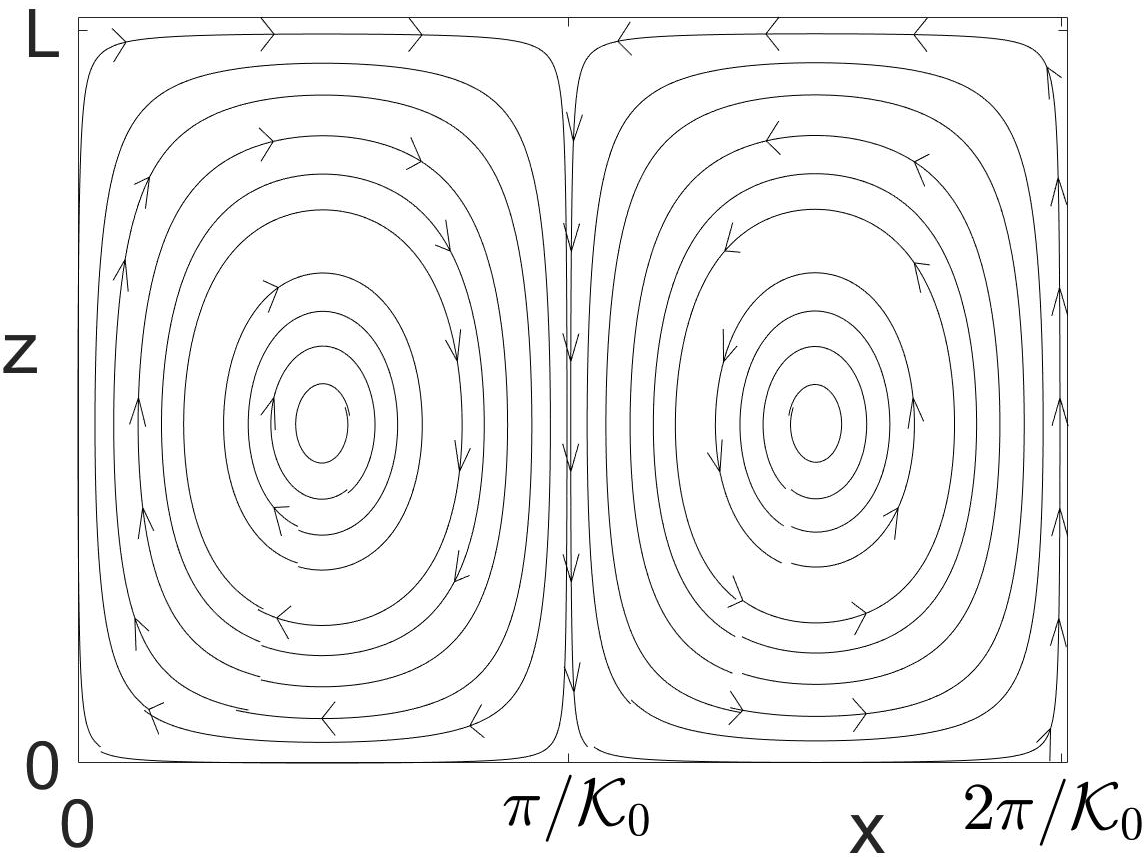}
\par\end{centering}
~

~
\centering{}c)\includegraphics[scale=0.09]{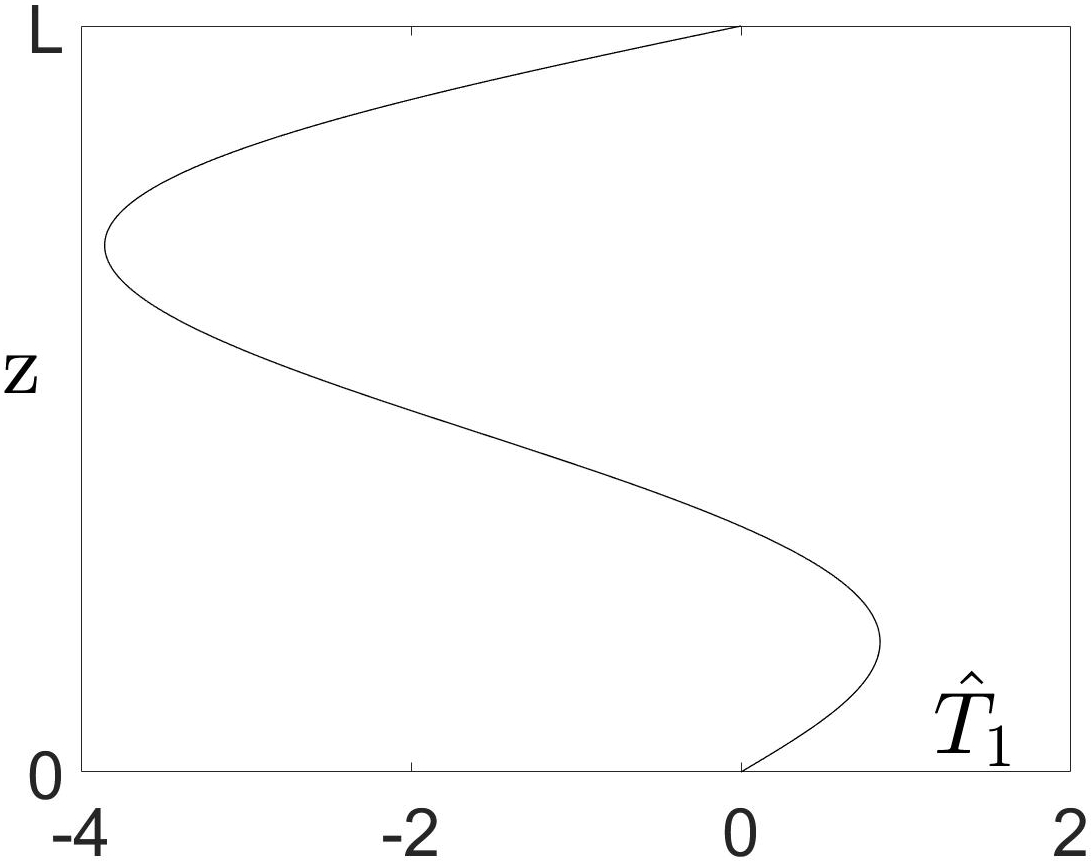}~~~d)\includegraphics[scale=0.09]{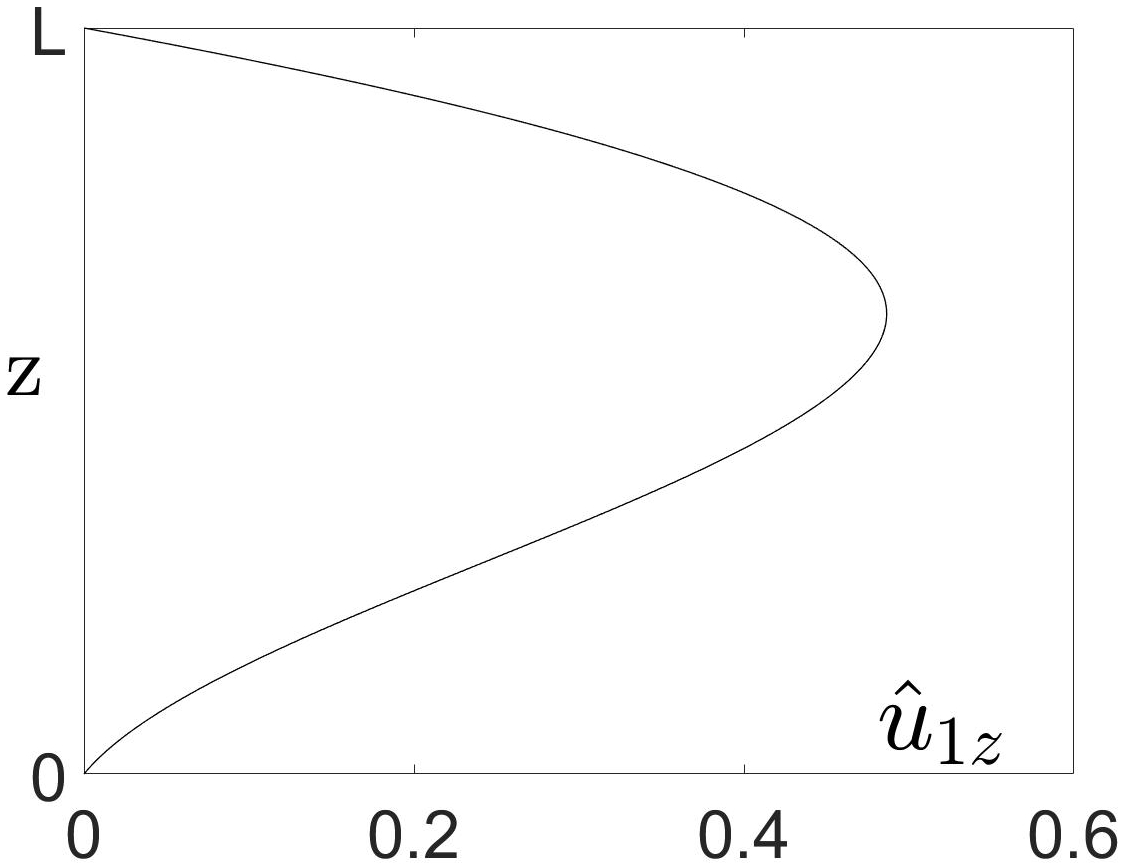}~~~e)\includegraphics[scale=0.09]{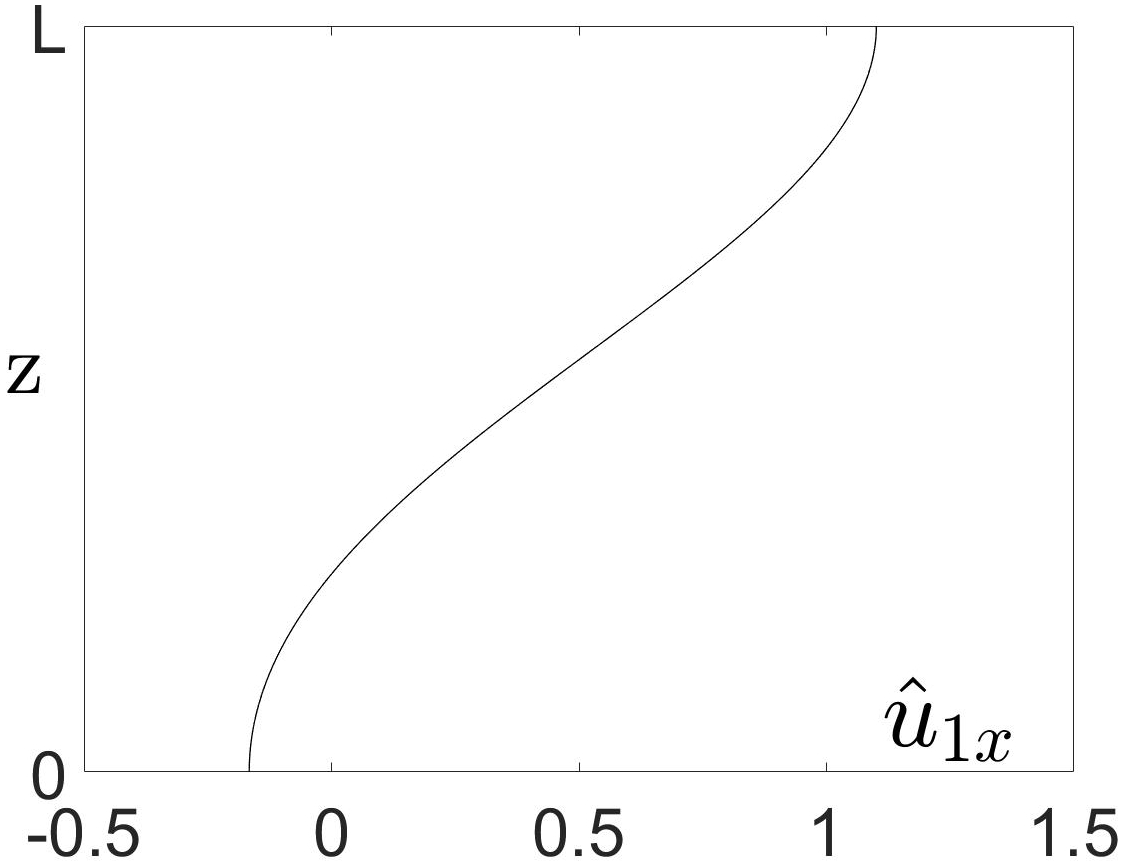}\caption{\label{fig:linear_anelastic_results}{\footnotesize{}Visual representation
of the linear solutions (\ref{eq:u0z_A_final}-\ref{eq:T1_A_final})
at the threshold of anelastic convection\index{SI}{threshold of convection} for an ideal gas with isothermal,
stress-free and impermeable boundaries, no heat sources $Q=0$, constant
$\nu$, $k$, $\mathbf{g}$ and $c_{p}$. Figure a) presents a colour
map and isolines of the temperature $T'=(\hat{T}_{0}+\theta\hat{T}_{1})\cos\mathcal{K}_{0}x$
and the streamlines of the flow at threshold, given by $u_{x}=(\hat{u}_{0x}+\theta\hat{u}_{1x})\sin\mathcal{K}_{0}x$
and $u_{z}=(\hat{u}_{0z}+\theta\hat{u}_{1z})\cos\mathcal{K}_{0}x$
are shown on figure b). To clearly visualize the effect of stratification
on linear solutions the stratification parameter was chosen $\theta=0.5$.
The vertical dependencies of the order $\theta$ corrections to temperature,
$\hat{T}_{1}(z)$, vertical velocity, $\hat{u}_{1z}(z)$ and horizontal
velocity, $\hat{u}_{1x}(z)$ are plotted on figures c), d) and e).}}
\end{figure}

We can now solve the first order problem (\ref{eq:uz1_A}-\ref{eq:BCs_order1_A_2})
with $\mathcal{K}_{1}$ set to zero in (\ref{eq:T1_A}). The homogeneous
equation for the vertical velocity correction $\hat{u}_{1z}$,
\begin{equation}
\frac{27}{8}\pi^{6}\hat{u}_{1z}^{HE}+\left(-\frac{\pi^{2}}{2}+\frac{\mathrm{d}^{2}}{\mathrm{d}z^{\sharp2}}\right)^{3}\hat{u}_{1z}^{HE}=0,\label{eq:e219}
\end{equation}
possesses a class of solutions
\begin{eqnarray}
\hat{u}_{1z}^{HE} & = & C_{1}\sin\left(q_{1}z^{\sharp}\right)\cosh\left(q_{2}z^{\sharp}\right)+C_{2}\cos\left(q_{1}z^{\sharp}\right)\sinh\left(q_{2}z^{\sharp}\right)\nonumber \\
 &  & +C_{3}\cos\left(q_{1}z^{\sharp}\right)\cosh\left(q_{2}z^{\sharp}\right)+C_{4}\sin\left(q_{1}z^{\sharp}\right)\sinh\left(q_{2}z^{\sharp}\right)\nonumber \\
 &  & +C_{5}\cos\left(\pi z^{\sharp}\right)+C_{6}\sin\left(\pi z^{\sharp}\right),\label{eq:u1_HE}
\end{eqnarray}
where $C_{j}$ with $j=1,\dots,6$ are constants and 
\begin{equation}
q_{1}=\frac{\pi}{2\sqrt{2}}\sqrt{\sqrt{52}-5},\quad q_{2}=\frac{\pi}{2\sqrt{2}}\sqrt{\sqrt{52}+5}.\label{eq:e220}
\end{equation}
The term $C_{6}\sin\left(\pi z^{\sharp}\right)$ is exactly of the
same type as the leading order solution, and hence the constant $C_{6}$
can be set to zero without loss of generality. The solution of the
homogeneous problem (\ref{eq:u1_HE}) must be accompanied by the specific
solution of the non-homogeneous equation (\ref{eq:uz1_A}), which can
be written in the form
\begin{equation}
A\left[\frac{1}{24}\left(19m-7\right)z^{\sharp}\sin\left(\pi z^{\sharp}\right)-\frac{\pi}{8}\left(m-1\right)z^{\sharp}\left(z^{\sharp}-1\right)\cos\left(\pi z^{\sharp}\right)\right].\label{eq:e221}
\end{equation}
One can now solve for the general form of the temperature from (\ref{eq:T1_A})
and apply the boundary conditions (\ref{eq:BCs_order_1_A_1},b) to
obtain the final formulae for the order $\theta$ corrections to the
vertical velocity and temperature. Once the form of $\hat{u}_{z}$
is known, the mass conservation equation can be used to calculate
the horizontal velocity component 
\begin{subequations}
\begin{equation}
\mathcal{K}_{0}^{\sharp}\hat{u}_{0x}=-\frac{\mathrm{d}\hat{u}_{0z}}{\mathrm{d}z^{\sharp}},\label{eq:e222}
\end{equation}
\begin{equation}
\mathcal{K}_{0}^{\sharp}\hat{u}_{1x}=-\frac{\mathrm{d}\hat{u}_{1z}}{\mathrm{d}z^{\sharp}}+m\theta\hat{u}_{0z}.\label{eq:e223}
\end{equation}
\end{subequations} 
The full solution in the dimensional coordinates,
together with the leading order terms invoked here again for the sake
of completeness, takes the following form
\begin{equation}
\hat{u}_{0z}=A\sin\left(\pi\frac{z}{L}\right),\label{eq:u0z_A_final}
\end{equation}
\begin{align}
\hat{u}_{1z}= & \,\,A\frac{2}{27\pi}\Bigg\{\left(m+2\right)\left[\cos\left(\pi\frac{z}{L}\right)-\cos\left(q_{1}\frac{z}{L}\right)\cosh\left(q_{2}\frac{z}{L}\right)\right]\nonumber \\
 & \qquad\qquad-\sqrt{3}\left(m-4\right)\sin\left(q_{1}\frac{z}{L}\right)\sinh\left(q_{2}\frac{z}{L}\right)\nonumber \\
 & \qquad\qquad+C_{1}\sin\left(q_{1}\frac{z}{L}\right)\cosh\left(q_{2}\frac{z}{L}\right)+C_{2}\cos\left(q_{1}\frac{z}{L}\right)\sinh\left(q_{2}\frac{z}{L}\right)\Bigg\}\nonumber \\
 & +\frac{A}{8}\left[\frac{1}{3}\left(19m-7\right)\frac{z}{L}\sin\left(\pi\frac{z}{L}\right)-\pi\left(m-1\right)\frac{z}{L}\left(\frac{z}{L}-1\right)\cos\left(\pi\frac{z}{L}\right)\right]\label{eq:u1z_A_final}
\end{align}
\begin{equation}
\hat{u}_{0x}=-A\sqrt{2}\cos\left(\pi\frac{z}{L}\right),\label{eq:u0x_A_final}
\end{equation}
\begin{align}
\hat{u}_{1x}= & -A\frac{2\sqrt{2}}{27\pi^{2}}\Bigg\{\left[q_{1}C_{1}+q_{2}C_{2}\right]\cos\left(q_{1}\frac{z}{L}\right)\cosh\left(q_{2}\frac{z}{L}\right)\nonumber \\
 & \qquad\qquad\quad+\left[q_{2}C_{1}-q_{1}C_{2}\right]\sin\left(q_{1}\frac{z}{L}\right)\sinh\left(q_{2}\frac{z}{L}\right)\nonumber \\
 & \qquad\qquad\quad+\left[q_{1}\left(m+2\right)-q_{2}\sqrt{3}\left(m-4\right)\right]\sin\left(q_{1}\frac{z}{L}\right)\cosh\left(q_{2}\frac{z}{L}\right)\nonumber \\
 & \qquad\qquad\quad-\left[q_{2}\left(m+2\right)+q_{1}\sqrt{3}\left(m-4\right)\right]\cos\left(q_{1}\frac{z}{L}\right)\sinh\left(q_{2}\frac{z}{L}\right)\Bigg\}\nonumber \\
 & -\frac{A\sqrt{2}}{8}\Bigg\{\left[\pi\left(m-1\right)\frac{z}{L}\left(\frac{z}{L}-1\right)-\frac{61m+95}{27\pi}\right]\sin\left(\pi\frac{z}{L}\right)\nonumber \\
 & \qquad\qquad+\left[\frac{1}{3}\left(13m-1\right)\frac{z}{L}+m-1\right]\cos\left(\pi\frac{z}{L}\right)\Bigg\}\label{eq:u1x_a_final}
\end{align}
\begin{equation}
\hat{T}_{0}=T_{B}\frac{\nu}{gL^{2}}A\frac{9}{2}\pi^{2}\sin\left(\pi\frac{z}{L}\right),\label{eq:T0_A_final}
\end{equation}
\begin{align}
\hat{T}_{1}= & \,\,T_{B}\frac{\nu}{gL^{2}}A\frac{\pi}{3}\Bigg\{-\left(m-7\right)\cos\left(q_{1}\frac{z}{L}\right)\cosh\left(q_{2}\frac{z}{L}\right)\nonumber \\
 & \qquad\qquad\qquad+\sqrt{3}\left(m-1\right)\sin\left(q_{1}\frac{z}{L}\right)\sinh\left(q_{2}\frac{z}{L}\right)\nonumber \\
 & \qquad\qquad\qquad+C_{3}\sin\left(q_{1}\frac{z}{L}\right)\cosh\left(q_{2}\frac{z}{L}\right)+C_{4}\cos\left(q_{1}\frac{z}{L}\right)\sinh\left(q_{2}\frac{z}{L}\right)\nonumber \\
 & \qquad\qquad\qquad+\frac{9\pi}{2}\left[\frac{3}{8}\left(m-5\right)\frac{z}{L}+\left(m-1\right)\right]\sin\left(\pi\frac{z}{L}\right)\nonumber \\
 & \qquad\qquad\qquad-\left[\frac{27}{16}\pi^{2}\left(m-1\right)\frac{z}{L}\left(\frac{z}{L}-1\right)-m+7\right]\cos\left(\pi\frac{z}{L}\right)\Bigg\},\label{eq:T1_A_final}
\end{align}
\begin{equation}
Ra_{crit}=\frac{27}{4}\pi^{4}\left[1+\frac{1}{2}\theta\left(m-1\right)\right]+\mathcal{O}\left(\theta^{2}\right),\label{eq:Ra_crit_final_A}
\end{equation}
\begin{equation}
\mathcal{K}_{crit}=\mathcal{K}_{0}+\mathcal{O}\left(\theta^{2}\right),\label{eq:Kcrit_A}
\end{equation}
where
\begin{equation}
C_{1}=\frac{\left(m+2\right)\sin q_{1}+\sqrt{3}\left(m-4\right)\sinh q_{2}}{\cosh q_{2}-\cos q_{1}},\label{eq:C1_A}
\end{equation}
\begin{equation}
C_{2}=\frac{\left(m+2\right)\sinh q_{2}-\sqrt{3}\left(m-4\right)\sin q_{1}}{\cosh q_{2}-\cos q_{1}},\label{eq:C2_A}
\end{equation}
\begin{equation}
C_{3}=\frac{\left(m-7\right)\sin q_{1}-\sqrt{3}\left(m-1\right)\sinh q_{2}}{\cosh q_{2}-\cos q_{1}},\label{eq:C3_A}
\end{equation}
\begin{equation}
C_{4}=\frac{\left(m-7\right)\sinh q_{2}+\sqrt{3}\left(m-1\right)\sin q_{1}}{\cosh q_{2}-\cos q_{1}}.\label{eq:C4_A}
\end{equation}
The solutions are depicted on figure \ref{fig:linear_anelastic_results}.
Note, that the centre of the convective cells is shifted downwards
with respect to the symmetric Boussinesq solutions and the horizontal
velocities are significantly stronger near the top than near the bottom.

The planform of the solutions was of course arbitrarily assumed here
as two-dimensional rolls. The weakly nonlinear analysis and the pattern
selection problem are very cumbersome in the case of anelastic convection,
since even the solutions of the linear problem (\ref{eq:u_z_equation_full_A-1})
are not known for arbitrary values of $\theta$ and the double asymptotic
calculations involving small departure from threshold and small $\theta$
become increasingly difficult at higher orders. Therefore the problem
of planform selection by anelastic convection near threshold remains
almost entirely unsolved as of yet.

\subsection{The growth rate of the convective instability in the limit $\theta\ll1$\label{subsec:The-growth-rate}}

Let us now derive the first order correction to the growth rate, which
results from vertical density stratification of the fluid layer, $\theta\ll1$.
We define the non-dimensional growth rate in a similar manner as in
the Boussinesq case
\begin{equation}
\sigma^{\sharp}=\frac{\sigma L^{2}}{\kappa_{B}},\label{eq:e224}
\end{equation}
with the use of the thermal diffusion time at the bottom $L^{2}/\kappa_{B}=L^{2}c_{p}\rho_{B}/k$.
The non-dimensional form of the full $\hat{u}_{z}$-equation (\ref{eq:u_z_equation_full_A}),
which constitutes the eigen problem for $\hat{u}_{z}$ and $\sigma^{\sharp}$
takes the form
\begin{eqnarray}
 &  & -\frac{\sigma^{\sharp2}}{Pr}\left(1-\theta z^{\sharp}\right)^{m+1}\mathfrak{D}^{\sharp2}\hat{u}_{z}\nonumber \\
 &  & +\sigma^{\sharp}\left[\left(1-\theta z^{\sharp}\right)^{m+1}\mathfrak{D}^{\sharp2}\mathfrak{D}^{\sharp2}\hat{u}_{z}+\frac{1}{Pr}\nabla^{\sharp2}\left[\left(1-\theta z^{\sharp}\right)\mathfrak{D}^{\sharp2}\hat{u}_{z}\right]+\frac{\theta}{Pr}\frac{\mathrm{d}}{\mathrm{d}z^{\sharp}}\mathfrak{D}^{\sharp2}\hat{u}_{z}\right]\nonumber \\
 &  & -\sigma^{\sharp}\mathcal{K}^{\sharp2}\hat{u}_{z}\frac{m\theta^{2}}{1-\theta z^{\sharp}}\left[\frac{2}{3}\left(m+3\right)\left(1-\theta z^{\sharp}\right)^{m}-\frac{1}{Pr}\right]\nonumber \\
 &  & \qquad=\mathcal{K}^{\sharp2}Ra\left(1-\theta z^{\sharp}\right)^{m}\hat{u}_{z}+\nabla^{\sharp2}\left[\left(1-\theta z^{\sharp}\right)\mathfrak{D}^{\sharp2}\mathfrak{D}^{\sharp2}\hat{u}_{z}\right]+\theta\frac{\mathrm{d}}{\mathrm{d}z^{\sharp}}\mathfrak{D}^{\sharp2}\mathfrak{D}^{\sharp2}\hat{u}_{z}\nonumber \\
 &  & \qquad\;\;\,+m\theta^{2}\mathcal{K}^{\sharp2}\left[\frac{4}{3}\mathfrak{D}^{\sharp2}\left(\frac{\hat{u}_{z}}{1-\theta z^{\sharp}}\right)-\frac{1}{1-\theta z^{\sharp}}\mathfrak{D}^{\sharp2}\hat{u}_{z}\right]\nonumber \\
 &  & \qquad\;\;\,-\frac{2}{3}m\left(m+3\right)\theta^{2}\mathcal{K}^{\sharp2}\left\{ \nabla^{\sharp2}\left(\frac{\hat{u}_{z}}{1-\theta z^{\sharp}}\right)+\theta\frac{\mathrm{d}}{\mathrm{d}z^{\sharp}}\left[\frac{\hat{u}_{z}}{\left(1-\theta z^{\sharp}\right)^{2}}\right]\right\} \nonumber \\
 &  & \qquad\;\;\,-2m\theta^{3}\mathcal{K}^{\sharp2}\frac{1}{\left(1-\theta z^{\sharp}\right)^{2}}\left[\frac{\mathrm{d}\hat{u}_{z}}{\mathrm{d}z^{\sharp}}-\frac{m\theta}{1-\theta z^{\sharp}}\hat{u}_{z}\right],\label{eq:u_z_equation_full_A-2}
\end{eqnarray}
where the Prandtl number
\begin{equation}
Pr=\frac{\nu}{\kappa_{B}},\label{eq:e225}
\end{equation}
is also defined with the use of the thermal diffusivity at the bottom.
Exactly at convection threshold the growth rate is zero,
therefore we introduce the parameter
\begin{equation}
\eta=\frac{Ra-Ra_{crit}}{Ra_{crit}}\ll1,\label{eq:e226}
\end{equation}
which measures the departure from critical state, where $Ra_{crit}=Ra_{0}+\theta Ra_{1}$
is defined in (\ref{eq:Ra_crit_final_A}). In other words, the state
we have in mind at present, is a flow, developed via the convective
instability at a Rayleigh number slightly above the threshold value
$Ra=Ra_{crit}(1+\eta)$. The eigen mode in such a state in general
differs from the threshold solution given in (\ref{eq:u0z_A_final})
and (\ref{eq:u1z_A_final}), hence we introduce the following double
asymptotic expansion in small parameters $\theta$ and $\eta$
\begin{equation}
\hat{u}_{z}=\hat{u}_{0z}+\theta\hat{u}_{1z}+\eta\hat{v}_{0z}+\eta\theta\hat{v}_{1z}+\mathcal{O}\left(\theta^{2},\,\eta^{2}\right),\label{eq:double_expansion_uz_A}
\end{equation}
where the corrections denoted by $\hat{v}_{0z}$ and $\hat{v}_{1z}$
result solely from the departure from threshold and are absent when
$Ra=Ra_{crit}$. Substituting the above formula (\ref{eq:double_expansion_uz_A})
into the eigen problem (\ref{eq:u_z_equation_full_A-2}) and balancing
the terms at the order $\theta^{0}$ (bearing in mind that $\sigma\sim\eta$) we obtain
\begin{align}
-\frac{\sigma_{0}^{\sharp2}}{Pr}\nabla^{\sharp2}\hat{u}_{0z}+\sigma_{0}^{\sharp}\frac{1+Pr}{Pr}\nabla^{\sharp4}\hat{u}_{0z}-\mathcal{K}^{\sharp2}Ra_{0}\eta\hat{u}_{0z}\nonumber \\
=\,\,\mathcal{K}^{\sharp2}Ra_{0}\hat{v}_{0z} & +\left(-\mathcal{K}^{\sharp2}+\frac{\mathrm{d}^{2}}{\mathrm{d}z^{\sharp2}}\right)^{3}\hat{v}_{0z},\label{eq:u0z_eq_sigma_0}
\end{align}
and hence
\begin{align}
\left[\frac{\sigma_{0}^{\sharp2}}{Pr}\left(\mathcal{K}^{\sharp2}+\pi^{2}\right)+\sigma_{0}^{\sharp}\frac{1+Pr}{Pr}\left(\mathcal{K}^{\sharp2}+\pi^{2}\right)^{2}-\mathcal{K}^{\sharp2}Ra_{0}\eta\right]A\sin\pi z^{\sharp}\qquad\qquad\qquad\nonumber \\
=\,\mathcal{K}^{\sharp2}Ra_{0}\hat{v}_{0z}+\left(-\mathcal{K}^{\sharp2}+\frac{\mathrm{d}^{2}}{\mathrm{d}z^{\sharp2}}\right)^{3}\hat{v}_{0z},\label{eq:u0z_eq_sigma_0-1}
\end{align}
where we have used the asymptotic form of the growth rate
\begin{equation}
\sigma^{\sharp}=\sigma_{0}^{\sharp}+\theta\sigma_{1}^{\sharp}+\mathcal{O}\left(\eta\theta^{2}\right).\label{eq:e227}
\end{equation}
The solvability condition for this equation (cf. Korn \& Korn 1961)
yields the same solution for the leading order term in the growth
rate expansion $\sigma_{0}^{\sharp}$ as that from the Boussinesq
case (\ref{eq:MUM_grate_B}). Thus the maximal growth rate is achieved
at $\mathcal{K}^{\sharp}=\mathcal{K}_{0}^{\sharp}=\pi/\sqrt{2}$ and
is of the order $\mathcal{O}(\eta)$.\footnote{Note, that this means, that the entire left hand side of the equation
(\ref{eq:u0z_eq_sigma_0-1}) vanishes, and therefore the solution
for the order $\eta$ correction to vertical velocity $\hat{v}_{0z}$
simply reproduces the leading order term $\hat{u}_{0z}$; consequently $\hat{v}_{0z}$
can be assumed zero without loss of generality.} 

We now proceed to the next order and gather all the terms of the order
$\theta^{1}$ in the equation (\ref{eq:u_z_equation_full_A-2}), which
yields
\begin{align}
 & -\frac{\sigma_{0}^{\sharp2}}{Pr}\nabla^{\sharp2}\hat{u}_{1z}-2\frac{\sigma_{0}^{\sharp}\sigma_{1}^{\sharp}}{Pr}\nabla^{\sharp2}\hat{u}_{0z}+m\frac{\sigma_{0}^{\sharp2}}{Pr}\frac{\mathrm{d}\hat{u}_{0z}}{\mathrm{d}z^{\sharp}}+\left(m+1\right)\frac{\sigma_{0}^{\sharp2}}{Pr}z^{\sharp}\nabla^{\sharp2}\hat{u}_{0z}\nonumber \\
 & +\sigma_{1}^{\sharp}\frac{1+Pr}{Pr}\nabla^{\sharp4}\hat{u}_{0z}+\sigma_{0}^{\sharp}\frac{1+Pr}{Pr}\nabla^{\sharp4}\hat{u}_{1z}-\sigma_{0}^{\sharp}\left(m+\frac{1+Pr}{Pr}\right)z^{\sharp}\nabla^{\sharp4}\hat{u}_{0z}\nonumber \\
 & -\left(2m+\frac{m+1}{Pr}\right)\sigma_{0}^{\sharp}\frac{\mathrm{d}}{\mathrm{d}z^{\sharp}}\nabla^{\sharp2}\hat{u}_{0z}\nonumber \\
 & -\mathcal{K}_{0}^{\sharp2}Ra_{1}\eta\hat{u}_{0z}-\mathcal{K}_{0}^{\sharp2}Ra_{0}\eta\left(\hat{u}_{1z}-mz^{\sharp}\hat{u}_{0z}\right)\nonumber \\
 & \qquad\qquad\qquad\qquad\qquad\qquad=\,\,\mathcal{K}_{0}^{\sharp2}Ra_{0}\hat{v}_{1z}+\left(-\mathcal{K}_{0}^{\sharp2}+\frac{\mathrm{d}^{2}}{\mathrm{d}z^{\sharp2}}\right)^{3}\hat{v}_{1z}.\label{eq:u_z_equation_order1_grate_A}
\end{align}
It can be anticipated, that just as the leading order term $\sigma_{0}^{\sharp}\sim\eta$
also the correction $\sigma_{1}$ is proportional to $\eta$, and
therefore all the terms proportional to $\sigma_{0}^{2}$ and $\sigma_{0}\sigma_{1}$
are of the order $\mathcal{O}(\eta^{2})$. Neglecting all the terms
$\mathcal{O}(\eta^{2})$ in the equation (\ref{eq:u_z_equation_order1_grate_A})
and leaving only terms of the order $\mathcal{O}(\eta)$ leads to
\begin{align}
\mathcal{K}_{0}^{\sharp2}Ra_{0}\hat{v}_{1z} & +\left(-\mathcal{K}_{0}^{\sharp2}+\frac{\mathrm{d}^{2}}{\mathrm{d}z^{\sharp2}}\right)^{3}\hat{v}_{1z}\nonumber \\
 & =-\frac{9}{4}\pi^{4}\left[\frac{3}{4}\pi^{2}\left(m-1\right)\eta-\sigma_{1}^{\sharp}\frac{1+Pr}{Pr}\right]A\sin\pi z^{\sharp}\nonumber \\
 & \quad\;-\frac{27}{8}\pi^{6}\eta\left(1-\frac{m}{1+Pr}\right)Az^{\sharp}\sin\pi z^{\sharp}\nonumber \\
 & \quad\;+\frac{9}{4}\pi^{5}\eta\frac{Pr}{1+Pr}\left(2m+\frac{m+1}{Pr}\right)A\cos\pi z^{\sharp}\nonumber \\
 & \quad\;-\frac{27}{8}\pi^{6}\eta\hat{u}_{1z}+\frac{3}{2}\pi^{2}\eta\left(-\mathcal{K}_{0}^{\sharp2}+\frac{\mathrm{d}^{2}}{\mathrm{d}z^{\sharp2}}\right)^{2}\hat{u}_{1z}.\label{eq:uz1_sigma1_A}
\end{align}
The solvability condition for the above equation can now be applied.
Integration by parts of the terms in the last row of (\ref{eq:uz1_sigma1_A})
and application of boundary conditions yields
\begin{equation}
\int_{0}^{1}\left[-\frac{27}{8}\pi^{6}\eta\hat{u}_{1z}+\frac{3}{2}\pi^{2}\eta\left(-\mathcal{K}_{0}^{\sharp2}+\frac{\mathrm{d}^{2}}{\mathrm{d}z^{\sharp2}}\right)^{2}\hat{u}_{1z}\right]\sin\pi z^{\sharp}\mathrm{d}z^{\sharp}=0,\label{eq:e228}
\end{equation}
hence by the use of
\[
\int_{0}^{1}\sin^{2}\pi z^{\sharp}\mathrm{d}z^{\sharp}=1/2,\quad\int_{0}^{1}z^{\sharp}\sin^{2}\pi z^{\sharp}\mathrm{d}z^{\sharp}=1/4,\quad\int_{0}^{1}\cos\pi z^{\sharp}\sin\pi z^{\sharp}\mathrm{d}z^{\sharp}=0
\]
one obtains the following equation for the correction to the growth
rate $\sigma_{1}$
\begin{equation}
\frac{3}{4}\pi^{2}\left(m-1\right)\eta-\sigma_{1}^{\sharp}\frac{1+Pr}{Pr}+\frac{3}{4}\pi^{2}\eta\left(1-\frac{m}{1+Pr}\right)=0.\label{eq:e229}
\end{equation}
The solution of the latter equation
\begin{equation}
\sigma_{1}^{\sharp}=\frac{3}{4}\pi^{2}m\left(\frac{Pr}{1+Pr}\right)^{2}\eta>0\label{eq:e230}
\end{equation}
is positive, thus the the convective instability in the anelastic
case with $\nu$ and $k$ constant is enhanced by the compressibility
of the fluid, since its growth rate\index{SI}{growth rate}
\begin{equation}
\sigma^{\sharp}=\frac{3}{2}\pi^{2}\frac{Pr}{1+Pr}\eta\left(1+\frac{1}{2}m\theta\frac{Pr}{1+Pr}\right),\label{eq:e231}
\end{equation}
is greater than the Boussinesq one. However, as in the case of critical
Rayleigh number, any generalization to other types of anelastic systems
should not be carried out without careful consideration of the full
equations; as we remarked above, a change in the physical properties
of the fluid, such as e.g. a change from constant $k$ to constant
$\kappa$ leads to an important change in the physical response of
the system to convective driving.

Finally, it is of interest to note, that since the solution slightly
above convection threshold differs from the marginal one (\ref{eq:u0z_A_final}),
(\ref{eq:u1z_A_final}) and contains corrections $\eta(\hat{v}_{0z}+\theta\hat{v}_{1z})$,
the convective flow in the exact form of the marginal solution at
$Ra=Ra_{crit}$, given in (\ref{eq:u0z_A_final}-\ref{eq:C4_A}),
might not be observable experimentally, since exactly at threshold
the amplitude of convection $A$ vanishes; to observe convective flow
one needs to exceed the threshold value $Ra_{crit}$ at least slightly,
so that $0<\eta=(Ra-Ra_{crit})/Ra_{crit}\ll1$, when the corrections
$\eta(\hat{v}_{0z}+\theta\hat{v}_{1z})$ already appear.

\section{Fully developed, stratified convection\label{sec:Fully-developed-convection-A}}

This section recalls the theory of turbulent stratified convection
of Jones \emph{et al.} (2020), with some amendments to present more
detailed clarifications. To make the description of the fully developed,
turbulent, stratified convection as clear as possible simplifications
need to be introduced. Therefore we assume the equation of state of
an ideal gas, uniform dynamic viscosity $\mu$ and uniform thermal
conductivity $k$, negligible bulk viscosity $\mu_{b}$ and internal
heating $Q=0$ and constant gravity $\mathbf{g}=-g\hat{\mathbf{e}}_{z}$, $g=\mathrm{const}$.
The equations of motion are expressed in the following way\index{SI}{anelastic!equations, perfect gas}
\begin{subequations}
\begin{equation}
\frac{\partial\mathbf{u}}{\partial t}+\left(\mathbf{u}\cdot\nabla\right)\mathbf{u}=-\nabla\frac{p'}{\tilde{\rho}}+\frac{s'}{c_{p}}g\hat{\mathbf{e}}_{z}+\frac{\mu}{\tilde{\rho}}\nabla^{2}\mathbf{u}+\frac{\mu}{3\tilde{\rho}}\nabla\left(\nabla\cdot\mathbf{u}\right)\label{eq:NS_Anonlin}
\end{equation}
\begin{equation}
\nabla\cdot\left(\tilde{\rho}\mathbf{u}\right)=0,\label{Cont_Anonlin}
\end{equation}
\begin{equation}
\tilde{\rho}\tilde{T}\left[\frac{\partial s'}{\partial t}+\mathbf{u}\cdot\nabla\left(\tilde{s}+s'\right)\right]=\nabla\cdot\left(k\nabla T\right)+2\mu\mathbf{G}^{s}:\mathbf{G}^{s}-\frac{2}{3}\mu\left(\nabla\cdot\mathbf{u}\right)^{2},\label{Energy_Anonlin}
\end{equation}
\begin{equation}
\frac{\rho'}{\tilde{\rho}}=-\frac{T'}{\tilde{T}}+\frac{p'}{\tilde{p}},\qquad s'=-R\frac{p'}{\tilde{p}}+c_{p}\frac{T'}{\tilde{T}},\label{State_eq_Anonlin}
\end{equation}
\end{subequations} 
where, as remarked, we have assumed
\begin{equation}
\mu=\mathrm{const},\quad k=\mathrm{const},\quad g=\mathrm{const},\quad\mu_{b}=0,\quad Q=0,\label{eq:uniform_material_props}
\end{equation}
\begin{equation}
p=\rho RT,\qquad s=c_{v}\ln\frac{p}{\rho^{\gamma}}.\label{eq:IG_state_eq}
\end{equation}

\subsection{The hydrostatic conduction reference state and the hydrostatic adiabatic
state}

The hydrostatic reference state determined by vertical conductive
heat transport and hydrostatic force balance is recalled here (cf.
(\ref{eq:hydrostatic_eq_A_pg}) and (\ref{eq:BS1}-c)) 
\begin{subequations}
\begin{equation}
\tilde{T}=\tilde{T}_{B}\left(1-\tilde{\theta}\frac{z}{L}\right),\qquad\tilde{\rho}=\tilde{\rho}_{B}\left(1-\tilde{\theta}\frac{z}{L}\right)^{m},\qquad\tilde{p}=\frac{gL\tilde{\rho}_{B}}{\tilde{\theta}\left(m+1\right)}\left(1-\tilde{\theta}\frac{z}{L}\right)^{m+1},\label{eq:BS1-1}
\end{equation}
\begin{equation}
\tilde{s}=c_{p}\frac{m+1-\gamma m}{\gamma}\ln\left(1-\tilde{\theta}\frac{z}{L}\right)+\mathrm{const},\label{eq:BS2-2}
\end{equation}
\end{subequations} 
where 
\begin{equation}
m=\frac{gL}{R\Delta\tilde T}-1,\qquad\tilde{\theta}=\frac{\Delta\tilde{T}}{\tilde{T}_{B}},\qquad\Delta\tilde{T}=\tilde{T}_{B}-\tilde{T}_{T}.\label{eq:m_and_theta}
\end{equation}
For the purpose of this section we make a clear distinction between
the top and bottom values of the temperature and density in the static reference
state and those of a convective state. The reason for doing so is
that two cases will be considered, that of isothermal boundaries,
when $\tilde{T}_{B}=T_{B}$, $\tilde{T}_{T}=T_{T}$ and that of isentropic
boundaries, when $\tilde{T}_{B}\neq T_{B}$, $\tilde{T}_{T}\neq T_{T}$.
In the latter case the total temperature at boundaries differs from
the reference state boundary values by terms of order $\delta$, i.e.
$T'(\mathbf{x}_{h},z=0,t)$ at the bottom and $T'(\mathbf{x}_{h},z=L,t)$
at the top.

We also recall the hydrostatic adiabatic state 
\begin{subequations}
\begin{align}
T_{ad} & =T_{ad\,B}\left(1-\frac{gz}{c_{p}T_{ad\,B}}\right),\quad\rho_{ad}=\rho_{ad\,B}\left(1-\frac{gz}{c_{p}T_{ad\,B}}\right)^{\frac{1}{\gamma-1}},\label{eq:adiabatic_profile_RS-1}\\
p_{ad} & =\rho_{ad\,B}RT_{ad\,B}\left(1-\frac{gz}{c_{p}T_{ad\,B}}\right)^{\frac{\gamma}{\gamma-1}},\quad s_{ad}=\mathrm{const}.\label{eq:adiabatic_profile_RS_1-1}
\end{align}
\end{subequations} 
Most of the time we will assume, that $T_{ad\,B}=\tilde{T}_{B}$,
whereas $\rho_{ad\,B}$ is related to $\tilde{\rho}_{B}$ through
(\ref{eq:rho_rel}) by the condition, that the total fluid mass is
contained in the hydrostatic reference state, either conduction state
or the adiabatic one. Furthermore we define the ratios of the bottom
to top temperature values
\begin{equation}
\tilde{\Gamma}=\frac{\tilde{T}_{B}}{\tilde{T}_{T}}=\frac{1}{1-\tilde{\theta}}>1,\qquad\Gamma_{ad}=\frac{T_{ad\,B}}{T_{ad\,T}}=\frac{1}{1-\frac{gL}{c_{p}T_{ad\,B}}}>1.\label{eq:gammas_def}
\end{equation}
On defining an alternative measure of small departure from adiabaticity
(cf. (\ref{eq:delta_intermsof_epsan}))
\begin{equation}
\epsilon_{a}=\frac{L}{\tilde{T}_{B}}\left(\frac{\Delta\tilde{T}}{L}-\frac{g}{c_{p}}\right)=\delta\frac{\tilde{\Gamma}-1}{\tilde{\Gamma}\ln\tilde{\Gamma}}>0,\qquad\epsilon_{a}\ll1,\label{eq:eps_a_def}
\end{equation}
where the subscript $a$ stands for '\emph{anelastic}' (to distinguish
this new parameter from the Boussinesq small parameter $\epsilon$
defined in (\ref{eq:epsilon_def_B})), for the case of 
\begin{equation}
T_{ad\,B}=\tilde{T}_{B}\label{eq:TBad_equals_TBref}
\end{equation}
 we get the following relations 
\begin{equation}
\tilde{\theta}=\frac{gL}{c_{p}\tilde{T}_{B}}+\epsilon_{a},\qquad\tilde{\Gamma}=\Gamma_{ad}+\epsilon_{a}\tilde{\Gamma}\Gamma_{ad}=\Gamma_{ad}+\epsilon_{a}\Gamma_{ad}^{2}+\mathcal{O}\left(\Gamma_{ad}^{3}\epsilon_{a}^{2}\right).\label{eq:theta_Gamma_rels}
\end{equation}
Finally we define the superadiabatic temperature fluctuation,
say $T_{S}(\mathbf{x},t)$, as a component of the total temperature
in the convective state $T(\mathbf{x},t)$, thus as a correction to
hydrostatic adiabatic state so that 
\begin{equation}
T\left(\mathbf{x},t\right)=\tilde{T}(z)+T'\left(\mathbf{x},t\right)=T_{ad}\left(z\right)+T_{S}\left(\mathbf{x},t\right).\label{eq:Tprime_Ts_def}
\end{equation}
By the use of the definitions of $\tilde{T}(z)$ and $T_{ad}\left(z\right)$
in (\ref{eq:BS1-1}) and (\ref{eq:adiabatic_profile_RS-1}) with $T_{ad\,B}=\tilde{T}_{B}$,
the superadiabatic temperature fluctuation $T_S(\mathbf{x},t)$ is related to the temperature fluctuation about the conduction reference
state $T'(\mathbf{x},t)$ in the following way
\begin{equation}
T'\left(\mathbf{x},t\right)=T_{S}\left(\mathbf{x},t\right)+\epsilon_{a}\frac{\tilde{T}_{B}}{L}z.\label{eq:Tprime_Ts_rel}
\end{equation}
Therefore the fluctuation $T'(\mathbf{x},t)$ is simply shifted with
respect to the superadiabatic temperature fluctuation by the superadiabatic
linear profile of the conduction reference state; of course both fluctuations
are of the order $\mathcal{O}(\epsilon_{a})=\mathcal{O}(\delta)$
by the second fundamental assumption of the anelastic approximation
(\ref{fluct_magnitude_A}).

\subsection{Relation between entropy and temperature jumps across the boundary
layers}

One of the major differences between compressible and Boussinesq convection
is that the temperature fluctuation and the entropy fluctuation are
not equivalent in the compressible case, contrary to the Boussinesq case (up to a constant,
cf. (\ref{eq:entropy_temperature_in_Boussinesq})). In the compressible
case the bulk of turbulent convection, where the fluid is efficiently
mixed by a vigorous flow, is characterized by an almost uniform mean
total entropy $\left\langle s\right\rangle _{h}$, as can be anticipated
from the energy equation (\ref{Energy_Anonlin}) which suggests efficient
advection of the mean entropy (cf. mean entropy profiles on figures
\ref{fig:DevAn_isothermal_profiles} and \ref{fig:DevAn_isentropic_profiles}
sketched for the cases of isothermal and isentropic boundaries).\footnote{\label{fn:fluct_over_mean_Anelastic}In a well-mixed bulk the fluctuations
about the horizontal means are expected to be small. Since the impermeability
of boundaries together with the continuity equation $\nabla\cdot(\tilde{\rho}\mathbf{u})=0$
imply $\left\langle u_{z}\right\rangle _{h}=0$, the horizontal average
of the stationary energy equation (\ref{Energy_Anonlin}) leaves a
dominant balance between the mean conduction and mean viscous heating;
in turn, by the use of the full equation, this leads to $u_{z}\partial_{z}\left\langle s\right\rangle _{h}\approx0$
in the well-mixed bulk of turbulent convection.} On the other hand the mean temperature, according to the equation
(\ref{Energy_eq1-1}) or (\ref{Energy_eq1-1-1}) for the considered
case of a perfect gas, is advected along with the mean pressure and
none of these two quantities alone needs to be homogenized in the
bulk (only the sum $-R\left\langle p'\right\rangle _{h}/\tilde{p}+c_{p}\left\langle T'\right\rangle _{h}/\tilde{T}+\tilde{s}=\left\langle s\right\rangle _{h}$,
which is the total mean entropy, as is clear from (\ref{State_eq_Anonlin})).
At the top and bottom of the fluid domain the boundary layers are
formed to adjust the bulk top and bottom values of the entropy and
temperature to their values at boundaries. However, these boundary
layers are not symmetric with respect to the mid plane, contrary to the symmetric Boussinesq case, and typically
the jumps of the total mean entropy and mean temperature fluctuation
across the top boundary layer, denoted by $\left(\Delta s\right)_{T}$
and $\left(\Delta T'\right)_{T}$ are considerably larger than the
relative jumps across the bottom boundary layer, $\left(\Delta s\right)_{B}$
and $\left(\Delta T'\right)_{B}$.\footnote{The jumps are positive by definition, hence e.g. the entropy jump
across the top boundary layer is defined as $(\Delta s)_{T}=\left\langle s\right\rangle _{h}(z=L-\delta_{th,T})-\left\langle s\right\rangle _{h}(z=L)$.} Note, that in the case of entropy both, the variations of the fluctuation $s'$ and
of the reference state entropy $\tilde{s}$ are of the order $\mathcal{O}(\epsilon_{a})$,
thus in developed convection the jump of $\tilde{s}$ across thin
boundary layers is negligible with respect to the jumps of $\left\langle s'\right\rangle _{h}$
across the layers. On the contrary, for the other thermodynamic variables,
such as temperature, pressure and density the reference state variables
are $\mathcal{O}(\epsilon_{a}^{-1})$ times stronger than the fluctuations
and thus jumps in the values of $\tilde{T}$, $\tilde{p}$ and $\tilde{\rho}$
across the boundary layers may significantly exceed the jumps of
the corresponding fluctuations.

Before we proceed to providing a dynamical picture of developed
compressible convection we note that in some cases it is also useful
to express the temperature jumps across the thermal boundary layers
by the entropy jumps. From the state equations (\ref{State_eq_Anonlin})
we obtain 
\begin{subequations}
\begin{align}
\frac{\left(\bigtriangleup\rho'\right)_{i}}{\tilde{\rho}_{i}} & =\frac{1}{\gamma-1}\left[\frac{\left(\bigtriangleup T'\right)_{i}}{\tilde{T}_{i}}-\gamma\frac{\left(\bigtriangleup s\right)_{i}}{c_{p}}\right]+\mathcal{O}\left(\epsilon_{a}^{2}\right),\label{eq:density_jump}\\
\frac{\left(\bigtriangleup p'\right)_{i}}{\tilde{p}_{i}} & =\frac{\gamma}{\gamma-1}\left[\frac{\left(\bigtriangleup T'\right)_{i}}{\tilde{T}_{i}}-\frac{\left(\bigtriangleup s\right)_{i}}{c_{p}}\right]+\mathcal{O}\left(\epsilon_{a}^{2}\right),\label{eq:pressure_jump}
\end{align}
\end{subequations} 
where the subscript $i$ stands either for $T$
or $B$, and similarly as for the entropy and temperature we denote
the jumps of the mean density and pressure fluctuations across the
top and bottom boundary layers by $\left(\bigtriangleup\rho'\right)_{T}$,
$\left(\bigtriangleup p'\right)_{T}$ and $\left(\bigtriangleup\rho'\right)_{B}$,
$\left(\bigtriangleup p'\right)_{B}$, respectively. Next we observe,
that in the boundary layers the mass conservation constraint $\nabla\cdot(\tilde{\rho}\mathbf{u})=0$
allows for the following estimate
\begin{equation}
u_{z,i}\sim\frac{\delta_{\nu,i}}{L},\label{eq:vertical_velocity_BLs}
\end{equation}
where $\delta_{\nu,i}$ denotes the thickness of either the top ($i=T$)
or the bottom ($i=B$) viscous boundary layer and $u_{z,i}$ the vertical
velocity in each of the layers. Consequently the $z$-component of
the nonlinear inertial term in the Navier-Stokes equation in a boundary
layer, i.e. $\mathbf{u}_{i}\cdot\nabla u_{z,i}$, is small, since
in fully developed turbulent convection the boundary layers are thin,
thus\begin{subequations}
\begin{equation}
\delta_{\nu,T}\ll L,\qquad\delta_{\nu,B}\ll L.\label{eq:deltas_small}
\end{equation}
\begin{equation}
\delta_{th,T}\ll L,\qquad\delta_{th,B}\ll L,\label{eq:deltas_small-1}
\end{equation}
\end{subequations}
where $\delta_{th,T}$ and $\delta_{th,B}$ denote
the thicknesses of the top and bottom thermal boundary layers respectively.
Therefore the balance on the $z$-component of the Navier-Stokes equation
(\ref{eq:NS_Anonlin}) in the boundary layers, occurs predominantly
between the pressure gradient and the buoyancy force, suggesting
\begin{equation}
\left(\bigtriangleup p'\right)_{i}\approx gL\tilde{\rho}_{i}\frac{\left(\bigtriangleup s\right)_{i}}{c_{p}}\frac{\delta_{th,i}}{L},\label{eq:Dp_BL}
\end{equation}
which implies, that the pressure jump across the thermal boundary
layers is rather small. Inserting (\ref{eq:Dp_BL}) into (\ref{eq:pressure_jump})
leads to
\begin{equation}
\frac{\left(\bigtriangleup T'\right)_{i}}{\tilde{T}_{i}}\approx\frac{\left(\bigtriangleup s\right)_{i}}{c_{p}}\left(1+\tilde{\theta}\frac{\delta_{th,i}}{L}\frac{\tilde{T}_{B}}{\tilde{T}_{i}}\right)+\mathcal{O}\left(\epsilon_{a}^{2}\frac{\delta_{th,i}}{L}\right).\label{eq:DT_rel_Ds_full}
\end{equation}
Typically the term $(\tilde{\theta}\delta_{th,i}\tilde{T}_{B})/(L\tilde{T}_{i})$
resulting from the pressure jump across the boundary layers is expected
to be small, however, it is not necessarily the case always. In fact,
as we will see in section \ref{subsec:estimates}, the thicknesses
of the thermal boundary layers depend in a complicated way on the
density scale height $D_{\rho}=-\tilde{\rho}/\mathrm{d}_{z}\tilde{\rho}$,
and can increase when the scale height decreases and the layer
becomes more strongly stratified. Consequently in strongly stratified
cases the boundary layers themselves may contain a few density scale
heights, i.e. we may have $D_{\rho}<\delta_{th,i}$, and thus the
boundary layers are not necessarily incompressible. Moreover, it is
clear, that especially the top boundary layer is more prone to become
compressible as the stratification $\tilde{\theta}$ is increased
(the scale heights decreased), as suggested by appearance of the factor
$\tilde{\Gamma}=\tilde{T}_{B}/\tilde{T}_{T}>1$ in the pressure-jump
correction at the top boundary layer in (\ref{eq:DT_rel_Ds_full})
(that is in the second term inside the brackets of that equation);
in strongly stratified systems this factor makes the top correction
significantly greater than the bottom one. 

Nevertheless, for clarity of the presentation it is of interest to
consider the simplest case when both the boundary layers are incompressible
and then it is allowed to use an approximate relation between the
jumps of the mean temperature fluctuation and of the mean total entropy
across the boundary layers,
\begin{equation}
\frac{\left(\bigtriangleup T'\right)_{i}}{\tilde{T}_{i}}\approx\frac{\left(\bigtriangleup s\right)_{i}}{c_{p}}.\label{eq:DT_rel_Ds}
\end{equation}
The other case, when the pressure-jump correction $(\tilde{\theta}\delta_{th,i}\tilde{T}_{B})/(L\tilde{T}_{i})$
in the equation (\ref{eq:DT_rel_Ds_full}) matters will be referred to
as the compressible boundary layer case.

\subsection{``Subadiabatic'' temperature gradient in the bulk\label{subsec:Subadiabatic-temperature-gradient}}

An important feature of the turbulent, (statistically) stationary
anelastic convection is that the magnitude of the mean temperature
gradient in the bulk of convection is weaker than $g/c_{p}$, i.e.
than that of the hydrostatic adiabatic state. This is because the
bulk is non-static, that is the mean force balance on the vertical direction in the
bulk is non-hydrostatic due to the influence of inertia. In other
words the vertical equilibrium of mean forces, obtained from horizontally
averaging the $z$-component of the Navier-Stokes equation (\ref{NS-Aderiv-1-1})
with the reference state balance $\mathrm{d}_{z}\tilde{p}=-\tilde{\rho}g$
incorporated back into it, reads
\begin{equation}
\frac{\mathrm{d}}{\mathrm{d}z}\left\langle \tilde{\rho}u_{z}^{2}\right\rangle _{h}+\frac{\mathrm{d}\left\langle p\right\rangle _{h}}{\mathrm{d}z}=-g\left\langle \rho\right\rangle _{h}.\label{eq:non-hydrostatic-balance-bulk}
\end{equation}
In obtaining the above equation we have used the fact, that $\left\langle u_{z}\right\rangle _{h}=0$
(cf. (\ref{eq:mean_h_uz_null_A})). As argued, the mean total entropy
is uniformly distributed by vigorous advection and thus almost constant
in the bulk of turbulent convection at high Rayleigh number. Therefore
we can assume that the bulk is approximately adiabatic, i.e. 
\begin{equation}
\left\langle s\right\rangle _{h}=\tilde{s}+\left\langle s'\right\rangle _{h}=\mathrm{const}\quad\textrm{in the bulk}.\label{eq:s_const_in_bulk}
\end{equation}
Consequently for the bulk of convection we may expect that the adiabatic
relation between mean pressure and density
\begin{equation}
\left\langle p\right\rangle _{h}=\mathrm{const}\left\langle \rho\right\rangle _{h}^{\gamma},\label{eq:p_rho_adiab}
\end{equation}
is satisfied. Moreover, the fact that the bulk is efficiently mixed
and the fluctuations about the means are assumed negligible implies
that approximately we may also expect 
\begin{equation}
\left\langle p\right\rangle _{h}=\left\langle \rho\right\rangle _{h}R\left\langle T\right\rangle _{h}.\label{eq:state_eq_bulk}
\end{equation}
Elimination of the mean total pressure $\left\langle p\right\rangle _{h}$
and density $\left\langle \rho\right\rangle _{h}$ form the equations
(\ref{eq:non-hydrostatic-balance-bulk}), (\ref{eq:p_rho_adiab})
and (\ref{eq:state_eq_bulk}) yields
\begin{equation}
\frac{d\left\langle T\right\rangle _{h}}{dz}=-\frac{g}{c_{p}}-\frac{1}{c_{p}\left\langle \rho\right\rangle _{h}}\frac{\mathrm{d}}{\mathrm{d}z}\left\langle \tilde{\rho}u_{z}^{2}\right\rangle _{h},\label{eq:temp_balance_bulk}
\end{equation}
where $\left\langle \rho\right\rangle _{h}$ has been retained in
the last term; this term, however, is of the order $\mathcal{O}(\epsilon_{a})$
(or equivalently $\mathcal{O}(\delta)$, cf (\ref{eq:vel_and_time_scales-1})),
and therefore keeping the order of accuracy at the level of the magnitude
of fluctuations, i.e. at $\mathcal{O}(\epsilon_{a})$, the mean total
density $\left\langle \rho\right\rangle _{h}$ can be replaced by
$\tilde{\rho}$, so that
\begin{align}
\frac{d\left\langle T\right\rangle _{h}}{dz}= & -\frac{g}{c_{p}}-\frac{1}{c_{p}\tilde{\rho}}\frac{\mathrm{d}}{\mathrm{d}z}\left\langle \tilde{\rho}u_{z}^{2}\right\rangle _{h}+\mathcal{O}\left(\epsilon_{a}^{2}\frac{g}{c_{p}}\right)\nonumber \\
= & -\frac{g}{c_{p}}+\frac{\left\langle u_{z}^{2}\right\rangle _{h}}{c_{p}D_{\rho}}-\frac{1}{c_{p}}\frac{\mathrm{d}\left\langle u_{z}^{2}\right\rangle _{h}}{\mathrm{d}z}+\mathcal{O}\left(\epsilon_{a}^{2}\frac{g}{c_{p}}\right),\label{eq:temp_balance_bulk-1}
\end{align}
where we have used the density scale height $D_{\rho}=-\tilde{\rho}/\mathrm{d}_{z}\tilde{\rho}$.
Integration of the latter equality across the bulk, hence from $z=\delta_{th,B}$
to $z=L-\delta_{th,T}$ allows to calculate the jump of the mean total
temperature across the bulk
\begin{align}
-\left(\Delta T\right)_{bulk}= & \left\langle T\right\rangle _{h}\left(z=L-\delta_{th,T}\right)-\left\langle T\right\rangle _{h}\left(z=\delta_{th,B}\right)\nonumber \\
= & -\frac{gL}{c_{p}}+\frac{g}{c_{p}}\left(\delta_{th,B}+\delta_{th,T}\right)+\int_{0}^{L}\frac{\left\langle u_{z}^{2}\right\rangle _{h}}{c_{p}D_{\rho}}\mathrm{d}z\nonumber \\
 & +\mathcal{O}\left(\epsilon_{a}\delta_{\nu,B}^{2}\frac{g}{c_{p}L}\right)+\mathcal{O}\left(\epsilon_{a}\delta_{\nu,T}^{2}\frac{g}{c_{p}L}\right)+\mathcal{O}\left(\epsilon_{a}^{2}\frac{gL}{c_{p}}\right).\label{eq:Temp_jump_bulk}
\end{align}
To obtain the above relation we have utilized the observation about
the magnitude of the vertical velocity in boundary layers made in
(\ref{eq:vertical_velocity_BLs}), which implies that at the top and
bottom of the bulk we have
\begin{equation}
\left\langle u_{z}^{2}\right\rangle _{h}\left(z=L-\delta_{th,T}\right)=\mathcal{O}\left(\epsilon_{a}\delta_{\nu,T}^{2}\right),\quad\left\langle u_{z}^{2}\right\rangle _{h}\left(z=\delta_{th,B}\right)=\mathcal{O}\left(\epsilon_{a}\delta_{\nu,B}^{2}\right).\label{eq:uz_squared_scale_TB}
\end{equation}
The term resulting from integration of the hydrostatic adiabatic gradient
across the boundary layers, i.e. $g(\delta_{th,B}+\delta_{th,T})/c_{p}$
must be retained, since it is of the order of non-dimensional thicknesses
of the boundary layers, $\delta_{th,B}/L$ and $\delta_{th,T}/L$,
which although small still have to be much greater than the small
anelastic parameter $\epsilon_{a}$ whose smallness guarantees validity
of the anelastic system of equations. The $\mathcal{O}(\epsilon_{a})$
correction to the hydrostatic adiabatic temperature jump across
the bulk in (\ref{eq:Temp_jump_bulk}) is also retained, because it
is of the same order as the thermodynamic fluctuations; it is denoted
by
\begin{equation}
\left(\Delta T\right)_{vel}=\int_{0}^{L}\frac{\left\langle u_{z}^{2}\right\rangle _{h}}{c_{p}D_{\rho}}\mathrm{d}z=\mathcal{O}\left(\epsilon_{a}\right)>0,\label{integral_positive}
\end{equation}
and is positive definite, hence convection reduces the magnitude of
the mean temperature jump across the bulk with respect to the magnitude
of the temperature jump between $z=\delta_{th,B}$ and $z=L-\delta_{th,T}$
in a hydrostatic adiabatic state, $gL/c_{p}+g(\delta_{th,B}+\delta_{th,T})/c_{p}$.
In other words the non-hydrostatic temperature gradient in the adiabatic
bulk is weaker than the hydrostatic adiabatic gradient $g/c_{p}$.

From the equation (\ref{eq:Temp_jump_bulk}) we can extract the jump
of the mean temperature fluctuation across the bulk
\begin{align}
0<\left(\Delta T'\right)_{bulk}= & \left\langle T'\right\rangle _{h}\left(z=L-\delta_{th,T}\right)-\left\langle T'\right\rangle _{h}\left(z=\delta_{th,B}\right)\nonumber \\
= & \epsilon_{a}\tilde{T}_{B}+\left(\Delta T\right)_{vel}\nonumber \\
 & +\mathcal{O}\left(\epsilon_{a}\delta_{th,B}\frac{g}{c_{p}}\right)+\mathcal{O}\left(\epsilon_{a}\delta_{th,T}\frac{g}{c_{p}}\right),\label{eq:Temp_jump_bulk-1}
\end{align}
where the term describing the differences between the temperature
jumps across the boundary layers for the hydrostatic reference and
adiabatic profiles
\begin{equation}
\epsilon_{a}\tilde{T}_{B}\frac{\delta_{th,B}+\delta_{th,T}}{L}\label{eq:rest_explicit}
\end{equation}
has been included in the remainders $\mathcal{O}(\epsilon_{a}\delta_{th,B}g/c_{p})$
and $\mathcal{O}(\epsilon_{a}\delta_{th,T}g/c_{p})$. The jump of
the mean temperature fluctuation is positive but defined in the opposite
way to the jump of the total mean temperature, that is the bottom
value is subtracted from the top value. This is because the mean
fluctuation $\left\langle T'\right\rangle _{h}(z)$ increases in the
bulk (cf. figures \ref{fig:DevAn_isothermal_profiles}b and \ref{fig:DevAn_isentropic_profiles}b).

\subsection{Vertical profiles of mean temperature and entropy}

In this section we schematically sketch the vertical profiles of the
total, horizontally averaged entropy and temperature and the horizontally
averaged temperature fluctuation. Justification of the important characteristics
of the profiles is provided. The reason for depicting separately the
total mean temperature and the mean temperature fluctuation is simply
clarity, because the total temperature is strongly dominated by the
contribution from the reference state, which by assumption is $\mathcal{O}(\epsilon_{a}^{-1})$
times greater than the fluctuation and hence details of the vertical
dependence of the total mean temperature are difficult to present;
at the same time, however, the general picture is also instructive.

\subsubsection{The case of isothermal boundaries\label{subsec:isothermal}}

\begin{figure}
\begin{centering}
a)\includegraphics[scale=0.17]{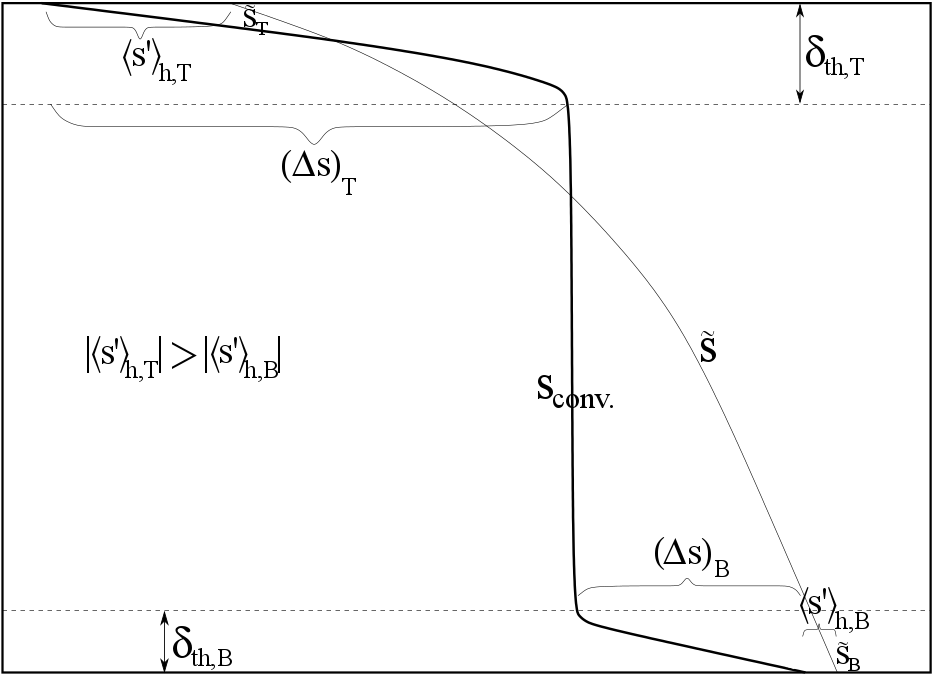}~~~b)\includegraphics[scale=0.17]{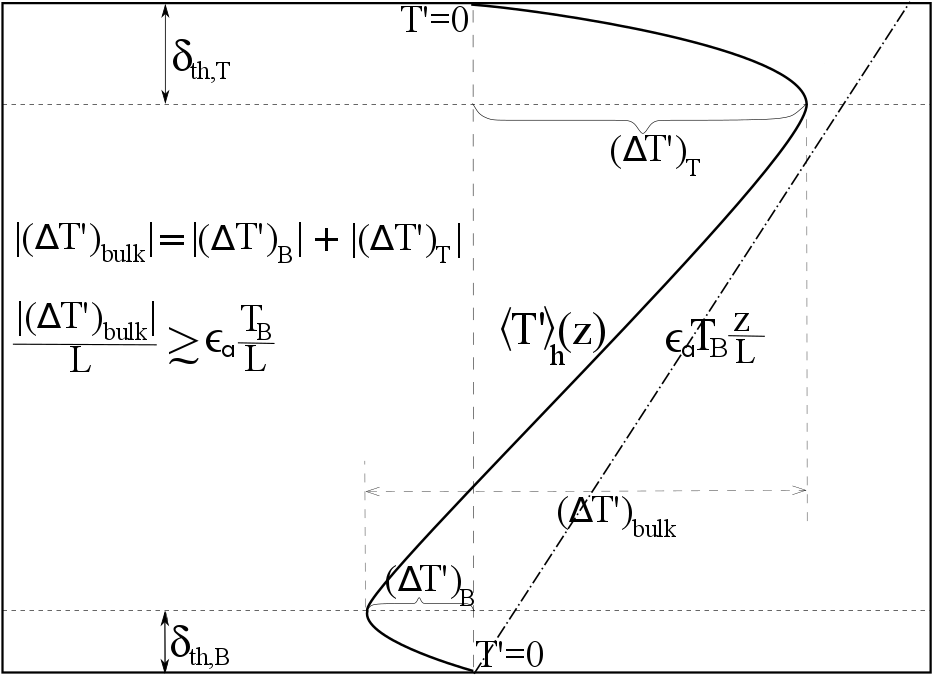}\vspace{3mm}
\par\end{centering}
\begin{centering}
c)\includegraphics[scale=0.17]{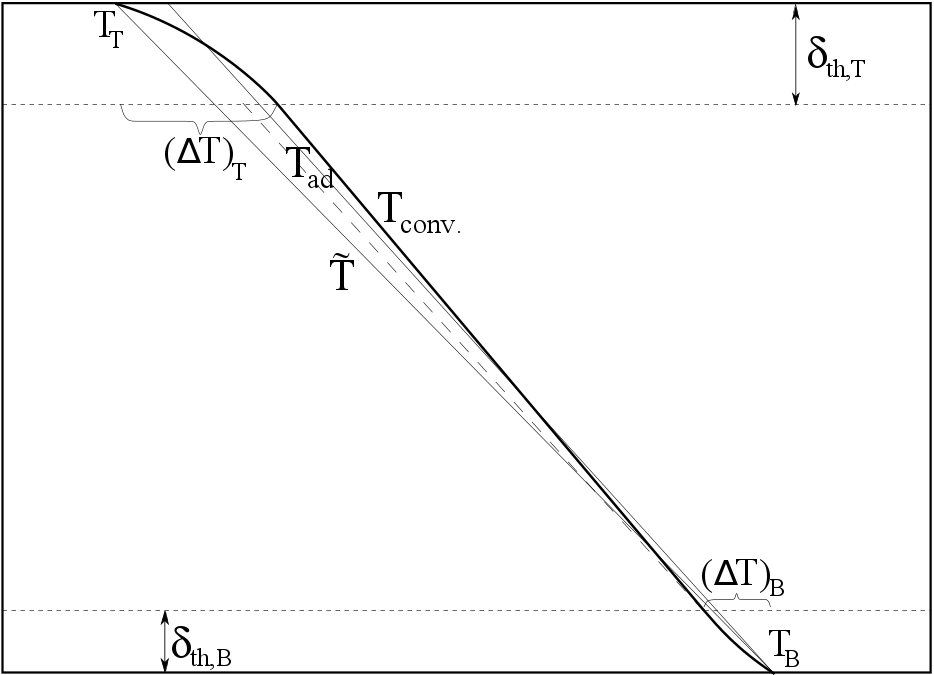}
\par\end{centering}
{\footnotesize{}\caption{{\footnotesize{}\label{fig:DevAn_isothermal_profiles}A schematic
picture of vertical profiles of the total entropy $s_{\mathrm{conv}}=\left\langle s\right\rangle _{h}=\tilde{s}+\left\langle s'\right\rangle _{h}$
a), the mean temperature fluctuation $\left\langle T'\right\rangle _{h}$
b), and the total temperature $T_{\mathrm{conv}}=\left\langle T\right\rangle _{h}=\tilde{T}+\left\langle T'\right\rangle _{h}$
c) (marked with bold lines) in developed convection with fixed temperature
at boundaries and $Q=0$. We stress, that the last figure c) is }\emph{\footnotesize{}not
in scale}{\footnotesize{}, and therefore may be somewhat misleading,
as the superadiabaticity $\delta$, i.e. the departure of the basic
profile gradient $\mathrm{d}_{z}\tilde{T}$ from the adiabatic one
had to be made significant for clarity of presentation; in particular
the jumps of the mean temperature fluctuation across the boundary
layers $(\Delta T')_{B}$ and $(\Delta T')_{T}$ are in general $\delta\ll1$
times smaller than corresponding jumps in the reference temperature
profile, i.e. $\tilde{T}(z=0)-\tilde{T}(\delta_{th,B})$ and $\tilde{T}(z=L-\delta_{th,T})-\tilde{T}(L)$
respectively. Moreover, the non-dimensional thicknesses of the boundary
layers, $\delta_{th,B}/L$ and $\delta_{th,T}/L$ must be much larger
than the superadiabaticity $\delta$ for consistency of the anelastic
approximation. The dashed line on figure c) represents the same gradient
as that of the hydrostatic adiabatic profile $T_{ad}=T_{B}-gz/c_{p}$,
but is shifted horizontally, so that it is hooked at the point $z=\delta_{B,th}$
and $T=\left\langle T\right\rangle _{h}(z=\delta_{th,B})$.}}
}{\footnotesize \par}
\end{figure}

We start by considering the case when the temperature at boundaries
is held fixed. The vertical profiles of $\left\langle s\right\rangle _{h}$,
$\left\langle T'\right\rangle _{h}$ and $\left\langle T\right\rangle _{h}$
are shown on figure \ref{fig:DevAn_isothermal_profiles} and the justification
of the negative sign of the entropy shifts at both boundaries with
respect to the reference state values $\tilde{s}_{B}$ and $\tilde{s}_{T}$
and their relative magnitudes is provided below. According to the
results of section \ref{subsec:Subadiabatic-temperature-gradient}
the gradient of the mean total temperature in the bulk on figure \ref{fig:DevAn_isothermal_profiles}c
is marked weaker than that of a hydrostatic adiabatic state. This
means, that the dashed line representing the horizontally shifted adiabatic
profile must intersect with the dashed horizontal line $z=L-\delta_{th,T}$
at a lower temperature than that of the mean total temperature profile
at the top of the bulk (the point of intersection of the two dashed
lines must be to the left of the point of intersection of the bold
line $T_{\mathrm{conv.}}$ and the horizontal line $z=L-\delta_{th,T}$).
We also note an interesting detail of the profiles. The relation between
the temperatures at the top of the bulk, i.e. at $z=L-\delta_{th,T}$
for the hydrostatic adiabatic profile 
\begin{equation}
T_{ad}\left(z=L-\delta_{th,T}\right)=T_{B}-\frac{g}{c_{p}}\left(L-\delta_{th,T}\right)\label{eq:T_ad_at_top_BL}
\end{equation}
hooked at $T_{B}$ (continuous line) and the mean total temperature
$\left\langle T\right\rangle _{h}(z=L-\delta_{th,T})$, which is suggested
by the figure to be $T_{ad}(z=L-\delta_{th,T})<\left\langle T\right\rangle _{h}(z=L-\delta_{th,T})$,
in fact remains unknown and either one could be greater. In other
words the point of intersection of the bold line $T_{\mathrm{conv}}$
with the horizontal dashed line $z=L-\delta_{th,T}$ could as well
be to the left of the point of intersection of the profile $T_{ad}(z)$
hooked at $T_{B}$ with the line $z=L-\delta_{th,T}$. The relation
between $T_{ad}(z=L-\delta_{th,T})$ and $\left\langle T\right\rangle _{h}(z=L-\delta_{th,T})$
is determined by the relation between $(\Delta T)_{T}$ and
\begin{align}
T_{ad}\left(z=L-\delta_{th,T}\right)-T_{T}= & \Delta T-\frac{g}{c_{p}}L+\frac{g}{c_{p}}\delta_{th,T}\nonumber \\
= & \left(\Delta T\right)_{T}+\left(\Delta T\right)_{B}+\left(\Delta T\right)_{bulk}-\frac{g}{c_{p}}L+\frac{g}{c_{p}}\delta_{th,T}\nonumber \\
= & \left(\Delta T\right)_{T}+\left(\Delta T\right)_{B}-\left(\Delta T\right)_{vel}-\frac{g}{c_{p}}\delta_{th,B}\nonumber \\
= & \left(\Delta T\right)_{T}+\left(\Delta T'\right)_{B}-\left(\Delta T\right)_{vel}+\epsilon_{a}\frac{T_{B}}{L}\delta_{th,B}\label{eq:T_ad_jump_top}
\end{align}
where we have used (\ref{eq:Temp_jump_bulk}) and (\ref{integral_positive}).
This comes down to the relation
\begin{equation}
\left(\Delta T\right)_{vel}\overset{?}{\gtrless}\left(\Delta T'\right)_{B}+\mathcal{O}\left(\epsilon_{a}\delta_{th,B}\frac{T_{B}}{L}\right),\label{eq:relation_aux}
\end{equation}
and which of these two quantities is greater is most likely dependent
on the Rayleigh number, that is on the strength of driving. 

\paragraph{Negative entropy shift at boundaries}

~

~

First we express the total mean entropy drop across the entire fluid
layer directly from the definition of the total entropy $s=\tilde{s}+s'$
\begin{equation}
\Delta\left\langle s\right\rangle _{h}=\tilde{s}_{B}+\left\langle s'\right\rangle _{h,B}-\tilde{s}_{T}-\left\langle s'\right\rangle _{h,T}.\label{Delta_s_1}
\end{equation}
However, because the mean entropy is constant in the bulk (\ref{eq:s_const_in_bulk})
the entropy drop can also be expressed by the entropy jumps across
the top and bottom boundary layers
\begin{equation}
\Delta\left\langle s\right\rangle _{h}=\left(\Delta s\right)_{B}+\left(\Delta s\right)_{T}.\label{Delta_s_2}
\end{equation}
From the two latter equations we get
\begin{equation}
\left\langle s'\right\rangle _{h\,T}-\left\langle s'\right\rangle _{h\,B}+\left(\Delta s\right)_{B}+\left(\Delta s\right)_{T}=\tilde{s}_{B}-\tilde{s}_{T}=\Delta\tilde{s},\label{entropy_deltas}
\end{equation}
and $\Delta\tilde{s}$ can be easily calculated from (\ref{eq:BS2-2})
\begin{equation}
\Delta\tilde{s}=\frac{c_{p}\epsilon_{a}}{\tilde{\theta}}\ln\tilde{\Gamma}=c_{p}\epsilon_{a}\frac{\tilde{\Gamma}}{\tilde{\Gamma}-1}\ln\tilde{\Gamma}.\label{eq:Delta_s_tilde}
\end{equation}
Similarly, we can write for the temperature drop
\begin{equation}
\left(\Delta T\right)_{B}+\left(\Delta T\right)_{T}+\left(\Delta T\right)_{bulk}=\Delta T,\label{eq:Delta_T_tilde}
\end{equation}
where of course $\left\langle T'\right\rangle _{h,T}=\left\langle T'\right\rangle _{h,T}=0$
because by assumption of this section \ref{subsec:isothermal} the
boundaries are held at constant temperature. Extracting the temperature
fluctuation jumps one obtains
\begin{equation}
\left(\Delta T'\right)_{B}+\left(\Delta T'\right)_{T}=\left(\Delta T'\right)_{bulk}.\label{eq:4-1-1-1}
\end{equation}
Introducing the ratio of temperature jumps across the boundary layers
\begin{equation}
r_{T}=\frac{\left(\Delta T'\right)_{T}}{\left(\Delta T'\right)_{B}}>1,\label{eq:rT_def}
\end{equation}
which in stratified developed convection is typically significantly
greater than one (cf. section \ref{subsec:estimates}), we can express
the jumps of the mean temperature fluctuation by the bulk jump in
the following way
\begin{equation}
\left(\Delta T'\right)_{B}=\frac{1}{1+r_{T}}\left(\Delta T'\right)_{bulk},\qquad\left(\Delta T'\right)_{T}=\frac{r_{T}}{1+r_{T}}\left(\Delta T'\right)_{bulk}.\label{Delta_Tprime_BT}
\end{equation}
Consequently by the use of (\ref{eq:DT_rel_Ds}) the mean entropy
jumps take the form 
\begin{subequations}
\begin{align}
\left(\bigtriangleup s\right)_{B}\approx & \frac{c_{p}\left(\bigtriangleup T'\right)_{B}}{T_{B}}=\frac{c_{p}}{T_{B}}\frac{1}{1+r_{T}}\left(\Delta T'\right)_{bulk},\label{Delta_s_B}\\
\left(\bigtriangleup s\right)_{T}\approx & \frac{c_{p}\left(\bigtriangleup T'\right)_{T}}{T_{T}}=\frac{c_{p}}{T_{B}}\frac{\tilde{\Gamma}r_{T}}{1+r_{T}}\left(\Delta T'\right)_{bulk};\label{Delta_s_T}
\end{align}
\end{subequations}
of course here $T_{B}=\tilde{T}_{B}$ and $T_{T}=\tilde{T}_{T}$
since the boundaries are isothermal. Substitution of the latter expressions
into (\ref{entropy_deltas}) yields
\begin{equation}
\left\langle s'\right\rangle _{h,T}-\left\langle s'\right\rangle _{h,B}+\frac{c_{p}}{T_{B}}\frac{1+\tilde{\Gamma}r_{T}}{1+r_{T}}\left(\Delta T'\right)_{bulk}=c_{p}\epsilon_{a}\frac{\tilde{\Gamma}}{\tilde{\Gamma}-1}\ln\tilde{\Gamma},\label{eq:4-2}
\end{equation}
and since $\left(\Delta T'\right)_{bulk}=\epsilon_{a}T_{B}+\left(\Delta T\right)_{vel}$
(cf. (\ref{eq:Temp_jump_bulk-1})) we get finally
\begin{equation}
\left\langle s'\right\rangle _{h,T}-\left\langle s'\right\rangle _{h,B}=c_{p}\epsilon_{a}\left(\frac{\tilde{\Gamma}\ln\tilde{\Gamma}}{\tilde{\Gamma}-1}-\frac{1+\tilde{\Gamma}r_{T}}{1+r_{T}}\right)-\frac{c_{p}}{\tilde{T}_{B}}\frac{1+\tilde{\Gamma}r_{T}}{1+r_{T}}\left(\Delta T\right)_{vel}.\label{eq:shT-shB}
\end{equation}
We will now show, that the right hand side of the equation (\ref{eq:shT-shB})
is negative. First we observe, that obviously
\begin{equation}
-\frac{c_{p}}{T_{B}}\frac{1+\tilde{\Gamma}r_{T}}{1+r_{T}}\left(\Delta T\right)_{vel}<0,\label{eq:Tvel_correction}
\end{equation}
so that we only need to demonstrate, that
\begin{equation}
\frac{\tilde{\Gamma}\ln\tilde{\Gamma}}{\tilde{\Gamma}-1}-\frac{1+\tilde{\Gamma}r_{T}}{1+r_{T}}<0.\label{aux_rel}
\end{equation}
In order to do this, based on the property $r_{T}>1$ let us express
the temperature jump ratio $r_{T}$ by a positive power of $\tilde{\Gamma}$, i.e.\footnote{In that way we imply a scaling law relation between $r_{T}$ and $\tilde{\Gamma}$,
which will be concretised in section \ref{subsec:estimates}},
\begin{equation}
r_{T}=\tilde{\Gamma}^{a},\qquad a\in\mathbb{R}_{+}.\label{eq:rT_Gamma}
\end{equation}
With the use of the latter relation the inequality (\ref{aux_rel})
may be cast in the following form
\begin{equation}
\frac{\left(\tilde{\Gamma}-1\right)\left(1+\tilde{\Gamma}^{1+a}\right)}{\tilde{\Gamma}+\tilde{\Gamma}^{1+a}}-\ln\tilde{\Gamma}>0.\label{aux_rel_2}
\end{equation}
Since at $\tilde{\Gamma}=1$ the left hand side (l.h.s.) of the inequality
(\ref{aux_rel_2}) vanishes it is enough if we prove, that the l.h.s.
is a monotonically increasing function of $\tilde{\Gamma}>1$. A straightforward
calculation leads to
\begin{align}
\frac{\mathrm{d}}{\mathrm{d}\tilde{\Gamma}}\left[\frac{\left(\tilde{\Gamma}-1\right)\left(1+\tilde{\Gamma}^{1+a}\right)}{\tilde{\Gamma}+\tilde{\Gamma}^{1+a}}-\ln\tilde{\Gamma}\right]\nonumber \\
=\frac{\tilde{\Gamma}\left(\tilde{\Gamma}-1\right)}{\left(\tilde{\Gamma}+\tilde{\Gamma}^{1+a}\right)^{2}} & \hspace{-1mm}\left[\tilde{\Gamma}^{2a+1}-1+\left(1+a\right)\tilde{\Gamma}^{a}\left(\tilde{\Gamma}-1\right)\right]>0\label{derivative}
\end{align}
for all $\tilde{\Gamma}>1$ which together with (\ref{eq:shT-shB})
and (\ref{eq:Tvel_correction}) proves, that
\begin{equation}
\left\langle s'\right\rangle _{h,T}-\left\langle s'\right\rangle _{h,B}<0.\label{eq:s_boundary_difference_neg}
\end{equation}
The final step is based on an exactly analogous argumentation as in
subsection \ref{subsec:isentropic_BCs_temp_shift}, which allows to
express the values of the mean entropy fluctuation at isothermal boundaries
by the values of the mean pressure fluctuation directly from (\ref{eq:no_pressure_jump})
and the second relation in (\ref{State_eq_Anonlin}) as follows
\begin{equation}
\left\langle s'\right\rangle _{h,B}=-\frac{\left\langle p'\right\rangle _{h,B}}{\tilde{\rho}_{B}T_{B}},\qquad\left\langle s'\right\rangle _{h,T}=-\frac{\left\langle p'\right\rangle _{h,B}}{\tilde{\rho}_{T}T_{T}}=\tilde{\Gamma}^{m+1}\left\langle s'\right\rangle _{h,B},\label{eq:BC_mean_entropy_isothermalBCs}
\end{equation}
where we have used $\tilde{\rho}_{B}/\tilde{\rho}_{T}=\tilde{\Gamma}^{m}>1$. Therefore
it is clear, that $\left\langle s'\right\rangle _{h,T}$ and $\left\langle s'\right\rangle _{h,B}$
are of the same sign and $\left|\left\langle s'\right\rangle _{h,T}\right|>\left|\left\langle s'\right\rangle _{h,B}\right|$,
which in light of (\ref{eq:s_boundary_difference_neg}) necessarily
implies
\begin{equation}
\left\langle s'\right\rangle _{h,T}<0,\qquad\left\langle s'\right\rangle _{h,B}<0.\label{eq:s_negative_at_boundaries}
\end{equation}

\subsubsection{The case of isentropic boundaries\label{subsec:isentropic_positve_shifts}}

\begin{figure}
\begin{centering}
a)\includegraphics[scale=0.17]{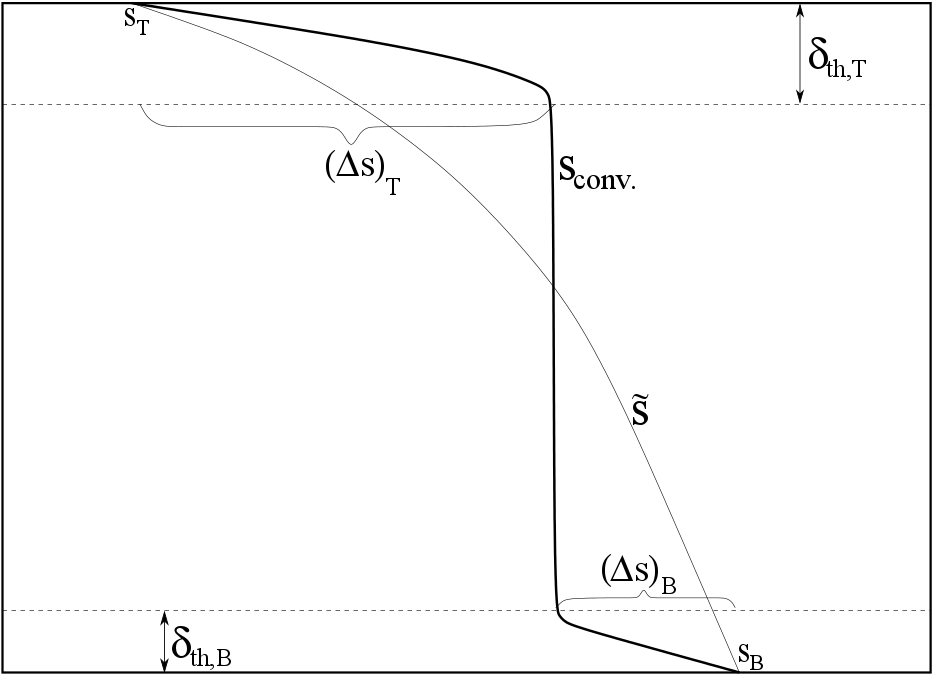}~~~b)\includegraphics[scale=0.17]{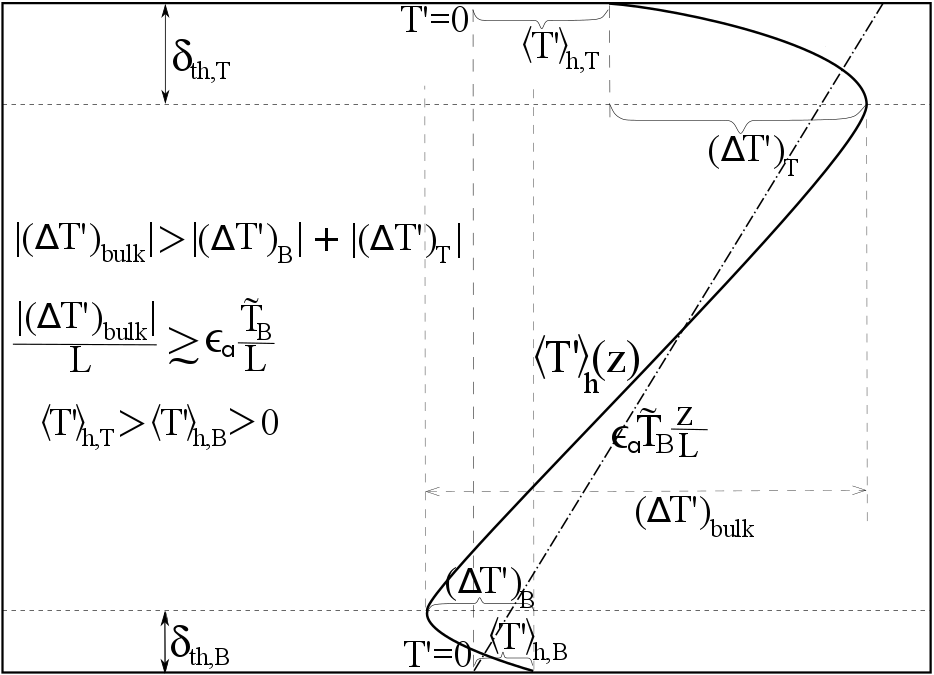}\vspace{3mm}
\par\end{centering}
\begin{centering}
c)\includegraphics[scale=0.17]{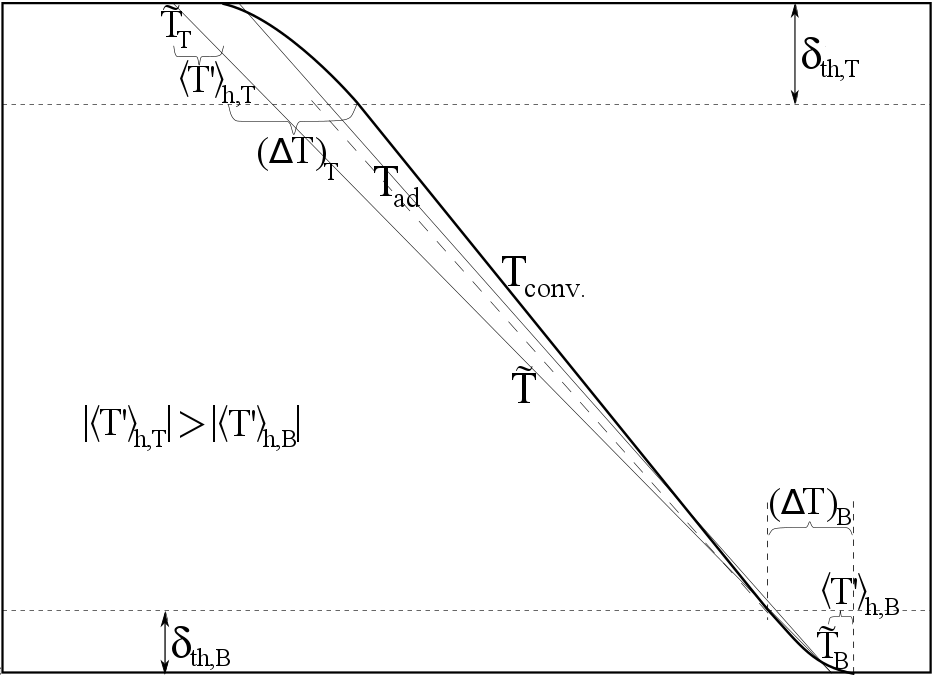}
\par\end{centering}
{\footnotesize{}\caption{{\footnotesize{}\label{fig:DevAn_isentropic_profiles}A schematic
picture of vertical profiles of the total entropy $s_{\mathrm{conv}}=\left\langle s\right\rangle _{h}=\tilde{s}+\left\langle s'\right\rangle _{h}$
a), the mean temperature fluctuation $\left\langle T'\right\rangle _{h}$
b) and the total temperature $T_{\mathrm{conv}}=\left\langle T\right\rangle _{h}=\tilde{T}+\left\langle T'\right\rangle _{h}$
c) (marked with bold lines) in developed convection with fixed entropy
at boundaries and $Q=0$. Note, that the last figure c) is }\emph{\footnotesize{}not
in scale}{\footnotesize{}, as the superadiabaticity $\delta$, i.e.
the departure of the basic profile gradient $\mathrm{d}_{z}\tilde{T}$
from the adiabatic one had to be made significant for clarity of presentation;
in particular the jumps of the mean temperature fluctuation across
the boundary layers $(\Delta T')_{B}$ and $(\Delta T')_{T}$ are
in general $\delta\ll1$ times smaller than corresponding jumps in
the reference temperature profile, i.e. $\tilde{T}(z=0)-\tilde{T}(\delta_{th,B})$
and $\tilde{T}(z=L-\delta_{th,T})-\tilde{T}(L)$ respectively. Moreover,
the non-dimensional thicknesses of the boundary layers, $\delta_{th,B}/L$
and $\delta_{th,T}/L$ must be much larger than the superadiabaticity
$\delta$ for consistency of the anelastic approximation. The dashed
line on figure c) represents the same gradient as that of the hydrostatic
adiabatic profile $T_{ad}$, but is shifted horizontally, so that
it is hooked at the point $z=\delta_{B,th}$ and $T=\left\langle T\right\rangle _{h}(z=\delta_{th,B})$.}}
}{\footnotesize \par}
\end{figure}

We now turn to the case when the entropy is fixed at boundaries for
which the dynamical description of fully developed stratified convection
will be further developed in the next few sections. The vertical profiles
of $\left\langle s\right\rangle _{h}$, $\left\langle T'\right\rangle _{h}$
and $\left\langle T\right\rangle _{h}$ are shown on figure \ref{fig:DevAn_isentropic_profiles}.
When the boundaries are isentropic the temperature at the boundaries
is shifted with respect to the reference state values $\tilde{T}_{B}$
and $\tilde{T}_{T}$ by $\left\langle T'\right\rangle _{h,B}$ and
$\left\langle T'\right\rangle _{h,T}$ respectively, but in this case
the shift is positive. We elaborate on this below. According to the observation made in
section \ref{subsec:Subadiabatic-temperature-gradient} the gradient
of the mean total temperature in the bulk on figure \ref{fig:DevAn_isentropic_profiles}c
is marked weaker than that of a hydrostatic adiabatic state. However,
we stress again, that also in the current case the relation between
the temperatures at the top of the bulk for the total mean $\left\langle T\right\rangle _{h}(z=L-\delta_{th,T})$
and the hydrostatic adiabatic profile hooked at $\tilde{T}_{B}$,
i.e. $T_{ad}(z=L-\delta_{th,T})$ remains unknown. On similar grounds
as in the case of isothermal boundaries explained at the beginning
of subsection \ref{subsec:isothermal} it can be shown, that the relation
between $\left\langle T\right\rangle _{h}(z=L-\delta_{th,T})$ and
$T_{ad}(z=L-\delta_{th,T})$ (with the latter hooked at $\tilde{T}_{B}$)
comes down to the relation between the positive quantities $(\Delta T)_{vel}$
and $(\Delta T')_{B}-\left\langle T'\right\rangle _{h,B}$, which
is most likely non-universal and depends on the Rayleigh number. Moreover,
it should be realized, that also the relations between the temperature
fluctuation jumps across the boundary layers and the temperature shifts
at boundaries, i.e. between $(\Delta T')_{B}$ and $\left\langle T'\right\rangle _{h,B}$
and likewise between $(\Delta T')_{T}$ and $\left\langle T'\right\rangle _{h,T}$,
can not be easily established; in particular this implies, that the
mean temperature fluctuation profile $\left\langle T'\right\rangle _{h}(z)$
on figure \ref{fig:DevAn_isentropic_profiles}b does not necessarily
cross the vertical dashed line indicating $T'=0$ and it is allowed,
that $\left\langle T'\right\rangle _{h}(z)>0$ for all $z$. 

\paragraph{Positive temperature shift at boundaries}

~

~

The mean total temperature drop across the entire fluid layer can
be expressed directly from the definition of the total temperature
$T=\tilde{T}+T'$ in the following way
\begin{equation}
\Delta\left\langle T\right\rangle _{h}=\tilde{T}_{B}+\left\langle T'\right\rangle _{h,B}-\tilde{T}_{T}-\left\langle T'\right\rangle _{h,T}.\label{eq:Delta_T_tot_1}
\end{equation}
At the same time we can express the total temperature drop by a sum
of the mean total temperature jumps across the bulk and both boundary
layers
\begin{equation}
\Delta\left\langle T\right\rangle _{h}=\left(\Delta T\right)_{B}+\left(\Delta T\right)_{T}+\left(\Delta T\right)_{bulk},\label{eq:Delta_T_tot_2}
\end{equation}
which in tandem with (\ref{eq:Delta_T_tot_1}) yields
\begin{equation}
\left\langle T'\right\rangle _{h,T}-\left\langle T'\right\rangle _{h,B}+\left(\Delta T\right)_{B}+\left(\Delta T\right)_{T}+\left(\Delta T\right)_{bulk}=\Delta\tilde{T}.\label{DeltaT_balance}
\end{equation}
Extraction of the mean temperature fluctuation jumps leads to
\begin{equation}
\left\langle T'\right\rangle _{h,T}-\left\langle T'\right\rangle _{h,B}+\left(\Delta T'\right)_{B}+\left(\Delta T'\right)_{T}=\left(\Delta T'\right)_{bulk}.\label{DeltaTprime_balance}
\end{equation}
The entropy is fixed at the boundaries by assumption, therefore $\left\langle s'\right\rangle _{h,B}=0$
and $\left\langle s'\right\rangle _{h,T}=0$ and consequently for
the entropy drops we may write
\begin{equation}
\left(\Delta s\right)_{B}+\left(\Delta s\right)_{T}=\Delta\tilde{s}=c_{p}\epsilon_{a}\frac{\tilde{\Gamma}}{\tilde{\Gamma}-1}\ln\tilde{\Gamma},\label{Delta_s_balance}
\end{equation}
where $(\Delta s)_{bulk}\approx0$ is negligibly small due to efficient
advection of entropy in the bulk (cf. (\ref{eq:s_const_in_bulk})).
Next we introduce the ratio of mean total entropy jumps across the
boundary layers,
\begin{equation}
r_{s}=\frac{\left(\Delta s\right)_{T}}{\left(\Delta s\right)_{B}}=\tilde{\Gamma}r_{T}>1,\qquad r_{T}=\frac{\left(\Delta T'\right)_{T}}{\left(\Delta T'\right)_{B}}>1,\label{r_s_def}
\end{equation}
which typically in developed stratified convection significantly exceeds
unity (cf. section \ref{subsec:estimates}) and which is related by
(\ref{eq:DT_rel_Ds}) to the temperature jump ratio $r_{T}$ already
introduced in (\ref{eq:rT_def}). By the use of (\ref{eq:DT_rel_Ds}),
(\ref{Delta_s_balance}) and (\ref{r_s_def}) the mean temperature
fluctuation jumps across the boundary layers $\left(\Delta T\right)_{B}$
and $\left(\Delta T\right)_{T}$ can be expressed as follows
\begin{subequations}
\begin{align}
\left(\bigtriangleup T'\right)_{B}\approx & \frac{\tilde{T}_{B}\left(\bigtriangleup s\right)_{B}}{c_{p}}=\epsilon_{a}\tilde{T}_{B}\frac{\tilde{\Gamma}\ln\tilde{\Gamma}}{\left(1+\tilde{\Gamma}r_{T}\right)\left(\tilde{\Gamma}-1\right)},\label{Delta_T_B}\\
\left(\bigtriangleup T'\right)_{T}\approx & \frac{\tilde{T}_{T}\left(\bigtriangleup s\right)_{T}}{c_{p}}=\epsilon_{a}\tilde{T}_{B}\frac{\tilde{\Gamma}r_{T}\ln\tilde{\Gamma}}{\left(1+\tilde{\Gamma}r_{T}\right)\left(\tilde{\Gamma}-1\right)}.\label{Delta_T_T}
\end{align}
\end{subequations}
Substituting the latter expressions into (\ref{DeltaTprime_balance})
and making use of $\left(\Delta T'\right)_{bulk}=\epsilon_{a}\tilde{T}_{B}+\left(\Delta T\right)_{vel}$
(cf. (\ref{eq:Temp_jump_bulk-1})) we obtain
\begin{equation}
\left\langle T'\right\rangle _{h,T}-\left\langle T'\right\rangle _{h,B}=\epsilon_{a}\tilde{T}_{B}\left[1-\frac{\tilde{\Gamma}\left(1+r_{T}\right)\ln\tilde{\Gamma}}{\left(1+\tilde{\Gamma}r_{T}\right)\left(\tilde{\Gamma}-1\right)}\right]+\left(\Delta T\right)_{vel}.\label{Tprime_at_Boundaries_difference}
\end{equation}
Since we already know from (\ref{aux_rel})-(\ref{derivative}) that
\begin{equation}
\frac{\tilde{\Gamma}\left(1+r_{T}\right)\ln\tilde{\Gamma}}{\left(1+\tilde{\Gamma}r_{T}\right)\left(\tilde{\Gamma}-1\right)}<1,\label{aux_rel_3}
\end{equation}
and since $\left(\Delta T\right)_{vel}>0$ (cf. (\ref{integral_positive}))
it is evident, that 
\begin{equation}
\left\langle T'\right\rangle _{h,T}-\left\langle T'\right\rangle _{h,B}>0.\label{eq:positive_temp_diff}
\end{equation}
Finally we make use of the observation made below (\ref{eq:BC_entropy_0L_fT-1}-b),
that at the isentropic boundaries $\left\langle T'\right\rangle _{h,T}$
and $\left\langle T'\right\rangle _{h,B}$ are of the same sign and
$\left|\left\langle T'\right\rangle _{h,T}\right|>\left|\left\langle T'\right\rangle _{h,B}\right|$,
which necessarily implies 
\begin{equation}
\left\langle T'\right\rangle _{h,T}>0,\qquad\left\langle T'\right\rangle _{h,B}>0.\label{eq:positive_temperature_shift}
\end{equation}
The actual values of $\left\langle T'\right\rangle _{h,T}$ and $\left\langle T'\right\rangle _{h,B}$
can be easily expressed in terms of the ratio $r_{T}$ (or alternatively
$r_{s}$), $\tilde{\Gamma}$, $\epsilon_{a}\tilde{T}_{B}$ and $\left(\Delta T\right)_{vel}$
by the use of (\ref{Tprime_at_Boundaries_difference}) and $\left\langle T'\right\rangle _{h,T}=\tilde{\Gamma}^{m}\left\langle T'\right\rangle _{h,B}$
(cf. (\ref{eq:BC_entropy_0L_fT-1-1})).

\subsection{The Nusselt, Rayleigh and Reynolds numbers}

At the top and bottom boundaries all the heat is carried
by conduction, thus the horizontally averaged superadiabatic heat
flux at the bottom boundary is
\begin{equation}
F_{S}(z=0)=-k\left.\frac{\mathrm{d}\left\langle T_{S}\right\rangle _{h}}{\mathrm{d}z}\right|_{z=0}.\label{eq:F_S_atB}
\end{equation}
Of course in a (statistically) stationary turbulent state the superadiabatic
flux at the bottom equals that at the top, as is clear from (\ref{eq:heat_flux_enter_exit})
and the fact that under current assumptions the superadiabatic flux
of the conduction reference state,
\begin{equation}
\tilde{F}_{S}=-k\frac{\mathrm{d}}{\mathrm{d}z}\left(\tilde{T}-T_{ad}\right)=k\epsilon_{a}\frac{\tilde{T}_{B}}{L},\label{eq:BS_superad}
\end{equation}
is independent of height. As a result
\begin{equation}
F_{S}(z=0)=-k\left.\frac{\mathrm{d}\left\langle T_{S}\right\rangle _{h}}{\mathrm{d}z}\right|_{z=0}=-k\left.\frac{\mathrm{d}\left\langle T_{S}\right\rangle _{h}}{\mathrm{d}z}\right|_{z=L}=F_{S}(z=L).\label{eq:top_and_bottom_fluxes}
\end{equation}
This allows to define the Nusselt number $Nu$ as the ratio of the
mean superadiabatic heat flux either at the top or at the bottom boundary
in a stationary convective state divided by the superadiabatic heat
flux in the conduction state (\ref{eq:BS_superad}), which yields\index{SI}{Nusselt number}
\begin{equation}
Nu=\frac{F_{S}(z=0)}{\tilde{F}_{S}}=\frac{-L\left.\frac{\mathrm{d}\left\langle T_{S}\right\rangle _{h}}{\mathrm{d}z}\right|_{z=0}}{\epsilon_{a}\tilde{T}_{B}}\approx\frac{\left(\bigtriangleup T_{S}\right)_{B}L}{\epsilon_{a}\tilde{T}_{B}\delta_{th,B}},\label{eq:Nu_def_anapp}
\end{equation}
or equivalently
\begin{equation}
Nu=\frac{F_{S}(z=L)}{\tilde{F}_{S}}=\frac{-L\left.\frac{\mathrm{d}\left\langle T_{S}\right\rangle _{h}}{\mathrm{d}z}\right|_{z=L}}{\epsilon_{a}\tilde{T}_{B}}\approx\frac{\left(\bigtriangleup T_{S}\right)_{T}L}{\epsilon_{a}\tilde{T}_{B}\delta_{th,T}}.\label{eq:Nu_def_anapp-2}
\end{equation}
The so-defined Nusselt number is unity at convection onset and large
in fully developed convection. Similarly as it is done in the theory
of Boussinesq turbulent convection, in the above definition of the
Nusselt number we have approximated the bottom and top values of the
superadiabatic temperature gradient with the ratios of the superadiabatic
temperature jumps across the boundary layers to the respective thicknesses
of boundary layers, so that
\begin{equation}
-\left.\frac{\mathrm{d}\left\langle T_{S}\right\rangle _{h}}{\mathrm{d}z}\right|_{z=0}\approx\frac{\left(\bigtriangleup T_{S}\right)_{B}}{\delta_{th,B}},\qquad-\left.\frac{\mathrm{d}\left\langle T_{S}\right\rangle _{h}}{\mathrm{d}z}\right|_{z=L}\approx\frac{\left(\bigtriangleup T_{S}\right)_{T}}{\delta_{th,T}}.\label{eq:bot_top_temp_grads}
\end{equation}
The Rayleigh number is a measure of the strength of the thermal driving.
For the case of isothermal boundaries it was defined in (\ref{eq:Ra_def-1}),
however, we now consider the case of isentropic boundaries, when the
flow is driven by the entropy jump across the fluid layer, that is
$\Delta\tilde{s}$ (cf. (\ref{eq:Delta_s_tilde})). Therefore we define
the Rayleigh number in the following way\index{SI}{Rayleigh number!anelastic}
\begin{equation}
Ra=\frac{g\bigtriangleup\tilde{s}L^{3}\tilde{\rho}_{B}^{2}}{\mu k}\approx\frac{c_{p}\bigtriangleup\tilde{s}\bigtriangleup\tilde{T}L^{2}\tilde{\rho}_{B}^{2}}{\mu k},\label{eq:Ra_def}
\end{equation}
where in obtaining the second expression we have used $g/c_{p}=\Delta\tilde{T}/L+\mathcal{O}(\epsilon_{a}g/c_{p})$.

Furthermore, the central idea of the theory of fully developed Boussinesq
convection is based on the assumption that the structure of turbulent
convective flow is always characterized by the presence of a large-scale
convective roll called the \emph{wind of turbulence}. This idea, which
in the non-stratified case stems from a vast numerical and experimental
evidence, is retained in the case of anelastic convection, however, it must be realized that
the significant stratification in the anelastic case breaks the Boussinesq
up-down symmetry. Thus we must distinguish between the magnitude
of the wind of turbulence near the bottom of the bulk and its magnitude
near the top of the bulk, denoted by $U_{B}$ and $U_{T}$ respectively,
which can now significantly differ (cf. e.g. Jones \emph{et al}. 2020
for numerical evidence of the 'wind of turbulence' with height-dependent
magnitude in stratified convection). With the use of the introduced
notation we define the bottom and top Reynolds numbers\index{SI}{Reynolds numbers}
\begin{equation}
Re_{B}=\frac{U_{B}L\tilde{\rho}_{B}}{\mu},\qquad Re_{T}=\frac{U_{T}L\tilde{\rho}_{T}}{\mu}.\label{eq:Re_numbers_def}
\end{equation}
We emphasize, that these Reynolds numbers are based on the large length
scale $L$, and such an approach is consistent with the up-to-date
available data from numerical simulations of strongly stratified convective
flows (cf. e.g. Verhoeven \emph{et al}. 2015, Jones \emph{et al}.
2020); this data, however, were obtained for a rather weak driving,
as large Rayleigh numbers are currently unachievable for large stratifications,
thus it is not clear if they yet correspond to the problem of fully
developed, strongly stratified convection. In fact it is possible
that at large stratifications both the horizontal and vertical length
scales of variation of the large scale convective flow $\left\langle \mathbf{u}(\mathbf{x},t)\right\rangle _{t}$
(where $\left\langle \cdot\right\rangle _{t}$ denotes a time average)
are determined by the scale heights $D_{p}$, $D_{\rho}$ and $D_{T}$;
these scale heights vary with height and are significantly smaller
at the top, than at the bottom of the fluid domain. It seems rather
likely that at larger stratifications the wind of turbulence becomes
splitted into two or more, still large scale convection rolls, whose
sizes are determined by the values of the scale heights. However,
since currently there is not enough evidence, neither numerical nor
experimental and observational for vertical splitting of large scale
rolls in strongly stratified convection, it seems more reasonable
to postulate the total depth of the layer as the typical length scale
of horizontal variation. 

\subsubsection{Thicknesses of thermal boundary layers, $\delta_{th,B}$ and $\delta_{th,T}$.}

It is possible to express the Nusselt number by the mean temperature
fluctuation jumps likewise by the entropy jumps across the boundary
layers, which in turn allows to directly relate the thicknesses of
the thermal boundary layers to the Nusselt number. By the use of (\ref{eq:Tprime_Ts_rel})
we get (cf. also figure \ref{fig:DevAn_isentropic_profiles}b)
\begin{subequations}
\begin{align}
\left(\bigtriangleup T_{S}\right)_{B}= & \left\langle T_{S}\right\rangle _{h}\left(z=0\right)-\left\langle T_{S}\right\rangle _{h}\left(z=\delta_{th,B}\right)\nonumber \\
= & \left\langle T'\right\rangle _{h}\left(z=0\right)-\left\langle T'\right\rangle _{h}\left(z=\delta_{th,B}\right)+\epsilon_{a}\tilde{T}_{B}\frac{\delta_{th,B}}{L}\nonumber \\
= & \left(\Delta T'\right)_{B}+\epsilon_{a}\frac{\delta_{th,B}}{L}\tilde{T}_{B},\label{eq:Delta_TSB_by_other_Deltas}
\end{align}
\begin{align}
\left(\bigtriangleup T_{S}\right)_{T}= & \left\langle T_{S}\right\rangle _{h}\left(z=L-\delta_{th,T}\right)-\left\langle T_{S}\right\rangle _{h}\left(z=L\right)\nonumber \\
= & \left\langle T'\right\rangle _{h}\left(z=L-\delta_{th,T}\right)-\left\langle T'\right\rangle _{h}\left(z=L\right)+\epsilon_{a}\tilde{T}_{B}\frac{\delta_{th,T}}{L}\nonumber \\
= & \left(\Delta T'\right)_{T}+\epsilon_{a}\frac{\delta_{th,T}}{L}\tilde{T}_{B},\label{eq:Delta_TST_by_other_Deltas}
\end{align}
\end{subequations}
hence the superadiabatic temperature jumps and
the temperature fluctuation jumps are equal at leading order, since
the corrections are much smaller. Next we can substitute for $\left(\Delta T'\right)_{B}$
and $\left(\Delta T'\right)_{T}$ from (\ref{eq:DT_rel_Ds}) and using
(\ref{Delta_s_balance}) and (\ref{r_s_def}) obtain
\begin{subequations}
\begin{equation}
\left(\bigtriangleup T_{S}\right)_{B}\approx\frac{\tilde{T}_{B}}{c_{p}}\frac{\Delta\tilde{s}}{1+r_{s}}+\epsilon_{a}\frac{\delta_{th,B}}{L}\tilde{T}_{B},\label{eq:Delta_TSB_by_other_Deltas-1}
\end{equation}
\begin{equation}
\left(\bigtriangleup T_{S}\right)_{T}\approx\frac{\tilde{T}_{T}}{c_{p}}\frac{r_{s}\Delta\tilde{s}}{1+r_{s}}+\epsilon_{a}\frac{\delta_{th,T}}{L}\tilde{T}_{B}.\label{eq:Delta_TST_by_other_Deltas-1}
\end{equation}
\end{subequations}
Finally we substitute for $\left(\bigtriangleup T_{S}\right)_{B}$
and $\left(\bigtriangleup T_{S}\right)_{T}$ from the latter equations
in the equations (\ref{eq:Nu_def_anapp}) and (\ref{eq:Nu_def_anapp-2}), which
with the aid of (\ref{eq:Delta_s_tilde}) yields\index{SI}{Nusselt number!anelastic}
\begin{equation}
Nu\approx\frac{\tilde{\Gamma}\ln\tilde{\Gamma}}{\left(1+r_{s}\right)\left(\tilde{\Gamma}-1\right)}\frac{L}{\delta_{th,B}}+1,\label{eq:Nu_def_delta_def_B}
\end{equation}
\begin{equation}
Nu\approx\frac{r_{s}\ln\tilde{\Gamma}}{\left(1+r_{s}\right)\left(\tilde{\Gamma}-1\right)}\frac{L}{\delta_{th,T}}+1.\label{eq:Nu_def_delta_def_T}
\end{equation}
This leads to the following expressions for thicknesses of thermal
boundary layers\index{SI}{boundary layer thickness!thermal boundary layer}
\begin{subequations}
\begin{equation}
\frac{\delta_{th,B}}{L}\approx\frac{\tilde{\Gamma}\ln\tilde{\Gamma}}{\left(1+r_{s}\right)\left(\tilde{\Gamma}-1\right)}\left(Nu+1\right)^{-1}\approx\frac{\tilde{\Gamma}\ln\tilde{\Gamma}}{\left(1+r_{s}\right)\left(\tilde{\Gamma}-1\right)}Nu^{-1},\label{eq:delta_th_B}
\end{equation}
\begin{equation}
\frac{\delta_{th,T}}{L}\approx\frac{r_{s}\ln\tilde{\Gamma}}{\left(1+r_{s}\right)\left(\tilde{\Gamma}-1\right)}\left(Nu+1\right)^{-1}\approx\frac{r_{s}\ln\tilde{\Gamma}}{\left(1+r_{s}\right)\left(\tilde{\Gamma}-1\right)}Nu^{-1},\label{eq:delta_th_T}
\end{equation}
\end{subequations}
where the last approximations concerning neglection
of unity with respect to the Nusselt number were made based on the
fact, that in fully developed convection the Nusselt number is large,
$Nu\gg1$ (they are of course equivalent to neglection of the small
corrections in (\ref{eq:Delta_TSB_by_other_Deltas-1},b)). The ratio
of the thicknesses of the thermal boundary layers is now easily expressed
by $r_{s}$
\begin{equation}
r_{\delta}\overset{\mathrm{def.}}{=}\frac{\delta_{th,T}}{\delta_{th,B}}\approx\tilde{\Gamma}^{-1}r_{s}=r_{T}.\label{eq:r_delta}
\end{equation}

\subsubsection{Thicknesses of the viscous boundary layers $\delta_{\nu,B}$ and
$\delta_{\nu,T}$.}

The viscous boundary layers are assumed laminar, as the intuition
developed from the Boussinesq theory suggests, that the Rayleigh numbers
necessary for the boundary layers to become turbulent are huge. We
therefore concentrate on the case, when the Rayleigh number is large
enough for convection to be already fully turbulent, but at the same
time does not exceed a critical much larger value above which the
boundary layers become turbulent. Consequently the thicknesses of
the viscous boundary layers are defined according to the standard
laminar Blasius theory\index{SI}{boundary layer thickness!Blasius boundary layer}
\begin{equation}
\frac{\delta_{\nu,B}}{L}=Re_{B}^{-1/2},\qquad\frac{\delta_{\nu,T}}{L}=Re_{T}^{-1/2}.\label{eq:viscous_BLs_thicknesses}
\end{equation}

\subsection{Estimates of the mean superadiabatic heat flux in a fully developed
state\label{subsec:estimates}}

The first formula for the mean superadiabatic heat flux
(\ref{eq:F_conv_superadiab_1}) for the analysed case of an ideal
gas with isentropic boundaries allows to write
\begin{align}
F_{S}(z=0)= & -k\left.\frac{\mathrm{d}}{\mathrm{d}z}\left(\tilde{T}+\left\langle T'\right\rangle _{h}-T_{ad}\right)\right|_{z=0}\nonumber \\
= & -k\frac{\mathrm{d}}{\mathrm{d}z}\left(\tilde{T}+\left\langle T'\right\rangle _{h}-T_{ad}\right)+\tilde{\rho}\tilde{T}\left\langle u_{z}s'\right\rangle _{h}-\int_{0}^{z}\tilde{\rho}\frac{\mathrm{d}\tilde{T}}{\mathrm{d}z}\left\langle u_{z}s'\right\rangle _{h}\mathrm{d}z\nonumber \\
 & -\mu\int_{0}^{z}\left\langle q\right\rangle _{h}\mathrm{d}z-2\mu\left[\frac{1}{2}\frac{\mathrm{d}\left\langle u_{z}^{2}\right\rangle _{h}}{\mathrm{d}z}-\frac{m\Delta\tilde{T}}{L\tilde{T}}\left\langle u_{z}^{2}\right\rangle _{h}\right],\label{eq:F_conv_superadiab_1-1}
\end{align}
whereas the second formula for the mean superadiabatic heat flux (\ref{eq:F_conv_superadiabatic_2})
implies
\begin{align}
F_{S}\left(z=0\right)= & -k\left.\frac{\mathrm{d}}{\mathrm{d}z}\left(\tilde{T}+\left\langle T'\right\rangle _{h}-T_{ad}\right)\right|_{z=0}\nonumber \\
= & -k\frac{\tilde{T}_{B}}{\tilde{T}}\frac{\mathrm{d}}{\mathrm{d}z}\left(\tilde{T}+\left\langle T'\right\rangle _{h}-T_{ad}\right)+\tilde{\rho}\tilde{T}_{B}\left\langle u_{z}s'\right\rangle _{h}-\mu\int_{0}^{z}\frac{\tilde{T}_{B}}{\tilde{T}}\left\langle q\right\rangle _{h}\mathrm{d}z\nonumber \\
 & -\epsilon_{a}k\frac{\tilde{T}_{B}}{L}\left(\frac{\tilde{T}_{B}}{\tilde{T}}-1\right)+k\tilde{T}_{B}\frac{\Delta\tilde{T}}{L}\left(\frac{\left\langle T'\right\rangle _{h}}{\tilde{T}^{2}}-\frac{\left\langle T'\right\rangle _{h,B}}{\tilde{T}_{B}^{2}}\right)\nonumber \\
 & -2k\tilde{T}_{B}\left(\frac{\Delta\tilde{T}}{L}\right)^{2}\int_{0}^{z}\frac{\left\langle T'\right\rangle _{h}}{\tilde{T}^{3}}\mathrm{d}z-2\mu\left[\frac{1}{2}\frac{\tilde{T}_{B}}{\tilde{T}}\frac{\mathrm{d}\left\langle u_{z}^{2}\right\rangle _{h}}{\mathrm{d}z}\right.\nonumber \\
 & \left.-\left(m+\frac{1}{2}\right)\frac{\tilde{T}_{B}\bigtriangleup\tilde{T}}{\tilde{T}^{2}L}\left\langle u_{z}^{2}\right\rangle _{h}+\left(m+1\right)\tilde{T}_{B}\left(\frac{\bigtriangleup\tilde{T}}{L}\right)^{2}\int_{0}^{z}\frac{\left\langle u_{z}^{2}\right\rangle _{h}}{\tilde{T}^{3}}\mathrm{d}z\right],\label{eq:F_conv_superadiabatic_2-2}
\end{align}
where we have introduced
\begin{equation}
q=\nabla\mathbf{u}:\nabla\mathbf{u}+\frac{1}{3}\left(\nabla\cdot\mathbf{u}\right)^{2}.\label{eq:q_def}
\end{equation}
Before we proceed let us first demonstrate that in fact
a lot of terms in (\ref{eq:F_conv_superadiab_1-1}) and (\ref{eq:F_conv_superadiabatic_2-2})
are negligible in comparison to the mean superadiabatic heat flux
$F_{S}\left(z=0\right)=F_{S}(z=L)$ entering the system at the bottom,
or leaving at the top; the latter two are equal in a stationary state
according to the equation (\ref{eq:heat_flux_enter_exit}) with subtracted constant
adiabatic gradient $-g/c_{P}$ from both sides or by (\ref{eq:F_conv_superadiab_1-1})
taken at $z=L$. We start with the sum of three temperature terms
in (\ref{eq:F_conv_superadiabatic_2-2}) 
\begin{align}
\Sigma_{T}\overset{\mathrm{def.}}{=} & -\epsilon_{a}k\frac{\tilde{T}_{B}}{L}\left(\frac{\tilde{T}_{B}}{\tilde{T}}-1\right)+k\tilde{T}_{B}\frac{\Delta\tilde{T}}{L}\left(\frac{\left\langle T'\right\rangle _{h}}{\tilde{T}^{2}}-\frac{\left\langle T'\right\rangle _{h,B}}{\tilde{T}_{B}^{2}}\right)\nonumber \\
 & -2k\tilde{T}_{B}\left(\frac{\Delta\tilde{T}}{L}\right)^{2}\int_{0}^{z}\frac{\left\langle T'\right\rangle _{h}}{\tilde{T}^{3}}\mathrm{d}z,\label{eq:3_thermal_terms}
\end{align}
which operatively was denoted by $\Sigma_{T}$. Examination of the
mean temperature fluctuation profile (\ref{fig:DevAn_isentropic_profiles}b)
allows to estimate the integral from above
\begin{align}
-2k\tilde{T}_{B}\left(\frac{\Delta\tilde{T}}{L}\right)^{2}\int_{0}^{z}\frac{\left\langle T'\right\rangle _{h}}{\tilde{T}^{3}}\mathrm{d}z\leq & -2k\tilde{T}_{B}\left(\frac{\Delta\tilde{T}}{L}\right)^{2}\left[\left\langle T'\right\rangle _{h,B}-\left(\Delta T'\right)_{B}\right]\int_{0}^{z}\frac{\mathrm{d}z}{\tilde{T}^{3}}\nonumber \\
 & =k\frac{\tilde{\theta}}{L}\left[\left(\Delta T'\right)_{B}-\left\langle T'\right\rangle _{h,B}\right]\left(\frac{\tilde{T}_{B}^{2}}{\tilde{T}^{2}}-1\right).\label{eq:estimate_temp_integral}
\end{align}
The second term in the sum (\ref{eq:3_thermal_terms}) can be bounded
from above by substituting the maximal value of the mean temperature
fluctuation, which according to the vertical profile of $\left\langle T'\right\rangle _{h}(z)$
sketched on figure (\ref{fig:DevAn_isentropic_profiles}b) is estimated
at $\left(\Delta T'\right)_{T}+\left\langle T'\right\rangle _{h,T}$.
Consequently the sum $\Sigma_{T}$ satisfies
\begin{align}
\Sigma_{T}\leq & -\epsilon_{a}k\frac{\tilde{T}_{B}}{L}\left(\frac{\tilde{T}_{B}}{\tilde{T}}-1\right)+k\frac{\tilde{\theta}}{L}\frac{\tilde{T}_{B}^{2}}{\tilde{T}^{2}}\left[\left(\Delta T'\right)_{T}+\left\langle T'\right\rangle _{h\,T}-\left\langle T'\right\rangle _{h,B}\right]\nonumber \\
 & +k\frac{\tilde{\theta}}{L}\left(\Delta T'\right)_{B}\left(\frac{\tilde{T}_{B}^{2}}{\tilde{T}^{2}}-1\right),\label{eq:Sigma_T_est11}
\end{align}
and with the aid of (\ref{DeltaTprime_balance}) and (\ref{eq:Temp_jump_bulk-1})
we can write
\begin{align}
\Sigma_{T}\leq & -\epsilon_{a}k\frac{\tilde{T}_{B}}{L}\left(\frac{\tilde{T}_{B}}{\tilde{T}}-1\right)+k\frac{\tilde{\theta}}{L}\frac{\tilde{T}_{B}^{2}}{\tilde{T}^{2}}\left(\Delta T'\right)_{bulk}-k\frac{\tilde{\theta}}{L}\left(\Delta T'\right)_{B}\nonumber \\
 & =\epsilon_{a}k\frac{\tilde{T}_{B}}{L}\left(\tilde{\theta}\frac{\tilde{T}_{B}^{2}}{\tilde{T}^{2}}-\frac{\tilde{T}_{B}}{\tilde{T}}+1\right)+k\frac{\tilde{\theta}}{L}\frac{\tilde{T}_{B}^{2}}{\tilde{T}^{2}}\left(\Delta T\right)_{vel}-k\tilde{\theta}\frac{\left(\Delta T'\right)_{B}}{\delta_{th,B}}\frac{\delta_{th,B}}{L}.\label{eq:Sigma_T_est2}
\end{align}
It follows from the definition (\ref{integral_positive}), that
\begin{equation}
\left(\Delta T\right)_{vel}=\int_{0}^{L}\frac{\left\langle u_{z}^{2}\right\rangle _{h}}{c_{p}D_{\rho}}\mathrm{d}z\leq\frac{m\left(\tilde{\Gamma}-1\right)}{c_{p}}\left\langle u_{z}^{2}\right\rangle ,\label{eq:Tvel_upper_bound}
\end{equation}
which in turn allows to write
\begin{align}
\Sigma_{T}\leq & \frac{\epsilon_{a}k\frac{\tilde{T}_{B}}{L}\left(\tilde{\Gamma}^{2}-\tilde{\Gamma}+1\right)}{k\frac{\left(\Delta T_{S}\right)_{B}}{\delta_{th,B}}}k\frac{\left(\Delta T_{S}\right)_{B}}{\delta_{th,B}}+\frac{mk}{c_{p}L}\tilde{\Gamma}\left(\tilde{\Gamma}-1\right)^{2}\left\langle u_{z}^{2}\right\rangle \nonumber \\
 & -\frac{\tilde{\Gamma}-1}{\tilde{\Gamma}}k\frac{\left(\Delta T_{S}\right)_{B}}{\delta_{th,B}}\frac{\delta_{th,B}}{L}\label{eq:Sigma_T_est22}
\end{align}
and finally
\begin{align}
\Sigma_{T}\leq & \frac{mk}{c_{p}L}\tilde{\Gamma}\left(\tilde{\Gamma}-1\right)^{2}\left\langle u_{z}^{2}\right\rangle +\left[\left(\tilde{\Gamma}^{2}-\tilde{\Gamma}+1\right)Nu^{-1}-\frac{\tilde{\Gamma}-1}{\tilde{\Gamma}}\frac{\delta_{th,B}}{L}\right]F_{S}(z=0)\nonumber \\
 & =\frac{mk}{c_{p}L}\tilde{\Gamma}\left(\tilde{\Gamma}-1\right)^{2}\left\langle u_{z}^{2}\right\rangle +\mathcal{O}\left(\frac{\delta_{th,B}}{L}F_{S}(z=0)\right),\label{eq:Sigma_T_est3}
\end{align}
where we have used (\ref{eq:Nu_def_anapp}) and (\ref{eq:delta_th_B})
to write 
\begin{equation}
\mathcal{O}\left(Nu^{-1}F_{S}(z=0)\right)=\mathcal{O}\left(\frac{\delta_{th,B}}{L}F_{S}(z=0)\right).\label{eq:equal_rests}
\end{equation}
It remains to prove, that the velocity term in (\ref{eq:Sigma_T_est3}),
\begin{equation}
\frac{mk}{c_{p}L}\tilde{\Gamma}\left(\tilde{\Gamma}-1\right)^{2}\left\langle u_{z}^{2}\right\rangle ,\label{eq:velocity_diss_term}
\end{equation}
is also negligibly small compared to the superadiabatic flux $F_{S}(z=0)$,
which we demonstrate along with negligibility of the viscous terms
in (\ref{eq:F_conv_superadiab_1-1}),
\begin{equation}
-2\mu\left[\frac{1}{2}\frac{\mathrm{d}\left\langle u_{z}^{2}\right\rangle _{h}}{\mathrm{d}z}-\frac{m\Delta\tilde{T}}{L\tilde{T}}\left\langle u_{z}^{2}\right\rangle _{h}\right]\ll F_{S}(z=0),\label{eq:viscous_terms_1}
\end{equation}
and in (\ref{eq:F_conv_superadiabatic_2-2})
\begin{align}
-2\mu\left[\frac{1}{2}\frac{\tilde{T}_{B}}{\tilde{T}}\frac{\mathrm{d}\left\langle u_{z}^{2}\right\rangle _{h}}{\mathrm{d}z}-\left(m+\frac{1}{2}\right)\frac{\tilde{T}_{B}\bigtriangleup\tilde{T}}{\tilde{T}^{2}L}\left\langle u_{z}^{2}\right\rangle _{h}\right.\nonumber \\
\left.+\left(m+1\right)\tilde{T}_{B}\left(\frac{\bigtriangleup\tilde{T}}{L}\right)^{2}\int_{0}^{z}\frac{\left\langle u_{z}^{2}\right\rangle _{h}}{\tilde{T}^{3}}\mathrm{d}z\right] & \ll F_{S}(z=0).\label{eq:viscous_terms_2}
\end{align}
The terms are either of order $\sim\mu\mathrm{d}\left\langle u_{z}^{2}\right\rangle _{h}/\mathrm{d}z$,
or $\sim\mu\left\langle u_{z}^{2}\right\rangle _{h}/L$, or $\sim k\left\langle u_{z}^{2}\right\rangle /c_{p}L$,
hence it is enough if we demonstrate that the squared vertical velocity
averaged over the horizontal plane and its first '$z$'-derivative
multiplied by the dissipative coefficients $\mu$ or $k/c_{p}$ are
negligible compared to the viscous dissipation terms
\begin{equation}
Q_{\nu}(z)\overset{\mathrm{def.}}{=}\mu\int_{0}^{z}\left\langle q\right\rangle _{h}\mathrm{d}z\qquad\textrm{and}\qquad Q_{\nu/T}(z)\overset{\mathrm{def.}}{=}\mu\int_{0}^{z}\frac{\tilde{T}_{B}}{\tilde{T}}\left\langle q\right\rangle _{h}\mathrm{d}z.\label{eq:Q1_and_Q2}
\end{equation}
First of all it is important to realize that in the viscous
boundary layers (of thicknesses $\delta_{\nu,B}$ and $\delta_{\nu,T}$)
and hence also at the bottom and at the top of the bulk, i.e. in the
vicinity of $z=\delta_{\nu,B}$ and $z=L-\delta_{\nu,T}$, the mass
conservation constraint $\nabla\cdot(\tilde{\rho}\mathbf{u})=0$ implies
that the vertical velocity $u_z$ must be very small, of the order of $\delta_{\nu,B}/L$
and $\delta_{\nu,T}/L$ at the bottom and top respectively (cf. (\ref{eq:vertical_velocity_BLs})). The viscous
dissipation is either dominated by the contributions from viscous
boundary layers or from the bulk, therefore in the former case it
is straightforward to see that the terms in (\ref{eq:viscous_terms_1}), likewise the terms in 
(\ref{eq:velocity_diss_term}) and (\ref{eq:viscous_terms_2}) are
$\mathcal{O}(\delta_{v,i})=\mathcal{O}(Re_{i}^{-1/2})$ times smaller
than 
\begin{equation}
Q_{\nu}\left(z\gtrsim\delta_{\nu,B}\right),\; Q_{\nu}\left(z\lesssim L-\delta_{\nu,T}\right);\qquad Q_{\nu/T}\left(z\gtrsim\delta_{\nu,B}\right),\; Q_{\nu/T}\left(z\lesssim L-\delta_{\nu,T}\right),\label{eq:Q_nus}
\end{equation}
respectively, because of smallness of $\left\langle u_{z}^{2}\right\rangle _{h}$ at the bottom and top of the bulk (see (\ref{eq:Q1_and_Q2}) for definitions of $Q_{\nu}(z)$ and $Q_{\nu/T}(z)$). More precisely
all the aforementioned terms can be estimated as follows 
\begin{subequations}
\begin{equation}
\frac{\mu\left\langle u_{z}^{2}\right\rangle _{h}}{L}\sim  \frac{\mu}{L}\frac{\delta_{\nu,i}^{2}}{L^{2}}U_{i}^{2}=\frac{\mu^{3}}{\tilde{\rho}_{i}^{2}L^{3}}Re_{i},\label{eq:estimate1}
\end{equation}
\begin{equation}
\mu\frac{\mathrm{d}\left\langle u_{z}^{2}\right\rangle _{h}}{\mathrm{d}z}\lesssim  \frac{\mu}{L}\frac{\delta_{\nu,i}}{L}U_{i}^{2}=\frac{\mu^{3}}{\tilde{\rho}_{i}^{2}L^{3}}Re_{i}^{3/2},\label{eq:estimate2}
\end{equation}
\begin{align}
\mu\tilde{T}_{B}\left(\frac{\bigtriangleup\tilde{T}}{L}\right)^{2}\int_{0}^{z}\frac{\left\langle u_{z}^{2}\right\rangle _{h}}{\tilde{T}^{3}}\mathrm{d}z\leq & \frac{\mu}{L}\tilde{\Gamma}^{3}\tilde{\theta}^{2}\frac{1}{L}\int_{0}^{z}\left\langle u_{z}^{2}\right\rangle _{h}\mathrm{d}z\nonumber \\
 & \sim\frac{\mu}{L}\tilde{\Gamma}^{3}\tilde{\theta}^{2}\frac{\delta_{\nu,i}^{3}}{L^{3}}U_{i}^{2}\nonumber \\
 & \sim\frac{\mu^{3}}{\tilde{\rho}_{i}^{2}L^{3}}\tilde{\Gamma}\left(\tilde{\Gamma}-1\right)^{2}Re_{i}^{1/2},\label{eq:estimate3}
\end{align}
\begin{align}
\frac{k\left\langle u_{z}^{2}\right\rangle }{c_{p}L}=Pr^{-1}\frac{\mu}{L}\left\langle u_{z}^{2}\right\rangle \sim & Pr^{-1}\frac{\mu}{L}\left(\frac{\delta_{\nu,B}^{3}}{L^{3}}U_{B}^{2}+\frac{\delta_{\nu,T}^{3}}{L^{3}}U_{T}^{2}\right)\nonumber \\
\sim & \frac{\mu^{3}}{\tilde{\rho}_{B}^{2}L^{3}}\tilde{\Gamma}\left(\tilde{\Gamma}-1\right)^{2}Pr^{-1}\left(Re_{B}^{1/2}+\frac{\tilde{\rho}_{B}^{2}}{\tilde{\rho}_{T}^{2}}Re_{T}^{1/2}\right),\label{eq:estimate4}
\end{align}
\end{subequations} 
where
\begin{equation}
Pr=\frac{c_{p}\mu}{k},\label{eq:Prandtl}
\end{equation}
is the Prandtl number, $U_{i}$ and $\tilde{\rho}_{i}$ are the maximal
horizontally averaged velocity and reference density either at the
top or the bottom of the bulk, whichever leads to a larger estimate
and $Re_{i}$ is the Reynolds number based on them. 

The second case, when the viscous dissipation takes place
predominantly in the bulk is a little bit more subtle. The viscous
dissipation terms $Q_{\nu}(L)$ and $Q_{\nu/T}(L)$ can be estimated
in a similar way as for the Boussinesq convection (cf. Grossman and
Lohse 2000), i.e. by the use of the fact, that in such a case the
dissipative effects are expected to balance the inertial effects in
the bulk, 
\begin{equation}
Q_{\nu}(L)\sim Q_{\nu/T}(L)\sim\tilde{\rho}_{i}U_{i}^{3}=\frac{\mu^{3}}{\tilde{\rho}_{i}^{2}L^{3}}Re_{i}^{3}.\label{eq:Qs_estimates_diss_in_bulk}
\end{equation}
The same estimate
can, in fact, be obtained by introducing the Kolmogorov cascade picture
and thus taking the Kolmogorov scale for velocity $u_{K}=U_{i}Re_{i}^{-1/4}$
and the dynamical length scale $l_{K}=LRe_{i}^{-3/4}$ to estimate dissipation, i.e. 
\begin{equation}
Q_{\nu}(L)\sim Q_{\nu/T}(L)\sim\mu L\frac{u_{K}^{2}}{l_{K}^{2}}=\frac{\mu^{3}}{\tilde{\rho}_{i}^{2}L^{3}}Re_{i}^{3}.\label{eq:Qs_Kologorov_estimates}
\end{equation}
This idea provides also estimates for 
\begin{subequations}
\begin{equation}
\frac{\mu\left\langle u_{z}^{2}\right\rangle _{h}}{L}\sim\mu\tilde{T}_{B}\left(\frac{\bigtriangleup\tilde{T}}{L}\right)^{2}\int_{0}^{z}\frac{\left\langle u_{z}^{2}\right\rangle _{h}}{\tilde{T}^{3}}\mathrm{d}z\sim \,\mu\frac{u_{K}^{2}}{L}=\frac{\mu^{3}}{\tilde{\rho}_{i}^{2}L^{3}}Re_{i}^{3/2},\label{eq:estimate1_bulk}
\end{equation}
\begin{equation}
\mu\frac{\mathrm{d}\left\langle u_{z}^{2}\right\rangle _{h}}{\mathrm{d}z}\sim \,\mu\frac{u_{K}^{2}}{l_{K}}=\frac{\mu^{3}}{\tilde{\rho}_{i}^{2}L^{3}}Re_{i}^{9/4},\label{eq:estimate2_bulk}
\end{equation}
\begin{equation}
\frac{k\left\langle u_{z}^{2}\right\rangle }{c_{p}L}\sim \,\frac{k}{c_{p}}\frac{u_{K}^{2}}{L}=\frac{\mu^{3}}{\tilde{\rho}_{i}^{2}L^{3}}Pr^{-1}Re_{i}^{3/2},\label{eq:estimate3_bulk}
\end{equation}
\end{subequations} 
in the bulk. Finally by the use of (\ref{eq:Ra_def}),
(\ref{eq:Nu_def_anapp}) and (\ref{eq:Delta_s_tilde}) one obtains
\begin{align}
\frac{\mu^{3}}{\tilde{\rho}_{i}^{2}L^{3}}\approx & \frac{\mu^{3}}{\tilde{\rho}_{i}^{2}L^{2}k\epsilon_{a}\tilde{T}_{B}}\frac{k\epsilon_{a}\tilde{T}_{B}\delta_{th,B}}{Lk\left(\Delta T_{S}\right)_{B}}\frac{k\left(\Delta T_{S}\right)_{B}}{\delta_{th,B}}\nonumber \\
\approx & \left(\frac{\tilde{\rho}_{B}}{\tilde{\rho}_{i}}\right)^{2}\frac{c_{p}^{2}\mu^{2}}{k^{2}}\frac{\mu k}{c_{p}^{2}\tilde{\rho}_{B}^{2}L^{2}\epsilon_{a}\Delta\tilde{T}}\tilde{\theta}Nu^{-1}F_{S}(z=0)\nonumber \\
\approx & \left(\frac{\tilde{\rho}_{B}}{\tilde{\rho}_{i}}\right)^{2}\ln\tilde{\Gamma}Pr^{2}Ra^{-1}Nu^{-1}F_{S}(z=0).\label{eq:factor_in_estimates}
\end{align}
It follows, that all the terms in (\ref{eq:viscous_terms_1}), (\ref{eq:viscous_terms_2})
and (\ref{eq:velocity_diss_term}) are always much smaller than the
dissipative terms $Q_{\nu}(z)$ and $Q_{\nu/T}(z)$ in (\ref{eq:Q1_and_Q2}).
Therefore the equations (\ref{eq:F_conv_superadiab_1-1}) and (\ref{eq:F_conv_superadiabatic_2-2})
can be written in a simpler, approximate form 
\begin{subequations}
\begin{align}
F_{S}(z=0)\approx & -k\frac{\mathrm{d}}{\mathrm{d}z}\left(\tilde{T}+\left\langle T'\right\rangle _{h}-T_{ad}\right)+\tilde{\rho}\tilde{T}\left\langle u_{z}s'\right\rangle _{h}\nonumber \\
 & +\frac{\Delta\tilde{T}}{L}\int_{0}^{z}\tilde{\rho}\left\langle u_{z}s'\right\rangle _{h}\mathrm{d}z-\mu\int_{0}^{z}\left\langle q\right\rangle _{h}\mathrm{d}z,\label{eq:F_superad_approx_rel}
\end{align}
\begin{equation}
F_{S}\left(z=0\right)\approx-k\frac{\tilde{T}_{B}}{\tilde{T}}\frac{\mathrm{d}}{\mathrm{d}z}\left(\tilde{T}+\left\langle T'\right\rangle _{h}-T_{ad}\right)+\tilde{\rho}\tilde{T}_{B}\left\langle u_{z}s'\right\rangle _{h}-\mu\int_{0}^{z}\frac{\tilde{T}_{B}}{\tilde{T}}\left\langle q\right\rangle _{h}\mathrm{d}z.\label{eq:F_superad_approx_rel_2}
\end{equation}
\end{subequations} 
up to 
\begin{equation}
\mathcal{O}\left(\frac{\delta_{th,B}}{L}F_{S}(z=0)\right)+\mathcal{O}\left(Ra^{-1}Nu^{-1}Re_{i}^{9/4}F_{S}(z=0)\right)\label{eq:rest_bulk}
\end{equation}
in the case of viscous dissipation dominated by the bulk contribution
or up to 
\begin{equation}
\mathcal{O}\left(\frac{\delta_{th,B}}{L}F_{S}(z=0)\right)+\mathcal{O}\left(Ra^{-1}Nu^{-1}Re_{i}^{3/2}F_{S}(z=0)\right)\label{eq:rest_BLs}
\end{equation}
in the case when viscous dissipation takes place predominantly in
the boundary layers.
It will be confirmed later in section \ref{subsec:Dynamical-description-ofA},
that the rests 
\begin{equation}
\mathcal{O}\left(Ra^{-1}Nu^{-1}Re_{i}^{9/4}F_{S}(z=0)\right)\quad\mathrm{and}\quad\mathcal{O}\left(Ra^{-1}Nu^{-1}Re_{i}^{3/2}F_{S}(z=0)\right)\label{eq:viscous_rests}
\end{equation}
are indeed negligibly small compared to the superadiabatic flux $F_{S}(z=0)$.
Finally we take the second relation, i.e. (\ref{eq:F_superad_approx_rel_2})
at $z=L$, which by the use of $F_{S}\left(z=0\right)=F_{S}(z=L)$
leads to
\begin{equation}
F_{S}\left(z=0\right)\left(\frac{1}{\tilde{T}_{T}}-\frac{1}{\tilde{T}_{B}}\right)=\mu\int_{0}^{L}\frac{1}{\tilde{T}}\left\langle q\right\rangle _{h}\mathrm{d}z=\frac{1}{\tilde{T}_{B}}Q_{\nu/T}(L).\label{eq:Entropy_production_integral}
\end{equation}

\subsection{Estimates of the ratios $r_{s}$, $r_{\delta}$, $r_{T}$ and $r_{U}$}

First we write down the leading order balance between inertia and
diffusion for the thermal and viscous boundary layers. In the simplest
case, when the viscous boundary layers are nested in the thermal ones,
$\delta_{th,T}>\delta_{\nu,T}$ and $\delta_{th,B}>\delta_{\nu,B}$
($Pr\lesssim1$), the inertia-diffusion balance takes the form
\begin{equation}
\frac{\tilde{\rho}_{B}U_{B}}{L}\approx\frac{k}{c_{p}\delta_{th,B}^{2}},\qquad\frac{\tilde{\rho}_{T}U_{T}}{L}\approx\frac{k}{c_{p}\delta_{th,T}^{2}},\label{eq:TEQ_balance_TBL-1}
\end{equation}
in the thermal layers, where (\ref{eq:DT_rel_Ds}) has been used, and
\begin{equation}
\frac{\tilde{\rho}_{B}U_{B}}{L}\approx\frac{\mu}{\delta_{\nu,B}^{2}},\qquad\frac{\tilde{\rho}_{T}U_{T}}{L}\approx\frac{\mu}{\delta_{\nu,T}^{2}},\label{eq:NS_horiz_balance_VBL}
\end{equation}
in the viscous layers. On dividing equations (\ref{eq:TEQ_balance_TBL-1})
by equations (\ref{eq:NS_horiz_balance_VBL}) respectively we get
\begin{equation}
\frac{\delta_{th,T}}{\delta_{\nu,T}}\approx Pr^{-1/2},\qquad\frac{\delta_{th,B}}{\delta_{\nu,B}}\approx Pr^{-1/2},\label{eq:delta_th_visc_1-1}
\end{equation}
thus
\begin{equation}
\frac{\delta_{th,T}}{L}\approx Re_{T}^{-1/2}Pr^{-1/2},\qquad\frac{\delta_{th,B}}{L}\approx Re_{B}^{-1/2}Pr^{-1/2}.\label{eq:delta_th_1-1}
\end{equation}
and hence also
\begin{equation}
r_{\delta}=\frac{\delta_{th,T}}{\delta_{th,B}}=\frac{\delta_{\nu,T}}{\delta_{\nu,B}}.\label{eq:r_delta_VT}
\end{equation}
In the case of thicker viscous layers $\delta_{th,T}<\delta_{\nu,T}$
and $\delta_{th,B}<\delta_{\nu,B}$ ($Pr\gtrsim1$) the velocity scale
in the thermal layers must be weakened with respect to the thermal
wind velocity by a factor $\delta_{th,i}/\delta_{\nu,i}$, which implies
the inertia-conduction balance in thermal layers in the form
\begin{equation}
\frac{\tilde{\rho}_{B}U_{B}}{L}\frac{\delta_{th,B}}{\delta_{\nu,B}}\approx\frac{k}{c_{p}\delta_{th,B}^{2}},\qquad\frac{\tilde{\rho}_{B}U_{T}}{L}\frac{\delta_{th,T}}{\delta_{\nu,T}}\approx\frac{K}{c_{p}\delta_{th,T}^{2}}.\label{eq:TEQ_balance_TBL-2}
\end{equation}
The dominant balance in the viscous boundary layers remains the same
(\ref{eq:NS_horiz_balance_VBL}), therefore on dividing equations
(\ref{eq:TEQ_balance_TBL-2}) by equations (\ref{eq:NS_horiz_balance_VBL})
respectively we get
\begin{equation}
\frac{\delta_{th,T}}{\delta_{\nu,T}}\approx Pr^{-1/3},\qquad\frac{\delta_{th,B}}{\delta_{\nu,B}}\approx Pr^{-1/3},\label{eq:delta_th_visc_1-2}
\end{equation}
thus
\begin{equation}
\frac{\delta_{th,T}}{L}\approx Re_{T}^{-1/2}Pr^{-1/3},\qquad\frac{\delta_{th,B}}{L}\approx Re_{B}^{-1/2}Pr^{-1/3}.\label{eq:delta_th_1-2}
\end{equation}
This clearly implies, that the ratios of the top to bottom thicknesses
of thermal boundary layers and top to bottom thicknesses of viscous
boundary layers are the same, cf. (\ref{eq:r_delta_VT}), no matter
the nesting between the thermal and viscous boundary layers. Moreover
from the equations (\ref{eq:delta_th_1-1}) and (\ref{eq:delta_th_1-2}),
supplied by the definitions of the Reynolds numbers in (\ref{eq:Re_numbers_def})
one obtains for both the cases, i.e. case 1: $\delta_{th,T}>\delta_{\nu,T}$,
$\delta_{th,B}>\delta_{\nu,B}$ ($Pr\lesssim1$) and case 2: $\delta_{th,T}<\delta_{\nu,T}$,
$\delta_{th,B}<\delta_{\nu,B}$ ($Pr\gtrsim1$) the following relation\footnote{We emphasize, that to estimate inertia in the boundary layers 
\[
\tilde{\rho}_{i}\left(\mathbf{u}_{h}\cdot\nabla+u_{z}\frac{\partial}{\partial z}\right)\mathbf{u}_{h}\approx\frac{\tilde{\rho}_{i}U_{i}^{2}}{L},
\]
the layer thickness $L$ was assumed as the horizontal length scale
of variation of velocity. This is suggested by results of numerical
simulations (cf. Verhoeven \emph{et al}. 2015, Jones \emph{et al}.
2020) and the reason for it may be, that although the dominant vertical
length scales in the bulk do scale with the pressure scale height,
the boundary layer wind of turbulence is selected by the longest horizontal
length scale over which the flow is coherent. With this approach the
thicknesses of the viscous boundary layers are simply given by (\ref{eq:viscous_BLs_thicknesses}),
but as remarked below (\ref{eq:Re_numbers_def}) it is possible that
at large stratifications $\tilde{\Gamma}\gg1$, the scale heights
determine both the vertical and horizontal length scales of variation
of the wind of turbulence; in such a case the inertial term in the boundary layers
can be estimated by $\tilde{\rho}_{i}U_{i}^{2}/D_{\rho}$, hence also
the definitions of the boundary layer thicknesses involve the scale
heights.}
\begin{equation}
r_{\delta}=\frac{\delta_{th,T}}{\delta_{th,B}}=\left(\frac{\tilde{\rho}_{B}U_{B}}{\tilde{\rho}_{T}U_{T}}\right)^{1/2}=\left(\frac{\tilde{\rho}_{B}}{\tilde{\rho}_{T}}\right)^{1/2}\frac{1}{r_{U}^{1/2}}=\left(\frac{\tilde{\Gamma}^{m}}{r_{U}}\right)^{1/2}.\label{eq:r_delta_rel}
\end{equation}
Gathering now the equations (\ref{eq:r_delta}) and (\ref{eq:r_delta_rel})
yields
\begin{equation}
r_{\delta}=r_{T},\quad r_{s}=\tilde{\Gamma}r_{T},\quad r_{U}r_{\delta}^{2}=\tilde{\Gamma}^{m},\label{eq:ratios}
\end{equation}
so that expressing things by $r_{U}$ we get
\begin{equation}
r_{\delta}=r_{T}=\frac{\tilde{\Gamma}^{m/2}}{r_{U}^{1/2}},\quad r_{s}=\frac{\tilde{\Gamma}^{m/2+1}}{r_{U}^{1/2}}.\label{eq:ratios_by_rU}
\end{equation}
The next step is evaluate somehow the velocity ratio $r_{U}$. It
should be made clear, that this step is the most speculative one in
the analysis of fully developed convection presented here. Nevertheless,
to obtain an estimate of $r_{U}$ it seems reasonable to consider
an analogue of the 'Deardorff' balance between mean inertia and mean
buoyancy (cf. Deardorff 1970). We therefore consider the stationary
Navier-Stokes equation multiplied by $\mathbf{u}$ and horizontally
averaged and assume, that the dominant terms in the resulting equation
are the inertial and buoyancy terms
\begin{equation}
\frac{1}{2\tilde{\rho}}\frac{\partial}{\partial z}\left(\tilde{\rho}\left\langle u_{z}u^{2}\right\rangle _{h}\right)\approx\frac{g}{c_{p}}\left\langle u_{z}s'\right\rangle _{h},\label{ff:1}
\end{equation}
which may also be rewritten in the form
\begin{equation}
-\frac{1}{2D_{\rho}}\left\langle u_{z}u^{2}\right\rangle _{h}+\frac{1}{2}\left\langle \frac{\partial u_{z}}{\partial z}u^{2}\right\rangle _{h}+\frac{1}{2}\left\langle u_{z}\frac{\partial u^{2}}{\partial z}\right\rangle _{h}\approx\frac{g}{c_{p}}\left\langle u_{z}s'\right\rangle _{h}.\label{ff:2}
\end{equation}
Assuming that the {\textit{vertical}} scale of variation of velocity outside
the boundary layers is determined by the density scale heights at
top and bottom we get
\begin{equation}
\frac{c_{p}}{g}\frac{U_{T}^{3}}{D_{\rho,T}}\approx\left[\left\langle u_{z}s'\right\rangle _{h}\right]_{T},\qquad\frac{c_{p}}{g}\frac{U_{B}^{3}}{D_{\rho,B}}\approx\left[\left\langle u_{z}s'\right\rangle _{h}\right]_{B}.\label{eq:1}
\end{equation}
It is postulated, that in turbulent convection the vertical velocity,
which is small in the boundary layers and in their vicinity in the bulk, is quickly amplified by strong
buoyancy which becomes important away from the boundary layers within upwelling large-scale convective currents (convection cells of the wind of turbulence). The vertical velocity is effectively assumed to become comparable with the horizontal one within the distances $D_{\rho,B}$ and $D_{\rho,T}$ away from the bottom and top boundaries respectively. Nevertheless we will still assume in this case, that the magnitudes of velocities can be approximated by $U_B$ and $U_T$ near the bottom and top respectively, which allows
to estimate the means $\left[\left\langle u_{z}\partial_{z}u^{2}\right\rangle _{h}\right]_{i}\sim\left[\left\langle u_{z}u^{2}\right\rangle _{h}\right]_{i}/D_{\rho}\sim U_{i}^{3}/D_{\rho}$,
where $i=B$ or $T$ for the bottom and top balance respectively. 

At this stage one needs to consider separately the different cases
determined by whether the dominant contributions to viscous and thermal
dissipation come from the bulk or boundary layers. For the sake of
simplicity and due to lack of sufficient experimental and numerical
data, we will consider only the two, perhaps simplest cases, when
the thermal dissipation takes place predominantly in the boundary
layers, but the viscous dissipation can be dominant either also in
the boundary layers or in the bulk. It is clear from figure (\ref{fig:PrRa_plane_diagram_GL2000}),
that at least for the Boussinesq convection the two aforementioned regimes
are the first to appear as the Rayleigh number increases and exceeds
a critical value for fully developed convection, whereas other regimes
appear for even much higher values of the Rayleigh number.

\subsubsection{Viscous and thermal dissipation predominantly in the boundary layers}

Since the dissipation in the bulk is negligible, estimates of the
superadiabatic heat flux at the top and bottom of the bulk, therefore
just above the bottom boundary layer and below the top one, according
to (\ref{eq:F_superad_approx_rel_2}) involve at the leading order
only advection. This is because conduction is negligible, and the
viscous dissipation integral at both locations, $z=\delta_{th,B}$
and $z=L-\delta_{th,T}$ (and in fact in the entire bulk) is dominated by the contribution from the
bottom boundary layer, which is approximately $-\mu\int_{0}^{\delta_{th,B}}\left\langle q\right\rangle _{h}\mathrm{d}z.$
This allows to write
\begin{equation}
\tilde{\rho}_{B}\left[\left\langle u_{z}s'\right\rangle _{h}\right]_{B}\approx\tilde{\rho}_{T}\left[\left\langle u_{z}s'\right\rangle _{h}\right]_{T},\label{eq:2}
\end{equation}
so that the latter together with (\ref{eq:1})\footnote{Once again, we stress, that the relation (\ref{eq:1}) is the weakest point of the presented analysis, despite the fact, that it leads to a rather satisfactory agreement with results of numerical simulations of Jones \emph{et al}. (2020)} produce
\begin{equation}
\frac{\left[\left\langle u_{z}s'\right\rangle _{h}\right]_{T}}{\left[\left\langle u_{z}s'\right\rangle _{h}\right]_{B}}\approx\frac{\tilde{\rho}_{B}}{\tilde{\rho}_{T}}=\tilde{\Gamma}^{m}\approx\frac{U_{T}^{3}}{U_{B}^{3}}\frac{D_{\rho,B}}{D_{\rho,T}},\label{eq:ratio_s}
\end{equation}
or equivalently
\begin{equation}
r_{U}=\tilde{\Gamma}^{(m-1)/3}.\label{eq:r_U1}
\end{equation}
It follows from (\ref{eq:ratios_by_rU}), that
\begin{equation}
r_{\delta}=r_{T}=\tilde{\Gamma}^{(2m+1)/6},\quad r_{s}=\tilde{\Gamma}^{(2m+7)/6}.\label{eq:ratios1}
\end{equation}

\subsubsection{Viscous dissipation predominantly in the bulk, thermal dissipation
dominated by contributions from boundary layers}

Now the dominant contributions to the mean superadiabatic
heat flux at the top and bottom of the bulk, by the use of the first
heat production formula (\ref{eq:F_superad_approx_rel}) are
\begin{equation}
F_{S}(z=0)\approx\tilde{\rho}_{T}\tilde{T}_{T}\left[\left\langle u_{z}s'\right\rangle _{h}\right]_{T}\approx\tilde{\rho}_{B}\tilde{T}_{B}\left[\left\langle u_{z}s'\right\rangle _{h}\right]_{B},\label{eq:top_bot_fluxes_c2-1}
\end{equation}
since at the bottom the work of the buoyancy force and viscous
dissipation integral are negligible and at the top, according to the
global balance $g\left\langle \tilde{\rho}u_{z}s'\right\rangle /c_{p}=\mu\left\langle q\right\rangle $
in (\ref{eq:e192}) they are approximately equal and thus cancel out.
Consequently
\begin{equation}
\frac{\left[\left\langle u_{z}s'\right\rangle _{h}\right]_{T}}{\left[\left\langle u_{z}s'\right\rangle _{h}\right]_{B}}\approx\frac{\tilde{\rho}_{B}\tilde{T}_{B}}{\tilde{\rho}_{T}\tilde{T}_{T}}=\tilde{\Gamma}^{m+1}\approx\frac{U_{T}^{3}}{U_{B}^{3}}\frac{D_{\rho,B}}{D_{\rho,T}},\label{eq:ratio_s-1}
\end{equation}
or equivalently
\begin{equation}
r_{U}=\tilde{\Gamma}^{m/3}.\label{eq:r_U2}
\end{equation}
Calculating the other ratios from (\ref{eq:ratios_by_rU}) we get
\begin{equation}
r_{\delta}=r_{T}=\tilde{\Gamma}^{m/3},\quad r_{s}=\tilde{\Gamma}^{m/3+1}.\label{eq:ratios2}
\end{equation}

\subsection{Scaling laws for fully developed stratified convection with isentropic
boundaries\label{subsec:Dynamical-description-ofA}}

We are now ready to derive the scaling laws for the Nusselt and Reynolds
numbers versus the driving force measured by the Rayleigh number.
We start by observing, that the relation between the Nusselt number
and the Reynolds number is now easily obtained. In the first case,
when thermal boundary layers are thicker than the viscous layers,
$\delta_{th,T}>\delta_{\nu,T}$ and $\delta_{th,B}>\delta_{\nu,B}$
($Pr\lesssim1$), from (\ref{eq:delta_th_B}) and (\ref{eq:delta_th_1-1})
we immediately get
\begin{equation}
Nu=\frac{\tilde{\Gamma}\ln\tilde{\Gamma}}{\left(1+r_{s}\right)\left(\tilde{\Gamma}-1\right)}Re_{B}^{1/2}Pr^{1/2}.\label{eq:Nu_ReB_2}
\end{equation}
For the second case of thermal layers nested in the viscous ones,
$\delta_{th,T}<\delta_{\nu,T}$ and $\delta_{th,B}<\delta_{\nu,B}$
($Pr\gtrsim1$), by (\ref{eq:delta_th_B}) and (\ref{eq:delta_th_1-2})
we immediately get
\begin{equation}
Nu=\frac{\tilde{\Gamma}\ln\tilde{\Gamma}}{\left(1+r_{s}\right)\left(\tilde{\Gamma}-1\right)}Re_{B}^{1/2}Pr^{1/3}.\label{eq:Nu_ReB_1}
\end{equation}
These results are due to the fact, that the thermal dissipation in
all the cases considered here is dominated by the contributions from
boundary layers and is independent of whether the viscous dissipation
takes place predominantly in the bulk or in the boundary layers. Let
us now turn to these cases separately.

\subsubsection{Viscous and thermal dissipation predominantly in the boundary layers}

We take the relation (\ref{eq:Entropy_production_integral}) and estimate
the term $Q_{\nu/T}(L)$ with a sum of the dominant contributions from boundary
layers
\begin{equation}
Q_{\nu/T}(L)\approx\mu\tilde{T}_{B}\left(\frac{U_{B}^{2}}{\delta_{\nu,B}\tilde{T}_{B}}+\frac{U_{T}^{2}}{\delta_{\nu,T}\tilde{T}_{T}}\right)=\frac{\mu U_{B}^{2}}{\delta_{\nu,B}}\left(1+\tilde{\Gamma}\frac{r_{U}^{2}}{r_{\delta}}\right),\label{eq:Qnu_est-1}
\end{equation}
which in light of (\ref{eq:Ra_def}) and
\begin{equation}
F_{S}\left(z=0\right)\approx k\frac{\left(\bigtriangleup T'\right)_{B}}{\delta_{th,B}}\approx k\frac{\tilde{T}_{B}}{c_{p}\left(1+r_{s}\right)}\frac{\Delta\tilde{s}}{\delta_{th,B}},\label{eq:F_S_by_Delta_s_tilde}
\end{equation}
(cf. (\ref{eq:bot_top_temp_grads}), (\ref{eq:Delta_TSB_by_other_Deltas})
and (\ref{eq:Delta_TSB_by_other_Deltas-1})) allows to write down
\begin{equation}
\frac{\tilde{\Gamma}}{\left(1+r_{s}\right)}RaPr^{-2}\approx\frac{\delta_{th,B}}{\delta_{\nu,B}}Re_{B}^{2}\left(1+\tilde{\Gamma}\frac{r_{U}^{2}}{r_{\delta}}\right).\label{eq:f1}
\end{equation}
The latter, by the use of (\ref{eq:r_U1}) and (\ref{eq:ratios1}) is equivalent to
\begin{equation}
Re_{B}^{2}\approx\frac{\delta_{\nu,B}}{\delta_{th,B}}\frac{\tilde{\Gamma}}{\left(1+\tilde{\Gamma}^{(2m+1)/6}\right)\left(1+\tilde{\Gamma}^{(2m+7)/6}\right)}RaPr^{-2}.\label{eq:Second_Nu_ReB_VBulk-1}
\end{equation}

\paragraph{case 1: thermal layers thicker than viscous layers, $\delta_{th,T}>\delta_{\nu,T}$
and $\delta_{th,B}>\delta_{\nu,B}$ ($Pr\lesssim1$)}

~

~

\noindent From (\ref{eq:Nu_ReB_2}), (\ref{eq:Second_Nu_ReB_VBulk-1}) and (\ref{eq:delta_th_visc_1-1})
we easily get the scaling laws\index{SI}{scaling laws}
\begin{align}
Nu & \approx\frac{\tilde{\Gamma}^{5/4}\ln\tilde{\Gamma}}{\left(1+\tilde{\Gamma}^{(2m+7)/6}\right)^{5/4}\left(1+\tilde{\Gamma}^{(2m+1)/6}\right)^{1/4}\left(\tilde{\Gamma}-1\right)}Pr^{1/8}Ra^{1/4}\nonumber \\
 & \overset{\tilde{\Gamma}\gg1}{\longrightarrow}\frac{\ln\tilde{\Gamma}}{\tilde{\Gamma}^{(2m+5)/4}}Pr^{1/8}Ra^{1/4},\label{eq:Nu_sc_1}
\end{align}
\begin{align}
Re_{B}=\tilde{\Gamma}^{(2m+1)/3}Re_{T} & \approx\left[\frac{\tilde{\Gamma}}{\left(1+\tilde{\Gamma}^{(2m+1)/6}\right)\left(1+\tilde{\Gamma}^{(2m+7)/6}\right)}\right]^{1/2}Pr^{-3/4}Ra^{1/2}\nonumber \\
 & \overset{\tilde{\Gamma}\gg1}{\longrightarrow}\tilde{\Gamma}^{-(2m+1)/6}Pr^{-3/4}Ra^{1/2},\label{eq:Re_sc_1}
\end{align}
supplied by (\ref{eq:delta_th_B},b), which now take the form 
\begin{subequations}
\begin{equation}
\frac{\delta_{th,B}}{L}\approx\frac{\tilde{\Gamma}\ln\tilde{\Gamma}}{\left(1+r_{s}\right)\left(\tilde{\Gamma}-1\right)}Nu^{-1}\overset{\tilde{\Gamma}\gg1}{\longrightarrow}\tilde{\Gamma}^{(2m+1)/12}Pr^{-1/8}Ra^{-1/4},\label{eq:delta_th_B-1}
\end{equation}
\begin{equation}
\frac{\delta_{th,T}}{\delta_{th,B}}\approx\frac{r_{s}}{\tilde{\Gamma}}\approx\tilde{\Gamma}^{(2m+1)/6}>1.\label{eq:delta_th_T-1}
\end{equation}
\end{subequations}
For completeness we recall (\ref{eq:Nu_ReB_2})
in the form
\begin{equation}
Nu=\frac{\tilde{\Gamma}\ln\tilde{\Gamma}}{\left(1+\tilde{\Gamma}^{(2m+7)/6}\right)\left(\tilde{\Gamma}-1\right)}Re_{B}^{1/2}Pr^{1/2}.\label{eq:Nu_ReB_2-1}
\end{equation}
The large $\tilde{\Gamma}$ limit requires care, because we have assumed
incompressibility of the boundary layers, in other words by assumption
the density stratification cannot exceed a critical large value, at
which the density scale height becomes comparable and less than the
thicknesses of the boundary layers; of course the smallest value of
$D_{\rho}(z)$ is achieved at the top, where also the boundary layer
is thicker than at the bottom, therefore the strongest constraint
on validity of the results of this section is obtained at the top.
More precisely from (\ref{eq:DT_rel_Ds_full}) it is clear, that the
term $(\tilde{\theta}\delta_{th,i}\tilde{T}_{B})/(L\tilde{T}_{i})$
must be small and the most restrictive constraint is, indeed, obtained
at the top, $\left(\tilde{\Gamma}-1\right)\delta_{th,T}/L\ll1$, so
that the large $\tilde{\Gamma}$ limit in the above formulae has an
upper bound, i.e. it corresponds to
\begin{equation}
1\ll\tilde{\Gamma}\ll Pr^{1/(4m+10)}Ra^{1/(2m+5)},\label{eq:large_Gamma_limit_1}
\end{equation}
which guarantees, that the boundary layers remain incompressible.

It remains to be verified, that the remainders (\ref{eq:estimate1}-d),
which were assumed negligible are indeed small. The expression (\ref{eq:estimate2})
with the largest power of the Reynolds number is most likely to be
the largest of the remainders, therefore consistency requires (cf. \ref{eq:factor_in_estimates})
\begin{equation}
\left(\frac{\tilde{\rho}_{B}}{\tilde{\rho}_{i}}\right)^{2}\ln\tilde{\Gamma}Pr^{2}Ra^{-1}Nu^{-1}Re_{i}^{3/2}\ll1,\label{eq:consistency1}
\end{equation}
which is largest at the top, that is for $i=T$, as long as $m>1/2$
or equivalently $\gamma<3$; since $\gamma>3$ for a fluid is highly
unlikely, we proceed with the assumption $m>1/2$. It follows from
(\ref{eq:consistency1}) taken at the top and the scaling laws (\ref{eq:Nu_sc_1}), (\ref{eq:Re_sc_1}), that the stratification
parameter $\tilde{\Gamma}$ must also satisfy $\tilde{\Gamma}\ll Ra^{1/(2m+1)}Pr^{-3/(4m+2)}$,
which is already satisfied by (\ref{eq:large_Gamma_limit_1}) since
\begin{equation}
\left(\frac{Ra}{Pr}\right)^{4/(2m+5)(2m+1)}>1,\label{eq:consistency2}
\end{equation}
in fully developed convection.

\paragraph{case 2: viscous layers thicker than thermal layers, $\delta_{th,T}<\delta_{\nu,T}$
and $\delta_{th,B}<\delta_{\nu,B}$ ($Pr\gtrsim1$)}

~

~

In this case we use the equations (\ref{eq:Nu_ReB_1}), (\ref{eq:Second_Nu_ReB_VBulk-1})
and (\ref{eq:delta_th_visc_1-2}) to obtain the scaling laws\index{SI}{scaling laws}
\begin{align}
Nu & \approx\frac{\tilde{\Gamma}^{5/4}\ln\tilde{\Gamma}}{\left(1+\tilde{\Gamma}^{(2m+7)/6}\right)^{5/4}\left(1+\tilde{\Gamma}^{(2m+1)/6}\right)^{1/4}\left(\tilde{\Gamma}-1\right)}Pr^{-1/12}Ra^{1/4}\nonumber \\
 & \overset{\tilde{\Gamma}\gg1}{\longrightarrow}\frac{\ln\tilde{\Gamma}}{\tilde{\Gamma}^{(2m+5)/4}}Pr^{-1/12}Ra^{1/4},\label{eq:Nu_sc_2}
\end{align}
\begin{align}
Re_{B}=\tilde{\Gamma}^{(2m+1)/3}Re_{T} & \approx\left[\frac{\tilde{\Gamma}}{\left(1+\tilde{\Gamma}^{(2m+1)/6}\right)\left(1+\tilde{\Gamma}^{(2m+7)/6}\right)}\right]^{1/2}Pr^{-5/6}Ra^{1/2}\nonumber \\
 & \overset{\tilde{\Gamma}\gg1}{\longrightarrow}\tilde{\Gamma}^{-(2m+1)/6}Pr^{-5/6}Ra^{1/2},\label{eq:Re_sc_2}
\end{align}
and from (\ref{eq:delta_th_B},b) it follows, that 
\begin{subequations}
\begin{equation}
\frac{\delta_{th,B}}{L}\approx\frac{\tilde{\Gamma}\ln\tilde{\Gamma}}{\left(1+r_{s}\right)\left(\tilde{\Gamma}-1\right)}Nu^{-1}\overset{\tilde{\Gamma}\gg1}{\longrightarrow}\tilde{\Gamma}^{(2m+1)/12}Pr^{1/12}Ra^{-1/4},\label{eq:delta_th_B-1-1}
\end{equation}
\begin{equation}
\frac{\delta_{th,T}}{\delta_{th,B}}\approx\frac{r_{s}}{\tilde{\Gamma}}\approx\tilde{\Gamma}^{(2m+1)/6}>1.\label{eq:delta_th_T-1-1}
\end{equation}
\end{subequations}
For completeness we recall (\ref{eq:Nu_ReB_1})
\begin{equation}
Nu=\frac{\tilde{\Gamma}\ln\tilde{\Gamma}}{\left(1+\tilde{\Gamma}^{(2m+7)/6}\right)\left(\tilde{\Gamma}-1\right)}Re_{B}^{1/2}Pr^{1/3}.\label{eq:Nu_ReB_1_fin}
\end{equation}
The large $\tilde{\Gamma}$ limit can only be taken up to
\begin{equation}
\tilde{\Gamma}\ll Pr^{-1/(6m+15)}Ra^{1/(2m+5)},\label{eq:large_Gamma_limit_2}
\end{equation}
so that the boundary layers remain incompressible. Consistency with
(\ref{eq:consistency1}) requires $\tilde{\Gamma}\ll Ra^{1/(2m+1)}Pr^{-5/(6m+3)}$,
which is satisfied by (\ref{eq:large_Gamma_limit_2}) in fully developed
convection at high Rayleigh numbers and moderately high Prandtl numbers,
\begin{equation}
\left(\frac{Ra}{Pr^{2m/3+2}}\right)^{4/(2m+5)(2m+1)}>1.\label{eq:consistency2-1}
\end{equation}

\subsubsection{Viscous dissipation predominantly in the bulk, thermal dissipation
dominated by contributions from boundary layers}

Again, we start with the relation (\ref{eq:Entropy_production_integral}).
First we make the previous estimate of $Q_{\nu/T}(L)$, provided in
(\ref{eq:Qs_estimates_diss_in_bulk}), somewhat more precise. The
viscous integral $Q_{\nu}(L)$ can be estimated from the Navier-Stokes
equation as follows
\begin{equation}
Q_{\nu}(L)=\mu\int_{0}^{L}\left\langle q\right\rangle _{h}\mathrm{d}z\approx\mu\int_{\delta_{th,B}}^{L-\delta_{th,T}}\left\langle q\right\rangle _{h}\mathrm{d}z\approx\tilde{\rho}_{B}U_{B}^{3}=\frac{\mu^{3}}{\tilde{\rho}_{B}^{2}L^{3}}Re_{B}^{3},\label{eq:Qnu_est}
\end{equation}
since the viscous dissipation is dominant in the bulk, where it is
expected to balance the nonlinear inertial term. Moreover, since $r_{U}^{3}=\tilde{\Gamma}^{m}$
in the current case, the maximal estimate of $Q_{\nu}(L)$ is obtained
either by taking $\tilde{\rho}_{B}U_{B}^{3}$ or equivalently $\tilde{\rho}_{T}U_{T}^{3}\approx\tilde{\rho}_{B}U_{B}^{3}$.
Next we observe, that
\begin{equation}
Q_{\nu/T}(L)\lesssim\tilde{\Gamma}Q_{\nu}(L),\label{eq:QQ}
\end{equation}
so that by (\ref{eq:Entropy_production_integral}), (\ref{eq:Qnu_est}),
(\ref{eq:QQ}) and (\ref{eq:factor_in_estimates}) we can finally
write 
\begin{equation}
\tilde{\Gamma}\ln\tilde{\Gamma}Pr^{2}Ra^{-1}Nu^{-1}Re_{B}^{3}\approx\tilde{\Gamma}-1,\label{eq:f3}
\end{equation}
or equivalently
\begin{equation}
RaNuPr^{-2}\approx\frac{\tilde{\Gamma}\ln\tilde{\Gamma}}{\tilde{\Gamma}-1}Re_{B}^{3}.\label{eq:Second_Nu_ReB_VBulk}
\end{equation}

\paragraph{case 1: thermal layers thicker than viscous layers, $\delta_{th,T}>\delta_{\nu,T}$
and $\delta_{th,B}>\delta_{\nu,B}$ ($Pr\lesssim1$)}

~

~

\noindent Using (\ref{eq:Nu_ReB_2}), (\ref{eq:Second_Nu_ReB_VBulk}) and (\ref{eq:ratios2})
one obtains the following scaling laws\index{SI}{scaling laws}
\begin{align}
Nu & \approx\frac{\tilde{\Gamma}\ln\tilde{\Gamma}}{\left(1+\tilde{\Gamma}^{m/3+1}\right)^{6/5}\left(\tilde{\Gamma}-1\right)}Pr^{1/5}Ra^{1/5}\nonumber \\
 & \overset{\tilde{\Gamma}\gg1}{\longrightarrow}\frac{\ln\tilde{\Gamma}}{\Gamma^{(2m+6)/5}}Pr^{1/5}Ra^{1/5},\label{eq:Nu_sc_3}
\end{align}
\begin{align}
Re_{B}=\tilde{\Gamma}^{2m/3}Re_{T} & \approx\frac{1}{\left(1+\tilde{\Gamma}^{m/3+1}\right)^{2/5}}Pr^{-3/5}Ra^{2/5}\nonumber \\
 & \overset{\tilde{\Gamma}\gg1}{\longrightarrow}\tilde{\Gamma}^{-(2m+6)/15}Pr^{-3/5}Ra^{2/5},\label{eq:Re_sc_3}
\end{align}
and consequently (\ref{eq:delta_th_B},b) implies 
\begin{subequations}
\begin{equation}
\frac{\delta_{th,B}}{L}\approx\frac{\tilde{\Gamma}\ln\tilde{\Gamma}}{\left(1+r_{s}\right)\left(\tilde{\Gamma}-1\right)}Nu^{-1}\overset{\tilde{\Gamma}\gg1}{\longrightarrow}\tilde{\Gamma}^{(m+3)/15}Pr^{-1/5}Ra^{-1/5},\label{eq:delta_th_B-1-2}
\end{equation}
\begin{equation}
\frac{\delta_{th,T}}{\delta_{th,B}}\approx\frac{r_{s}}{\tilde{\Gamma}}\approx\tilde{\Gamma}^{m/3}>1.\label{eq:delta_th_T-1-2}
\end{equation}
\end{subequations}
For completeness (\ref{eq:Nu_ReB_2}) is recalled
in an explicit form
\begin{equation}
Nu=\frac{\tilde{\Gamma}\ln\tilde{\Gamma}}{\left(1+\tilde{\Gamma}^{m/3+1}\right)\left(\tilde{\Gamma}-1\right)}Re_{B}^{1/2}Pr^{1/2}.\label{eq:Nu_ReB_2-2}
\end{equation}
Due to the assumption of incompressibility of the boundary layers (cf.
(\ref{eq:DT_rel_Ds_full}) which implies $\tilde{\theta}\tilde{\Gamma}\delta_{th,T}/L\ll1$)
the large $\tilde{\Gamma}$ limit can only be taken up to
\begin{equation}
\tilde{\Gamma}\ll\left(PrRa\right)^{1/(2m+6)},\label{eq:large_Gamma_limit_2-1}
\end{equation}
so that the boundary layers remain incompressible. 

On the other hand neglection
of the terms (\ref{eq:estimate1_bulk}-c) in the process of derivation
of the scaling laws requires verification, that these terms are indeed
small in comparison with the total superadiabatic heat flux $F_{S}(z=0)$.
The expression (\ref{eq:estimate2_bulk}) with the largest power of
the Reynolds number is the largest, therefore consistency requires
(cf. \ref{eq:factor_in_estimates})
\begin{equation}
\left(\frac{\tilde{\rho}_{B}}{\tilde{\rho}_{i}}\right)^{2}\ln\tilde{\Gamma}Pr^{2}Ra^{-1}Nu^{-1}Re_{i}^{9/4}\ll1,\label{eq:consistency1-1}
\end{equation}
and the strongest restriction is obtained at the top, that is for
$i=T$. It follows, that the stratification parameter $\tilde{\Gamma}$
must also satisfy $\tilde{\Gamma}\ll Ra^{1/(2m+1)}Pr^{-3/(4m+2)}$,
which is already satisfied by (\ref{eq:large_Gamma_limit_1}) since
\begin{equation}
\left(\frac{Ra}{Pr^{m+2}}\right)^{5/(2m+6)(2m+1)}>1,\label{eq:consistency2-2}
\end{equation}
in fully developed convection.

\paragraph{case 2: viscous layers thicker than thermal layers, $\delta_{th,T}<\delta_{\nu,T}$
and $\delta_{th,B}<\delta_{\nu,B}$ ($Pr\gtrsim1$)}

~

~

\noindent From (\ref{eq:Nu_ReB_1}), (\ref{eq:Second_Nu_ReB_VBulk}) and (\ref{eq:ratios2})
we get the scaling laws\index{SI}{scaling laws}
\begin{align}
Nu & \approx\frac{\tilde{\Gamma}\ln\tilde{\Gamma}}{\left(1+\tilde{\Gamma}^{m/3+1}\right)^{6/5}\left(\tilde{\Gamma}-1\right)}Ra^{1/5}\nonumber \\
 & \overset{\tilde{\Gamma}\gg1}{\longrightarrow}\frac{\ln\tilde{\Gamma}}{\Gamma^{(2m+6)/5}}Ra^{1/5},\label{eq:Nu_sc_4}
\end{align}
\begin{align}
Re_{B}=\tilde{\Gamma}^{2m/3}Re_{T} & \approx\frac{1}{\left(1+\tilde{\Gamma}^{m/3+1}\right)^{2/5}}Pr^{-2/3}Ra^{2/5}\nonumber \\
 & \overset{\tilde{\Gamma}\gg1}{\longrightarrow}\tilde{\Gamma}^{-(2m+6)/15}Pr^{-2/3}Ra^{2/5},\label{eq:Re_sc_4}
\end{align}
and by the use of (\ref{eq:delta_th_B},b) 
\begin{subequations}
\begin{equation}
\frac{\delta_{th,B}}{L}\approx\frac{\tilde{\Gamma}\ln\tilde{\Gamma}}{\left(1+r_{s}\right)\left(\tilde{\Gamma}-1\right)}Nu^{-1}\overset{\tilde{\Gamma}\gg1}{\longrightarrow}\tilde{\Gamma}^{(m+3)/15}Ra^{-1/5},\label{eq:delta_th_B-1-2-1}
\end{equation}
\begin{equation}
\frac{\delta_{th,T}}{\delta_{th,B}}\approx\frac{r_{s}}{\tilde{\Gamma}}\approx\tilde{\Gamma}^{m/3}>1.\label{eq:delta_th_T-1-2-1}
\end{equation}
\end{subequations}
For completeness we provide (\ref{eq:Nu_ReB_1})
in the form
\begin{equation}
Nu=\frac{\tilde{\Gamma}\ln\tilde{\Gamma}}{\left(1+\tilde{\Gamma}^{m/3+1}\right)\left(\tilde{\Gamma}-1\right)}Re_{B}^{1/2}Pr^{1/3}.\label{eq:Nu_ReB_1-1}
\end{equation}
The large $\tilde{\Gamma}$ limit can only be taken up to
\begin{equation}
\tilde{\Gamma}\ll Ra^{1/(2m+6)},\label{eq:large_Gamma_limit_2-1-1}
\end{equation}
so that the boundary layers remain incompressible. Consistency with
(\ref{eq:consistency1-1}) requires $\tilde{\Gamma}\ll Ra^{1/(2m+1)}Pr^{-5/(6m+3)}$,
which is satisfied by (\ref{eq:large_Gamma_limit_2-1-1}) in fully
developed convection at high Rayleigh numbers and moderately high
Prandtl numbers,
\begin{equation}
\left(\frac{Ra^{1/(2m+6)}}{Pr^{1/3}}\right)^{5/(2m+1)}>1.\label{eq:consistency2-2-1}
\end{equation}

\subsection{Discussion}

\noindent The influence of stratification on the dynamics of fully
developed convection is, indeed, substantial. In a stationary state
the heat flux entering at the bottom is equal to the heat flux leaving
the system at the top as in the Boussinesq case, but the work done
by the buoyancy force and the viscous heating\index{SI}{viscous heating} are no longer negligible
and are of the same order as the total heat flux in the system. Therefore
the total heat flux passing through every plane $z=\textrm{const}.$
within the fluid domain is no longer the same. In the case when most
of the viscous dissipation takes place in the boundary layers the
heat flux entering at the bottom is increased by the viscous heating in the bottom boundary
layer, then the work done by the buoyancy force reduces
the flux in the bulk so that it falls below the value at the enter
and then it is again boosted by the viscous heating in the top boundary
layer to reach the same value at the top boundary as at the bottom
boundary (cf. (\ref{eq:heta_flux_balance_BULK_AN}) and (\ref{eq:e192})). 

\noindent It should also be noted that direct quantitative comparisons
of the results obtained in here taken at $\tilde{\theta}=0$ with
the Boussinesq case cannot not be carried out for two reasons. Firstly
no experimental data on compressible convection is available to provide
the prefactors in the relations used (e.g. the estimates of the viscous
dissipation in the system, the final scaling laws). More importantly,
however, in most of our analysis it is assumed that the viscous heating
and work of buoyancy are of comparable magnitude with the total heat
flux and hence that the kinetic energy is comparable with the thermal
energy, which is not true in the Boussinesq case. Hence the analysis
presented here is valid only for finite values of the compressibility
$\tilde{\theta}$.

\noindent Furthermore, we note that the top boundary layer is always
thicker than the bottom one and that the entropy jump across the boundary
layer is always greater at the top, that is $r_{\delta}>1$ and $r_{s}>1$.
Moreover, the thicknesses of the boundary layers generally increase
with the stratification parameter $\tilde{\Gamma}$, in other words
the thermal boundary layers thicken as the density scale height decreases.
This means, that for strong enough stratifications the boundary layers
become compressible and since the top boundary layer is thicker, it
is also much more prone to such a transition. 

\noindent Next we may observe, that since typically $r_{U}$ also
exceeds unity (always when $m>1$), the convective velocities are
also larger close to the top of the flow domain than in the bottom
region. The latter fact, together with $\delta_{th,T}>\delta_{th,B}$
mean that the top boundary layer is also more prone to instability,
i.e. is more likely to become turbulent at high Rayleigh numbers. 

\noindent The validity restrictions for the presented approach, that
is the upper bounds in (\ref{eq:large_Gamma_limit_1}), (\ref{eq:large_Gamma_limit_2}),
(\ref{eq:large_Gamma_limit_2-1}) and (\ref{eq:large_Gamma_limit_2-1-1}),
which result from the assumed incompressibility of the boundary layers
indicate, that for moderately high Rayleigh numbers, about $10^{7}$
and $10^{8}$, even relatively weak stratifications, with $\tilde{\Gamma}\approx10$
lead to compressible boundary layers (at least the top one), thus
fall out of the regime of validity for the presented theory. Smaller
values of the polytropic index $m$ allow for larger stratifications
at which $\delta_{th,T}\ll D_{\rho}(L)$ is still satisfied and the
theory remains valid, but generally speaking the large stratification
limit with incompressible boundary layers is expected for larger values
of $Ra\gtrsim10^{10}$.

\section{Validity of the approximation and summary\label{sec:Validity-A}}

The anelastic approximation is designed for description of systems
with significant stratification
\begin{equation}
L\sim D_{\rho},\,D_{T},\,D_{p},\label{eq:e232}
\end{equation}
such as e.g. planetary atmospheres or planetary and stellar interiors.
The shear and bulk dynamical viscosities $\mu(x,y,z)$ and $\mu_{b}(x,y,z)$,
the specific heats $c_{v}(x,y,z)$ and $c_{p}(x,y,z)$ are allowed
any spatial variation, whereas the thermal conduction $k(z)$ and
the gravitational acceleration $\mathbf{g}=-g(z)\hat{\mathbf{e}}_{z}$
are typically assumed only depth-dependent. If the boundary conditions
are assumed time-independent the hydrostatic reference state satisfies
the following equations
\begin{equation}
\frac{\mathrm{d}\tilde{p}}{\mathrm{d}z}=-\tilde{\rho}g,\quad\frac{\mathrm{d}}{\mathrm{d}z}\left(k\frac{\mathrm{d}\tilde{T}}{\mathrm{d}z}\right)=-\tilde{Q},\quad\tilde{\rho}=\rho(\tilde{p},\tilde{T}),\quad\tilde{s}=s(\tilde{p},\tilde{T}).\label{eq:RefState_eqs_A_summary}
\end{equation}
The actual form of the reference state depends on the functions $g(z)$
and $k(z)$, which have to be known beforehand. It must be emphasized,
that a comparison between two anelastic systems, which is often necessary
when two different numerical codes are expected to produce the same
results, must be done with great care and inclusion of the form of
the reference state. We elaborate on this issue later. 

The \emph{first fundamental assumption} leading to the anelastic
approximation is that the convective system is only slightly superadiabatic, that is
\begin{equation}
0<\delta\equiv\left\langle \frac{L}{\tilde{T}}\Delta_{S}\right\rangle =-\left\langle \frac{L}{\tilde{T}}\left(\frac{\mathrm{d}\tilde{T}}{\mathrm{\mathrm{d}z}}+\frac{g\tilde{\alpha}\tilde{T}}{\tilde{c}_{p}}\right)\right\rangle =-\left\langle \frac{L}{\tilde{c}_{p}}\frac{\mathrm{d}\tilde{s}}{\mathrm{\mathrm{d}z}}\right\rangle \ll1.\label{eq:e233}
\end{equation}
\emph{Secondly}, one must require that the weak superadiabaticity
which drives convection implies weak fluctuations of thermodynamic
variables
\begin{equation}
\left|\frac{\rho'}{\tilde{\rho}}\right|\sim\left|\frac{T'}{\tilde{T}}\right|\sim\left|\frac{p'}{\tilde{p}}\right|\sim\left|\frac{s'}{\tilde{s}}\right|\sim\left|\frac{\psi'}{\tilde{\psi}}\right|\sim\mathcal{O}(\delta)\ll1,\label{fluct_magnitude_A-1}
\end{equation}
Furthermore, the convective velocity and time scales are
\begin{equation}
\mathscr{U}\sim\delta^{1/2}\sqrt{\bar{g}L},\qquad\mathscr{T}\sim\delta^{-1/2}\sqrt{\frac{L}{\bar{g}}},\label{eq:vel_and_time_scales-1-1}
\end{equation}
and hence consistency of the approximation requires, that the scales viscosity
and thermal conductivity also have to be small,
\begin{equation}
\mu_{b}/\bar{\rho}\,\lesssim\,\nu\sim\delta^{1/2}\sqrt{\bar{g}L}L,\qquad k\sim \delta^{1/2}\bar{\rho}\bar{c}_{p}\sqrt{\bar{g}L}L.\label{eq:visc_scale-1-1}
\end{equation}
The \emph{third assumption} involves the equation of state, that is
to say we require, that the derivative $\partial p/\partial\rho$
is of the same order of magnitude in terms of the small parameter
$\delta$ as $p/\rho$, which is typically satisfied by fluids and
thus does not impose a strong restriction on the system. The latter
assumption, together with (\ref{eq:vel_and_time_scales-1-1}) imply
small Mach number $Ma^{2}=\mathscr{U}^{2}/\left\langle \left(\frac{\partial p}{\partial\rho}\right)_{s}\right\rangle =\mathcal{O}\left(\delta\right)\ll1.$

The full system of anelastic equations reads 
\begin{subequations}
\begin{align}
\tilde{\rho}\left[\frac{\partial\mathbf{u}}{\partial t}+\left(\mathbf{u}\cdot\nabla\right)\mathbf{u}\right]= & -\nabla p'+\rho'\mathbf{g}-\tilde{\rho}\nabla\psi'+\mu\nabla^{2}\mathbf{u}+\left(\frac{\mu}{3}+\mu_{b}\right)\nabla\left(\nabla\cdot\mathbf{u}\right)\nonumber \\
 & +2\nabla\mu\cdot\mathbf{G}^{s}+\nabla\left(\mu_{b}-\frac{2}{3}\mu\right)\nabla\cdot\mathbf{u},\label{NS-Aderiv-1-1-2}
\end{align}
\begin{equation}
\nabla\cdot\left(\tilde{\rho}\mathbf{u}\right)=0,\qquad\nabla^{2}\psi'=4\pi G\rho',\label{Cont_Aderiv-1-1-3}
\end{equation}
\begin{align}
\tilde{\rho}\tilde{T}\left(\frac{\partial s'}{\partial t}+\mathbf{u}\cdot\nabla s'\right)-\tilde{\rho}\tilde{c}_{p}u_{z}\Delta_{S}= & \nabla\cdot\left(k\nabla T'\right)+2\mu\mathbf{G}^{s}:\mathbf{G}^{s}\nonumber \\
 & +\left(\mu_{b}-\frac{2}{3}\mu\right)\left(\nabla\cdot\mathbf{u}\right)^{2}+Q',\label{Energy_Aderiv-1-1-2}
\end{align}
\begin{equation}
\frac{\rho'}{\tilde{\rho}}=-\tilde{\alpha}T'+\tilde{\beta}p',\qquad s'=-\tilde{\alpha}\frac{p'}{\tilde{\rho}}+\tilde{c}_{p}\frac{T'}{\tilde{T}}.\label{State_eq_Aderiv-1-1-2}
\end{equation}
\end{subequations} 
The entropy equation may be replaced by the temperature
equation, for which the most general form is
\begin{align}
\tilde{\rho}\tilde{c}_{p}\left(\frac{\partial T'}{\partial t}+\mathbf{u}\cdot\nabla T'\right)-\tilde{\alpha}\tilde{T}\left(\frac{\partial p'}{\partial t}+\mathbf{u}\cdot\nabla p'\right)+\rho c_{p}u_{z}\frac{\mathrm{d}\tilde{T}}{\mathrm{d}z}-\alpha Tu_{z}\frac{\mathrm{d}\tilde{p}}{\mathrm{d}z}\qquad\nonumber \\
=\nabla\cdot\left(k\nabla T'\right)+2\mu\mathbf{G}^{s}:\mathbf{G}^{s}+\left(\mu_{b}-\frac{2}{3}\mu\right)\left(\nabla\cdot\mathbf{u}\right)^{2} & +Q'.\quad\label{Energy_eq1-1-2}
\end{align}

We also recall here the results of section \ref{subsec:Conservarion-of-mass-A}.
In the case of uniform gravity the conservation of mass implies, that
in order for the thermodynamic fluctuations to be correctly resolved,
the jump of the mean pressure fluctuation across the depth of the
fluid layer must vanish at all times, i.e. $\left\langle p'\right\rangle _{h}(z=L)-\left\langle p'\right\rangle _{h}(z=0)=0$.
This constitutes a boundary condition, which must be imposed on the
pressure field.

The system of anelastic equations can be expressed solely in terms
of the velocity field and the pressure and entropy fluctuations, which
is known as the ``entropy formulation''. Under the assumptions that
the fluid satisfies the ideal gas equation, the volume cooling can
be modelled by $\tilde{Q}=\kappa g\mathrm{d}_{z}\tilde{\rho}$,
the gravity $g=\mathrm{const}$, thermal diffusivity
$\kappa=\mathrm{const}$ and specific heat $\tilde{c}_{p}=\mathrm{const}$
are uniform, the entropy formulation takes the form 
\begin{subequations}
\begin{align}
\frac{\partial\mathbf{u}}{\partial t}+\left(\mathbf{u}\cdot\nabla\right)\mathbf{u}= & -\nabla\left(\frac{p'}{\tilde{\rho}}\right)+\frac{gs'}{c_{p}}\hat{\mathbf{e}}_{z}+\frac{\mu}{\tilde{\rho}}\nabla^{2}\mathbf{u}+\left(\frac{\mu}{3\tilde{\rho}}+\frac{\mu_{b}}{\tilde{\rho}}\right)\nabla\left(\nabla\cdot\mathbf{u}\right)\nonumber \\
 & +\frac{2}{\tilde{\rho}}\nabla\mu\cdot\mathbf{G}^{s}+\frac{1}{\tilde{\rho}}\nabla\left(\mu_{b}-\frac{2}{3}\mu\right)\nabla\cdot\mathbf{u},\label{NS-Aderiv-1-1-1-1-1-1}
\end{align}
\begin{equation}
\nabla\cdot\left(\tilde{\rho}\mathbf{u}\right)=0,\label{Cont_Aderiv-1-1-1-1-1-1}
\end{equation}
\begin{equation}
\tilde{\rho}\tilde{T}\left(\frac{\partial s'}{\partial t}+\mathbf{u}\cdot\nabla s'\right)-\tilde{\rho}c_{p}u_{z}\Delta_{S}=\nabla\cdot\left(\kappa\tilde{\rho}\tilde{T}\nabla s'\right)+\mathcal{J}+Q',\label{eq:energy_final_constkappa-1}
\end{equation}
\end{subequations} 
where $\mathcal{J}$ is given in (\ref{eq:J_def_A}).
In such a way the pressure fluctuation is entirely eliminated from
the energy balance and it appears only in the momentum equation. The
pressure problem is easily removable by taking a curl of the momentum
balance. It follows, that the boundary conditions for the pressure which
physically cannot be controlled at the boundaries are no longer necessary,
if one is searching for the velocity and entropy fields only (however, imposition of explicit boundary conditions on the entropy, not temperature is necessary). Such
a two-variable approach gives an adventage from the point of view
of effectiveness of numerical simulations of anelastic convection
under the aforementioned assumptions.

The energetic properties of the anelastic systems are presented in
section \ref{subsec:Energetic-properties-of-A}. We recall here \emph{the
second formula for superadiabatic heat flux}
\begin{align}
F_{S}\left(z=0\right)= & -k\left.\frac{\mathrm{d}}{\mathrm{d}z}\left(\tilde{T}+\left\langle T'\right\rangle _{h}-T_{ad}\right)\right|_{z=0}\nonumber \\
= & -k\frac{T_{B}}{\tilde{T}}\frac{\mathrm{d}}{\mathrm{d}z}\left(\tilde{T}+\left\langle T'\right\rangle _{h}-T_{ad}\right)\nonumber \\
 & -\left[\left.k\frac{\mathrm{d}\left(\tilde{T}-T_{ad}\right)}{\mathrm{d}z}\right|_{z=0}-\frac{T_{B}}{\tilde{T}}k\frac{\mathrm{d}\left(\tilde{T}-T_{ad}\right)}{\mathrm{d}z}\right]\nonumber \\
 & -T_{B}k\frac{\mathrm{d}\tilde{T}}{\mathrm{d}z}\left(\frac{\left\langle T'\right\rangle _{h}}{\tilde{T}^{2}}-\frac{\left.\left\langle T'\right\rangle _{h}\right|_{z=0}}{T_{B}^{2}}\right)-2T_{B}k\frac{\mathrm{d}\tilde{T}}{\mathrm{d}z}\int_{0}^{z}\frac{\left\langle T'\right\rangle _{h}}{\tilde{T}^{3}}\frac{\mathrm{d}\tilde{T}}{\mathrm{d}z}\mathrm{d}z\nonumber \\
 & +\tilde{\rho}T_{B}\left\langle u_{z}s'\right\rangle _{h}\nonumber \\
 & -2T_{B}\int_{0}^{z}\left\langle \frac{\mu}{\tilde{T}}\mathbf{G}^{s}:\mathbf{G}^{s}\right\rangle _{h}\mathrm{d}z+T_{B}\int_{0}^{z}\left\langle \frac{2\mu-3\mu_{b}}{3\tilde{T}}\left(\nabla\cdot\mathbf{u}\right)^{2}\right\rangle _{h}\mathrm{d}z.\label{eq:F_conv_superadiabatic_2-1}
\end{align}
which is one of two provided general formulae for the superadiabatic
heat flux, satisfied by any anelastic system. If for simplicity one
assumes, that the fluid satisfies the equation of state of an ideal
gas, isothermal boundaries, uniform fluid properties $k=\mathrm{const.},$
$\mu=\mathrm{const}.$, $\mu_{b}=\mathrm{const}.$, $c_{p}=\mathrm{const.}$,
no radiation $Q=0$ and $g=\textrm{const}.$ and $\mathbf{g}'=0$,
the latter formula implies
\begin{eqnarray}
\frac{k^{2}}{\rho_{B}^{2}c_{p}^{2}L^{4}}Ra\left(Nu-1\right)\left(\Gamma-1\right) & = & 2\Delta T\left\langle \frac{1}{\tilde{T}}\mathbf{G}^{s}:\mathbf{G}^{s}\right\rangle \nonumber \\
 &  & -\Delta T\left(\frac{2}{3}-\frac{\mu_{b}}{\mu}\right)\left\langle \frac{\left(\nabla\cdot\mathbf{u}\right)^{2}}{\tilde{T}}\right\rangle \nonumber \\
 &  & +2c_{p}Pr^{-1}L\left(\frac{\Delta T}{L}\right)^{3}\left\langle \frac{T'}{\tilde{T}^{3}}\right\rangle ,\qquad\label{eq:rel_flux_VD-1-1-1}
\end{eqnarray}
where $Pr=\mu c_{p}/k$, $\Gamma=T_{B}/T_{T}$ and the Nusselt and
Rayleigh numbers are defined as follows
\begin{equation}
Nu=\frac{F_{S}\left(z=0\right)}{k\Delta_{S}}=\frac{-k\left.\frac{\mathrm{d}}{\mathrm{d}z}\left(\tilde{T}+\left\langle T'\right\rangle _{h}-T_{ad}\right)\right|_{z=0}}{k\Delta_{S}},\label{eq:Nu_def_anapp-1-1}
\end{equation}
\begin{equation}
Ra=\frac{g\bigtriangleup_{S}L^{4}\rho_{B}^{2}c_{p}}{T_{B}\mu k}=\frac{g\Delta T\Delta\tilde{s}L^{3}\rho_{B}^{2}}{\ln\Gamma T_{B}\mu k}.\label{eq:Ra_def-1-2}
\end{equation}
In the above $k\Delta_{S}=k(\Delta T/L-g/c_{p})$ is the superadiabatic
conductive heat flux in the hydrostatic basic state.

On the other hand \emph{the first formula for the total heat flux}
\begin{align}
F_{total}\left(z=0\right)= & -k\left.\frac{\mathrm{d}}{\mathrm{d}z}\left(\tilde{T}+\left\langle T'\right\rangle _{h}\right)\right|_{z=0}\nonumber \\
= & -k\frac{\mathrm{d}}{\mathrm{d}z}\left(\tilde{T}+\left\langle T'\right\rangle _{h}\right)\nonumber \\
 & +\tilde{\rho}\tilde{T}\left\langle u_{z}s'\right\rangle _{h}-\int_{0}^{z}\tilde{\rho}\frac{\mathrm{d}\tilde{T}}{\mathrm{d}z}\left\langle u_{z}s'\right\rangle _{h}\mathrm{d}z\nonumber \\
 & -2\int_{0}^{z}\left\langle \mu\mathbf{G}^{s}:\mathbf{G}^{s}\right\rangle _{h}\mathrm{d}z+\int_{0}^{z}\left\langle \left(\frac{2}{3}\mu-\mu_{b}\right)\left(\nabla\cdot\mathbf{u}\right)^{2}\right\rangle _{h}\mathrm{d}z,\label{eq:F_conv_total_1-1}
\end{align}
comes useful when the heat flux is kept fixed at the boundaries, which
under the same assumptions of constant $k$, $\mu$, $\mu_{b}$, $c_{p}$,
$Q$, $g$ and $\mathbf{g}'= 0$, implies
\begin{align}
\frac{k^{2}}{\rho_{B}^{2}c_{p}^{2}L^{4}}Ra\left(Nu_{Q}-\frac{k\frac{\Delta\left\langle T'\right\rangle _{h}}{2L}}{k\Delta_{S}}\right)\frac{\Gamma}{\Gamma-1}= & \left\langle \left(\frac{\tilde{T}}{\Delta T}-1\right)\mathbf{G}^{s}:\mathbf{G}^{s}\right\rangle \nonumber \\
 & -\left(\frac{1}{3}-\frac{\mu_{b}}{2\mu}\right)\left\langle \left(\frac{\tilde{T}}{\Delta T}-1\right)\left(\nabla\cdot\mathbf{u}\right)^{2}\right\rangle .\label{eq:e204-1}
\end{align}
where the Rayleigh number is still defined as in (\ref{eq:Ra_def-1-2})
but a new definition of the Nusselt number is required
\begin{equation}
Nu_{Q}=\frac{\left\langle \tilde{\rho}\tilde{T}u_{z}s'\right\rangle }{k\Delta_{S}},\label{eq:Nu_Q_anelastic-1}
\end{equation}
and $\Delta\left\langle T'\right\rangle _{h}=\left.\left\langle T'\right\rangle _{h}\right|_{z=L}-\left.\left\langle T'\right\rangle _{h}\right|_{z=0}.$

It is important to note, that contrary to the Boussinesq case, it
is clear from (\ref{eq:F_conv_total_1-1}), that
\begin{align}
F_{total}\left(z\right)= & F_{total}\left(z=0\right)-\int_{0}^{z}\frac{\tilde{\alpha}\tilde{T}g\tilde{\rho}}{\tilde{c}_{p}}\left\langle u_{z}s'\right\rangle _{h}\mathrm{d}z\nonumber \\
 & +2\int_{0}^{z}\left\langle \mu\mathbf{G}^{s}:\mathbf{G}^{s}\right\rangle _{h}\mathrm{d}z-\int_{0}^{z}\left\langle \left(\frac{2}{3}\mu-\mu_{b}\right)\left(\nabla\cdot\mathbf{u}\right)^{2}\right\rangle _{h}\mathrm{d}z,\;\label{eq:heta_flux_balance_BULK_AN-1}
\end{align}
therefore anelastic systems are \emph{not} characterized by constant
heat flux at every $z$, but the heat flux is strongly influenced
by the viscous heating\index{SI}{viscous heating} and the work of the buoyancy force.

Linear stability of an anelastic ideal gas at constant $\nu$, $k$,
$\mathbf{g}$ and $c_{p}$, $\mathbf{g}'=0$, $Q=0$, with isothermal,
stress-free and impermeable boundaries is characterized by the critical
Rayleigh number
\begin{equation}
Ra_{crit}\approx\frac{27}{4}\pi^{4}\left[1+\frac{1}{2}\theta\left(m-1\right)\right],\label{eq:Ra_crit_A-1}
\end{equation}
and the growth rate
\begin{equation}
\sigma=\frac{\kappa_{B}}{L^{2}}\frac{3}{2}\pi^{2}\frac{Pr}{1+Pr}\eta\left(1+\frac{1}{2}m\theta\frac{Pr}{1+Pr}\right),\label{eq:e236}
\end{equation}
where $Ra=g\Delta_{S}L^{4}/T_{B}\kappa_{B}\nu$ and $\eta=(Ra-Ra_{crit})/Ra_{crit}$;
both the growth rate and the Rayleigh number are greater than in the
Boussinesq, that is non-stratified (very weakly compressible)
case. The total heat per unit mass accumulated in the fluid layer
(between top and bottom boundaries) in the marginal state, $c_{p}\kappa_{B}\nu T_{B}Ra_{crit}/gL^{3}$,
is also greater in the stratified case, than in the Boussinesq one,
but it is known, that the assumptions regarding the depth dependence
of transport coefficients, in particular the thermal diffusivity, can
strongly influence the latter results concerning the threshold of
convection. 

\subsection{Comparison of anelastic systems and relation to the adiabatic reference
state formulation\label{subsec:Comparison-of-anelastic}}

There are different forms of anelastic equations used in the literature,
and the differences result from various possible assumptions regarding
the fluid properties $k$, $\mu$, $\mu_{b}$, $c_{p}$, $c_v$, the form
of the heating source term $Q$ and whether or not the gravity is
assumed uniform or influenced by the density fluctuations. Moreover
the different formulations can result from two most common possibilities
for the choice of the reference state, that is either a weakly superadiabatic
hydrostatic reference state, satisfying the boundary conditions can
be chosen, as e.g. in (\ref{eq:BS1}-c) or the adiabatic state can
serve as the reference one (see e.g. (\ref{eq:adiabatic_profile_RS})),
but then the boundary conditions on convective fluctuations are non-uniform. However, the reference
state, can in fact be chosen in an arbitrary way, even as a time-dependent
state, but in the latter case its time evolution must be included
in the set of equations. 

It is often necessary to compare the results of different anelastic
formulations, e.g. in computing, when testing numerical codes against
some benchmarking solutions. The choice of the reference state does
not matter in the sense, that two formulations with different reference
state can obviously still lead to the same results. What matters,
is first of all, that the physical assumptions are the same, thus
in particular that the fluid properties $k$, $\mu$, $\mu_{b}$,
$c_{p}$, $c_v$, the heating source $Q$ and gravity have the same spatial
(and possibly temporal) dependence, the equation of state for the fluid is the same, and the boundary conditions are
of the same type. The latter implies, that the departure from adiabatic
state, which drives the convective flow and physically can be realized
by heating up the bottom boundary and cooling the upper one to generate
superadiabatic temperature gradients, must also be the same, if two
anelastic systems are expected to produce the same results. In other
words one needs to take care so that the excess in temperature jump
across the layer over the temperature jump
in the adiabatic state relative to the bottom temperature $T_{B}$ 
\begin{equation}
\frac{\Delta T}{T_{B}}-\int_{0}^{L}\frac{g\tilde{\alpha}\tilde{T}}{\tilde{c}_{p}T_B}\mathrm{d}z,\label{Tjump_excess}
\end{equation}
(equal to $\Delta T/T_{B}-gL/c_{p}T_{B}$
for a perfect gas when $g$ and $c_{p}$ are uniform), is the same
in both compared anelastic formulations.  The thermodynamic fluctuations about the conduction and adiabatic reference states by the use of (\ref{full_therm_vars}) are related through
\begin{subequations}\label{therm_vars_transforms}
\begin{equation}
T_S(\mathbf{x},t)=T'(\mathbf{x},t)+\tilde{T}(z)-T_{ad}(z),
\end{equation}
\begin{equation}
\rho_S(\mathbf{x},t)=\rho'(\mathbf{x},t)+\tilde{\rho}(z)-\rho_{ad}(z),
\end{equation}
\begin{equation}
p_S(\mathbf{x},t)=p'(\mathbf{x},t)+\tilde{p}(z)-p_{ad}(z),
\end{equation}
\begin{equation}
s_S(\mathbf{x},t)=s'(\mathbf{x},t)+\tilde{s}(z)+\mathrm{const},
\end{equation}
\end{subequations}
where the subscript $S$ denotes the superadiabatic fluctuation about the adiabatic state. When the total mass of the fluid is the same in both formulations direct correspondence is achieved, see section \ref{subsec:The-reference-state} and equations (\ref{eq:M_cond},b). It is a matter of choice whether or not the total mass is contained in the reference state or not, but the former option is certainly useful and more clear. In such a case the bottom densities $\rho_B$ and $\rho_{ad\,B}$ in the conduction and adiabatic reference states are related through
\begin{equation}
\int_{0}^{L}\tilde{\rho}\mathrm{d}z=\int_{0}^{L}\rho_{ad}\mathrm{d}z,\label{eq:Equal_masses_formulations}
\end{equation}
(cf. (\ref{eq:rho_rel}) for an explicit  relation between $\rho_B$ and $\rho_{ad\,B}$ in the case of a perfect gas with uniform material properties). The relation (\ref{eq:Equal_masses_formulations}) involves corrections of the order $\mathcal{O}(\delta\rho_B)$, which are important in the transformations (\ref{therm_vars_transforms}b,c) for the density and pressure fluctuations, as the bottom values of density $\rho_B$ and $\rho_{ad\,B}$ have to explicitly appear in the expressions for $\tilde{\rho}(z)$, $\rho_{ad}(z)$, $\tilde{p}(z)$ and $p_{ad}(z)$. For example in the case when all the fluid properties and gravity are uniform, and the fluid is described by the equation of state of a perfect gas, by the use of (\ref{eq:rhotilde_expanded_RS}), (\ref{eq:adiabatic_profile_RS}) and (\ref{eq:rho_rel}) the density fluctuation transformation between the two anelastic formulations takes the form (\ref{rho_transform}) with the ratio $\rho_{B}/\rho_{ad\,B}$ given by (\ref{rho_ratio_formulations}), cf. section \ref{subsec:The-reference-state}.

The anelastic numerical codes are often constructed based on a non-dimensional form of the dynamical equations. But here again the comparison requires
extra care, since the non-dimensional variables are often defined with the
use of different scales, such as the density scale, temperature scale,
etc. E.g., it is vital to compare Rayleigh numbers, which utilize the same
scale definitions, e.g. scales defined by the bottom values of density, temperature, etc. Nevertheless, comparison and full correspondence between two anelastic formulations, with conduction and adiabatic reference state is possible, and in fact not very difficult. Let us provide an example of a non-dimensional form of the dynamical equations under the anelastic approximation, where for simplicity we assume that the viscosities are uniform, $\mu=\mathrm{const}$ and $\mu_b=\mathrm{const}$; this assumption is by no means necessary, and it is used only to make the Navier-Stokes equation somewhat simpler, as it allows to remove some viscous dissipation terms. 
We introduce the following non-dimensional variables (which can, of course be chosen differently)
\begin{subequations}\label{nondim_vars_form}
\begin{equation}
\mathbf{x}=L\mathbf{x}^{\sharp},\quad t=\frac{L^2}{\kappa_B}t^{\sharp},\quad \mathbf{u}=\frac{\kappa_B}{L}\mathbf{u}^{\sharp},
\end{equation}
\begin{equation}
\rho=\rho_B\rho^{\sharp},\quad T=T_B T^{\sharp},\quad p=\frac{\rho_B \kappa_B^2}{L^2} p^{\sharp},\quad s=c_{p\,B}s^{\sharp},\quad \psi= g_B L\psi^{\sharp},
\end{equation}
\begin{equation} 
\alpha=\frac{\alpha^{\sharp}}{T_B},\quad \beta=\frac{L^2}{\rho_B \kappa_B^2} \beta^{\sharp},
\end{equation}
\begin{equation} 
\mathbf{g}=g_B\mathbf{g}^{\sharp},\quad Q=Q_B Q^{\sharp},\quad c_p=c_{p\,B} c_p^{\sharp},\quad k=k_B k^{\sharp}.
\end{equation}
\end{subequations}
It follows, that the non-dimensional dynamical equations take the form
\begin{subequations}
\begin{align}
\rho_r^{\sharp}\left[\frac{\partial\mathbf{u}^{\sharp}}{\partial t^{\sharp}}+\left(\mathbf{u}^{\sharp}\cdot\nabla^{\sharp}\right)\mathbf{u}^{\sharp}\right] = &\, -\nabla p^{\prime\sharp}+RaPr\rho^{\prime\sharp}\mathbf{g}_r^{\sharp}-RaPr\rho_r^{\sharp}\nabla\psi^{\prime\sharp}\nonumber \\
 &\, +Pr\nabla^{\sharp2}\mathbf{u}^{\sharp}+Pr\left(\frac{1}{3}+\frac{\mu_{b}}{\mu}\right)\nabla^{\sharp}\left(\nabla^{\sharp}\cdot\mathbf{u}^{\sharp}\right),\label{NS-nondim_form}
\end{align}
\begin{equation}
\nabla^{\sharp}\cdot\left(\rho^{\sharp}_r\mathbf{u}^{\sharp}\right)=0,\qquad
\nabla^{\sharp2}\psi^{\prime\sharp}=\frac{4\pi G\rho_B L}{g_{B}}\rho^{\prime\sharp},\label{cont_grav_nondim_form}
\end{equation}
\begin{align}
\rho_r^{\sharp}T_r^{\sharp}\left(\frac{\partial s^{\prime\sharp}}{\partial t^{\sharp}}+\mathbf{u}^{\sharp}\cdot\nabla^{\sharp} s^{\prime\sharp}\right)-\rho_r^{\sharp}c_{p\,r}^{\sharp}u_{z}^{\sharp}\frac{\Delta_{S}}{\langle\Delta_S\rangle}= \,\nabla^{\sharp}\cdot\left(k^{\sharp}\nabla T^{\prime\sharp}\right) &\qquad\qquad\qquad\qquad\nonumber \\
 +\frac{2g_B L}{c_{p\,B}T_B Ra}\left[\mathbf{G}^{s}:\mathbf{G}^{s} +\left(\frac{\mu_{b}}{\mu}-\frac{2}{3}\right)\left(\nabla^{\sharp}\cdot\mathbf{u}^{\sharp}\right)^{2}\right] &+\frac{Q_B L}{\kappa_B \rho_B c_{p\,B}\Delta_S}Q^{\prime\sharp}.\label{Energy_nondim_form}
\end{align}
\begin{equation}
\frac{\rho^{\prime\sharp}}{\rho_r^{\sharp}}=-\alpha_r^{\sharp}T^{\prime\sharp}+\beta_r^{\sharp}p^{\prime\sharp},\qquad s^{\prime\sharp}=-\frac{\nu_B\kappa_B}{g_B L^3}\frac{g_B L}{c_{p\,B}T_B Pr}\alpha_r^{\sharp}\frac{p^{\prime\sharp}}{\rho_r^{\sharp}}+c_{p\,r}^{\sharp}\frac{T^{\prime\sharp}}{T_r^{\sharp}},\label{State_eq_nondim_form}
\end{equation}
\end{subequations} 
where the subscript $r$ denotes a reference state variable, either the conduction reference state ($\rho_r=\tilde{\rho}$, $T_r=\tilde{T}$, etc.) or the adiabatic reference state ($\rho_r=\rho_{ad}$, $T_r=T_{ad}$, etc.), depending on the formulation used; the term $-\rho_r^{\sharp}c_{p\,r}^{\sharp}u_{z}^{\sharp}\Delta_{S}/\langle\Delta_S\rangle$ in the energy equation (\ref{Energy_nondim_form}) is absent in the case $r=ad$, i.e. when the reference state is adiabatic. In the above non-dimensional equations the Rayleigh number, which measures the relative strength of
buoyancy with respect to dissipation (and provides a useful measure of departure from the critical
state), and the Prandtl number (which both must have the same values for the two compared formulations) are defined as follows
\begin{equation}
Ra=\frac{g_{B}L^{3}}{\nu_{B}\kappa_{B}}\left(\frac{\Delta T}{T_{B}}-\frac{L}{T_{B}}\left\langle \frac{g\tilde{\alpha}\tilde{T}}{\tilde{c}_{p}}\right\rangle \right),\qquad Pr=\frac{\nu_B}{\kappa_B}=\frac{\mu c_{p\,B}}{k_B}.\label{eq:e237}
\end{equation}
The expression in the brackets in the Rayleigh number definition is the aforementioned
superadiabatic excess in the temperature jump across the layer, which as remarked must be the same in the both considered formulations of the anelastic approximation. It follows that the
non-dimensional parameter $g_{B}L^{3}/\nu_{B}\kappa_{B}$ must also
be the same in two anelastic, physically equivalent systems. Finally
the non-dimensional measure of stratification (compressibility) $\theta=\Delta T/T_{B}$
(or equivalently $\Gamma=T_{B}/T_{T}$ or $g_B L/c_{p\,B}T_B$ instead) 
must coincide as well. Eventually an exemplary set of parameters which
need to be compared between two different anelastic formulations in
order to expect from them the same physical results (i.e. describing the same physical system) can be written
as follows
\begin{subequations}
\begin{equation}
\frac{\Delta T}{T_{B}}-\frac{L}{T_{B}}\left\langle \frac{g\tilde{\alpha}\tilde{T}}{\tilde{c}_{p}}\right\rangle ,\quad\frac{g_{B}L^{3}}{\nu_{B}\kappa_{B}},\quad\theta=\frac{\Delta T}{T_{B}},\quad Pr=\frac{\nu_{B}}{\kappa_{B}},\label{eq:e238}
\end{equation}
\begin{equation}
\frac{\mu_{b}}{\mu},\quad \tilde{\alpha}_B T_B,\quad \frac{Q_B L}{\kappa_B \rho_B c_{p\,B}\Delta_S},\quad \frac{G\rho_B L}{g_B},\label{eq:e238and_a_half}
\end{equation}
\end{subequations}
where, as mentioned $\left\langle g\tilde{\alpha}\tilde{T}/\tilde{c}_{p}\right\rangle =g/c_{p}$
in the simplest case of an ideal gas and constant $g$ and $c_{p}$. Of course the parameter $\theta$ could be replaced by another measure of
stratification, e.g. $\Gamma=T_{B}/T_{T}$ or the non-dimensional measure of the adiabatic gradient $g_B L/c_{p\,B}T_B$. $\tilde{\alpha}$, $\tilde{T}$ and $\tilde{c}_p$ could be replaced by their profiles (values) in the adiabatic state; in any case the conduction reference state is determined by the boundary conditions, which must be the same in both formulations, since they are responsible for the driving and consequently $\tilde{\alpha}$, $\tilde{T}$ and $\tilde{c}_p$ and therefore all the above parameters  in (\ref{eq:e238},b) can be evaluated no matter the formulation. The three bottom-row parameters in (\ref{eq:e238and_a_half}) are irrelevant in the case of a perfect gas with negligible bulk viscosity, no heating source, $Q=0$ and uniform gravity. Note, that the scales
of thermodynamic variables do not need to be defined as their values
at the bottom, and e.g. the mid-plane values are often used instead.

\subsection{Anelastic vs Boussinesq approximations\label{subsec:Anelastic-vs-Boussinesq}}

The transformation of the anelastic system of equations under the
Boussinesq limit $\mathrm{d}\tilde{\rho}/\mathrm{d}z\ll\bar{\rho}/L$,
$\mathrm{d}\tilde{T}/\mathrm{d}z\ll\bar{T}/L$ and $\mathrm{d}\tilde{p}/\mathrm{d}z\ll\bar{p}/L$
has been thoroughly described in section \ref{subsec:Boussinesq-limit-A}.
Since $\theta=\Delta T/T_{B}=\mathcal{O}(\epsilon)$, where $\epsilon=\Delta\dbtilde{\rho}/\bar{\rho}$,
the Boussinesq limit simply corresponds to taking the formal limit
$\theta\rightarrow0$ (or equivalently $\Gamma\rightarrow1$) in the
anelastic equations, but one must bare in mind, that this also implies
$\delta\lesssim\mathcal{O}(\theta)$, $c_{p}=\mathcal{O}(\theta^{-1}gL/\tilde{T})$
likewise $\mathscr{U}=\mathcal{O}(\theta^{1/2}\sqrt{gL})$ and $\mathscr{T}=\mathcal{O}(\theta^{-1/2}\sqrt{L/g})$.
The thermodynamic variables then scale as in (\ref{eq:rho_scaling_B}-d),
where $\epsilon$ may be replaced by $\theta$, that is $T'/\tilde{T}=\mathcal{O}(\theta)$,
$\rho'/\tilde{\rho}=\mathcal{O}(\theta)$, but the mean pressure is
boosted so that $p'/\tilde{p}=\mathcal{O}(\theta T'/\tilde{T})=\mathcal{O}(\theta^{2})$; moreover
$s'/c_{p}=\mathcal{O}(T'/\tilde{T})=\mathcal{O}(\theta)$ likewise
$\dbtilde{s}/c_{p}=\mathcal{O}(\theta)$, but $\tilde{s}/c_{p}=\mathcal{O}(1)$. Note, that the scalings for the
thermodynamic variables are fully consistent with the reference state
solutions for perfect gas given in (\ref{eq:BS1}-c) and with (\ref{eq:delta_intermsof_epsan}). 
\begin{figure}
a)\includegraphics[scale=0.17]{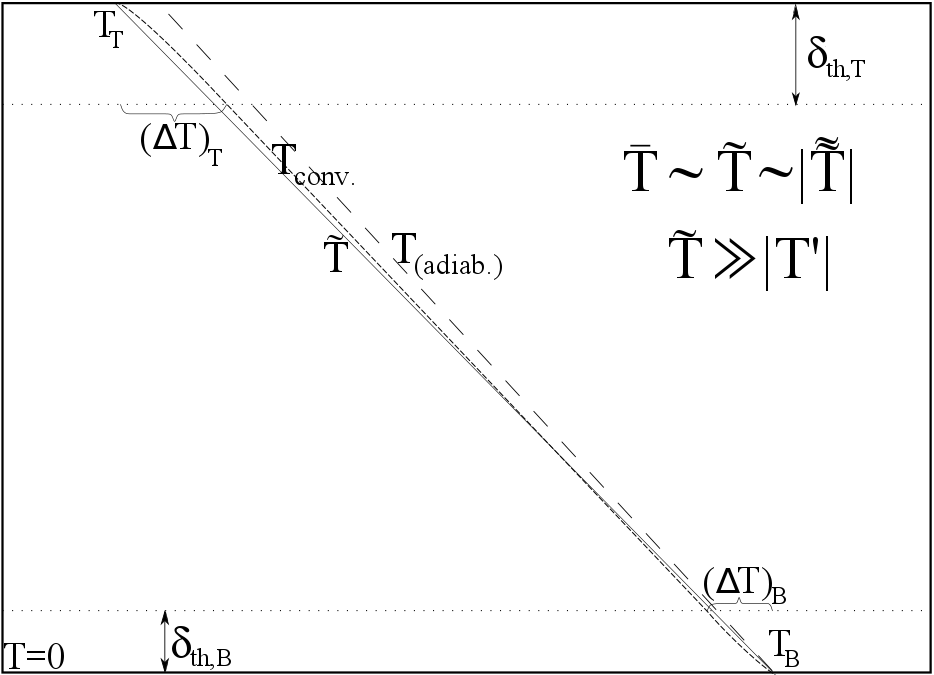}~~b)\includegraphics[scale=0.17]{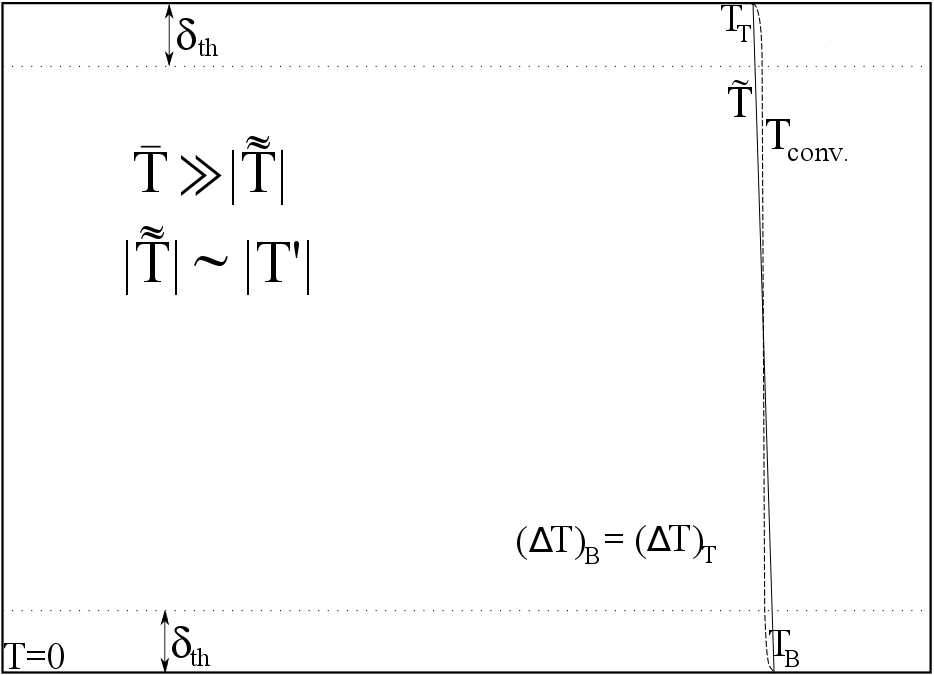}

~

~

c)\includegraphics[scale=0.17]{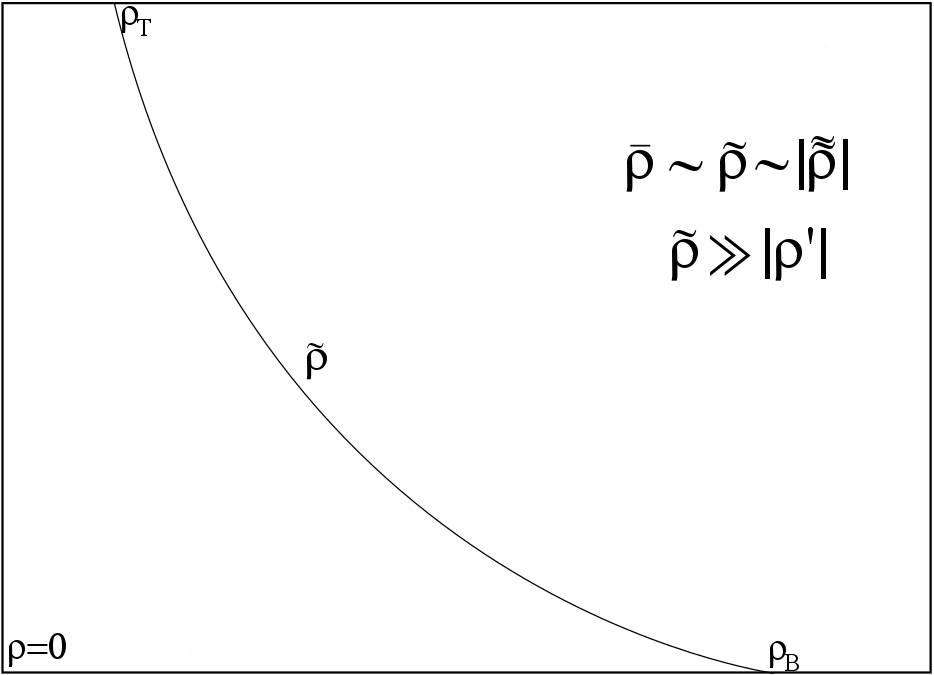}~~d)\includegraphics[scale=0.17]{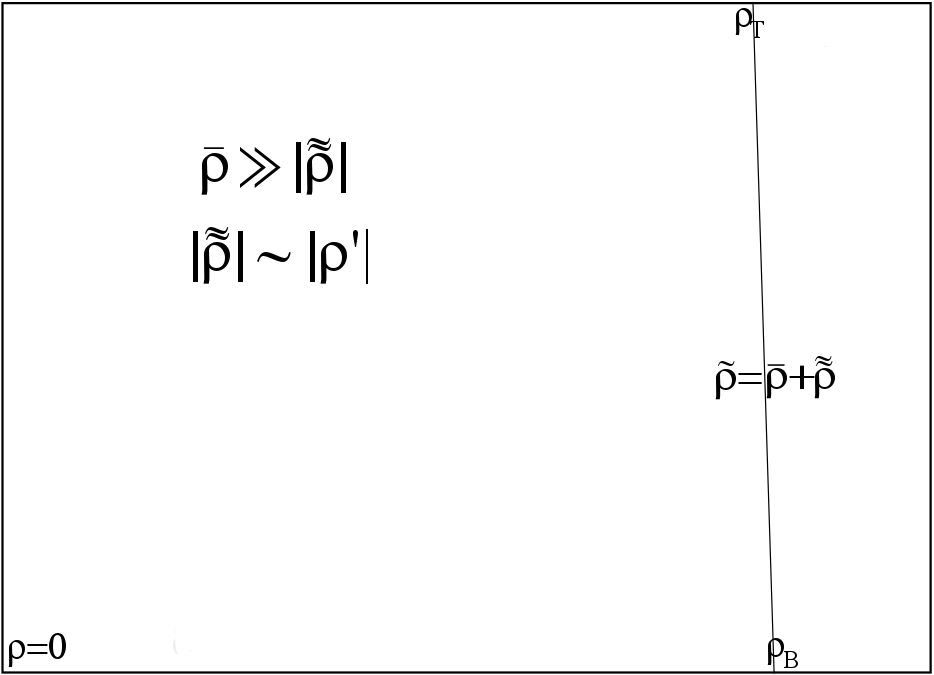}

~

~

e)\includegraphics[scale=0.17]{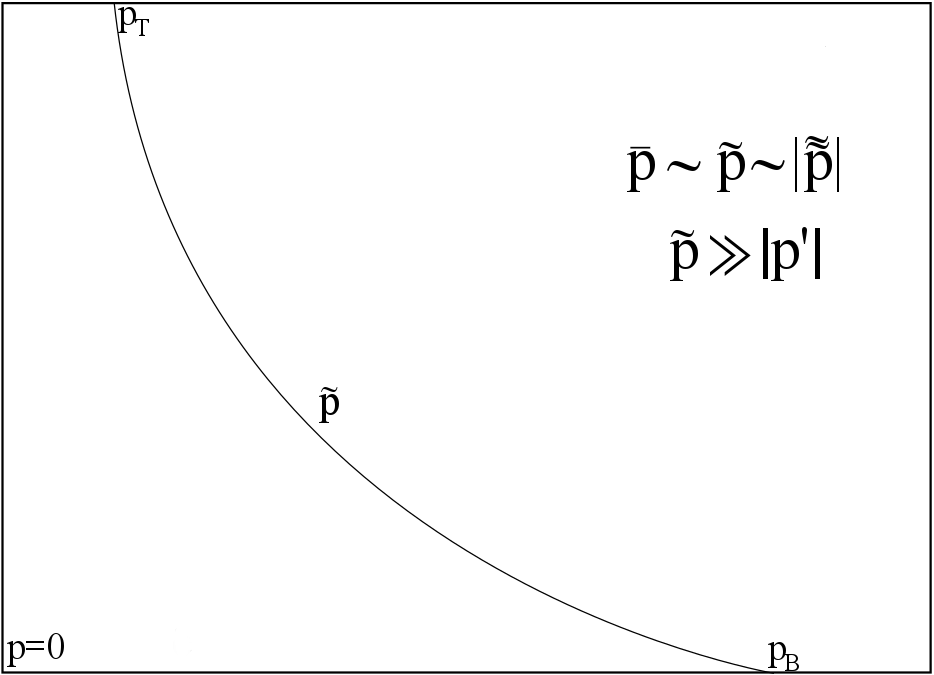}~~f)\includegraphics[scale=0.17]{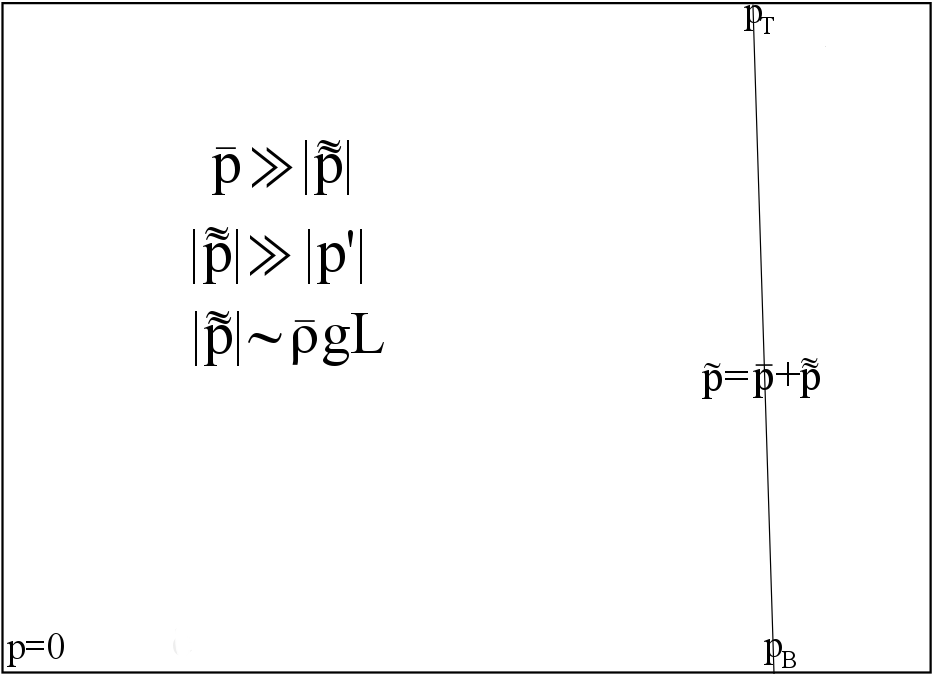}\caption{\label{fig:An_vs_Bsq_comparison}{\footnotesize{}A schematic comparison
of physical situations described by anelastic (left column) and Boussinesq
(right column) approximations. The Boussinesq approximation is characterized
by the fact, that the total variations of temperature, density and
pressure are very small compared to their means, whereas in the
anelastic approximation only the fluctuations induced by the convective
flow are small. Moreover, within the Boussinesq approximation the
vertical profiles of temperature, density and pressure in the hydrostatic
state are approximately linear (\ref{eq:T0_example}-c) and the mean
pressure is extremely high, $\bar{p}\sim\epsilon^{-1}\bar{\rho}gL$.}}
\end{figure}

\begin{figure}
\begin{centering}
\includegraphics[scale=0.2]{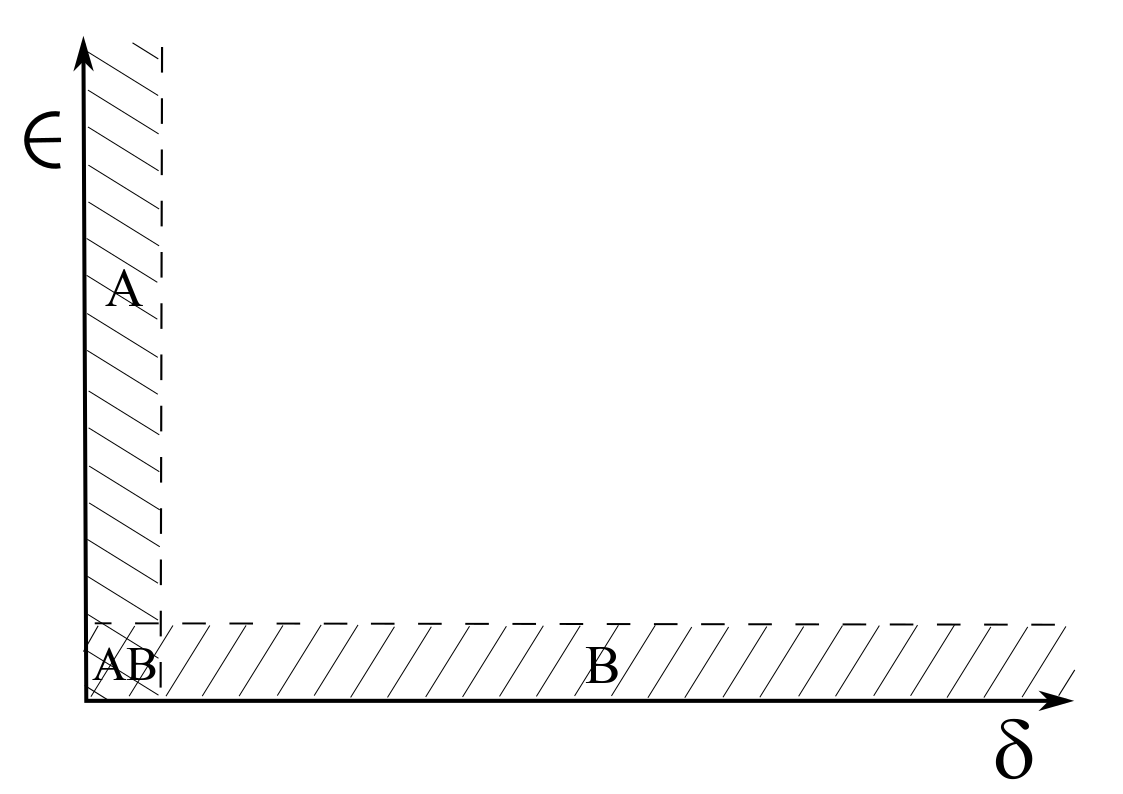}
\par\end{centering}
\caption{\label{fig:An_vs_Bou_limits}Schematic comparison of the anelastic
($\delta\ll1$, cf. (\ref{eq:delta_definition}); marked A) and Boussinesq ($\epsilon\ll1$,
cf. (\ref{eq:epsilon_def_B}); marked B) approximations on the $\delta\epsilon$
plane. The small region where $\delta\ll1$ and $\epsilon\ll1$ marked
AB is where both the approximations are valid.}
\end{figure}

The significant difference between the hydrostatic reference states
in the anelastic and Boussinesq approximations can be seen on figure
\ref{fig:An_vs_Bsq_comparison}. The characteristic feature of Boussinesq
systems is that the mean temperature, density and pressure are large
compared to the static state variations, whereas it is not the case
in the anelastic systems. Thus the latter are stratified and the density
and temperature stratification can be strong, which means, that the
convective flow is compressible $\nabla\cdot\mathbf{u}\neq0$, contrary
to the Boussinesq case. Moreover, the leading-order hydrostatic balance
within the Boussinesq approximation, $\mathrm{d}_{z}\dbtilde{p}\approx-\bar{\rho}g$,
requires relatively strong hydrostatic pressure variations, $\dbtilde{p}\sim\bar{\rho}gL$,
and this aided by the fundamental assumption $|\dbtilde{p}|/\bar{p}\sim\epsilon\ll1$
implies that the mean pressure has to be extremely high, of the order
$\bar{p}\sim\epsilon^{-1}\bar{\rho}gL$. A particular consequence
is that within the Boussinesq approximation the entropy fluctuation
is equivalent to the temperature fluctuation up to a constant factor,
$s'\approx\bar{c}_{p}T'/\bar{T}$; the latter coincidence of $s'$
and $T'$ is \emph{not} satisfied under the anelastic approximation
and in the anelastic formulation the mean pressure is of the same
order as pressure in the static reference state. It is also important
to realize, that Boussinesq convection is not necessarily weakly superadiabatic
and the ratio of the temperature gradient in the static reference
state to the adiabatic gradient, $\left|c_{p}\mathrm{d}_{z}\dbtilde{T}/g\alpha T\right|$,
can be much greater than unity for Boussinesq systems. A schematic
figure \ref{fig:An_vs_Bou_limits} depicts regions of validity of
the anelastic and Boussinesq approximations on the $\delta\epsilon$
plane; the limit marked AB obtained for $\delta\ll1$ and $\epsilon\ll1$
is where the two approximations merge. 

Finally, a very important difference between the two approximations
is that in Boussinesq convection the thermal energy strongly dominates
over the kinetic energy (even the part of the thermal energy associated
with temperature fluctuation only, is still $\epsilon^{-1}$ times greater
than the kinetic energy). Consequently the viscous heating in the
energy balance is negligible and thus in the absence of radiation the
only process contributing to the heat exchange between a fluid parcel
and its surroundings within the Boussinesq approximation is the molecular
heat conduction, 
\begin{equation}
\frac{\dbar q_{Bq}}{\mathrm{d}t}=T\frac{\mathrm{d}s}{\mathrm{d}t}=\frac{1}{\rho}\nabla\cdot(k\nabla T),\label{eq:e239}
\end{equation}
where $Bq$ stands for \emph{Boussinesq}. On the contrary, in anelastic
convection the kinetic energy is comparable with the thermal one and
thus the viscous heating\index{SI}{viscous heating} strongly contributes to the total heat flux
in the system. Therefore in the absence of radiation the contributions
to heat exchanged by a fluid parcel with its surroundings come from
both, the molecular conduction and viscous friction,
\begin{equation}
\frac{\dbar q_{Ac}}{\mathrm{d}t}=T\frac{\mathrm{d}s}{\mathrm{d}t}=\frac{1}{\rho}\nabla\cdot(k\nabla T)+2\frac{\mu}{\rho}\mathbf{G}^{s}:\mathbf{G}^{s}+\left(\frac{\mu_{b}}{\rho}-\frac{2\mu}{3\rho}\right)\left(\nabla\cdot\mathbf{u}\right)^{2},\label{eq:e240}
\end{equation}
where $Ac$ stands for \emph{Anelastic}. Furthermore, the work done
by the buoyancy force is also comparable with the thermal energy.
Therefore the effects of viscous heating\index{SI}{viscous heating} and of the work done by buoyancy
have a substantial influence on the total, horizontally averaged convective
heat flux, which varies with height, contrary to Boussinesq systems,
where the mean heat flux is uniform.

\section*{Review exercises}

{\textbf{Exercise 1.}} \\
Solve for a hydrostatic reference state in a perfect gas under presence of uniform heat sink $Q=-Q_0$, $Q_0=\mathrm{const}>0$. Assume $g=\mathrm{const}$, $k=\mathrm{const}$ and that the bottom and top temperatures $T_B$ and $T_T$ are held constant; consider the values of $\rho_B$, $T_B$ and $T_T$ as given. Determine the small anelastic parameter $\delta$.

\noindent\emph{Hint}: use equations (\ref{eq:hydrostatic_eq_A_pg}) with the second one modified by the presence of a heat sink $\mathrm{d}^2 \tilde{T}/\mathrm{d}z^2=Q_0/k$.
\\

\noindent{\textbf{Exercise 2.}} \\
For the previous problem determine the explicit transformations between the superadiabatic variables $T_S=T-T_{ad}$, $\rho_S=\rho-\rho_{ad}$ and the fluctuations about the calculated reference state $T'$, $\rho'$.

\noindent\emph{Hint}: cf. (\ref{TS_Tprime_transform}) and (\ref{rho_transform}).
\\

\noindent{\textbf{Exercise 3.}} \\
For the problem of Ex. 1, calculate the total pressure jump across the fluid layer $p(z=0)-p(z=L)$.

\noindent\emph{Hint}: cf. section \ref{subsec:Conservarion-of-mass-A}.
\\

\noindent{\textbf{Exercise 4.}} \\
Assuming that in some anelastic, convective system the viscous dissipation takes place predominantly in the upper, turbulent boundary layer and is negligible in the rest of the fluid domain, calculate the mid value (at $z=L/2$) of the work of the buoyancy force. The bottom value $F_B$ and the mid value $F_M$ of the total heat flux are given.

\noindent\emph{Hint}: utilize the first formula for the total heat flux and (\ref{eq:e192}).
\\

\noindent{\textbf{Exercise 5.}} \\
Assuming the physical setting of section 3.5 (ideal gas, $Q=0$, isothermal, stress-free and impermeable boundaries, constant $\nu$, $k$, $\mathbf{g}$ and $c_{p}$) and given that the critical Rayleigh number is equal to $Ra_{crit}=660$, the polytropic index $m=1.49$ and the Prandtl number $Pr=5$ estimate the growth rate of convection at $Ra=670$.

\noindent\emph{Answer}: $\sigma\approx0.189L^2c_p\rho_B/k$, $(\theta\approx0.015)$.

\chapter{Inclusion of compositional effects\label{chap:Inclusion-of-compositional}}

Thermal convection considered up to now is not the only type of buoyancy-driven
flows. A very common type of convective flows occurring in natural systems
is the compositional convection in binary alloys, composed of light
and heavy constituents, when the buoyancy is generated via local increase
in the concentration of the light constituent. This type of driving
is e.g. an important energy source for convective motions in the Earth's
liquid core, where at the inner-outer core boundary the process of
iron solidification leads to excess in the concentration of the light
constituents in the core alloy, composed mainly of iron.

We now turn to the fundamental ideas regarding buoyancy generation
in liquids via thermal and compositional effects. A comprehensive
derivation of the full set of equations describing the dynamics of
convection driven by the compositional and thermal effects can also
be found e.g. in Landau and Lifshitz (1987) or Braginsky and Roberts
(1995); the latter is done in the Earth's core context. The case,
when driving is solely thermal has been explained in section \ref{subsec:Adiabatic-gradient}.
Let us start by introducing the symbol
\begin{equation}
\xi=\frac{\rho_{l}}{\rho},\label{eq:e241}
\end{equation}
to denote the mass fraction of the light component in a binary alloy,
where $\rho_{l}$ is the mass of the light constituent in a unitary
volume and $\rho$ is the total mass of that volume; in other words $\rho$
is the density field of the entire liquid alloy, $\rho=\rho_{l}+\rho_{h}$,
where subscript $h$ denotes the heavy constituent. In the following
we assume that there are \emph{no chemical reactions} between the light and
heavy components of the alloy. In a static state, i.e. when no motions
are present the thermodynamic fields $\rho$, $T$, $p$, $s$ and
$\xi$ are stationary but height-dependent. Stability of a fluid layer,
with temperature and mass fraction gradients, can be studied by considering
two infinitesimally spaced horizontal (perpendicular to gravity) fluid
layers situated at heights $z_{0}$ and $z_{0}+\mathrm{d}z$. A very
similar type of reasoning as in section \ref{subsec:Adiabatic-gradient}
can be put forward now. A fluid volume of unitary mass $V=1/\rho$
taken from the level $z_{0}$ and placed slightly higher at $z_{0}+\mathrm{d}z$
experiences a thermodynamic transformation, since its temperature,
pressure and chemical potential\footnote{\label{fn:chem_pot}Because we have used $\mu$ to denote the shear
dynamic viscosity throughout the previous chapters, the chemical potential\index{SI}{chemical potential}
will be denoted here by $\mu_{c}$; moreover, to avoid any confusion
we will only use the symbols $\nu$ and $\nu_{b}$ to denote kinematic
shear and bulk viscosities, whereas the dynamic ones will be simply
denoted by $\tilde{\rho}\nu$ and $\tilde{\rho}\nu_{b}$.

The introduction of one chemical potential $\mu_{c}$ for a binary
mixture composed of light and heavy constituents follows from writing
down the differential of the internal energy
\[
\mathrm{d}\mathcal{E}=T\mathrm{d}S-p\mathrm{d}V+\mu^{(l)}\mathrm{d}N^{(l)}+\mu^{(h)}\mathrm{d}N^{(h)},
\]
where $\mathcal{E}$ and $S$ are the actual internal energy and entropy
of the alloy (as opposed to the mass densities $\varepsilon$ and
$s$), $V$ is the volume, $(\mu^{(l)},\,N^{(l)})$ and $(\mu^{(h)},\,N^{(h)})$
are the pairs of chemical potentials and the numbers of particles of
the light and heavy constituents respectively; division by the total
mass $M=m_{l}N^{(l)}+m_{h}N^{(h)}$, where $m_{l}$ and $m_{h}$ denote
the molecular masses for both constituents, allows to transform the
above into the differential for the internal energy mass density,
which takes the form
\[
\mathrm{d}\varepsilon=T\mathrm{d}s-p\mathrm{d}\left(\frac{1}{\rho}\right)+\mu_{c}\mathrm{d}\xi,
\]
where 
\[
\mu_{c}=\frac{\mu^{(l)}}{m_{l}}-\frac{\mu^{(h)}}{m_{h}},\quad\xi=\frac{m_{l}N^{(l)}}{M}=\frac{\rho_{l}}{\rho}.
\]
}, denoted here by $\mu_{c}$, will start adjusting to the environment at the higher level. If the
fluid volume after the transformation becomes denser than the surroundings,
the gravity will act to put it back at the original level $z_{0}$
and then the situation is stable; in the opposite case the buoyancy
is non-zero and the system looses stability. 

Let us take the pressure $p(z)$, the entropy $s(z)$ and the mass
fraction $\xi(z)$ as the system parameters, then the fluid volume,
which initially is 
\begin{equation}
V\left(p(z_{0}),s(z_{0}),\xi(z_{0})\right),\label{eq:e242}
\end{equation}
after the shift and the thermodynamic transformation which adjusts
the pressure to the value $p(z_{0}+\mathrm{d}z)$ of the surrounding fluid at
the higher level changes to 
\begin{equation}
V\left(p(z_{0}+\mathrm{d}z),s(z_{0})+\mathrm{D}s,\xi(z_{0})+\mathrm{D}\xi\right),\label{eq:e243}
\end{equation}
where $s(z_{0})+\mathrm{D}s$ and $\xi(z_{0})+\mathrm{D}\xi$ denote
the entropy and mass fraction after the thermodynamic process ($\mathrm{D}s$
is the entropy change and $\mathrm{D}\xi$ is the mass fraction change
in the process). We can now simplify things slightly, by considering
the most constrained case of adiabatic transformation of the fluid
volume at constant $s$ and $\xi$, thus with $\mathrm{D}s=0$ and
$\mathrm{D}\xi=0$; in that way we will obtain a sufficient and strongest
stability restriction. The hydrostatic state is stable as long as
the fluid parcel after the transformation is denser than the surroundings
at the new level $z_{0}+\mathrm{d}z$, i.e. when
\begin{equation}
V\left(p(z_{0}+\mathrm{d}z),s(z_{0}),\xi(z_{0})\right)<V\left(p(z_{0}+\mathrm{d}z),s(z_{0}+\mathrm{d}z_{0}),\xi(z_{0}+\mathrm{d}z_{0})\right).\label{eq:e244}
\end{equation}
Expanding the right hand side of the latter inequality in the entropy
and mass fraction about the values at the level $z_{0}$ we get
\begin{equation}
0<\left(\frac{\partial V}{\partial s}\right)_{p,\xi}\mathrm{d}s+\left(\frac{\partial V}{\partial\xi}\right)_{p,s}\mathrm{d}\xi,\label{eq:stab_cond_comp}
\end{equation}
where $\mathrm{d}s=s(z_{0}+\mathrm{d}z_{0})-s(z_{0})$ and $\mathrm{d}\xi=\xi(z_{0}+\mathrm{d}z_{0})-\xi(z_{0})$. Furthermore, the coefficient of thermal expansion $\alpha$ (\ref{eq:alpha})
likewise the coefficient of isentropic compositional expansion\index{SI}{compositional expansion!isentropic} 
\begin{equation}
\chi=\frac{1}{V}\left(\frac{\partial V}{\partial\xi}\right)_{p,s}=-\frac{1}{\rho}\left(\frac{\partial\rho}{\partial\xi}\right)_{p,s},\label{eq:e245}
\end{equation}
are both positive for all standard fluids/binary alloys, therefore
\begin{equation}
\left(\frac{\partial V}{\partial s}\right)_{p,\xi}=\frac{\left(\frac{\partial V}{\partial T}\right)_{p,\xi}}{\left(\frac{\partial s}{\partial T}\right)_{p,\xi}}=\frac{\alpha T}{\rho c_{p,\xi}}>0,\label{eq:e246}
\end{equation}
\begin{equation}
\left(\frac{\partial V}{\partial\xi}\right)_{p,s}=\frac{\chi}{\rho}>0.\label{eq:e247}
\end{equation}
which further implies, that the stability condition (\ref{eq:stab_cond_comp})
can be expressed in the following way
\begin{equation}
0<\frac{\alpha T}{\rho c_{p,\xi}}\frac{\mathrm{d}s}{\mathrm{d}z}+\frac{\chi}{\rho}\frac{\mathrm{d}\xi}{\mathrm{d}z}.\label{eq:stab_cond_comp_xi_s}
\end{equation}
This means that an adiabatic well-mixed state\index{SI}{well-mixed state} characterized by uniform entropy
$\mathrm{d}_{z}s=0$ and mass fraction $\mathrm{d}_{z}\xi=0$ corresponds
to instability threshold in the absence of dissipative effects; in
particular if both the entropy and mass fraction gradients become
negative in the absence of diffusion, the system becomes unstable.
Note however, that in fact the stability problem is governed by a
sum of the two gradients and the system may become unstable when only
one, say the mass fraction gradient, is negative and overcomes the
effect of positive entropy gradient. 

It can be seen now, that anelastic convection driven by both mechanisms
- thermal and compositional, has to be based on the requirement, that
the convective state departures only weakly from the well-mixed, adiabatic
state and therefore both the gradients $\mathrm{d}_{z}s$ and $\mathrm{d}_{z}\xi$
must be small. Next let us obtain the expressions for vertical gradients
of the intensive parameters $T$, $p$ and $\mu_{c}$ in the well-mixed,
adiabatic, hydrostatic state. The pressure gradient is easily obtained
from the hydrostatic momentum balance, $\mathrm{d}_{z}p=-\rho g$.
This allows to express the entropy gradient in terms of the temperature,
pressure and mass fraction gradients in the following way,
\begin{equation}
0=\frac{\mathrm{d}s}{\mathrm{d}z}=\left(\frac{\partial s}{\partial T}\right)_{p,\xi}\frac{\mathrm{d}T}{\mathrm{d}z}+\left(\frac{\partial s}{\partial p}\right)_{T,\xi}\frac{\mathrm{d}p}{\mathrm{d}z}+\left(\frac{\partial s}{\partial\xi}\right)_{T,p}\frac{\mathrm{d}\xi}{\mathrm{d}z}=\frac{c_{p,\xi}}{T}\frac{\mathrm{d}T}{\mathrm{d}z}+\alpha g,\label{eq:dsbydz_comp}
\end{equation}
where we have used the Maxwell identity $(\partial s/\partial p)_{T}=-(\partial V/\partial T)_{p}$
and $\mathrm{d}_{z}\xi=0$ for the well-mixed state. From the latter
we can infer, that the standard formula for the adiabatic gradient
still applies, that is the adiabatic, well-mixed state is characterized
by
\begin{equation}
\frac{\mathrm{d}T}{\mathrm{d}z}=-\frac{\alpha Tg}{c_{p,\xi}}.\label{eq:e248}
\end{equation}
Expressing now, the mass fraction gradient in terms of the temperature,
pressure and chemical potential gradients
\begin{align}
0=\frac{\mathrm{d}\xi}{\mathrm{d}z}= & \left(\frac{\partial\xi}{\partial T}\right)_{p,\mu_{c}}\frac{\mathrm{d}T}{\mathrm{d}z}+\left(\frac{\partial\xi}{\partial p}\right)_{T,\mu_{c}}\frac{\mathrm{d}p}{\mathrm{d}z}+\left(\frac{\partial\xi}{\partial\mu_{c}}\right)_{p,T}\frac{\mathrm{d}\mu_{c}}{\mathrm{d}z}\nonumber \\
= & -\left(\frac{\partial\xi}{\partial T}\right)_{p,\mu_{c}}\frac{\alpha Tg}{c_{p,\xi}}-\left(\frac{\partial\xi}{\partial p}\right)_{T,\mu_{c}}\rho g+\left(\frac{\partial\xi}{\partial\mu_{c}}\right)_{p,T}\frac{\mathrm{d}\mu_{c}}{\mathrm{d}z},\label{eq:dxibydz_comp}
\end{align}
leads to
\begin{equation}
\frac{\mathrm{d}\mu_{c}}{\mathrm{d}z}=\varUpsilon g\left[\left(\frac{\partial\xi}{\partial p}\right)_{T,\mu_{c}}\rho+\left(\frac{\partial\xi}{\partial T}\right)_{p,\mu_{c}}\frac{\alpha T}{c_{p,\xi}}\right],\label{eq:mu_grad_1}
\end{equation}
where we have introduced
\begin{equation}
\varUpsilon=\left(\frac{\partial\mu_{c}}{\partial\xi}\right)_{p,T}.\label{eq:e249}
\end{equation}
However, utilizing the implicit function theorem we may write
\begin{equation}
\left(\frac{\partial\xi}{\partial p}\right)_{T,\mu_{c}}=-\frac{\left(\frac{\partial\mu_{c}}{\partial p}\right)_{T,\xi}}{\left(\frac{\partial\mu_{c}}{\partial\xi}\right)_{p,T}}=-\frac{\chi_{T}}{\rho\varUpsilon},\label{eq:e250}
\end{equation}
\begin{equation}
\left(\frac{\partial\xi}{\partial T}\right)_{p,\mu_{c}}=-\frac{\left(\frac{\partial\mu_{c}}{\partial T}\right)_{p,\xi}}{\left(\frac{\partial\mu_{c}}{\partial\xi}\right)_{p,T}}=\frac{h_{p,T}}{T\varUpsilon},\label{eq:e251}
\end{equation}
where
\begin{equation}
\chi_{T}=-\frac{1}{\rho}\left(\frac{\partial\rho}{\partial\xi}\right)_{p,T}=\rho\left(\frac{\partial\mu_{c}}{\partial p}\right)_{T,\xi},\label{eq:e252}
\end{equation}
is the coefficient of isothermal compositional expansion\index{SI}{compositional expansion!isothermal} (the last
equality is simply a Maxwell relation) and the coefficient\index{SI}{heat of reaction}
\begin{equation}
h_{p,T}=T\left(\frac{\partial s}{\partial\xi}\right)_{p,T}=-T\left(\frac{\partial\mu_{c}}{\partial T}\right)_{p,\xi},\label{eq:h_pT}
\end{equation}
(again, the last equality is a Maxwell relation) is a thermodynamic
property of the fluid, which is related (but not directly !\footnote{The relation is not
direct, since as we will show in section \ref{sec:Compositional-and-heat-fluxes}
in a binary alloy composed of the light and heavy constituents the
total heat delivered to the unitary mass in an infinitesimal process
can not be expressed solely by $T\mathrm{d}s$, and there are contributions
from the process of chemical potential equilibration.}) to the
amount of heat delivered to the system when the mass fraction of the
light constituent is increased at constant temperature and pressure
(sometimes termed ``the heat of reaction'').

Equation (\ref{eq:mu_grad_1}) can now be rewritten in the form
\begin{equation}
\frac{\mathrm{d}\mu_{c}}{\mathrm{d}z}=-g\left[\chi_{T}-\frac{\alpha h_{p,T}}{c_{p,\xi}}\right].\label{eq:mu_grad_2}
\end{equation}
Furthermore, the coefficients of isentropic and isothermal compositional
expansions are related through the following equation
\begin{align}
\chi= & -\frac{1}{\rho}\left(\frac{\partial\rho}{\partial\xi}\right)_{p,s}=-\frac{1}{\rho}\left[\left(\frac{\partial\rho}{\partial\xi}\right)_{p,T}+\left(\frac{\partial\rho}{\partial T}\right)_{p,\xi}\left(\frac{\partial T}{\partial\xi}\right)_{p,s}\right]\nonumber \\
= & \chi_{T}-\frac{\alpha h_{p,T}}{c_{p,\xi}},\label{eq:e253}
\end{align}
where we have utilized the implicit function theorem to obtain $\left(\partial T/\partial\xi\right)_{p,s}=-h_{p,T}/c_{p,\xi}$,
which allows to express the chemical potential gradient in the well-mixed,
adiabatic and hydrostatic state in the most compact form
\begin{equation}
\frac{\mathrm{d}\mu_{c}}{\mathrm{d}z}=-\chi g.\label{eq:e254}
\end{equation}
The full expressions for the entropy and mass fraction gradient in
(\ref{eq:dsbydz_comp}) and in (\ref{eq:dxibydz_comp}) yield, 
\begin{subequations}
\begin{equation}
\frac{\mathrm{d}s}{\mathrm{d}z}=\frac{c_{p,\xi}}{T}\left(\frac{\mathrm{d}T}{\mathrm{d}z}+\frac{\alpha Tg}{c_{p,\xi}}\right)+\frac{h_{p,T}}{T}\frac{\mathrm{d}\xi}{\mathrm{d}z},\label{eq:entropy_grad_comp}
\end{equation}
\begin{align}
\frac{\mathrm{d}\xi}{\mathrm{d}z}= & \frac{h_{p,T}}{T\varUpsilon}\frac{\mathrm{d}T}{\mathrm{d}z}+\frac{\chi_{T}g}{\varUpsilon}+\frac{1}{\varUpsilon}\frac{\mathrm{d}\mu_{c}}{\mathrm{d}z}\nonumber \\
= & \frac{h_{p,T}}{T\varUpsilon}\left(\frac{\mathrm{d}T}{\mathrm{d}z}+\frac{\alpha Tg}{c_{p,\xi}}\right)+\frac{1}{\varUpsilon}\left(\chi g+\frac{\mathrm{d}\mu_{c}}{\mathrm{d}z}\right),\label{eq:xi_grad_comp}
\end{align}
\end{subequations} 
hence the stability condition (\ref{eq:stab_cond_comp_xi_s}),
can now be cast in the form
\begin{align}
0< & \frac{\alpha T}{\rho c_{p,\xi}}\frac{\mathrm{d}s}{\mathrm{d}z}+\frac{\chi}{\rho}\frac{\mathrm{d}\xi}{\mathrm{d}z}=\frac{\alpha}{\rho}\left(\frac{\mathrm{d}T}{\mathrm{d}z}+\frac{\alpha Tg}{c_{p,\xi}}\right)+\frac{\chi_{T}}{\rho}\frac{\mathrm{d}\xi}{\mathrm{d}z}\nonumber \\
 & =\left(\frac{\alpha}{\rho}+\frac{\chi_{T}h_{p,T}}{\rho T\varUpsilon}\right)\left(\frac{\mathrm{d}T}{\mathrm{d}z}+\frac{\alpha Tg}{c_{p,\xi}}\right)+\frac{\chi_{T}}{\rho\varUpsilon}\left(\frac{\mathrm{d}\mu_{c}}{\mathrm{d}z}+\chi g\right).\label{eq:gen_stab}
\end{align}
We recall, that this condition does not involve the effect of heat
exchange and material diffusion between a fluid parcel and surroundings,
thus providing a strongest stability restriction, which is typically
weakened by the presence of dissipation. Since $\alpha$ and $\chi$
are positive for standard binary alloys we can easily formulate
the following sufficient (but \emph{not necessary}) conditions for
stability
\begin{equation}
\frac{\mathrm{d}s}{\mathrm{d}z}>0,\quad\textrm{and}\quad\frac{\mathrm{d}\xi}{\mathrm{d}z}>0.\label{eq:stab_cond_s_xi}
\end{equation}
Additionally, another sufficient (but \emph{not necessary}) set of
stability conditions can be formulated, based on $\chi_{T}>0$,
\begin{equation}
-\frac{\mathrm{d}T}{\mathrm{d}z}<\frac{g\alpha T}{c_{p,\xi}},\quad\textrm{and}\quad\frac{\mathrm{d}\xi}{\mathrm{d}z}>0,\label{eq:stab_cond_xi_T}
\end{equation}
but we emphasize, that it is not directly equivalent to the former
one (\ref{eq:stab_cond_s_xi}), since the former can be transformed
to
\begin{equation}
-\frac{\mathrm{d}T}{\mathrm{d}z}<\frac{g\alpha T}{c_{p,\xi}}+\frac{\chi_{T}-\chi}{\alpha}\frac{\mathrm{d}\xi}{\mathrm{d}z},\quad\textrm{and}\quad\frac{\mathrm{d}\xi}{\mathrm{d}z}>0,\label{eq:e255}
\end{equation}
and whether it is a stronger or a weaker restriction on the temperature
gradient than (\ref{eq:stab_cond_xi_T}) depends on the sign of the
``heat of reaction'' $h_{p,T}=c_{p,\xi}(\chi_{T}-\chi)/\alpha$.
Nevertheless, we can conclude that convection does not develop when
the negative temperature gradient is below the adiabatic one and the
mass fraction gradient is positive, i.e. exceeds that of the ``well-mixed''
uniform profile. Thus the buoyancy forces may only start to appear
when the temperature gradient exceeds that of the adiabatic profile
and/or the mass fraction gradient exceeds zero (as explained above,
convection can develop if only one of the gradients exceeds the threshold
value strongly enough, to overcome the stabilizing effect of the other
one). 

The last line of (\ref{eq:gen_stab}) expresses the stability condition
in terms of the temperature and chemical potential gradients. It is
of interest to observe, that imposing simultaneously $-\mathrm{d}_{z}T<g\alpha T/c_{p,\xi}$
and $-\mathrm{d}_{z}\mu_{c}<\chi g$ does not guarantee stability,
because the sign of the coefficient $h_{p,T}$ is not specified. Therefore
the stability condition (\ref{eq:gen_stab}) in terms of temperature
and chemical potential gradients is simply expressed in the following
way
\begin{equation}
-\frac{\mathrm{d}\mu_{c}}{\mathrm{d}z}<\chi g+\left(\frac{\alpha\varUpsilon}{\chi_{T}}+\frac{h_{p,T}}{T}\right)\left(\frac{\mathrm{d}T}{\mathrm{d}z}+\frac{\alpha Tg}{c_{p,\xi}}\right),\label{eq:mu_stab_cond}
\end{equation}
where we have utilized the fact, that $\varUpsilon>0$ (cf. Landau
and Lifschitz 1980, p. 288, eq. (96.7), chapter on ``Thermodynamic
inequalities for solutions'', where this property of binary alloys
is derived directly from the minimal work principle).

However, in natural systems convection is typically driven by only
weak departure from the adiabatic, well-mixed state\index{SI}{well-mixed state}. In such a case,
which corresponds to the first fundamental assumption of the anelastic
approximation, all the following quantities are small
\begin{equation}
-\frac{L}{c_{p,\xi}}\frac{\mathrm{d}s}{\mathrm{d}z}\ll1,\quad-L\frac{\mathrm{d}\xi}{\mathrm{d}z}\ll1,\quad-\alpha L\left(\frac{\mathrm{d}T}{\mathrm{d}z}+\frac{g\alpha T}{c_{p,\xi}}\right)\ll1,\label{eq:small_departure_requirements}
\end{equation}
and as long as $\varUpsilon$ remains an order unity quantity, also
\begin{equation}
-\frac{\chi L}{\varUpsilon}\left(\frac{\mathrm{d}\mu_{c}}{\mathrm{d}z}+\chi g\right)\ll1\label{eq:e256}
\end{equation}
must be small (which is not true in weak solutions defined by $\xi\ll1$
when $\varUpsilon\gg c_{p,\xi}T$ is large, cf. section \ref{sec:Weak-solution-limit}).

We end this section by listing a full set of thermodynamic
properties of binary alloys which are used in this chapter, for an easy reference\index{SI}{compositional expansion!isentropic}\index{SI}{compositional expansion!isothermal}\index{SI}{heat of reaction}\index{SI}{thermal expansion coefficient}\index{SI}{isothermal compressibility coefficient}
\begin{subequations}
\begin{equation}
h_{p,T}=T\left(\frac{\partial s}{\partial\xi}\right)_{p,T}=-T\left(\frac{\partial\mu_{c}}{\partial T}\right)_{p,\xi},\label{eq:h_pT_list}
\end{equation}
\begin{equation}
\chi_{T}=-\frac{1}{\rho}\left(\frac{\partial\rho}{\partial\xi}\right)_{p,T}=\rho\left(\frac{\partial\mu_{c}}{\partial p}\right)_{T,\xi}>0,\label{eq:chi_T_list}
\end{equation}
\begin{equation}
\chi=-\frac{1}{\rho}\left(\frac{\partial\rho}{\partial\xi}\right)_{p,s}=\chi_{T}-\frac{\alpha h_{p,T}}{c_{p,\xi}}>0,\label{eq:chi_list}
\end{equation}
\begin{equation}
\varUpsilon=\left(\frac{\partial\mu_{c}}{\partial\xi}\right)_{p,T}>0,\quad c_{p,\xi}=T\left(\frac{\partial s}{\partial T}\right)_{p,\xi}>0,\quad c_{v,\xi}=T\left(\frac{\partial s}{\partial T}\right)_{\rho,\xi}>0,\label{eq:ups_cp_cv_list}
\end{equation}
\begin{equation}
\alpha=-\frac{1}{\rho}\left(\frac{\partial\rho}{\partial T}\right)_{p,\xi}>0,\quad\beta=\frac{1}{\rho}\left(\frac{\partial\rho}{\partial p}\right)_{T,\xi}>0.\label{eq:alpha_beta_list}
\end{equation}
\end{subequations}

\section{Compositional and heat fluxes\label{sec:Compositional-and-heat-fluxes}}

It is essential to derive the formulae for the compositional flux
and the heat flux; in the case of convection
driven by both, compositional and thermal effects the expression for the heat flux is significantly modified with respect to that for a single-component fluid. We start by
introducing the notation for the total molecular flux of entropy and the total molecular compositional
flux (without the contribution from the effect of advection by the
flow), which will be denoted by $\mathbf{j}_{s,\mathrm{mol}}$ and
$\mathbf{j}_{\xi,\mathrm{mol}}$ respectively. The total heat flux
will be denoted by $\mathbf{j}_{q}$. Therefore the general forms
of the entropy and mass fraction equations are
\begin{equation}
\rho\left(\frac{\partial s}{\partial t}+\mathbf{u}\cdot\nabla s\right)+\nabla\cdot\mathbf{j}_{s,\mathrm{mol}}=\sigma_{s},\label{eq:entropy_balance_gen_comp}
\end{equation}
\begin{equation}
\rho\left(\frac{\partial\xi}{\partial t}+\mathbf{u}\cdot\nabla\xi\right)+\nabla\cdot\mathbf{j}_{\xi,\mathrm{mol}}=0,\label{eq:mass_fraction_balance_gen_comp}
\end{equation}
where the mass conservation law $\partial_{t}\rho+\nabla\cdot(\rho\mathbf{u})=0$
has been used\footnote{The law of conservation of mass has the standard form for a continuous
medium, with the density $\rho=\rho_{l}+\rho_{h}$ and the flow velocity
of the alloy being the centre of mass velocity $\mathbf{u}=\xi\mathbf{u}_{l}+(1-\xi)\mathbf{u}_{h}$
for a fluid element, where $\mathbf{u}_{l}$ and $\mathbf{u}_{h}$
denote the velocities of motion of each of the two constituents of
the alloy. Therefore the law of mass conservation for the alloy, $\partial_{t}\rho+\nabla\cdot(\rho\mathbf{u})=0$,
results in a straightforward way from summing up the equations describing
the mass balance for each of the constituents, $\partial_{t}\rho_{l}+\nabla\cdot(\rho_{l}\mathbf{u}_{l})=0,$
and $\partial_{t}\rho_{h}+\nabla\cdot(\rho_{h}\mathbf{u}_{h})=0,$
i.e. 
\[
\partial_{t}(\rho_{l}+\rho_{h})+\nabla\cdot\left[\rho\left(\xi\mathbf{u}_{l}+(1-\xi)\mathbf{u}_{h}\right)\right]=0.
\]
}, $\sigma_{s}$ are the volume entropy sources and there are no volume
sources of the light constituent. In order to obtain the formulae
for the fluxes $\mathbf{j}_{s,\mathrm{mol}}$ and $\mathbf{j}_{\xi,\mathrm{mol}}$,
we must first identify the ``thermodynamic forces''\footnote{Also called ``affinities''.},
that is the causative thermodynamic stimuli of the fluxes, which take
the form of gradients of intensive thermodynamic variables, such as
e.g. the temperature or the chemical potential. To that end, we need
to find an explicit expression for the entropy production, which in
general terms is known to take the bilinear form of a sum of products
of the fluxes and thermodynamic forces associated with irreversible
processes, $\sigma_{s}=\sum_{j}\mathbf{j}_{j}\cdot\mathbf{X}_{j}$,
where the symbol $\mathbf{X}$ is commonly used to denote the thermodynamic
forces (cf. Landau and Lifshitz 1987, Glansdorff and Prigogine 1971,
de Groot and Mazur 1984).

The total energy density per unit mass in a fluid volume $V$ is a
sum of the kinetic energy density, the potential energy resulting
from the presence of conservative body forces, $\mathbf{F}=-\nabla\psi$,
which we will assume stationary and the internal energy,
\begin{equation}
e=\frac{1}{2}\mathbf{u}^{2}+\psi+\varepsilon.\label{eq:tot_en_comp}
\end{equation}
 The evolution of the total energy is described by the following law
(cf. (\ref{eq:gen_evolution_law_local}))
\begin{equation}
\rho\left(\frac{\partial e}{\partial t}+\mathbf{u}\cdot\nabla e\right)+\nabla\cdot\mathbf{j}_{e,\mathrm{mol}}=Q,\label{eq:energy_evolution_gen-2}
\end{equation}
where $\rho e\mathbf{u}$ is the advective energy flux, $\mathbf{j}_{e,\mathrm{mol}}$
is the flux of energy from molecular mechanical, thermal and compositional
effects and $Q$ is the energy source (absorbed heat per unit mass
per unit volume), here unspecified, which e.g. may describe the effects
of radioactive heating, thermal radiation, etc. From section \ref{sec:The-energy-equation}
we know, that the molecular energy flux is composed of two factors,
that is the heat transferred between a fluid volume and the rest of
the fluid and the total work done on the volume by the stresses,
\begin{equation}
\mathbf{j}_{e,\mathrm{mol}}=-\boldsymbol{\tau}\cdot\mathbf{u}+\mathbf{j}_{q},\label{eq:je_vs_jq_comp}
\end{equation}
where $\mathbf{j}_{q}$ denotes the total molecular heat flux. From
the expression for the total differential of the internal energy per
unit mass
\begin{equation}
\mathrm{d}\varepsilon=T\mathrm{d}s+\frac{p}{\rho^{2}}\mathrm{d}\rho+\mu_{c}\mathrm{d}\xi,\label{eq:e257}
\end{equation}
supplied by the mass conservation equation $\partial_{t}\rho+\mathbf{u}\cdot\nabla\rho=-\rho\nabla\cdot\mathbf{u}$
and the mass fraction balance (\ref{eq:mass_fraction_balance_gen_comp})
one obtains
\begin{equation}
\rho\left(\frac{\partial\varepsilon}{\partial t}+\mathbf{u}\cdot\nabla\varepsilon\right)=\rho T\left(\frac{\partial s}{\partial t}+\mathbf{u}\cdot\nabla s\right)-p\nabla\cdot\mathbf{u}-\mu_{c}\nabla\cdot\mathbf{j}_{\xi,\mathrm{mol}}.\label{eq:int_en_comp}
\end{equation}
The variation of the kinetic and potential energies is described by
the same equations as in section \ref{sec:The-energy-equation}, thus
\begin{equation}
\rho\left(\frac{\partial}{\partial t}+\mathbf{u}\cdot\nabla\right)\frac{1}{2}\mathbf{u}^{2}=\nabla\cdot\left(\boldsymbol{\tau}\cdot\mathbf{u}\right)-\boldsymbol{\tau}:\mathbf{G}^{s}+\rho\mathbf{u}\cdot\mathbf{F},\label{eq:kin_en_comp}
\end{equation}
\begin{equation}
\rho\left(\frac{\partial}{\partial t}+\mathbf{u}\cdot\nabla\right)\psi=-\rho\mathbf{u}\cdot\mathbf{F},\label{eq:pot_en_comp}
\end{equation}
where
\begin{equation}
\boldsymbol{\tau}:\mathbf{G}^{s}=-p\nabla\cdot\mathbf{u}+2\rho\nu\mathbf{G}^{s}:\mathbf{G}^{s}+\left(\rho\nu_{b}-\frac{2}{3}\rho\nu\right)\left(\nabla\cdot\mathbf{u}\right)^{2}.\label{eq:e258}
\end{equation}
Substitution of (\ref{eq:tot_en_comp}), (\ref{eq:int_en_comp}),
(\ref{eq:kin_en_comp}) and (\ref{eq:pot_en_comp}) into the total
energy balance (\ref{eq:energy_evolution_gen-2}) leads to
\begin{equation}
\rho T\left(\frac{\partial s}{\partial t}+\mathbf{u}\cdot\nabla s\right)=-\nabla\cdot\left(\mathbf{j}_{e,\mathrm{mol}}+\boldsymbol{\tau}\cdot\mathbf{u}-\mu_{c}\mathbf{j}_{\xi,\mathrm{mol}}\right)+\boldsymbol{\tau}_{\nu}:\mathbf{G}^{s}-\mathbf{j}_{\xi,\mathrm{mol}}\cdot\nabla\mu_{c}+Q,\label{eq:entropy_working_comp}
\end{equation}
where the expression
\begin{equation}
\boldsymbol{\tau}_{\nu}:\mathbf{G}^{s}=2\rho\nu\mathbf{G}^{s}:\mathbf{G}^{s}+\left(\rho\nu_{b}-\frac{2}{3}\rho\nu\right)\left(\nabla\cdot\mathbf{u}\right)^{2},\label{eq:e259}
\end{equation}
involves now only the dissipative part of the stress tensor, $\boldsymbol{\tau}_{\nu}$.
Comparison of the equation (\ref{eq:entropy_working_comp}) with the
entropy balance (\ref{eq:entropy_balance_gen_comp}) multiplied by
the temperature $T$, which reads
\begin{equation}
\rho T\left(\frac{\partial s}{\partial t}+\mathbf{u}\cdot\nabla s\right)=-\nabla\cdot\left(T\mathbf{j}_{s,\mathrm{mol}}\right)+\mathbf{j}_{s,\mathrm{mol}}\cdot\nabla T+T\sigma_{s},\label{eq:e260}
\end{equation}
allows to express the molecular flux of the total energy\index{SI}{flux!of the total energy} and the entropy
sources in terms of the entropy and mass fraction fluxes
\begin{equation}
\mathbf{j}_{e,\mathrm{mol}}=-\boldsymbol{\tau}\cdot\mathbf{u}+\mu_{c}\mathbf{j}_{\xi,\mathrm{mol}}+T\mathbf{j}_{s,\mathrm{mol}},\label{eq:je_vs_js_and_jxi}
\end{equation}
\begin{equation}
T\sigma_{s}=\boldsymbol{\tau}_{\nu}:\mathbf{G}^{s}-\mathbf{j}_{\xi,\mathrm{mol}}\cdot\nabla\mu_{c}-\mathbf{j}_{s,\mathrm{mol}}\cdot\nabla T+Q.\label{eq:entropy_source}
\end{equation}
Additionally, by the use of (\ref{eq:je_vs_jq_comp}) and (\ref{eq:je_vs_js_and_jxi})
we get an expression for the total molecular heat flux\index{SI}{heat flux!total}
\begin{equation}
\mathbf{j}_{q}=\mu_{c}\mathbf{j}_{\xi,\mathrm{mol}}+T\mathbf{j}_{s,\mathrm{mol}}.\label{eq:heat_flux_vs_jxi_js}
\end{equation}
We can now identify form (\ref{eq:entropy_source}), that the causative
thermodynamic stimuli of the compositional and entropy fluxes are
the chemical potential and temperature gradients divided by the temperature.
Although the actual stimulus of the entropy flow is the temperature
gradient and the material diffusion is stimulated by the chemical
potential differences, in a binary alloy the material diffusion \emph{can
not} take place without simultaneous entropy transfer and thermal
conduction is \emph{necessarily} associated with simultaneous concentration
flow. Thus each of the fluxes of $\xi$ and $s$ must depend on both
of the gradients, i.e. the gradient of the chemical potential and
the temperature gradient. If the gradients are small we may suppose
the fluxes are linearly related to $\nabla\mu_{c}$ and $\nabla T$,
\begin{equation}
\mathbf{j}_{\xi,\mathrm{mol}}=L_{\xi\xi}\frac{\nabla\mu_{c}}{T}+L_{\xi s}\frac{\nabla T}{T},\label{eq:jxi_gen}
\end{equation}
\begin{equation}
\mathbf{j}_{s,\mathrm{mol}}=L_{s\xi}\frac{\nabla\mu_{c}}{T}+L_{ss}\frac{\nabla T}{T},\label{eq:js_gen}
\end{equation}
where $L_{\xi\xi}$, $L_{\xi s}$, $L_{s\xi}$, $L_{ss}$ are called
phenomenological coefficients\footnote{Also termed ``kinetic coefficients''.},
independent of any thermodynamic forces and fluxes (the first lower
index in the coefficients $L_{ij}$ is associated with the type of
flux expressed by linear combination of $\nabla\mu_{c}/T$ and $\nabla T/T$
and the second lower index with the quantity, whose flux is multiplied
by the relevant stimulus in the expression for $\sigma_{s}$). In
the case when two or more fluxes share the same causative stimuli,
the Onsager's reciprocity principle applies (cf. e.g. Landau and Lifshitz
1980, p. 365, chapter ``The symmetry of the kinetic coefficients'',
or de Groot and Mazur 1984, p. 100), which stems from kinetic theory
and utilizes the time reversal symmetry of the equations of motion
of individual particles and the assumption of only small departures
from thermodynamic equilibrium\footnote{In the current case the equations (\ref{eq:jxi_gen})-(\ref{eq:js_gen})
can be rewritten in the more general form utilized e.g by Landau and
Lifshitz 1987, Glansdorff and Prigogine 1971 or de Groot and Mazur
1984, $\mathrm{d}_{t}x_{i}=L_{ij}X_{j}$, where $\mathrm{d}_{t}x_{1}=\mathrm{d}_{t}\mathring{m}_{l}=\mathbf{j}_{\xi,\mathrm{mol}}$,
$\mathrm{d}_{t}x_{2}=\mathrm{d}_{t}\mathring{S}=\mathbf{j}_{s,\mathrm{mol}}$,
and $\mathring{m}_{l}$ expressed in $[kg/m^{2}]$ is the mass of
the light constituent transported through a unit surface of a fluid
element in a time unit near thermodynamic equilibrium, whereas $\mathring{S}$
is the total entropy transported through a unit surface of a fluid
element in a time unit near thermodynamic equilibrium. Inspection
of the expression for the entropy sources (\ref{eq:entropy_source})
allows to write $\mathrm{d}_{t}\int_{V_{0}}\rho s\mathrm{d}V=-\int_{V_{0}}\mathbf{j}_{\xi,\mathrm{mol}}\cdot(\nabla\mu_{c}/T)\mathrm{d}V-\int_{V_{0}}\mathbf{j}_{s,\mathrm{mol}}\cdot(\nabla T/T)\mathrm{d}V+\dots$,
where the flux term, likewise the viscous and radiogenic terms have
been omitted for brevity and $V_{0}$ denotes here the entire fluid
volume. Thus we recognize $\rho\partial s/\partial\mathring{S}=\nabla T/T$
and $\rho\partial s/\partial\mathring{m}_{l}=\nabla\mu/T$, and the
Onsager's reciprocity principle applies (cf. e.g. Landau and Lifshitz
1987, p. 230, chapter ``Coefficients of mass transfer and thermal
diffusion''). }; it states, that the kinetic coefficients are symmetric
\begin{equation}
L_{\xi s}=L_{s\xi}.\label{eq:e261}
\end{equation}
The gradient of the chemical potential as a function of the pressure,
the temperature and the mass fraction, $\mu_{c}(p,\,T,\,\xi)$,
can be easily cast in the form
\begin{equation}
\nabla\mu_{c}=\frac{\chi_{T}}{\rho}\nabla p-\frac{h_{p,T}}{T}\nabla T+\varUpsilon\nabla\xi.\label{eq:grad_muc}
\end{equation}
On inserting the latter formula into expressions for the molecular compositional and entropy fluxes (\ref{eq:jxi_gen})-(\ref{eq:js_gen})
one obtains\index{SI}{flux!compositional (material)}\index{SI}{entropy!flux}
\begin{equation}
\mathbf{j}_{\xi,\mathrm{mol}}=-K\left(\nabla\xi+\frac{k_{T}}{T}\nabla T+\frac{k_{p}}{p}\nabla p\right),\label{eq:j_xi_gen_comp}
\end{equation}
\begin{equation}
T\mathbf{j}_{s,\mathrm{mol}}=-k\nabla T+\Lambda\mathbf{j}_{\xi,\mathrm{mol}},\label{eq:j_s_gen_comp}
\end{equation}
where the following new set of phenomenological coefficients was introduced
\begin{equation}
K=-L_{\xi\xi}\frac{\varUpsilon}{T},\qquad k=\frac{\Lambda}{T}L_{\xi s}-L_{ss},\label{eq:Kk}
\end{equation}
\begin{equation}
k_{T}=\frac{\Lambda-h_{p,T}}{\varUpsilon},\quad k_{p}=\frac{p\chi_{T}}{\rho\varUpsilon},\quad\Lambda=T\frac{L_{\xi s}}{L_{\xi\xi}}.\label{eq:kT_kp_muprime}
\end{equation}
In the above $k$ is the coefficient of thermal conductivity, $K$
denotes the coefficient of material conductivity and
\begin{equation}
D=K/\rho\label{eq:e262}
\end{equation}
will be used to denote the material diffusion coefficient. Furthermore,
$k_{T}$ is the Soret coefficient, which describes the effect of temperature
gradient on the material flux, whereas $k_{p}$ describes an analogous
effect of the pressure gradient. The coefficient $\Lambda$ (which
has the units of energy mass density, $J/kg$) can be called the Dufour
coefficient, since it describes the effect of concentration gradient
on the entropy flux. It is also of interest to observe, that the chemical
potential gradient can be expressed by the material flux in the following
way
\begin{equation}
\nabla\mu_{c}=-\frac{\varUpsilon}{K}\mathbf{j}_{\xi,\mathrm{mol}}-\frac{\Lambda}{T}\nabla T.\label{eq:chem_pot_grad}
\end{equation}
We are now ready to write down the entropy balance (\ref{eq:entropy_balance_gen_comp})
in a more explicit form
\begin{equation}
\rho T\left(\frac{\partial s}{\partial t}+\mathbf{u}\cdot\nabla s\right)=\nabla\cdot\left(k\nabla T\right)-\nabla\cdot\left(\Lambda\mathbf{j}_{\xi,\mathrm{mol}}\right)+\boldsymbol{\tau}_{\nu}:\mathbf{G}^{s}-\mathbf{j}_{\xi,\mathrm{mol}}\cdot\nabla\mu_{c}+Q,\label{eq:entropy_balance_gen_explicit_comp}
\end{equation}
where the flux of the light constituent is given by (\ref{eq:j_xi_gen_comp});
the mass fraction balance, on the other hand, is given in (\ref{eq:mass_fraction_balance_gen_comp}).

Let us now utilize formula (\ref{eq:heat_flux_vs_jxi_js}) for the
total heat flux to write down the heat balance. By the use of the
general local evolution law (\ref{eq:gen_evolution_law_local}) the
rate of variation of the heat exchanged via molecular processes between
a fluid parcel and its surroundings is expressed by the heat flux
divergence $-\nabla\cdot\mathbf{j}_{q}$ and the heat sources, which
contain the viscous heating\index{SI}{viscous heating} $\boldsymbol{\tau}_{\nu}:\mathbf{G}^{s}$
and other possible sources $Q$, e.g. radiogenic. Therefore the formulae
(\ref{eq:heat_flux_vs_jxi_js}), (\ref{eq:entropy_balance_gen_explicit_comp})
and (\ref{eq:mass_fraction_balance_gen_comp}) yield
\begin{align}
\frac{\dbar q_{\mathrm{tot}}}{\mathrm{d}t}= & -\nabla\cdot\mathbf{j}_{q}+\boldsymbol{\tau}_{\nu}:\mathbf{G}^{s}+Q\nonumber \\
= & -\nabla\cdot\left(\mu_{c}\mathbf{j}_{\xi,\mathrm{mol}}\right)-\nabla\cdot\left(T\mathbf{j}_{s,\mathrm{mol}}\right)+\boldsymbol{\tau}_{\nu}:\mathbf{G}^{s}+Q\nonumber \\
= & -\mu_{c}\nabla\cdot\mathbf{j}_{\xi,\mathrm{mol}}+\rho T\frac{Ds}{\partial t}\nonumber \\
= & \rho T\frac{Ds}{\partial t}+\rho\mu_{c}\frac{D\xi}{\partial t}.\label{eq:e263}
\end{align}
An important conclusion can be drawn from the latter equation, that
the total infinitesimal heat delivered to the unit volume can not
be expressed solely by $T\rho\mathrm{d}s$, contrary to single-component
fluids. In a binary alloy the effects of compositional variation (chemical
potential equilibration) contribute significantly to the total molecular
heat flux.

A significant simplification of the entropy balance (\ref{eq:entropy_balance_gen_explicit_comp})
can be achieved through neglection of the Soret effect (the effect
of temperature gradient on the material flux) and the effect of the
pressure gradient on the mass fraction flux, so that effectively the
binary alloy satisfies the Fick's law\index{SI}{Fick's law}
\begin{equation}
\mathbf{j}_{\xi,\mathrm{mol}}=-K\nabla\xi.\label{eq:Ficks_law}
\end{equation}
This can be obtained by assuming e.g. $\Lambda\approx h_{p,T}$, so
that $k_{T}\approx0$ and only small values of the relative pressure
gradient, $L\nabla p/p\ll1$ (such as in the Boussinesq approximation).
The entropy balance and the mass fraction equation then read
\begin{equation}
\rho T\left(\frac{\partial s}{\partial t}+\mathbf{u}\cdot\nabla s\right)=\nabla\cdot\left(k\nabla T\right)+\nabla\cdot\left(\Lambda K\nabla\xi\right)+\boldsymbol{\tau}_{\nu}:\mathbf{G}^{s}+K\nabla\xi\cdot\nabla\mu_{c}+Q,\label{eq:entropy_eq_no_Soret}
\end{equation}
\begin{equation}
\rho\left(\frac{\partial\xi}{\partial t}+\mathbf{u}\cdot\nabla\xi\right)=\nabla\cdot\left(K\nabla\xi\right),\label{eq:mass_frac_eq_no_Soret}
\end{equation}
where $\nabla\mu_{c}$ is given in (\ref{eq:grad_muc}). Further simplification
is still possible when the Dufour coefficient $\Lambda$ is small,
compared to some typical energy scale in the system, say the gravitational
energy scale $\bar{g}L$ or thermal energy scale $c_{v,\xi}T$. Then
the Dufour effect, that is the influence of the mass concentration
flux on the the entropy variation, can be neglected yielding
\begin{equation}
\rho T\left(\frac{\partial s}{\partial t}+\mathbf{u}\cdot\nabla s\right)=\nabla\cdot\left(k\nabla T\right)+\boldsymbol{\tau}_{\nu}:\mathbf{G}^{s}+K\nabla\xi\cdot\nabla\mu_{c}+Q.\label{eq:e264}
\end{equation}
It might be tempting to also provide some justification for neglection
of the term $K\nabla\xi\cdot\nabla\mu_{c}$, so that the entropy balance
directly corresponds to that of a single-component fluid. Unfortunately,
it is hard to justify neglection of this term unless the parameter
$\varUpsilon$ is assumed small. However, if $\varUpsilon$ were assumed
small then the coefficients $k_{T}$ and $k_{p}$ describing the Soret
effect and the pressure effect on the material flux become significant,
unless $\chi_{T}$ and thus the compositional buoyancy is assumed
negligible. The latter assumption means that only the thermal buoyancy
plays a dynamical role (cf. section \ref{sec:Buoyancy-force})
and the problem is reduced to purely thermal convection studied in
previous chapters.

Therefore the only sensible way to neglect the term $K\nabla\xi\cdot\nabla\mu_{c}$
is on the grounds, that the ratio of material to thermal diffusion
coefficients
\begin{equation}
\frac{D}{\kappa}=\frac{Pr}{Sc}\ll1,\label{eq:e265}
\end{equation}
is small, where 
\begin{equation}
Pr=\frac{\nu}{\kappa},\qquad Sc=\frac{\nu}{D},\label{eq:e266}
\end{equation}
are the Prandtl and Schmidt numbers\index{SI}{Schmidt number}, respectively. Indeed, the ratios
of the term $K\nabla\xi\cdot\nabla\mu_{c}$ to the thermal diffusion
and entropy advection can be estimated as follows
\begin{equation}
\frac{\left|K\nabla\xi\cdot\nabla\mu_{c}\right|}{\left|\nabla\cdot\left(k\nabla T\right)\right|}\sim\frac{K\varUpsilon\left|\nabla\xi\right|^{2}}{k\left|\nabla^{2}T\right|}\sim\frac{Pr}{Sc}\frac{\varUpsilon\left|\xi'\right|^{2}}{c_{p,\xi}\left|T'\right|},\label{eq:e267}
\end{equation}
\begin{equation}
\frac{\left|K\nabla\xi\cdot\nabla\mu_{c}\right|}{\left|\rho T\mathbf{u}\cdot\nabla s\right|}\sim\frac{K\varUpsilon\left|\nabla\xi\right|^{2}}{\left|\rho\right|\left|T\right|\mathscr{U}\left|\nabla s\right|}\sim\frac{1}{ReSc}\frac{\varUpsilon\left|\xi'\right|^{2}}{T_{B}\left|s'\right|},\label{eq:e268}
\end{equation}
hence the assumptions $D/\kappa\ll1$ and $ReSc\gg1$, or simply $Sc\gg1$
allow to obtain the entropy equation in the single-component form
(note, that in order for the thermal and compositional buoyancy forces
to be comparable in magnitude we must require for the magnitudes of
fluctuations $\left|\xi'\right|\sim\left|s'/c_{p,\xi}\right|$, which
also implies $\left|\xi'\right|\sim\left|\alpha T'\right|$; cf. the
following sections \ref{sec:Hydrostatic-reference-state} and \ref{sec:Buoyancy-force}).
However, assumption such as $Sc\gg1$ affects also the mass fraction
equation (\ref{eq:mass_frac_eq_no_Soret}). A non-dimensional form
of that equation requires introduction of some time scale, say the
viscous time scale $L^{2}/\nu$ or the inertial time scale $L/\mathscr{U}$.
This leads to appearance of small parameter in front of the diffusive
term which contains the highest-order derivative in the mass fraction
equation, introducing compositional boundary layers into the dynamics
(alternative choice of $L^{2}/D$ for the time scale leads to small,
or large parameters in the remaining dynamical equations).

Perhaps a simplest way to neglect the Soret effect and the effect of the pressure gradient on the compositional flux is to assume large $\varUpsilon\gg c_{v,\xi}\bar T$ (but \emph{without} assuming the weak solution limit $\xi\ll1$ explained in section \ref{sec:Weak-solution-limit}). In such a case the coefficients $k_T\approx 0$, $k_p\approx0$ are negligibly small (cf. (\ref{eq:kT_kp_muprime})), whereas the expression for the chemical potential gradient (\ref{eq:grad_muc}) simplifies to $\nabla\mu_c\approx\varUpsilon\nabla\xi$. Additional assumption of a weak Dufour effect, $\Lambda\ll c_{v,\xi}\bar T$ allows to achieve the mass fraction and energy equations in greatly simplified forms of (\ref{eq:mass_frac_eq_no_Soret}) and 
\begin{equation}
\rho T\left(\frac{\partial s}{\partial t}+\mathbf{u}\cdot\nabla s\right)=\nabla\cdot\left(k\nabla T\right)+\boldsymbol{\tau}_{\nu}:\mathbf{G}^{s}+K\varUpsilon\left(\nabla\xi\right)^2+Q.\label{eq:ff23}
\end{equation}
Finally, it is of interest to note, that from the mass fraction equation
(\ref{eq:mass_frac_eq_no_Soret}) we may anticipate already, that
the anelastic approximation for binary alloys requires
\begin{equation}
D\sim\delta^{1/2}\sqrt{\bar{g}L}L,\label{eq:e269}
\end{equation}
just as in the case of viscous and thermal diffusion coefficients,
cf. (\ref{eq:visc_scale-1}) and (\ref{eq:kappa_scale}).

\subsection{Boundary conditions on the mass fraction\label{subsec:Boundary-conditions-on-xi}}

The standard conditions that the mass fraction can be sensibly assumed
to satisfy at boundaries are similar to those often assumed for temperature.
Most frequently the flux of the light constituent through the boundaries
is required to be constant, i.e.
\begin{equation}
K\left.\left(\frac{\partial\xi}{\partial z}+\frac{k_{T}}{T}\frac{\partial T}{\partial z}+\frac{k_{p}}{\tilde{p}}\frac{\partial p}{\partial z}\right)\right|_{z=0,L}=\mathrm{const}.\label{eq:e270}
\end{equation}
Such a boundary condition corresponds to an exemplary situation, when
the heavy constituent solidifies at the bottom, and the light constituent
solidifies at the top of the domain; that way the light constituent
is effectively supplied at the bottom and expelled at the top. Of
course, when the Soret effect is negligible, $k_{T}=0$, and the flux
is independent of the pressure gradient, $k_{p}=0$, this boundary
condition simplifies to
\begin{equation}
K\left.\frac{\partial\xi}{\partial z}\right|_{z=0,L}=\mathrm{const}.\label{eq:e271}
\end{equation}
Note, that in such a case the compositional convection at constant
temperature, thus in the absence of thermal driving, is directly relevant
to the Boussinesq Rayleigh-B$\acute{\mathrm{e}}$nard problem with fixed heat flux at
the boundaries.

Another possibility is to assume, that the boundaries dissolve in
the fluid (binary alloy) in such a way, that the saturation concentration
$\xi_{0}$ (say) is rapidly established near their surfaces. In such
a case the mass fraction can be assumed constant at the boundaries,
\begin{equation}
\left.\xi\right|_{z=0,L}=\mathrm{const}.\label{eq:e272}
\end{equation}

\subsection{Production of the total entropy and the second law of thermodynamics\label{subsec:Entropy_production_comp}}\index{SI}{entropy!production}

The second law of thermodynamics states, that in a closed and adiabatically
insulated system the production of the total entropy is always positive
or null. The assumption of a closed system means, that there are no
material interactions with the surroundings, i.e. the material (compositional)
flux at the boundaries is zero
\begin{equation}
\left.\mathbf{j}_{\xi,\mathrm{mol}}\cdot\hat{\mathbf{e}}_{z}\right|_{z=0,L}=0.\label{eq:j_xi_BC_EP_comp}
\end{equation}
Adiabatic insulation on the other hand implies that the total heat flux
(\ref{eq:heat_flux_vs_jxi_js}), composed of the material and entropy
fluxes, needs to vanish at the boundaries. By the use of the expression
for the total entropy flux provided in (\ref{eq:j_s_gen_comp}) and
the fact, that the material flux has already been assumed to vanish
at the boundaries, the assumption of adiabatic insulation reduces
to
\begin{equation}
-\left.k\frac{\partial T}{\partial z}\right|_{z=0,L}=0.\label{eq:HF_BC_EP_comp}
\end{equation}
Impermeability of the boundaries additionally implies
\begin{equation}
\left.\mathbf{u}\cdot\hat{\mathbf{e}}_{z}\right|_{z=0,L}=0,\label{eq:IMP_BC_EP_comp}
\end{equation}
and to fix ideas let us assume periodicity in the horizontal directions.
We can now divide the energy equation (\ref{eq:entropy_balance_gen_explicit_comp})
by $T$ to get the entropy balance and rearrange the flux terms in
the following way 
\begin{subequations}
\begin{equation}
\frac{1}{T}\nabla\cdot\left(k\nabla T\right)=\nabla\cdot\left(k\frac{\nabla T}{T}\right)+k\frac{\left(\nabla T\right)^{2}}{T^{2}},\label{eq:e273}
\end{equation}
\begin{equation}
-\frac{1}{T}\nabla\cdot\left(\Lambda\mathbf{j}_{\xi,\mathrm{mol}}\right)=-\nabla\cdot\left(\Lambda\frac{\mathbf{j}_{\xi,\mathrm{mol}}}{T}\right)-\Lambda\frac{\mathbf{j}_{\xi,\mathrm{mol}}\cdot\nabla T}{T^{2}}.\label{eq:e274}
\end{equation}
\end{subequations} 
On integration of the entropy balance over the
entire horizontally periodic volume and application of the boundary
conditions (\ref{eq:j_xi_BC_EP_comp}), (\ref{eq:HF_BC_EP_comp})
and (\ref{eq:IMP_BC_EP_comp}) one obtains
\begin{align}
\frac{\partial}{\partial t}\int_{V}\rho s\mathrm{d}V= & \int_{V}k\frac{\left(\nabla T\right)^{2}}{T^{2}}\mathrm{d}V-\int_{V}\Lambda\frac{\mathbf{j}_{\xi,\mathrm{mol}}\cdot\nabla T}{T^{2}}\mathrm{d}V+\int_{V}\frac{1}{T}\boldsymbol{\tau}_{\nu}:\mathbf{G}^{s}\mathrm{d}V\nonumber \\
 & -\int_{V}\frac{\mathbf{j}_{\xi,\mathrm{mol}}\cdot\nabla\mu_{c}}{T}\mathrm{d}V+\int_{V}Q\mathrm{d}V,\label{eq:entropy_prod_comp-1}
\end{align}
where the mass conservation equation $\partial_{t}\rho+\nabla\cdot(\rho\mathbf{u})=0$
was used on the left hand side and
\begin{equation}
\int_{V}\nabla\cdot(\rho\mathbf{u}s)\mathrm{d}V=\int_{\partial V}\rho s\mathbf{u}\cdot\hat{\mathbf{n}}\mathrm{d}\Sigma=0\label{eq:e275}
\end{equation}
by the impermeability (and horizontal periodicity) conditions. Finally,
introducing the expression (\ref{eq:chem_pot_grad}) for the chemical
potential gradient into the first term in the second line of (\ref{eq:entropy_prod_comp-1})
leads to
\begin{align}
\frac{\partial}{\partial t}\int_{V}\rho s\mathrm{d}V= & \int_{V}k\frac{\left(\nabla T\right)^{2}}{T^{2}}\mathrm{d}V+\int_{V}\frac{1}{T}\boldsymbol{\tau}_{\nu}:\mathbf{G}^{s}\mathrm{d}V\nonumber \\
 & +\int_{V}\frac{\varUpsilon}{K}\frac{\mathbf{j}_{\xi,\mathrm{mol}}^{2}}{T}\mathrm{d}V+\int_{V}Q\mathrm{d}V.\label{eq:entropy_prod_comp}
\end{align}
It was shown in section \ref{subsec:Total-entropy-production}, that
the viscous term
\begin{equation}
\int_{V}\frac{1}{T}\boldsymbol{\tau}_{\nu}:\mathbf{G}^{s}\mathrm{d}V\geq0\label{eq:e276}
\end{equation}
is positive definite and since by the second law of thermodynamics
the entire production of the total entropy must be positive definite
\begin{equation}
\frac{\partial}{\partial t}\int_{V}\rho s\mathrm{d}V\geq0,\label{eq:e277}
\end{equation}
this implies positivity of the material conductivity coefficient
\begin{equation}
K>0.\label{eq:e278}
\end{equation}

\section{Hydrostatic reference state\label{sec:Hydrostatic-reference-state}}\index{SI}{reference (basic) state}

At this stage we introduce the concept of the \emph{anelastic approximation}
for binary alloys; i.e. we list the additional assumptions with respect
to those made for single-component fluids, that have to be made when
binary alloys are considered. We proceed to derive the equations for
the hydrostatic reference state. As in chapter \ref{chap:Anelastic-convection}
we assume time independent boundary conditions and decompose the thermodynamic
variables into the reference state contributions and the fluctuations
induced by the convective flow (\ref{eq:rho_gen_A_deriv}-e), thus
the mass fraction variable also takes the form
\begin{equation}
\xi(\mathbf{x},t)=\tilde{\xi}(z)+\xi'(\mathbf{x},t).\label{eq:p_gen_A_deriv-1-1}
\end{equation}
Within the anelastic approximation the convection is driven thermally
by only small departures from the adiabatic profile, $s=\mathrm{const}$,
and compositionally through small departures from the well-mixed profile
$\xi=\mathrm{const}$ (the \emph{first fundamental assumption} of
the anelastic approximation). Therefore the assumption (\ref{eq:delta_definition})
still holds, but additionally we need to require\index{SI}{anelastic!assumption (1)}
\begin{equation}
-L\left\langle \frac{\mathrm{d}\tilde{\xi}}{\mathrm{\mathrm{d}z}}\right\rangle =\Delta\xi\sim\delta\ll1,\label{eq:delta_definition-1}
\end{equation}
where $\Delta\xi=\xi_{B}-\xi_{T}$ denotes the mass concentration
jump across the fluid layer. Note, that similarly as in the case of
the entropy variable (cf. discussion below (\ref{fluct_magnitude_A})),
the static state function $\tilde{\xi}$ can be splitted into two
parts, $\tilde{\xi}_{0}+\dbtilde{\xi}(z)$, where $\tilde{\xi}_{0}=\mathrm{const}$
is the uniform value of the mass concentration in the well-mixed,
adiabatic state and $\dbtilde{\xi}(z)=\mathcal{O}(\delta)$ is the $z$-dependent
part describing the departure of the reference state from the well-mixed
one (i.e. in the hydrostatic equilibrium the variations in the mass
fraction are only a weak correction to the mean, similarly as in
the case of the entropy).

The \emph{second fundamental assumption} of the anelastic approximation
consists of (\ref{fluct_magnitude_A}) and an additional requirement,
that the convective fluctuations of the mass fraction are small\index{SI}{anelastic!assumption (2)}
\begin{equation}
\left|\xi'\right|\sim\mathcal{O}(\delta)\ll1.\label{xi_fluct_magnitude_comp}
\end{equation}
It is important to realize, that the mass fraction fluctuation need
not be smaller than the equilibrium value $\tilde{\xi}$ and both
$\left|\xi'\right|$ and $\tilde{\xi}$ can, in fact, be small (which
means that the weak solution limit, $\xi\ll1$, for a binary
alloy is formally allowed, cf. section \ref{subsec:Weak-solution-limit_RS}).

It is now clear, that by inspection of the mass fraction equation
(\ref{eq:mass_fraction_balance_gen_comp}) and the formula for the
concentration flux (\ref{eq:j_xi_gen_comp}) (or the simplified version
(\ref{eq:mass_frac_eq_no_Soret})) we must require for consistency
\begin{equation}
K=\tilde{\rho}D=\mathcal{O}\left(\delta^{1/2}\rho_{B}\sqrt{\bar{g}L}L\right).\label{eq:e279}
\end{equation}
In consequence we obtain
\begin{equation}
\tilde{j}_{\xi}=\mathcal{O}\left(\delta K/\bar{\chi}L\right)=\mathcal{O}\left(\delta^{3/2}\rho_{B}\sqrt{\bar{g}L}/\bar{\chi}\right).\label{eq:e280}
\end{equation}
The general expression for the gradient of the chemical potential
(\ref{eq:grad_muc}) implies
\begin{align}
\frac{\mathrm{d}\tilde{\mu}_{c}}{\mathrm{d}z}= & \tilde{\varUpsilon}\frac{\mathrm{d}\tilde{\xi}}{\mathrm{d}z}-\frac{\tilde{h}_{p,T}}{\tilde{T}}\frac{\mathrm{d}\tilde{T}}{\mathrm{d}z}+\frac{\tilde{\chi}_{T}}{\tilde{\rho}}\frac{\mathrm{d}\tilde{p}}{\mathrm{d}z}\nonumber \\
= & -\tilde{\chi}\tilde{g}+\frac{\tilde{h}_{p,T}}{\tilde{T}}\Delta_{S}+\tilde{\varUpsilon}\frac{\mathrm{d}\tilde{\xi}}{\mathrm{d}z},\label{eq:e281}
\end{align}
where 
\begin{equation}
\Delta_{S}=-\left(\frac{\mathrm{d}\tilde{T}}{\mathrm{d}z}+\frac{\tilde{\alpha}\tilde{T}\tilde{g}}{\tilde{c}_{p,\xi}}\right),\label{eq:e282}
\end{equation}
therefore
\begin{equation}
\frac{L}{\tilde{h}_{p,T}}\left(\frac{\mathrm{d}\tilde{\mu}_{c}}{\mathrm{d}z}+\tilde{\chi}\tilde{g}\right)=\frac{\tilde{\varUpsilon}L}{\tilde{h}_{p,T}}\frac{\mathrm{d}\tilde{\xi}}{\mathrm{d}z}+\frac{L}{\tilde{T}}\Delta_{S}.\label{eq:e283}
\end{equation}
From the latter equation and the assumptions (\ref{eq:delta_definition})
and (\ref{eq:delta_definition-1}) we see, that as long as $\tilde{\varUpsilon}$
remains comparable with other thermodynamic properties of the system,
thus e.g. $\tilde{\varUpsilon}\sim\tilde{c}_{p,\xi}\tilde{\chi}/\tilde{\alpha}\sim\tilde{\chi}\tilde{h}_{p,T}$,
then the quantity $L\left(\mathrm{d}_{z}\tilde{\mu}_{c}+\tilde{\chi}\tilde{g}\right)/\tilde{h}_{p,T}$
must also be small, i.e.\footnote{However, as we have already remarked below the equation (\ref{eq:small_departure_requirements}),
it is to be emphasized, that in the case when the binary alloy is
a weak solution of the light constituent, $\xi\ll1$, the coefficient
$\tilde{\varUpsilon}\sim\tilde{c}_{p,\xi}\tilde{T}/\tilde{\xi}$ becomes
large (cf. Landau and Lifschitz 1980, eqs 87.4-5, chapter ``Weak
solutions'' and a comment below eq. (96.7), chapter ``Thermodynamic
inequalities for solutions''). Then $\tilde{\varUpsilon}/\tilde{h}_{p,T}\gg1$
and the quantity $L\left(\mathrm{d}_{z}\tilde{\mu}_{c}+\tilde{\chi}\tilde{g}\right)/\tilde{h}_{p,T}$
is \emph{not} small, despite the fact, that the assumptions (\ref{eq:delta_definition})
and (\ref{eq:delta_definition-1}) still hold. The weak solutions
are treated in section \ref{sec:Weak-solution-limit}.}
\begin{equation}
\frac{\mathrm{d}\tilde{\mu}_{c}}{\mathrm{d}z}=-\tilde{\chi}\tilde{g}+\mathcal{O}\left(\delta\bar{g}\right).\label{eq:grad_muc_BS}
\end{equation}
The equations of the hydrostatic equilibrium, which include the hydrostatic
momentum, mass fraction and entropy balances, the gravitational potential
equation and the equations of state take the following, general form
\begin{subequations}
\begin{equation}
\frac{\mathrm{d}\tilde{p}}{\mathrm{d}z}=-\tilde{\rho}\tilde{g},\label{eq:hydrostatic_eq_A_1-1}
\end{equation}
\begin{equation}
\frac{\mathrm{d}}{\mathrm{d}z}\left[K\left(\frac{\mathrm{d}\tilde{\xi}}{\mathrm{d}z}+\frac{\tilde{k}_{T}}{\tilde{T}}\frac{\mathrm{d}\tilde{T}}{\mathrm{d}z}+\frac{\tilde{k}_{p}}{\tilde{p}}\frac{\mathrm{d}\tilde{p}}{\mathrm{d}z}\right)\right]=0,\label{eq:mass_fraction_hydrostatic_comp}
\end{equation}
\begin{align}
\frac{\mathrm{d}}{\mathrm{d}z}\left[\left(k+\frac{\Lambda}{\tilde{T}}\tilde{k}_{T}K\right)\frac{\mathrm{d}\tilde{T}}{\mathrm{d}z}+\Lambda K\frac{\mathrm{d}\tilde{\xi}}{\mathrm{d}z}+\frac{\Lambda}{\tilde{p}}\tilde{k}_{p}K\frac{\mathrm{d}\tilde{p}}{\mathrm{d}z}\right]\qquad\qquad\qquad\nonumber \\
+K\left(\frac{\mathrm{d}\tilde{\xi}}{\mathrm{d}z}+\frac{\tilde{k}_{T}}{\tilde{T}}\frac{\mathrm{d}\tilde{T}}{\mathrm{d}z}+\frac{\tilde{k}_{p}}{\tilde{p}}\frac{\mathrm{d}\tilde{p}}{\mathrm{d}z}\right)\left(\tilde{\varUpsilon}\frac{\mathrm{d}\tilde{\xi}}{\mathrm{d}z}-\frac{\tilde{h}_{p,T}}{\tilde{T}}\frac{\mathrm{d}\tilde{T}}{\mathrm{d}z}+\frac{\tilde{\chi}_{T}}{\tilde{\rho}}\frac{\mathrm{d}\tilde{p}}{\mathrm{d}z}\right) & =-\tilde{Q},\label{eq:eq:hydrostatic_eq_A_2-1}
\end{align}
\begin{equation}
\frac{\mathrm{d}^{2}\tilde{\psi}}{\mathrm{d}z^{2}}=4\pi G\left[\tilde{\rho}(z)\left(\theta_{H}(z)-\theta_{H}(z-L)\right)+\rho_{in}(z)\theta_{H}(-z)\right],\label{eq:grav_pot_hydrostatic_comp}
\end{equation}
\begin{equation}
\tilde{\rho}=\rho(\tilde{p},\tilde{T},\tilde{\xi}),\quad\tilde{s}=s(\tilde{p},\tilde{T},\tilde{\xi}),\label{eq:hydrostatic_eq_A_3-1}
\end{equation}
\end{subequations}
 where, $\tilde{g}=-\nabla\tilde{\psi}$ is the
gravitational acceleration in the reference state, which consists
of the contributions from the entire mass below the fluid layer of
density, $\rho_{in}(z)$, and from the fluid layer itself, where the
density is $\tilde{\rho}$. The assumption of small departures from
the well-mixed state, (\ref{eq:delta_definition-1}) affects the above
system of reference state equations. In the most general case, when
the coefficients $\tilde{\varUpsilon}$, $\tilde{k}_{T}$ and $\tilde{k}_{p}$
are finite, all the terms containing the mass fraction gradient $\mathrm{d}_{z}\tilde{\xi}$
constitute $\mathcal{O}(\delta)$ corrections to the remaining terms.
It follows from the equation (\ref{eq:mass_fraction_hydrostatic_comp})
that
\begin{equation}
K\left(\frac{\tilde{k}_{T}}{\tilde{T}}\frac{\mathrm{d}\tilde{T}}{\mathrm{d}z}+\frac{\tilde{k}_{p}}{\tilde{p}}\frac{\mathrm{d}\tilde{p}}{\mathrm{d}z}\right)=\mathrm{const}+\mathcal{O}\left(\delta\frac{K}{L}\right),\label{eq:mass_fraction_balance_eq}
\end{equation}
and thus the entropy balance (\ref{eq:eq:hydrostatic_eq_A_2-1})
supplied by the force balance (\ref{eq:hydrostatic_eq_A_1-1}) imply
\begin{equation}
\frac{\mathrm{d}}{\mathrm{d}z}\left(k\frac{\mathrm{d}\tilde{T}}{\mathrm{d}z}+\Lambda\,\mathrm{const}\right)-\mathrm{const}\left(\frac{\tilde{h}_{p,T}}{\tilde{T}}\frac{\mathrm{d}\tilde{T}}{\mathrm{d}z}+\tilde{\chi}_{T}\tilde{g}\right)=-\tilde{Q}+\mathcal{O}(\delta);\label{eq:e284}
\end{equation}
the latter determines the temperature gradient at leading order. On
the other hand the pressure gradient can be expressed by the density,
temperature and mass fraction gradients, 
\begin{align}
\frac{\mathrm{d}\tilde{p}}{\mathrm{d}z}= & \frac{1}{\tilde{\rho}\tilde{\beta}}\frac{\mathrm{d}\tilde{\rho}}{\mathrm{d}z}+\frac{\tilde{\alpha}}{\tilde{\beta}}\frac{\mathrm{d}\tilde{T}}{\mathrm{d}z}+\frac{\chi_{T}}{\beta}\frac{\mathrm{d}\tilde{\xi}}{\mathrm{d}z}\nonumber \\
 & \frac{1}{\tilde{\rho}\tilde{\beta}}\frac{\mathrm{d}\tilde{\rho}}{\mathrm{d}z}+\frac{\tilde{\alpha}}{\tilde{\beta}}\frac{\mathrm{d}\tilde{T}}{\mathrm{d}z}+\mathcal{O}\left(\delta\bar{\rho}\bar{g}\right),\label{eq:e285}
\end{align}
and substitution of the latter to the hydrostatic force balance (\ref{eq:hydrostatic_eq_A_1-1})
yields
\begin{equation}
\frac{1}{\tilde{\rho}}\frac{\mathrm{d}\tilde{\rho}}{\mathrm{d}z}=-\tilde{\beta}\tilde{\rho}\tilde{g}-\tilde{\alpha}\frac{\mathrm{d}\tilde{T}}{\mathrm{d}z}+\mathcal{O}\left(\frac{\delta}{L}\right),\label{eq:e286}
\end{equation}
which in turn determines the density gradient at leading order. Finally
the state equation allows to determine the pressure. However, since
the mass fraction gradient drops out of the leading order analysis,
the so-determined reference state profiles $\tilde{\rho}$, $\tilde{T}$
and $\tilde{p}$ are in general inconsistent with the compositional
flux balance (\ref{eq:mass_fraction_balance_eq}), which is an independent
equation that the temperature and pressure have to satisfy at leading
order. In other words a height-dependent only hydrostatic state satisfying
the assumption of weak mass fraction gradients does not exist. The
situation may be rescued by assuming a slowly time-dependent basic
state\footnote{The dependence on time of the reference state can only be slow, in
order not to violate the fundamental for the anelastic approximation
leading order property $\nabla\cdot\left(\tilde{\rho}\mathbf{u}\right)=0$,
i.e. not to introduce the fast sound waves into the problem. } and/or a basic state non-uniform in all three directions (that is
dependent on $x$, $y$ and $z$), however, this introduces additional
complications; nevertheless, slowly varying in time basic states are
relevant to natural systems and commonly used in their quantitative
description, since astrophysical objects such as planets and stars
cool down and change chemical composition on long time scales. Furthermore,
the well-mixed adiabatic state could also be chosen for the reference
state, and then the convective flow would have to be driven by only
weak departures from that state on boundaries. Since the reference
state has to satisfy the equations (momentum, energy, mass fraction
and mass balances and the gravitational and state equations), in general
also in this case the adiabatic, well-mixed reference state would
be required to be non-stationary.

The situation becomes greatly simplified, when the Soret effect and
the term involving the pressure gradient in the expression for the
compositional flux (\ref{eq:j_xi_gen_comp}) are negligibly small
compared to the mass fraction gradient. In such a case the material
flux is given by the Fick's law (\ref{eq:Ficks_law}) and the equations
for the reference state\index{SI}{reference (basic) state} take the following solvable form at the
leading order 
\begin{subequations}
\begin{equation}
\frac{\mathrm{d}\tilde{p}}{\mathrm{d}z}=\tilde{\rho}\nabla\tilde{\psi},\quad\frac{\mathrm{d}}{\mathrm{d}z}\left(K\frac{\mathrm{d}\tilde{\xi}}{\mathrm{d}z}\right)=0,\quad\frac{\mathrm{d}}{\mathrm{d}z}\left(k\frac{\mathrm{d}\tilde{T}}{\mathrm{d}z}\right)=-\tilde{Q},\label{eq:hydrostatic_eq_AC_NoSoret_1}
\end{equation}
\begin{equation}
\tilde{\rho}=\rho(\tilde{p},\tilde{T},\tilde{\xi}),\quad\tilde{s}=s(\tilde{p},\tilde{T},\tilde{\xi}),\label{eq:hydrostatic_eq_AC_NoSoret_2}
\end{equation}
\begin{equation}
\frac{\mathrm{d}^{2}\tilde{\psi}}{\mathrm{d}z^{2}}=4\pi G\left[\tilde{\rho}(z)\left(\theta_{H}(z)-\theta_{H}(z-L)\right)+\rho_{in}(z)\theta_{H}(-z)\right],\label{eq:grav_pot_hydrostatic_AC_NoSoret_3}
\end{equation}
\end{subequations} 
where $\theta_H(z)$ is the Heaviside step function. 

Finally, let us derive explicit relations between
the thermodynamic fluctuations assuming, that the binary alloy is not a weak solution, which implies
\begin{equation}
\frac{\left|\xi'\right|}{\tilde{\xi}}\ll1,\label{eq:not_weak_solution}
\end{equation}
i.e. that the mean value of the concentration of the light constituent is much greater than its variations. The equations of state $\rho=\rho(p,\,T,\,\xi)$ and $s=s(p,\,T,\,\xi)$
now include the dependence on the mass fraction; expansion about the
hydrostatic equilibrium at every height, similarly as in (\ref{eq:Taylor_exp_rho_A},b)
results in 
\begin{subequations}
\begin{equation}
\frac{\rho'}{\tilde{\rho}}=-\tilde{\alpha}T'-\tilde{\chi}_{T}\xi'+\tilde{\beta}p'+\mathcal{O}\left(\delta^{2}\right),\label{eq:Taylor_exp_rho_A-1-1}
\end{equation}
\begin{equation}
s'=-\tilde{\alpha}\frac{p'}{\tilde{\rho}}+\tilde{c}_{p,\xi}\frac{T'}{\tilde{T}}+\tilde{h}_{p,T}\frac{\xi'}{\tilde{T}}+\mathcal{O}\left(\tilde{c}_{p,\xi}\delta^{2}\right),\label{eq:Taylor_exp_entropy_A-1-1}
\end{equation}
\end{subequations} 
which will be later utilized to cast the dynamical
equations in a most suitable form. Additionally we may expand the
chemical potential $\mu_{c}=\mu_{c}(p,\,T,\,\xi)$ about the reference
state at every height, which yields
\begin{equation}
\mu_{c}^{\prime}=\frac{\tilde{\chi}_{T}}{\tilde{\rho}}p'-\frac{\tilde{h}_{p,T}}{\tilde{T}}T'+\tilde{\varUpsilon}\xi'.\label{chem_pot_pert}
\end{equation}
However, we stress, that the assumption (\ref{eq:not_weak_solution})
is crucial in order for the expansions (\ref{eq:Taylor_exp_rho_A-1-1},b)
and (\ref{chem_pot_pert}) to be valid; in other words these expansions are valid only for solutions which are \emph{not} weak. In particular, in the limit
of a weak solution, when both $\tilde{\xi}\ll1$ and $\xi'\ll1$ are
small the fluctuation $\xi'$ can be of comparable magnitude with
the magnitude of the mass fraction in the reference state , i.e. $\xi'/\tilde{\xi}=\mathcal{O}(1)$.
In such a case $\tilde{\varUpsilon}$ is of the same order of magnitude
as $\xi^{\prime-1}$ and the expressions for thermodynamic fluctuations
need to be modified (cf. section \ref{sec:Weak-solution-limit} on
the weak solution limit).

\subsection{Mixture of ideal gases\label{subsec:Mixture-of-ideal-gases}}

Let us now describe in detail one possible, and perhaps the simplest
example of a binary alloy, which is formed when two ideal gases are
mixed. In particular, this is the simplest example of what is termed
an ``ideal solution'' in chemistry, that is a type of solution for
which the entropy and internal energy are additive, i.e. the sum of
entropies (or internal energies) of each constituent equals the total
entropy (internal energy) of the solution; this means, that in the process of mixing of the constituents no heat
is released. The heavy and light constituents, which
occupy the same volume $V$ and have the same temperature $T$ are
both described by the equations of state of a perfect gas\footnote{see e.g. Gumi\'{n}ski (1974), p. 167.},
\begin{subequations}
\begin{equation}
p^{(h)}V=N^{(h)}k_{B}T,\qquad p^{(l)}V=N^{(l)}k_{B}T,\label{eq:ideal_gas_1}
\end{equation}
\begin{align}
S^{(h)} & =\frac{N^{(h)}}{N_{A}}\left(C_{p}^{(h)}\ln T-k_{B}N_{A}\ln p^{(h)}+S_{0}^{(h)}\right),\label{eq:ideal_gas_2}\\
S^{(l)} & =\frac{N^{(l)}}{N_{A}}\left(C_{p}^{(l)}\ln T-k_{B}N_{A}\ln p^{(l)}+S_{0}^{(l)}\right),\label{eq:ideal_gas_2b}
\end{align}
\begin{equation}
\mathcal{E}^{(h)}=\frac{N^{(h)}}{N_{A}}\left(C_{v}^{(h)}T+\mathcal{E}_{0}^{(h)}\right),\qquad\mathcal{E}^{(l)}=\frac{N^{(l)}}{N_{A}}\left(C_{v}^{(l)}T+\mathcal{E}_{0}^{(l)}\right),\label{eq:ideal_gas_3}
\end{equation}
\end{subequations} 
where $p^{(h)}$ and $p^{(l)}$ are the partial
pressures of the heavy and light constituents, $k_{B}$ is the Boltzmann
constant, $N_{A}$ is the Avogadro constant, uppercase $C_{p}$ and
$C_{v}$ denote the molar specific heats ($C_{p}-C_{v}=k_{B}N_{A}$)
and $S_{0}^{(h)}$, $S_{0}^{(l)}$, $\mathcal{E}_{0}^{(h)}$ and $\mathcal{E}_{0}^{(l)}$
are constants. In the above we have adopted for the time being the
``canonical'' thermodynamic variables, such as the volume $V$, the
number of particles $N^{(h)}+N^{(l)}=N$, the actual entropy $S$
and internal energy $\mathcal{E}$\footnote{As opposed to the mass densities of the entropy ($s$) and the internal
energy ($\varepsilon$).}, pressure and temperature. To obtain the equations of state in terms
of thermodynamic parameters of the mixture $\rho$, $p$, $T$, $s$
and $\varepsilon$ we first sum the two equations in (\ref{eq:ideal_gas_1})
and use the Dalton's law to write
\begin{equation}
p^{(h)}+p^{(l)}=p=\left(\frac{N^{(h)}m_{h}}{V}R^{(h)}+\frac{N^{(l)}m_{l}}{V}R^{(l)}\right)T,\label{eq:p_state_eq_aux}
\end{equation}
where we have introduced the molecular masses of the light ($m_{l}$)
and heavy ($m_{h}$) constituents and the specific gas constants for
both constituents $R^{(h)}=k_{B}/m_{h}$ and $R^{(l)}=k_{B}/m_{l}$.
Since the partial densities are
\begin{equation}
\rho^{(h)}=\frac{N^{(h)}m_{h}}{V},\qquad\rho^{(l)}=\frac{N^{(l)}m_{l}}{V},\label{eq:e287}
\end{equation}
and the total density
\begin{equation}
\rho=\frac{N^{(h)}m_{h}+N^{(l)}m_{l}}{V},\label{eq:e288}
\end{equation}
whereas the mass fraction of the light constituent is defined as follows
\begin{equation}
\xi=\frac{\rho^{(l)}}{\rho}=\frac{N^{(l)}m_{l}}{N^{(h)}m_{h}+N^{(l)}m_{l}},\label{eq:xi_def}
\end{equation}
we may rewrite the state equation (\ref{eq:p_state_eq_aux}) in the
form
\begin{equation}
p=\rho R^{(h)}\left[1+\xi\left(r_{m}-1\right)\right]T,\label{eq:p_state_eq}
\end{equation}
where
\begin{equation}
r_{m}=\frac{m_{h}}{m_{l}}\label{eq:e289}
\end{equation}
is the ratio of molecular masses of the heavy to light constituent.
Next we turn to the state equations for the entropy. According to
the Gibbs law the total entropy of a mixture of ideal gases equals
the sum of entropies of the individual constituents at the same temperature
and volume; dividing by the total mass and using again the Dalton's
law
\begin{equation}
p^{(h)}=\frac{N^{(h)}}{N}p,\qquad p^{(l)}=\frac{N^{(l)}}{N}p,\label{eq:e290}
\end{equation}
one obtains the formula for the mass density of the entropy of the
mixture 
\begin{align}
s= & \,\,\frac{S^{(h)}+S^{(l)}}{N^{(h)}m_{h}+N^{(l)}m_{l}}\nonumber \\
= & \left[\xi c_{p}^{(l)}+\left(1-\xi\right)c_{p}^{(h)}\right]\ln T-R^{(h)}\left[1+\xi\left(r_{m}-1\right)\right]\ln p\nonumber \\
 & -R^{(h)}\left(1-\xi\right)\ln\frac{N^{(h)}}{N}-r_{m}R^{(h)}\xi\ln\frac{N^{(l)}}{N}+s_{0}^{(l)}\xi+s_{0}^{(h)}\left(1-\xi\right),\label{eq:entropy_comp}
\end{align}
where
\begin{equation}
c_{p}^{(h)}=\frac{C_{p}^{(h)}}{m_{h}N_{A}},\qquad c_{p}^{(l)}=\frac{C_{p}^{(l)}}{m_{l}N_{A}},\label{eq:e291}
\end{equation}
\begin{equation}
s_{0}^{(l)}=\frac{S_{0}^{(l)}}{m_{l}N_{A}},\qquad s_{0}^{(h)}=\frac{S_{0}^{(h)}}{m_{h}N_{A}}.\label{eq:e292}
\end{equation}
The ratios $N^{(h)}/N$ and $N^{(h)}/N$ can be expressed in terms
of the mass fraction $\xi$ by the use of (\ref{eq:xi_def}),
\begin{equation}
\frac{N^{(h)}}{N}=\frac{1-\xi}{1+\xi\left(r_{m}-1\right)},\quad\frac{N^{(l)}}{N}=\frac{r_{m}\xi}{1+\xi\left(r_{m}-1\right)}.\label{eq:e293}
\end{equation}
Finally, in a mixture of ideal gases the internal energy is also additive,
thus the mass density of the internal energy of the mixture can be
calculated in a straightforward way 
\begin{equation}
\varepsilon=\frac{\mathcal{E}^{(h)}+\mathcal{E}^{(l)}}{N^{(h)}m_{h}+N^{(l)}m_{l}}=\left[\xi c_{v}^{(l)}+\left(1-\xi\right)c_{v}^{(h)}\right]T+\varepsilon_{0}^{(l)}\xi+\varepsilon_{0}^{(h)}\left(1-\xi\right),\label{eq:int_en_comp-1}
\end{equation}
where
\begin{equation}
c_{v}^{(h)}=\frac{C_{v}^{(h)}}{m_{h}N_{A}},\qquad c_{v}^{(l)}=\frac{C_{v}^{(l)}}{m_{l}N_{A}},\label{eq:e294}
\end{equation}
\begin{equation}
\varepsilon_{0}^{(l)}=\frac{\mathcal{E}_{0}^{(l)}}{m_{l}N_{A}},\qquad\varepsilon_{0}^{(h)}=\frac{\mathcal{E}_{0}^{(h)}}{m_{h}N_{A}}.\label{eq:e295}
\end{equation}
Directly from (\ref{eq:entropy_comp}) and (\ref{eq:int_en_comp-1})
and the definitions of the specific heats at constant volume and pressure
for the mixture, it follows that
\begin{equation}
c_{p,\xi}=T\left(\frac{\partial s}{\partial T}\right)_{p,\xi}=\xi c_{p}^{(l)}+\left(1-\xi\right)c_{p}^{(h)},\label{eq:e296}
\end{equation}
\begin{equation}
c_{v,\xi}=T\left(\frac{\partial s}{\partial T}\right)_{\rho,\xi}=\left(\frac{\partial\varepsilon}{\partial T}\right)_{\rho,\xi}=\xi c_{v}^{(l)}+\left(1-\xi\right)c_{v}^{(h)}\label{eq:e297}
\end{equation}
therefore the specific heats of a binary alloy are $\xi$-dependent.
Note also, that for each constituent we have 
\begin{equation}
c_{p}^{(h)}-c_{v}^{(h)}=R^{(h)}=\frac{k_{B}}{m_{h}},\quad c_{p}^{(l)}-c_{v}^{(l)}=R^{(l)}=r_{m}R^{(h)}=\frac{k_{B}}{m_{l}}.\label{eq:e298}
\end{equation}
It is also useful to provide a formula for the chemical potential
of the mixture of ideal gases. By the use of the general expression
for the total differential of the internal energy 
\begin{equation}
\mathrm{d}\varepsilon=T\mathrm{d}s+\frac{p}{\rho^{2}}\mathrm{d}\rho+\mu_{c}\mathrm{d}\xi,\label{eq:e299}
\end{equation}
the chemical potential can be calculated directly from the definition
\begin{align}
\mu_{c}=\left(\frac{\partial\varepsilon}{\partial\xi}\right)_{\rho,s}= & \left(\frac{\partial\varepsilon}{\partial\xi}\right)_{\rho,T}+\left(\frac{\partial\varepsilon}{\partial T}\right)_{\rho,\xi}\left(\frac{\partial T}{\partial\xi}\right)_{\rho,s}\nonumber \\
 & \left(c_{v}^{(l)}-c_{v}^{(h)}\right)T+\varepsilon_{0}^{(l)}-\varepsilon_{0}^{(h)}-T\left(\frac{\partial s}{\partial\xi}\right)_{\rho,T},\label{eq:e300}
\end{align}
where we have used the implicit function theorem to get 
\begin{equation}
\left(\frac{\partial T}{\partial\xi}\right)_{\rho,s}=-\frac{T}{c_{v,\xi}}\left(\frac{\partial s}{\partial\xi}\right)_{\rho,T}.\label{eq:e301}
\end{equation}
By analogy with $h_{p,T}$ defined in (\ref{eq:h_pT}), the thermodynamic
property of the alloy, $T\left(\partial_{\xi}s\right)_{\rho,T}$ can
be denoted by $h_{v,T}$ and 
\begin{equation}
\frac{h_{v,T}}{T}=\frac{h_{p,T}}{T}+\left(\frac{\partial s}{\partial p}\right)_{T,\xi}\left(\frac{\partial p}{\partial\xi}\right)_{\rho,T}.\label{eq:e302}
\end{equation}
With the aid of the Maxwell relation $\left(\partial_{p}s\right)_{T,\xi}=-\alpha/\rho$
and the equation of state (\ref{eq:p_state_eq}), the latter equation
can be transformed to
\begin{equation}
\frac{h_{v,T}}{T}=\frac{h_{p,T}}{T}-R^{(h)}\left(r_{m}-1\right),\label{eq:e303}
\end{equation}
therefore the chemical potential can be alternatively expressed as
follows
\begin{equation}
\mu_{c}=\left(c_{p}^{(l)}-c_{p}^{(h)}\right)T-h_{p,T}+\varepsilon_{0}^{(l)}-\varepsilon_{0}^{(h)}.\label{eq:chem_pot_expression}
\end{equation}
Note, that $h_{p,T}=T(\partial_{\xi}s)_{p,T}$ is a function of $p$,
$T$ and $\xi$. Summarizing, we collect the expressions for $p(\rho,T,\xi)$,
$s(p,T,\xi)$, $\varepsilon(T,\xi)$ and $\mu_{c}(p,T,\xi)$ and provide
a complete set of equations of state for a mixture of ideal gases,
\begin{subequations}
\begin{equation}
p=\rho R^{(h)}\left[1+\xi\left(r_{m}-1\right)\right]T,\label{eq:st_eq_idgm_1}
\end{equation}
\begin{align}
s= & c_{p,\xi}\ln T-R^{(h)}\left[1+\xi\left(r_{m}-1\right)\right]\ln p\nonumber \\
 & -R^{(h)}\left(1-\xi\right)\ln\frac{1-\xi}{1+\xi\left(r_{m}-1\right)}-r_{m}R^{(h)}\xi\ln\frac{r_{m}\xi}{1+\xi\left(r_{m}-1\right)}\nonumber \\
 & +s_{0}^{(l)}\xi+s_{0}^{(h)}\left(1-\xi\right),\label{eq:st_eq_idgm_2}
\end{align}
\begin{equation}
\varepsilon=c_{v,\xi}T++\varepsilon_{0}^{(l)}\xi+\varepsilon_{0}^{(h)}\left(1-\xi\right),\label{st_eq_idgm_3}
\end{equation}
\begin{equation}
\mu_{c}=\left(c_{p}^{(l)}-c_{p}^{(h)}\right)T-h_{p,T}+\varepsilon_{0}^{(l)}-\varepsilon_{0}^{(h)}.\label{st_eq_idgm4}
\end{equation}
\end{subequations} 
Formulae for any other thermodynamic potential,
such as the free energy, the enthalpy, etc. can now be easily derived,
if necessary. Having obtained the equations of state we can write
down explicitly the expressions for all thermodynamic properties of
interest, cf. their definitions in (\ref{eq:h_pT_list}-e) and (\ref{eq:kT_kp_muprime}),
\begin{subequations}
\begin{align}
h_{p,T}= & \left(c_{p}^{(l)}-c_{p}^{(h)}\right)T\ln T-\left(r_{m}-1\right)R^{(h)}T\ln p\nonumber \\
 & +R^{(h)}T\ln\frac{1-\xi}{1+\xi\left(r_{m}-1\right)}-r_{m}R^{(h)}T\ln\frac{r_{m}\xi}{1+\xi\left(r_{m}-1\right)}\nonumber \\
 & +T\left(s_{0}^{(l)}-s_{0}^{(h)}\right),\label{h_pt_idgm_1}
\end{align}
\begin{equation}
c_{p,\xi}=\xi c_{p}^{(l)}+\left(1-\xi\right)c_{p}^{(h)},\quad c_{v,\xi}=\xi c_{v}^{(l)}+\left(1-\xi\right)c_{v}^{(h)},\label{eq:c_idgm_2}
\end{equation}
\begin{equation}
c_{p,\xi}-c_{v,\xi}=R^{(h)}\left[1+\xi\left(r_{m}-1\right)\right]\label{R_idgm_3}
\end{equation}
\begin{equation}
\chi_{T}=\frac{r_{m}-1}{1+\xi\left(r_{m}-1\right)},\quad\chi=\frac{r_{m}-1}{1+\xi\left(r_{m}-1\right)}-\frac{h_{p,T}}{c_{p,\xi}T},\label{chi_idgm_4}
\end{equation}
\begin{equation}
\varUpsilon=\frac{r_{m}R^{(h)}T}{\xi\left(1-\xi\right)\left[1+\xi\left(r_{m}-1\right)\right]},\quad\alpha=\frac{1}{T},\quad\beta=\frac{1}{p},\label{upsilon_idgm_5}
\end{equation}
\begin{equation}
k_{p}=\frac{r_{m}-1}{r_{m}}\xi\left(1-\xi\right)\left[1+\xi\left(r_{m}-1\right)\right],\label{eq:k_p_idgm_6}
\end{equation}
\end{subequations} 
We recall, that $c_{p}^{(h)}-c_{v}^{(h)}=R^{(h)}=k_{B}/m_{h}$
and $c_{p}^{(l)}-c_{v}^{(l)}=r_{m}R^{(h)}=R^{(l)}=k_{B}/m_{l}$. 

\section{Buoyancy force\label{sec:Buoyancy-force}}

In a similar manner as it was done in chapter \ref{chap:Anelastic-convection} (cf. equations (\ref{eq:Buoyancy_entropy_A_1}) and (\ref{eq:Buoyancy_entropy_A_2}))
we will now express the buoyancy force in the Navier-Stokes equation
solely in terms of the entropy and the mass fraction fluctuations.
Elimination of the temperature fluctuation $T'$ from the equations
(\ref{eq:Taylor_exp_rho_A-1-1},b) leads to
\begin{equation}
\frac{\rho'}{\tilde{\rho}}=-\frac{\tilde{\alpha}\tilde{T}}{\tilde{c}_{p,\xi}}s'-\left(\tilde{\chi}_{T}-\frac{\tilde{\alpha}\tilde{h}_{p,T}}{\tilde{c}_{p,\xi}}\right)\xi'+\tilde{\beta}\frac{\tilde{c}_{v,\xi}}{\tilde{c}_{p,\xi}}p',\label{eq:e304}
\end{equation}
therefore by the use of the isentropic compositional expansion coefficient
(\ref{eq:chi_list}) one obtains
\begin{align}
-\frac{1}{\tilde{\rho}}\nabla p'-\nabla\psi'+\frac{\rho'}{\tilde{\rho}}\tilde{\mathbf{g}}= & -\frac{1}{\tilde{\rho}}\nabla p'-\nabla\psi'+\left(\frac{\tilde{\alpha}\tilde{T}}{\tilde{c}_{p,\xi}}s'+\tilde{\chi}\xi'-\tilde{\beta}\frac{\tilde{c}_{v,\xi}}{\tilde{c}_{p,\xi}}p'\right)\tilde{g}\hat{\mathbf{e}}_{z}\nonumber \\
= & -\nabla\left(\frac{p'}{\tilde{\rho}}+\psi'\right)+\frac{\tilde{\alpha}\tilde{T}}{\tilde{c}_{p,\xi}}s'\tilde{g}\hat{\mathbf{e}}_{z}+\tilde{\chi}\xi'\tilde{g}\hat{\mathbf{e}}_{z}\nonumber \\
 & -\left(\tilde{g}\tilde{\beta}\frac{\tilde{c}_{v,\xi}}{\tilde{c}_{p,\xi}}+\frac{1}{\tilde{\rho}^{2}}\frac{\mathrm{d}\tilde{\rho}}{\mathrm{d}z}\right)p'\hat{\mathbf{e}}_{z}.\label{eq:Buoyancy_entropy_A_1-1}
\end{align}
Next, by the use of
\begin{equation}
\frac{\mathrm{d}\tilde{\rho}}{\mathrm{d}z}=\tilde{\rho}\tilde{\beta}\frac{\mathrm{d}\tilde{p}}{\mathrm{d}z}-\tilde{\rho}\tilde{\alpha}\frac{\mathrm{d}\tilde{T}}{\mathrm{d}z}-\tilde{\rho}\tilde{\chi}_{T}\frac{\mathrm{d}\tilde{\xi}}{\mathrm{d}z},\label{eq:rho_derivative_in_z_A-2}
\end{equation}
the hydrostatic pressure balance $\mathrm{d}_{z}\tilde{p}=-\tilde{\rho}\tilde{g}$,
the definition of the parameter $\delta$ in (\ref{eq:delta_definition})
together with (\ref{eq:delta_definition-1}) and $c_{p,\xi}-c_{v,\xi}=\alpha^{2}T/\beta\rho$
(cf. (\ref{eq:cp-cv_B})), one can neglect the entire term proportional
to the pressure fluctuation $p'$, which is equal to
\begin{equation}
\left(-\tilde{\alpha}\Delta_{S}+\tilde{\chi}_{T}\frac{\mathrm{d}\tilde{\xi}}{\mathrm{d}z}\right)\frac{p'}{\tilde{\rho}}\hat{\mathbf{e}}_{z}=\left(\frac{\tilde{\alpha}\tilde{T}}{\tilde{c}_{p,\xi}}\frac{\mathrm{d}\tilde{s}}{\mathrm{d}z}+\tilde{\chi}\frac{\mathrm{d}\tilde{\xi}}{\mathrm{d}z}\right)\frac{p'}{\tilde{\rho}}\hat{\mathbf{e}}_{z}=\mathcal{O}\left(\bar{g}\delta^{2}\right);\label{eq:e305}
\end{equation}
in the above we have also used the expression (\ref{eq:entropy_grad_comp})
for the basic entropy vertical gradient and we recall here $\Delta_{S}=-(\mathrm{d}_{z}\tilde{T}+\tilde{\alpha}\tilde{T}\tilde{g}/\tilde{c}_{p,\xi})$.
This yields
\begin{equation}
-\frac{1}{\tilde{\rho}}\nabla p'-\nabla\psi'+\frac{\rho'}{\tilde{\rho}}\tilde{\mathbf{g}}=-\nabla\left(\frac{p'}{\tilde{\rho}}+\psi'\right)+\frac{\tilde{\alpha}\tilde{T}}{\tilde{c}_{p,\xi}}s'\tilde{g}\hat{\mathbf{e}}_{z}+\tilde{\chi}\xi'\tilde{g}\hat{\mathbf{e}}_{z},\label{eq:buoy_force_simpl_comp}
\end{equation}
at the leading order.

\section{Final set of equations\label{sec:Final-set-of-eqs}}

In the most general case, when the reference state is arbitrary, the
thermodynamic variables take the form 
\begin{subequations}
\begin{align}
\rho\left(\mathbf{x},t\right) & =\tilde{\rho}\left(\mathbf{x},\varepsilon t\right)+\rho'\left(\mathbf{x},t\right),\quad p\left(\mathbf{x},t\right)=\tilde{p}\left(\mathbf{x},\varepsilon t\right)+p'\left(\mathbf{x},t\right),\label{eq:e306}\\
T\left(\mathbf{x},t\right) & =\tilde{T}\left(\mathbf{x},\varepsilon t\right)+T'\left(\mathbf{x},t\right),\quad s\left(\mathbf{x},t\right)=\tilde{s}\left(\mathbf{x},\varepsilon t\right)+s'\left(\mathbf{x},t\right),\label{eq:e307}\\
\xi\left(\mathbf{x},t\right) & =\tilde{\xi}\left(\mathbf{x},\varepsilon t\right)+\xi'\left(\mathbf{x},t\right),\quad\psi\left(\mathbf{x},t\right)=\tilde{\psi}\left(\mathbf{x},\varepsilon t\right)+\psi'\left(\mathbf{x},t\right),\label{eq:e308}
\end{align}
\end{subequations} 
where $\varepsilon=\delta^{n}\ll1$, $n\geq1$
was introduced to stress, that within the anelastic approximation,
which in particular requires $\nabla\cdot\left(\tilde{\rho}\mathbf{u}\right)=0$
to be satisfied, the time dependence of the reference state can only
be slow. The reference state could be chosen in various ways, e.g. as adiabatic, $\mathrm{d}_{z}\tilde{T}=-\tilde{\alpha}\tilde{T}\tilde{g}/\tilde{c}_{p,\xi}$
and well-mixed, $\mathrm{d}_{z}\tilde{\xi}=0$, thus not satisfying
the temperature and mass fraction boundary conditions which drive
the flow or time-dependent and/or fully spatially inhomogeneous. Although
a fully spatially dependent reference state may seem rather complicated
at first sight, it can still be useful, since through satisfaction
of the boundary conditions it can allow to impose homogeneous boundary
conditions on temperature fluctuations (or entropy fluctuations if
the entropy is fixed at boundaries) and mass fraction fluctuations. The equations for the fluctuations
of the mass fraction and the entropy, in such a general case, can
not be expressed in a simple, compact way. Hence we provide here only
the most general form of those equations, however, we adopt the standard
assumption that the hydrostatic force balance $\nabla\tilde{p}=-\tilde{\rho}\tilde{\mathbf{g}}$
holds in the reference state (thus flows in the basic state, even
if present, are weak), which allows to simplify the Navier-Stokes
equation, \index{SI}{anelastic!equations, thermal and compositional driving}
\begin{subequations}
\begin{align}
\tilde{\rho}\left[\frac{\partial\mathbf{u}}{\partial t}+\left(\mathbf{u}\cdot\nabla\right)\mathbf{u}\right]= & -\nabla p'-\tilde{\rho}\nabla\psi'+\rho'\tilde{\mathbf{g}}+\tilde{\rho}\nu\nabla^{2}\mathbf{u}+\tilde{\rho}\left(\frac{\nu}{3}+\nu_{b}\right)\nabla\left(\nabla\cdot\mathbf{u}\right)\nonumber \\
 & +2\nabla\left(\tilde{\rho}\nu\right)\cdot\mathbf{G}^{s}+\nabla\left(\tilde{\rho}\nu_{b}-\frac{2}{3}\tilde{\rho}\nu\right)\nabla\cdot\mathbf{u},\label{NS-Aderiv-1-1-3-3}
\end{align}
\begin{equation}
\nabla\cdot\left(\tilde{\rho}\mathbf{u}\right)=0,\label{Cont_Aderiv-1-1-4-2}
\end{equation}
\begin{equation}
\nabla^{2}\psi'=4\pi G\rho',\label{eq:grav_pot_comp-2}
\end{equation}
\begin{equation}
\tilde{\rho}\left(\frac{\partial\xi}{\partial t}+\mathbf{u}\cdot\nabla\xi\right)=-\nabla\cdot\mathbf{j}_{\xi,\mathrm{mol}},\label{eq:mass_fraction_balance_gen_comp-1}
\end{equation}
\begin{equation}
\tilde{\rho}\tilde{T}\left(\frac{\partial s}{\partial t}+\mathbf{u}\cdot\nabla s\right)=\nabla\cdot\left(k\nabla T\right)-\nabla\cdot\left(\Lambda\mathbf{j}_{\xi,\mathrm{mol}}\right)+\boldsymbol{\tau}_{\nu}:\mathbf{G}^{s}-\mathbf{j}_{\xi,\mathrm{mol}}\cdot\nabla\mu_{c}+Q,\label{eq:entropy_balance_gen_explicit_comp-1}
\end{equation}
\begin{equation}
\frac{\rho'}{\tilde{\rho}}=-\tilde{\alpha}T'-\tilde{\chi}_{T}\xi'+\tilde{\beta}p',\qquad s'=-\tilde{\alpha}\frac{p'}{\tilde{\rho}}+\tilde{c}_{p,\xi}\frac{T'}{\tilde{T}}+\tilde{h}_{p,T}\frac{\xi'}{\tilde{T}},\label{State_eq_Aderiv-pert_comp}
\end{equation}
\begin{equation}
\mu_{c}^{\prime}=\frac{\tilde{\chi}_{T}}{\tilde{\rho}}p'-\frac{\tilde{h}_{p,T}}{\tilde{T}}T'+\tilde{\varUpsilon}\xi'.\label{eq:muc_prime_final}
\end{equation}
\end{subequations} where
\begin{equation}
\mathbf{j}_{\xi,\mathrm{mol}}=-K\left(\nabla\xi+\frac{k_{T}}{T}\nabla T+\frac{k_{p}}{p}\nabla p\right),\label{eq:j_xi_gen_comp-1}
\end{equation}
\begin{align}
\nabla\mu_{c}= & \frac{\chi_{T}}{\rho}\nabla p-\frac{h_{p,T}}{T}\nabla T+\varUpsilon\nabla\xi\nonumber \\
= & -\frac{\varUpsilon}{K}\mathbf{j}_{\xi,\mathrm{mol}}-\frac{\Lambda}{T}\nabla T.\label{eq:grad_muc-1}
\end{align}
The list of definitions of the thermodynamic properties is provided
in (\ref{eq:h_pT_list}-e) and in (\ref{eq:kT_kp_muprime}) and the
total molecular heat flux possesses contributions from both the thermal and
compositional fluxes i.e. is given by\index{SI}{heat flux!total}
\begin{equation}
\mathbf{j}_{q}=-k\nabla T+\left(\Lambda+\mu_{c}\right)\mathbf{j}_{\xi,\mathrm{mol}}.\label{eq:total_heat_flux_final}
\end{equation}
We note, that the equations for thermodynamic fluctuations take the
form provided in (\ref{State_eq_Aderiv-pert_comp}) and (\ref{eq:muc_prime_final})
only when (cf. equations (\ref{eq:Taylor_exp_rho_A-1-1},b) and (\ref{chem_pot_pert})
and the comment below)
\begin{equation}
\frac{\left|\xi'\right|}{\tilde{\xi}}\ll1,\label{eq:small_xiprime}
\end{equation}
and the limit of a weak solution, when $|\xi'|/\tilde{\xi}=\mathcal{O}(1)$
is allowed, is considered later in section \ref{sec:Weak-solution-limit}.
A significant simplification is achieved when the Soret effect and
the effect of the pressure gradient on the flux of the light constituent
are weak,
\begin{equation}
\frac{k_{T}}{T}\nabla T+\frac{k_{p}}{p}\nabla p\ll\nabla\xi,\label{eq:noSoret_nogradp}
\end{equation}
so that the Fick's law for the material flux is satisfied
\begin{equation}
\mathbf{j}_{\xi,\mathrm{mol}}=-K\nabla\xi.\label{Ficks_law_Final}
\end{equation}
This allows to introduce a time independent hydrostatic reference
state (cf. section \ref{sec:Hydrostatic-reference-state} and the
discussion below (\ref{eq:hydrostatic_eq_AC_NoSoret_1}-c)), so that
\begin{subequations}
\begin{align}
\rho\left(\mathbf{x},t\right) & =\tilde{\rho}\left(z\right)+\rho'\left(\mathbf{x},t\right),\quad p\left(\mathbf{x},t\right)=\tilde{p}\left(z\right)+p'\left(\mathbf{x},t\right),\label{eq:e309}\\
T\left(\mathbf{x},t\right) & =\tilde{T}\left(z\right)+T'\left(\mathbf{x},t\right),\quad s\left(\mathbf{x},t\right)=\tilde{s}\left(z\right)+s'\left(\mathbf{x},t\right),\label{eq:e310}\\
\xi\left(\mathbf{x},t\right) & =\tilde{\xi}\left(z\right)+\xi'\left(\mathbf{x},t\right),\quad\psi\left(\mathbf{x},t\right)=\tilde{\psi}\left(z\right)+\psi'\left(\mathbf{x},t\right).\label{eq:e311}
\end{align}
\end{subequations} 
Note, that just like for the thermal conductivity
coefficient $k$ we assume, that the material conductivity coefficient
is also a function of height only, $K=K(z)$ and the coefficients
are of the order $k=\mathcal{O}(\delta^{1/2}\rho_{B}\bar{c}_{p,\xi}\sqrt{\bar{g}L}L)$
and $K=\mathcal{O}(\delta^{1/2}\rho_{B}\sqrt{\bar{g}L}L)$. Under
the above assumptions the material flux can be approximated as follows
\begin{subequations}
\begin{equation}
\tilde{\mathbf{j}}_{\xi,\mathrm{mol}}\cdot\hat{\mathbf{e}}_{z}=-K\frac{\mathrm{d}\tilde{\xi}}{\mathrm{d}z}=\mathrm{const,}\label{eq:e312}
\end{equation}
\begin{equation}
\tilde{\mathbf{j}}_{\xi,\mathrm{mol}}\cdot\hat{\mathbf{e}}_{x}=\tilde{\mathbf{j}}_{\xi,\mathrm{mol}}\cdot\hat{\mathbf{e}}_{y}=0\label{eq:e313}
\end{equation}
\begin{equation}
\mathbf{j}_{\xi,\mathrm{mol}}^{\prime}=-K\nabla\xi'.\label{eq:e314}
\end{equation}
\end{subequations} 
Furthermore, under the anelastic approximation
it is demanded, that the material flux is weak, i.e.
\begin{equation}
-K\frac{\mathrm{d}\tilde{\xi}}{\mathrm{d}z}=\mathcal{O}\left(\delta^{3/2}\rho_{B}\sqrt{\bar{g}L}\right),\qquad-K\nabla\xi'=\mathcal{O}\left(\delta^{3/2}\rho_{B}\sqrt{\bar{g}L}\right),\label{eq:e315}
\end{equation}
so that the departure from the well-mixed state, which drives convection
is small (cf. the definition of $\delta\ll1$ in (\ref{eq:delta_definition})
and (\ref{eq:delta_definition-1})). In consequence the heat flux
in the reference state is dominated by the thermal flux $-k\mathrm{d}_{z}\tilde{T}=\mathcal{O}(\delta^{1/2}\rho_{B}\bar{c}_{p,\xi}\Delta T\sqrt{\bar{g}L})$,
which is $\delta^{-1}$ times greater than the material flux contribution
$\Lambda K\mathrm{d}_{z}\tilde{\xi}=\mathcal{O}(\delta^{3/2}\rho_{B}\bar{c}_{p,\xi}\Delta T\sqrt{\bar{g}L})$.

Therefore in the light of (\ref{eq:grad_muc_BS}) we can make the
following estimate
\begin{align}
-\mathbf{j}_{\xi,\mathrm{mol}}\cdot\nabla\mu_{c}= & -\tilde{\chi}\tilde{g}K\frac{\mathrm{d}\tilde{\xi}}{\mathrm{d}z}-\tilde{\chi}\tilde{g}K\frac{\partial\xi'}{\partial z}+K\frac{\mathrm{d}\tilde{\xi}}{\mathrm{d}z}\nabla\mu_{c}^{\prime}+\mathcal{O}\left(\delta^{5/2}\rho_{B}\bar{g}\sqrt{\bar{g}L}\right)\nonumber \\
= & -\tilde{\chi}\tilde{g}K\frac{\mathrm{d}\tilde{\xi}}{\mathrm{d}z}-\tilde{\chi}\tilde{g}K\frac{\partial\xi'}{\partial z}+\mathcal{O}\left(\delta^{5/2}\rho_{B}\bar{g}\sqrt{\bar{g}L}\right),\label{eq:jgradmuc_term}
\end{align}
which is valid as long as the assumption of a \emph{not} weak solution (\ref{eq:small_xiprime})
holds and consequently the convective fluctuation of the chemical
potential $\mu_{c}^{\prime}$ and its gradient are small ($\varUpsilon=\mathcal{O}(\bar{g}L)$),
so that
\begin{align}
K\frac{\mathrm{d}\tilde{\xi}}{\mathrm{d}z}\nabla\mu_{c}^{\prime}= & K\frac{\mathrm{d}\tilde{\xi}}{\mathrm{d}z}\nabla\left(\frac{\tilde{\chi}_{T}}{\tilde{\rho}}p'-\frac{\tilde{h}_{p,T}}{\tilde{T}}T'+\tilde{\varUpsilon}\xi'\right)\nonumber \\
= & K\mathcal{O}\left(\frac{\delta}{L}\right)\mathcal{O}\left(\delta\bar{g}\right)=\mathcal{O}\left(\delta^{5/2}\rho_{B}\bar{g}\sqrt{\bar{g}L}\right).\label{eq:e316}
\end{align}
Of course the term $-\tilde{\chi}\tilde{g}K\mathrm{d}_{z}\tilde{\xi}$
in (\ref{eq:jgradmuc_term}) belongs to the basic state energy (entropy) balance,
hence it does not contribute to the dynamical equation for the evolution
of the entropy fluctuation. 

Finally by the use of (\ref{eq:buoy_force_simpl_comp}) the Navier-Stokes
equation can be written in the form which emphasizes, that the buoyancy
force is created by departures from the isentropic and well-mixed
state. Therefore the leading-order anelastic equations under the additional
assumptions that the mean concentration of the light constituent greatly
exceeds its convective fluctuation (\ref{eq:small_xiprime}) and that
the Soret and pressure gradient effects on the material flux are
negligible (\ref{eq:noSoret_nogradp}) can be cast in the following,
simplified form\index{SI}{anelastic!equations, thermal and compositional driving}
\begin{subequations}
\begin{align}
\frac{\partial\mathbf{u}}{\partial t}+\left(\mathbf{u}\cdot\nabla\right)\mathbf{u}= & -\nabla\left(\frac{p'}{\tilde{\rho}}+\psi'\right)+\frac{\tilde{\alpha}\tilde{T}}{\tilde{c}_{p,\xi}}s'\tilde{g}\hat{\mathbf{e}}_{z}+\tilde{\chi}\xi'\tilde{g}\hat{\mathbf{e}}_{z}\nonumber \\
 & +\nu\nabla^{2}\mathbf{u}+\left(\frac{\nu}{3}+\nu_{b}\right)\nabla\left(\nabla\cdot\mathbf{u}\right)\nonumber \\
 & +\frac{2}{\tilde{\rho}}\nabla\left(\tilde{\rho}\nu\right)\cdot\mathbf{G}^{s}+\frac{1}{\tilde{\rho}}\nabla\left(\tilde{\rho}\nu_{b}-\frac{2}{3}\tilde{\rho}\nu\right)\nabla\cdot\mathbf{u},\label{NS-Aderiv-comp_final}
\end{align}
\begin{equation}
\nabla\cdot\left(\tilde{\rho}\mathbf{u}\right)=0,\label{Cont_Aderiv-1-1-4-2-1}
\end{equation}
\begin{equation}
\nabla^{2}\psi'=4\pi G\rho',\label{eq:grav_pot_comp-2-1}
\end{equation}
\begin{equation}
\tilde{\rho}\left(\frac{\partial\xi'}{\partial t}+\mathbf{u}\cdot\nabla\xi'\right)+\tilde{\rho}u_{z}\frac{\mathrm{d}\tilde{\xi}}{\mathrm{d}z}=\nabla\cdot\left(K\nabla\xi'\right),\label{eq:mass_fraction_comp}
\end{equation}
\begin{align}
\tilde{\rho}\tilde{T}\left(\frac{\partial s'}{\partial t}+\mathbf{u}\cdot\nabla s'\right) & -\tilde{\rho}\tilde{c}_{p,\xi}u_{z}\Delta_{S}+\tilde{\rho}\tilde{h}_{p,T}u_{z}\frac{\mathrm{d}\tilde{\xi}}{\mathrm{d}z}\nonumber \\
 & =\nabla\cdot\left(k\nabla T'\right)+\nabla\cdot\left(\Lambda K\nabla\xi'\right)-\tilde{\chi}\tilde{g}K\frac{\partial\xi'}{\partial z}\nonumber \\
 & \quad\;+2\tilde{\rho}\nu\mathbf{G}^{s}:\mathbf{G}^{s}+\tilde{\rho}\left(\nu_{b}-\frac{2}{3}\nu\right)\left(\nabla\cdot\mathbf{u}\right)^{2}+Q',\label{eq:entropy_balance_noSoret}
\end{align}
\begin{equation}
\frac{\rho'}{\tilde{\rho}}=-\tilde{\alpha}T'-\tilde{\chi}_{T}\xi'+\tilde{\beta}p',\qquad s'=-\tilde{\alpha}\frac{p'}{\tilde{\rho}}+\tilde{c}_{p,\xi}\frac{T'}{\tilde{T}}+\tilde{h}_{p,T}\frac{\xi'}{\tilde{T}},\label{State_eq_Aderiv-pert_comp-1}
\end{equation}
\end{subequations} 
where we have used
\begin{equation}
\frac{\mathrm{d}\tilde{s}}{\mathrm{d}z}=\frac{\tilde{c}_{p,\xi}}{\tilde{T}}\left(\frac{\mathrm{d}\tilde{T}}{\mathrm{d}z}+\frac{\tilde{\alpha}\tilde{T}\tilde{g}}{\tilde{c}_{p,\xi}}\right)+\frac{\tilde{h}_{p,T}}{\tilde{T}}\frac{\mathrm{d}\tilde{\xi}}{\mathrm{d}z},\label{eq:e317}
\end{equation}
and $\Delta_{S}=-(\mathrm{d}_{z}\tilde{T}+\tilde{\alpha}\tilde{T}\tilde{g}/\tilde{c}_{p,\xi})$.
Once the thermodynamic properties of the binary alloy $\tilde{\alpha}$,
$\tilde{\beta}$, $\tilde{\chi}_{T}$, $\tilde{c}_{p,\xi}$ and $\tilde{h}_{p,T}$,
together with $r_{m}$ and the density profile for the bottom body
$\rho_{in}(z)$ are specified, the system of equations (\ref{NS-Aderiv-comp_final}-f),
supplied by adequate boundary conditions becomes fully determined
and can be solved. One of the simplest examples of an equation of
state for a mixture of two constituents corresponds to a mixture of
ideal gases for which all the thermodynamic properties were provided
in section \ref{subsec:Mixture-of-ideal-gases}.

\subsection{Global balance}\label{subsec:Global_balance_comp}

In order to obtain a global force balance we multiply the Navier-Stokes
equation (\ref{NS-Aderiv-comp_final}) by $\tilde{\rho}\mathbf{u}$,
average over the entire volume (with assumed periodicity in the '$x$'
and '$y$' directions) and for a (statistically) stationary state
this yields the following balance between the total work per unit
volume of the buoyancy forces averaged over the horizontal directions
and the total viscous dissipation in the fluid volume
\begin{equation}
\left\langle \frac{\tilde{\alpha}\tilde{T}\tilde{g}\tilde{\rho}}{\tilde{c}_{p,\xi}}u_{z}s'\right\rangle +\left\langle \tilde{\chi}\tilde{g}\tilde{\rho}u_{z}\xi'\right\rangle =2\left\langle \mu\mathbf{G}^{s}:\mathbf{G}^{s}\right\rangle -\left\langle \left(\frac{2}{3}\mu-\mu_{b}\right)\left(\nabla\cdot\mathbf{u}\right)^{2}\right\rangle .\label{eq:W_BF_and_VD-1}
\end{equation}
To derive the latter relation we have used the impermeable and either
no-slip or stress-free boundary conditions, as in (\ref{eq:impermeable},b).

Next we investigate the mean energy (entropy) and material fluxes
and for simplicity we assume no additional (such as e.g. radiogenic
or radiational) heat sources $Q=0$. First we average over a horizontal
plane and integrate from $0$ to $z$ the stationary energy equation
(\ref{eq:entropy_balance_gen_explicit_comp-1}) and the stationary
mass fraction equation (\ref{eq:mass_fraction_balance_gen_comp-1}),
\begin{subequations}
\begin{eqnarray}
0 & = & \int_{0}^{z}\tilde{\rho}\frac{\mathrm{d}\tilde{T}}{\mathrm{d}z}\left\langle u_{z}s'\right\rangle _{h}\mathrm{d}z-\tilde{\rho}\tilde{T}\left\langle u_{z}s'\right\rangle _{h}-\left.k\frac{\mathrm{d}\left\langle T\right\rangle _{h}}{\mathrm{d}z}\right|_{z=0}+k\frac{\mathrm{d}\left\langle T\right\rangle _{h}}{\mathrm{d}z}\nonumber \\
 &  & +\left.\left(\Lambda+\mu_{c}\right)\left\langle \mathbf{j}_{\xi,\mathrm{mol}}\cdot\hat{\mathbf{e}}_{z}\right\rangle _{h}\right|_{z=0}-\left(\Lambda+\mu_{c}\right)\left\langle \mathbf{j}_{\xi,\mathrm{mol}}\cdot\hat{\mathbf{e}}_{z}\right\rangle _{h}\nonumber \\
 &  & +2\int_{0}^{z}\left\langle \mu\mathbf{G}^{s}:\mathbf{G}^{s}\right\rangle _{h}\mathrm{d}z-\int_{0}^{z}\left\langle \left(\frac{2}{3}\mu-\mu_{b}\right)\left(\nabla\cdot\mathbf{u}\right)^{2}\right\rangle _{h}\mathrm{d}z\nonumber \\
 &  & +\int_{0}^{z}\left\langle \mu_{c}\nabla\cdot\mathbf{j}_{\xi,\mathrm{mol}}\right\rangle _{h}\mathrm{d}z,\label{eq:Global_energy_comp}
\end{eqnarray}
\begin{equation}
\tilde{\rho}\left\langle u_{z}\xi'\right\rangle _{h}=\left.\left\langle \mathbf{j}_{\xi,\mathrm{mol}}\cdot\hat{\mathbf{e}}_{z}\right\rangle _{h}\right|_{z=0}-\left\langle \mathbf{j}_{\xi,\mathrm{mol}}\cdot\hat{\mathbf{e}}_{z}\right\rangle _{h},\label{eq:global_xi_comp}
\end{equation}
\end{subequations} 
where we have used the fact, that the impermeable
boundary conditions at $z=0,\,L$ and the mass conservation imply
$\left\langle u_{z}\right\rangle _{h}=0$ in the fluid volume (cf.
(\ref{eq:mean_h_uz_null_A})).

By the use of stationary form of the equation (\ref{eq:mass_fraction_balance_gen_comp-1})
and the fact that $\nabla\xi=\mathcal{O}(\delta/L)$
\begin{align}
\int_{0}^{z}\left\langle \mu_{c}\nabla\cdot\mathbf{j}_{\xi,\mathrm{mol}}\right\rangle _{h}\mathrm{d}z= & -\int_{0}^{z}\left\langle \tilde{\mu}_{c}\tilde{\rho}\mathbf{u}\cdot\nabla\xi'\right\rangle _{h}\mathrm{d}z+\mathcal{O}\left(\delta^{5/2}\frac{k\Delta T}{L}\right)\nonumber \\
= & -\int_{0}^{z}\left\langle \tilde{\mu}_{c}\nabla\cdot\left(\tilde{\rho}\mathbf{u}\xi'\right)\right\rangle _{h}\mathrm{d}z+\mathcal{O}\left(\delta^{5/2}\frac{k\Delta T}{L}\right)\nonumber \\
= & -\tilde{\mu}_{c}\tilde{\rho}\left\langle u_{z}\xi'\right\rangle _{h}+\int_{0}^{z}\frac{\mathrm{d}\tilde{\mu}_{c}}{\mathrm{d}z}\tilde{\rho}\left\langle u_{z}\xi'\right\rangle _{h}\mathrm{d}z+\mathcal{O}\left(\delta^{5/2}\frac{k\Delta T}{L}\right)\label{eq:global_mucgradj}
\end{align}
where we have used again $\left\langle u_{z}\right\rangle _{h}=0$,
the anelastic mass balance $\nabla\cdot(\tilde{\rho}\mathbf{u})=0$
and finally the impermeability condition at the bottom boundary. On
introduction of (\ref{eq:global_mucgradj}) into the global balance
(\ref{eq:Global_energy_comp}), together with $\mathrm{d}_{z}\tilde{T}=-\tilde{g}\tilde{\alpha}\tilde{T}/\tilde{c}_{p}+\mathcal{O}(\delta\Delta T/L)$
and $\mathrm{d}_{z}\tilde{\mu}_{c}=-\tilde{g}\tilde{\chi}+\mathcal{O}(\delta\bar{g})$
we get the following expression for the total horizontally averaged heat flux
entering the system at the bottom\index{SI}{heat flux!total}
\begin{align}
F_{total}(z=0)= & -\left.k\frac{\mathrm{d}}{\mathrm{d}z}\left(\tilde{T}+\left\langle T'\right\rangle _{h}\right)\right|_{z=0}+\left.\left(\Lambda+\mu_{c}\right)\left\langle \mathbf{j}_{\xi,\mathrm{mol}}\cdot\hat{\mathbf{e}}_{z}\right\rangle _{h}\right|_{z=0}\nonumber \\
= & -k\frac{\mathrm{d}}{\mathrm{d}z}\left(\tilde{T}+\left\langle T'\right\rangle _{h}\right)+\left(\Lambda+\mu_{c}\right)\left\langle \mathbf{j}_{\xi,\mathrm{mol}}\cdot\hat{\mathbf{e}}_{z}\right\rangle _{h}\nonumber \\
 & +\tilde{\rho}\tilde{T}\left\langle u_{z}s'\right\rangle _{h}+\tilde{\mu}_{c}\tilde{\rho}\left\langle u_{z}\xi'\right\rangle _{h}\nonumber \\
 & +\int_{0}^{z}\frac{\tilde{\rho}\tilde{g}\tilde{\alpha}\tilde{T}}{\tilde{c}_{p,\xi}}\left\langle u_{z}s'\right\rangle _{h}\mathrm{d}z+\int_{0}^{z}\tilde{g}\tilde{\chi}\tilde{\rho}\left\langle u_{z}\xi'\right\rangle _{h}\mathrm{d}z\nonumber \\
 & -2\int_{0}^{z}\left\langle \mu\mathbf{G}^{s}:\mathbf{G}^{s}\right\rangle _{h}\mathrm{d}z+\int_{0}^{z}\left\langle \left(\frac{2}{3}\mu-\mu_{b}\right)\left(\nabla\cdot\mathbf{u}\right)^{2}\right\rangle _{h}\mathrm{d}z.\label{eq:e318}
\end{align}
The latter formula expresses the fact, that the horizontally averaged
vertical heat flux at any height $z$ is significantly influenced
(could be either increased or decreased) by the work of the buoyancy
forces and the viscous heating\index{SI}{viscous heating}, i.e.
\begin{align}
F_{total}(z)= & -k\frac{\mathrm{d}}{\mathrm{d}z}\left(\tilde{T}+\left\langle T'\right\rangle _{h}\right)+\left(\Lambda+\mu_{c}\right)\left\langle \mathbf{j}_{\xi,\mathrm{mol}}\cdot\hat{\mathbf{e}}_{z}\right\rangle _{h}\nonumber \\
 & +\tilde{\rho}\tilde{T}\left\langle u_{z}s'\right\rangle _{h}+\tilde{\mu}_{c}\tilde{\rho}\left\langle u_{z}\xi'\right\rangle _{h}\nonumber \\
= & F_{total}(z=0)-\int_{0}^{z}\frac{\tilde{\rho}\tilde{g}\tilde{\alpha}\tilde{T}}{\tilde{c}_{p,\xi}}\left\langle u_{z}s'\right\rangle _{h}\mathrm{d}z-\int_{0}^{z}\tilde{g}\tilde{\chi}\tilde{\rho}\left\langle u_{z}\xi'\right\rangle _{h}\mathrm{d}z\nonumber \\
 & +2\int_{0}^{z}\left\langle \mu\mathbf{G}^{s}:\mathbf{G}^{s}\right\rangle _{h}\mathrm{d}z-\int_{0}^{z}\left\langle \left(\frac{2}{3}\mu-\mu_{b}\right)\left(\nabla\cdot\mathbf{u}\right)^{2}\right\rangle _{h}\mathrm{d}z.\label{eq:e319}
\end{align}
In an analogous way, from (\ref{eq:global_xi_comp}) we can derive
the expression for the total horizontally averaged material flux,
denoted by $G_{total}(z)$, entering the system at the bottom, \index{SI}{flux!compositional (material)}
\begin{equation}
G_{total}(z=0)=\left.\left\langle \mathbf{j}_{\xi,\mathrm{mol}}\cdot\hat{\mathbf{e}}_{z}\right\rangle _{h}\right|_{z=0}=\tilde{\rho}\left\langle u_{z}\xi'\right\rangle _{h}+\left\langle \mathbf{j}_{\xi,\mathrm{mol}}\cdot\hat{\mathbf{e}}_{z}\right\rangle _{h}=G_{total}(z),\label{eq:G_total}
\end{equation}
which is the same at every horizontal plane, thus independent of height,
$G_{total}(z)=G_{total}(z=0)$ \footnote{Similarly as the heat flux in the Boussinesq approximation; see section
\ref{subsec:Energetic-properties-of_B}, equation (\ref{eq:stationary_flux_B}).}.

Next, by setting the upper limit of the vertical integration in (\ref{eq:Global_energy_comp})
and (\ref{eq:global_mucgradj}) to $z=L$ (i.e. integrating over the
entire fluid volume), applying the boundary condition of impermeability
at the top $u_{z}(z=L)=0$ and utilizing (\ref{eq:W_BF_and_VD-1})
one obtains
\begin{align}
L\left\langle \frac{\tilde{g}\tilde{\alpha}\tilde{T}\tilde{\rho}}{\tilde{c}_{p}}u_{z}s'\right\rangle  & +L\left\langle \tilde{\rho}\frac{\mathrm{d}\tilde{T}}{\mathrm{d}z}u_{z}s'\right\rangle +L\left\langle \tilde{\chi}\tilde{g}\tilde{\rho}u_{z}\xi'\right\rangle +L\left\langle \frac{\mathrm{d}\tilde{\mu}_{c}}{\mathrm{d}z}\tilde{\rho}u_{z}\xi'\right\rangle \nonumber \\
= & -\left.k\frac{\mathrm{d}\left\langle T\right\rangle _{h}}{\mathrm{d}z}\right|_{z=L}+\left.k\frac{\mathrm{d}\left\langle T\right\rangle _{h}}{\mathrm{d}z}\right|_{z=0}+\left.\left(\Lambda+\mu_{c}\right)\left\langle \mathbf{j}_{\xi,\mathrm{mol}}\cdot\hat{\mathbf{e}}_{z}\right\rangle _{h}\right|_{z=L}\nonumber \\
 & -\left.\left(\Lambda+\mu_{c}\right)\left\langle \mathbf{j}_{\xi,\mathrm{mol}}\cdot\hat{\mathbf{e}}_{z}\right\rangle _{h}\right|_{z=0}.\label{eq:e320}
\end{align}
Of course the left hand side of the latter equation vanishes at leading
order since $\mathrm{d}_{z}\tilde{T}=-\tilde{g}\tilde{\alpha}\tilde{T}/\tilde{c}_{p}+\mathcal{O}(\delta\Delta T/L)$
and $\mathrm{d}_{z}\tilde{\mu}_{c}=-\tilde{g}\tilde{\chi}+\mathcal{O}(\delta\bar{g})$.
Therefore we can conclude, that in a stationary state the total heat
flux which enters the system at the bottom must be equal to the total
heat flux which leaves the system at the top,\index{SI}{heat flux balance}
\begin{align}
-\left.k\frac{\mathrm{d}\left\langle T\right\rangle _{h}}{\mathrm{d}z}\right|_{z=0}+\left.\left(\Lambda+\mu_{c}\right)\left\langle \mathbf{j}_{\xi,\mathrm{mol}}\cdot\hat{\mathbf{e}}_{z}\right\rangle _{h}\right|_{z=0}\nonumber \\
=-\left.k\frac{\mathrm{d}\left\langle T\right\rangle _{h}}{\mathrm{d}z}\right|_{z=L}+\left(\Lambda+\mu_{c}\right) & \hspace{-0.5mm}\left.\left\langle \mathbf{j}_{\xi,\mathrm{mol}}\cdot\hat{\mathbf{e}}_{z}\right\rangle _{h}\right|_{z=L}.\label{eq:heat_flux_enter_exit-1}
\end{align}
The same is true for the total compositional flux\index{SI}{material flux balance}
\begin{equation}
\left.\left\langle \mathbf{j}_{\xi,\mathrm{mol}}\cdot\hat{\mathbf{e}}_{z}\right\rangle _{h}\right|_{z=0}=\left.\left\langle \mathbf{j}_{\xi,\mathrm{mol}}\cdot\hat{\mathbf{e}}_{z}\right\rangle _{h}\right|_{z=L},\label{eq:global_xi_comp-2}
\end{equation}
which has been obtained by setting $z=L$ in (\ref{eq:global_xi_comp})
and application of the impermeability condition at the top boundary.

\subsection{Definitions of the Rayleigh and Nusselt numbers }

For simplicity let us assume, that the influence of the temperature
and pressure gradients on the material flux is negligible, i.e. $\mathbf{j}_{\xi,\mathrm{mol}}=-K\nabla\xi$.
Then the total local heat flux is given by (cf. equation (\ref{eq:total_heat_flux_final}))
\begin{equation}
\mathbf{j}_{q}=-k\nabla T-\left(\Lambda+\mu_{c}\right)K\nabla\xi.\label{eq:e321}
\end{equation}
To properly define the Nusselt number we must first identify the heat
flux associated with the physical driving of the convective flow.
Since buoyancy effects are allowed only when the adiabatic temperature
gradient is exceeded and/or when the gradient of the mass fraction
$\xi$ becomes negative, the driving comes from the superadiabatic
excess in the temperature gradient and the magnitude of the negative
mass fraction gradient, both imposed by the boundary conditions. Therefore
to describe heat transfer by convection it is suitable to use the
total superadiabatic heat flux composed of contributions from the
superadiabatic thermal molecular flux, compositional molecular flux
and the heat flux resulting from advection of the entropy and concentration,
defined in the following way
\begin{align}
F_{S}(z)= & -k\frac{\mathrm{d}}{\mathrm{d}z}\left(\tilde{T}+\left\langle T'\right\rangle _{h}-T_{ad}\right)-\left(\Lambda+\mu_{c}\right)K\frac{\mathrm{d}}{\mathrm{d}z}\left(\tilde{\xi}+\left\langle \xi'\right\rangle _{h}\right)\nonumber \\
 & +\tilde{\rho}\tilde{T}\left\langle u_{z}s'\right\rangle _{h}+\tilde{\mu}_{c}\tilde{\rho}\left\langle u_{z}\xi'\right\rangle _{h}.\label{eq:e322}
\end{align}
In such a way the heat flux conducted down the adiabat, unrelated
to the convective motions, is excluded. This allows to define the
Nusselt number $Nu$ as a ratio of the total superadiabatic heat flux
which enters the system at the bottom in a convective state, $F_{S}(z=0)$
to the total superadiabatic heat flux in the hydrostatic reference
state, for which the temperature $\tilde{T}$ and mass fraction $\tilde{\xi}$
satisfy the non-homogeneous boundary conditions responsible for driving
(the fluctuations $T'$ and $\xi'$ satisfy homogeneous boundary conditions),\index{SI}{Nusselt number!anelastic}
\begin{align}
Nu= & \frac{F_{S}\left(z=0\right)}{k\Delta_{S}-\left(\Lambda+\mu_{c}\right)K\frac{\mathrm{d}\tilde{\xi}}{\mathrm{d}z}}\nonumber \\
= & \frac{-k\left.\frac{\mathrm{d}}{\mathrm{d}z}\left(\tilde{T}+\left\langle T'\right\rangle _{h}-T_{ad}\right)\right|_{z=0}-\left.\left(\Lambda+\mu_{c}\right)K\frac{\mathrm{d}}{\mathrm{d}z}\left(\tilde{\xi}+\left\langle \xi'\right\rangle _{h}\right)\right|_{z=0}}{k\Delta_{S}-\left(\Lambda+\mu_{c}\right)K\frac{\mathrm{d}\tilde{\xi}}{\mathrm{d}z}}.\label{eq:Nu_def_an_comp}
\end{align}
Furthermore, the most convenient definition of the Rayleigh numbers
associated with thermal and compositional buoyancies depends on a
particular application. If the fluid properties are spatially non-uniform
a useful Rayleigh number definition can involve spatial averaging.
However, when one assumes uniform dynamical viscosity $\mu$, thermal
conductivity $k$, specific heat $c_{p,\xi}$ and gravity $g=\textrm{const}$
(which is allowed e.g. in the case of a weak solution considered in
section \ref{sec:Weak-solution-limit}), the following definitions
of the thermal and compositional Rayleigh numbers can be proposed\index{SI}{Rayleigh number!anelastic}
\begin{equation}
Ra_{th}=\frac{g\bigtriangleup_{S}L^{4}\rho_{B}^{2}c_{p,\xi}}{T_{B}\mu k},\label{eq:Ra_def-1-3}
\end{equation}
\begin{equation}
Ra_{comp}=\frac{g\left(-\frac{\mathrm{d}\tilde{\xi}}{\mathrm{d}z}\right)L^{4}\rho_{B}^{2}c_{p,\xi}}{\mu k},\label{eq:Ra_def-1-3-1}
\end{equation}
where $k\Delta_{S}=k(\Delta T/L-g/c_{p})$ is the superadiabatic thermal conductive
heat flux in the hydrostatic basic state.

\subsection{Boussinesq equations\label{subsec:Boussinesq-equations_comp}}\index{SI}{Boussinesq!limit}

In order to reduce the system of anelastic dynamical equations into
the Boussinesq one it is necessary to assume that the scale heights associated
with density, temperature and pressure are large compared to the fluid
layer thickness, i.e.
\begin{equation}
\frac{\mathrm{d}\tilde{\rho}}{\mathrm{d}z}\ll\frac{\bar{\rho}}{L},\quad\frac{\mathrm{d}\tilde{T}}{\mathrm{d}z}\ll\frac{\bar{T}}{L},\quad\frac{\mathrm{d}\tilde{p}}{\mathrm{d}z}\ll\frac{\bar{p}}{L}.\label{eq:e323}
\end{equation}
Since this limit is mainly applicable to laboratory systems we assume
for simplicity, that the gravitational acceleration is constant, $g=\mathrm{const}$.
The Boussinesq approximation is characterized by small departures
of the thermodynamic variables from their mean values, thus we split
the hydrostatic basic state into the mean and the small vertically
varying correction, e.g. $\tilde{T}=\bar{T}+\dbtilde{T}$ and $\dbtilde{T}/\bar{T}\ll1$,
cf. chapter \ref{chap:The-Boussinesq-convection} on the Boussinesq
convection. Recall, that the small Boussinesq parameter was defined
as $\epsilon=\Delta\dbtilde{\rho}/\bar{\rho}\ll1$ and therefore the
Boussinesq limit of the anelastic approximation implies $\delta\lesssim\mathcal{O}(\epsilon)$ and
$\mathrm{d}_{z}\tilde{\xi}\leq\mathcal{O}(\epsilon)$. In the equations
we can substitute $\tilde{\rho}\approx\bar{\rho}$, $\tilde{T}\approx\bar{T}$,
$\tilde{p}\approx\bar{p}$, $\mathrm{d}\tilde{\rho}/\mathrm{d}z=\mathrm{d}\dbtilde{\rho}/\mathrm{d}z$,
$\mathrm{d}\tilde{T}/\mathrm{d}z=\mathrm{d}\dbtilde{T}/\mathrm{d}z$,
$\mathrm{d}\tilde{p}/\mathrm{d}z=\mathrm{d}\dbtilde{p}/\mathrm{d}z$. 

It can already be seen, that in the continuity equation the term $u_{z}\mathrm{d}\tilde{\rho}/\mathrm{d}z=\mathcal{O}(\bar{\rho}\sqrt{g/L}\epsilon^{3/2})$
becomes negligible compared to $\bar{\rho}\nabla\cdot\mathbf{u}=\mathcal{O}(\bar{\rho}\sqrt{g/L}\epsilon^{1/2})$
and hence the mass conservation at leading order is expressed by $\nabla\cdot\mathbf{u}=0$;
we recall, that in the Boussinesq limit $|\mathbf{u}|=\mathcal{O}(\epsilon^{1/2}\sqrt{gL})$.

Next, by the use of the formula
\begin{equation}
\frac{\partial s}{\partial z}=\frac{c_{p,\xi}}{T}\frac{\partial T}{\partial z}-\frac{\alpha}{\rho}\frac{\partial p}{\partial z}+\frac{h_{p,T}}{T}\frac{\partial\xi}{\partial z},\label{eq:e324}
\end{equation}
and with the aid of the hydrostatic force balance $\partial_{z}p\sim-g\rho$
we can provide the following estimates
\begin{equation}
c_{p,\xi}=\mathcal{O}\left(-\frac{\bar{\alpha}\bar{T}gL}{L\frac{\mathrm{d}\dbtilde{T}}{\mathrm{d}z}}\right)=\mathcal{O}\left(\epsilon^{-1}gL/\bar{T}\right),\qquad h_{p,T}=\mathcal{O}\left(\epsilon^{-1}gL\right).\label{eq:e325}
\end{equation}
The compositional and thermal conduction and diffusion coefficients
satisfy
\begin{equation}
D=\frac{K}{\bar{\rho}}=\mathcal{O}\left(\epsilon^{1/2}\sqrt{gL}L\right),\quad\kappa=\frac{k}{\bar{\rho}\bar{c}_{p,\xi}}=\mathcal{O}\left(\epsilon^{1/2}\sqrt{gL}L\right),\label{eq:e326}
\end{equation}
hence for the compositional heat flux we get
\begin{equation}
\mathbf{j}_{\xi,\mathrm{mol}}\sim-K\nabla\xi=\mathcal{O}\left(\epsilon^{3/2}\bar{\rho}\sqrt{gL}\right).\label{eq:e327}
\end{equation}
Since the full expression for the compositional flux (\ref{eq:j_xi_gen_comp-1})
involves gradients of temperature and pressure and the total heat
flux (\ref{eq:total_heat_flux_final}) involves a contribution from
the material flux, we require for consistency
\begin{equation}
\Lambda=\mathcal{O}\left(\epsilon^{-1}gL\right),\qquad\varUpsilon=\mathcal{O}\left(\epsilon^{-1}gL\right),\label{eq:e328}
\end{equation}
\begin{equation}
k_{T}=\frac{\Lambda-h_{p,T}}{\varUpsilon}\lesssim\mathcal{O}\left(1\right)\qquad k_{p}=\frac{p\chi_{T}}{\rho\varUpsilon}=\chi_{T}\frac{p}{\rho gL}\frac{gL}{\varUpsilon}=\mathcal{O}(1),\label{eq:e329}
\end{equation}
where we have utilized (\ref{eq:p_vs_rho_B}) to estimate $p/\rho gL$.
Furthermore, by the use of (\ref{eq:p_scaling_B}) the pressure fluctuation
in the Boussinesq limit is so small, that the effect of $\nabla p'$
on the material flux can be neglected to yield
\begin{align}
\mathbf{j}_{\xi,\mathrm{mol}}= & -K\left(\nabla\xi+\frac{k_{T}}{T}\nabla T+\frac{k_{p}}{p}\nabla p\right)\nonumber \\
 & -K\left(\frac{\mathrm{d}\dbtilde{\xi}}{\mathrm{d}z}+\frac{\bar{k}_{T}}{\bar{T}}\frac{\mathrm{d}\dbtilde{T}}{\mathrm{d}z}+\frac{\bar{k}_{p}}{\bar{p}}\frac{\mathrm{d}\dbtilde{p}}{\mathrm{d}z}\right)-K\left(\nabla\xi'+\frac{\bar{k}_{T}}{\bar{T}}\nabla T'\right).\label{eq:e330}
\end{align}
On the basis of (\ref{eq:grad_muc-1}) the gradient of the chemical
potential is of the order $\epsilon^{0}$, i.e. $\nabla\mu_{c}=\mathcal{O}(g)$
and this implies
\begin{equation}
-\frac{1}{\bar{\rho}\bar{c}_{p,\xi}}\mathbf{j}_{\xi,\mathrm{mol}}\cdot\nabla\mu_{c}=\mathcal{O}\left(\epsilon^{5/2}\bar{T}\sqrt{g/L}\right),\label{eq:e331}
\end{equation}
which will allow to neglect this term in the temperature equation,
since it is of the same order of magnitude as the viscous heating,
cf. (\ref{eq:visc_heat_negligible_B}). The following expression for
the entropy differential
\begin{equation}
\mathrm{d}s=\frac{c_{p,\xi}}{T}\mathrm{d}T-\frac{\alpha}{\rho}\mathrm{d}p+\frac{h_{p,T}}{T}\mathrm{d}\xi,\label{eq:e332}
\end{equation}
supplied by the mass fraction balance (\ref{eq:mass_fraction_balance_gen_comp-1})
allows to transform the entropy balance (\ref{eq:entropy_balance_gen_explicit_comp-1})
into the temperature equation\index{SI}{temperature equation!for binary alloy}
\begin{align}
\bar{\rho}\bar{c}_{p,\xi}\left(\frac{\partial T}{\partial t}+\mathbf{u}\cdot\nabla T\right) & -\bar{\alpha}\bar{T}\left(\frac{\partial p}{\partial t}+\mathbf{u}\cdot\nabla p\right)\qquad\qquad\qquad\qquad\qquad\qquad\qquad\qquad\quad\nonumber \\
=\nabla\cdot\left(k\nabla T\right)- & \nabla\cdot\left[\left(\Lambda+\bar{h}_{p,T}K\right)\mathbf{j}_{\xi,\mathrm{mol}}\right]+\boldsymbol{\tau}_{\nu}:\mathbf{G}^{s}-\mathbf{j}_{\xi,\mathrm{mol}}\cdot\nabla\mu_{c}+Q.\label{eq:entropy_balance_gen_explicit_comp-h1}
\end{align}
The pressure material derivative can be greatly simplified using the
fact, that the pressure fluctuation is small, as in (\ref{eq:p_scaling_B}),
so that
\begin{equation}
\frac{\partial p}{\partial t}+\mathbf{u}\cdot\nabla p=\frac{\partial p'}{\partial t}+u_{z}\frac{\mathrm{d}\dbtilde{p}}{\mathrm{d}z}+\mathbf{u}\cdot\nabla p'=-\bar{\rho}gu_{z}+\mathcal{O}\left(\epsilon^{3/2}\bar{\rho}\sqrt{g^{3}L}\right).\label{eq:e333}
\end{equation}
Finally the smallness of the pressure fluctuation implies also $\left|\tilde{\beta}p'\right|\ll\left|\tilde{\alpha}T'\right|$
(cf. discussion below (\ref{eq:Buoyancy_simpl_B})), thus the buoyancy
force in the Navier-Stokes equation (\ref{NS-Aderiv-1-1-3-3}) takes
the leading order form
\begin{equation}
\mathbf{g}\frac{\rho'}{\bar{\rho}}=-\mathbf{g}\bar{\alpha}T'-\mathbf{g}\bar{\chi}_{T}\xi'+\mathcal{O}\left(\mathbf{g}\epsilon^{2}\right).\label{rho_prime_comp_B}
\end{equation}
We are now ready to write down the final form of the dynamical equations
under the Boussinesq approximation which reads\index{SI}{Boussinesq!equations, thermal and compositional driving}
\begin{subequations}
\begin{equation}
\frac{\partial\mathbf{u}}{\partial t}+\left(\mathbf{u}\cdot\nabla\right)\mathbf{u}=-\frac{1}{\bar{\rho}}\nabla p'+g\bar{\alpha}T'\hat{\mathbf{e}}_{z}+g\bar{\chi}_{T}\xi'\hat{\mathbf{e}}_{z}+\nu\nabla^{2}\mathbf{u}+2\nabla\nu\cdot\mathbf{G}^{s},\label{eq:NS-comp_Blimit}
\end{equation}
\begin{equation}
\nabla\cdot\mathbf{u}=0,\label{Cont_comp_Blimit}
\end{equation}
\begin{equation}
\frac{\partial\xi'}{\partial t}+\mathbf{u}\cdot\nabla\xi'+u_{z}\frac{\mathrm{d}\dbtilde{\xi}}{\mathrm{d}z}=\nabla\cdot\left[D\left(\nabla\xi'+\frac{k_{T}}{\bar{T}}\nabla T'\right)\right],\label{eq:mass_fraction_comp_Blimit}
\end{equation}
\begin{align}
\frac{\partial T'}{\partial t}+\mathbf{u}\cdot\nabla T'+u_{z}\left(\frac{\mathrm{d}\dbtilde{T}}{\mathrm{d}z}+\frac{g\bar{\alpha}\bar{T}}{\bar{c}_{p,\xi}}\right)= & \nabla\cdot\left[\left(\kappa+D\frac{\bar{\varUpsilon}k_{T}^{2}}{\bar{c}_{p,\xi}\bar{T}}\right)\nabla T'\right]\nonumber \\
 & +\frac{\bar{\varUpsilon}}{\bar{c}_{p,\xi}}\nabla\cdot\left(k_{T}D\nabla\xi'\right)+\frac{Q'}{\bar{\rho}\bar{c}_{p,\xi}}.\label{eq:temp_comp_Blimit}
\end{align}
\end{subequations} 
Note, that we did not utilize the smallness of
the ratios $\xi'/\bar{\xi}$ nor $s'/\bar{s}$ at any point in the
derivation of the Boussinesq equation and in fact in the Boussinesq
limit those ratios need not be small.

Moreover, we observe that when the Soret coefficient $k_{T}$ is negligibly
small both the Soret and Dufour effects are excluded at once and the
mass fraction and temperature equations simplify to
\begin{equation}
\frac{\partial\xi'}{\partial t}+\mathbf{u}\cdot\nabla\xi'+u_{z}\frac{\mathrm{d}\dbtilde{\xi}}{\mathrm{d}z}=\nabla\cdot\left(D\nabla\xi'\right),\label{eq:mass_fraction_comp_Blimit-1}
\end{equation}
\begin{equation}
\frac{\partial T'}{\partial t}+\mathbf{u}\cdot\nabla T'+u_{z}\left(\frac{\mathrm{d}\dbtilde{T}}{\mathrm{d}z}+\frac{g\bar{\alpha}\bar{T}}{\bar{c}_{p,\xi}}\right)=\nabla\cdot\left(\kappa\nabla T'\right)+\frac{Q'}{\bar{\rho}\bar{c}_{p,\xi}}.\label{eq:temp_comp_Blimit-1}
\end{equation}
When there are no heating sources, $Q=0$, the equations for the mass
fraction and the temperature are of exactly the same type and since
the compositional and thermal contributions to the total buoyancy force
$g\bar{\alpha}T'\hat{\mathbf{e}}_{z}+g\bar{\chi}_{T}\xi'\hat{\mathbf{e}}_{z}$
are alike, if additionally the boundary conditions for $T'$ and $\xi'$
are of the same type, the physical effect of both is qualitatively
the same. If, however, some radiogenic or radioactive heat sources
are present the effect of thermal driving significantly differs from
that of the compositional driving. In general the boundary conditions
for $T'$ and $\xi'$ are also not of the same type, making their
effects distinguishable.

\section{Weak solution limit, $\xi\ll1$.\label{sec:Weak-solution-limit}}\index{SI}{weak solution limit}

When the solution of the light constituent is weak, that is the mass
fraction
\begin{equation}
\xi\ll1\label{eq:e334}
\end{equation}
is small, the coefficient $\varUpsilon$, which describes the variation
of the chemical potential with concentration at constant pressure
and temperature satisfies
\begin{equation}
\varUpsilon\approx\frac{k_{B}T}{\xi m_{l}}=\mathcal{O}\left(\bar{c}_{p,\xi}\bar{T}/\bar{\xi}\right)\gg1,\label{eq:e335}
\end{equation}
where $k_{B}$ is the Boltzmann constant (cf. Landau and Lifschitz
1980, eqs 87.4-5, chapter ``Weak solutions'' and chapter 96 on ``Thermodynamic
inequalities for solutions'').\footnote{\label{fn:varUpsilon}Note, that in the book of Landau and Lifschitz
1980 the temperature is expressed in energy units, thus their temperature
is in fact $k_{B}T$, where $T$ is expressed in degrees Kelvin. Since
in the limit of $\xi\ll1$ and in the notation of the footnote (\ref{fn:chem_pot})
we get $\xi\approx N^{(l)}m_{l}/N^{(h)}m_{h}$, therefore Landau's
$c=N^{(l)}/N^{(h)}\approx\xi m_{h}/m_{l}$ is equivalent to $\xi$
up to a constant factor and hence the parameter $\varUpsilon\approx k_{B}T/\xi m_{l}$. } Therefore $\alpha\varUpsilon/\chi_{T}\gg c_{p,\xi}\sim h_{p,T}/T$
is large and by (\ref{eq:mu_stab_cond}) the quantity
\begin{equation}
-\frac{L}{Tc_{p,\xi}}\left(\frac{\mathrm{d}\mu_{c}}{\mathrm{d}z}+\chi g\right)=\mathcal{O}(1)\label{eq:muc_tilde_grad_small_xi}
\end{equation}
is, in general \emph{not} small and much greater than the superadiabatic
gradient $\alpha L(\mathrm{d}_{z}T+\alpha Tg/c_{p,\xi})$, despite
the smallness of $\mathrm{d}_{z}\xi$.

A hydrostatic reference state for anelastic convection in the case
of a weak solution, $\tilde{\xi}\ll1$ and $\xi'\ll1$, can be obtained
in the following way. In such a case the coefficient $\varUpsilon\gg\chi c_{p,\xi}/\alpha$
is large in comparison with other standard thermodynamic properties
of the binary alloy, which by the use of (\ref{eq:kT_kp_muprime})
implies that the coefficients $k_{T}$ and $k_{p}$ are small, hence
the Soret effect and $\nabla p$-dependence of the mass concentration
flux are weak. The following assumption
\begin{equation}
k_{T}\sim k_{p}\sim\frac{\bar{p}}{\bar{\rho}\varUpsilon}\sim\delta,\label{eq:e336}
\end{equation}
where $\delta$ is defined in (\ref{eq:delta_definition}) (cf. also
(\ref{eq:delta_definition-1}), allows to simplify the reference state
equations, so that at the leading order they read 
\begin{subequations}
\begin{equation}
\frac{\mathrm{d}\tilde{p}}{\mathrm{d}z}=-\tilde{\rho}\tilde{g},\label{eq:hydrostatic_eq_A_1-1-1}
\end{equation}
\begin{equation}
\frac{\mathrm{d}}{\mathrm{d}z}\left[K\left(\frac{\mathrm{d}\tilde{\xi}}{\mathrm{d}z}+\frac{\tilde{k}_{T}}{\tilde{T}}\frac{\mathrm{d}\tilde{T}}{\mathrm{d}z}+\frac{\tilde{k}_{p}}{\tilde{p}}\frac{\mathrm{d}\tilde{p}}{\mathrm{d}z}\right)\right]=0,\label{eq:mass_fraction_hydrostatic_comp-1}
\end{equation}
\begin{equation}
\frac{\mathrm{d}}{\mathrm{d}z}\left(k\frac{\mathrm{d}\tilde{T}}{\mathrm{d}z}\right)=-\tilde{Q},\label{eq:hydrostatic_eq_A_2-1-1}
\end{equation}
\begin{equation}
\frac{\mathrm{d}^{2}\tilde{\psi}}{\mathrm{d}z^{2}}=4\pi G\left[\tilde{\rho}(z)\left(\theta_{H}(z)-\theta_{H}(z-L)\right)+\rho_{in}(z)\theta_{H}(-z)\right],\label{eq:grav_pot_hydrostatic_comp-1}
\end{equation}
\begin{equation}
\tilde{\rho}=\rho(\tilde{p},\tilde{T},\tilde{\xi}),\quad\tilde{s}=s(\tilde{p},\tilde{T},\tilde{\xi}).\label{eq:hydrostatic_eq_A_3-1-1}
\end{equation}
\end{subequations} 
In the above system the reference temperature,
density and pressure profiles are determined from the equations (\ref{eq:hydrostatic_eq_A_1-1-1}),
(\ref{eq:hydrostatic_eq_A_2-1-1}) and (\ref{eq:hydrostatic_eq_A_3-1-1})
and then the equation (\ref{eq:mass_fraction_hydrostatic_comp-1})
allows to calculate the mass fraction reference profile. $\theta_H (z)$ is the Heaviside step function.

\subsection{Mixture of ideal gases in the weak solution limit\label{subsec:Weak-solution-limit_RS}}

It is of interest to consider the simplifying case when the binary
alloy is a mixture of two ideal gases (cf. section \ref{subsec:Mixture-of-ideal-gases}),
but it remains a weak solution of the light constituent, more precisely
\begin{equation}
\xi\sim\delta\ll1.\label{eq:e337}
\end{equation}
Under this assumption the thermodynamic properties of the alloy significantly
simplify at the leading order to 
\begin{subequations}
\begin{align}
h_{p,T}\approx & \left(c_{p}^{(l)}-c_{p}^{(h)}\right)T\ln T-\left(r_{m}-1\right)R^{(h)}T\ln p-r_{m}R^{(h)}T\ln\left(r_{m}\xi\right)\nonumber \\
 & +T\left(s_{0}^{(l)}-s_{0}^{(h)}\right),\label{eq:h_pt_weak}
\end{align}
\begin{equation}
c_{p,\xi}\approx c_{p}^{(h)},\qquad c_{v,\xi}\approx c_{v}^{(h)},\qquad c_{p,\xi}-c_{v,\xi}\approx R^{(h)},\label{eq:c_v_c_p_weak}
\end{equation}
\begin{equation}
\chi_{T}\approx r_{m}-1,\quad\varUpsilon\approx\frac{r_{m}R^{(h)}T}{\xi},\qquad k_{p}\approx\frac{r_{m}-1}{r_{m}}\xi,\label{eq:chi_T_Ups_k_p_weak}
\end{equation}
\begin{align}
\chi\approx & \left(1-\frac{c_{p}^{(l)}}{c_{p}^{(h)}}\right)\ln T+\left(r_{m}-1\right)\frac{\gamma^{(h)}-1}{\gamma^{(h)}}\ln p+r_{m}\frac{\gamma^{(h)}-1}{\gamma^{(h)}}\ln\left(r_{m}\xi\right)\nonumber \\
 & +r_{m}-1-\frac{s_{0}^{(l)}-s_{0}^{(h)}}{c_{p}^{(h)}}.\label{eq:chi_weak}
\end{align}
\end{subequations} 
Therefore $\varUpsilon\sim R^{(h)}T\delta^{-1}$
is large, whereas $k_{p}\sim\delta$ likewise $k_{T}=(\Lambda-h_{p,T})/\varUpsilon\sim\delta$
are small\footnote{\label{fn:xilogxi_small}Note, that $h_{p,T}$ has a logarithmic dependence
on the mass fraction $\xi$, however, since $\xi\ln\xi\overset{\xi\rightarrow0}{\longrightarrow}-\xi$
($\Lambda$ is a phenomenological material property), the assumption
$k_{T}\sim\delta$ is justified.}. In other words we can write
\begin{equation}
\xi=\mathcal{O}(\delta)\;\Rightarrow\;\varUpsilon=\mathcal{O}\left(\delta^{-1}\bar{g}L\right),\quad k_{T}=\mathcal{O}(\delta),\quad k_{p}=\mathcal{O}(\delta),\label{eq:xi_small_and_coseq-1}
\end{equation}
where $\delta\ll1$ is the measure of the departure of the system
from the adiabatic well-mixed state, defined in (\ref{eq:delta_definition})
and (\ref{eq:delta_definition-1}). The equations describing the hydrostatic
reference state (\ref{eq:hydrostatic_eq_A_1-1-1}-e), in the current
situation take the form \index{SI}{reference (basic) state}
\begin{subequations}
\begin{equation}
\frac{\mathrm{d}\tilde{p}}{\mathrm{d}z}=-\tilde{\rho}\tilde{g},\label{eq:hydrostatic_eq_A_1-1-1-1}
\end{equation}
\begin{equation}
\frac{\mathrm{d}}{\mathrm{d}z}\left[K\left(\frac{\mathrm{d}\tilde{\xi}}{\mathrm{d}z}+\frac{\tilde{k}_{T}}{\tilde{T}}\frac{\mathrm{d}\tilde{T}}{\mathrm{d}z}-\frac{\left(r_{m}-1\right)\tilde{g}\tilde{\xi}}{r_{m}R^{(h)}\tilde{T}}\right)\right]=0,\label{eq:mass_fraction_hydrostatic_comp-1-1}
\end{equation}
\begin{equation}
\frac{\mathrm{d}}{\mathrm{d}z}\left(k\frac{\mathrm{d}\tilde{T}}{\mathrm{d}z}\right)=-\tilde{Q},\label{eq:hydrostatic_eq_A_2-1-1-1}
\end{equation}
\begin{equation}
\frac{\mathrm{d}^{2}\tilde{\psi}}{\mathrm{d}z^{2}}=4\pi G\left[\tilde{\rho}(z)\left(\theta_{H}(z)-\theta_{H}(z-L)\right)+\rho_{in}(z)\theta_{H}(-z)\right],\label{eq:grav_pot_hydrostatic_comp-1-1}
\end{equation}
\begin{equation}
\tilde{p}=\tilde{\rho}R^{(h)}\tilde{T},\quad\tilde{s}=c_{p}^{(h)}\ln\tilde{T}-R^{(h)}\ln\tilde{p}+s_{0}^{(h)}.\label{eq:hydrostatic_eq_A_3-1-1-1}
\end{equation}
\end{subequations} 
The temperature in the reference state $\tilde{T}$
can now be explicitly calculated from (\ref{eq:hydrostatic_eq_A_2-1-1-1}),
whereas the pressure $\tilde{p}$ and density $\tilde{\rho}$ from
the hydrostatic force balance (\ref{eq:hydrostatic_eq_A_1-1-1-1})
and the first of the equations of state in (\ref{eq:hydrostatic_eq_A_3-1-1-1}).
Let us now consider the special case when the transport coefficients
$K$ and $k$, the Soret coefficient $k_{T}$ and the specific heats
$c_{p,\xi}\approx c_{p}^{(h)}$ and $c_{v,\xi}\approx c_{v}^{(h)}$
can be assumed constant, and moreover the total mass of the fluid
layer is negligibly small compared to the mass of the body below the
layer, which implies $\mathrm{d}_{z}^{2}\tilde{\psi}=4\pi G\rho_{in}(z)\theta_{H}(-z)$
and consequently $g\approx\tilde{g}=\mathrm{const}$ within the region
of the fluid. Additionally we assume, that the radiogenic heat can
be neglected $\tilde{Q}=0$. The temperature, density, pressure and
entropy in the reference state are then described at the leading order
by the equations (\ref{eq:BS1}-c) with $c_{p}$, $\gamma=c_{p}/c_{v}$
and $R$ replaced by $c_{p}^{(h)}$, $\gamma^{(h)}=c_{p}^{(h)}/c_{v}^{(h)}$
and $R^{(h)}$ respectively, i.e. the equations of state of a single-component
ideal gas, that is the heavy constituent only. These expressions are
now supplied by the solution of the mass fraction equation (\ref{eq:hydrostatic_eq_A_2-1-1-1})
which in the case at hand takes the following, fairly simple form\index{SI}{reference (basic) state}
\begin{equation}
\tilde{\xi}=\frac{C_{0}}{\left(1-\theta\frac{z}{L}\right)^{l}}+\frac{\tilde{j}_{\xi}}{K}\frac{L}{\theta}\frac{1}{l+1}\left(1-\theta\frac{z}{L}\right)-\frac{k_{T}}{l},\label{eq:e338}
\end{equation}
where $\theta=\Delta T/T_{B}$, $C_{0}=\mathrm{const}$,
\begin{equation}
\tilde{j}_{\xi}=-K\left(\frac{\mathrm{d}\tilde{\xi}}{\mathrm{d}z}+\frac{\tilde{k}_{T}}{\tilde{T}}\frac{\mathrm{d}\tilde{T}}{\mathrm{d}z}-\frac{\left(r_{m}-1\right)g\tilde{\xi}}{r_{m}R^{(h)}\tilde{T}}\right),\label{eq:e339}
\end{equation}
is the constant flux of the mass concentration and
\begin{equation}
l=\frac{r_{m}-1}{r_{m}}\frac{gL}{R^{(h)}\Delta T}=\frac{r_{m}-1}{r_{m}}\left(m+1\right).\label{eq:e340}
\end{equation}
The constant $\mathrm{C_{0}}$ can be determined by application of
the boundary conditions when the values of the mass fraction are specified
at the boundaries, however, when the mass concentration flux is held
fixed at the boundaries, this constant remains undetermined. In the
limit of weak stratification, $\theta\ll1$, the $\tilde{\xi}$ profile
becomes linear.

It is sometimes useful to possess also the explicit formula for the
chemical potential in the reference state, which in the limit $\xi\ll1$
reads (cf. (\ref{eq:chem_pot_expression}) and (\ref{eq:h_pt_weak}))
\begin{align}
\tilde{\mu}_{c}= & \left(c_{p}^{(l)}-c_{p}^{(h)}\right)\tilde{T}\left(1-\ln\tilde{T}\right)+\left(r_{m}-1\right)R^{(h)}\tilde{T}\ln\tilde{p}\nonumber \\
 & +r_{m}R^{(h)}\tilde{T}\ln\left(r_{m}\tilde{\xi}\right)-\tilde{T}\left(s_{0}^{(l)}-s_{0}^{(h)}\right)+\varepsilon_{0}^{(l)}-\varepsilon_{0}^{(h)}.\label{eq:muc_tilde}
\end{align}
Let us stress, that there is, in fact, a large degree of freedom in
the choice of the reference state, which depends on the particular
application. The reference state must satisfy the dynamical equations,
and therefore may be required to be non-stationary if e.g. the thermal
diffusion in the basic state does not vanish $\nabla\cdot\left(k\nabla\tilde{T}\right)\neq0$.
However, when specifying a reference state one must take good care
to specify also the boundary conditions for the fluctuations and realize
precisely what does the assumption of small departures from the adiabatic
well-mixed state mean in the particular situation. For example, as
already noted, the well-mixed adiabatic state itself can be chosen
for the reference state, in which case the boundary conditions on
the fluctuations \emph{must be} non-zero in order to drive the flow;
in such a case the boundary conditions define the departure from adiabaticity
and the well-mixed state, which must be weak.

Next we proceed to derive the final form of the dynamical equations
in the considered limit. A significant simplification is achieved
when the Dufour effect is weak
\begin{equation}
\Lambda\lesssim\mathcal{O}\left(\delta\bar{g}L\right),\label{eq:Xi_small-1}
\end{equation}
where $\bar{g}L$ is the gravitational energy scale, since the chemical
potential gradient, which is an order unity quantity, by the use of
(\ref{eq:grad_muc-1}) and (\ref{eq:Xi_small-1}) is proportional
to the material flux
\begin{equation}
\nabla\mu_{c}=-\frac{\varUpsilon}{K}\mathbf{j}_{\xi,\mathrm{mol}}+\mathcal{O}\left(\delta\bar{g}\right).\label{eq:grad_muc_weak_sol}
\end{equation}
It is important to emphasize, that in the weak solution limit both,
the mass fraction in the reference state, $\tilde{\xi}$ and its fluctuation
$\xi'$ are small and of the same order of magnitude in terms of the
small parameter $\delta$, that is $\tilde{\xi}=\mathcal{O}(\delta)$
and $\xi'=\mathcal{O}(\delta)$, but the compositional expansion coefficients
$\chi$ and $\chi_{T}$ are both of the order unity. Under the above
assumptions the terms in the material flux can be approximated as
follows
\begin{align}
\mathbf{j}_{\xi,\mathrm{mol}}= & -K\left(\nabla\xi'+\frac{\mathrm{d}\tilde{\xi}}{\mathrm{d}z}\hat{\mathbf{e}}_{z}+\frac{\tilde{k}_{T}}{\tilde{T}}\frac{\mathrm{d}\tilde{T}}{\mathrm{d}z}\hat{\mathbf{e}}_{z}+\frac{\tilde{k}_{p}}{\tilde{p}}\frac{\mathrm{d}\tilde{p}}{\mathrm{d}z}\hat{\mathbf{e}}_{z}\right)+\mathcal{O}\left(\delta^{5/2}\rho_{B}\sqrt{\bar{g}L}\right)\nonumber \\
 & -K\nabla\xi'+\tilde{j}_{\xi}\hat{\mathbf{e}}_{z}+\mathcal{O}\left(\delta^{5/2}\rho_{B}\sqrt{\bar{g}L}\right),\label{eq:j_xi_weak_sol}
\end{align}
where the material flux in the reference state 
\begin{equation}
\tilde{j}_{\xi}=-K\left(\frac{\mathrm{d}\tilde{\xi}}{\mathrm{d}z}+\frac{\tilde{k}_{T}}{\tilde{T}}\frac{\mathrm{d}\tilde{T}}{\mathrm{d}z}+\frac{\tilde{k}_{p}}{\tilde{p}}\frac{\mathrm{d}\tilde{p}}{\mathrm{d}z}\right)=\mathrm{const}\label{eq:uniform_jxi_weak_sol}
\end{equation}
is uniform (recall, that the material conductivity coefficient is
a function of height only, $K=K(z)$). In the light of (\ref{eq:grad_muc_weak_sol})
and (\ref{eq:j_xi_weak_sol}) the term $-\mathbf{j}_{\xi,\mathrm{mol}}\cdot\nabla\mu_{c}$
in the energy equation takes the form
\begin{align}
-\mathbf{j}_{\xi,\mathrm{mol}}\cdot\nabla\mu_{c}=\frac{\varUpsilon}{K}\mathbf{j}_{\xi,\mathrm{mol}}^{2}= & \frac{\varUpsilon}{K}\left(K\nabla\xi'-\tilde{j}_{\xi}\hat{\mathbf{e}}_{z}\right)^{2}\nonumber \\
= & \varUpsilon K\left(\nabla\xi'\right)^{2}-2\tilde{j}_{\xi}\varUpsilon\frac{\partial\xi'}{\partial z}-\frac{\tilde{j}_{\xi}^{2}}{K}\frac{r_{m}R^{(h)}\tilde{T}}{\tilde{\xi}+\xi'}\frac{\xi'}{\tilde{\xi}}\nonumber \\
 & +\frac{\tilde{j}_{\xi}^{2}}{K}\frac{r_{m}R^{(h)}\tilde{T}}{\tilde{\xi}}+\mathcal{O}\left(\delta^{5/2}\rho_{B}\bar{g}\sqrt{\bar{g}L}\right),\label{eq:jgradmuc_term-1}
\end{align}
where in the last term we have substituted
\begin{equation}
\varUpsilon=\frac{r_{m}R^{(h)}\left(\tilde{T}+T'\right)}{\tilde{\xi}+\xi'}=\frac{r_{m}R^{(h)}\tilde{T}}{\tilde{\xi}+\xi'}+\mathcal{O}\left(\bar{g}L\right)=\frac{r_{m}R^{(h)}\tilde{T}}{\tilde{\xi}}-\frac{r_{m}R^{(h)}\tilde{T}}{\tilde{\xi}+\xi'}\frac{\xi'}{\tilde{\xi}}+\mathcal{O}\left(\bar{g}L\right),\label{eq:Upsilon_small_xi}
\end{equation}
to clearly separate the last term $\tilde{j}_{\xi}^{2}r_{m}R^{(h)}\tilde{T}/K\tilde{\xi}$
in (\ref{eq:jgradmuc_term-1}) which belongs to the basic state energy (entropy)
balance. It can now be clearly seen, that as a result of (\ref{eq:muc_tilde_grad_small_xi}),
in the limit of a weak solution the term $-\mathbf{j}_{\xi,\mathrm{mol}}\cdot\nabla\mu_{c}$
takes a significantly different form than (\ref{eq:jgradmuc_term})
obtained for the case $|\xi'|\ll\tilde{\xi}$. Consequently the entropy
equation (\ref{eq:entropy_balance_noSoret}) at the leading order
is modified to
\begin{align}
\tilde{\rho}\tilde{T}\left(\frac{\partial s'}{\partial t}+\mathbf{u}\cdot\nabla s'\right)-\tilde{\rho}\tilde{c}_{p,\xi}u_{z}\Delta_{S}+\tilde{\rho}\tilde{h}_{p,T}u_{z}\frac{\mathrm{d}\tilde{\xi}}{\mathrm{d}z}\qquad\qquad\qquad\qquad\qquad\nonumber \\
=\nabla\cdot\left(k\nabla T'\right)+K\varUpsilon\left(\nabla\xi'\right)^{2}-2\tilde{j}_{\xi}\varUpsilon\frac{\partial\xi'}{\partial z}-\frac{\tilde{j}_{\xi}^{2}}{K}\frac{r_{m}R^{(h)}\tilde{T}}{\tilde{\xi}+\xi'}\frac{\xi'}{\tilde{\xi}}\nonumber \\
+2\tilde{\rho}\nu\mathbf{G}^{s}:\mathbf{G}^{s}+\tilde{\rho}\left(\nu_{b}-\frac{2}{3}\nu\right)\left(\nabla\cdot\mathbf{u}\right)^{2}+Q',\label{eq:entropy_comp-1-1}
\end{align}
where $\varUpsilon$ is given by the full expression in (\ref{eq:Upsilon_small_xi}).
The equations (\ref{Cont_Aderiv-1-1-4-2-1}-d) remain unaltered. The
Navier-Stokes equation can be conveniently left in the most general
form (\ref{NS-Aderiv-1-1-3-3}). The full equations of state of a
mixture of ideal gases and the thermodynamic properties are provided
in (\ref{eq:st_eq_idgm_1}-d) and (\ref{h_pt_idgm_1}-f) whereas the
leading order form of the thermodynamic properties in the limit $\xi\ll1$
can be found in (\ref{eq:h_pt_weak}-d). Since the parameters $\alpha$,
$\beta$ and $\chi_{T}$ for weak solutions remain order unity quantities in terms of
$\delta$, the density fluctuation is still expressed
by
\begin{equation}
\frac{\rho'}{\tilde{\rho}}=\frac{p'}{\tilde{p}}-\frac{T'}{\tilde{T}}-\left(r_{m}-1\right)\xi',\label{eq:rhoprime_weak_sol}
\end{equation}
and is of the order $\rho'=\mathcal{O}\left(\delta\tilde{\rho}\right)$.
The entropy fluctuation on the other hand can no longer be expressed
as in (\ref{State_eq_Aderiv-pert_comp-1}) because $\xi'/\tilde{\xi}$
is no longer required to be small and the parameter $h_{p,T}$ is
irregular when $\xi\rightarrow0$. Hence the entropy fluctuation must
be calculated explicitly from (\ref{eq:st_eq_idgm_2}) and (\ref{h_pt_idgm_1}-c)
\begin{align}
s'= & \,\,s-\tilde{s}\nonumber \\
= & \,\,c_{p}^{(h)}\frac{T'}{\tilde{T}}-\frac{p'}{\tilde{\rho}\tilde{T}}+\frac{\tilde{h}_{p,T}}{\tilde{T}}\xi'\nonumber \\
 & +r_{m}R^{(h)}\tilde{\xi}\left[\frac{\xi'}{\tilde{\xi}}-\left(1+\frac{\xi'}{\tilde{\xi}}\right)\ln\left(1+\frac{\xi'}{\tilde{\xi}}\right)\right],\label{sprime_weak_sol}
\end{align}
where $\tilde{h}_{p,T}$ is taken from (\ref{eq:h_pt_weak}). However,
$s'$ still remains an order $\mathcal{O}\left(\delta\bar{c}_{p,\xi}\right)$
quantity because $\tilde{\xi}=\mathcal{O}(\delta)$ and as remarked
in the footnote (\ref{fn:xilogxi_small}), $\tilde{h}_{p,T}\xi'/\tilde{T}=\mathcal{O}(\delta\tilde{h}_{p,T}/\tilde{T})=\mathcal{O}(\delta\bar{c}_{p,\xi})$.
The two latter expressions for the density and entropy fluctuations can be used to express the buoyancy
force in terms of the entropy and mass fraction fluctuations, similarly
as in section \ref{sec:Buoyancy-force}; in an analogous way one obtains
at the leading order
\begin{align}
-\frac{1}{\tilde{\rho}}\nabla p'-\nabla\psi'+\frac{\rho'}{\tilde{\rho}}\tilde{\mathbf{g}}= & -\nabla\left(\frac{p'}{\tilde{\rho}}+\psi'\right)+\frac{s'}{c_{p}^{(h)}}\tilde{g}\hat{\mathbf{e}}_{z}+\tilde{\chi}\xi'\tilde{g}\hat{\mathbf{e}}_{z}\nonumber \\
 & -r_{m}g\frac{\gamma^{(h)}-1}{\gamma^{(h)}}\tilde{\xi}\left[\frac{\xi'}{\tilde{\xi}}-\left(1+\frac{\xi'}{\tilde{\xi}}\right)\ln\left(1+\frac{\xi'}{\tilde{\xi}}\right)\right]\hat{\mathbf{e}}_{z},\label{eq:buoy_force_simpl_weak_sol}
\end{align}
where $\tilde{\chi}$ can be taken from (\ref{eq:chi_weak}). Such
an expression for the buoyancy force, although not as simple as for
the case of $|\xi'|/\tilde{\xi}\ll1$ may still be useful, since it
does not involve the density and temperature fluctuations and the
pressure fluctuation could be easily removed from the Navier-Stokes
equation by taking its curl. 

For the sake of completeness we also provide the expressions for the
chemical potential in the current case. The leading order form of
the chemical potential in the reference state was already derived
in (\ref{eq:muc_tilde}). The convective fluctuation has to be calculated
directly from (\ref{st_eq_idgm4}) and (\ref{h_pt_idgm_1})
\begin{align}
\mu_{c}^{\prime}= & \mu_{c}-\tilde{\mu}_{c}\nonumber \\
= & \left(r_{m}-1\right)R^{(h)}\tilde{T}\frac{p'}{\tilde{p}}-\tilde{h}_{p,T}\frac{T'}{\tilde{T}}\nonumber \\
 & +r_{m}R^{(h)}\tilde{T}\left(\frac{T'}{\tilde{T}}+1\right)\ln\left(1+\frac{\xi'}{\tilde{\xi}}\right).\label{eq:e341}
\end{align}
Observe, that the fluctuation of the chemical potential is an order
unity quantity in terms of the small parameter $\delta$, i.e. $\mu_{c}^{\prime}=\mathcal{O}\left(\bar{g}L\right)$
and thus the magnitude of the reference state profile $\tilde{\mu}_{c}(z)$
is not expected to exceed the magnitude of the fluctuation.

\subsubsection{Entropy formulation with compositional effects for systems with volume
cooling\label{subsec:Entropy-formulation-CE}}\index{SI}{entropy!formulation}

When the binary alloy is a mixture of ideal, light and heavy gases,
but the solution of the light constituent is weak, i.e. we hold $\xi=\mathcal{O}(\delta)\ll1$,
and there are volume heat sinks which can be modelled by $\tilde{Q}=\kappa g\mathrm{d}_{z}\tilde{\rho}<0$,
it is possible to express the dynamical equations in terms of only
three thermodynamic variables describing fluctuations, namely $p'$,
$s'$ and $\xi'$. Moreover, the pressure then, appears solely in
the Navier-Stokes equation and can be easily removed by taking a curl
of this equation. The remaining variables $\rho'$ and $T'$ are entirely
eliminated and if necessary can be calculated afterwards, when the
system of dynamical equations is solved and $\mathbf{u}$, $p'$,
$s'$ and $\xi'$ are determined. This is called the \emph{entropy
formulation} (cf. section \ref{subsec:Simplifications-through-entropy}
for comparison with the case of a single-component fluid). This formulation
can be achieved only, when additionally constant thermal diffusivity
is assumed and that the fluid's contributions to the total gravity
are negligible, therefore the gravitational acceleration is effectively
constant within the fluid, i.e.
\begin{equation}
\kappa=\frac{k}{\tilde{\rho}c_{p}^{(h)}}=\mathrm{const},\qquad g=\mathrm{const},\label{eq:e342}
\end{equation}
where $c_{p}^{(h)}$ is the specific heat at constant pressure of
the heavy constituent alone; the specific heats $c_{p}^{(h)}$ and
$c_{v}^{(h)}$ are also assumed constant. The properties of the binary
alloy are described by (\ref{eq:h_pt_weak}-d) and we assume, that
the Dufour effect is negligible
\begin{equation}
\Lambda\lesssim\mathcal{O}\left(\delta\bar{g}L\right),\label{eq:e343}
\end{equation}
thus the Soret coefficient simplifies to
\begin{equation}
\tilde{k}_{T}=-\frac{\tilde{h}_{p,T}}{\tilde{\varUpsilon}}.\label{eq:e344}
\end{equation}
The hydrostatic reference state is determined by the equations (\ref{eq:hydrostatic_eq_A_1-1-1-1}-c,e)
with $g=\mathrm{const}$ (the chemical potential is given in (\ref{eq:muc_tilde})). 

The equation (\ref{sprime_weak_sol}) allows to express the temperature
fluctuations in terms of the entropy, pressure and mass fraction fluctuations
\begin{align}
T'= & \frac{\tilde{T}}{c_{p}^{(h)}}s'+\frac{p'}{\tilde{\rho}c_{p}^{(h)}}-\frac{\tilde{h}_{p,T}}{c_{p}^{(h)}}\xi'\nonumber \\
 & -r_{m}\frac{\gamma^{(h)}-1}{\gamma^{(h)}}\tilde{T}\tilde{\xi}\left[\frac{\xi'}{\tilde{\xi}}-\left(1+\frac{\xi'}{\tilde{\xi}}\right)\ln\left(1+\frac{\xi'}{\tilde{\xi}}\right)\right],\label{eq:e345}
\end{align}
where $\gamma^{(h)}=c_{p}^{(h)}/c_{v}^{(h)}$. On inserting the latter
expression into the thermal diffusion term in the entropy equation
one obtains
\begin{align}
\nabla\cdot\left(k\nabla T'\right)= & \kappa\nabla\cdot\left[\tilde{\rho}\nabla\left(\tilde{T}s'\right)\right]+\kappa\nabla\cdot\left(\tilde{\rho}\nabla\frac{p'}{\tilde{\rho}}\right)-\kappa\nabla\cdot\left[\tilde{\rho}\nabla\left(\tilde{h}_{p,T}\mathcal{Z}\right)\right]\nonumber \\
= & \kappa\nabla\cdot\left[\tilde{\rho}\tilde{T}\nabla s'\right]-\frac{\kappa g}{c_{p}^{(h)}}\frac{\partial}{\partial z}\left(\tilde{\rho}s'\right)+\kappa\nabla\cdot\left(\tilde{\rho}\nabla\frac{p'}{\tilde{\rho}}\right)\nonumber \\
 & -\kappa\nabla\cdot\left[\tilde{\rho}\nabla\left(\tilde{h}_{p,T}\mathcal{Z}\right)\right],\label{eq:div_grad_T}
\end{align}
where we have used $\nabla\tilde{T}=\mathrm{d}_{z}\tilde{T}\hat{\mathbf{e}}_{z}=-g/c_{p}^{(h)}\hat{\mathbf{e}}_{z}+\mathcal{O}(\delta\Delta T/L)$
and we have defined
\begin{equation}
\mathcal{Z}=\xi'+\frac{r_{m}R\tilde{T}\tilde{\xi}}{\tilde{h}_{p,T}}\left[\frac{\xi'}{\tilde{\xi}}-\left(1+\frac{\xi'}{\tilde{\xi}}\right)\ln\left(1+\frac{\xi'}{\tilde{\xi}}\right)\right].\label{eq:Z}
\end{equation}
On the other hand the Navier-Stokes equation by the use of (\ref{eq:buoy_force_simpl_comp})
can be cast in the form
\begin{align}
\frac{\partial\mathbf{u}}{\partial t}+\frac{1}{\tilde{\rho}}\nabla\cdot\left(\tilde{\rho}\mathbf{u}\mathbf{u}\right)= & -\nabla\left(\frac{p'}{\tilde{\rho}}\right)+\frac{s'}{c_{p}^{(h)}}g\hat{\mathbf{e}}_{z}+\tilde{\chi}\xi'g\hat{\mathbf{e}}_{z}+\frac{1}{\tilde{\rho}}\nabla\cdot\left(2\tilde{\rho}\nu\mathbf{G}^{s}\right)\nonumber \\
 & +\frac{1}{\tilde{\rho}}\nabla\left[\left(\nu_{b}-\frac{2}{3}\nu\right)\tilde{\rho}\nabla\cdot\mathbf{u}\right],\label{eq:NS_aux_comp}
\end{align}
hence multiplying the latter equation by $\tilde{\rho}$, taking its
divergence and utilizing $\nabla\cdot\left(\tilde{\rho}\mathbf{u}\right)=0$
we get
\begin{align}
\nabla\cdot\left(\tilde{\rho}\nabla\frac{p'}{\tilde{\rho}}\right)= & \frac{g}{c_{p}^{(h)}}\frac{\partial}{\partial z}\left(\tilde{\rho}s'\right)+g\frac{\partial}{\partial z}\left(\tilde{\rho}\tilde{\chi}\xi'\right)+\nabla\cdot\left[\nabla\cdot\left(2\tilde{\rho}\nu\mathbf{G}^{s}-\tilde{\rho}\mathbf{u}\mathbf{u}\right)\right]\nonumber \\
 & +\nabla^{2}\left[\left(\nu_{b}-\frac{2}{3}\nu\right)\tilde{\rho}\nabla\cdot\mathbf{u}\right].\label{eq:div_grad_p}
\end{align}
Substitution of the formula (\ref{eq:div_grad_p}) into (\ref{eq:div_grad_T})
allows to express the thermal diffusion solely in terms of the entropy
$s'$ and the mass fraction $\xi'$. Therefore we can now write down
the final set of dynamical equations formulated in terms of the entropy and the mass fraction fluctuations
in the following form 
\begin{subequations}
\begin{align}
\frac{\partial\mathbf{u}}{\partial t}+\left(\mathbf{u}\cdot\nabla\right)\mathbf{u}= & -\nabla\left(\frac{p'}{\tilde{\rho}}\right)+\frac{s'}{c_{p}^{(h)}}g\hat{\mathbf{e}}_{z}+\tilde{\chi}\xi'g\hat{\mathbf{e}}_{z}+\nu\nabla^{2}\mathbf{u}+\left(\frac{\nu}{3}+\nu_{b}\right)\nabla\left(\nabla\cdot\mathbf{u}\right)\nonumber \\
 & +\frac{2}{\tilde{\rho}}\nabla\left(\tilde{\rho}\nu\right)\cdot\mathbf{G}^{s}+\frac{1}{\tilde{\rho}}\nabla\left(\tilde{\rho}\nu_{b}-\frac{2}{3}\tilde{\rho}\nu\right)\nabla\cdot\mathbf{u},\label{NS-Aderiv-1-1-3-2}
\end{align}
\begin{equation}
\nabla\cdot\left(\tilde{\rho}\mathbf{u}\right)=0,\label{Cont_Aderiv-1-1-1-1-1-2}
\end{equation}
\begin{equation}
\tilde{\rho}\left(\frac{\partial\xi'}{\partial t}+\mathbf{u}\cdot\nabla\xi'\right)+\tilde{\rho}u_{z}\frac{\mathrm{d}\tilde{\xi}}{\mathrm{d}z}=\nabla\cdot\left(K\nabla\xi'\right),\label{eq:mass_fraction_comp-1}
\end{equation}
\begin{equation}
\tilde{\rho}\tilde{T}\left(\frac{\partial s'}{\partial t}+\mathbf{u}\cdot\nabla s'\right)-\tilde{\rho}c_{p}^{(h)}u_{z}\Delta_{S}+\tilde{\rho}\tilde{h}_{p,T}u_{z}\frac{\mathrm{d}\tilde{\xi}}{\mathrm{d}z}=\kappa\nabla\cdot\left(\tilde{\rho}\tilde{T}\nabla s'\right)+\Xi+\mathcal{J}+Q'.\label{eq:energy_final_constkappa_comp}
\end{equation}
\end{subequations} 
In the above the term
\begin{align}
\Xi= & -\kappa\nabla\cdot\left[\tilde{\rho}\nabla\left(\tilde{h}_{p,T}\mathcal{Z}\right)\right]+\kappa g\frac{\partial}{\partial z}\left(\tilde{\chi}\tilde{\rho}\xi'\right)+K\varUpsilon\left(\nabla\xi'\right)^{2}\nonumber \\
 & -2\tilde{j}_{\xi}\varUpsilon\frac{\partial\xi'}{\partial z}-\frac{\tilde{j}_{\xi}^{2}}{K}\frac{r_{m}R^{(h)}\tilde{T}}{\tilde{\xi}+\xi'}\frac{\xi'}{\tilde{\xi}},\label{eq:e346}
\end{align}
(where $\mathcal{Z}$ is given in (\ref{eq:Z})), depends on the mass
fraction fluctuation only, i.e. no other type of thermodynamic fluctuations
such as $\rho'$, $p'$, $T'$ nor $s'$ contributes to the above
expression for $\Xi$, and
\begin{align}
\mathcal{J}= & \kappa\nabla\cdot\left[\nabla\cdot\left(2\tilde{\rho}\nu\mathbf{G}^{s}-\tilde{\rho}\mathbf{u}\mathbf{u}\right)\right]+\kappa\nabla^{2}\left[\left(\nu_{b}-\frac{2}{3}\nu\right)\tilde{\rho}\nabla\cdot\mathbf{u}\right]\nonumber \\
 & +2\tilde{\rho}\nu\mathbf{G}^{s}:\mathbf{G}^{s}+\tilde{\rho}\left(\nu_{b}-\frac{2}{3}\nu\right)\left(\nabla\cdot\mathbf{u}\right)^{2},\label{eq:J_def_A-1}
\end{align}
is the same as in the case of entropy formulation for a single-component
fluid, cf. (\ref{eq:J_def_A}). The coefficients $\tilde{h}_{p,T}$
and $\tilde{\chi}$ are provided in (\ref{eq:h_pt_weak},d), with
$T$, $p$, $\xi$ replaced by $\tilde{T}$, $\tilde{p}$, $\tilde{\xi}$;
the parameter $\varUpsilon$ at the leading order was given in (\ref{eq:Upsilon_small_xi}).
Once the dynamical equations (\ref{NS-Aderiv-1-1-3-2}-d) are solved
and in particular the fluctuations of pressure $p'$, entropy $s'$
and the mass fraction $\xi'$ are determined, the density and temperature
fluctuations can be found from
\begin{equation}
\frac{\rho'}{\tilde{\rho}}=-\frac{s'}{c_{p}^{(h)}}+\frac{p'}{\gamma^{(h)}\tilde{p}}-\tilde{\chi}\xi',\qquad\frac{T'}{\tilde{T}}=\frac{s'}{c_{p}^{(h)}}+\frac{\gamma^{(h)}-1}{\gamma^{(h)}}\frac{p'}{\tilde{p}}-\frac{\tilde{h}_{p,T}}{c_{p}^{(h)}\tilde{T}}\xi'.\label{State_eq_Aderiv-1-1-3-2-1}
\end{equation}

\section*{Review exercises}

{\textbf{Exercise 1.}} \\
In the case of a weak solution of two perfect gases the reference state takes the form provided in section \ref{subsec:Weak-solution-limit_RS}  (cf. (\ref{eq:e338}) and (\ref{eq:BS1}-c)). Calculate the compositional Rayleigh number and the total heat flux for this reference state.

\noindent\emph{Hint}: utilize (\ref{eq:Ra_def-1-3-1}) (why?), (\ref{eq:total_heat_flux_final}) and (\ref{eq:j_xi_gen_comp-1}).
\\

\noindent{\textbf{Exercise 2.}} \\
For the case of Ex. 1 demonstrate explicitly, that $\mathrm{d}\tilde{\mu}_c/\mathrm{d}z+g\tilde{\chi}$ is an $\mathcal{O}(1)$ quantity in terms of the small parameter $\delta$, which under the anelastic approximation is a unique feature of weak solutions. 

\noindent\emph{Hint}: utilize (\ref{eq:muc_tilde}).
\\

\noindent{\textbf{Exercise 3.}} \\
Under the assumptions $k_T=0$, $k_p=0$, $g=\mathrm{const}$ and $\chi=\mathrm{const}$ calculate the mean vertical molecular material flux $\langle\mathbf{j}_{\xi,\mathrm{mol}}\cdot\hat{\mathbf{e}}_z\rangle$, given the total material flux through the system $G_0$, the total viscous dissipation rate
\[
D_{\nu}=2\left\langle \mu\mathbf{G}^{s}:\mathbf{G}^{s}\right\rangle -\left\langle \left(\frac{2}{3}\mu-\mu_{b}\right)\left(\nabla\cdot\mathbf{u}\right)^{2}\right\rangle,
\]
and the mean work of the thermal buoyancy force
\[
W_{th}=\left\langle \frac{\tilde{\alpha}\tilde{T}\tilde{g}\tilde{\rho}}{\tilde{c}_{p,\xi}}u_{z}s'\right\rangle.
\]

\noindent\emph{Hint}: utilize the results of section \ref{subsec:Global_balance_comp}.

\subsection*{Acknowledgements}
The author wishes to thank Professor Chris Jones for many fruitfull discussions on the topic of compressible convection, which have greatly helped to improve the third chapter of the book. The support of the National Science Centre of Poland (grant no. 2017/26/E/ST3/00554) is
gratefully acknowledged.

\chapter*{Notation and definitions}
\markboth{}{}

~

\begin{longtable}{|c|c|c|}
\hline 
{\scriptsize{}Symbol} & {\scriptsize{}Mathematical definition} & {\scriptsize{}Explanation}\tabularnewline
\hline 
\hline 
{\scriptsize{}$L$} & {\scriptsize{}-} & {\scriptsize{}vertical span of the fluid layer}\tabularnewline
\hline 
{\scriptsize{} $L_{x}$, $L_{y}$} & {\scriptsize{}-} & {\scriptsize{}horizontal periods of the domain}\tabularnewline
\hline 
{\scriptsize{}$\left\langle f\right\rangle $, $\bar{f}$} & {\scriptsize{}$\frac{1}{L_{x}L_{y}L}\int_{-L_{x}/2}^{L_{x}/2}\int_{-L_{y}/2}^{L_{y}/2}\int_{-L/2}^{L/2}f\mathrm{d}x\mathrm{d}y\mathrm{d}z$} & {\scriptsize{}spatial average}\tabularnewline
\hline 
{\scriptsize{}$\left\langle f\right\rangle _{h}$} & {\scriptsize{}$\frac{1}{L_{x}L_{y}}\int_{-L_{x}/2}^{L_{x}/2}\int_{-L_{y}/2}^{L_{y}/2}f\mathrm{d}x\mathrm{d}y$} & {\scriptsize{}average over a horizontal plane}\tabularnewline
\hline 
{\scriptsize{}$\tilde{f}$} & {\scriptsize{}-} & {\scriptsize{}variable in the hydrostatic reference state}\tabularnewline
\hline 
{\scriptsize{}$\dbtilde{f}$ } & {\scriptsize{}$\tilde{f}-\bar{f}$} & {\scriptsize{}variation above the mean in hydrostatic state}\tabularnewline
\hline 
{\scriptsize{}$f_{ad}$} & {\scriptsize{}-} & {\scriptsize{}variable in the hydrostatic adiabatic state}\tabularnewline
\hline 
{\scriptsize{}$f^{\sharp}$} & {\scriptsize{}-} & {\scriptsize{}non-dimensional variable $f$}\tabularnewline
\hline 
{\scriptsize{}$f_{B}$, $f_{T}$} & {\scriptsize{}$f(z=0)$, $f(z=L)$} & {\scriptsize{}bottom and top values of the field $f$}\tabularnewline
\hline 
{\scriptsize{}$\left(\frac{\partial f}{\partial A}\right)_{B,C}$} & {\scriptsize{}-} & {\scriptsize{}derivative with respect to $A$ at constant $B$, $C$}\tabularnewline
\hline 
{\scriptsize{}$\mathbf{u}$} & {\scriptsize{}-} & {\scriptsize{}velocity field}\tabularnewline
\hline 
{\scriptsize{}$\boldsymbol{\zeta}$} & {\scriptsize{}$\nabla\times\mathbf{u}$} & {\scriptsize{}vorticity field}\tabularnewline
\hline 
{\scriptsize{}$\rho$} & {\scriptsize{}-} & {\scriptsize{}density }\tabularnewline
\hline 
{\scriptsize{}$V$} & {\scriptsize{}-} & {\scriptsize{}volume of the fluid}\tabularnewline
\hline 
{\scriptsize{}$p$} & {\scriptsize{}-} & {\scriptsize{}pressure }\tabularnewline
\hline 
{\scriptsize{}$s$} & {\scriptsize{}-} & {\scriptsize{}entropy per unit mass}\tabularnewline
\hline 
{\scriptsize{}$S$} & {\scriptsize{}-} & {\scriptsize{}entropy}\tabularnewline
\hline 
{\scriptsize{}$T$} & {\scriptsize{}-} & {\scriptsize{}temperature }\tabularnewline
\hline 
{\scriptsize{}$\xi$} & {\scriptsize{}$\frac{m_{l}N^{(l)}}{m_{l}N^{(l)}+m_{h}N^{(h)}}$} & {\scriptsize{}mass fraction of a light constituent in a binary alloy}\tabularnewline
\hline 
{\scriptsize{}$N$} & {\scriptsize{}-} & {\scriptsize{}number of particles in the fluid}\tabularnewline
\hline 
{\scriptsize{}$N^{(l)}$} & {\scriptsize{}-} & {\scriptsize{}number of particles of light constituent in alloy}\tabularnewline
\hline 
{\scriptsize{}$N^{(h)}$} & {\scriptsize{}-} & {\scriptsize{}number of particles of heavy constituent in alloy}\tabularnewline
\hline 
{\scriptsize{}$\mu_{c}$} & {\scriptsize{}$\frac{\mu_{l}}{m_{l}}-\frac{\mu_{h}}{m_{h}}$} & {\scriptsize{}chemical potential for a binary alloy}\tabularnewline
\hline 
{\scriptsize{}$\mu_{l}$} & {\scriptsize{}-} & {\scriptsize{}chemical potential of light constituent in alloy}\tabularnewline
\hline 
{\scriptsize{}$\mu_{h}$} & {\scriptsize{}-} & {\scriptsize{}chemical potential of heavy constituent in alloy}\tabularnewline
\hline 
{\scriptsize{}$\varepsilon$} & {\scriptsize{}-} & {\scriptsize{}internal energy per unit mass}\tabularnewline
\hline 
{\scriptsize{}$\mathcal{E}$} & {\scriptsize{}-} & {\scriptsize{}internal energy}\tabularnewline
\hline 
{\scriptsize{}$\psi$} & {\scriptsize{}-} & {\scriptsize{}potential energy per unit mass from external forcing}\tabularnewline
\hline 
{\scriptsize{}$\mathbf{F}$} & {\scriptsize{}$-\nabla\psi$} & {\scriptsize{}external force per unit mass}\tabularnewline
\hline 
{\scriptsize{}$e$} & {\scriptsize{}$e=\frac{1}{2}\mathbf{u}^{2}+\psi+\varepsilon$} & {\scriptsize{}total energy per unit mass}\tabularnewline
\hline 
{\scriptsize{}$m_{m}$} & {\scriptsize{}-} & {\scriptsize{}molecular mass of fluid particles}\tabularnewline
\hline 
{\scriptsize{}$m_{l}$} & {\scriptsize{}-} & {\scriptsize{}molecular mass of light constituent in alloy}\tabularnewline
\hline 
{\scriptsize{}$m_{h}$} & {\scriptsize{}-} & {\scriptsize{}molecular mass of heavy constituent in alloy}\tabularnewline
\hline 
{\scriptsize{}$r_{m}$} & {\scriptsize{}$\frac{m_{h}}{m_{l}}$} & {\scriptsize{}molecular mass ratio}\tabularnewline
\hline 
{\scriptsize{}$\alpha$} & {\scriptsize{}$-\frac{1}{\rho}\left(\frac{\partial\rho}{\partial T}\right)_{p,\xi}$} & {\scriptsize{}coefficient of thermal expansion}\tabularnewline
\hline 
{\scriptsize{}$\alpha_{a}$} & {\scriptsize{}$(\ref{eq:radiative_flux_governing_eq})$} & {\scriptsize{}radiation absorption coefficient per unit volume}\tabularnewline
\hline 
{\scriptsize{}$\beta$} & {\scriptsize{}$\frac{1}{\rho}\left(\frac{\partial\rho}{\partial p}\right)_{T,\xi}$} & {\scriptsize{}coefficient of isothermal compressibility}\tabularnewline
\hline 
{\scriptsize{}$c_{p,\xi}$} & {\scriptsize{}$T\left(\frac{\partial s}{\partial T}\right)_{p,\xi}$} & {\scriptsize{}specific heat at constant pressure}\tabularnewline
\hline 
{\scriptsize{}$c_{v,\xi}$} & {\scriptsize{}$T\left(\frac{\partial s}{\partial T}\right)_{\rho,\xi}$} & {\scriptsize{}specific heat at constant volume}\tabularnewline
\hline 
{\scriptsize{}$\gamma$} & {\scriptsize{}$\frac{c_{p}}{c_{v}}>1$} & {\scriptsize{}specific heat ratio}\tabularnewline
\hline 
{\scriptsize{}$C$} & {\scriptsize{}$\sqrt{\left(\frac{\partial p}{\partial\rho}\right)_{s,\xi}}$} & {\scriptsize{}speed of sound}\tabularnewline
\hline 
{\scriptsize{}$C_{T}$} & {\scriptsize{}$\sqrt{\left(\frac{\partial p}{\partial\rho}\right)_{T,\xi}}$} & {\scriptsize{}isothermal speed of sound}\tabularnewline
\hline 
{\scriptsize{}$\Gamma$} & {\scriptsize{}$\frac{T_{B}}{T_{T}}$} & {\scriptsize{}temperature bottom to top ratio}\tabularnewline
\hline 
{\scriptsize{}$D_{\rho}$} & {\scriptsize{}$\left|\frac{1}{\rho}\frac{\mathrm{d}\rho}{\mathrm{d}z}\right|^{-1}$} & {\scriptsize{}density scale height}\tabularnewline
\hline 
{\scriptsize{}$D_{p}$} & {\scriptsize{}$\left|\frac{1}{p}\frac{\mathrm{d}p}{\mathrm{d}z}\right|^{-1}$} & {\scriptsize{}pressure scale height}\tabularnewline
\hline 
{\scriptsize{}$D_{T}$} & {\scriptsize{}$\left|\frac{1}{T}\frac{\mathrm{d}T}{\mathrm{d}z}\right|^{-1}$} & {\scriptsize{}temperature scale height}\tabularnewline
\hline 
{\scriptsize{}$\Delta_{S}$} & {\scriptsize{}$-\left(\frac{\partial T}{\partial z}+\frac{g\alpha T}{c_{p,\xi}}\right)$} & {\scriptsize{}superadiabatic excess}\tabularnewline
\hline 
{\scriptsize{}$\Delta\tilde{s}$} & {\scriptsize{}$\tilde{s}_{B}-\tilde{s}_{T}$} & {\scriptsize{}basic entropy jump across the fluid layer}\tabularnewline
\hline 
{\scriptsize{}$\Delta T$} & {\scriptsize{}$T_{B}-T_{T}$} & {\scriptsize{}temperature jump across the fluid layer}\tabularnewline
\hline 
{\scriptsize{}$\Delta\tilde{T}$} & {\scriptsize{}$\tilde{T}_{B}-\tilde{T}_{T}$} & {\scriptsize{}basic temperature jump across the fluid layer}\tabularnewline
\hline 
{\scriptsize{}$\left(\Delta f\right)_{B}$} & {\scriptsize{}$\left\langle f\right\rangle _{h}\left(z=\delta_{th,B}\right)-\left\langle f\right\rangle _{h}\left(z=0\right)$} & {\scriptsize{}jump in value of $\left\langle f\right\rangle _{h}$
across bottom boundary layer}\tabularnewline
\hline 
{\scriptsize{}$\left(\Delta f\right)_{T}$} & {\scriptsize{}$\left\langle f\right\rangle _{h}\left(z=L\right)-\left\langle f\right\rangle _{h}\left(z=L-\delta_{th,T}\right)$} & {\scriptsize{}jump in value of $\left\langle f\right\rangle _{h}$
across top boundary layer}\tabularnewline
\hline 
{\scriptsize{}$\left(\Delta T\right)_{bulk}$} & {\scriptsize{}$\left\langle T\right\rangle _{h}\left(z=L-\delta_{th,T}\right)-\left\langle T\right\rangle _{h}\left(z=\delta_{th,B}\right)>0$} & {\scriptsize{}mean temperature jump across the bulk}\tabularnewline
\hline 
{\scriptsize{}$\left(\Delta T'\right)_{bulk}$} & {\scriptsize{}$\left\langle T'\right\rangle _{h}\left(z=\delta_{th,B}\right)-\left\langle T'\right\rangle _{h}\left(z=L-\delta_{th,T}\right)>0$} & {\scriptsize{}mean temperature fluctuation jump across the bulk}\tabularnewline
\hline 
{\scriptsize{}$\left(\Delta T\right)_{vel}$} & {\scriptsize{}$(\ref{integral_positive})$} & {\scriptsize{}convective correction to bulk temperature jump}\tabularnewline
\hline 
{\scriptsize{}$\delta$} & {\scriptsize{}$\left\langle \frac{L}{\tilde{T}}\Delta_{S}\right\rangle $} & {\scriptsize{}non-dimensional superadiabatic excess}\tabularnewline
\hline 
{\scriptsize{}$\delta_{th}$} & {\scriptsize{}$\sim Nu^{-1}$} & {\scriptsize{}thermal boundary layer thickness}\tabularnewline
\hline 
{\scriptsize{}$\delta_{\nu}$} & {\scriptsize{}$Re^{-1/2}$} & {\scriptsize{}viscous boundary layer thickness}\tabularnewline
\hline 
{\scriptsize{}$\epsilon$} & {\scriptsize{}$\frac{\Delta\dbtilde{\rho}}{\bar{\rho}}\ll1$} & {\scriptsize{}small dernsity stratification (Boussinesq)}\tabularnewline
\hline 
{\scriptsize{}$\epsilon_{a}$} & {\scriptsize{}$\frac{L}{\tilde{T}_{B}}\left(\frac{\Delta\tilde{T}}{L}-\frac{g}{c_{p}}\right)\ll1$} & {\scriptsize{}small departure from adiabaticity (anelastic)}\tabularnewline
\hline 
{\scriptsize{}$F_{total}$} & {\scriptsize{}$(\ref{eq:F_total})$} & {\scriptsize{}total, horizontally averaged heat flux}\tabularnewline
\hline 
{\scriptsize{}$F_{conv.}$} & {\scriptsize{}(\ref{eq:F_conv})} & {\scriptsize{}convective, horizontally averaged heat flux}\tabularnewline
\hline 
{\scriptsize{}$F_{S}$} & {\scriptsize{}(\ref{eq:F_conv_superadiab_1})} & {\scriptsize{}superadiabatic, horizontally averaged heat flux}\tabularnewline
\hline 
{\scriptsize{}$G_{total}$} & {\scriptsize{}$(\ref{eq:G_total})$} & {\scriptsize{}total, horizontally averaged material flux}\tabularnewline
\hline 
{\scriptsize{}$\mathbf{G}$} & {\scriptsize{}$\frac{\partial u_{i}}{\partial x_{j}}$} & {\scriptsize{}velocity gradient tensor}\tabularnewline
\hline 
{\scriptsize{}$G$} & {\scriptsize{}$6.67\times10^{-11}\,m^{3}/s^{2}kg$} & {\scriptsize{}gravitational constant}\tabularnewline
\hline 
{\scriptsize{}$\mathbf{g}$} & {\scriptsize{}-} & {\scriptsize{}acceleration of gravity}\tabularnewline
\hline 
{\scriptsize{}$\eta$} & {\scriptsize{}$(\ref{eq:eta_def_WNT})$} & {\scriptsize{}departure from threshold of convection}\tabularnewline
\hline 
{\scriptsize{}$h_{p,T}$} & {\scriptsize{}$T\left(\frac{\partial s}{\partial\xi}\right)_{p,T}$} & {\scriptsize{}heat of reaction}\tabularnewline
\hline 
{\scriptsize{}$\theta$} & {\scriptsize{}$\frac{\Delta T}{T_{B}}$} & {\scriptsize{}temperature stratification parameter}\tabularnewline
\hline 
{\scriptsize{}$\mathbf{I}$} & {\scriptsize{}$\delta_{ij}$} & {\scriptsize{}unitary matrix}\tabularnewline
\hline 
{\scriptsize{}$\mathbf{j}_{A}$} & {\scriptsize{}-} & {\scriptsize{}flux of quantity $A$}\tabularnewline
\hline 
{\scriptsize{}$\mathbf{j}_{A,\mathrm{mol}}$} & {\scriptsize{}-} & {\scriptsize{}molecular flux of quantity $A$}\tabularnewline
\hline 
{\scriptsize{}$k$} & {\scriptsize{}-} & {\scriptsize{}coefficient of heat conduction}\tabularnewline
\hline 
{\scriptsize{}$\kappa$} & {\scriptsize{}$\frac{k}{\rho c_{p,\xi}}$} & {\scriptsize{}coefficient of thermal diffusion}\tabularnewline
\hline 
{\scriptsize{}$K$} & - & {\scriptsize{}coefficient of material conductivity}\tabularnewline
\hline 
{\scriptsize{}$D$} & {\scriptsize{}$\frac{K}{\rho}$} & {\scriptsize{}coefficient of material diffusion}\tabularnewline
\hline 
{\scriptsize{}$k_{T}$} & {\scriptsize{}$(\ref{eq:kT_kp_muprime})$} & {\scriptsize{}Soret coefficient}\tabularnewline
\hline 
{\scriptsize{}$k_{p}$} & {\scriptsize{}$(\ref{eq:kT_kp_muprime})$} & {\scriptsize{}pressure gradient coefficient in material flux}\tabularnewline
\hline 
{\scriptsize{}$k_{B}$} & {\scriptsize{}$1.38\times10^{-23}\,J/K$} & {\scriptsize{}Boltzmann constant}\tabularnewline
\hline 
{\scriptsize{}$\Lambda$} & {\scriptsize{}$(\ref{eq:kT_kp_muprime})$} & {\scriptsize{}Dufour coefficient}\tabularnewline
\hline 
{\scriptsize{}$m$} & {\scriptsize{}$(\ref{eq:m_exp1})$} & {\scriptsize{}polytropic index}\tabularnewline
\hline 
{\scriptsize{}$\mu$} & {\scriptsize{}-} & {\scriptsize{}dynamic shear viscosity}\tabularnewline
\hline 
{\scriptsize{}$\mu_{b}$} & {\scriptsize{}-} & {\scriptsize{}dynamic bulk viscosity}\tabularnewline
\hline 
{\scriptsize{}$\nu$} & {\scriptsize{}$\frac{\mu}{\rho}$} & {\scriptsize{}kinematic shear viscosity}\tabularnewline
\hline 
{\scriptsize{}$\nu_{b}$} & {\scriptsize{}$\frac{\mu_{b}}{\rho}$} & {\scriptsize{}kinematic bulk viscosity}\tabularnewline
\hline 
{\scriptsize{}$Q$} & {\scriptsize{}-} & {\scriptsize{}heat sources other than viscous friction}\tabularnewline
\hline 
{\scriptsize{}$q_{tot}$} & {\scriptsize{}-} & {\scriptsize{}heat delivered to fluid parcel from surroundings}\tabularnewline
\hline 
{\scriptsize{}$r_{\delta}$} & {\scriptsize{}$\frac{\delta_{th,T}}{\delta_{th,B}}$} & {\scriptsize{}top to bottom thermal boundary layer thickness ratio}\tabularnewline
\hline 
{\scriptsize{}$r_{s}$} & {\scriptsize{}$\frac{\left(\Delta s\right)_{T}}{\left(\Delta s\right)_{B}}$} & {\scriptsize{}ratio of jumps of $\left\langle s\right\rangle _{h}$
across boundary layers }\tabularnewline
\hline 
{\scriptsize{}$r_{T}$} & {\scriptsize{}$\frac{\left(\Delta T'\right)_{T}}{\left(\Delta T'\right)_{B}}$} & {\scriptsize{}ratio of jumps of $\left\langle T'\right\rangle _{h}$
across boundary layers }\tabularnewline
\hline 
{\scriptsize{}$r_{U}$} & {\scriptsize{}$\frac{U_{T}}{U_{B}}$} & {\scriptsize{}top to bottom thermal wind magnitude ratio}\tabularnewline
\hline 
{\scriptsize{}$R$} & {\scriptsize{}$\frac{k_{B}}{m_{m}}$} & {\scriptsize{}specific gas constant}\tabularnewline
\hline 
{\scriptsize{}$\sigma_{A}$} & {\scriptsize{}-} & {\scriptsize{}volume sources of quantity $A$}\tabularnewline
\hline 
{\scriptsize{}$\sigma$} & {\scriptsize{}-} & {\scriptsize{}growth rate of convective instability}\tabularnewline
\hline 
{\scriptsize{}$\sigma_{rad}$} & {\scriptsize{}$5.67\times10^{-8}\,W/m^{2}K^{4}$} & {\scriptsize{}Stefa-Boltzmann constant}\tabularnewline
\hline 
{\scriptsize{}$\boldsymbol{\tau}$} & {\scriptsize{}$(\ref{eq:Newtonian_stress_tensor})$} & {\scriptsize{}stress tensor}\tabularnewline
\hline 
{\scriptsize{}$\mathscr{T}$} & {\scriptsize{}$(\ref{eq:vel_and_time_scales})$, $(\ref{eq:vel_and_time_scales-1})$} & {\scriptsize{}time scale}\tabularnewline
\hline 
{\scriptsize{}$\mathscr{U}$} & {\scriptsize{}$(\ref{eq:vel_and_time_scales})$, $(\ref{eq:vel_and_time_scales-1})$} & {\scriptsize{}velocity scale}\tabularnewline
\hline 
{\scriptsize{}$\varUpsilon$} & {\scriptsize{}$\left(\frac{\partial\mu_{c}}{\partial\xi}\right)_{p,T}$} & {\scriptsize{}compositional derivative of chemical potential}\tabularnewline
\hline 
{\scriptsize{}$\chi$} & {\scriptsize{}$-\frac{1}{\rho}\left(\frac{\partial\rho}{\partial\xi}\right)_{p,s}$} & {\scriptsize{}compositional expansion cofficient}\tabularnewline
\hline 
{\scriptsize{}$\chi_{T}$} & {\scriptsize{}$-\frac{1}{\rho}\left(\frac{\partial\rho}{\partial\xi}\right)_{p,T}$} & {\scriptsize{}isothermal compositional expansion coefficient}\tabularnewline
\hline 
{\scriptsize{}$\omega$} & {\scriptsize{}-} & {\scriptsize{}frequency of oscillations of fluctuations}\tabularnewline
\hline 
{\scriptsize{}$\Omega$} & {\scriptsize{}-} & {\scriptsize{}background rotation rate}\tabularnewline
\hline 
{\scriptsize{}$Nu$} & {\scriptsize{}$\left(\ref{eq:Nu_def_B}\right)$, $\left(\ref{eq:Nu_def_anapp-1}\right)$,
$\left(\ref{eq:Nu_def_an_comp}\right)$} & {\scriptsize{}Nusselt number}\tabularnewline
\hline 
{\scriptsize{}$Nu_{Q}$} & {\scriptsize{}$\left(\ref{eq:Nu_Q}\right)$, $\left(\ref{eq:Nu_Q_anelastic}\right)$} & {\scriptsize{}Nusselt number for fixed heat flux at boundaries}\tabularnewline
\hline 
{\scriptsize{}$Ra$} & {\scriptsize{}$\left(\ref{eq:Ra_def_B}\right)$, $\left(\ref{eq:Ra_def-1}\right)$} & {\scriptsize{}Rayleigh number}\tabularnewline
\hline 
{\scriptsize{}$Ra_{comp}$} & {\scriptsize{} $\left(\ref{eq:Ra_def-1-3-1}\right)$} & {\scriptsize{}compositional Rayleigh number}\tabularnewline
\hline 
{\scriptsize{}$Ra_{R}$} & {\scriptsize{}$(\ref{eq:radiative_Ra})$} & {\scriptsize{}radiative Rayleigh number}\tabularnewline
\hline 
{\scriptsize{}$Re$} & {\scriptsize{}$\frac{\mathscr{U}L}{\nu}$} & {\scriptsize{}Reynolds number}\tabularnewline
\hline 
{\scriptsize{}$Pr$} & {\scriptsize{}$\frac{\nu}{\kappa}$} & {\scriptsize{}Prandtl number}\tabularnewline
\hline 
{\scriptsize{}$E$} & {\scriptsize{}$\frac{\nu}{2\Omega L^{2}}$} & {\scriptsize{}Ekman number}\tabularnewline
\hline 
\end{longtable}

\printindex{SI}{Subject Index}

\end{document}